\documentclass[12pt]{article}
\usepackage[utf8]{inputenc}
\usepackage[export]{adjustbox}
\usepackage{authblk}
\usepackage{bbm}
\usepackage{pgfpages}
\usepackage{graphicx}
\usepackage{multicol}
\usepackage{csquotes}
\usepackage{multirow}
\usepackage{graphicx} 
\usepackage{booktabs} 
\usepackage{authblk}
\usepackage{caption}
\usepackage{subcaption}
\usepackage{stackengine}
\usepackage{amsmath}

\usepackage[maxfloats=256]{morefloats}

\usepackage{comment}
\usepackage{lscape}
\usepackage{hyperref}
\hypersetup{
    colorlinks=true,
    linkcolor=blue,
    filecolor=blue,      
    urlcolor=blue,
    citecolor=blue,
}
\usepackage[capposition=top]{floatrow}

\usepackage[spanish,english]{babel}

\usepackage{amssymb}
\usepackage{stackrel}
\usepackage{natbib} 

\usepackage[a4paper, margin=0.85in]{geometry}

\title{\textbf{Spillovers of US Interest Rates} \\ Monetary Policy \& Information Effects\footnote{I would like to thank the Monetary and Fiscal History of Latin America project of the Becker Friedman Institute for their generous financial support. Particularly, I would like to thank Edward R. Allen, for supporting my research.}}

\author[1]{Santiago Camara}

\affil[1]{McGill University \& Red-NIE}
\date{\normalsize This Draft \today, First Draft: August 2nd 2021}

\begin{document}
    
\maketitle

\begin{abstract}
    This paper quantifies the international spillovers of US interest rates by explicitly controlling for the ``Fed Information Effect’’. I use multiple identification strategies that identify two components of monetary policy surprises around FOMC meetings: a pure US monetary policy shock component and a ``Fed Information Effect'' component. On the one hand, a US tightening caused by a pure US monetary policy component leads to an economic recession, an exchange rate depreciation and tighter financial conditions. On the other hand, a tightening of US monetary policy caused by the ``Fed Information Effect'' leads to an economic expansion, an exchange rate appreciation and looser financial conditions. Ignoring the ``Fed Information Effect'' biases the impact of US interest rates and may explain recent atypical findings which suggest an expansionary impact of US monetary policy shocks on the rest of the world.  
    

    \medskip
    
    \medskip

\end{abstract}

\newpage
\section{Introduction} \label{sec:introduction}

The international spillovers of US monetary policy is a classic question in international macroeconomics, going back to \cite{fleming1962domestic}, \cite{mundell1963capital}, \cite{dornbusch1976expectations} and \cite{frenkel1983monetary}. The status of the US dollar as the global reserve currency, unit of account, invoice of international trade and its dominant role in financial markets implies that the Federal Open Market Committee's (FOMC) policy decisions have spillover effects on the rest of the world. In recent years, the conventional view that a US monetary tightening leads to negative international spillovers such as a recession, exchange rate depreciation and financial distress (\cite{svensson1989excess}, \cite{obstfeld1995exchange}, \cite{betts2000exchange}), has been challenged. In fact, a recent literature has found opposite empirical results, an increase in the US monetary policy rate has been associated with a depreciation of the US dollar and an economic boom and looser financial conditions in the rest of the world (\cite{stavrakeva2019dollar}, \cite{ilzetzki2021puzzling}). In this paper, I show that the atypical dynamics documented in this recent literature can be explained by the disclosure of and response by the Federal Reserve to information released around FOMC announcements, known as the ``Fed Information Effect''. I argue that this ``Fed Information Effect'' contaminates the identification of monetary policy shocks and biases the estimates of the international spillovers of US interest rates. Controlling for this ``Fed Information Effect'' re-establishes the conventional view that a US monetary tightening leads to a recession, an exchange rate depreciation and tighter financial conditions. 

The recent literature estimating atypical effects of US monetary policy shocks has identified them using the standard high frequency identification strategy, i.e. as unexpected movements in interest rates around FOMC announcements, as in \cite{gertler2015monetary} and \cite{nakamura2018high}. However, FOMC announcements not only convey decisions about policy rates, but also about the future path of policy rates (see \cite{miranda2021transmission}), its views about the state of the US economy (see \cite{nakamura2018identification,jarocinski2020central}), and its response to macroeconomic and financial news (see \cite{cieslak2019non,bauer2023alternative}). Both Advanced (AE) and Emerging Market (EME) economies depend heavily on the US business cycle (for instance, because of its international trade with the US or because of the impact of the US economy on commodity goods' prices) or on the conditions in US financial markets (for example, due to the appetite for AEs and EMEs' sovereign and corporate bonds and/or equity markets). As a result, these economies are affected by both the FOMC's policy decisions and its disclosure and reaction to macroeconomic and financial information. Therefore, separately identifying monetary policy shocks from information effects is essential to study the spillovers of US monetary policy on AE and EME economies. 

In line with this argument, I estimate the international spillovers of the Federal Reserve's rates for a panel of 50 countries for the period 1988-2019 using five identification schemes that allow me to separate two components of monetary policy surprises around FOMC meetings: a pure US monetary policy shock component and a ``Fed Information Effect' component. In my benchmark analysis, I separate these two components using the identification strategy introduced by \cite{bauer2023alternative} which regress monetary policy surprises on macroeconomic and financial news that predate the FOMC announcements. The orthogonalized residuals are the pure monetary policy shock (MP) component, while the component of monetary policy surprises systematically correlated with macroeconomic news represents the ``Fed Information Effect'' (FIE) component.\footnote{In \cite{bauer2023reassessment} and \cite{bauer2023alternative}, the authors show that disregarding the component of interest rate surprises systematically correlated with macroeconomic and financial news lead to significant biases in the estimates of the effects of monetary policy on the US economy.} By separately identifying these two components of the Federal Reserve's monetary policy surprises into both panel local projections and SVAR models, I find that a MP component produces conventional international spillovers, i.e. a recession, exchange rate depreciation and financial
distress. On the contrary, the FIE component of interest rate surprises produce an economic expansion, an exchange rate appreciation and looser financial conditions. 

However, several other papers have also documented the existence of the ``Fed Information Effect'' and have proposed several alternative interpretations of it. An first alternative interpretation is presented by \cite{miranda2021transmission} which shows that monetary policy surprises are both auto-correlated and correlated with the Federal Reserve's Greenbook forecasts. The authors interpret these empirical results as evidence of a ``signalling channel'' of US monetary policy, through which the FOMC conveys information about stronger than expected economic fundamentals to which it endogenously responds to. Furthermore, the authors show that this ``signalling channel'' significantly biases the identification of conventional monetary policy shocks and its impact in the US economy. A second alternative interpretation of the ``Fed Information Effect'' is proposed by \cite{nakamura2018high} and \cite{jarocinski2020deconstructing} which present evidence that the Federal Reserve systematically discloses information about the state of the US economy during FOMC announcements.\footnote{In addition, \cite{caicamara2021} provide evidence on the information content of FOMC announcements by using text from both the Federal Reserve and New York Times articles to compute differences in expectations.} The authors argue that this systematic disclosure of information by the Federal Reserve around FOMC meetings affect private sector beliefs about both monetary policy and other economic fundamentals. In this paper, I show that following four alternative identification strategies with different interpretations and different methodologies to control for the ``Fed Information Effect'', lead to qualitatively similar estimates of my benchmark results.\footnote{In particular, I estimate the impulse response functions of pure monetary policy shocks and ``Fed Information Effect'' components following the methodologies introduced by \cite{jarocinski2020deconstructing}, \cite{miranda2021transmission}, \cite{bu2021unified}, and \cite{miranda2022tale}.} I interpret this result as evidence that purging monetary policy surprises of the ``Fed Information Effect'' is key to accurately estimate the international spillovers of US interest rates. 

Moreover, I argue that the recently found atypical dynamics can be attributed to not controlling for the ``Fed Information Effect'' around FOMC meetings. I show that following the standard high frequency identification scheme to identify US interest rate shocks leads to dynamics which are an average of those arising from the MP and FIE components. In particular, this leads to a US interest rate tightening associated with a significant expansion of industrial production and equity indexes for both AE and EME economies. Overall, I argue that not controlling for the ``Fed Information Effect'' around FOMC meetings biases the estimation and quantification of the international spillovers of US interest rates. 

Lastly, I show that the benchmark results survive  a battery of robustness checks. The results are present for both AEs and EMEs and even within sub-categories of these countries, such as across different geographical regions. Additionally, results are present across different policy regimes such as exchange rate regimes and degrees of current account and/or trade openness. Finally, results are robust to alternative estimation methods, such as country-by-country and panel SVAR models.

\noindent 
\textbf{Related literature.} This paper relates to three main strands of literature. First, this paper contributes to a long strand of literature which has focused on identifying and quantifying the international spillovers of US monetary policy shocks and their transmission channels. A significant share of this literature has found that a US monetary policy tightening is associated with negative international spillovers such as an economic recession, an exchange rate depreciation or fall of the value of the country's currency and tighter overall financial conditions. Examples of this literature are \cite{eichenbaum1995some} and \cite{uribe2006country} using data from the 1980s, 1990s, and more recently \cite{dedola2017if}, \cite{vicondoa2019monetary} using data up to the late 2000s. During the rest of the paper I will refer to these results as the conventional view or impact of a US monetary policy tightening. My contribution to this literature is threefold. First, this paper innovates  by introducing an identification scheme that clearly purges any information content included in US monetary policy decisions and finds that conventional results still hold for a time sample of the 2000s and 2010s. Second, this paper contributes to the literature by showing evidence that it is crucial to purge US monetary policy surprises of any informational content to accurately identify the international spillovers of pure US monetary policy shocks. Third, this paper contributes to this literature by providing evidence that the real and financial international spillovers of US monetary policy shocks may be increasing across time.

Second, this paper also relates to a more recent literature in international economics which has found an atypical association between the US interest rates and AE and EME dynamics. Particularly, \cite{ilzetzki2021puzzling} argue that there has been a significant change over time in the transmission of US monetary policy shocks on the rest of the world. The authors find that while in the 1980s and early 1990s a US monetary policy tightening lead to the conventional results described in the previous paragraph, in the last two decades there has been a shift whereby increases in US interest rates depreciate the US dollar but stimulate the rest of the world economy. The authors label this shift as a puzzling change in the transmission of US interest rates. Consequently, in the rest of the paper I will refer to these responses of AEs and EMEs to a US monetary policy tightening as atypical dynamics. Another example of these atypical dynamics is \cite{canova2005transmission} which finds that after a US monetary policy tightening, Latin American currencies appreciate while the conventional view would expect a currency depreciation.\footnote{Evidence of atypical dynamics can also be found in an influential paper of the conventional view as \cite{uribe2006country}. In the working paper, the authors explore an identification scheme different from the one presented in the actual paper, which allows for real domestic variables to react contemporaneously to innovations in the US interest rate. Under this alternative identification strategy, the point estimate of the impact of a US-interest-rate shock on output and investment is slightly positive. This lead to adoption of a different identification scheme.} I contribute to this literature by presenting evidence that these atypical dynamics are caused by the ``Fed Information Effect''. To this end, I introducing multiple identification schemes which deconstruct monetary policy surprises into a pure monetary policy shock component and a ``Fed Information Effect'' component. I show that for all these identification schemes, the ``Fed Information Effect'' component entirely explains the atypical dynamics found by this recent literature. Additionally, for these identification schemes the pure monetary policy shock component re-establishes the results presented by the conventional view. Consequently, by deconstructing monetary policy surprises I am able to match both the conventional and atypical results.\footnote{Additionally, \cite{camara2023international} has recently found significant biases from ``information effects'' from ECB interest rates in an international setting. }

Third, this paper relates to a recent literature which has studied the spillovers of US interest rates over the rest of the world by using identification strategies that control for possible informational effects around FOMC meetings.  For instance, \cite{jarocinski2022central} uses the identification strategy introduced by \cite{jarocinski2020deconstructing} to demonstrate that central bank information effects are an important channel of the transatlantic spillover of monetary policy, as they account for a significant share of the co-movement of German and US government bond yields around ECB and FOMC policy announcements. Another example is \cite{degasperi2020global} which uses the identification strategy constructed by \cite{miranda2021transmission} which controls for potential ``signalling information'' effects around FOMC meetings.\footnote{Another example of this literature is \cite{camararamirez2022} which uses the identification strategy constructed by \cite{bauer2023reassessment} which controls for the Federal Reserve's ``responding to news'' informational effect to decompose the different transmission channels of US monetary policy rates in Chile.} My contribution to this literature is that it actively seeks to document the spillovers of the US interest rates originated by the ``Fed Information Effect''. While the main focus of \cite{degasperi2020global} is to study the different transmission channels of US interest rates, it disregard the transmission of US interest rates through information effects. In this paper, I show that changes in US interest rates originated in the ``Fed Information Effect'' lead to large spillovers on the rest of the world.\footnote{While this paper innovates by presenting an empirical analysis of the international spillovers of the FOMC's disclosure of information, the theoretical literature has already studied its potential impacts, see \cite{ahmed2021us}.} While \cite{degasperi2020global} suggests that informational effects may explain these recent atypical dynamics, it does not seek to answer this question.

\section{Country Sample, Variables \& Methodology} \label{sec:data_methodology_identification}

In this section, I describe the construction of my dataset, describe the local projection and SVAR methodologies used and delineate the identification strategies used across the paper.

\noindent
\textbf{Country sample.} For the purpose of comprehensively estimating the international spillovers of US interest rates I construct a sample of both macroeconomic and financial variables for the largest possible time period and largest possible number of countries. To do so, I select all the countries that have data on the following five macroeconomic and financial variables from the either the IMF or OECD datasets for the period January 1988 to December 2019: (i) the nominal exchange rate with respect to the US dollar; (ii) a long-term interest rate; (iii) a consumer price index; (iv) an industrial production index; and (v) an equity index. This variable choice is crucial to measure the international spillovers of US interest rates. This variable choice is crucial as their response to US interest rates is completely opposite depending on the component of monetary policy surprises. I choose to include both long term interest rates and the equity index as proxies of the financial conditions, as higher interest rates are not always a signal of tighter financial conditions in response to US interest rate hikes (see \cite{hoek2022higher}).

\begin{figure}[ht]
    \centering
    \includegraphics[width=16cm, height=8cm]{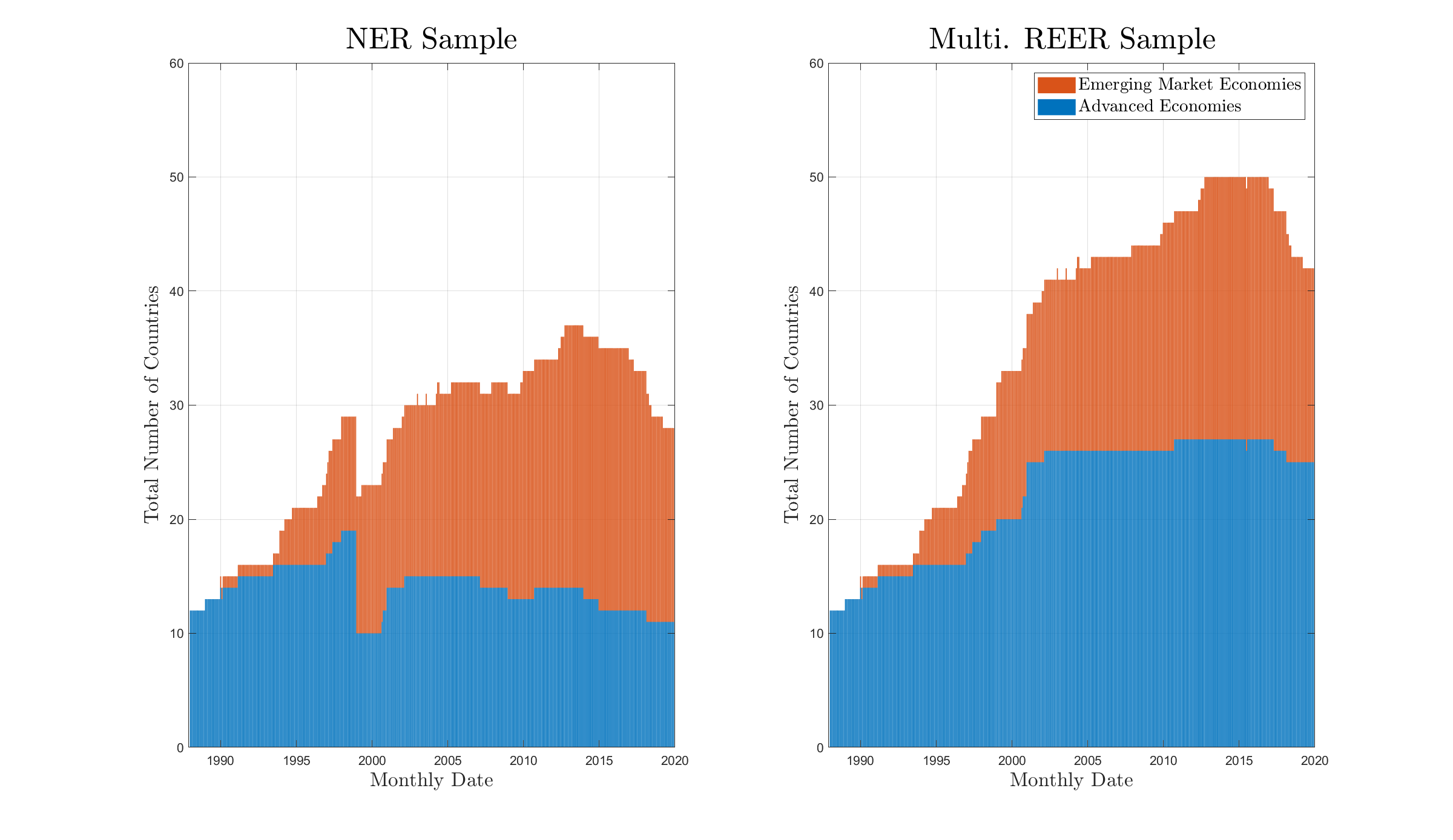}
    \caption{Number of Countries by Type of Sample}
    \label{fig:Samples}
    \floatfoot{\textbf{Note:} The significant drop in the amount of Advanced Economies in the ``NER Sample'' in the year 1999 is related to the introduction of the euro. As described in the text, once a country adopts the euro I drop it from the sample. In doing so, I prevent my results to be driven by over-representing the Euro Area. }
\end{figure}

Countries are grouped into Advanced Economies (AE) and Emerging Market Economies (EME) following the classification used by the IMF in the year 2019, the last year of my sample. Figure \ref{fig:Samples} presents the number of countries in the sample by group and by date. The resulting sample is comprised of an unbalanced panel of 50 countries, 28 AEs and 22 EMEs. The list of countries by group and additional sample details can be found in Appendix \ref{sec:appendix_data_details}. The time frame and number of countries included is significantly larger than those included in previous studies in the literature which studies the international spillovers of US interest rates at the monthly frequency. For instance, \cite{ilzetzki2021puzzling} includes 21 countries, 12 AEs and 9 EMEs, \cite{hoek2022higher} includes data on 20 EMEs, and \cite{degasperi2020global} uses an unbalanced panel of 30 economies, 15 AEs and 15 EMEs.\footnote{Few studies at the quarterly frequency have a similar number of countries, such as \cite{lin2018international} which includes 61 countries, or \cite{kalemli2019us} which includes 55 countries; but have a shorter time window, usually starting in the mid or late 1990s.} Note that the sample includes the specific time window for which \cite{ilzetzki2021puzzling} find atypical results.\footnote{In Section \ref{sec:robustness_checks_additional_results}, I show that my results are robust to different sub-samples in terms of both countries and time. More importantly, I show that my results are robust to using the same country and time sample used by \cite{ilzetzki2021puzzling}.}

I source all of the variables from IMF \& OECD datasets in order to construct a harmonized dataset of macroeconomic and financial variables for both AE and EME economies. This guarantees that the variables in the dataset are constructed following closely related methodologies.\footnote{Data is sourced from the IMF's International Financial Statistics dataset and the OECD Statistics website (\url{https://stats.oecd.org/}). The OECD dataset constructs long-term interest rates for most of the countries in its dataset. For the countries without or with a significantly short time series of a long term interest rate, I construct the variable by summing the country's EMBI Spread (constructed and produced by JP Morgan) and Kenneth French's short term US free risk rate (sourced from \url{https://mba.tuck.dartmouth.edu/pages/faculty/ken.french/data_library.html}. This is based following the approach by \cite{camara2024international}. For additional details on the construction of the dataset see Appendix \ref{sec:appendix_data_details}. In Section \ref{subsec:robustness_checks} I show that all of the results are robust to replacing the constructed long term interest rate with the EMBI spreads.} To rule out the possibility that the results are driven by countries in the Euro Area, I carry out two modifications of the sample. First, as countries join the Euro Area I drop them from the sample and replace them with a single Euro Area country-unit. Second, I estimate the results replacing the nominal exchange rate with the US dollar with a trade weighted real exchange rate.\footnote{This dataset is constructed following the methodology introduced by \cite{darvas2012real} and can be sourced from \url{https://www.bruegel.org/publications/datasets/real-effective-exchange-rates-for-178-countries-a-new-database}. I normalize the real exchange rate index such that an increase in the index implies a depreciation of the home currency.} While countries in a currency union share a nominal exchange rate, there is evidence of persistent differences in real exchange rate dynamics across countries in the Euro Area.\footnote{See, for instance, \cite{crucini2005understanding} and \cite{fidora2021real}.} 

In Section \ref{sec:main_results}, I show that results are robust across two important sub-samples. First, I show that results are present for both AE and EME samples separately. Second, I show that results are present across several different time samples. In Section \ref{subsec:additional_results}, I show that the benchmark results are robust to both removing and introducing several additional macroeconomic, labor and/or financial variables to the benchmark specification. In Section \ref{subsec:robustness_checks}, I show that the benchmark results are robust to using the exact sample and variable specification as in \cite{ilzetzki2021puzzling}, and across different sub-samples according to a country's continent, exchange rate regime and current account and trade openness. 

\noindent
\textbf{Methodology.} I quantify the international spillovers of US interest rates by estimating both local projection regressions \'a la \cite{jorda2005estimation} and Bayesian SVAR models. 

First, local projection regressions allow me to estimate the impact of the two components of monetary policy surprises for a large unbalanced panels of countries. This exercise allows me to estimate the international spillovers of US interest rates for the largest possible sample in terms of the time and country dimensions. The main local projection regression estimated is given by
\begin{align} 
    y_{i,t+h} = \beta^{MP}_{h} i^{\text{MP}}_t + \beta^{FIE}_{h} i^{\text{FIE}}_t + \sum^{J_y}_{j=1} \delta^{j}_i y_{i,t-j} + \sum^{J_x}_{j=1} \alpha^{j}_i x_{i,t-j} + \sum^{J_i}_{j=1} \left( \phi^{j}_i i^{\text{MP}}_{t-j} + \varphi^{j}_i i^{\text{FIE}}_{t-j} \right) + \epsilon_{i,t} \label{eq:LP_pooled}
\end{align}
where $y_{i,t+h}$ represents the outcome variable from country $i$'s at time horizon of $h$ months from date $t$. Coefficients $\beta^{MP}_{h}$ and $\beta^{FIE}_{h}$ give the impulse response of outcome variable $y$ at horizon $h$ of a pure monetary policy shock (MP) component and of the ``Fed Information Effect'' (FIE) component, respectively. The specification includes $J_y$ lags of the outcome variable, $y_{i}$, $J_x$ lags of the other ``country specific'' variables $x_{i}$ and $J_i$ lags of the monetary policy surprises components. For the benchmark sample I set the lags on the dependent and other control variables at 2, i.e., $J_y = J_x = 2$; and the amount of lags on the components of monetary policy components at 4, i.e., $J_i=4$. Introducing lagged values of the dependent variable controls for any potential seasonality in the monthly variables. The choice of two lags is set considering the Bayesian Information Criterion (BIC) and the Hannan-Quinn (HQ) information criteria. I also introduce lagged values of the monetary policy components for efficiency reasons as suggested by \cite{jorda2023local}. In Section \ref{subsec:robustness_checks}, I show that  results are robust to introducing several alternative lag structures of dependent variables and components of monetary policy surprises. Additionally, I estimate an alternative specification of Equation \ref{eq:LP_pooled} in which I follow the standard high-frequency identification assumption (``Standard HFI") and replace the MP and FIE component with the monetary policy surprise. The standard errors are clustered at the time level, which takes into account the fact that all countries are hit by the shocks simultaneously. In Section \ref{subsec:robustness_checks} I show that all of the benchmark results are robust to alternative specifications of Equation \ref{eq:LP_pooled} which include time trend and/or country fixed effects.

Second, I estimate Bayesian SVAR models to study potential sources of heterogeneity across countries and to provide a robustness check for my local projection regression analysis. First, I estimate country-by-country standard Bayesian SVAR model introducing the MP and FIE components as exogenous variables. Bayesian methods permit me to estimate the model for each country separately and study potentially heterogeneous impacts of US interest rates across countries, even with a limited time series. I show that results are prevalent across the vast majority of countries in my sample and not driven by any outlier. Second, I estimate several specifications of pooled and mean-group Bayesian panel SVAR model. These exercises provide a robustness check to the local projection regression result given the well known bias-variance trade off between the two methodologies, see \cite{plagborg2021local}. The details of these Bayesian SVAR models are described in Appendix \ref{sec:appendix_model_details}. 

\noindent
\textbf{Identification strategies.} The benchmark results in this paper are estimated using the identification strategy introduced by \cite{bauer2023reassessment}. The authors construct monetary policy surprise by taking the first principal component of the changes in the first four quarterly Eurodollar futures contracts around FOMC announcements. The authors extract the pure monetary policy shock component by orthogonalizing monetary policy surprises of publicly available macroeconomic and financial information released close to the FOMC meeting:
\begin{equation} \label{eq:mps_orth}
    mps_t=\alpha+\beta X_t +u_t
\end{equation}
where $mps_t$ is the monetary policy surprise, $X_t$ denotes a vector of controls and $u_t$ is the regression residual.\footnote{This equation relates to equation 14 from \cite{bauer2023reassessment}.}$^{,}$\footnote{The vector of controls are (i) Non-farm payrolls surprise: the surprise component of the most recent non-farm payrolls release prior to the FOMC announcement, measured as the difference between the released value of the statistic minus the median expectation for that release from the Money Market Services survey; (ii) Employment growth: the log change in nonfarm payroll employment from 1 year earlier to the most recent release before the FOMC announcement; (iii) S\&P 500: the log change in the S\&P 500 stock market index from 3 months (65 trading days) before the FOMC announcement to the day before the FOMC announcement; (iv) Yield curve slope: the change in the slope of the yield curve from 3 months before the FOMC announcement to the day before the FOMC announcement, measured as the second principal component of 1-to-10-year zero-coupon Treasury yields; (v) Commodity prices: the log change in the Bloomberg Commodity Spot Price index from 3 months before the FOMC announcement to the day before the FOMC announcement; (vi) Treasury skewness: the implied skewness of the 10-year Treasury yield, measured using options on 10-year Treasury note futures with expirations in 1–3 months, averaged over the preceding month, from \cite{bauer2024interest}.} The authors define the monetary policy component of monetary policy surprises as
\begin{align}
    \text{MP component} = mps_t - \hat{\alpha} - \hat{\beta} X_t
\end{align}
where $ \hat{\alpha}, \hat{\beta}, X_t$ correspond to the estimated coefficients and predictors from Equation \ref{eq:mps_orth}. I define the ``Fed Information Effect'' component of monetary policy surprises as
\begin{align}
    \text{FIE component} = \hat{\alpha} + \hat{\beta} X_t
\end{align}
Note that this method of constructing the FIE component of monetary policy surprises is analogous to the method used by \cite{miranda2021transmission} to isolate the Federal Reserve's ``signalling channel''.

This identification strategy has two main advantages. First, it provides a decomposition of interest rate surprises into MP and FIE components for the longest possible time sample, February 1988 to December 2019. This large time sample is of particular importance to estimate the international spillovers of US interest rates as there is evidence that US monetary policy has been conducted more systematically in recent years, see \cite{ramey2016macroeconomic}, making monetary policy surprises smaller and less informative. Second, as argued in \cite{bauer2023reassessment} and \cite{bauer2023alternative}, the authors present evidence of their interpretation of the ``Fed Information Effect'' over alternative previous interpretations.\footnote{However, it is noteworthy to mention that \cite{acosta2022perceived} has presented evidence against \cite{bauer2023alternative} and highlighted the importance of the disclosure of information and information mis-perceptions around FOMC meetings.}

Additionally, I show that the benchmark results are robust to alternative identification strategies which provide alternative explanations for the ``Fed Information Effect''. In particular, I estimate the international spillovers of US interest rates using the methodologies introduced by \cite{jarocinski2020deconstructing}, \cite{miranda2021transmission} and \cite{miranda2022tale}. These methodologies emphasize the role of the Federal Reserve in disclosing information about the state of the US economy around FOMC meetings and provide a decomposition of interests rate surprises into MP and FIE components for shorter time periods. Furthermore, I show that benchmark results are robust to following the methodology introduced by \cite{bu2021unified} which identify a pure monetary policy shock cleansed of the ``Fed Information Effect'' following a Fama-MacBeth two-step procedure. Lastly, I show that the recent atypical results presented by \cite{ilzetzki2021puzzling} arise when following the methodology introduced by \cite{nakamura2018high} which highlights the importance of the ``Fed Information Effects'' in the transmission of monetary policy in the US. While this paper does not provide evidence in favour or against any of the identification strategies, it does stress the importance of purging monetary policy surprises of the ``Fed Information Effect''. I want to stress that it is not the intention of this paper to opine in the validity of these different methodological approaches. I take these different methodologies as the state of the art in the empirical literature of identification of monetary policy shocks. The core insight of this paper is that, regardless of the interpretation or methodology to cleanse monetary policy surprises, controlling for the ``Fed Information Effect'' is crucial to accurately estimate the international spillovers of US interest rates.

\section{Spillovers of Monetary Policy \& Information Effects} \label{sec:main_results}

This section presents the main results of this paper. I estimate and quantify the impact of the two components of monetary policy surprises: the pure monetary policy shock (MP) and the ``Fed Information Effects'' (FIE) component. I compare the resulting impulse response functions with those arising from following the standard identification strategy (``Standard HFI'') of only using the high-frequency monetary policy surprises. My first main result is that the MP and FIE components have completely opposite international spillovers. My second main result is that the presence of the ``Fed Information Effects'' explains the recently found atypical dynamics arising from following the standard identification strategy.

\begin{figure}[ht]
    \centering
    \includegraphics[scale=0.4]{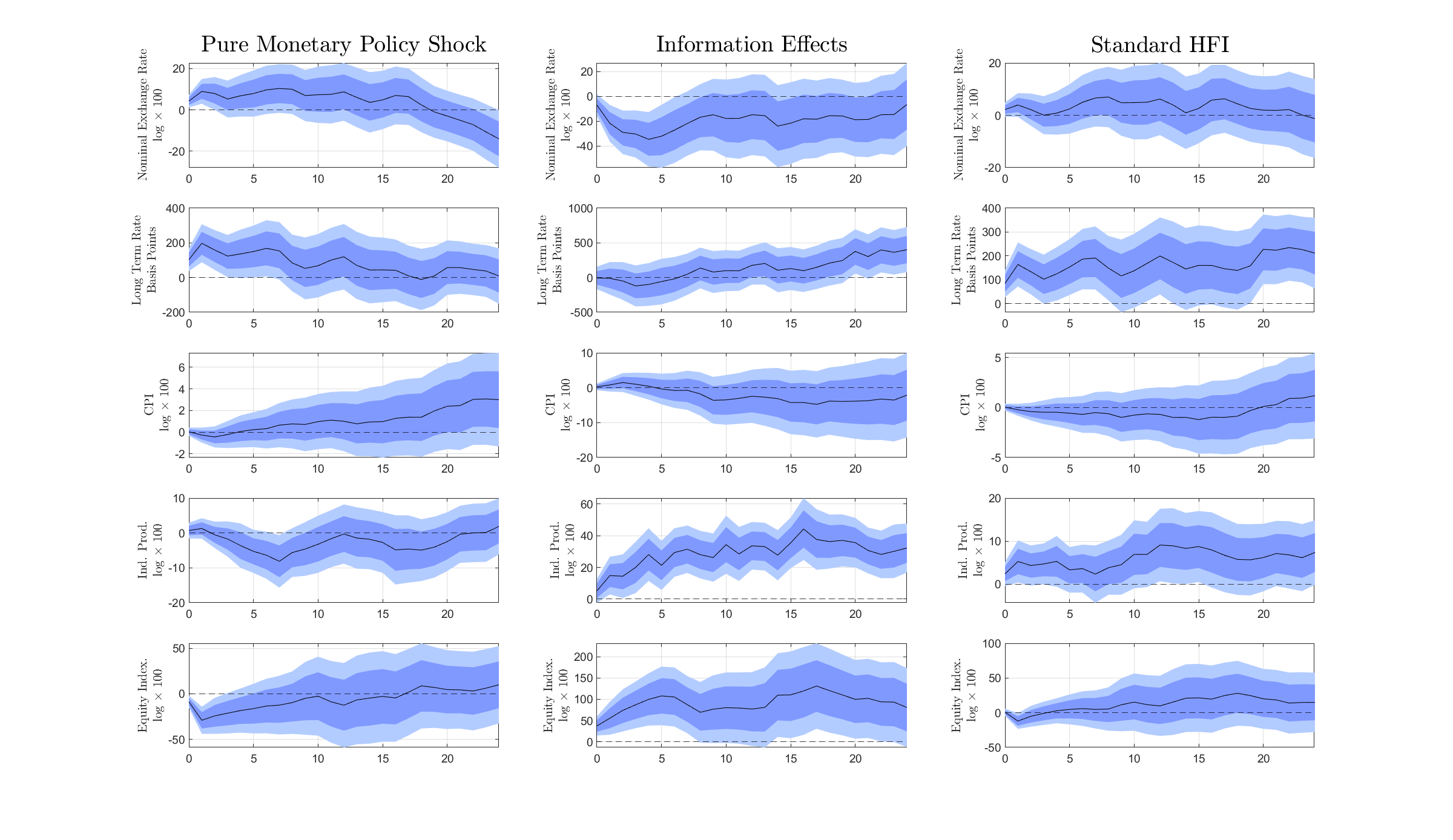}
    \caption{Impulse Response Functions \\ Benchmark Specification}
    \label{fig:Figure_1}
    \floatfoot{\textbf{Note:} The figure is comprised of 15 sub-figures ordered in three columns and five rows. The left column relates to the estimates of $\beta^{MP}$ in Equation \ref{eq:LP_pooled}, the middle column relates to the estimate of $\beta^{FIE}$ in Equation \ref{eq:LP_pooled}, while the right column relates to estimating Equation \ref{eq:LP_pooled}, replacing the MP and FIE components with the un-orthogonalized monetary policy surprise. The rows represent the impact on (i) the nominal exchange rate with the US dollar (in logs times 100); (ii) long term interest rates in basis points; (iii) the consumer price index (in logs times 100); (iv) the industrial production index (in logs times 100); (v) the equity index (in logs times 100). The solid black line represents the point estimate, the dark blue area represents the 68\% confidence interval, and the light blue area represents the 90\% confidence interval. In the text, when referring to Panel $(i,j)$, $i$ refers to the row and $j$ to the column of the figure. Each variable, in its own transformation, is demeaned at the country level. }
\end{figure}

First, I start by testing whether the deconstruction of US monetary policy surprises into two distinct components matter for identifying the international spillovers of US interest rates. The first and second columns of Figure \ref{fig:Figure_1} present the impulse response functions of the macroeconomic and financial variables to a MP and FIE component of monetary policy surprises.
Comparing the responses across these two columns lead to a first important conclusion. This is, that the controlling for the ``Fed Information Effect'' around FOMC meetings deconstructs policy monetary surprises into two distinct components. Comparing the result presented in Figure \ref{fig:Figure_1} it is straightforward to conclude that the impulse response functions of these two components have opposite effects in macroeconomic and financial variables.  For instance, the behavior of the nominal exchange rate and the industrial production index is completely opposite across figures, with a MP component leading to an exchange rate depreciation and a persistent decline in industrial output and a FIE component leading significant nominal exchange rate appreciation and increase in the industrial production index. Hence, correctly identifying pure monetary policy shocks is crucial to accurately quantify the international spillovers of US interest rates.

Next, I describe in greater detail the impulse response functions. The first column of Figure \ref{fig:Figure_1} shows the responses of domestic macro and financial variables to a MP component. Panel (1.1) shows that a MP component leads to a significant depreciation on impact of the nominal exchange rate. The nominal exchange rate further depreciates during the first year after the shock, being statistically different from zero at the 90\% for the first four months. Panel (2.1) shows that on impact, a MP component leads to an increase in long term interest rates of 100 basis points, which reaches a peak of 200 basis points two months after the initial shock. This increase in long term interest rates is statistically significant at the 90\% up to a year after the initial shock. Panel (3.1) shows that the MP component leads to a slow and sluggish increase in consumer prices, which only becomes statistically different from zero at the 68\% a year after the initial shock. Panel (4.1) shows that the MP component leads to an economic recession shown by a drop in industrial production index. This fall in industrial output reaches it trough between 7 to 8 months after the initial shock. Lastly, Panel (5.1) shows that a MP component leads to a significant and persistent drop in equity indexes. After an initial drop close to 10\%, equity indexes keep falling for the next two months, reaching a trough close to 25\%. Moreover, the equity index only recovers back to the pre-shock level a year after the initial shock.\footnote{A possible concern arising from the dynamics of the equity index on Panel 5.1 of Figure \ref{fig:Figure_1} is that the drop in the equity index is driven by the depreciation of the nominal exchange rate, shown in Panel 1.1. Note that, as described in Appendix \ref{sec:appendix_data_details} the equity index is defined in domestic currency. Furthermore, the drop in the equity index is between 50\% and 100\% greater in magnitude than the increase in the nominal exchange rate. Thus, Panel 5.1 shows that the equity index drops in value both in domestic currency and in US dollars.}

The spillovers of the FIE component of monetary policy surprises are opposite to those of the MP component. The
second column of Figure \ref{fig:Figure_1} presents the estimates of $\beta^{FIE}$ in Equation \ref{eq:LP_pooled}. Panel (1.2) shows that the FIE component leads to a significant and persistent appreciation of the exchange rate. This appreciation is hump-shaped and persistent, being statistically different from zero at the 68\% level up to 18 months after the initial shock. Panel (2.2) shows that long term interest rates do not react on impact to a FIE component, and only increase persistently above pre-shock levels between 6 and 9 months after the initial shock. Panel (3.2) shows that the consumer price index slightly increases on impact, but seems to fall below pre-shock levels after 9 months. Panel (4.2) shows that the FIE component of US monetary policy surprises leads to a persistent increase in industrial production. This expansion in industrial output is statistically different from zero at the 90\% for all the time horizons in the figure, reaching a peak between 16 to 18 months after the initial shock. Lastly, Panel (5.2) shows that the FIE component also leads to a sizable expansion of the equity index, remaining significantly above pre-shock levels up to 24 months after the initial shock. Overall, the first and second columns of Figure \ref{fig:Figure_1} show that the two components of monetary policy surprises shocks lead to almost completely opposite impulse response functions for the full sample of countries. 

Next, I argue that following the ``Standard HFI'' strategy, using only the high frequency monetary policy surprise may lead to biased impulse response functions. I replace the benchmark specification in Equation \ref{eq:LP_pooled} by replacing the MP and FIE components with the un-orthogonalized monetary policy surprise. The third column of Figure \ref{eq:LP_pooled} exhibits the impulse response functions under this identification strategy. Across the different variables, the impulse responses are an average of the responses presented for the MP and FIE shocks in the left and middle columns. Most shockingly, the ``Standard HF'' strategy leads to a qualitatively different response of the industrial production index from that arising after a MP shock. Under the ``Standard HFI'' strategy, the industrial production production increases significantly after a monetary policy surprise. This expansion is statistically different zero on impact and remains persistently above pre-shock levels for more than 12 months after the initial shock. Similarly, after exhibiting a short fall on impact, the response of the equity index turns positive and remains above pre-shock levels 9 months after the initial shock. In addition to this qualitative differences, there are notable quantitative differences in the impulse response functions of other variables. Panel 1.3 shows that under the ``Standard HFI'' strategy, the depreciation of the nominal exchange rate is not statistically different from zero. Consequently, following the ``Standard HFI'' strategy may lead to underestimating the depreciation of the exchange rate after a pure US monetary policy shock.

Following this analysis, I test whether results are robust to using the expanded sample which replaces the nominal exchange rate with the multilateral trade-weighted real exchange rate index, as described in Section \ref{sec:data_methodology_identification}. 
\begin{figure}[ht]
    \centering
    \includegraphics[scale=0.4]{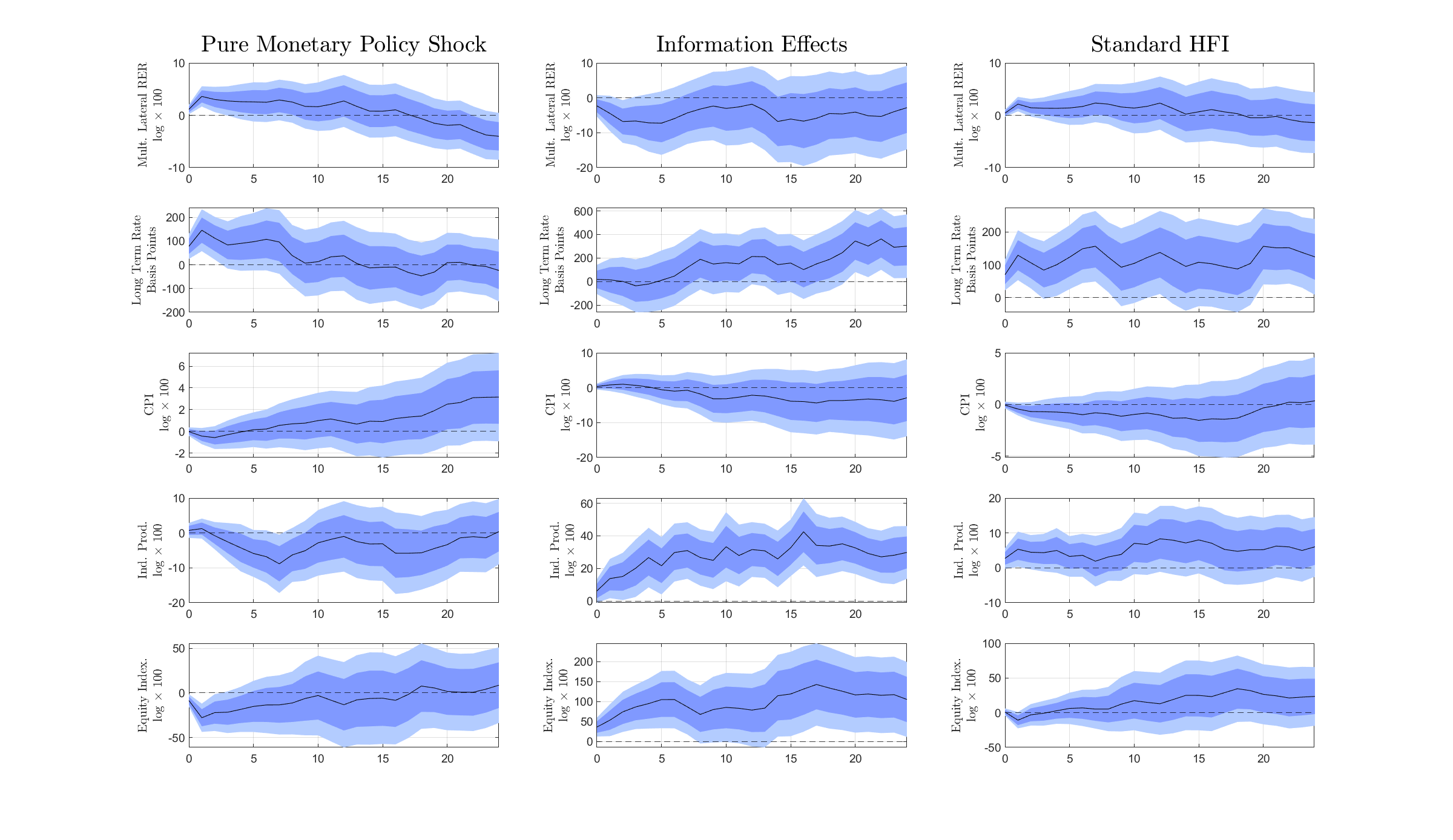}
    \caption{Impulse Response Functions \\ Multilateral Real Exchange Rate Specification}
    \label{fig:Figure_2}
    \floatfoot{\textbf{Note:} The figure is comprised of 15 sub-figures ordered in three columns and five rows. The left column relates to the estimates of $\beta^{MP}$ in Equation \ref{eq:LP_pooled}, the middle column relates to the estimate of $\beta^{FIE}$ in Equation \ref{eq:LP_pooled}, while the right column relates to estimating Equation \ref{eq:LP_pooled}, replacing the MP and FIE components with the un-orthogonalized monetary policy surprise. The rows represent the impact on (i) the multilateral trade weighted real exchange rate index (in logs times 100); (ii) long term interest rates in basis points; (iii) the consumer price index (in logs times 100); (iv) the industrial production index (in logs times 100); (v) the equity index (in logs times 100). The solid black line represents the point estimate, the dark blue area represents the 68\% confidence interval, and the light blue area represents the 90\% confidence interval. In the text, when referring to Panel $(i,j)$, $i$ refers to the row and $j$ to the column of the figure. Each variable, in its own transformation, is demeaned at the country level. }
\end{figure}
Figure \ref{fig:Figure_2} shows the result of this exercise. The first row of Figure \ref{fig:Figure_2} presents the impulse response function of the real exchange rate in response to the different components of US monetary policy surprises. On the one hand, Panel (1.1) shows that in response to the MP component, countries' real exchange rate depreciates. On the other hand, Panel (1.2) shows that in response to the FIE component the real exchange rate appreciates. Panel (1.3) shows that following the ``Standard HFI'' approach leads to a response an average of the previous two impulse response functions, exhibiting a muted and not statistically different from zero response. The response of the remaining four variables in Figure \ref{fig:Figure_2} are qualitatively and quantitatively in line with those in Figure \ref{fig:Figure_1}. Thus, the main results presented in Figure \ref{fig:Figure_1} are robust to this alternative variable selection.

\begin{landscape} 
\begin{figure}[ht]
    \centering
    \caption{Impulse Response Functions \\ \footnotesize Separate Samples for Adv. \& Emerging Economies}
    \label{fig:Figure_3}
     \centering
     \begin{subfigure}[b]{0.495\textwidth}
         \centering
         \includegraphics[width=\textwidth,height=9.5cm]{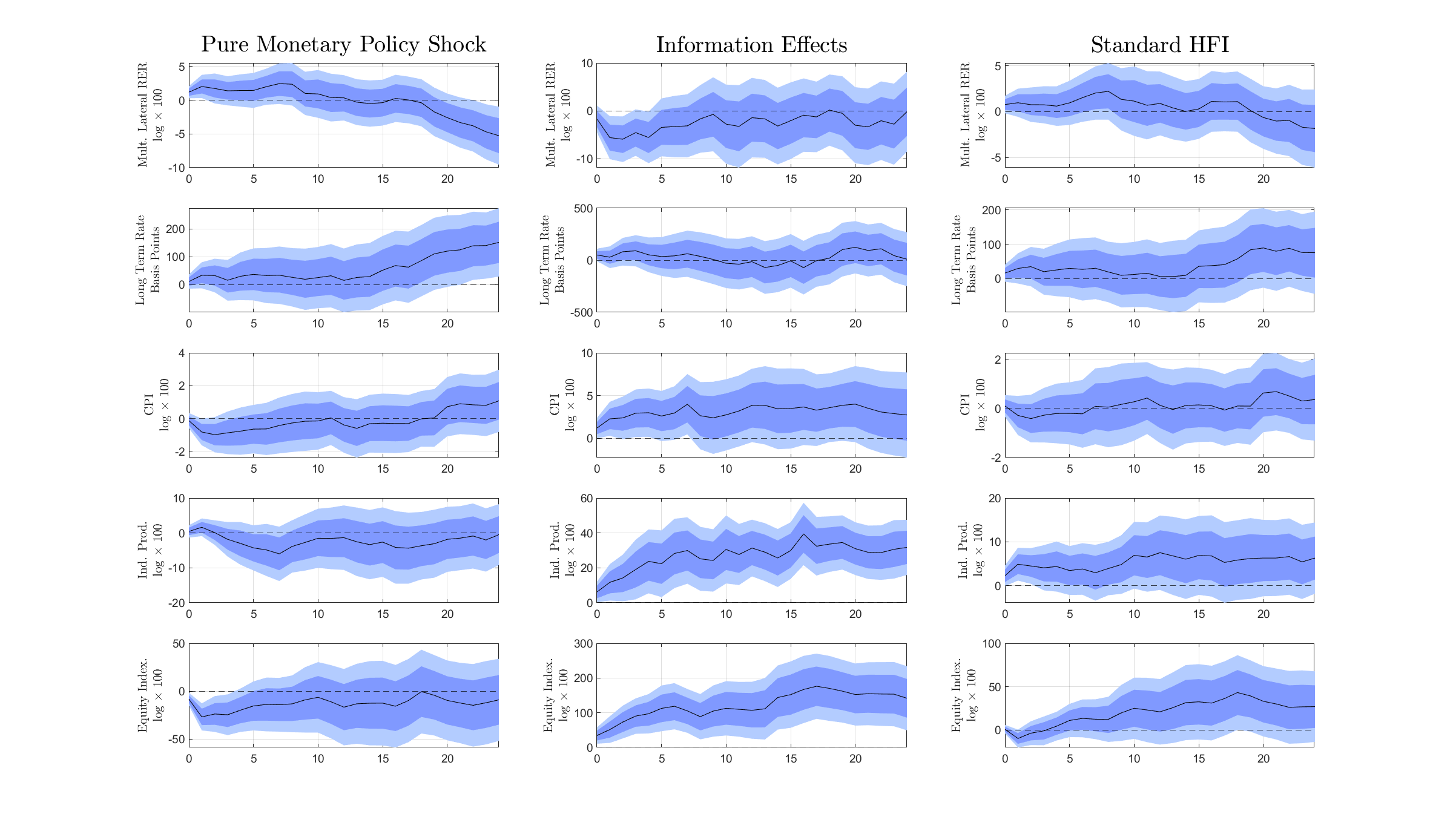}
         \caption{Advanced Economies}
         \label{fig:Figure_3_1}
     \end{subfigure}
     \hfill
     \begin{subfigure}[b]{0.495\textwidth}
         \centering
         \includegraphics[width=\textwidth,height=9.5cm]{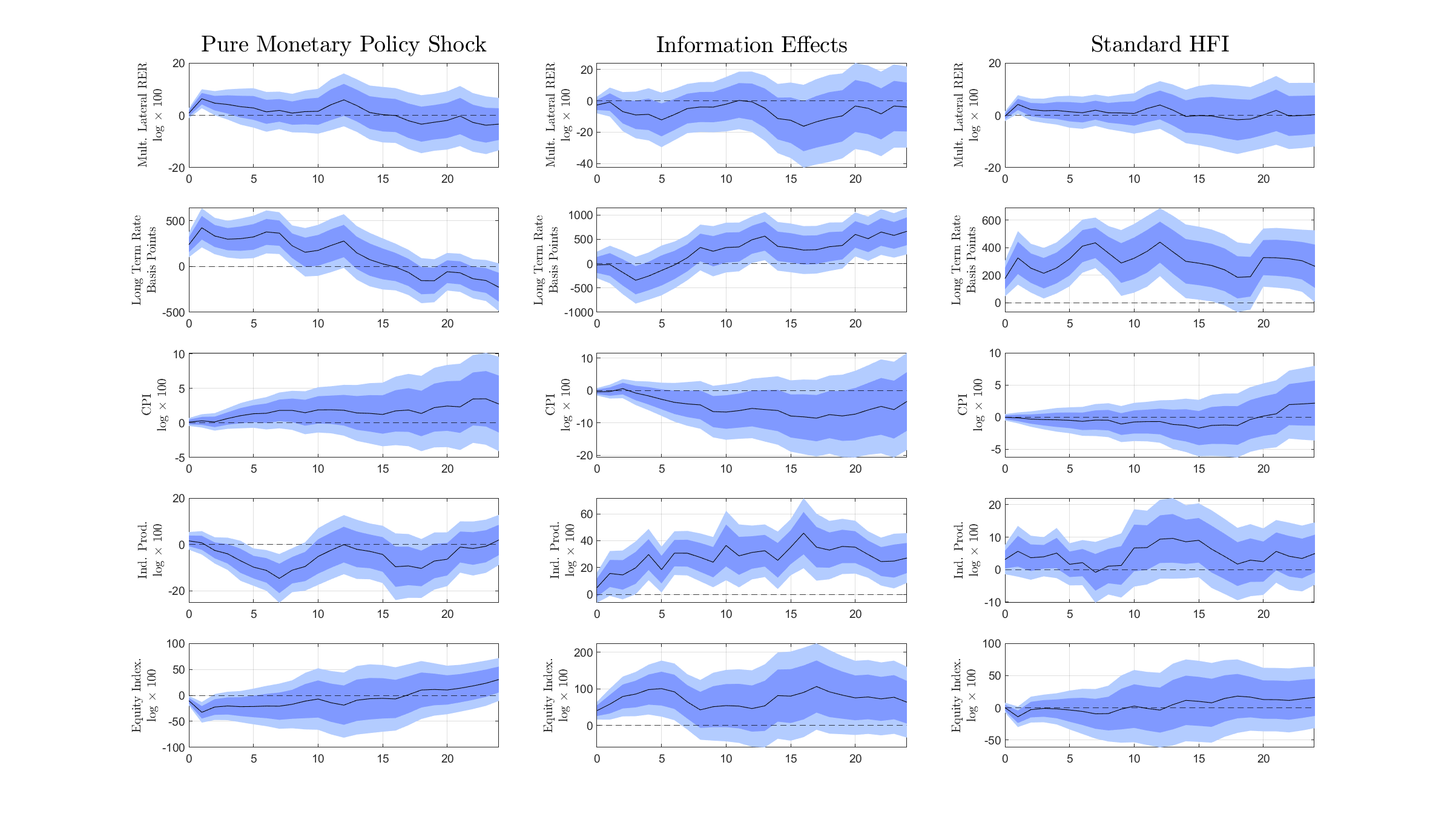}
         \caption{Emerging Market Economies}
         \label{fig:Figure_3_2}
     \end{subfigure} 
     \floatfoot{\scriptsize \textbf{Note:} Each of the two figures are comprised of 15 sub-figures ordered in three columns and five rows. The first figure to the left is constructed by estimating Equation \ref{eq:LP_pooled} for the sample of Advanced Economies. The second figure to the right is constructed by estimating Equation \ref{eq:LP_pooled} for the sample of Emerging Market Economies. See Appendix \ref{sec:appendix_data_details} for the list of countries in each sample. The left column relates to the estimates of $\beta^{MP}$ in Equation \ref{eq:LP_pooled}, the middle column relates to the estimate of $\beta^{FIE}$ in Equation \ref{eq:LP_pooled}, while the right column relates to estimating Equation \ref{eq:LP_pooled}, replacing the MP and FIE components with the un-orthogonalized monetary policy surprise. The rows represent the impact on (i) the nominal exchange rate with respect to the US dollar (in logs times 100); (ii) long term interest rates in basis points; (iii) the consumer price index (in logs times 100); (iv) the industrial production index (in logs times 100); (v) the equity index (in logs times 100). The solid black line represents the point estimate, the dark blue area represents the 68\% confidence interval, and the light blue area represents the 90\% confidence interval. In the text, when referring to Panel $(i,j)$, $i$ refers to the row and $j$ to the column of the figure. Each variable, in its own transformation, is demeaned at the country level. }
\end{figure}
\end{landscape}

To conclude the main results of this paper, I carry out two robustness checks in terms of the composition of countries in the sample and in terms of different time samples. I begin by estimating the impulse response functions for the MP and FIE components and following the ``Standard HFI'' for the panel of Advanced Economies (Figure \ref{fig:Figure_3_1}) and for the panel of Emerging Market Economies (Figure \ref{fig:Figure_3_2}), separately. 
Across the two sub-samples, the main results presented in Figure \ref{fig:Figure_1} hold, with the two components of monetary policy surprises leading to completely opposite dynamics and the ``Standard HFI'' leading to a weighted average of them. Still, there are some noteworthy quantitative differences. First, after a MP component, Emerging Market economies exhibit a persistent increase in the consumer price index while Advanced Economies exhibit a significant drop. As argued by \cite{garcia2020revisiting} and \cite{auclert2021exchange}, EME economies depend relatively more in imported goods for both consumer and input goods. Thus, one would expect an exchange rate depreciation to lead to a greater impact on consumer prices in Emerging Markets relative to Advanced Economies, all else equal. Similarly, the impact of the MP component on long term interest rates is faster and quantitatively larger larger for EMEs compared to AEs. While EMEs exhibit an increase in long term interest of 250 basis points on impact, AEs experience a slower and quantitatively smaller increase in long term rates. This difference across samples may be driven by EMEs having relatively less sophisticated and smaller domestic financial markets and, thus, exhibiting a relatively greater dependence on international financial markets than AEs (see \cite{dages2000foreign,broner2013emerging,cortina2018corporate,abraham2020growth}). Figures \ref{fig:AEs_REER} and \ref{fig:EMs_REER} in Appendix \ref{sec:appendix_figures} present this robustness check for the sample which replaces the nominal exchange rate with the multilateral exchange rate.

\begin{landscape} 
\begin{figure}[ht]
    \centering
    \caption{Impulse Response Functions \\ \footnotesize Samples starting in January 1998 \& January 2008}
    \label{fig:Figure_4}
     \centering
     \begin{subfigure}[b]{0.495\textwidth}
         \centering
         \includegraphics[width=\textwidth,height=9.5cm]{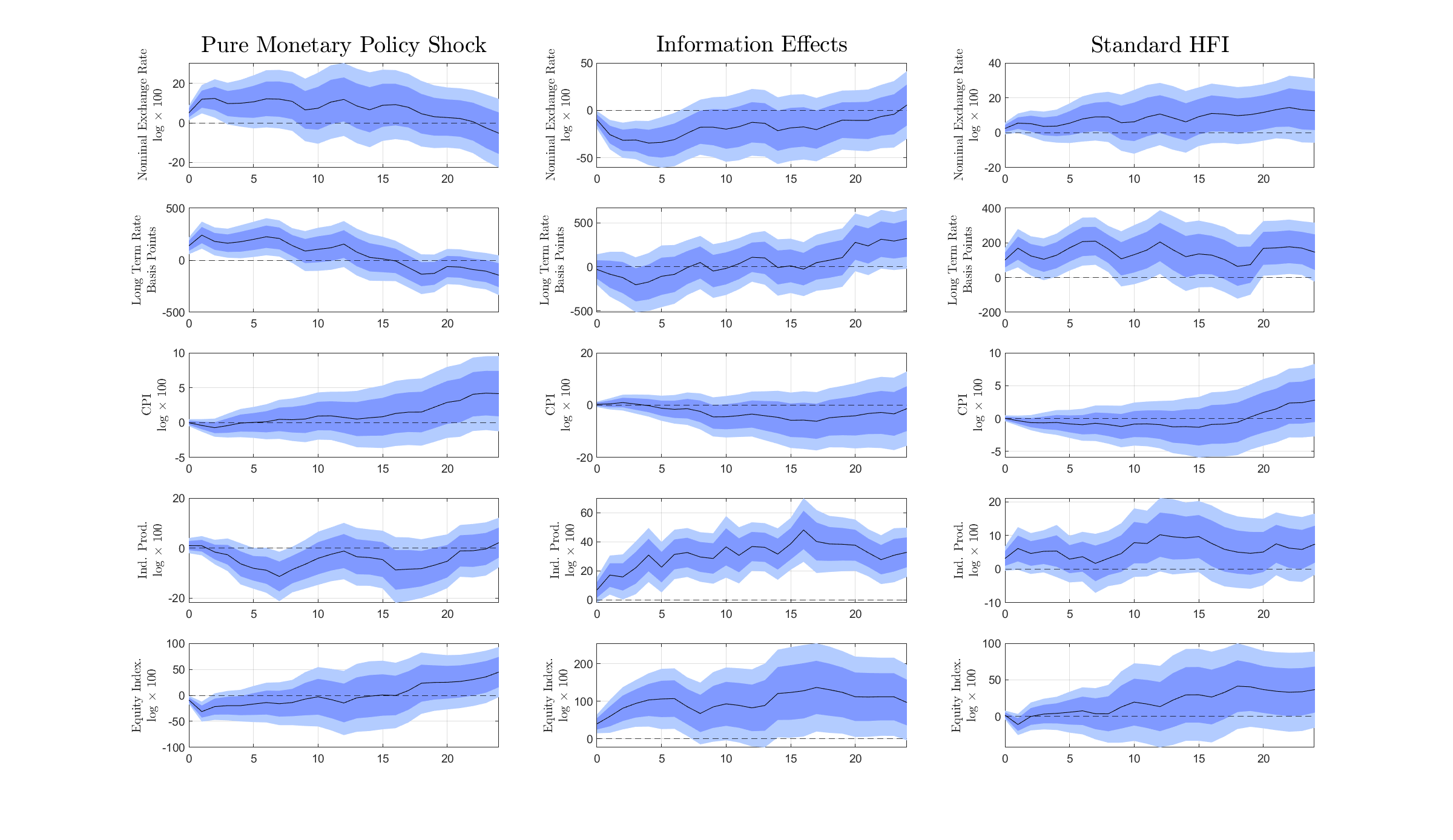}
         \caption{Sample January 1998 - December 2019}
         \label{fig:Figure_4_1}
     \end{subfigure}
     \hfill
     \begin{subfigure}[b]{0.495\textwidth}
         \centering
         \includegraphics[width=\textwidth,height=9.5cm]{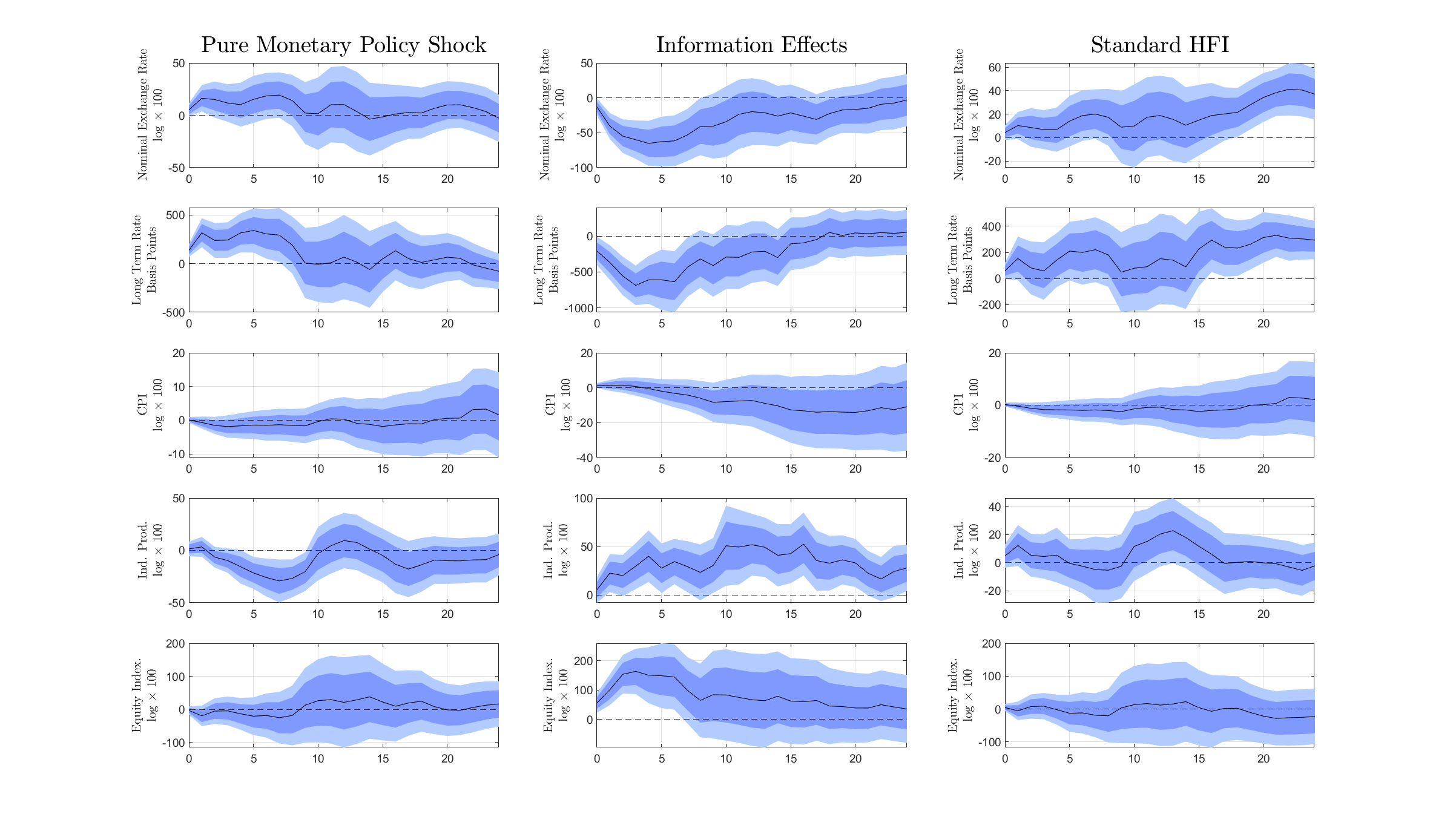}
         \caption{Sample January 2008 - December 2019}
         \label{fig:Figure_4_2}
     \end{subfigure} 
     \floatfoot{\scriptsize \textbf{Note:} Each of the two figures are comprised of 15 sub-figures ordered in three columns and five rows. The first figure to the left is constructed by estimating Equation \ref{eq:LP_pooled} for the sample starting in January 1998 and ending in December 2019. The second figure to the right is constructed by estimating Equation \ref{eq:LP_pooled} for the sample starting in January 2008 and ending in December 2019. The left column relates to the estimates of $\beta^{MP}$ in Equation \ref{eq:LP_pooled}, the middle column relates to the estimate of $\beta^{FIE}$ in Equation \ref{eq:LP_pooled}, while the right column relates to estimating Equation \ref{eq:LP_pooled}, replacing the MP and FIE components with the un-orthogonalized monetary policy surprise. The rows represent the impact on (i) the nominal exchange rate with respect to the US dollar (in logs times 100); (ii) long term interest rates in basis points; (iii) the consumer price index (in logs times 100); (iv) the industrial production index (in logs times 100); (v) the equity index (in logs times 100). The solid black line represents the point estimate, the dark blue area represents the 68\% confidence interval, and the light blue area represents the 90\% confidence interval. In the text, when referring to Panel $(i,j)$, $i$ refers to the row and $j$ to the column of the figure. Each variable, in its own transformation, is demeaned at the country level. }
\end{figure}
\end{landscape}

Lastly, I estimate the impulse response functions for the MP and FIE components and following the``Standard HFI'' for a sample starting in January 1998 and for a sample starting in January 2008, 10 and 20 years after the start of the benchmark sample, respectively. This exercise allows me to test whether the international spillovers of US interests have varied significantly in recent years as suggested by \cite{ilzetzki2021puzzling}. The results are presented in Figure \ref{fig:Figure_4}. The resulting impulse response functions in the two time sub-samples are in line with those obtained for the benchmark sample presented in Figure \ref{fig:Figure_1}. One quantitative difference worth mentioning is that the impact of the MP component on industrial production seems to have increased across time. For the benchmark sample, Figure \ref{fig:Figure_1} shows a drop in industrial production below 10\%, while Figure \ref{fig:Figure_4} show an impact close to 12\% and to 20\% for the sample starting in 1998 and 2008 respectively. Figures \ref{fig:Figure_AE_Across_Time} and \ref{fig:Figure_EM_Across_Time} in Appendix \ref{sec:appendix_figures} repeats the exercise for the sample of AE and EME respectively to show that this trend is present across countries.\footnote{Additionally, Figures \ref{fig:Figure_AE_Across_Time_REER} and \ref{fig:Figure_EM_Across_Time_REER} in Appendix \ref{sec:appendix_figures} show that this result is also present when using the sample that replaces the nominal exchange rate with the multilateral real exchange rate.}

To summarize, by introducing an identification scheme which deconstruct US monetary policy surprises into two different components I am able to show that an increase in US interest rates may have completely different international spillovers. While a pure monetary policy shock leads to conventional results, a monetary policy surprise driven by the ``Fed Information Effect'' leads to an economic expansion, an exchange rate appreciation and looser financial conditions. Moreover, I argued that following a ``Standard HFI'' strategy underestimates the spillovers of a US monetary policy shock and leads to atypical dynamics, such as an expansion of economic activity. Lastly, I show that these results are present for both the sub-samples of Advanced and Emerging Market economies with only minor caveats, and across different time samples. 

\section{Additional Results \& Robustness Checks} \label{sec:robustness_checks_additional_results}

This section presents additional results and robustness checks that complement and validate the findings presented in Section \ref{sec:main_results}. In Subsection \ref{subsec:additional_results}, I present three additional results: (i) I estimate the international spillovers of US interest rates on samples with alternative variable selections; (ii) I show that alternative identification schemes that control for the ``Fed Information Effect'' through different methodologies and interpretations lead to results qualitatively similar to those presented in Section \ref{sec:main_results}; (iii) I estimate the international spillovers country-by-country using a SVAR model. In Subsection \ref{subsec:robustness_checks}, I show that the main results of this paper are robust to alternative econometric specifications and across different sub-sample exercises. Figures are presented in Appendix \ref{sec:appendix_figures}.

\subsection{Additional Results} \label{subsec:additional_results}

\noindent
\textbf{Additional variables.} I extend the benchmark sample with extra variables to provide supporting evidence that the MP and FIE components have distinct international spillovers. Figures \ref{fig:LendingRate_NER}, \ref{fig:Real_Variables_NER}, \ref{fig:URATE_NER} and \ref{fig:Trade_Variables} in Appendix \ref{sec:appendix_figures} present the impact on additional variables related to countries' financial conditions and economic performance. 

Figure \ref{fig:LendingRate_NER} presents evidence of the pass-through of US interest rates into domestic monetary policy and short term private sector lending rates. The left column shows that the MP component leads to worsening financial conditions as there is an immediate increase in monetary policy and lending rates. The middle column shows that the FIE component lead to slower spillovers into domestic rates, with rates increasing above pre-shock levels between 6 and 9 months after the initial impact. This is line with the dynamics estimated for the long term rates in the benchmark sample. Following the Standard HFI would lead one to conclude that a US interest rate hike leads to persistent increases in international policy rates for up to 2 years after the initial shock. 

Figure \ref{fig:Real_Variables_NER} presents the impact on three additional real variables related to economic performance: (i) a construction index, (ii) a mining index, (iii) a retail sales index. While these variables generally have a lower coverage than industrial production indexes, they represents shares of GDP comparable or larger to total manufacturing. The left column of Figure \ref{fig:Real_Variables_NER} shows that a US monetary policy tightening caused by the MP component leads to a persistent drop in the construction, mining and retail sales indexes. On the contrary, the middle column  of Figure \ref{fig:Real_Variables_NER} shows that a FIE component of monetary policy surprises lead to a significant increase of these three indexes. The right column of Figure \ref{fig:Real_Variables_NER} shows that following the ``Standard HFI'' approach would lead an econometrician to conclude that a US monetary policy tightening leads to an economic expansion. This is in line with the recent atypical dynamics found by the literature.

Figure \ref{fig:URATE_NER} presents the impact of the different components of US monetary policy rate surprises in countries'  unemployment rates. While a MP component (left panel) causes a hump shaped increase in the unemployment rate, the FIE component (middle panel) causes a persistent decline in the unemployment rate up to 18 months after the initial impact. Once more, following the ``Standard HFI'' approach would lead to the conclusion that a US monetary policy tightening positively affects the rest of the world, shown by a significantly lower unemployment rate. 

Figure \ref{fig:Trade_Variables} presents the impact of US interest rates into international trade variables: (i) a country specific commodity export price index; (ii) a country specific commodity import price index; (iii) the total export value in FOB USD dollars; (iv) the total import value in CIF USD dollars. On the one hand, the left column shows that the MP component leads to a decrease in international trade, showing a decrease in export and import commodity prices indexes and a reduction in the total value traded. On the other hand, the middle column shows that the FIE component leads to an overall expansion of international trade in terms of commodity prices and total value traded. The right column shows that following the ``Standard HFI'' approach would lead to the conclusion of a US monetary policy tightening leading to an increase in commodity prices and non discernible impact in value traded. 

I carry out a final robustness check in terms of variables by removing variables from the benchmark specification. Reducing the number of variables in the econometric specification increases the number of countries and the number of observations per country, but increases the risk of model misspecification. I estimate impulse response functions for a sample that removes the long term interest rate (see Figures \ref{fig:Removing_LT} to \ref{fig:Removing_LT_REER_08} in Appendix \ref{sec:appendix_figures}), and for a sample that removes both the long term interest rate and the equity index (see Figures \ref{fig:Removing_LT_Equity} to \ref{fig:Removing_LT_Equity_REER_08} in Appendix \ref{sec:appendix_figures}).\footnote{Removing the long term interest and/or the equity index does not increase significantly the number of countries in the sample, from 50 to 58.} The estimated impulse response functions are in line with those presented in Section \ref{sec:main_results}, yet confidence intervals are somewhat larger. 

In summary, introducing extra variables to the benchmark sample supports the main results presented in Section \ref{fig:Figure_1}. The MP component of US monetary policy surprises leads to an immediate worsening of financial conditions, measured as higher domestic interest rates, and a persistent drop in economic activity, in international trade and an increase in unemployment. Conversely, the FIE component of US monetary policy surprises lead to no immediate impact on domestic rates and a significant expansion of economic output, international trade and a persistent reduction in the unemployment rate. Figures \ref{fig:LendingRate_REER} to \ref{fig:Trade_Variables_REER} in Appendix \ref{sec:appendix_figures} show that these results are robust to using the sample which replaces the nominal exchange rate with the trade-weighted multilateral real exchange rate. 

\noindent
\textbf{Alternative identification schemes which control for the ``Fed Information Effect''.} I turn to showing that alternative identification strategies that control for the ``Fed Information Effect'' lead to results in line with those presented in Section \ref{sec:main_results}. I use the identification strategies introduced by \cite{bu2021unified}, \cite{jarocinski2020deconstructing}, \cite{miranda2021transmission} and \cite{miranda2022tale}, which have alternative interpretations and methods to control for the ``Fed Information Effect''. 

First, I estimate the international spillovers of US interest rates using the identification strategy proposed by \cite{bu2021unified}. The authors use the Fama-MacBeth two-step procedure to extract a pure monetary policy shocks, analog of the MP component defined in my benchmark scenario, for the period January 1994 to September 2019. The authors show that this shock is largely unpredictable, and contains no significant ``Fed Information Effect''. Figure \ref{fig:BenchmarkNER_BU} in Appendix \ref{sec:appendix_figures} present the results of estimating the impulse response functions by replacing the MP and FIE components in Equation \ref{eq:LP_pooled} with \cite{bu2021unified}'s shock. This monetary policy shock purged of the ``Fed Information Effect'' leads to an exchange rate depreciation, a persistent increase in long term rates, and a drop in both industrial production and equity indexes.\footnote{Figure \ref{fig:BenchmarkREER_BU} in Appendix \ref{sec:appendix_figures} shows that these results are also present for the sample which replaces the nominal exchange rate with the trade weighted multilateral real exchange rate.}  

Second, I estimate the international spillovers of US interest rates using the identification strategy proposed by \cite{jarocinski2020deconstructing}. The authors separately identify two structural shocks, a pure monetary policy shock and information disclosure shock, by exploiting the co-movement of monetary policy rates and equity prices surprises around FOMC meetings for the period January 1990 to December 2016. The authors impose sign-restriction moment conditions on these surprises to recover structural shocks. Figure \ref{fig:JK_Comparison_MP} and \ref{fig:JK_Comparison_CBI} in Appendix \ref{sec:appendix_figures} present the results for the authors' pure monetary policy and information disclosure shock respectively, under the median rotational angle that satisfies the moment conditions. Comparing the results presented in the left column of Figure \ref{fig:Figure_1} with those presented in Figure \ref{fig:JK_Comparison_MP}, it is clear that the MP component lead to qualitatively similar spillovers to \cite{jarocinski2020deconstructing}'s pure monetary policy shock. For instance, a nominal exchange rate depreciation, higher long-term interest rates and lower equity indexes. However, the impact on industrial production is muted under this alternative identification strategy for the full sample.\footnote{Nevertheless, the drop in industrial production becomes statistically different from zero for the sample starting in January 2008, as can be seen in the right column of Figure \ref{fig:JK_Comparison_MP}.} Similarly, comparing the results presented in the middle column of Figure \ref{fig:Figure_1} with those presented in Figure \ref{fig:JK_Comparison_CBI}, it is clear that the FIE component lead to qualitatively similar spillovers to \cite{jarocinski2020deconstructing}'s information disclosure shock. For example, an exchange rate appreciation, no initial impact on long term rates, and an expansion of both the industrial production and the equity index.\footnote{Figures \ref{fig:JK_Comparison_MP_REER} and \ref{fig:JK_Comparison_CBI_REER} in Appendix \ref{sec:appendix_figures} show that these results are also present for the sample which replaces the nominal exchange rate with the trade weighted multilateral real exchange rate.}

Third, I estimate the international spillovers of US interest rates using the identification strategy proposed by \cite{miranda2021transmission}. The authors separately identify two components of US monetary policy surprises by projecting the surprises on their own lags, and on the Federal Reserve's own
information set, as summarised by Greenbook forecasts, for the period January 1991 to December 2015. The authors' interpret the orthogonalized component as pure monetary policy shock and interpret the predictable component as an information component or ``signalling effect''. Figures \ref{fig:MAR_Comparison_MP} and \ref{fig:MAR_Comparison_CBI} in Appendix \ref{sec:appendix_figures} present the results for the authors' pure monetary policy shock component and the information component respectively. Comparing the results presented in the left column of Figure \ref{fig:Figure_1} with those presented in Figure \ref{fig:MAR_Comparison_MP}, it is clear that the MP component lead to qualitatively similar spillovers to \cite{miranda2021transmission}'s pure monetary policy shock. For instance, a nominal exchange rate depreciation, higher long-term interest rates and a drop in industrial production and equity indexes. Similarly, comparing the results presented in the middle column of Figure \ref{fig:Figure_1} with those presented in Figure \ref{fig:MAR_Comparison_CBI}, it is clear that the FIE component lead to qualitatively similar spillovers to \cite{miranda2021transmission}'s information or ``signalling effect'' component. This is, an exchange rate appreciation, an initial drop on long term rates, and an expansion of both the industrial production and the equity indexes.\footnote{Figures \ref{fig:MAR_Comparison_MP_REER} and \ref{fig:MAR_Comparison_CBI_REER} in Appendix \ref{sec:appendix_figures} show that these results are also present for the sample which replaces the nominal exchange rate with the trade weighted multilateral real exchange rate.} 

Fourth, I estimate the international spillovers of US interest rates using the identification strategy proposed by \cite{miranda2022tale}. Monetary policy surprises are constructed following \cite{swanson2021measuring}'s methodology, recovering a target surprise (that loads predominantly on the overnight policy rate), a path surprise (that has higher loading on 1 to 2-year maturity rates), and an quantitative easing or QE surprise (that mostly captures the variation at the long end of the yield curve). For each FOMC meeting, \cite{miranda2022tale} classify monetary policy surprises into pure monetary policy components and information component by the co-movement between the interest rates and the S\&P 500.\footnote{This is analog to \cite{jarocinski2020deconstructing}'s ``poor man's'' sign-restriction identification.} Figures \ref{fig:Target_BenchmarkNER}, \ref{fig:Path_BenchmarkNER} and \ref{fig:QE_BenchmarkNER} in Appendix \ref{sec:appendix_figures} present the international spillovers for both the pure monetary policy and information component for the target, path and QE shocks, respectively. Across the three shocks, the pure monetary policy components lead to an exchange rate depreciation, higher long-term interest rates and a drop in equity indexes. The impact on industrial production is slightly muted for the target shock, but statistically lower than zero for the path and QE shock.\footnote{Similar to the results under the \cite{jarocinski2020deconstructing} identification strategy, the impact of the target shock becomes larger in magnitude and statistical significance for later samples. This quantitatively larger effect is present across all three shocks as can be seen in Figures \ref{fig:Target_Comparison_MP}, \ref{fig:Target_Comparison_CBI}, \ref{fig:Target_Comparison_MP_REER}, \ref{fig:Target_Comparison_CBI_REER} for the target shock, Figures \ref{fig:Path_Comparison_MP}, \ref{fig:Path_Comparison_CBI}, \ref{fig:Path_Comparison_MP_REER}, \ref{fig:Path_Comparison_CBI_REER} for the path shock and Figures \ref{fig:QE_Comparison_MP}, \ref{fig:QE_Comparison_CBI}, \ref{fig:QE_Comparison_MP_REER}, \ref{fig:QE_Comparison_CBI_REER} for the QE shock.} To the contrary, the information components lead to an exchange rate appreciation, an initial reduction in long-term interest rates and an expansion of industrial and equity indexes.\footnote{Figures \ref{fig:Target_BenchmarkREER}, \ref{fig:Path_BenchmarkREER} and \ref{fig:QE_BenchmarkREER} in Appendix \ref{sec:appendix_figures} show that these results are also present for the sample which replaces the nominal exchange rate with the trade weighted multilateral real exchange rate.}

Fifth, I show that using the identification strategy proposed by \cite{nakamura2018identification} which highlights the role of the ``Fed Information Effect'' in monetary policy transmission leads to the same atypical dynamics as following the ``Standard HFI'' approach.\footnote{The vast majority of the literature which follows the ``Standard HFI'' uses the high-frequency surprises in the one quarter ahead Fed Funds Futures as the relevant monetary policy surprise. \cite{nakamura2018identification} uses a policy indicator constructed using the principal component of the surprises on: the Fed funds rate immediately following the FOMC meeting, the expected Fed funds rate immediately following the next FOMC meeting, and expected three-month eurodollar interest rates at horizons of two, three, and four quarters.} Figure \ref{fig:BenchmarkNER_NS} in Appendix \ref{sec:appendix_figures} presents the results. Following this identification strategy, a US monetary policy shock causes no exchange rate depreciation on impact, a slow increase on long term rates and a persistent increase in both industrial production and equity indexes. These results are in line with the recent atypical dynamics which suggest that a US interest rate hike leads to looser financial conditions and an economic expansion. I interpret this result as further evidence that recent atypical dynamics are caused by the ``Fed Information Effect''.\footnote{Figure \ref{fig:BenchmarkREER_NS} in Appendix \ref{sec:appendix_figures} show that these results are also present for the sample which replaces the nominal exchange rate with the trade weighted multilateral real exchange rate.}

In short, I show that four alternative identification strategies that control for and have alternative interpretations for the ``Fed Information Effect'' lead to international spillovers of US interest rates qualitatively in line with those presented in Section \ref{sec:main_results}. The different measures that capture a pure monetary policy shock component lead to an exchange rate depreciation, tighter financial conditions and a drop in industrial production. The different measures that capture a ``Fed Information Effect'' lead to an exchange rate appreciation, looser financial conditions and an industrial production expansion. I take this as supporting evidence that controlling for the ``Fed Information Effect'' is key to accurately estimate the international spillovers of US interest rates.

\noindent
\textbf{SVAR results \& country-specific heterogeneity.} So far, the impulse response functions are estimated exploiting the panel structure of the dataset. This approach produces an average of the impact across countries and does not allow for any country-specific heterogeneity. I relax this condition by estimating a series of country-by-country Bayesian SVAR models. Bayesian SVAR models are an appropriate method given the low number of observations for several of the countries in my sample. The model is a Bayesian SVAR model with standard Minnesota priors.\footnote{The model is estimated with 12 lags, an auto-regressive coefficient of 0.8, an overall tightness parameter of 0.1, a cross-variable weighting parameter of 0.5, and a lag decay parameter of 1. Each of the MP components is introduced to the model as an exogenous variable one at a time. The prior on the impact of this exogenous variable is of a mean of 0 and a variance of 100. All of these parameter values are standard in the literature. Changes on these parameter values do not lead to any substantial changes in results.} See Section \ref{sec:appendix_model_details} of the Appendix for additional details on this model. The MP and FIE components are introduced as exogenous variables. Each model is estimated for the largest possible time sample for each country, and includes the same country specific variables as in the benchmark specification, plus a set of US specific variables: (i) the two year treasury rate, (ii) the Federal Reserve's real trade weighted dollar index, (iii) the PCE price index, (iv) the dollar value of US exports, (v) the dollar value of US imports, (vi) the excess bond premium by \cite{gilchrist2012credit}, and (vii) industrial production index.\footnote{For each country's specific start and end sample see Appendix \ref{sec:appendix_data_details}.} The use of US variables in country-by-country SVAR models is standard in the literature as it minimizes the risk of model misspecification.\footnote{See \cite{degasperi2020global} and \cite{camara2024international}. The choice of US variables is also standard with the literature and with the literature which studies the impact of US monetary policy in the US itself, see \cite{bauer2023alternative}.}

Figures \ref{fig:CbC_NER_AE_NER} to \ref{fig:CbC_NER_EM_LT} in Appendix \ref{sec:appendix_figures} present the impulse response functions for the five benchmark  variables for both AE and EMEs. First, Figures \ref{fig:CbC_NER_AE_NER} and \ref{fig:CbC_NER_EM_NER} present the impact of the MP and FIE components to the nominal exchange rate. On the one hand, in response to the MP component of monetary policy surprises 90\% of the countries exhibited a depreciation of the nominal exchange rate (see Figures \ref{fig:CbC_NER_AE_NER} and \ref{fig:CbC_NER_EM_NER}), 82.5\% of the countries exhibited a fall in industrial production (see Figures \ref{fig:CbC_NER_AE_IND} and \ref{fig:CbC_NER_EM_IND}), and 92.5\% of the countries exhibited a fall in the equity index, respectively (see Figures \ref{fig:CbC_NER_AE_EQUITY} and \ref{fig:CbC_NER_EM_Equity}).\footnote{The countries which exhibited either an exchange appreciation or a non-significant response are China, Finland, India, Venezuela and Ukraine. Out of the 40 countries in this sample, only 4 did not exhibit an exchange rate depreciation}$^{,}$\footnote{The countries which exhibited either an industrial production expansion or a non-significant response are Denmark, Poland, Slovenia, Switzerland, UK, Croatia and Romania. Out of the 40 countries in this sample, only 7 did not exhibit a drop in industrial production.}$^{,}$\footnote{The countries which exhibited either an equity index expansion or a non-significant response are Slovenia, India and Pakistan. Out of the 40 countries in this sample, only 3 did not exhibit a drop in the equity index.} On the other hand, in response to the FIE component of monetary policy surprises 95\% of the countries experience an appreciation of the nominal exchange rate (see Figures \ref{fig:CbC_NER_AE_NER} and \ref{fig:CbC_NER_EM_NER}), 95\% of the countries experience an expansion of industrial production (see Figures \ref{fig:CbC_NER_AE_IND} and \ref{fig:CbC_NER_EM_IND}), and 97.5\% of the countries experience an expansion of the equity index.\footnote{The countries which exhibited either an exchange depreciation or a non-significant response are Finland and Japan. Out of the 40 countries in this sample, only 2 did not exhibit an exchange rate appreciation}$^{,}$\footnote{The countries which exhibited either an industrial production fall or a non-significant response are Norway and Iceland. Out of the 40 countries in this sample, only 2 did not exhibit an increase in industrial production.}$^{,}$\footnote{The only country which exhibited either an equity index fall or a non-significant response are Slovenia. Out of the 40 countries in this sample, only 1 did not exhibit an increase in the equity index.}  To stress this result, Figure \ref{fig:NER_MedianResponse} in Appendix \ref{sec:appendix_figures} pools the median impulse response function across countries and plots the interquartile range for both the MP and FIE components of monetary policy surprises, confirming the previous results. While there is heterogeneity in magnitude across countries, the vast majority of countries exhibit responses in line with those presented in Section \ref{sec:main_results}. This is, the MP and FIE components of monetary policy surprises lead to opposite international spillovers, highlighting the need to control for the ``Fed Information Effect'' to accurately estimate the impact of US interest rates. 

I carry out several additional country-by-country Bayesian SVAR exercises. Figures \ref{fig:CbC_REER_AE_REER} to \ref{fig:CbC_REER_EM_LT} in Appendix \ref{sec:appendix_figures} presents the results of carrying out the same exercise but for the sample that replaces the nominal exchange rate with the US dollar with the trade weighted multilateral real exchange rate. I present the results for the set of countries present for the full time sample 1988 to 2019, see Figures \ref{fig:MeanMedian_88_NER_Modify} and \ref{fig:MeanMedian_88_REER_Modify}; for the sample of 1998 to 2019, see Figures \ref{fig:MeanMedian_98_NER_Modify} and \ref{fig:MeanMedian_98_REER_Modify}; and for the sample 2008 to 2019, see Figures \ref{fig:MeanMedian_08_NER_Modify} and \ref{fig:MeanMedian_08_REER_Modify}. Across both time and variable sample specification, the resulting impulse response functions are in line with those presented in Section \ref{sec:main_results}. 

\subsection{Robustness Checks} \label{subsec:robustness_checks}

\noindent
\textbf{Alternative econometric specifications.} I start by showing that the main results in Section \ref{sec:main_results} are robust to econometric specifications different from Equation \ref{eq:LP_pooled}. In particular, I consider the following two alternative regressions to estimate the impulse response functions. 
\begin{align}
    y_{i,t+h} &= \beta^{MP}_{h} i^{\text{MP}}_t + \beta^{FIE}_{h} i^{\text{FIE}}_t + \nonumber \\
    & \qquad + \sum^{J_y}_{j=1} \delta^{j}_i y_{i,t-j} + \sum^{J_x}_{j=1} \alpha^{j}_i x_{i,t-j} + \sum^{J_i}_{j=1} \left( \phi^{j}_i i^{\text{MP}}_{t-j} + \varphi^{j}_i i^{\text{FIE}}_{t-j} \right) + \mu_1 \times t + \mu_2 \times t^{2} + \epsilon_{i,t} \label{eq:LP_Trend}  \\
    \nonumber \\
    y_{i,t+h} &= \beta^{MP}_{h} i^{\text{MP}}_t + \beta^{FIE}_{h} i^{\text{FIE}}_t + \nonumber \\
    & \qquad + \sum^{J_y}_{j=1} \delta^{j}_i y_{i,t-j} + \sum^{J_x}_{j=1} \alpha^{j}_i x_{i,t-j} + \sum^{J_i}_{j=1} \left( \phi^{j}_i i^{\text{MP}}_{t-j} + \varphi^{j}_i i^{\text{FIE}}_{t-j} \right) + \mu_1 \times t + \mu_2 + \phi_i \times t^{2} + \Gamma_i + \epsilon_{i,t} \label{eq:LP_Trend_FE}
\end{align}
The specification in Equation \ref{eq:LP_Trend} includes a linear and quadratic time trend, while Equation \ref{eq:LP_Trend_FE} adds country fixed effects.\footnote{Note that the specification given by Equation \ref{eq:LP_Trend_FE} introduces both lagged values of the independent variable and country fixed effects. As shown by \cite{nickell1981biases}, this could lead to biased estimates. However, I also estimate the impulse response functions using this specification as it was the method followed by \cite{ilzetzki2021puzzling}.} Figure \ref{fig:Trend_NER} and \ref{fig:TrendFE_NER} in Appendix \ref{sec:appendix_figures} present the results of estimating the impulse response functions following the specifications in Equations \ref{eq:LP_Trend} and \ref{eq:LP_Trend_FE}. From these figures it is clear that the results presented in Section \ref{sec:main_results} are robust to the alternative econometric specifications.\footnote{Furthermore, Figure \ref{fig:Trend_REER} and \ref{fig:TrendFE_REER} in Appendix \ref{sec:appendix_figures} show similar results for the sample that replaces the nominal exchange rate with the trade weighted real exchange rate.}

\noindent
\textbf{Panel SVAR econometric specification.} I carry out an additional econometric specification robustness check by estimating impulse response functions through a two panel SVAR models: (i) a pooled model; (ii) a mean-group model. Details on these models are found in Appendix \ref{sec:appendix_model_details}. Analogous to the country-by-country estimated in the previous section I introduce the MP and FIE components and the complete monetary policy surprise as exogenous variables.\footnote{This is a point of difference with previous versions of this paper. In previous versions of this paper the different components of monetary policy surprises where introduced as an additional variable in the SVAR for each country. To identify the impact of the different components of monetary policy I imposed an ordering assumption. However,  this approach relies on the assumption that structural shocks are uncorrelated across countries. This is not a plausible assumption under the study of the international spillovers of US interest rates.} Model details are left for Appendix \ref{sec:appendix_model_details}.

I begin by describing the results from the pooled panel SVAR model. I estimate the impulse response functions of the MP component, FIE component and following the Standard HFI approach for samples starting in January 1988, 1998 and 2008. Figures \ref{fig:Pool_88_NER} through \ref{fig:Pool_08_NER} in Appendix \ref{sec:appendix_figures} presents the results.\footnote{In addition, I estimate the impulse response functions for all three samples for the sample which replaces the nominal exchange rate with respect to the US dollar with the multilateral trade weighted real exchange rate, see Figures \ref{fig:Pool_88_REER} through \ref{fig:Pool_08_REER} in Appendix \ref{sec:appendix_figures}} In line with the country-by-country SVAR results, the panel SVAR impulse response functions show that the MP and FIE components of monetary policy surprises lead to opposite international spillovers. In consequence, the main results in Section \ref{sec:main_results} are robust to this alternative SVAR econometric specification.

Second, I describe the results from the mean-group panel SVAR model. I demean the variables and average them across countries for each moment in the data. I estimate a one unit SVAR, such as the one for each country in the previous exercise, on the averaged data. Once again, I estimate the impulse response functions of the MP component, FIE component and following the Standard HFI approach for samples starting in January 1988, 1998 and 2008. In line with the country-by-country SVAR results, and the pooled SVAR model, impulse response functions show that the MP and FIE components of monetary policy surprises lead to opposite international spillovers. I interpret these results as supporting evidence that the main results are robust to two alternative panel SVAR estimation methodologies. Moreover, all the results hold for the alternative sample with the trade weighted multilateral real exchange rate.\footnote{See Figures \ref{fig:Average_1988_NER} through \ref{fig:Average_2008_REER} in Appendix \ref{sec:appendix_figures} presents the results. }

\noindent
\textbf{Results including US variables in the specification.} The following econometric exercise augments the benchmark econometric specification in Equation \ref{eq:LP_pooled} with a set of US control variables. This alternative econometric specification becomes
\begin{align}
    y_{i,t+h} &= \beta^{MP}_{h} i^{\text{MP}}_t + \beta^{FIE}_{h}  i^{\text{FIE}}_t + \nonumber \\
    & \quad + \sum^{J_y}_{j=1} \delta^{j}_i y_{i,t-j} + \sum^{J_x}_{j=1} \alpha^{j}_i x_{i,t-j} + \sum^{J_i}_{j=1} \left( \phi^{j}_i i^{\text{MP}}_{t-j} + \varphi^{j}_i i^{\text{FIE}}_{t-j} \right) + \sum^{J_{US}}_{s=i} \alpha^{US}_i x_{US,t-j}+ \epsilon_{i,t} \label{eq:LP_pooled_USA}
\end{align}
where $x_{US,t-j}$ represents a vector of US control variables.\footnote{In particular, I include the ``Market Yield on U.S. Treasury Securities at 2-Year Constant Maturity, Quoted on an Investment Basis'', the real dollar index constructed by the Federal Reserve, the PCE price index, the industrial production index constructed by the Federal Reserve, the US' total export and imports in current US dollars and the excess bond premium (see \cite{gilchrist2012credit}). As in Section \ref{sec:main_results}, I include 2 lags of these US variables into the estimation.} Figure \ref{fig:US_BenchmarkNER} and \ref{fig:US_BenchmarkREER} in Appendix \ref{sec:appendix_figures} show that the main results in Section \ref{sec:main_results} are robust to this alternative econometric specification for both the benchmark sample and the sample with the trade weighted real exchange rate. Even more, impact of the MP and FIE components on the exchange rates and industrial production are slightly greater in terms of magnitude and in terms of statistical significance. 

\noindent
\textbf{Results across exchange rate regimes.} I show that the results in Section \ref{sec:main_results} are not driven by countries with different exchange rate regimes. To do so, I classify countries into four different groups based on \cite{ilzetzki2017country}'s coarse classification which goes from 1 or ``de facto peg'' to the US dollar, to 6 or ``Dual market in which parallel market data is missing''. Given that only 10\% of the observations are classified in groups 4 (``freely floating''), 5 (``freely falling'') or 6 (``dual market with parallel market is missing''), I combine countries under these classifications into one group. 

The first econometric exercise computes the impulse response functions for each of the four groups according the country's classification in each month. Figures \ref{fig:BenchmarkNER_ERA1} to \ref{fig:BenchmarkNER_ERA4} in Appendix \ref{sec:appendix_figures} present the results for each of the four groups, respectively. The second econometric exercise computes the impulse response functions according to the median coarse classification during the period of time each country is present in my benchmark sample. Figures \ref{fig:BenchmarkNER_Median_ERA1} to \ref{fig:BenchmarkNER_Median_ERA4} in Appendix \ref{sec:appendix_figures} present the results for each of the four groups, respectively. Across both econometric exercises and all country groups, a MP component leads to an exchange rate depreciation, higher long-term rates and a drop in industrial production and equity indexes. On the other hand, the FIE component leads to an exchange rate appreciation, no initial impact on long term rates, and an expansion of the industrial production and equity indexes. Following the ``Standard HFI'' approach leads to an expansion of the industrial production and the equity indexes for all four groups. Intuitively, the impact on nominal exchange rates is quantitatively lower for countries categorized in the ``de facto peg'' classification group. Overall, these results provide supporting evidence that the main results in Section \ref{sec:main_results} are present across countries with different exchange rate regimes.

\noindent
\textbf{Results across levels of openness of the current account.} Countries differ significantly in their openness in cross-border financial transactions, as shown by \cite{chinn2008new}. This heterogeneity might matter for the transmission of US interest rate shocks as there is indirect evidence that current account openness might be associated with a lower probability of financial crises.\footnote{For instance, see \cite{aizenman1999capital}.} Thus, it could be the case that the international spillovers of US interest rates differ systematically with an economy's degree of current account openness. 

To test this potential heterogeneity of the international spillovers of US interest rates I classify countries according to the current account openness index constructed by \cite{chinn2008new}.\footnote{An updated index can be sourced from the authors' website.} I compute the mean index across the country's sample and then group countries into four quartiles. I compute the international spillovers of US interest rates by estimating Equation \ref{eq:LP_pooled} for each group separately. Figures \ref{fig:BenchmarkNER_KA1} to \ref{fig:BenchmarkNER_KA4} in Appendix \ref{sec:appendix_figures} present the impulse response functions for the four groups of countries. The impulse response functions across all four groups are qualitatively inline with those presented in Section \ref{sec:main_results}, presenting completely opposite spillovers for the MP and FIE components. Thus, the main results in Section \ref{sec:main_results} are not driven by a country's degree of current account openness.

\noindent
\textbf{Results across levels of GDP per capita.} I expand on the results between AEs and EMEs in Figure \ref{fig:Figure_3} by estimating the impulse response functions across countries' level of GDP per capita. I group countries into GDP per capital quartiles using data from the year 2004.\footnote{Data is sourced from the World Bank's dataset.} The year 2004 is picked as its the midpoint of the full time sample. There is significant differences in GDP per capita across the four groups, with the average GDP per capita in constant 2015 USD: \$7848, \$15 907, \$27 697 and \$40 106. The study of the international spillovers of US interest rates across these four groups is of interest given this significant heterogeneity in income levels. Figures \ref{fig:BenchmarkNER_GDPpc_1} to \ref{fig:BenchmarkNER_GDPpc_4} in Appendix \ref{sec:appendix_figures} present the results for the four groups. Results are in line with the benchmark results in Figure \ref{fig:Figure_1} and those in Figures \ref{fig:Figure_3_1} and \ref{fig:Figure_3_2}.\footnote{Additionally, Figures \ref{fig:BenchmarkREER_GDPpc_1} to \ref{fig:BenchmarkREER_GDPpc_4} in Appendix \ref{sec:appendix_data_details}, carries out the same exercise for the sample that replaces the nominal exchange rate with respect to the US with the multilateral trade weighted real exchange rate. Results do not vary significantly.}

\noindent
\textbf{Results across trade openness.} I show that results are robust across countries with different degrees of trade openness. \cite{camara2024international} argues that the main transmission channel of US monetary policy operates through international trade. Furthermore, there is evidence that countries which are relatively more open to trade experience less frequent and less severe Sudden Stop crises, see \cite{cavallo2008does}. Consequently, it could be the case that the main results in Section \ref{sec:main_results} are driven by countries which are relatively more exposed to international trade flows.

To test this hypothesis I classify countries according to their total trade to GDP ratio.\footnote{Data on the total trade to GDP ratio are sourced from the World Bank.} I compute the mean total trade to GDP ratio across the country's sample and then group countries into four quartiles. There is significant variation across countries' total trade to GDP. The average ratio across the four groups are: 39.6\%, 59.8\%, 79\% and 140.3\%, respectively. I compute the international spillovers of US interest rates by estimating Equation \ref{eq:LP_pooled} for each group separately. Figures \ref{fig:BenchmarkNER_TT_1} to \ref{fig:BenchmarkNER_TT_4} in Appendix \ref{sec:appendix_figures} present the impulse response functions for the four groups of countries. The impulse response functions across all four groups are qualitatively inline with those presented in Section \ref{sec:main_results}, presenting completely opposite spillovers for the MP and FIE components.\footnote{Additionally, Figures \ref{fig:BenchmarkNER_TT_1} to \ref{fig:BenchmarkNER_TT_1} in Appendix \ref{sec:appendix_data_details}, carries out the same exercise for the sample that replaces the nominal exchange rate with respect to the US with the multilateral trade weighted real exchange rate. Results do not vary significantly.}

\noindent
\textbf{Results using \cite{ilzetzki2021puzzling} dataset.} Next, I test whether my results are driven by the choice of my specific benchmark sample. To do so, I estimate the international spillovers of US interest rates on the data set used by \cite{ilzetzki2021puzzling} which found the recent atypical dynamics.\footnote{The dataset can be downloaded from the authors' website.}  I estimate the impulse response functions by estimating Equations \ref{eq:LP_pooled}, \ref{eq:LP_Trend} and \ref{eq:LP_Trend_FE}. Note that this last specification is the empirical specification used in \cite{ilzetzki2021puzzling}.

Figure \ref{fig:Results_1} to \ref{fig:Results_1_08} present the results of estimating Equation \ref{eq:LP_pooled} for the full sample, for the sample starting in January 1998 and the sample starting in January 2008. Across the three figures, a US interest rate hike caused by the MP component leads to an exchange rate depreciation and a drop in industrial production. On the contrary, a US interest rate hike caused by the FIE component leads to an exchange rate appreciation and expansion of the industrial production index. In other words, a monetary policy surprise cleansed of the ``Fed Information Effect'' leads to the conventional results. Similarly, the component of monetary policy surprises explained by the ``Fed Information Effect'' leads to the recent atypical results presented by \cite{ilzetzki2021puzzling}. Figures \ref{fig:Results_Trend} to \ref{fig:Results_Trend_08} present the results for estimating Equation \ref{eq:LP_Trend} for the three time samples. Figures \ref{fig:Results_Trend_FE} to \ref{fig:Results_Trend_FE_08} present the results for estimating Equation \ref{eq:LP_Trend_FE} for the three time samples. I take these results as evidence that my main empirical findings are robust across time samples and econometric specifications.

\noindent
\textbf{Results across continents.} For the last geographical robustness check discussed in this paper, I present results estimated across different continents In particular, I classify countries into Africa, Asia, Europe, Latin America, North America and Oceania. Figures \ref{fig:BenchmarkNER_AF} through \ref{fig:BenchmarkREER_OC} in Appendix \ref{sec:appendix_figures} present the results across all the continents for both the benchmark sample and the sample which replaces the nominal exchange rate with the trade weighted real exchange rate. Across the continent sub-samples in response to a MP component, the exchange rate depreciates and the industrial production and equity indexes fall; whilst in response to a FIE component the exchange rate appreciates and the industrial production and equity indexes increase persistently. Following the ``Standard HFI'' approach across the continent sub-samples leads to a muted response of the exchange rate and an expansion in the industrial production index. I take these results as evidence that the main results presented in Figure \ref{sec:main_results} are present and robust across different country samples, with only minor quantitative differences, and not driven by any outlier. 

\noindent
\textbf{Results using EMBI spreads.} I show that results are robust to replacing the long interest rate used in the benchmark specification with the EMBI spreads. These spreads have been widely used in the economic literature as a measure of an economy's risk of default and as a proxy of the interest rate its economic agents would face in global financial markets.\footnote{For example, see \cite{garcia2006role}, \cite{uribe2006country} and \cite{ozatay2009emerging}.} Thus, I estimate the benchmark specification in Equation \ref{eq:LP_pooled} replacing the long-term interest rate with the EMBI spread.

Figure \ref{fig:EMBI_BenchmarkNER} to \ref{fig:EMBI_Regressions_08NER} present the results for the full sample, for the sample starting in January 1998 and the sample starting in January 2008. Overall, results are in line with those presented in Section \ref{sec:main_results}. On the one hand, the MP component of monetary policy surprises lead to a persistent increase in EMBI spreads across all samples. On the other hand, the FIE component of EMBI spreads leads to a persistent and significant drop in EMBI spreads. Following the Standard HFI approach leads to a smaller and less significant increase in EMBI spreads. As an additional robustness check, Figures \ref{fig:EMBI_BenchmarkREER} to \ref{fig:EMBI_Regressions_08REER} present the results by also replacing the nominal exchange rate with the trade-weighted real exchange rate. All results hold. In summary, this robustness check highlights the fact that to accurately measure the impact of US monetary policy on EMEs' financial conditions it is crucial to control for the ``Fed Information Effect''.

\noindent
\textbf{Additional robustness checks presented in the Appendix.} Appendix \ref{sec:appendix_figures} presents additional results using alternative data transformations, see Figures \ref{fig:Levels_BenchmarkNER} to \ref{fig:Levels_BenchmarkREER_08}; dropping the few countries which leave and re-enter the sample, see Figures \ref{fig:Exiters_BenchmarkNER} to \ref{fig:Exiters_BenchmarkREER_08}. I also show that results are robust to alternative choices of lags on the control variables, such as setting $J_y = J_x = 12$, (see Figures \ref{fig:Regressions_NER_deplag12} to \ref{fig:Regressions_08REER_deplag12}); and setting the number of lags on the components of monetary policy surprises to zero, $J_i = 0$, see Figures \ref{fig:Regressions_NER_shock0lag} to \ref{fig:Regressions_08REER_shock0lag}).

In summary, this section carried out and presented robustness checks showing that the main results in Section \ref{sec:main_results} are robust to alternative econometric specifications, alternative identification strategies that control for the ``Fed Information Effect'' and across different time and country samples. I interpret this results as supporting evidence that controlling for the ``Fed Information Effect'' is crucially to accurately estimate the international spillovers of US interest rates. Furthermore, following the ``Standard HFI'' approach leads to atypical dynamics such as an expansion of industrial production in response to an increase in US interest rates. 

\section{Conclusion} \label{sec:conclusion}

In this paper I argue that the international spillovers of a US interest rates hike depend critically on the underlying shock that causes this tightening. Using the identification strategy proposed by \cite{bauer2023alternative}, I deconstruct monetary policy surprises into a pure monetary policy shock component (MP) and a ``Fed Information Effect'' component (FIE). Introducing these two components of monetary policy surprises into a panel of 50 countries, both Advanced and Emerging Market economies, I find that the two components of monetary policy surprises lead to qualitatively opposite spillovers. On the one hand, a MP shock component leads to a nominal exchange rate depreciation, a persistent drop in industrial output and overall tighter financial conditions. On the other hand, a FIE component leads to an exchange rate appreciation, a persistent expansion in industrial output and overall looser financial conditions.

Furthermore, I argue that following the standard high-frequency identification of US monetary policy leads to atypical spillover dynamics as it does not control for the systematic disclosure of information about the state of the US economy around FOMC announcements. I show that this identification strategy leads to impulse response functions which are an average of those resulting from the MP and FIE components. Under this identification strategy, a US interest rate tightening leads to a significant expansion of the industrial production and equity index. Moreover, by not controlling for the ``Fed Information Effects'', this identification strategy introduces an attenuates bias in terms of the quantitative impact US monetary policy shocks on the variables which do not show atypical dynamics.

To conclude, I show that the main results are robust to a battery of different model and sample specifications. Results hold when considering Advanced and Emerging Market economies separately, when estimating impulse response functions using alternative local projection methodologies and SVAR models, alternative variable selection, across countries' level of GDP per capita, exchange rate regime, current account openness, countries' geographic location, and even when using the sample of other papers who found the recent atypical dynamics. Most interestingly, results are robust to four alternative identification strategies that control for the ``Fed Information Effect'', although with alternative interpretations and methodologies. This result highlights the importance of controlling for the ``Fed Information Effect'' to accurately identify the international spillovers of US interest rates.

\newpage

\large
\noindent
\textbf{Acknowledgements:} \normalsize

\noindent
I owe an unsustainable debt of gratitude to Giorgio Primiceri for his guidance and advice. I would also like to thank the comments by two anonymous referees, Martin Eichenbaum, Matt Rognlie, Guido Lorenzoni, Larry Christiano, Husnu Dalgic, Fabian Lange, Fernando Saltiel, Francisco Ruge-Murcia, Francesco Amodio, Rui Castro and Markus Poschke. I would like to thank the Monetary and Fiscal History of Latin America project of the Becker Friedman Institute for their generous financial support. Particularly, I would like to thank Edward R. Allen, for supporting my research.  I would like to thank the attendees of Northwestern University's macro lunch seminar, Javier Garcia Cicco, and Yong Cai. Susan Belles provided helpful comments.

\newpage
\bibliography{main.bib}

\newpage
\appendix

\section{Data Details} \label{sec:appendix_data_details}

In this Appendix, I present additional details on the construction of the different datasets used across the paper. 

\subsection{Benchmark Sample} \label{subsec:appendix_data_details_benchmark}

First, I describe in greater detail the countries in the benchmark sample. To group countries into Advanced Economies and Emerging Market Economies I use the IMF's classification in the year 2019. Tables \ref{tab:AE_Sample} and \ref{tab:EME_Sample} describe the sample of 28 Advanced Economies (29 if adding the Euro Area unit), and the 22 Emerging Market economies, respectively. 
\begin{table}[ht]
    \centering
    \footnotesize
    \begin{tabular}{c|c|c|c|c|c}
\textbf{Entry Date}	&	\textbf{Country}	&	\textbf{Continent}	&	\textbf{Exit Date}	&	\textbf{Re-Enters}	&	\textbf{\# Obs.}	\\ \hline \hline
2001m1	&	Australia	&	OC	&	2019m12	&	No	&	228	\\
1990m1	&	Austria	&	EU	&	2019m12	&	No	&	360	\\
1988m1	&	Belgium	&	EU	&	2019m12	&	No	&	384	\\
1988m1	&	Canada	&	NA	&	2019m12	&	No	&	384	\\
1988m1	&	Denmark	&	EU	&	2019m12	&	No	&	384	\\
1999m1	&	Euro Area	&	EU	&	2019m12	&	No	&	291	\\
1988m1	&	Finland	&	EU	&	2019m12	&	No	&	383	\\
1988m1	&	France	&	EU	&	2019m12	&	No	&	384	\\
1988m1	&	Germany	&	EU	&	2019m12	&	No	&	384	\\
1997m6	&	Greece	&	EU	&	2015m6	&	Yes	&	270	\\
1998m1	&	Iceland	&	EU	&	2018m2	&	No	&	242	\\
1988m1	&	Ireland	&	EU	&	2019m12	&	No	&	384	\\
1997m1	&	Israel	&	AS	&	2019m12	&	No	&	276	\\
1991m3	&	Italy	&	EU	&	2019m12	&	No	&	346	\\
1989m1	&	Japan	&	AS	&	2019m12	&	No	&	372	\\
2001m1	&	Latvia	&	EU	&	for NER 2013m12	&	No	&	228	\\
2001m1	&	Lithuania	&	EU	&	for NER 2014m12	&	No	&	196	\\
1999m1	&	Luxembourg	&	EU	&	2019m12	&	No	&	252	\\
1988m1	&	Norway	&	EU	&	2019m12	&	No	&	384	\\
1994m10	&	Poland	&	EU	&	2019m12	&	No	&	303	\\
1993m7	&	Portugal	&	EU	&	2019m12	&	No	&	318	\\
2000m9	&	Slovak Rep.	&	EU	&	for NER 2008M12	&	No	&	232	\\
2002m3	&	Slovenia	&	EU	&	for NER 2007m2	&	No	&	214	\\
2000m10	&	South Korea	&	AS	&	2019m12	&	No	&	231	\\
1988m1	&	Spain	&	EU	&	2019m12	&	No	&	384	\\
1988m1	&	Sweden	&	EU	&	2019m12	&	No	&	384	\\
2010m10	&	Switzerland	&	EU	&	2019m12	&	No	&	111	\\
1988m1	&	The Netherlands	&	EU	&	2019m12	&	No	&	384	\\
1988m1	&	United Kingdom	&	EU	&	2019m12	&	No	&	384	\\ \hline \hline
    \end{tabular}
    \caption{Emerging Market Economies in the Benchmark Sample}
    \label{tab:AE_Sample}
\end{table}
\begin{table}[ht]
    \centering
    \footnotesize
    \begin{tabular}{c|c|c|c|c|c}
\textbf{Entry Date}	&	\textbf{Country}	&	\textbf{Continent}	&	\textbf{Exit Date}	&	\textbf{Leaves \& Re-Enters} & \# Obs. \\ \hline \hline
1994m4	&	Brazil	&	LA	&	2019m12	&	No	&	309	\\
2003m1	&	Bulgaria	&	EU	&	2019m12	&	No	&	204	\\
1999m5	&	Chile	&	LA	&	2019m12	&	No	&	248	\\
2010m1	&	China	&	AS	&	2019m12	&	No	&	120	\\
1997m2	&	Colombia	&	LA	&	2019m12	&	No	&	275	\\
2012m7	&	Costa Rica	&	LA	&	2019m12	&	No	&	90	\\
1998m1	&	Croatia	&	EU	&	2019m12	&	Yes	&	200	\\
1999m1	&	Hungary	&	EU	&	2019m12	&	No	&	252	\\
2012m10	&	India	&	AS	&	2019m12	&	No	&	87	\\
2004m5	&	Indonesia	&	AS	&	2019m12	&	No	&	188	\\
1996m10	&	Malaysia	&	AS	&	2019m3	&	No	&	270	\\
1993m12	&	Mexico	&	LA	&	2018m2	&	No	&	291	\\
2012m5	&	Mongolia	&	AS	&	2019m12	&	No	&	92	\\
2001m6	&	Pakistan	&	AS	&	2017m4	&	Yes	&	178	\\
1997m3	&	Peru	&	LA	&	2018m4	&	No	&	254	\\
1993m12	&	Philippines	&	AS	&	2019m12	&	No	&	313	\\
2005m4	&	Romania	&	EU	&	2019m12	&	No	&	177	\\
1999m1	&	Russian Federation	&	EU	&	2018m6	&	No	&	234	\\
1990m1	&	South Africa	&	AF	&	2019m12	&	No	&	360	\\
1996m6	&	Turkiye	&	EU	&	2019m12	&	No	&	283	\\
2002m1	&	Ukraine	&	EU	&	2019m12	&	No	&	216	\\
2007m12	&	Venezuela	&	LA	&	2016m12	&	No	&	109	\\ \hline \hline
    \end{tabular}
    \caption{Emerging Market Economies in the Benchmark Sample}
    \label{tab:EME_Sample}
\end{table}
Estonia satisfies the condition of reporting data on the five variables macroeconomic and financial variables described in Section \ref{sec:data_methodology_identification}. However, it is excluded from the sample as it satisfies this condition for less than 3 years. 

In terms of continent decomposition, there is one country from Africa (AF), 10 countries from Asia (AS), 30 countries from Europe (EU), 7 countries from Latin America (LA), one country from North America and one country from Oceania (OC). In Section \ref{subsec:robustness_checks} show that the benchmark results are robust across the different continents. 

\noindent
\textbf{Data set sources.} The benchmark dataset is constructed sourcing data from the OECD and IMF-IFS datasets. I establish a priority across the two datasets, with the OECD preferred with respect to the IMF-IFS dataset. This is IMF-IFS dataset systematically has lower coverage within each country for those countries that are present in both datasets. For instance, for the case of Chile data on industrial production, equity index and the CPI are jointly present in the IMF-IFS data set starts in 1991m5 while in the OECD dataset, the data starts in 1990m1. Thus, if the country is present in either only in the OECD dataset or in both datasets, I use data from the OECD dataset.

The OECD data set produces the variable ``Growth Industrial Production'' which produces an index of industrial production at the monthly frequency. The IMF-IFS dataset produces three potential measures of industrial production:
       \begin{itemize}
        \item Economic Activity, Industrial Production, Index
        
        \item Economic Activity, Industrial Production, Seasonally Adjusted, Index
        
        \item Economic Activity, Industrial Production, Manufacturing, Index
    \end{itemize}
    Ideally, I would  construct the variable ``Industrial Production Index'' by choosing only one of the variables mentioned above. However, this is impossible as countries do not report to the IMF all three of these variables for the benchmark time sample. For instance, Peru provides neither the ``Economic Activity, Industrial Production, Index'' nor the ``Economic Activity, Industrial Production, Seasonally Adjusted, Index'', but does provide the ``Economic Activity, Industrial Production, Manufacturing, Index''. Visiting Peru's Central Bank statistics website, there is no ``Industrial Production Index'', but there is an ``Industrial Production, Manufacturing Index'', which coincides with the variable reported as ``Economic Activity, Industrial Production, Manufacturing, Index'' to the IMF. In order to deal with this, I establish the following priority between the three IMF IFS variables: (i) ``Economic Activity, Industrial Production, Seasonally Adjusted, Index'' (ii) ``Economic Activity, Industrial Production, Index'', (iii) ``Economic Activity, Industrial Production, Manufacturing, Index''.

Similarly, the OECD produces one measure of equity index at the monthly frequency. However, the IMF-IFS dataset produces two measures and coverage across countries is heterogeneous. In order to construct countries' ``Equity Index'' I rely on two variables of the IMF IFS' dataset:
    \begin{itemize}
        \item Monetary and Financial Accounts, Financial Market Prices, Equities, Index
        
        \item Monetary and Financial Accounts, Financial Market Prices, Equities, End of Period, Index
    \end{itemize}
    
    I establish a priority: (i) ``Monetary and Financial Accounts, Financial Market Prices, Equities, Index'' (ii) ``Monetary and Financial Accounts, Financial Market Prices, Equities, End of Period, Index''.

The IMF-IFS dataset produces CPI data for all countries except Australia is constructed using the variable ``Prices, Consumer Price Index, All items, Index'' from IMF IFS data set. Australia does not report a monthly CPI series to the IMF-IFS data set. Furthermore, the Australian Bureau of Statistics provides only quarterly data on their consumer price index.\footnote{See \url{https://www.abs.gov.au/statistics/economy/price-indexes-and-inflation/consumer-price-index-australia/jun-2022}.} Thus, for the case of Australia I proxy the monthly consumer price index by using the ``Prices, Producer Price Index, All Commodities, Index''. Once again, given that the paper's main results are robust to the different exercises that partition the sample, I believe that using this proxy variable does not guide any of the of the results presented in the paper. 

Lastly, the benchmark variable nominal exchange rate is sourced completely from the IMF-IFS dataset. The variable's full name at the IMF IFS data set is ``Exchange Rates, National Currency Per U.S. Dollar, Period Average, Rate''. 

\subsection{Additional Samples} \label{subsec:appendix_data_details_additional}

Next, I turn to describing the additional samples used across the paper, mainly in Section \ref{sec:robustness_checks_additional_results}. I present the countries and their entry date into the sample. 

\begin{table}[ht]
    \centering
    \footnotesize
    \caption{Sample of Countries - Construction Index}
    \label{tab:appendix_sample_construction}
    \begin{tabular}{c|c|c|c|c}
Entry Date	&	AEs	&	&	Entry Date	&	EMEs	\\ \hline \hline
2001m1	&	Australia	&	&	1994m4	&	Brazil	\\
1996m1	&	Austria	&	&	2003m1	&	Bulgaria	\\
1988m1	&	Belgium	&	&	1999m5	&	Chile	\\
1988m1	&	Canada	&	&	2012m7	&	Costa Rica	\\
2000m1	&	Denmark	&	&	2000m1	&	Croatia, Rep. of	\\
1999m1	&	Euro Area	&	&	1999m1	&	Hungary	\\
1995m1	&	Finland	&	&	2010m1	&	Malaysia	\\
1988m1	&	France	&	&	1993m12	&	Mexico	\\
1988m1	&	Germany	&	&	2003m1	&	Peru	\\
1998m1	&	Iceland	&	&	2005m4	&	Romania	\\
1995m1	&	Italy	&	&	2010m1	&	Russian Federation	\\
1993m1	&	Japan	&	&	1993m1	&	South Africa	\\
2000m10	&	Korea, Rep. of	&	&	2005m1	&	Türkiye, Rep of	\\
1993m10	&	Luxembourg	&	&		&		\\
2000m1	&	Netherlands, The	&	&		&		\\
1994m10	&	Poland, Rep. of	&	&		&		\\
2000m1	&	Portugal	&	&		&		\\
2000m9	&	Slovak Rep.	&	&		&		\\
2002m3	&	Slovenia, Rep. of	&	&		&		\\
1988m1	&	Spain	&	&		&		\\
1994m1	&	Sweden	&	&		&		\\
2010m1	&	United Kingdom	&	&		&		\\
    \end{tabular}
\end{table}

\newpage
\begin{table}[ht]
    \centering
    \footnotesize
    \caption{Sample of Countries - Retail Index}
    \label{tab:appendix_sample_retail}
    \begin{tabular}{c|c|c|c|c}
Entry Date	&	AEs	&	&	Entry Date	&	EMEs	\\ \hline \hline
1990m1	&	Austria	&	&	2000m1	&	Brazil	\\
1988m1	&	Belgium	&	&	2003m1	&	Bulgaria	\\
1991m1	&	Canada	&	&	2005m1	&	Chile	\\
1988m1	&	Denmark	&	&	2013m1	&	Colombia	\\
1999m1	&	Euro Area	&	&	2012m7	&	Costa Rica	\\
1988m1	&	Finland	&	&	2000m1	&	Croatia, Rep. of	\\
1988m1	&	France	&	&	1999m1	&	Hungary	\\
1988m1	&	Germany	&	&	1993m12	&	Mexico	\\
1997m6	&	Greece	&	&	2005m4	&	Romania	\\
1998m1	&	Iceland	&	&	1999m1	&	Russian Federation	\\
1988m1	&	Ireland	&	&	1990m1	&	South Africa	\\
1997m1	&	Israel	&	&	2010m1	&	Türkiye, Rep of	\\
1991m3	&	Italy	&	&		&		\\
1989m1	&	Japan	&	&		&		\\
2000m10	&	Korea, Rep. of	&	&		&		\\
2001m1	&	Latvia	&	&		&		\\
2001m1	&	Lithuania	&	&		&		\\
1995m1	&	Luxembourg	&	&		&		\\
1994m1	&	Netherlands, The	&	&		&		\\
1988m1	&	Norway	&	&		&		\\
1994m10	&	Poland, Rep. of	&	&		&		\\
1993m7	&	Portugal	&	&		&		\\
2000m9	&	Slovak Rep.	&	&		&		\\
2002m3	&	Slovenia, Rep. of	&	&		&		\\
1995m1	&	Spain	&	&		&		\\
1988m1	&	Sweden	&	&		&		\\
2010m10	&	Switzerland	&	&		&		\\
1988m1	&	United Kingdom	&	&		&		\\
    \end{tabular}
\end{table}

\newpage
\begin{table}[ht]
    \centering
    \footnotesize
    \caption{Sample of Countries - Mining Index}
    \label{tab:appendix_sample_mining}
    \begin{tabular}{c|c|c|c|c}
Entry Date	&	AEs	&	&	Entry Date	&	EMEs	\\ \hline \hline
1996m1	&	Austria	&	&	2003m1	&	Bulgaria	\\
2000m1	&	Belgium	&	&	1999m5	&	Chile	\\
2000m1	&	Denmark	&	&	2012m7	&	Costa Rica	\\
2000m1	&	Euro Area	&	&	1998m1	&	Croatia, Rep. of	\\
1995m1	&	Finland	&	&	2000m1	&	Hungary	\\
1990m1	&	France	&	&	2012m10	&	India	\\
1991m1	&	Germany	&	&	2010m1	&	Malaysia	\\
2000m1	&	Greece	&	&	1997m12	&	Mexico	\\
1991m3	&	Italy	&	&	2012m5	&	Mongolia	\\
1998m1	&	Japan	&	&	2003m1	&	Peru	\\
2000m10	&	Korea, Rep. of	&	&	2005m4	&	Romania	\\
2001m1	&	Latvia	&	&	2000m1	&	Russian Federation	\\
2001m1	&	Lithuania	&	&	1990m1	&	South Africa	\\
2000m1	&	Luxembourg	&	&	1996m6	&	Türkiye, Rep of	\\
1990m1	&	Netherlands, The	&	&		&		\\
1995m1	&	Norway	&	&		&		\\
2000m1	&	Poland, Rep. of	&	&		&		\\
2000m1	&	Portugal	&	&		&		\\
2000m9	&	Slovak Rep.	&	&		&		\\
2002m3	&	Slovenia, Rep. of	&	&		&		\\
1988m1	&	Spain	&	&		&		\\
1988m1	&	Sweden	&	&		&		\\
1997m1	&	United Kingdom	&	&		&		\\
    \end{tabular}
\end{table}

\newpage
\begin{table}[ht]
    \centering
    \footnotesize
    \caption{Sample of Countries - Commodity Prices}
    \label{tab:appendix_sample_PComm}
    \begin{tabular}{c|c|c|c|c}
Entry Date	&	AEs	&	&	Entry Date	&	EMEs	\\ \hline \hline
2001m1	&	Australia	&	&	1994m4	&	Brazil	\\
1990m1	&	Austria	&	&	2003m1	&	Bulgaria	\\
1999m1	&	Belgium	&	&	1999m5	&	Chile	\\
1988m1	&	Canada	&	&	2010m1	&	China, P.R.: Mainland	\\
1988m1	&	Denmark	&	&	1997m2	&	Colombia	\\
1988m1	&	Finland	&	&	2012m7	&	Costa Rica	\\
1988m1	&	France	&	&	1999m1	&	Hungary	\\
1988m1	&	Germany	&	&	2012m10	&	India	\\
1997m6	&	Greece	&	&	2004m5	&	Indonesia	\\
1998m1	&	Iceland	&	&	1996m10	&	Malaysia	\\
1988m1	&	Ireland	&	&	1993m12	&	Mexico	\\
1997m1	&	Israel	&	&	2012m5	&	Mongolia	\\
1991m3	&	Italy	&	&	2001m6	&	Pakistan	\\
1989m1	&	Japan	&	&	1997m3	&	Peru	\\
2001m1	&	Latvia	&	&	1993m12	&	Philippines	\\
2001m1	&	Lithuania	&	&	2005m4	&	Romania	\\
1999m1	&	Luxembourg	&	&	1999m1	&	Russian Federation	\\
1988m1	&	Norway	&	&	1990m1	&	South Africa	\\
1993m7	&	Portugal	&	&	2002m1	&	Ukraine	\\
1988m1	&	Spain	&	&		&		\\
1988m1	&	Sweden	&	&		&		\\
2010m10	&	Switzerland	&	&		&		\\
1988m1	&	United Kingdom	&	&		&		\\
    \end{tabular}
\end{table}

\newpage
\begin{table}[ht]
    \centering
    \footnotesize
    \caption{Sample of Countries - Exports \& Imports Values}
    \label{tab:appendix_sample_Exports}
    \begin{tabular}{c|c|c|c|c}
Entry Date	&	AEs	&	&	Entry Date	&	EMEs	\\ \hline \hline
2006m1	&	Austria	&	&	2006m1	&	Brazil	\\
2006m1	&	Belgium	&	&	2006m1	&	Bulgaria	\\
2006m1	&	Canada	&	&	2006m1	&	Chile	\\
2006m1	&	Denmark	&	&	2006m1	&	Colombia	\\
2006m1	&	Finland	&	&	2012m7	&	Costa Rica	\\
2006m1	&	France	&	&	2009m11	&	Croatia, Rep. of	\\
2006m1	&	Germany	&	&	2006m1	&	Hungary	\\
2006m1	&	Greece	&	&	2012m10	&	India	\\
2006m1	&	Iceland	&	&	2006m1	&	Indonesia	\\
2006m1	&	Ireland	&	&	2006m1	&	Malaysia	\\
2006m1	&	Israel	&	&	2013m1	&	Mongolia	\\
2006m1	&	Italy	&	&	2006m1	&	Peru	\\
2006m1	&	Japan	&	&	2006m1	&	Philippines	\\
2006m1	&	Latvia	&	&	2006m1	&	South Africa	\\
2006m1	&	Luxembourg	&	&	2006m1	&	Türkiye, Rep of	\\
2006m1	&	Norway	&	&	2006m1	&	Ukraine	\\
2006m1	&	Poland, Rep. of	&	&		&		\\
2006m1	&	Portugal	&	&		&		\\
2002m3	&	Slovenia, Rep. of	&	&		&		\\
2010m10	&	Switzerland	&	&		&		\\
    \end{tabular}
\end{table}

\newpage
\begin{table}[ht]
    \centering
    \footnotesize
    \caption{Sample of Countries - Lending Rates}
    \label{tab:appendix_sample_LR}
    \begin{tabular}{c|c|c|c|c}
Entry Date	&	AEs	&	&	Entry Date	&	EMEs	\\ \hline \hline
1997m1	&	Brazil	&	&	2001m1	&	Australia	\\
2003m1	&	Bulgaria	&	&	1988m1	&	Canada	\\
1999m5	&	Chile	&	&	1998m1	&	Iceland	\\
2010m1	&	China, P.R.: Mainland	&	&	2013m4	&	Israel	\\
1997m2	&	Colombia	&	&	1991m3	&	Italy	\\
2012m7	&	Costa Rica	&	&	1993m4	&	Japan	\\
1998m1	&	Croatia, Rep. of	&	&	2000m10	&	Korea, Rep. of	\\
1999m1	&	Hungary	&	&	1999m1	&	Netherlands, The	\\
2012m10	&	India	&	&	2013m12	&	Norway	\\
2004m5	&	Indonesia	&	&	1992m12	&	Sweden	\\
1996m10	&	Malaysia	&	&	2010m10	&	Switzerland	\\
1993m12	&	Mexico	&	&	1988m1	&	United Kingdom	\\
2012m5	&	Mongolia	&	&		&		\\
2004m4	&	Pakistan	&	&		&		\\
1997m3	&	Peru	&	&		&		\\
1993m12	&	Philippines	&	&		&		\\
2005m4	&	Romania	&	&		&		\\
1999m1	&	Russian Federation	&	&		&		\\
1990m1	&	South Africa	&	&		&		\\
2002m1	&	Ukraine	&	&		&		\\
2007m12	&	Venezuela, Rep. Bolivariana de	&	&		&		\\
    \end{tabular}
\end{table}

\newpage
\begin{table}[ht]
    \centering
    \footnotesize
    \caption{Sample of Countries - Monetary Policy Rate}
    \label{tab:appendix_sample_MP}
    \begin{tabular}{c|c|c|c|c}
Entry Date	&	AEs	&	&	Entry Date	&	EMEs	\\ \hline \hline
2001m1	&	Australia	&	&	1999m4	&	Brazil	\\
1999m1	&	Austria	&	&	2005m2	&	Bulgaria	\\
1999m1	&	Belgium	&	&	1999m5	&	Chile	\\
1992m12	&	Canada	&	&	2016m2	&	China, P.R.: Mainland	\\
1988m1	&	Denmark	&	&	1997m2	&	Colombia	\\
1999m1	&	Euro Area	&	&	2012m7	&	Costa Rica	\\
1999m1	&	Finland	&	&	1999m1	&	Hungary	\\
1999m1	&	France	&	&	2012m10	&	India	\\
1999m1	&	Germany	&	&	2005m7	&	Indonesia	\\
1999m1	&	Greece	&	&	2001m12	&	Malaysia	\\
2001m12	&	Iceland	&	&	2001m12	&	Mexico	\\
1999m1	&	Ireland	&	&	2012m5	&	Mongolia	\\
1997m1	&	Israel	&	&	2003m9	&	Peru	\\
1999m1	&	Italy	&	&	2001m12	&	Philippines	\\
2006m3	&	Japan	&	&	2005m4	&	Romania	\\
2000m10	&	Korea, Rep. of	&	&	2001m12	&	Russian Federation	\\
2001m1	&	Latvia	&	&	1998m3	&	South Africa	\\
2001m1	&	Lithuania	&	&	1996m6	&	Türkiye, Rep of	\\
1999m1	&	Luxembourg	&	&		&		\\
1999m1	&	Netherlands, The	&	&		&		\\
1988m1	&	Norway	&	&		&		\\
1998m1	&	Poland, Rep. of	&	&		&		\\
1999m1	&	Portugal	&	&		&		\\
2000m9	&	Slovak Rep.	&	&		&		\\
2002m3	&	Slovenia, Rep. of	&	&		&		\\
1999m1	&	Spain	&	&		&		\\
2002m7	&	Sweden	&	&		&		\\
2010m10	&	Switzerland	&	&		&		\\
1988m1	&	United Kingdom	&	&		&		\\
    \end{tabular}
\end{table}

\begin{table}[ht]
    \centering
    \footnotesize
    \caption{Sample of Countries - PPI}
    \label{tab:appendix_sample_PPI}
    \begin{tabular}{c|c|c|c|c}
Entry Date	&	AEs	&	&	Entry Date	&	EMEs	\\ \hline \hline
2001m1	&	Australia	&	&	1994m8	&	Brazil	\\
1990m1	&	Austria	&	&	2003m1	&	Bulgaria	\\
1988m1	&	Belgium	&	&	2009m1	&	Chile	\\
1988m1	&	Canada	&	&	2010m1	&	China, P.R.: Mainland	\\
1988m1	&	Denmark	&	&	1997m2	&	Colombia	\\
1999m1	&	Euro Area	&	&	2012m7	&	Costa Rica	\\
1988m1	&	Finland	&	&	1998m1	&	Croatia, Rep. of	\\
1995m1	&	France	&	&	1999m1	&	Hungary	\\
1988m1	&	Germany	&	&	2012m10	&	India	\\
1997m6	&	Greece	&	&	1996m10	&	Malaysia	\\
2006m1	&	Iceland	&	&	1993m12	&	Mexico	\\
1988m1	&	Ireland	&	&	1993m12	&	Philippines	\\
1997m1	&	Israel	&	&	2005m4	&	Romania	\\
1991m3	&	Italy	&	&	1999m1	&	Russian Federation	\\
1989m1	&	Japan	&	&	1990m1	&	South Africa	\\
2000m10	&	Korea, Rep. of	&	&	1996m6	&	Türkiye, Rep of	\\
2001m1	&	Latvia	&	&	2002m1	&	Ukraine	\\
2001m1	&	Lithuania	&	&		&		\\
1993m10	&	Luxembourg	&	&		&		\\
1988m1	&	Netherlands, The	&	&		&		\\
1988m1	&	Norway	&	&		&		\\
1994m10	&	Poland, Rep. of	&	&		&		\\
2000m1	&	Portugal	&	&		&		\\
2000m9	&	Slovak Rep.	&	&		&		\\
2002m3	&	Slovenia, Rep. of	&	&		&		\\
1988m1	&	Spain	&	&		&		\\
1988m1	&	Sweden	&	&		&		\\
2010m10	&	Switzerland	&	&		&		\\
1988m1	&	United Kingdom	&	&		&		\\
    \end{tabular}
\end{table}

\begin{table}[ht]
    \centering
    \footnotesize
    \caption{Sample of Countries - Unemployment Rate}
    \label{tab:appendix_sample_URATE}
    \begin{tabular}{c|c|c|c|c}
Entry Date	&	AEs	&	&	Entry Date	&	EMEs	\\ \hline \hline
2001m1	&	Australia	&	&	2012m3	&	Brazil	\\
1994m1	&	Austria	&	&	2003m1	&	Bulgaria	\\
1988m1	&	Belgium	&	&	2004m1	&	Chile	\\
1988m1	&	Canada	&	&	2019m1	&	China, P.R.: Mainland	\\
1988m1	&	Denmark	&	&	2001m7	&	Colombia	\\
1988m1	&	Finland	&	&	2012m7	&	Costa Rica	\\
2003m1	&	France	&	&	2000m1	&	Croatia, Rep. of	\\
2007m1	&	Germany	&	&	1999m1	&	Hungary	\\
1998m4	&	Greece	&	&	2009m1	&	Malaysia	\\
2003m1	&	Iceland	&	&	2001m4	&	Peru	\\
1988m1	&	Ireland	&	&	2005m4	&	Romania	\\
1991m3	&	Italy	&	&	2010m1	&	Russian Federation	\\
1989m1	&	Japan	&	&	2005m1	&	Türkiye, Rep of	\\
2000m10	&	Korea, Rep. of	&	&		&		\\
2001m1	&	Latvia	&	&		&		\\
2001m1	&	Lithuania	&	&		&		\\
1993m10	&	Luxembourg	&	&		&		\\
1988m1	&	Netherlands, The	&	&		&		\\
1989m1	&	Norway	&	&		&		\\
1997m1	&	Poland, Rep. of	&	&		&		\\
1998m2	&	Portugal	&	&		&		\\
2000m9	&	Slovak Rep.	&	&		&		\\
2002m3	&	Slovenia, Rep. of	&	&		&		\\
1988m1	&	Spain	&	&		&		\\
1988m1	&	Sweden	&	&		&		\\
2010m10	&	Switzerland	&	&		&		\\
1988m1	&	United Kingdom	&	&		&		\\
    \end{tabular}
\end{table}

\newpage
\section{Model Details} \label{sec:appendix_model_details}

In this section of the Appendix I describe the details of the Bayesian SVAR models used to produce the results in Section \ref{sec:robustness_checks_additional_results}.

First, I describe the Bayesian SVAR model used to estimate the country-by-country SVAR model. In compact form, a general VAR model with $n$ endogenous variables, $p$ lags, and $m$ exogenous variables can be written as
\begin{align} \label{eq:VAR_CompactForm}
    y_t = A_1 y_{t-1} + \ldots + A_p y_{t-p} + C x_t + \epsilon_t
\end{align}
where $y_t = \left(y_{1,t}, \ldots, y_{n,t}\right)$ is a $n\times 1$ vector of endogenous data, $A_1,\ldots,A_p$ are matrices of dimension $n \times n$, $C$ is a $n\times m$  matrix, $x_t$ is a vector of exogenous regressors, $\epsilon_t = \left(\epsilon_{1,t},\ldots,\epsilon_{n,t}\right)$ is a vector of residuals following a multivariate normal distribution:
\begin{align*}
    \epsilon_t \sim \mathcal{N} \left(0, \Sigma\right)
\end{align*}
which satisfies the standard assumptions in the literature. The model is estimated imposing a Minnesota Prior as proposed by \cite{litterman1986forecasting}. All variables are assumed to follow a unit root random walk. Additionally, for the case of exogenous variables, the most reasonable strategy is to assume that they have no impact on endogenous variables, and thus that their coefficients are equal to zero.

The country-by-country models estimated are composed of 11 variables: the five country specific variables in the benchmark sample plus 6 US specific variables: (i)  two year treasury rate, (ii) the Federal Reserve's real trade weighted dollar index, (iii) the PCE price index, (iv) the dollar value of US exports, (v) the dollar value of US imports, (vi) the excess bond premium by \cite{gilchrist2012credit}. As argued in Section \ref{sec:robustness_checks_additional_results}, introducing US variables is common in the literature and allows for more precise estimates of the impact of US monetary policy shocks. Thus, the number of endogenous variables is set equal to $n=11$. The number of lags is $p=12$. In order to estimate the impact of the MP component, the FIE component and when following the Standard HFI, I introduce each of these variables one at a time as an exogenous variable $x_t$. This implies that the number of exogenous variables in any specification of the model is always $m=1$. The model is estimated using the BEAR toolbox from the ECB. See \cite{dieppe2016bear}.

Second, I describe the Bayesian pooled panel SVAR model estimated to measure the impact of US monetary policy for several panels of countries at a time. In its most general form, a panel SVAR model comprises of N countries or units, $n$ endogenous variables, $p$ lagged values and $T$ time periods. The pooled panel SVAR model can be written as
\begin{align} \label{eq:pooled_estimator}
\begin{pmatrix}
y_{1,t} \\
y_{2,t} \\
\vdots  \\
y_{N,t}
\end{pmatrix}
&=
\begin{pmatrix}
A^1 \quad 0 \quad \cdots \quad 0 \\
0 \quad  A^1 \quad \cdots \quad 0 \\
\vdots \quad \vdots \quad \ddots \quad \vdots \\
0 \quad 0 \quad \cdots \quad A^1 
\end{pmatrix}
\begin{pmatrix}
y_{1,t-1} \\
y_{2,t-1} \\
\vdots  \\
y_{N,t-1}
\end{pmatrix}
+ \cdots \nonumber \\
\\
&+ \nonumber
\begin{pmatrix}
A^p \quad 0 \quad \cdots \quad 0 \\
0 \quad  A^p \quad \cdots \quad 0 \\
\vdots \quad \vdots \quad \ddots \quad \vdots \\
0 \quad 0 \quad \cdots \quad A^p 
\end{pmatrix}
\begin{pmatrix}
y_{1,t-p} \\
y_{2,t-p} \\
\vdots  \\
y_{N,t-p}
\end{pmatrix}
+
\begin{pmatrix}
C \\
C \\
\vdots \\
C
\end{pmatrix}
x_t + 
\begin{pmatrix}
\epsilon_{1,t} \\
\epsilon_{2,t} \\
\vdots \\
\epsilon_{N,t}
\end{pmatrix}
\end{align}
where $y_{i,t}$ denotes an $n \times 1$ vector of $n$ endogenous variables of country $i$ at time $t$ and $A^{j}$ is an $n \times n$ matrix of coefficients providing the response of country $i$ to the $j^{th}$ lag at period $t$. Note that by assuming that $A^j_1 = A^j_n = A^j$ for $j=1,\ldots,n$ implies the assumption that the estimated coefficients are common across countries. $x_t$ is a vector $m \times 1$ of exogenous variables, $C$ is a $Nn\times1$ vector of constant terms which are also assumed to be common across countries. Lastly, $\epsilon_{i,t}$ is an $n \times 1$ vector of residuals for the variables of country $i$, such that
\begin{align*}
    \epsilon_{i,t} \sim \mathcal{N}\left(0,\Sigma_{ii,t}\right)
\end{align*}
with 
\begin{align*}
    \epsilon_{ii,t} &= \mathbb{E} \left(\epsilon_{i,t} \epsilon_{i,t}' \right) = \Sigma_c \quad \forall i \\
    \epsilon_{ij,t} &= \mathbb{E} \left(\epsilon_{i,t} \epsilon_{j,t}' \right) = 0 \quad \text{for } i \neq j
\end{align*}
The last two equations imply that, as for the model's auto-regressive coefficients, the innovation's variance is equal across countries. 

The model described by Equation \ref{eq:pooled_estimator} is estimated using Bayesian methods. In order to carry out the estimation of this model I first re-write the model. In particular, the model can be reformulated in compact form as

\begin{align} \label{eq:model_compact_form}
\underbrace{\begin{pmatrix}
y_{1,t}' \\
y_{2,t}' \\
\vdots  \\
y_{N,t}'
\end{pmatrix}}_{Y_t, \quad N \times n}
&=
\underbrace{\begin{pmatrix}
y_{1,t-1}' \ldots y_{1,t-p}',  x_t'\\
y_{2,t-1}' \ldots y_{2,t-p}', x_t'\\
\vdots  \ddots \vdots \\
y_{N,t-1}' \ldots y_{N,t-p},  x_t'
\end{pmatrix}}_{\mathcal{X}_t, \quad N \times \left( np + m\right)}
\underbrace{\begin{pmatrix}
\left(A^{1}\right)' \\
\left(A^{2}\right)' \\
\vdots \\
\left(A^{N}\right)' \\
C 
\end{pmatrix}}_{B, \quad \left( np + m\right) \times n}
+
\underbrace{\begin{pmatrix}
\epsilon_{1,t}' \\
\epsilon_{2,t}' \\
\vdots \\
\epsilon_{N,t}'
\end{pmatrix}}_{\mathcal{E}_t, \quad N \times n}
\end{align}
or
\begin{align}
    Y_t = X_t \mathcal{B} + \mathcal{E}_t
\end{align}
Even more, the model can be written in vectorised form by stacking over the $T$ time periods 
\begin{align}
    \underbrace{vec\left(Y\right)}_{NnT \times 1} = \underbrace{\left(I_n \otimes X \right)}_{NnT \times n np} \quad \underbrace{vec\left(\mathcal{B}\right)}_{n np \times 1} \quad + \quad \underbrace{vec\left(\mathcal{E}\right)}_{NnT \times 1}
\end{align}
or
\begin{align}
    y = \Bar{X} \beta + \epsilon
\end{align}
where $\epsilon \sim \mathcal{N}\left(0, \Bar{\Sigma}\right)$, with $\Bar{\Sigma} = \Sigma_c \otimes I_{NT}$. Note, I do not rely in the matrix $\Sigma_c$ for the identification of the US monetary policy shocks. As it was the case for the country-by-country SVAR model, the components of monetary policy surprises are introduced as exogenous variables in the vector $x_t$. Once again, I introduce the MP, FIE and the Standard HFI approach one at a time. In consequence, $m=1$ for every panel SVAR model estimated. 

The model described above is just a conventional VAR model. Furthermore, a single SVAR model is estimated for the whole set of units. Thus, the traditional Normal-Wishart identification strategy is carried out to estimate it. The likelihood function is given by
\begin{align}
    f\left(y | \Bar{X} \right) \propto |\Bar{\Sigma}|^{-\frac{1}{2}} \exp \left(-\frac{1}{2} \left(y - \Bar{X}\beta\right)' \Bar{\Sigma}^{-1} \left(y - \Bar{X}\beta\right) \right)
\end{align}
As for the Normal-Wishart, the prior of $\beta$ is assumed to be multivariate normal and the prior for $\Sigma_c$ is inverse Wishart. For further details, see \cite{dieppe2016bear}. All of the panel SVAR model computations are carried out using the BEAR Toolbox version 5.1. In particular, once estimated the model, and in particular $C$, impulse response functions can be directly computed.

\newpage
\section{Additional Results \& Robustness Check Figures} \label{sec:appendix_figures}

In this section of the Appendix I present the figures described and discussed in Sections \ref{sec:main_results} and \ref{sec:robustness_checks_additional_results}. 

\subsection{Additional Result Figures}
\newpage
\begin{figure}[ht]
    \centering
    \includegraphics[scale=0.4]{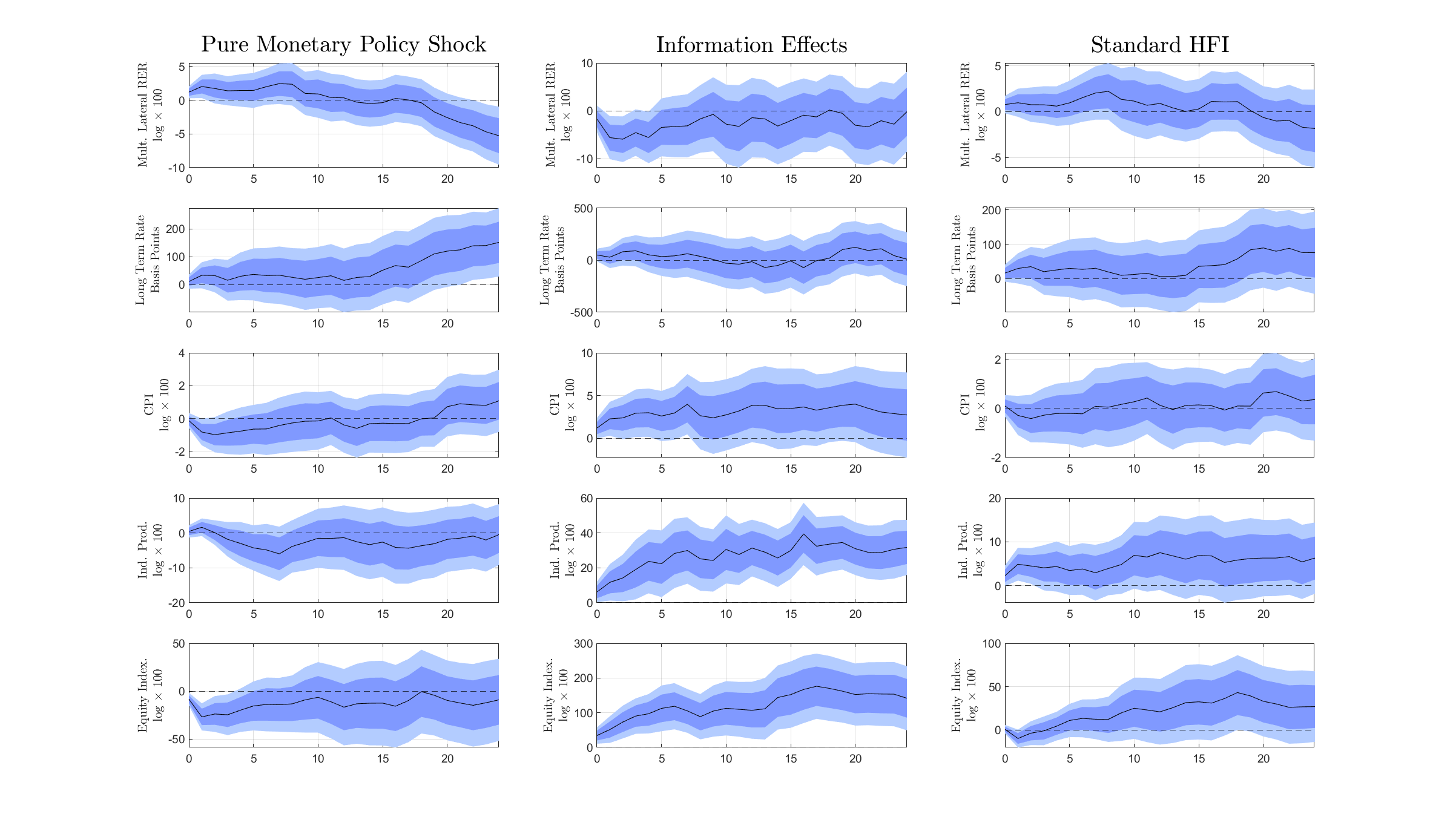}
    \caption{Impulse Response Functions - Advanced Economies \\ Multilateral Real Exchange Rate Specification}
    \label{fig:AEs_REER}
    \floatfoot{\textbf{Note:} The figure is comprised of 15 sub-figures ordered in three columns and five rows. The impulse response functions are computed for the sample of Advanced Economies alone. The left column relates to the estimates of $\beta^{MP}$ in Equation \ref{eq:LP_pooled}, the middle column relates to the estimate of $\beta^{FIE}$ in Equation \ref{eq:LP_pooled}, while the right column relates to estimating Equation \ref{eq:LP_pooled}, replacing the MP and FIE components with the un-orthogonalized monetary policy surprise. The rows represent the impact on (i) the multilateral trade weighted real exchange rate index (in logs times 100); (ii) long term interest rates in basis points; (iii) the consumer price index (in logs times 100); (iv) the industrial production index (in logs times 100); (v) the equity index (in logs times 100). The solid black line represents the point estimate, the dark blue area represents the 68\% confidence interval, and the light blue area represents the 90\% confidence interval. In the text, when referring to Panel $(i,j)$, $i$ refers to the row and $j$ to the column of the figure. Each variable, in its own transformation, is demeaned at the country level. }
\end{figure}

\newpage
\begin{figure}[ht]
    \centering
    \includegraphics[scale=0.4]{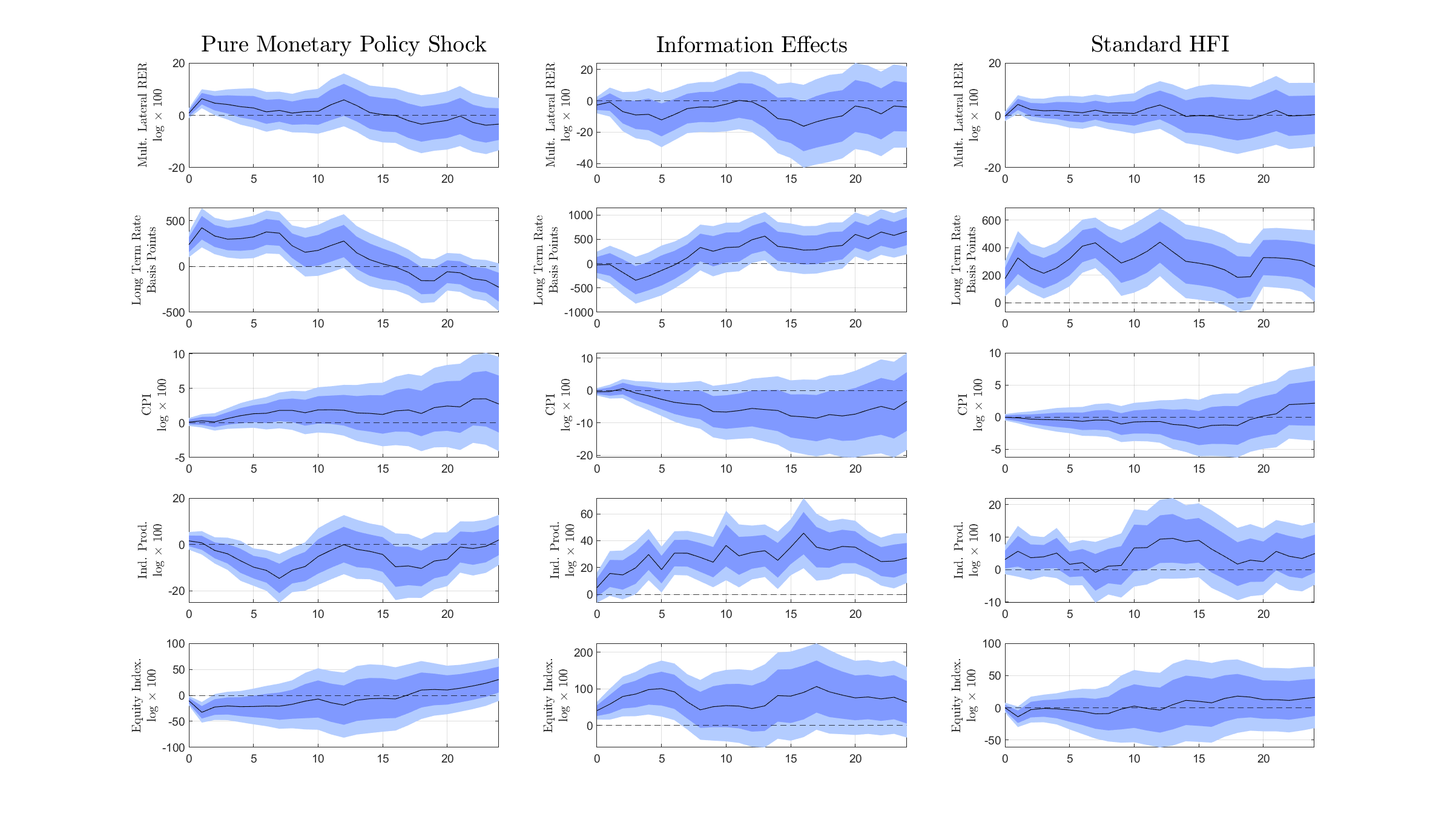}
    \caption{Impulse Response Functions - Emerging Market Economies \\ Multilateral Real Exchange Rate Specification}
    \label{fig:EMs_REER}
    \floatfoot{\textbf{Note:} The figure is comprised of 15 sub-figures ordered in three columns and five rows. The impulse response functions are computed for the sample of Advanced Economies alone. The left column relates to the estimates of $\beta^{MP}$ in Equation \ref{eq:LP_pooled}, the middle column relates to the estimate of $\beta^{FIE}$ in Equation \ref{eq:LP_pooled}, while the right column relates to estimating Equation \ref{eq:LP_pooled}, replacing the MP and FIE components with the un-orthogonalized monetary policy surprise. The rows represent the impact on (i) the multilateral trade weighted real exchange rate index (in logs times 100); (ii) long term interest rates in basis points; (iii) the consumer price index (in logs times 100); (iv) the industrial production index (in logs times 100); (v) the equity index (in logs times 100). The solid black line represents the point estimate, the dark blue area represents the 68\% confidence interval, and the light blue area represents the 90\% confidence interval. In the text, when referring to Panel $(i,j)$, $i$ refers to the row and $j$ to the column of the figure. Each variable, in its own transformation, is demeaned at the country level. }
\end{figure}

\newpage
\begin{landscape} 
\begin{figure}[ht]
    \centering
    \caption{Impulse Response Functions - Advanced Economies \\ \footnotesize Samples starting in January 1998 \& January 2008}
    \label{fig:Figure_AE_Across_Time}
     \centering
     \begin{subfigure}[b]{0.495\textwidth}
         \centering
         \includegraphics[width=\textwidth,height=9.5cm]{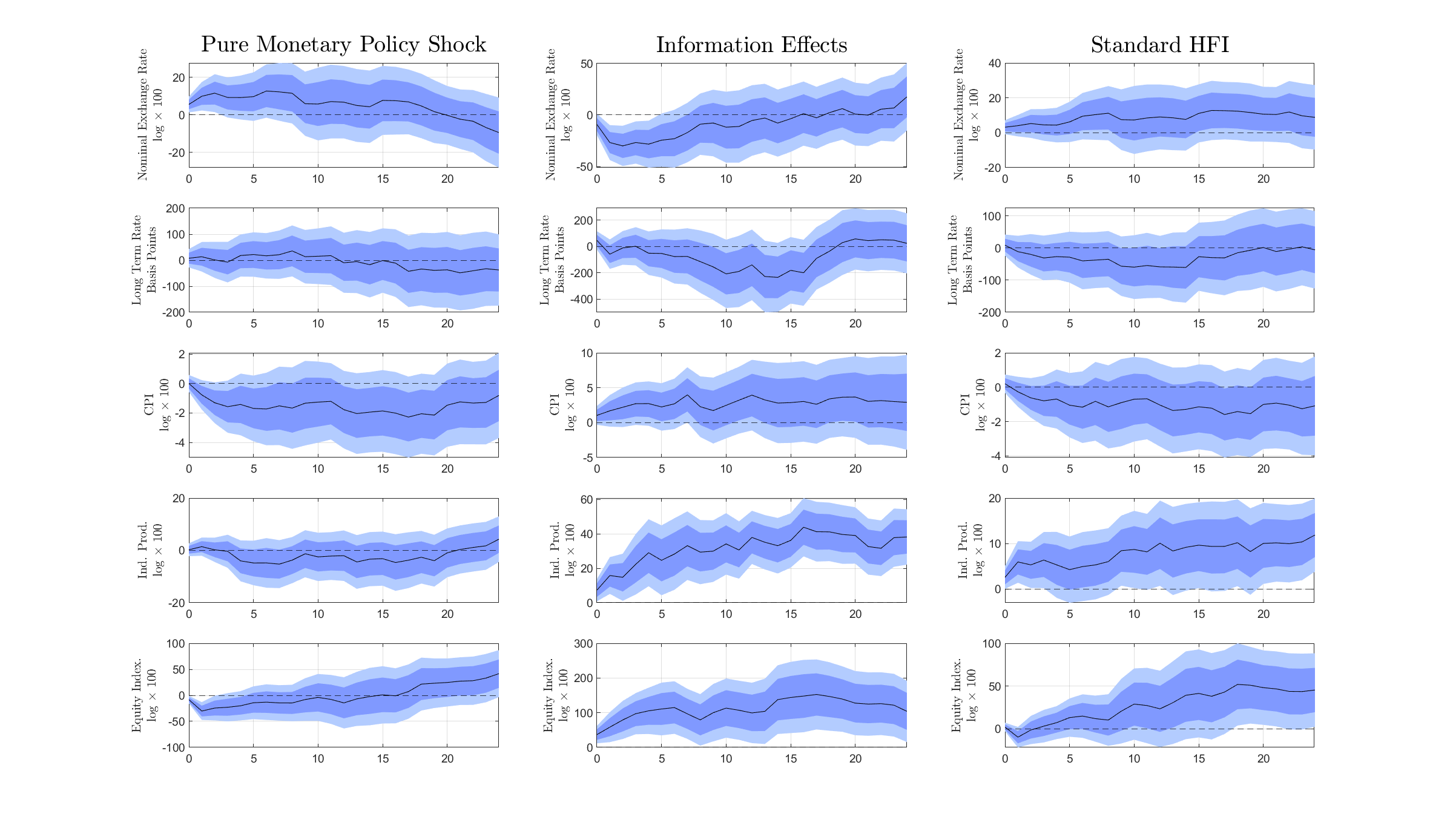}
         \caption{Sample January 1998 - December 2019}
         \label{fig:Figure_Appendix_1_1}
     \end{subfigure}
     \hfill
     \begin{subfigure}[b]{0.495\textwidth}
         \centering
         \includegraphics[width=\textwidth,height=9.5cm]{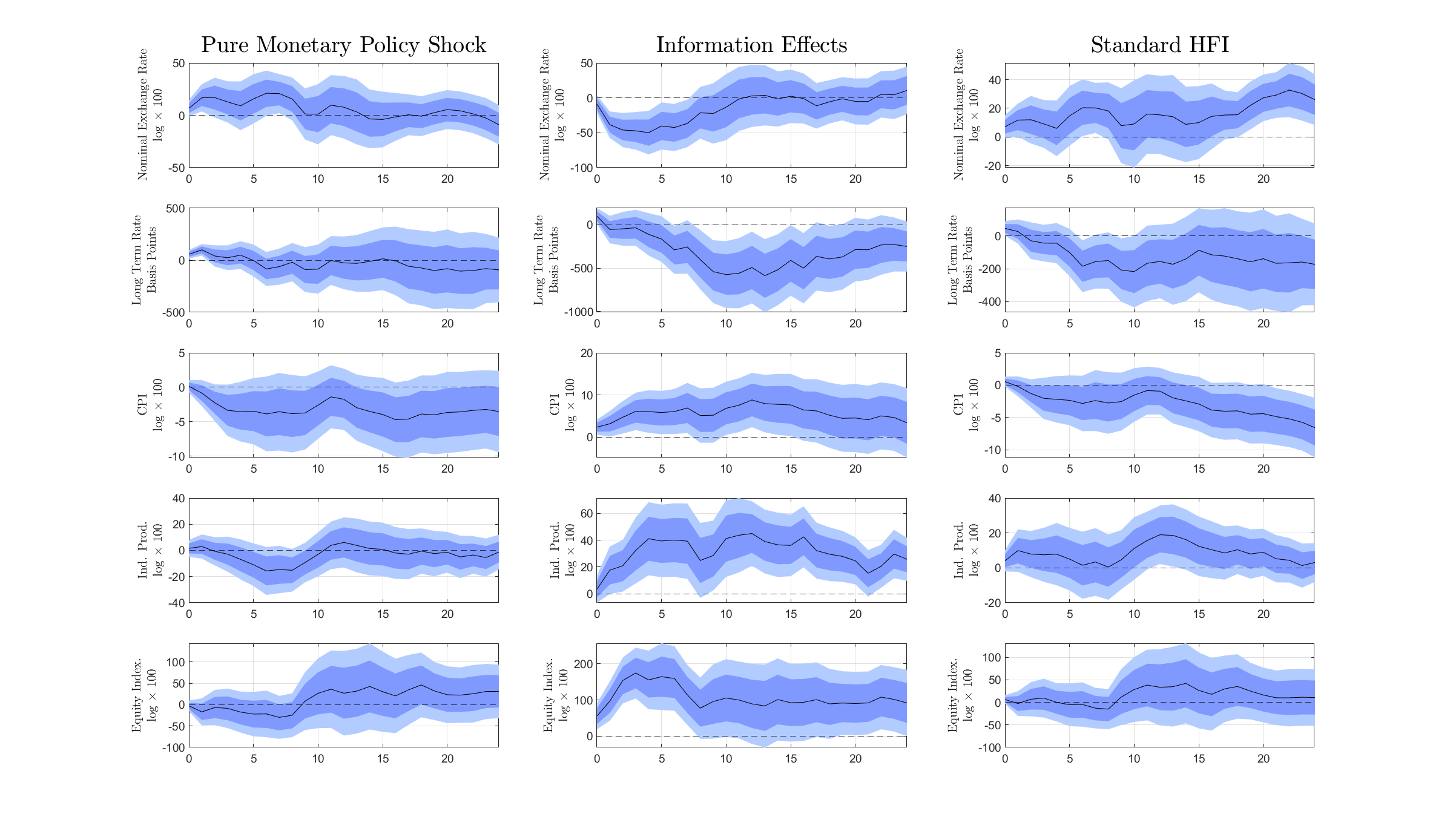}
         \caption{Sample January 2008 - December 2019}
         \label{fig:Figure_Appendix_1_2}
     \end{subfigure} 
     \floatfoot{\scriptsize \textbf{Note:} Each of the two figures are comprised of 15 sub-figures ordered in three columns and five rows. Both are constructed for the sample of Advanced Economies alone. The first figure to the left is constructed by estimating Equation \ref{eq:LP_pooled} for the sample starting in January 1998 and ending in December 2019. The second figure to the right is constructed by estimating Equation \ref{eq:LP_pooled} for the sample starting in January 2008 and ending in December 2019. The left column relates to the estimates of $\beta^{MP}$ in Equation \ref{eq:LP_pooled}, the middle column relates to the estimate of $\beta^{FIE}$ in Equation \ref{eq:LP_pooled}, while the right column relates to estimating Equation \ref{eq:LP_pooled}, replacing the MP and FIE components with the un-orthogonalized monetary policy surprise. The rows represent the impact on (i) the nominal exchange rate with respect to the US dollar (in logs times 100); (ii) long term interest rates in basis points; (iii) the consumer price index (in logs times 100); (iv) the industrial production index (in logs times 100); (v) the equity index (in logs times 100). The solid black line represents the point estimate, the dark blue area represents the 68\% confidence interval, and the light blue area represents the 90\% confidence interval. In the text, when referring to Panel $(i,j)$, $i$ refers to the row and $j$ to the column of the figure. Each variable, in its own transformation, is demeaned at the country level. }
\end{figure}
\end{landscape}

\newpage
\begin{landscape} 
\begin{figure}[ht]
    \centering
    \caption{Impulse Response Functions - Advanced Economies \\ \footnotesize Samples starting in January 1998 \& January 2008}
    \label{fig:Figure_EM_Across_Time}
     \centering
     \begin{subfigure}[b]{0.495\textwidth}
         \centering
         \includegraphics[width=\textwidth,height=9.5cm]{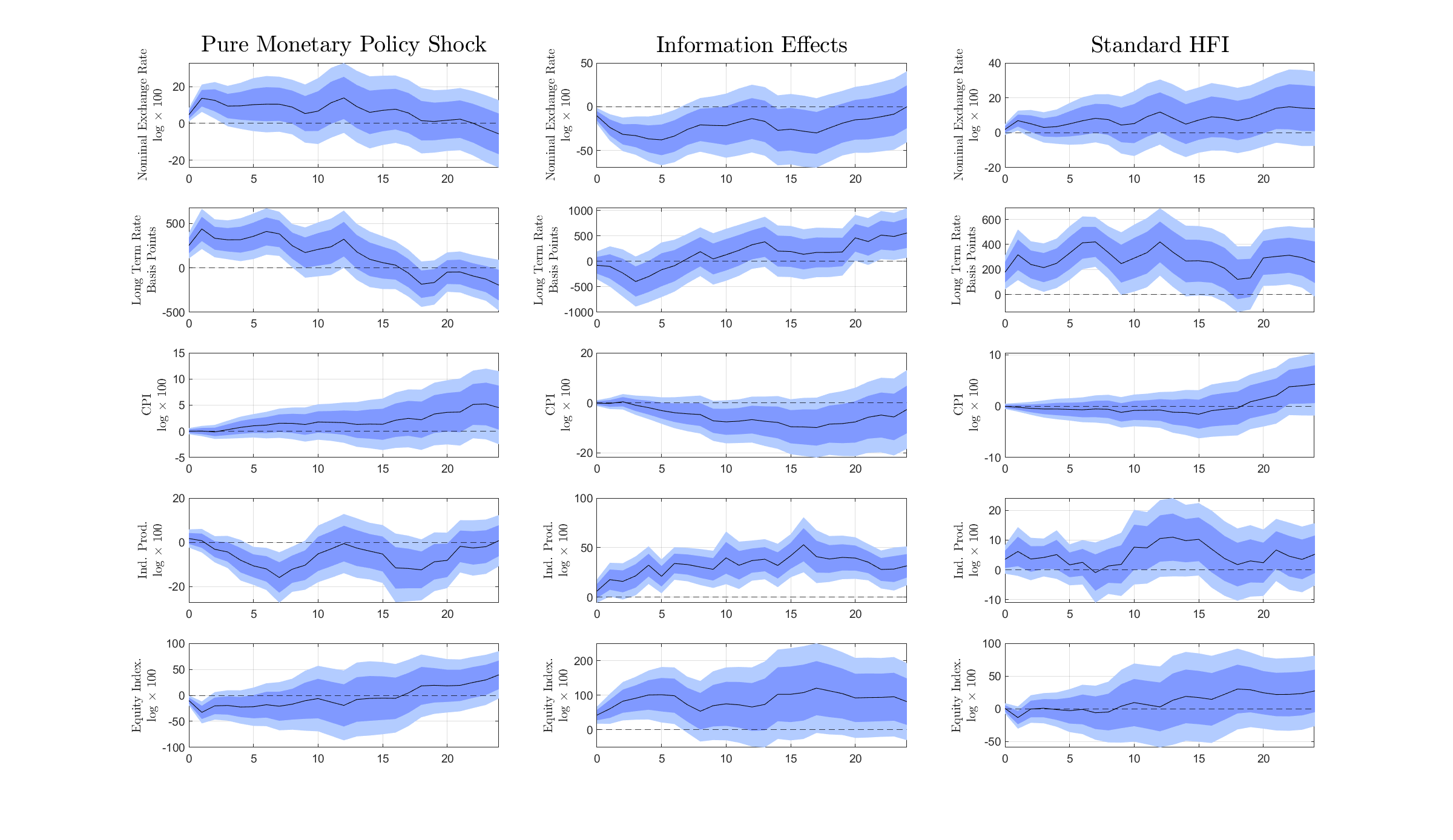}
         \caption{Sample January 1998 - December 2019}
         \label{fig:EMs_NER_98}
     \end{subfigure}
     \hfill
     \begin{subfigure}[b]{0.495\textwidth}
         \centering
         \includegraphics[width=\textwidth,height=9.5cm]{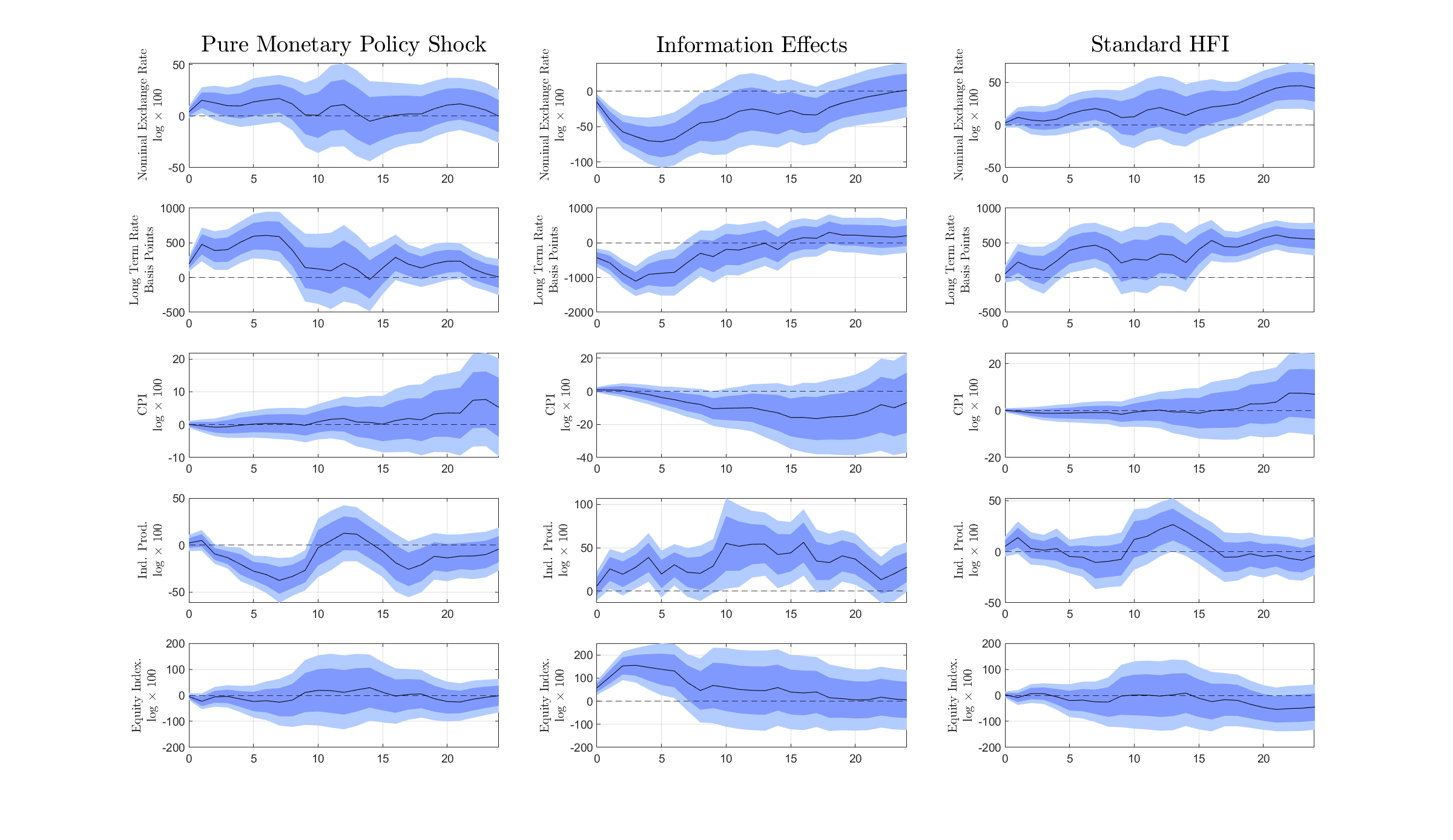}
         \caption{Sample January 2008 - December 2019}
         \label{fig:EMs_NER_08}
     \end{subfigure} 
     \floatfoot{\scriptsize \textbf{Note:} Each of the two figures are comprised of 15 sub-figures ordered in three columns and five rows. Both are constructed for the sample of Advanced Economies alone. The first figure to the left is constructed by estimating Equation \ref{eq:LP_pooled} for the sample starting in January 1998 and ending in December 2019. The second figure to the right is constructed by estimating Equation \ref{eq:LP_pooled} for the sample starting in January 2008 and ending in December 2019. The left column relates to the estimates of $\beta^{MP}$ in Equation \ref{eq:LP_pooled}, the middle column relates to the estimate of $\beta^{FIE}$ in Equation \ref{eq:LP_pooled}, while the right column relates to estimating Equation \ref{eq:LP_pooled}, replacing the MP and FIE components with the un-orthogonalized monetary policy surprise. The rows represent the impact on (i) (i) the multilateral trade weighted real exchange rate index (in logs times 100); (ii) long term interest rates in basis points; (iii) the consumer price index (in logs times 100); (iv) the industrial production index (in logs times 100); (v) the equity index (in logs times 100). The solid black line represents the point estimate, the dark blue area represents the 68\% confidence interval, and the light blue area represents the 90\% confidence interval. In the text, when referring to Panel $(i,j)$, $i$ refers to the row and $j$ to the column of the figure. Each variable, in its own transformation, is demeaned at the country level. }
\end{figure}
\end{landscape}

\newpage
\begin{landscape} 
\begin{figure}[ht]
    \centering
    \caption{Impulse Response Functions - Advanced Economies \\ \footnotesize Samples starting in January 1998 \& January 2008}
    \label{fig:Figure_AE_Across_Time_REER}
     \centering
     \begin{subfigure}[b]{0.495\textwidth}
         \centering
         \includegraphics[width=\textwidth,height=9.5cm]{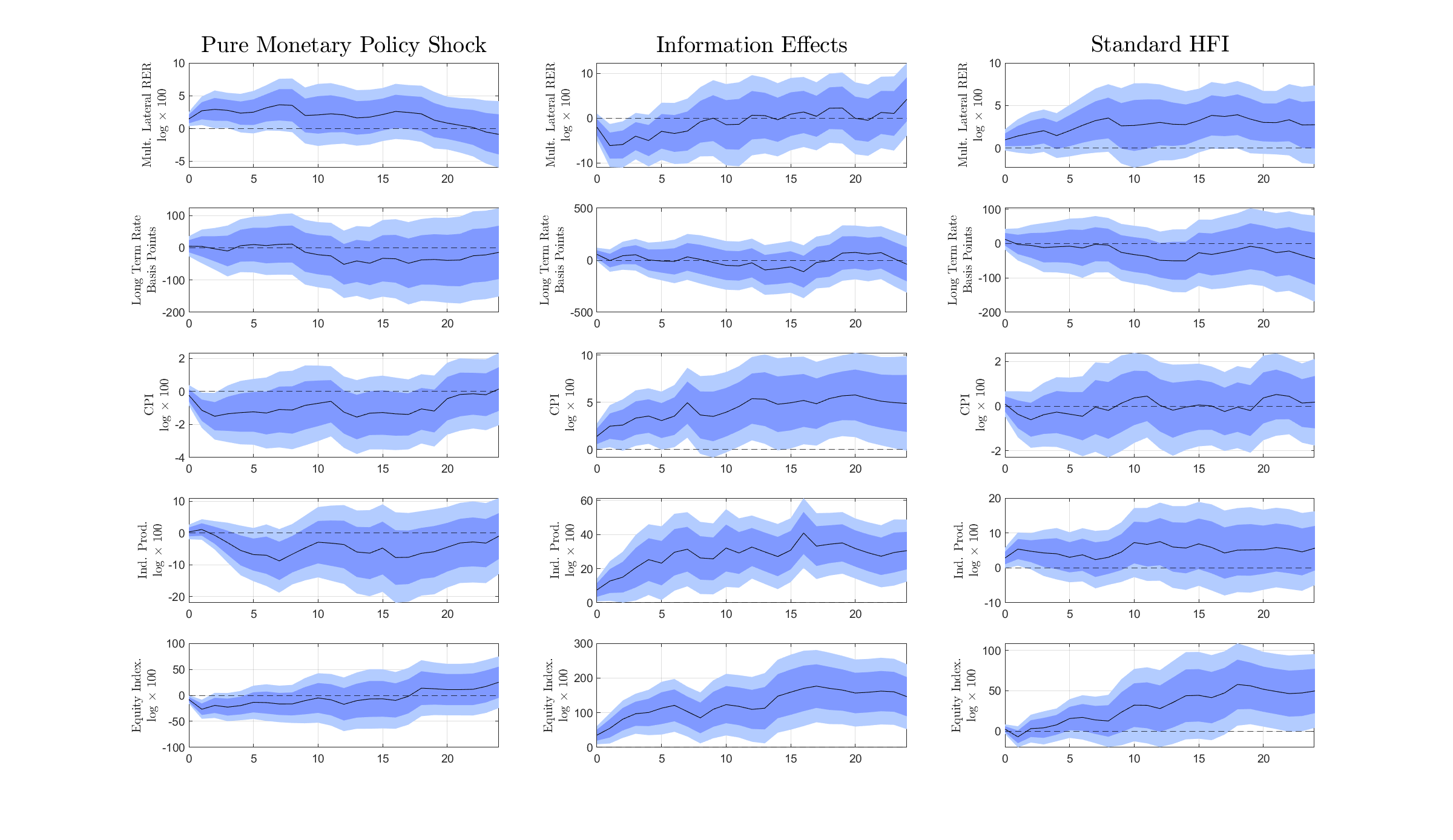}
         \caption{Sample January 1998 - December 2019}
         \label{fig:AEs_REER_98}
     \end{subfigure}
     \hfill
     \begin{subfigure}[b]{0.495\textwidth}
         \centering
         \includegraphics[width=\textwidth,height=9.5cm]{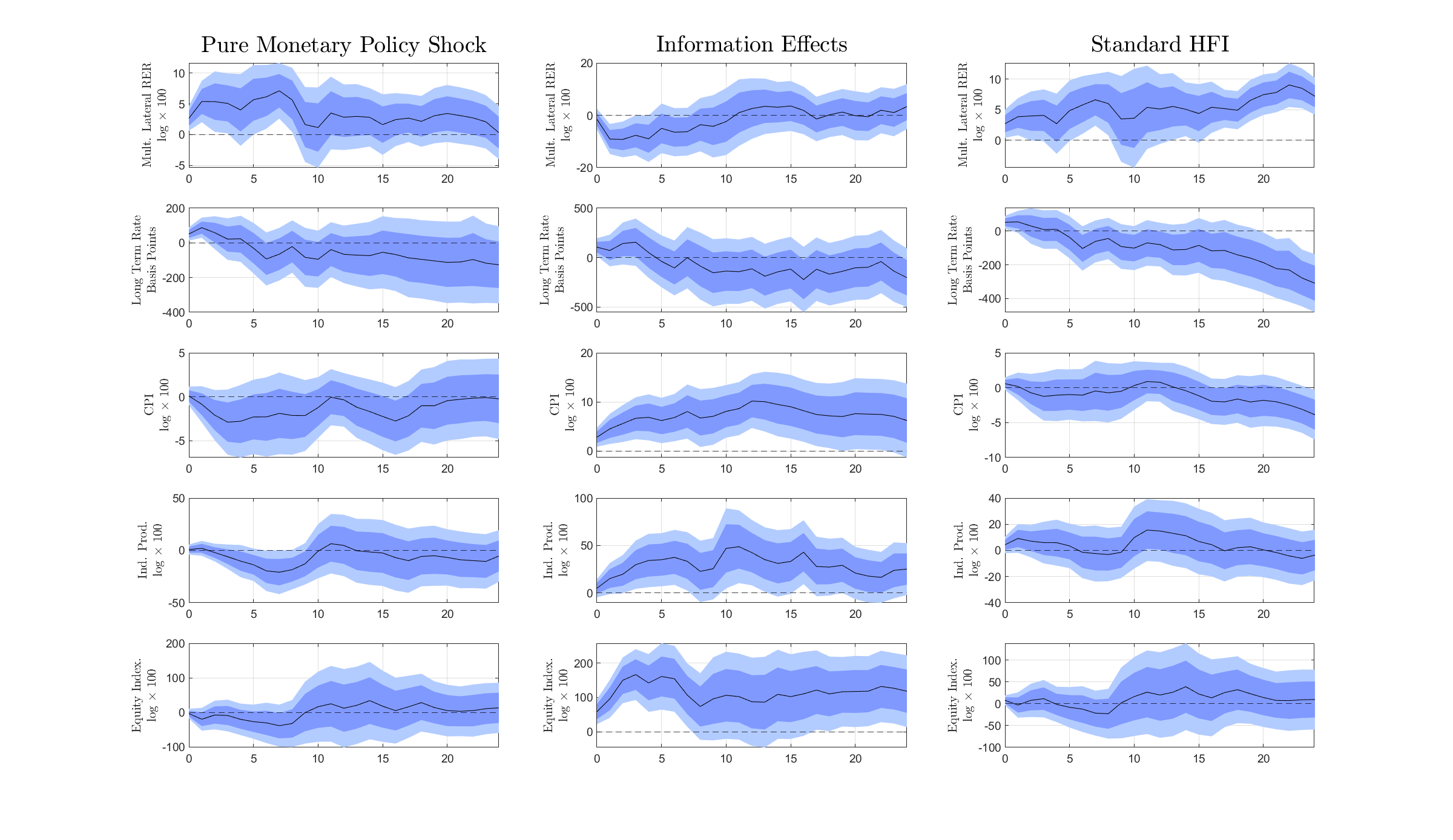}
         \caption{Sample January 2008 - December 2019}
         \label{fig:AEs_REER_08}
     \end{subfigure} 
     \floatfoot{\scriptsize \textbf{Note:} Each of the two figures are comprised of 15 sub-figures ordered in three columns and five rows. Both are constructed for the sample of Advanced Economies alone. The first figure to the left is constructed by estimating Equation \ref{eq:LP_pooled} for the sample starting in January 1998 and ending in December 2019. The second figure to the right is constructed by estimating Equation \ref{eq:LP_pooled} for the sample starting in January 2008 and ending in December 2019. The left column relates to the estimates of $\beta^{MP}$ in Equation \ref{eq:LP_pooled}, the middle column relates to the estimate of $\beta^{FIE}$ in Equation \ref{eq:LP_pooled}, while the right column relates to estimating Equation \ref{eq:LP_pooled}, replacing the MP and FIE components with the un-orthogonalized monetary policy surprise. The rows represent the impact on (i) (i) the multilateral trade weighted real exchange rate index (in logs times 100); (ii) long term interest rates in basis points; (iii) the consumer price index (in logs times 100); (iv) the industrial production index (in logs times 100); (v) the equity index (in logs times 100). The solid black line represents the point estimate, the dark blue area represents the 68\% confidence interval, and the light blue area represents the 90\% confidence interval. In the text, when referring to Panel $(i,j)$, $i$ refers to the row and $j$ to the column of the figure. Each variable, in its own transformation, is demeaned at the country level. }
\end{figure}
\end{landscape}

\newpage
\begin{landscape} 
\begin{figure}[ht]
    \centering
    \caption{Impulse Response Functions - Advanced Economies \\ \footnotesize Samples starting in January 1998 \& January 2008}
    \label{fig:Figure_EM_Across_Time_REER}
     \centering
     \begin{subfigure}[b]{0.495\textwidth}
         \centering
         \includegraphics[width=\textwidth,height=9.5cm]{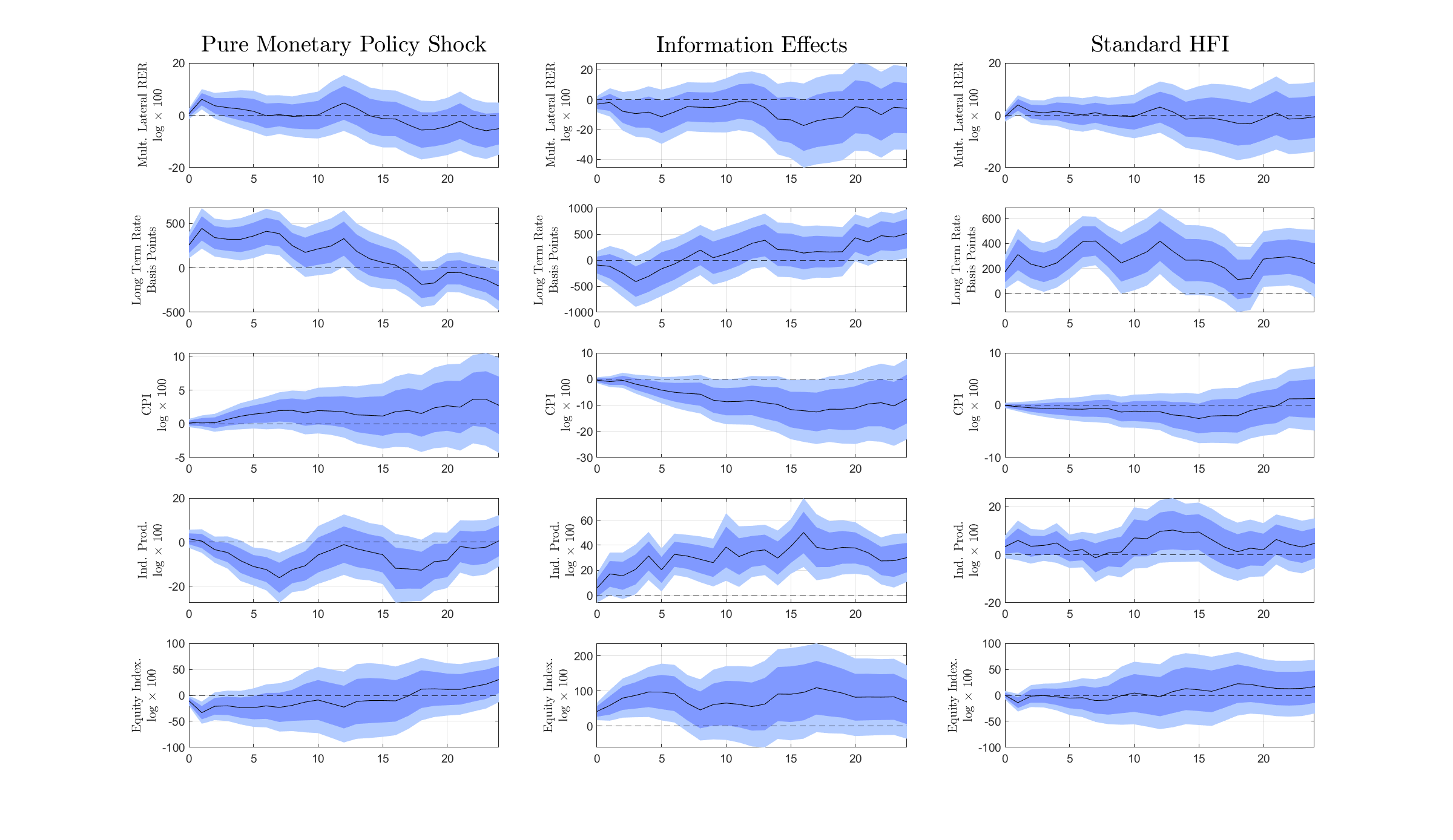}
         \caption{Sample January 1998 - December 2019}
         \label{fig:EMs_REER_98}
     \end{subfigure}
     \hfill
     \begin{subfigure}[b]{0.495\textwidth}
         \centering
         \includegraphics[width=\textwidth,height=9.5cm]{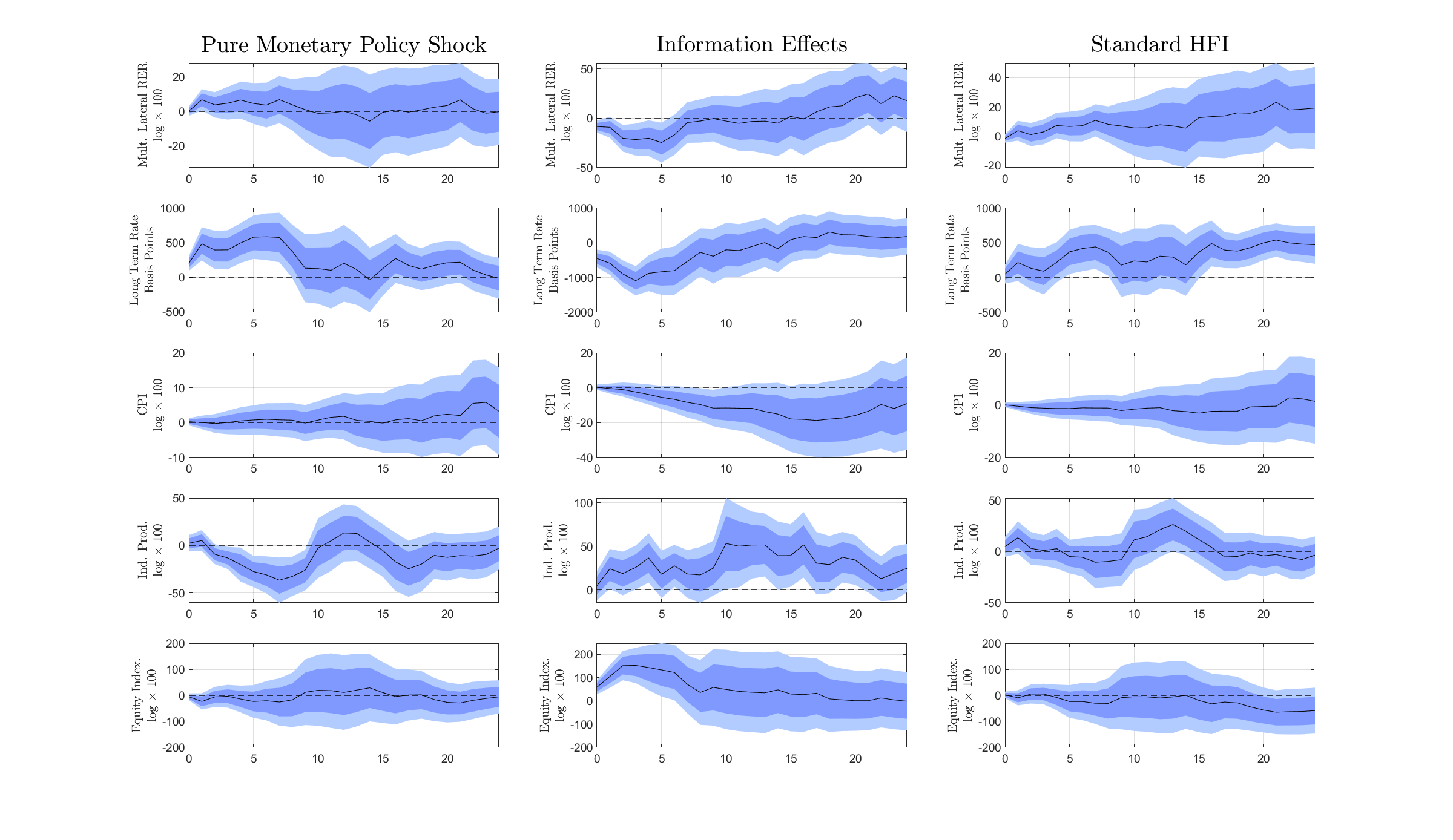}
         \caption{Sample January 2008 - December 2019}
         \label{fig:EMs_REER_08}
     \end{subfigure} 
     \floatfoot{\scriptsize \textbf{Note:} Each of the two figures are comprised of 15 sub-figures ordered in three columns and five rows. Both are constructed for the sample of Advanced Economies alone. The first figure to the left is constructed by estimating Equation \ref{eq:LP_pooled} for the sample starting in January 1998 and ending in December 2019. The second figure to the right is constructed by estimating Equation \ref{eq:LP_pooled} for the sample starting in January 2008 and ending in December 2019. The left column relates to the estimates of $\beta^{MP}$ in Equation \ref{eq:LP_pooled}, the middle column relates to the estimate of $\beta^{FIE}$ in Equation \ref{eq:LP_pooled}, while the right column relates to estimating Equation \ref{eq:LP_pooled}, replacing the MP and FIE components with the un-orthogonalized monetary policy surprise. The rows represent the impact on (i) the nominal exchange rate with respect to the US dollar (in logs times 100); (ii) long term interest rates in basis points; (iii) the consumer price index (in logs times 100); (iv) the industrial production index (in logs times 100); (v) the equity index (in logs times 100). The solid black line represents the point estimate, the dark blue area represents the 68\% confidence interval, and the light blue area represents the 90\% confidence interval. In the text, when referring to Panel $(i,j)$, $i$ refers to the row and $j$ to the column of the figure. Each variable, in its own transformation, is demeaned at the country level. }
\end{figure}
\end{landscape}

\newpage
\begin{figure}[ht]
    \centering
    \includegraphics[scale=0.4]{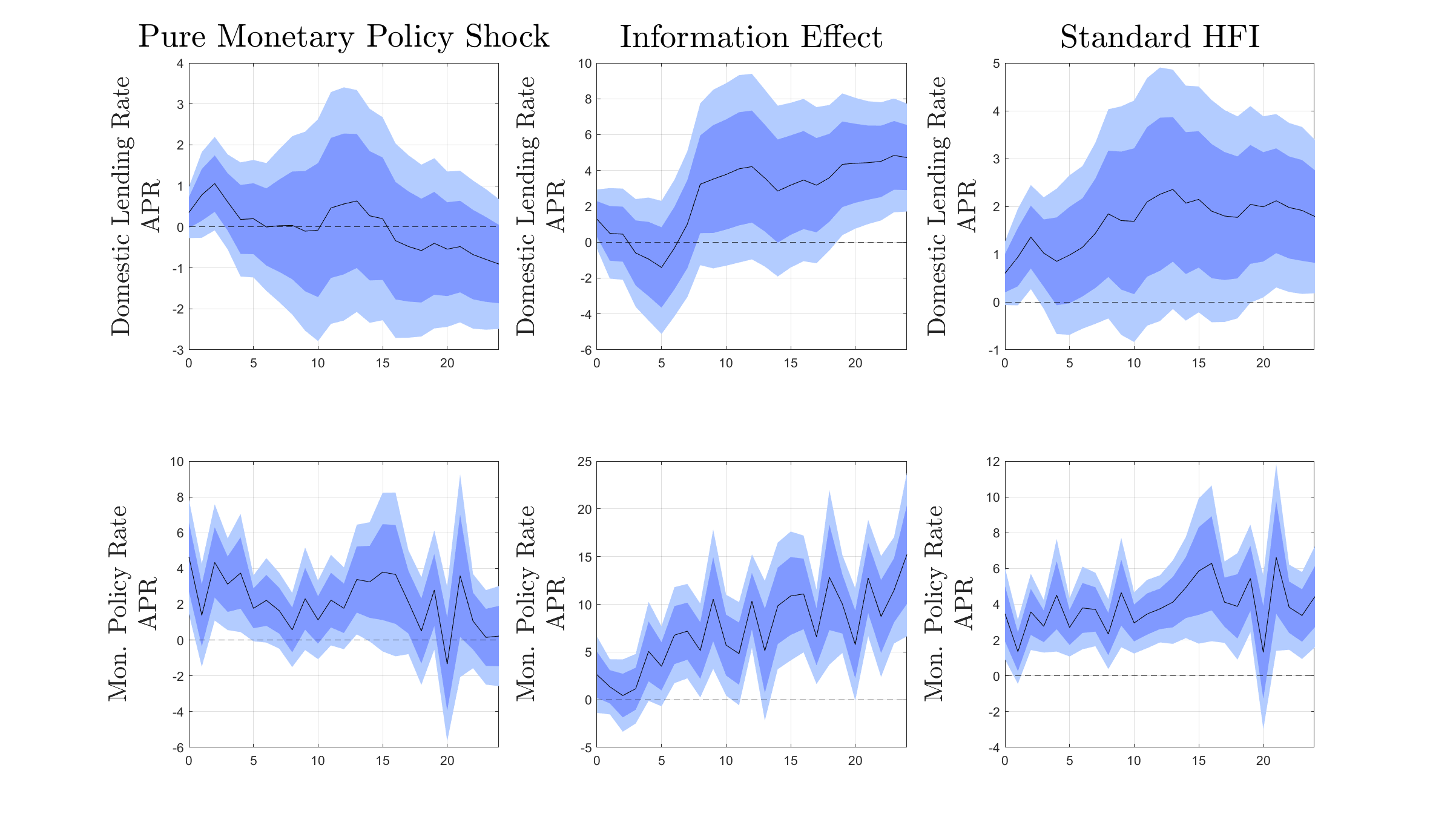}
    \caption{Impulse Response Functions \\ Additional Financial Variables}
    \label{fig:LendingRate_NER}
    \floatfoot{\textbf{Note:} The figure is comprised of 6 sub-figures ordered in three columns and two rows. The left column relates to the estimates of $\beta^{MP}$ in Equation \ref{eq:LP_pooled}, the middle column relates to the estimate of $\beta^{FIE}$ in Equation \ref{eq:LP_pooled}, while the right column relates to estimating Equation \ref{eq:LP_pooled}, replacing the MP and FIE components with the un-orthogonalized monetary policy surprise. The rows represent the impact on (i) domestic lending rates (APR); (ii) monetary policy rate (APR). The response of these variables are estimated by estimating Equation \ref{eq:LP_pooled} and adding each extra variable separately. The solid black line represents the point estimate, the dark blue area represents the 68\% confidence interval, and the light blue area represents the 90\% confidence interval. In the text, when referring to Panel $(i,j)$, $i$ refers to the row and $j$ to the column of the figure. Each variable, in its own transformation, is demeaned at the country level.}
\end{figure}

\newpage
\begin{figure}
    \centering
    \includegraphics[scale=0.4]{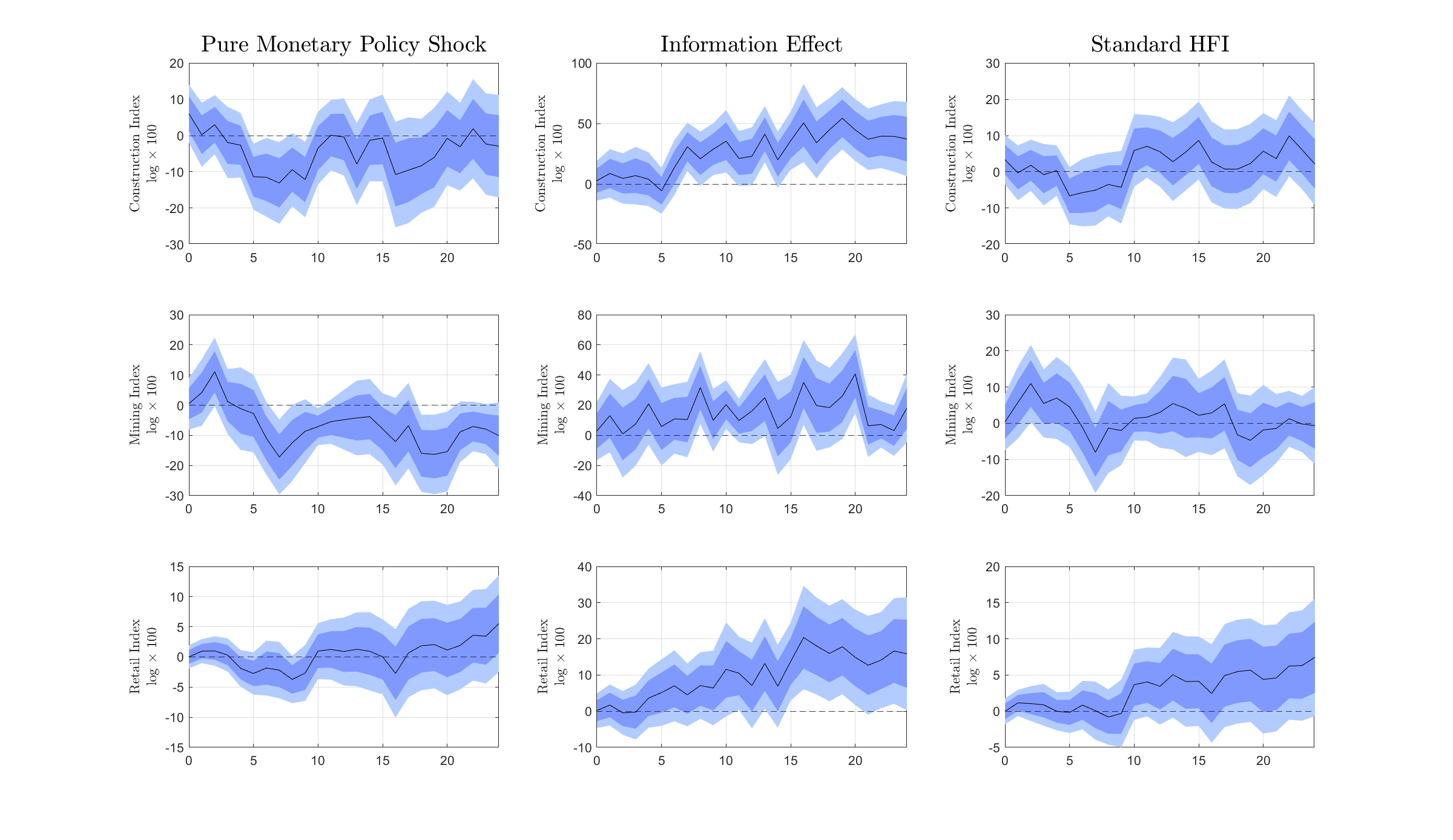}
    \caption{Impulse Response Functions \\ Additional Real Variables}
    \label{fig:Real_Variables_NER}
    \floatfoot{\textbf{Note:} The figure is comprised of 9 sub-figures ordered in three columns and three rows. The left column relates to the estimates of $\beta^{MP}$ in Equation \ref{eq:LP_pooled}, the middle column relates to the estimate of $\beta^{FIE}$ in Equation \ref{eq:LP_pooled}, while the right column relates to estimating Equation \ref{eq:LP_pooled}, replacing the MP and FIE components with the un-orthogonalized monetary policy surprise. The rows represent the impact on (i) construction index (in logs times 100); (ii) mining index (in logs times 100); (iii) retail sales index (in logs times 100). The response of these variables are estimated by estimating Equation \ref{eq:LP_pooled} and adding each extra variable separately. The solid black line represents the point estimate, the dark blue area represents the 68\% confidence interval, and the light blue area represents the 90\% confidence interval. In the text, when referring to Panel $(i,j)$, $i$ refers to the row and $j$ to the column of the figure. Each variable, in its own transformation, is demeaned at the country level.}
\end{figure}

\newpage
\begin{figure}
    \centering
    \includegraphics[width=14cm,height=8cm]{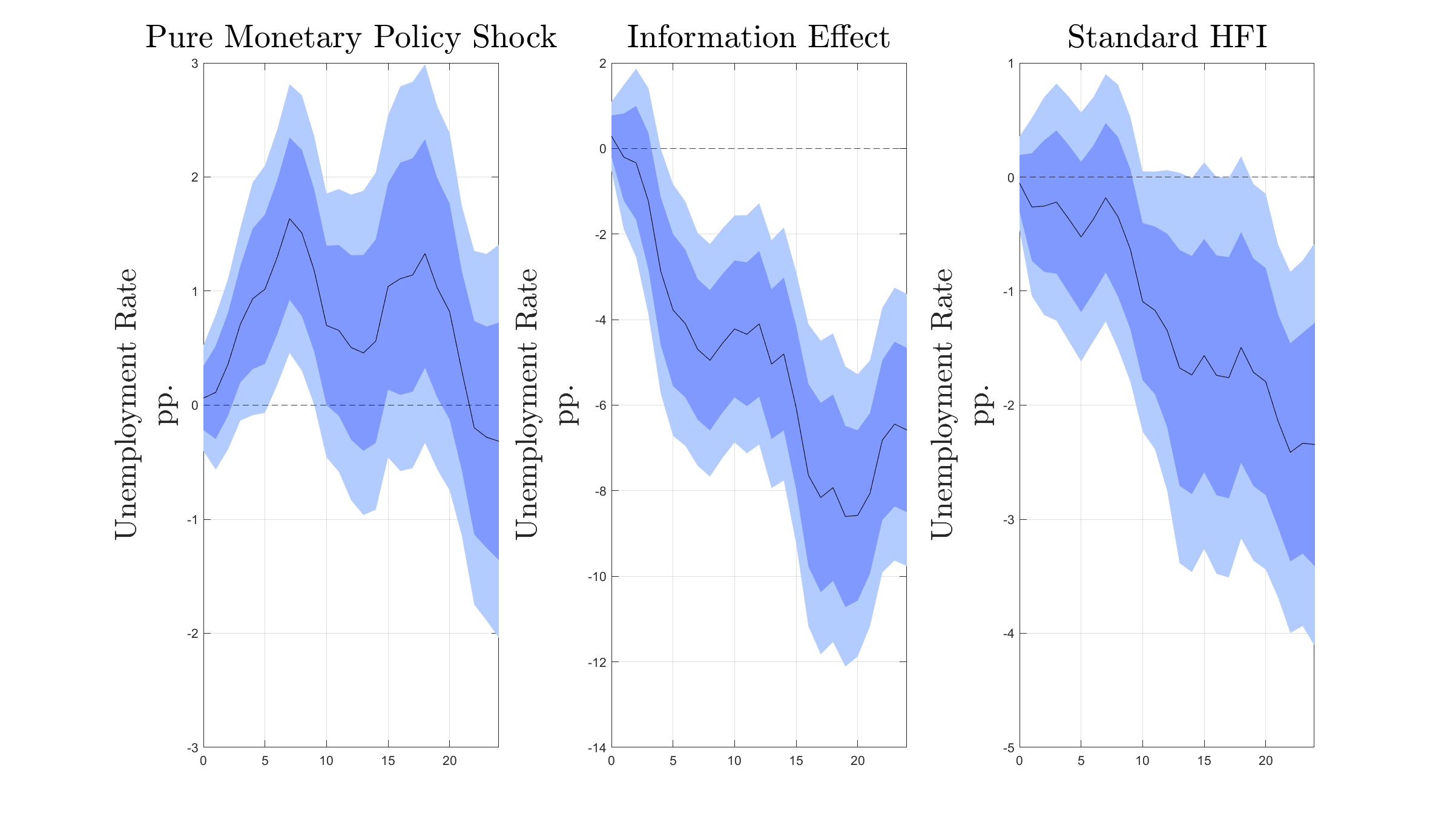}
    \caption{Impulse Response Functions \\ Unemployment Rate}
    \label{fig:URATE_NER}
    \floatfoot{\textbf{Note:} The figure is comprised of 3 sub-figures ordered in three columns and one row. The left column relates to the estimates of $\beta^{MP}$ in Equation \ref{eq:LP_pooled}, the middle column relates to the estimate of $\beta^{FIE}$ in Equation \ref{eq:LP_pooled}, while the right column relates to estimating Equation \ref{eq:LP_pooled}, replacing the MP and FIE components with the un-orthogonalized monetary policy surprise. The row represents the impact on the unemployment rate in percentage points. The response of this variable is estimated by estimating Equation \ref{eq:LP_pooled} and the unemployment variable in addition to the benchmark sample. The solid black line represents the point estimate, the dark blue area represents the 68\% confidence interval, and the light blue area represents the 90\% confidence interval. In the text, when referring to Panel $(i,j)$, $i$ refers to the row and $j$ to the column of the figure. Each variable, in its own transformation, is demeaned at the country level.}
\end{figure}

\newpage
\begin{figure}
    \centering
    \includegraphics[scale=0.4]{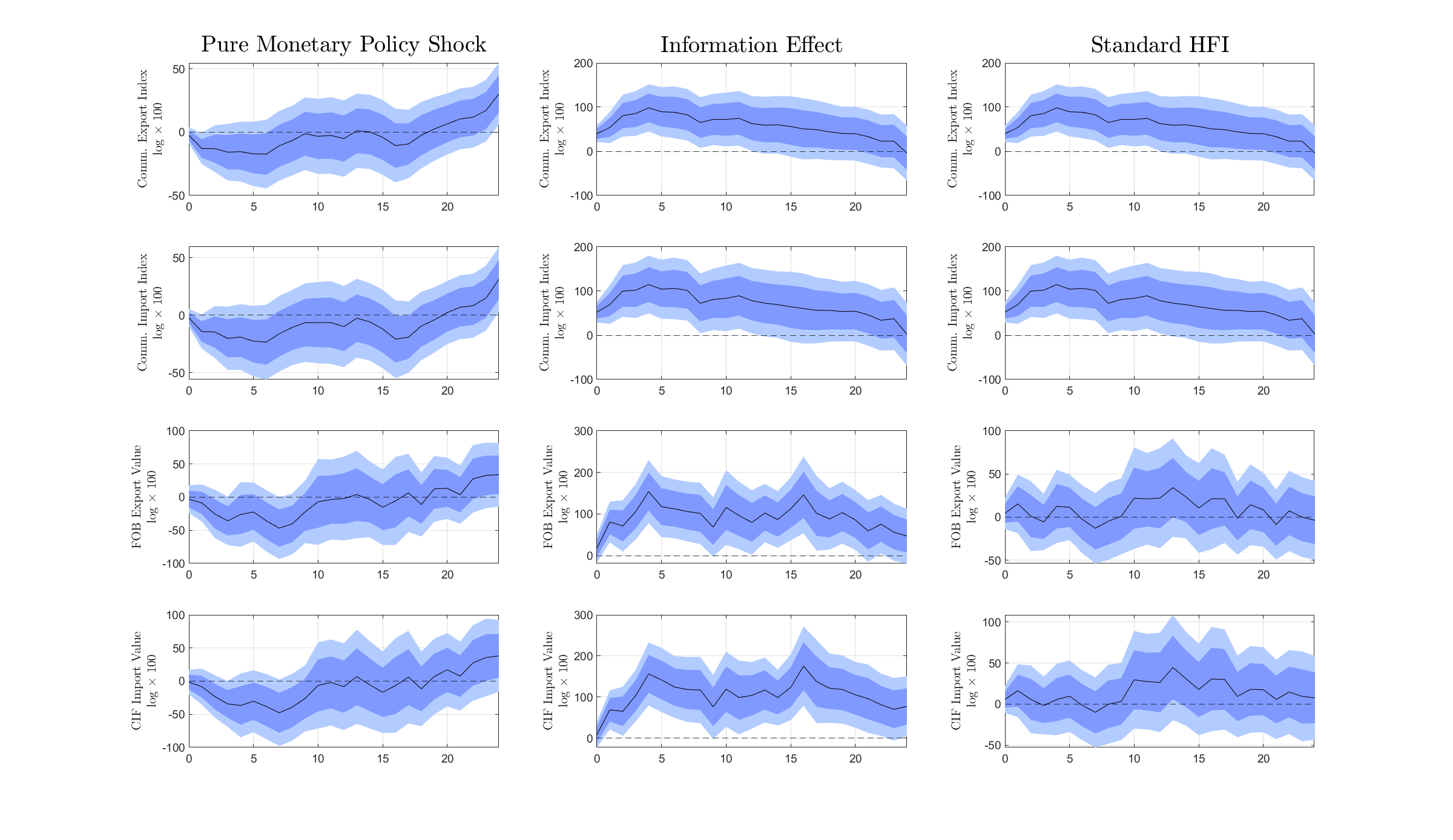}
    \caption{Impulse Response Functions \\ Additional Trade Related Variables}
    \label{fig:Trade_Variables}
    \floatfoot{\textbf{Note:} The figure is comprised of 12 sub-figures ordered in three columns and four rows. The left column relates to the estimates of $\beta^{MP}$ in Equation \ref{eq:LP_pooled}, the middle column relates to the estimate of $\beta^{FIE}$ in Equation \ref{eq:LP_pooled}, while the right column relates to estimating Equation \ref{eq:LP_pooled}, replacing the MP and FIE components with the un-orthogonalized monetary policy surprise. The rows represent the impact on (i) country specific commodity export price index (in logs times 100); (ii) country specific commodity import price index (in logs times 100); (iii) total export value in FOB USD dollars (in logs times 100); (iv) total import value in CIF USD dollars (in logs times 100). The response of these variables are estimated by estimating Equation \ref{eq:LP_pooled} by first adding variables (i) and (ii) to the benchmark sample, and then adding variables (iii) and (iv) to the benchmark sample. The solid black line represents the point estimate, the dark blue area represents the 68\% confidence interval, and the light blue area represents the 90\% confidence interval. In the text, when referring to Panel $(i,j)$, $i$ refers to the row and $j$ to the column of the figure. Each variable, in its own transformation, is demeaned at the country level.}
\end{figure}

\newpage
\begin{figure}
    \centering
    \includegraphics[scale=0.4]{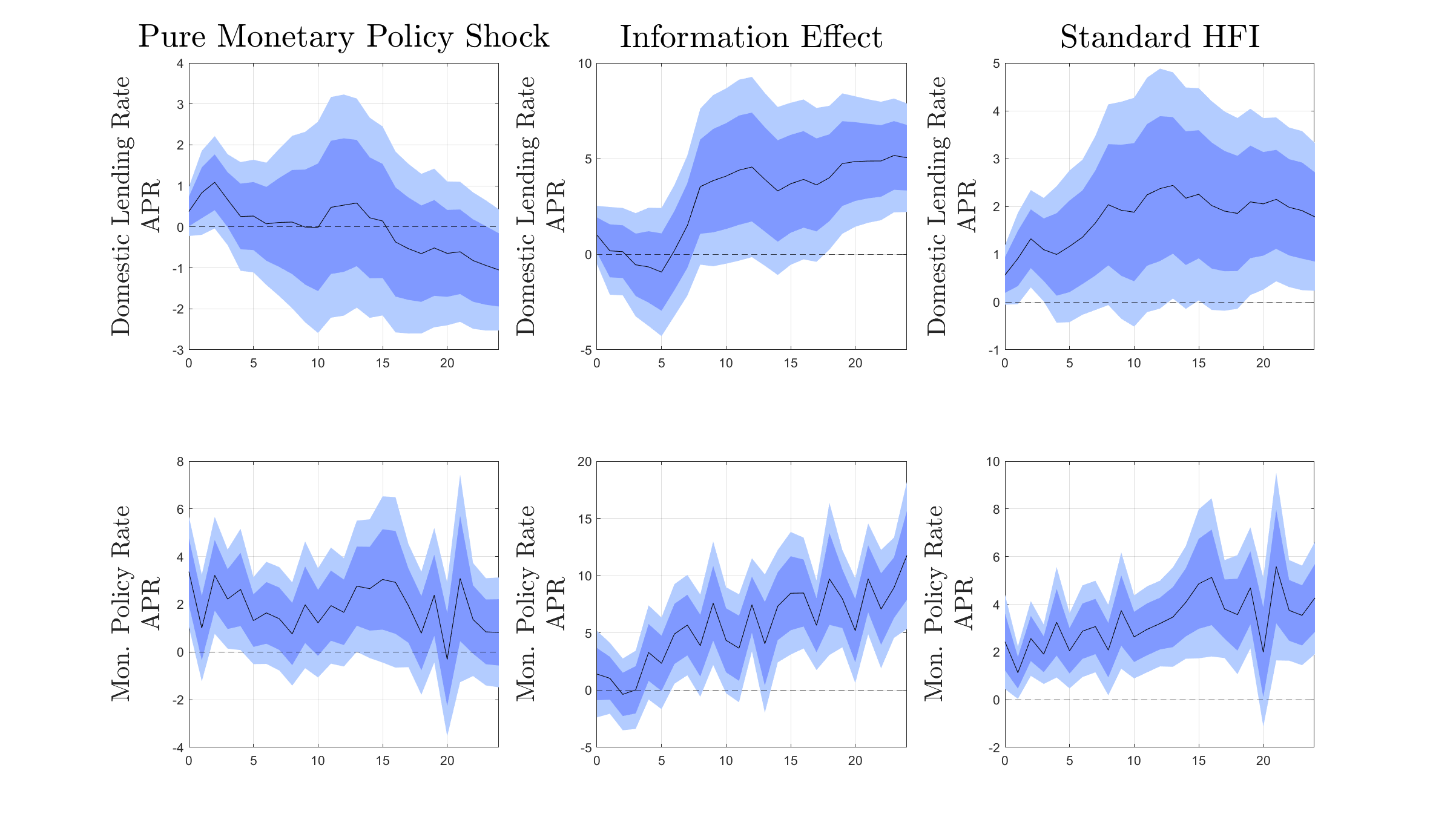}
    \caption{Impulse Response Functions \\ Additional Financial Variables - REER Sample}
    \label{fig:LendingRate_REER}
    \floatfoot{\textbf{Note:} The figure is comprised of 6 sub-figures ordered in three columns and two rows. The left column relates to the estimates of $\beta^{MP}$ in Equation \ref{eq:LP_pooled}, the middle column relates to the estimate of $\beta^{FIE}$ in Equation \ref{eq:LP_pooled}, while the right column relates to estimating Equation \ref{eq:LP_pooled}, replacing the MP and FIE components with the un-orthogonalized monetary policy surprise. The rows represent the impact on (i) domestic lending rates (APR); (ii) monetary policy rate (APR). The response of these variables are estimated by estimating Equation \ref{eq:LP_pooled} and adding each extra variable separately. The nominal exchange rate is replaced by the trade weighted multilateral real exchange rate. The solid black line represents the point estimate, the dark blue area represents the 68\% confidence interval, and the light blue area represents the 90\% confidence interval. In the text, when referring to Panel $(i,j)$, $i$ refers to the row and $j$ to the column of the figure. Each variable, in its own transformation, is demeaned at the country level.}
\end{figure}

\newpage
\begin{figure}
    \centering
    \includegraphics[scale=0.4]{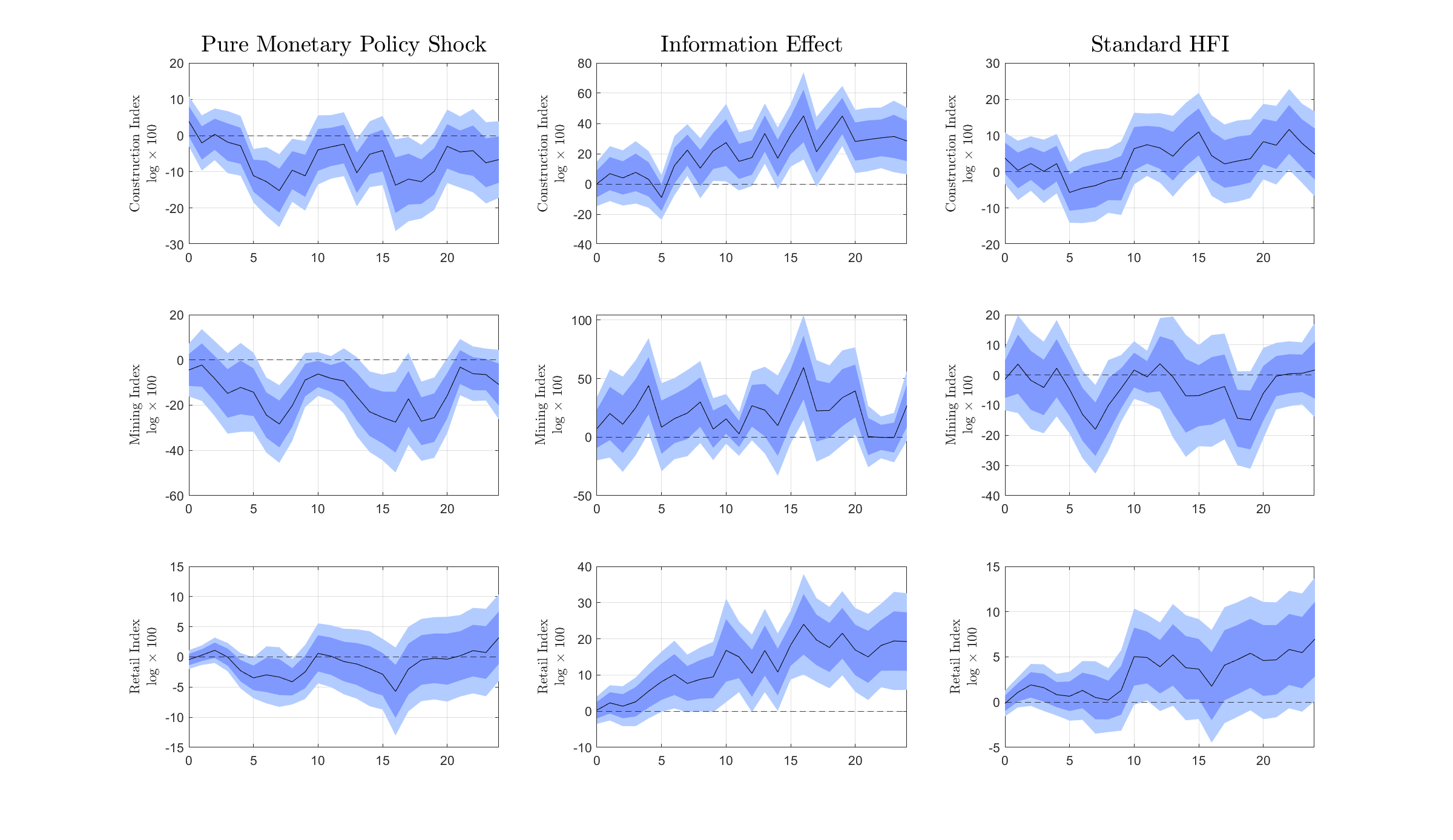}
    \caption{Impulse Response Functions \\ Additional Real Variables - REER Sample}
    \label{fig:Real_Variables_REER}
    \floatfoot{\textbf{Note:} The figure is comprised of 9 sub-figures ordered in three columns and three rows. The left column relates to the estimates of $\beta^{MP}$ in Equation \ref{eq:LP_pooled}, the middle column relates to the estimate of $\beta^{FIE}$ in Equation \ref{eq:LP_pooled}, while the right column relates to estimating Equation \ref{eq:LP_pooled}, replacing the MP and FIE components with the un-orthogonalized monetary policy surprise. The rows represent the impact on (i) construction index (in logs times 100); (ii) mining index (in logs times 100); (iii) retail sales index (in logs times 100). The response of these variables are estimated by estimating Equation \ref{eq:LP_pooled} and adding each extra variable separately.  The nominal exchange rate is replaced by the trade weighted multilateral real exchange rate.The solid black line represents the point estimate, the dark blue area represents the 68\% confidence interval, and the light blue area represents the 90\% confidence interval. In the text, when referring to Panel $(i,j)$, $i$ refers to the row and $j$ to the column of the figure. Each variable, in its own transformation, is demeaned at the country level.}
\end{figure}

\newpage
\begin{figure}
    \centering
    \includegraphics[width=14cm,height=8cm]{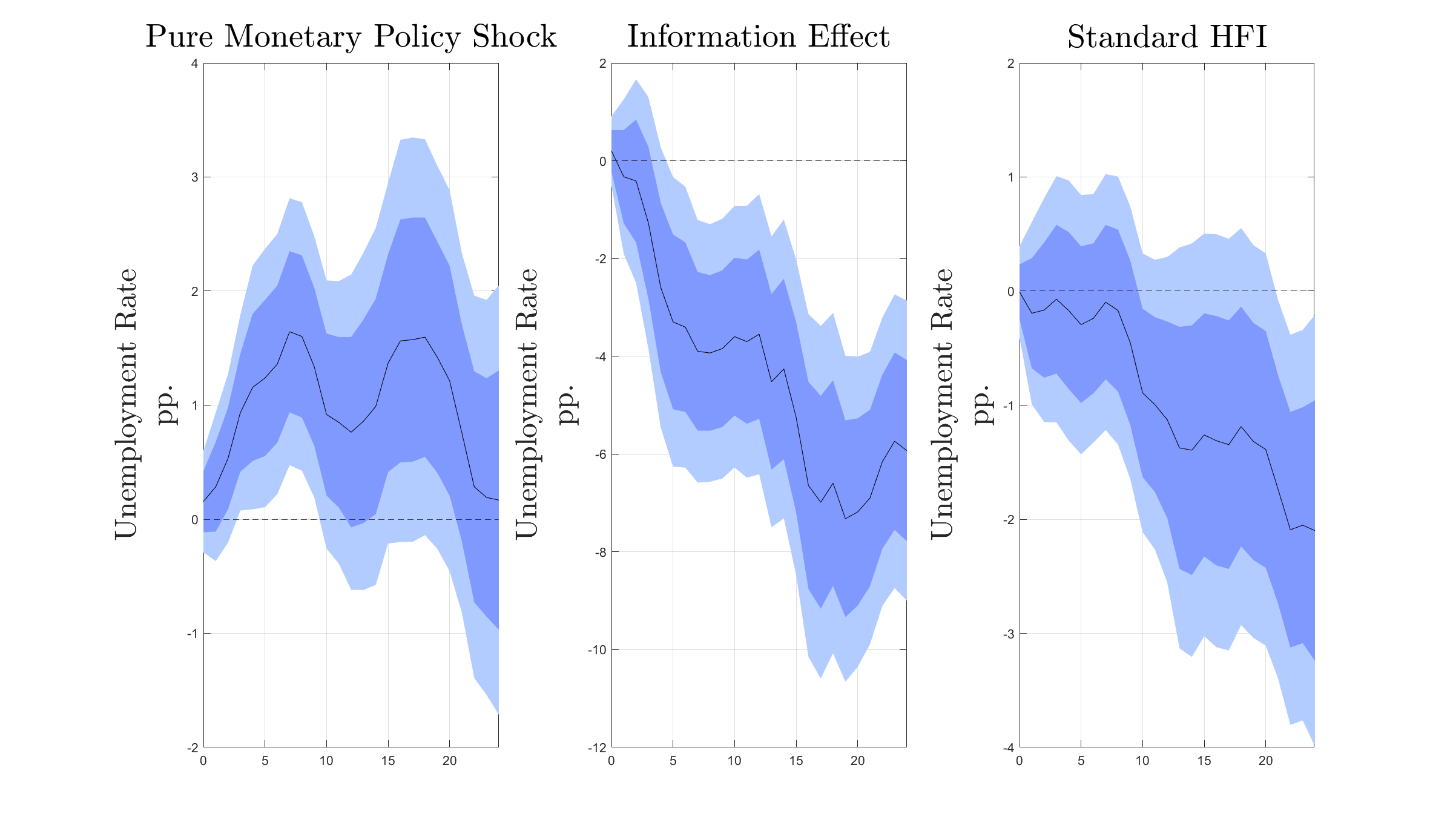}
    \caption{Impulse Response Functions \\ Unemployment Rate  - REER Sample}
    \label{fig:URATE_REER}
    \floatfoot{\textbf{Note:} The figure is comprised of 3 sub-figures ordered in three columns and one row. The left column relates to the estimates of $\beta^{MP}$ in Equation \ref{eq:LP_pooled}, the middle column relates to the estimate of $\beta^{FIE}$ in Equation \ref{eq:LP_pooled}, while the right column relates to estimating Equation \ref{eq:LP_pooled}, replacing the MP and FIE components with the un-orthogonalized monetary policy surprise. The row represents the impact on the unemployment rate in percentage points. The response of this variable is estimated by estimating Equation \ref{eq:LP_pooled} and the unemployment variable in addition to the sample where the nominal exchange rate is replaced by the trade weighted multilateral real exchange rate. The solid black line represents the point estimate, the dark blue area represents the 68\% confidence interval, and the light blue area represents the 90\% confidence interval. In the text, when referring to Panel $(i,j)$, $i$ refers to the row and $j$ to the column of the figure. Each variable, in its own transformation, is demeaned at the country level.}
\end{figure}

\newpage
\begin{figure}
    \centering
    \includegraphics[scale=0.4]{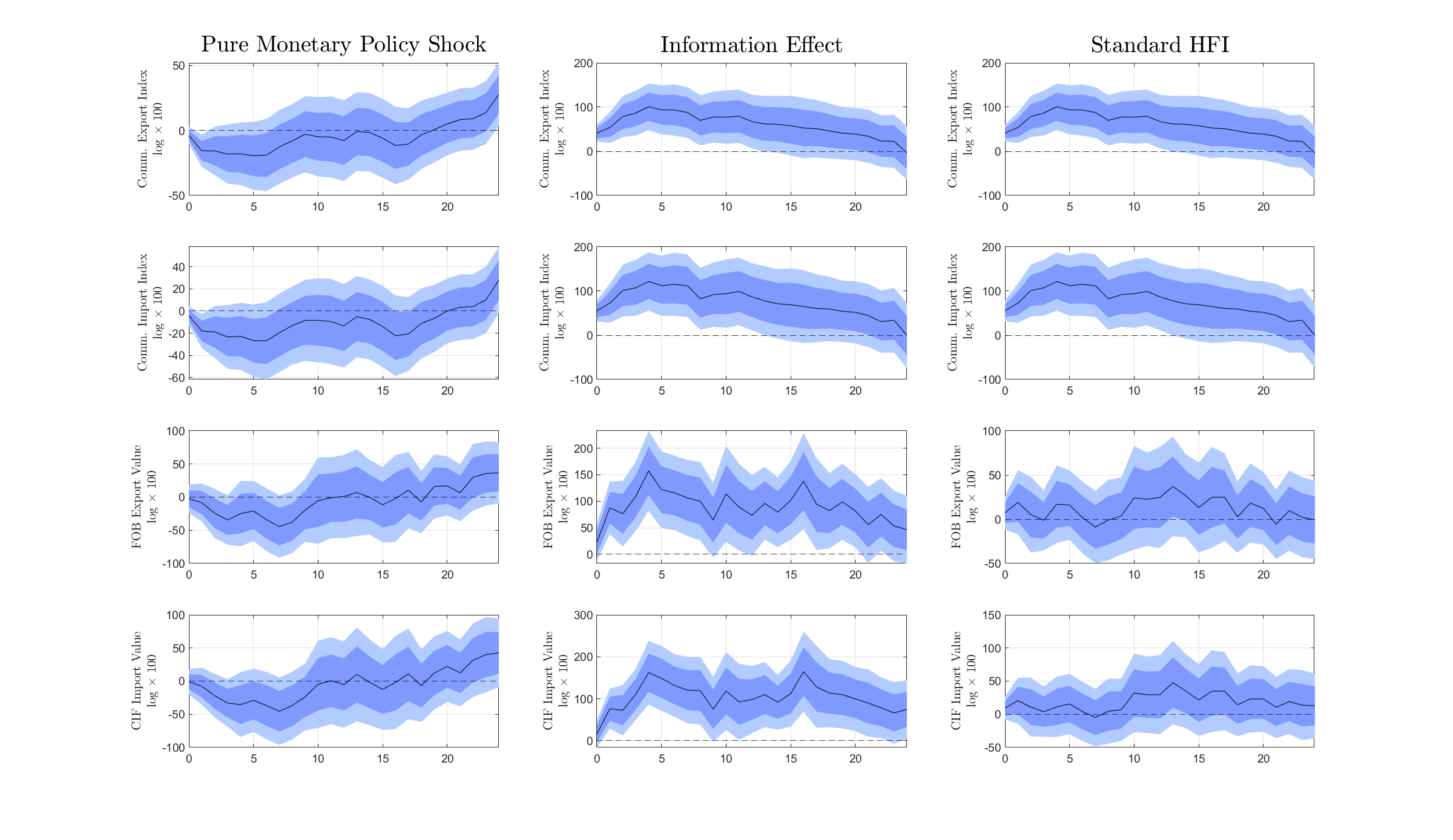}
    \caption{Impulse Response Functions \\ Additional Trade Related Variables - REER Sample}
    \label{fig:Trade_Variables_REER}
    \floatfoot{\textbf{Note:} The figure is comprised of 12 sub-figures ordered in three columns and four rows. The left column relates to the estimates of $\beta^{MP}$ in Equation \ref{eq:LP_pooled}, the middle column relates to the estimate of $\beta^{FIE}$ in Equation \ref{eq:LP_pooled}, while the right column relates to estimating Equation \ref{eq:LP_pooled}, replacing the MP and FIE components with the un-orthogonalized monetary policy surprise. The rows represent the impact on (i) country specific commodity export price index (in logs times 100); (ii) country specific commodity import price index (in logs times 100); (iii) total export value in FOB USD dollars (in logs times 100); (iv) total import value in CIF USD dollars (in logs times 100). The response of these variables are estimated by estimating Equation \ref{eq:LP_pooled} by first adding variables (i) and (ii) to the sample where the nominal exchange rate is replaced by the trade weighted multilateral real exchange rate, and then adding variables (iii) and (iv) to the sample where the nominal exchange rate is replaced by the trade weighted multilateral real exchange rate. The solid black line represents the point estimate, the dark blue area represents the 68\% confidence interval, and the light blue area represents the 90\% confidence interval. In the text, when referring to Panel $(i,j)$, $i$ refers to the row and $j$ to the column of the figure. Each variable, in its own transformation, is demeaned at the country level.}
\end{figure}

\newpage
\begin{figure}[ht]
    \centering
    \includegraphics[scale=0.4]{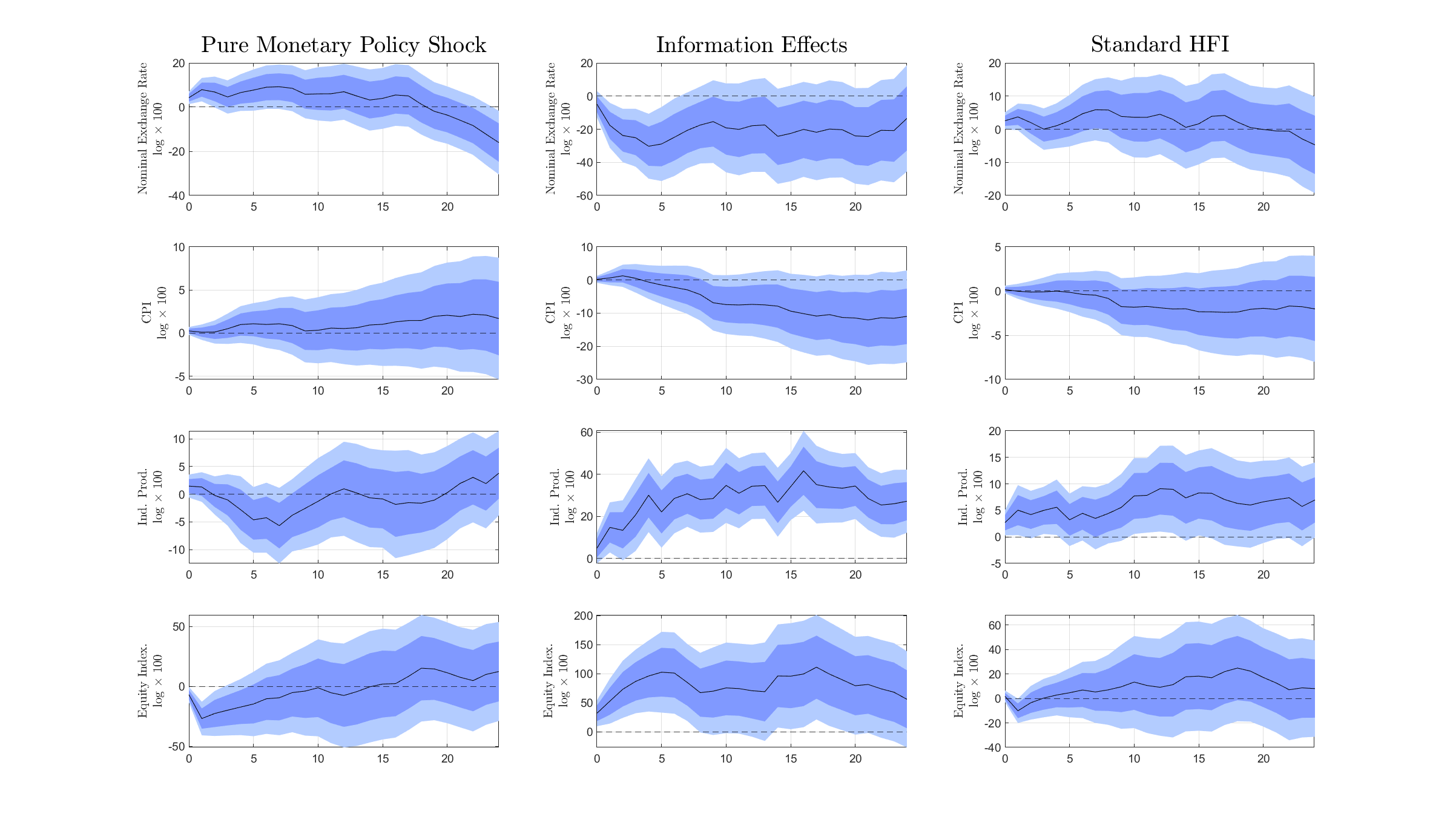}
    \caption{Impulse Response Functions - Alternative Variables \\ Removing Long Term Rates }
    \label{fig:Removing_LT}
    \floatfoot{\textbf{Note:} The figure is comprised of 12 sub-figures ordered in three columns and four rows. The left column relates to the estimates of $\beta^{MP}$ in Equation \ref{eq:LP_pooled}, the middle column relates to the estimate of $\beta^{FIE}$ in Equation \ref{eq:LP_pooled}, while the right column relates to estimating Equation \ref{eq:LP_pooled}, replacing the MP and FIE components with the un-orthogonalized monetary policy surprise. The rows represent the impact on (i) the nominal exchange rate with the US dollar (in logs times 100); (ii) the consumer price index (in logs times 100); (iii) the industrial production index (in logs times 100); (iv) the equity index (in logs times 100). The solid black line represents the point estimate, the dark blue area represents the 68\% confidence interval, and the light blue area represents the 90\% confidence interval. In the text, when referring to Panel $(i,j)$, $i$ refers to the row and $j$ to the column of the figure. Each variable, in its own transformation, is demeaned at the country level.}
\end{figure}

\newpage
\begin{figure}[ht]
    \centering
    \includegraphics[scale=0.4]{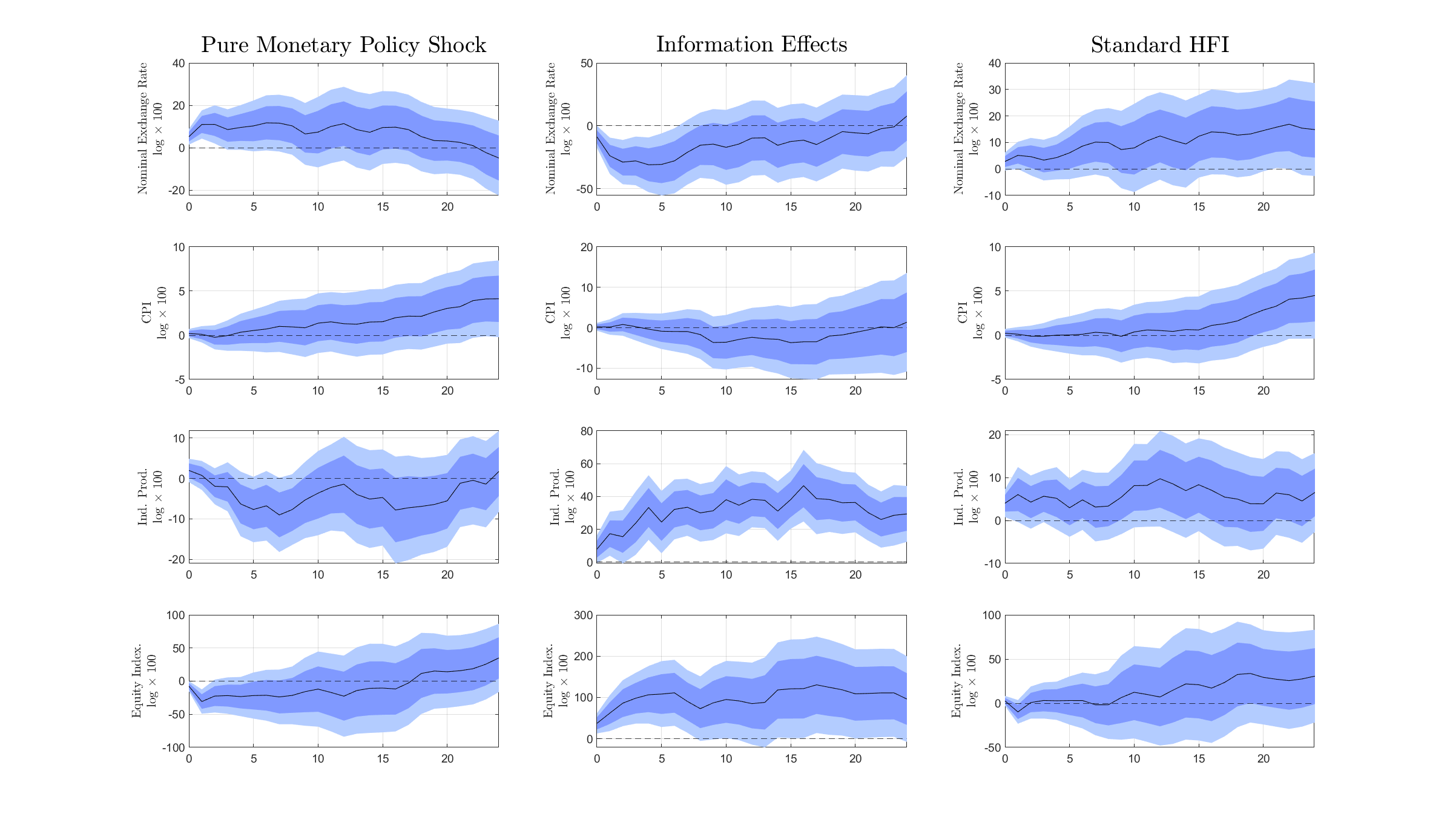}
    \caption{Impulse Response Functions - Alternative Variables \\ Removing LT. Rates - Starting January 1998}
    \label{fig:Removing_LT_98}
    \floatfoot{\textbf{Note:} The figure is comprised of 12 sub-figures ordered in three columns and four rows. The left column relates to the estimates of $\beta^{MP}$ in Equation \ref{eq:LP_pooled}, the middle column relates to the estimate of $\beta^{FIE}$ in Equation \ref{eq:LP_pooled}, while the right column relates to estimating Equation \ref{eq:LP_pooled}, replacing the MP and FIE components with the un-orthogonalized monetary policy surprise. Sample from January 1998 to December 2019. The rows represent the impact on (i) the nominal exchange rate with the US dollar (in logs times 100); (ii) the consumer price index (in logs times 100); (iii) the industrial production index (in logs times 100); (iv) the equity index (in logs times 100). The solid black line represents the point estimate, the dark blue area represents the 68\% confidence interval, and the light blue area represents the 90\% confidence interval. In the text, when referring to Panel $(i,j)$, $i$ refers to the row and $j$ to the column of the figure. Each variable, in its own transformation, is demeaned at the country level.}
\end{figure}

\newpage
\begin{figure}[ht]
    \centering
    \includegraphics[scale=0.4]{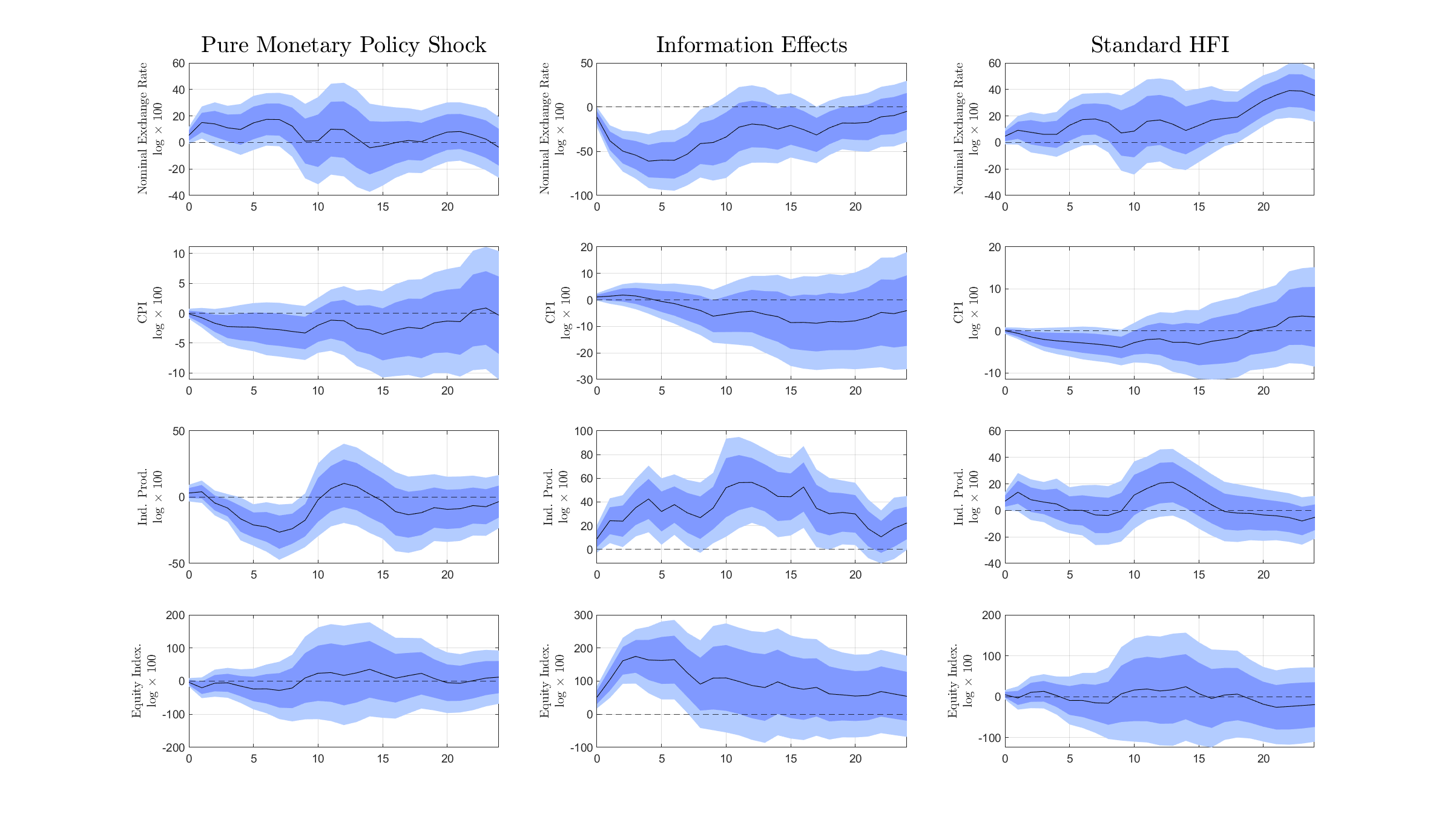}
    \caption{Impulse Response Functions - Alternative Variables \\ Removing LT. Rates - Starting January 2008}
    \label{fig:Removing_LT_08}
    \floatfoot{\textbf{Note:} The figure is comprised of 12 sub-figures ordered in three columns and four rows. The left column relates to the estimates of $\beta^{MP}$ in Equation \ref{eq:LP_pooled}, the middle column relates to the estimate of $\beta^{FIE}$ in Equation \ref{eq:LP_pooled}, while the right column relates to estimating Equation \ref{eq:LP_pooled}, replacing the MP and FIE components with the un-orthogonalized monetary policy surprise. Sample from January 2008 to December 2019. The rows represent the impact on (i) the nominal exchange rate with the US dollar (in logs times 100); (ii) the consumer price index (in logs times 100); (iii) the industrial production index (in logs times 100); (iv) the equity index (in logs times 100). The solid black line represents the point estimate, the dark blue area represents the 68\% confidence interval, and the light blue area represents the 90\% confidence interval. In the text, when referring to Panel $(i,j)$, $i$ refers to the row and $j$ to the column of the figure. Each variable, in its own transformation, is demeaned at the country level.}
\end{figure}

\newpage
\begin{figure}[ht]
    \centering
    \includegraphics[scale=0.4]{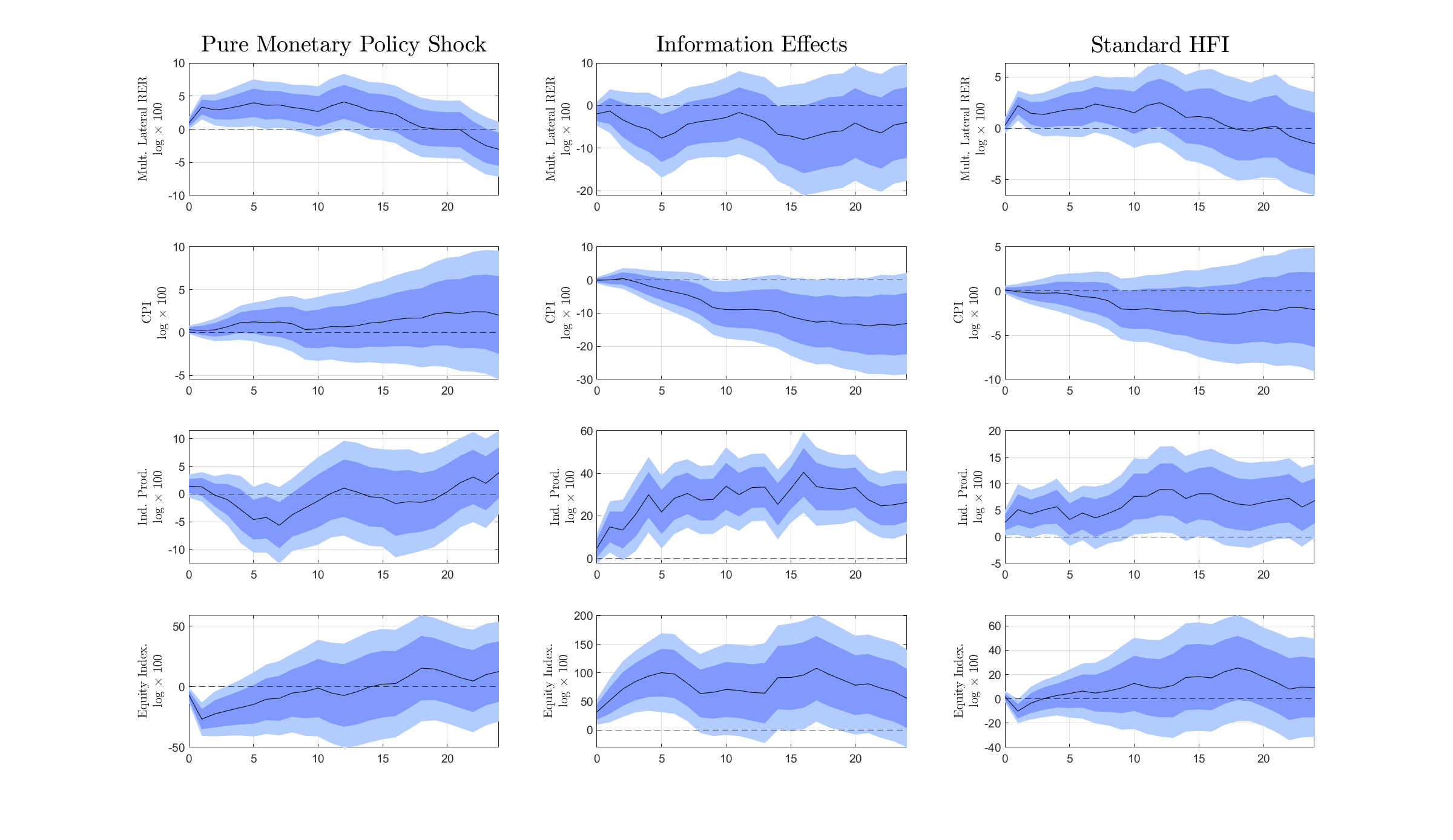}
    \caption{Impulse Response Functions - Alternative Variables \\ Multi. REER Sample - Removing Long Term Rates }
    \label{fig:Removing_LT_REER}
    \floatfoot{\textbf{Note:} The figure is comprised of 12 sub-figures ordered in three columns and four rows. The left column relates to the estimates of $\beta^{MP}$ in Equation \ref{eq:LP_pooled}, the middle column relates to the estimate of $\beta^{FIE}$ in Equation \ref{eq:LP_pooled}, while the right column relates to estimating Equation \ref{eq:LP_pooled}, replacing the MP and FIE components with the un-orthogonalized monetary policy surprise. The rows represent the impact on (i) the multilateral trade weighted real exchange rate index (in logs times 100); (ii) the consumer price index (in logs times 100); (iii) the industrial production index (in logs times 100); (iv) the equity index (in logs times 100). The solid black line represents the point estimate, the dark blue area represents the 68\% confidence interval, and the light blue area represents the 90\% confidence interval. In the text, when referring to Panel $(i,j)$, $i$ refers to the row and $j$ to the column of the figure. Each variable, in its own transformation, is demeaned at the country level.}
\end{figure}

\newpage
\begin{figure}[ht]
    \centering
    \includegraphics[scale=0.4]{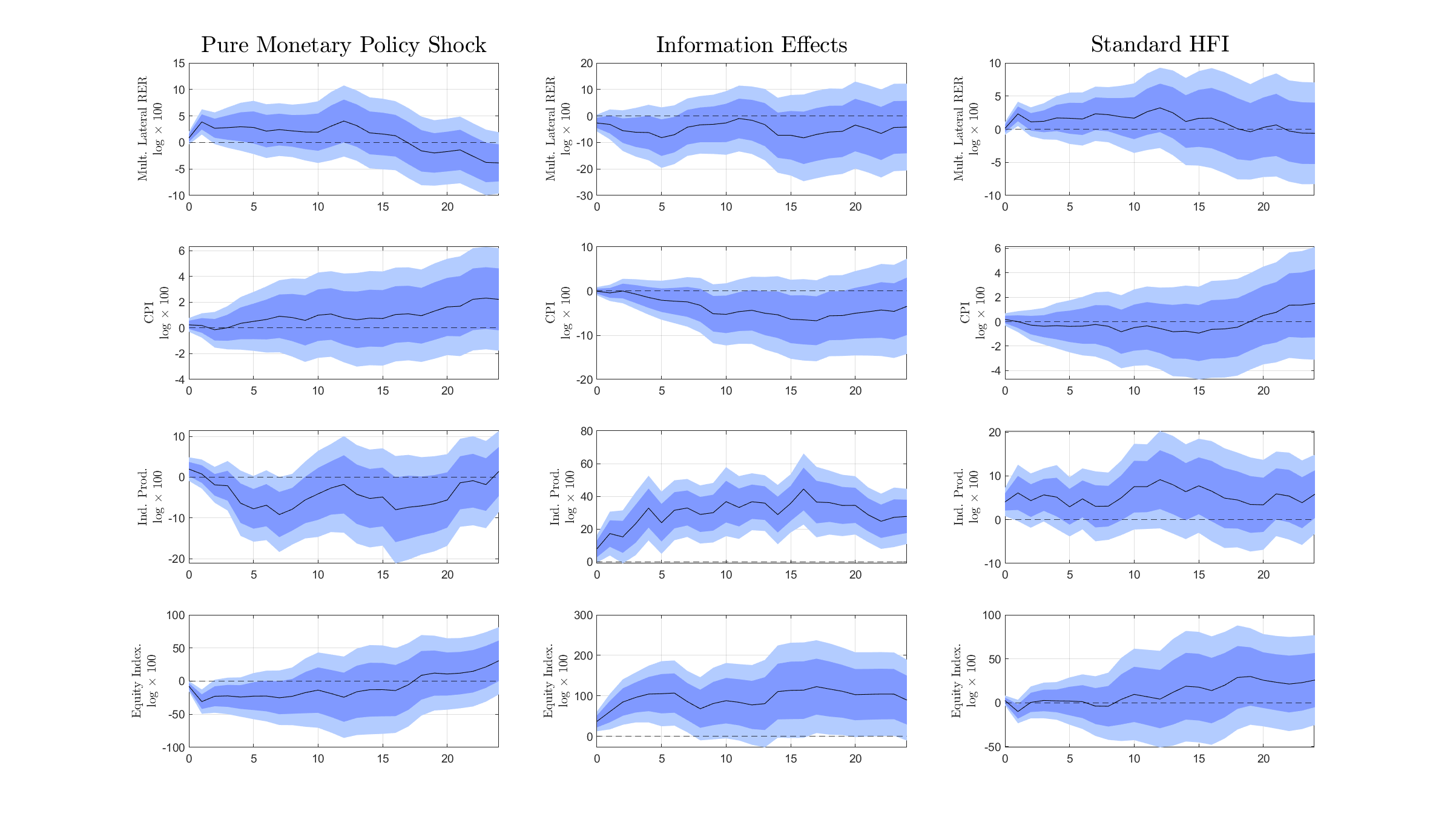}
    \caption{Impulse Response Functions - Alternative Variables \\ Multi. REER Sample - Removing LT. Rates - Starting January 1998}
    \label{fig:Removing_LT_REER_98}
    \floatfoot{\textbf{Note:} The figure is comprised of 12 sub-figures ordered in three columns and four rows. The left column relates to the estimates of $\beta^{MP}$ in Equation \ref{eq:LP_pooled}, the middle column relates to the estimate of $\beta^{FIE}$ in Equation \ref{eq:LP_pooled}, while the right column relates to estimating Equation \ref{eq:LP_pooled}, replacing the MP and FIE components with the un-orthogonalized monetary policy surprise. Sample from January 1998 to December 2019. The rows represent the impact on (i) the multilateral trade weighted real exchange rate index (in logs times 100); (ii) the consumer price index (in logs times 100); (iii) the industrial production index (in logs times 100); (iv) the equity index (in logs times 100). The solid black line represents the point estimate, the dark blue area represents the 68\% confidence interval, and the light blue area represents the 90\% confidence interval. In the text, when referring to Panel $(i,j)$, $i$ refers to the row and $j$ to the column of the figure. Each variable, in its own transformation, is demeaned at the country level.}
\end{figure}

\newpage
\begin{figure}[ht]
    \centering
    \includegraphics[scale=0.4]{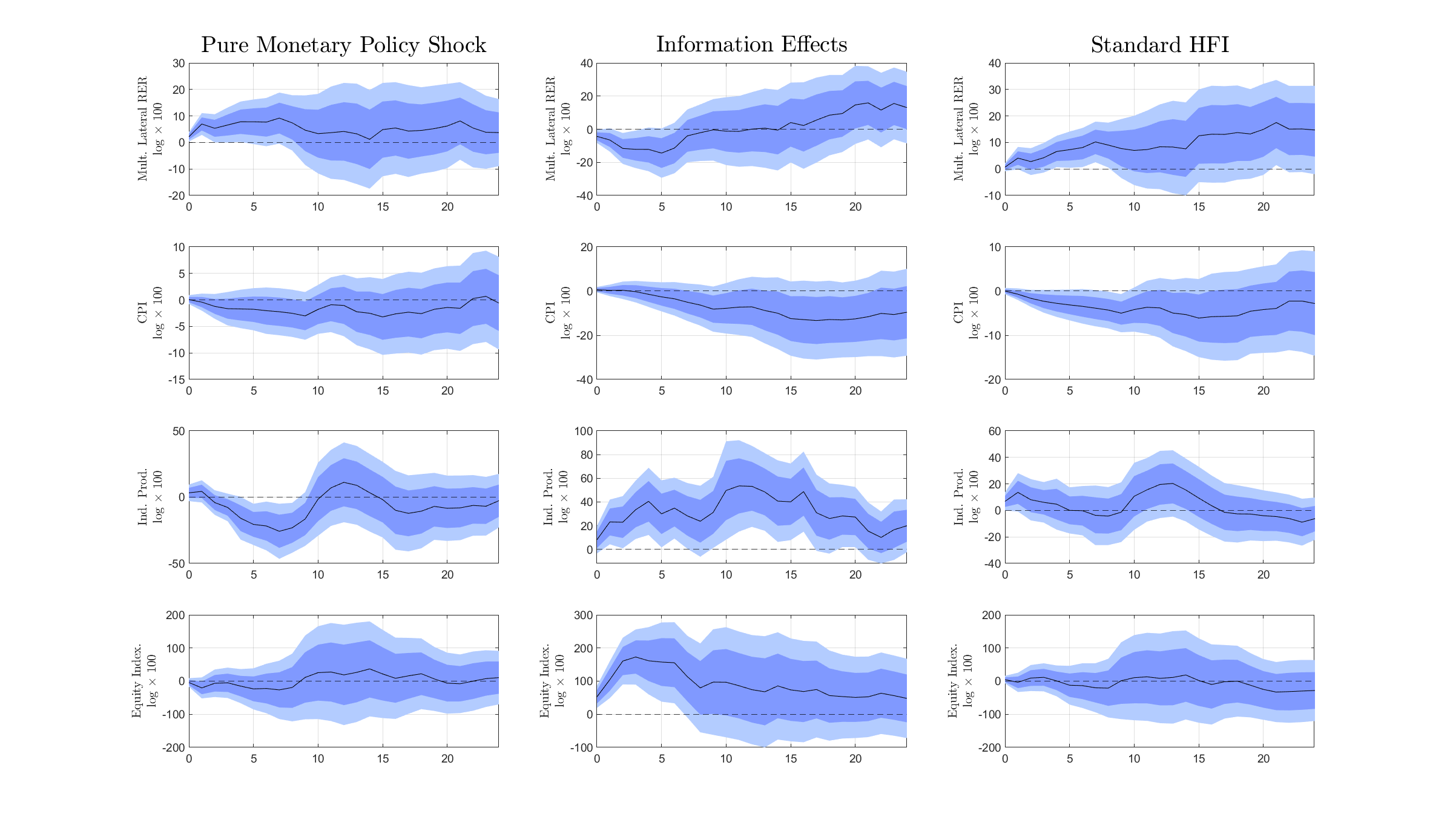}
    \caption{Impulse Response Functions - Alternative Variables \\ Multi. REER Sample - Removing LT. Rates - Starting January 2008}
    \label{fig:Removing_LT_REER_08}
    \floatfoot{\textbf{Note:} The figure is comprised of 12 sub-figures ordered in three columns and four rows. The left column relates to the estimates of $\beta^{MP}$ in Equation \ref{eq:LP_pooled}, the middle column relates to the estimate of $\beta^{FIE}$ in Equation \ref{eq:LP_pooled}, while the right column relates to estimating Equation \ref{eq:LP_pooled}, replacing the MP and FIE components with the un-orthogonalized monetary policy surprise. Sample from January 2008 to December 2019. The rows represent the impact on (i) the multilateral trade weighted real exchange rate index (in logs times 100); (ii) the consumer price index (in logs times 100); (iii) the industrial production index (in logs times 100); (iv) the equity index (in logs times 100). The solid black line represents the point estimate, the dark blue area represents the 68\% confidence interval, and the light blue area represents the 90\% confidence interval. In the text, when referring to Panel $(i,j)$, $i$ refers to the row and $j$ to the column of the figure. Each variable, in its own transformation, is demeaned at the country level.}
\end{figure}

\newpage
\begin{figure}[ht]
    \centering
    \includegraphics[scale=0.4]{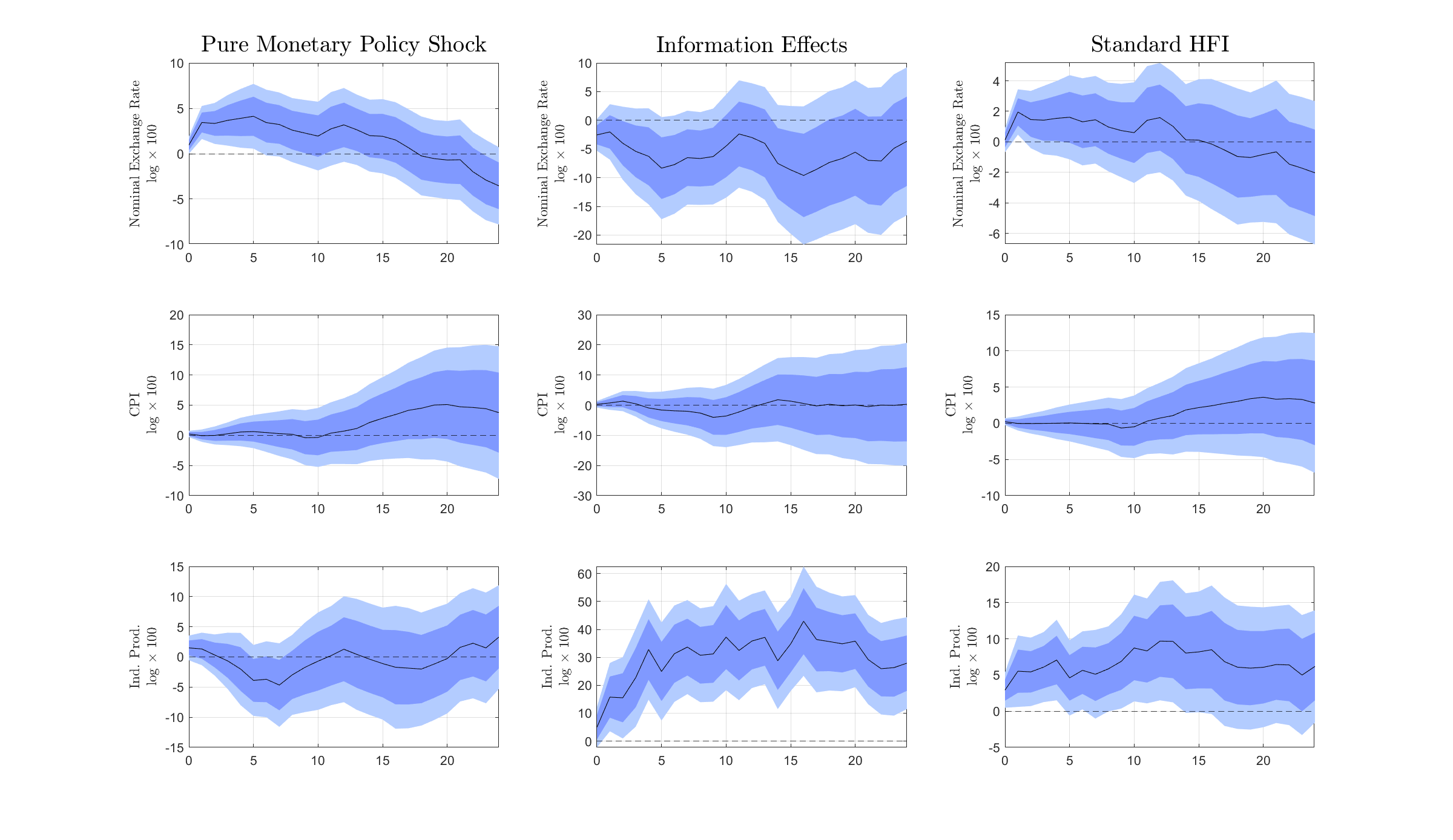}
    \caption{Impulse Response Functions - Alternative Variables \\ Removing LT Rates \& Equity Index }
    \label{fig:Removing_LT_Equity}
    \floatfoot{\textbf{Note:} The figure is comprised of 9 sub-figures ordered in three columns and three rows. The left column relates to the estimates of $\beta^{MP}$ in Equation \ref{eq:LP_pooled}, the middle column relates to the estimate of $\beta^{FIE}$ in Equation \ref{eq:LP_pooled}, while the right column relates to estimating Equation \ref{eq:LP_pooled}, replacing the MP and FIE components with the un-orthogonalized monetary policy surprise. The rows represent the impact on (i) the nominal exchange rate with the US dollar (in logs times 100); (ii) the consumer price index (in logs times 100); (iii) the industrial production index (in logs times 100). The solid black line represents the point estimate, the dark blue area represents the 68\% confidence interval, and the light blue area represents the 90\% confidence interval. In the text, when referring to Panel $(i,j)$, $i$ refers to the row and $j$ to the column of the figure. Each variable, in its own transformation, is demeaned at the country level.}
\end{figure}

\newpage
\begin{figure}[ht]
    \centering
    \includegraphics[scale=0.4]{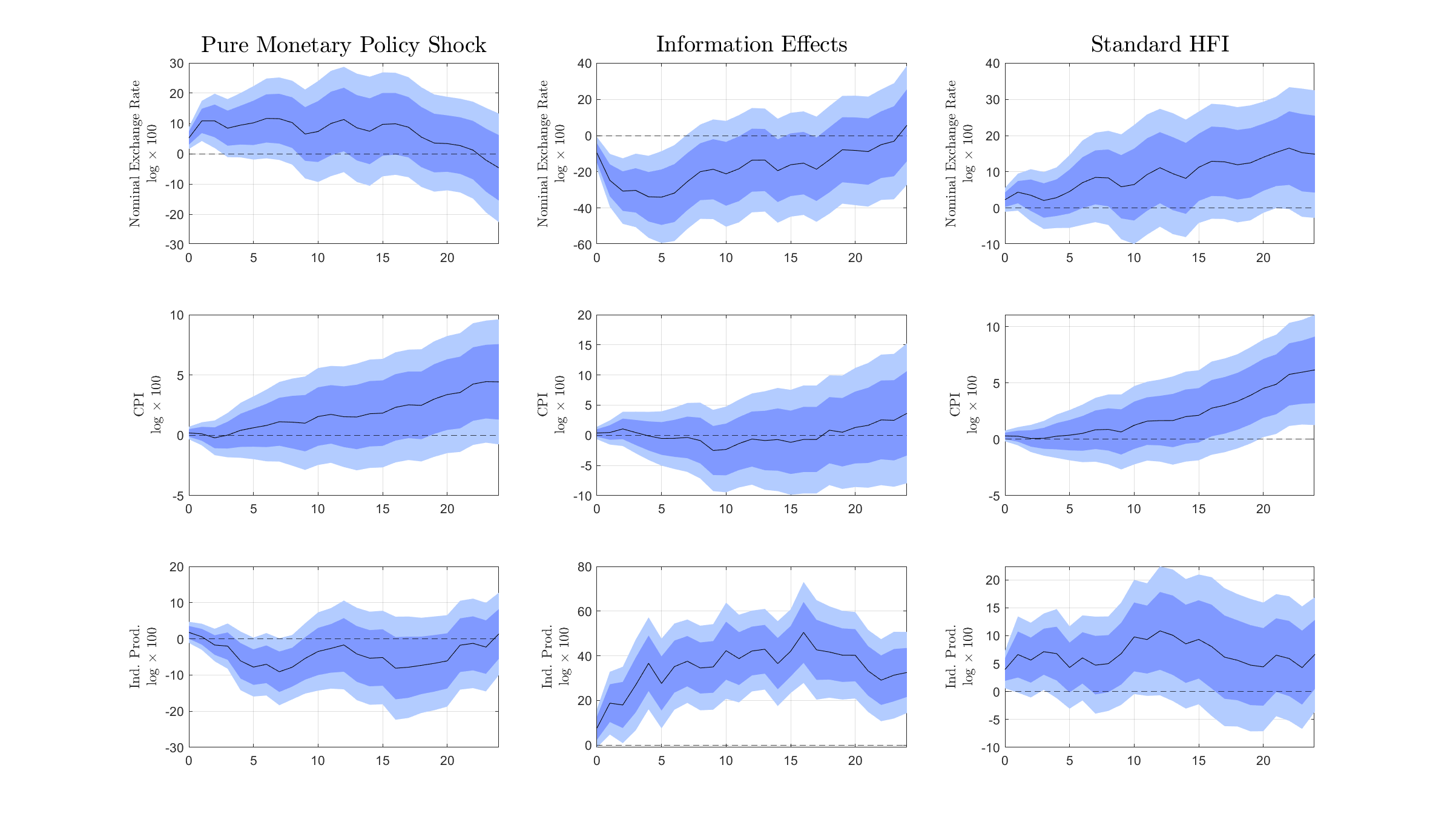}
    \caption{Impulse Response Functions - Alternative Variables \\ Removing LT Rates \& Equity Index - Starting Jan 1998}
    \label{fig:Removing_LT_Equity_98}
    \floatfoot{\textbf{Note:} The figure is comprised of 9 sub-figures ordered in three columns and three rows. The left column relates to the estimates of $\beta^{MP}$ in Equation \ref{eq:LP_pooled}, the middle column relates to the estimate of $\beta^{FIE}$ in Equation \ref{eq:LP_pooled}, while the right column relates to estimating Equation \ref{eq:LP_pooled}, replacing the MP and FIE components with the un-orthogonalized monetary policy surprise. The rows represent the impact on (i) the nominal exchange rate with the US dollar (in logs times 100); (ii) the consumer price index (in logs times 100); (iii) the industrial production index (in logs times 100). The solid black line represents the point estimate, the dark blue area represents the 68\% confidence interval, and the light blue area represents the 90\% confidence interval. In the text, when referring to Panel $(i,j)$, $i$ refers to the row and $j$ to the column of the figure. Each variable, in its own transformation, is demeaned at the country level.}
\end{figure}

\newpage
\begin{figure}[ht]
    \centering
    \includegraphics[scale=0.4]{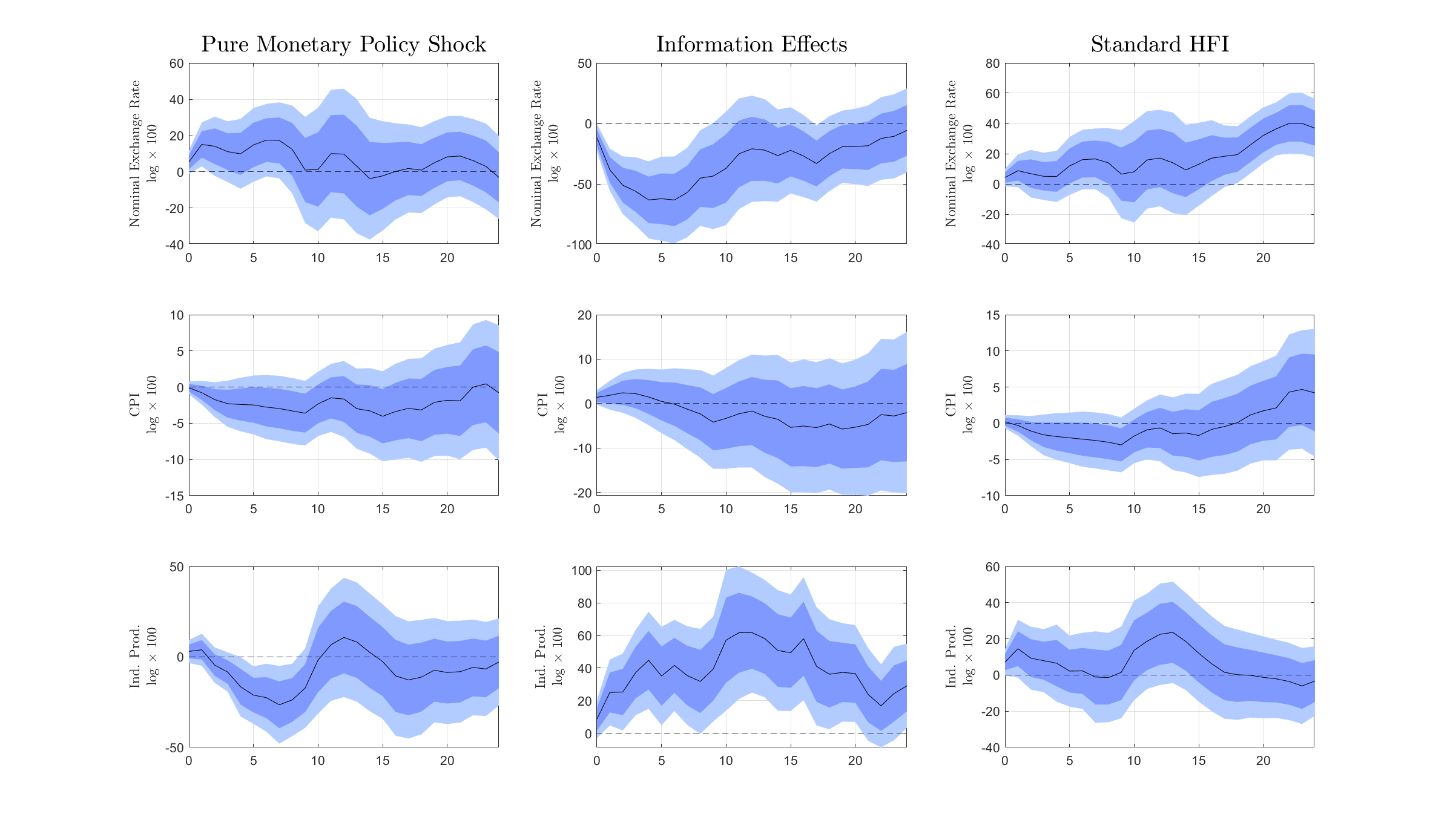}
    \caption{Impulse Response Functions - Alternative Variables \\ Removing LT Rates \& Equity Index - Starting Jan 2008}
    \label{fig:Removing_LT_Equity_08}
    \floatfoot{\textbf{Note:} The figure is comprised of 9 sub-figures ordered in three columns and three rows. The left column relates to the estimates of $\beta^{MP}$ in Equation \ref{eq:LP_pooled}, the middle column relates to the estimate of $\beta^{FIE}$ in Equation \ref{eq:LP_pooled}, while the right column relates to estimating Equation \ref{eq:LP_pooled}, replacing the MP and FIE components with the un-orthogonalized monetary policy surprise. The rows represent the impact on (i) the nominal exchange rate with the US dollar (in logs times 100); (ii) the consumer price index (in logs times 100); (iii) the industrial production index (in logs times 100). The solid black line represents the point estimate, the dark blue area represents the 68\% confidence interval, and the light blue area represents the 90\% confidence interval. In the text, when referring to Panel $(i,j)$, $i$ refers to the row and $j$ to the column of the figure. Each variable, in its own transformation, is demeaned at the country level.}
\end{figure}

\newpage
\begin{figure}[ht]
    \centering
    \includegraphics[scale=0.4]{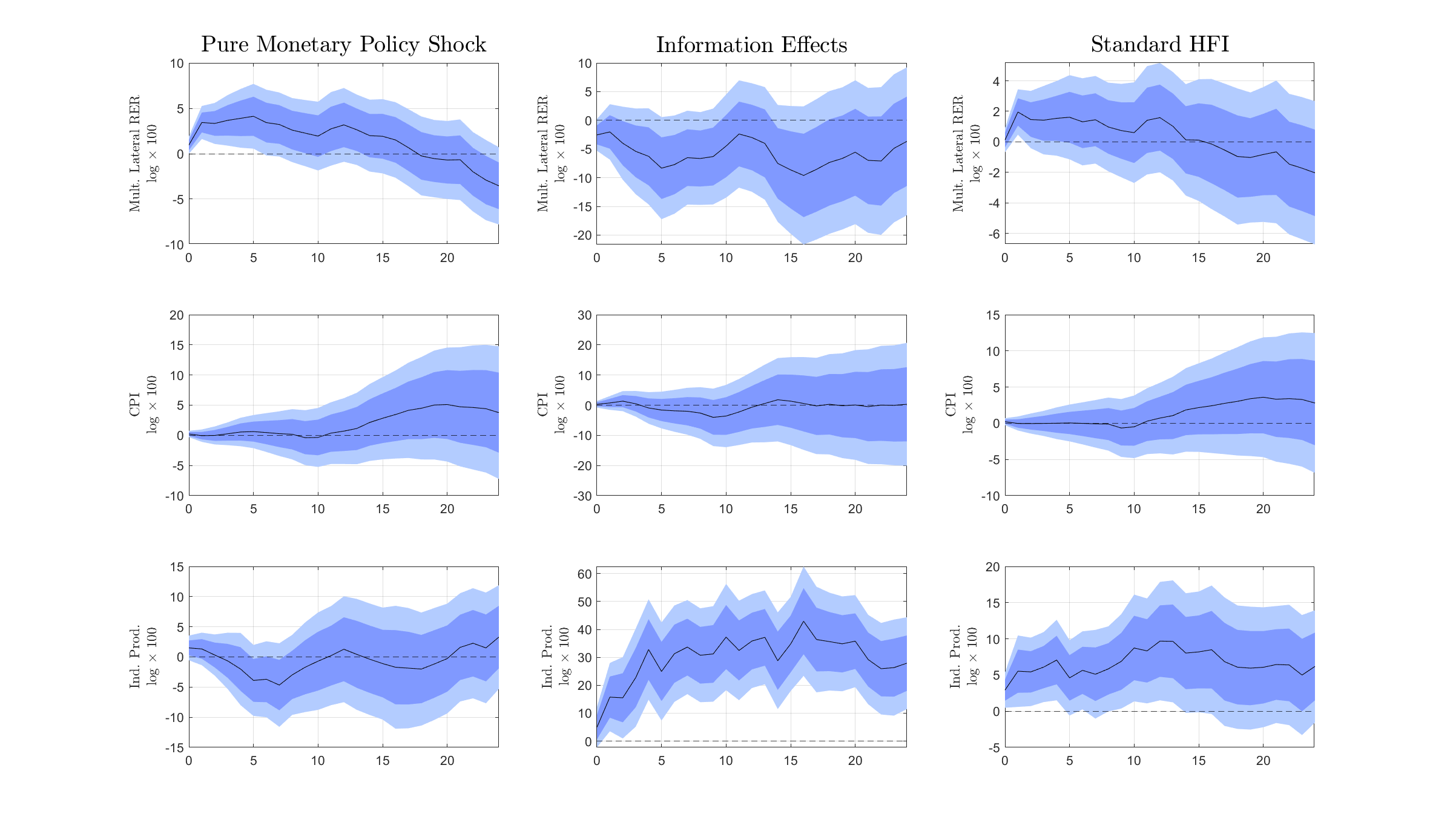}
    \caption{Impulse Response Functions - Alternative Variables \\ Multi. REER Sample - Removing LT Rates \& Equity Index }
    \label{fig:Removing_LT_Equity_REER}
    \floatfoot{\textbf{Note:} The figure is comprised of 9 sub-figures ordered in three columns and three rows. The left column relates to the estimates of $\beta^{MP}$ in Equation \ref{eq:LP_pooled}, the middle column relates to the estimate of $\beta^{FIE}$ in Equation \ref{eq:LP_pooled}, while the right column relates to estimating Equation \ref{eq:LP_pooled}, replacing the MP and FIE components with the un-orthogonalized monetary policy surprise. The rows represent the impact on (i) the multilateral trade weighted real exchange rate index (in logs times 100); (ii) the consumer price index (in logs times 100); (iii) the industrial production index (in logs times 100). The solid black line represents the point estimate, the dark blue area represents the 68\% confidence interval, and the light blue area represents the 90\% confidence interval. In the text, when referring to Panel $(i,j)$, $i$ refers to the row and $j$ to the column of the figure. Each variable, in its own transformation, is demeaned at the country level.}
\end{figure}

\newpage
\begin{figure}[ht]
    \centering
    \includegraphics[scale=0.4]{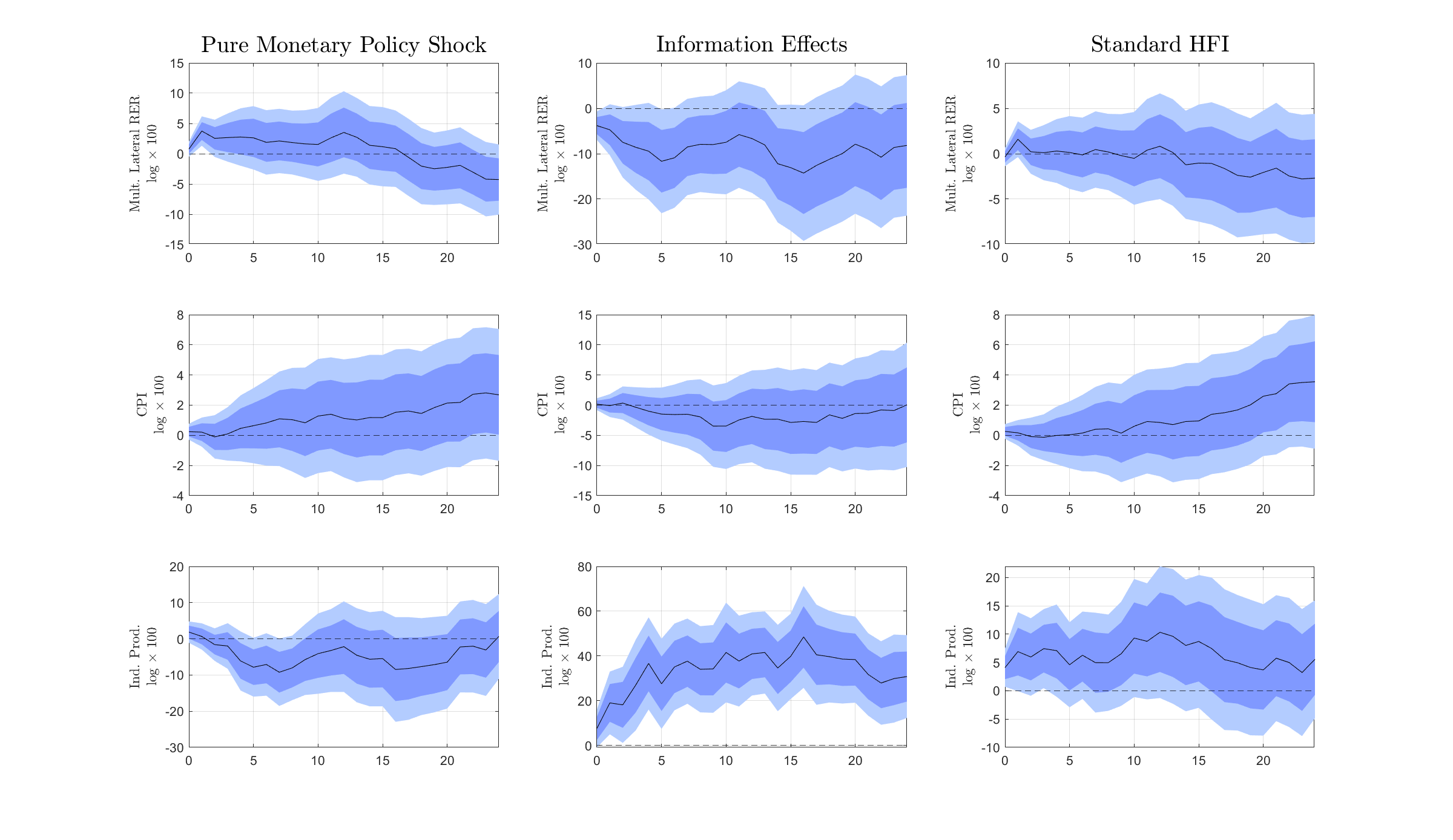}
    \caption{Impulse Response Functions - Alternative Variables \\ Multi. REER Sample - Removing LT Rates \& Equity Index - Starting Jan 1998}
    \label{fig:Removing_LT_Equity_REER_98}
    \floatfoot{\textbf{Note:} The figure is comprised of 9 sub-figures ordered in three columns and three rows. The left column relates to the estimates of $\beta^{MP}$ in Equation \ref{eq:LP_pooled}, the middle column relates to the estimate of $\beta^{FIE}$ in Equation \ref{eq:LP_pooled}, while the right column relates to estimating Equation \ref{eq:LP_pooled}, replacing the MP and FIE components with the un-orthogonalized monetary policy surprise. The rows represent the impact on (i) the multilateral trade weighted real exchange rate index (in logs times 100); (ii) the consumer price index (in logs times 100); (iii) the industrial production index (in logs times 100). The solid black line represents the point estimate, the dark blue area represents the 68\% confidence interval, and the light blue area represents the 90\% confidence interval. In the text, when referring to Panel $(i,j)$, $i$ refers to the row and $j$ to the column of the figure. Each variable, in its own transformation, is demeaned at the country level.}
\end{figure}

\newpage
\begin{figure}[ht]
    \centering
    \includegraphics[scale=0.4]{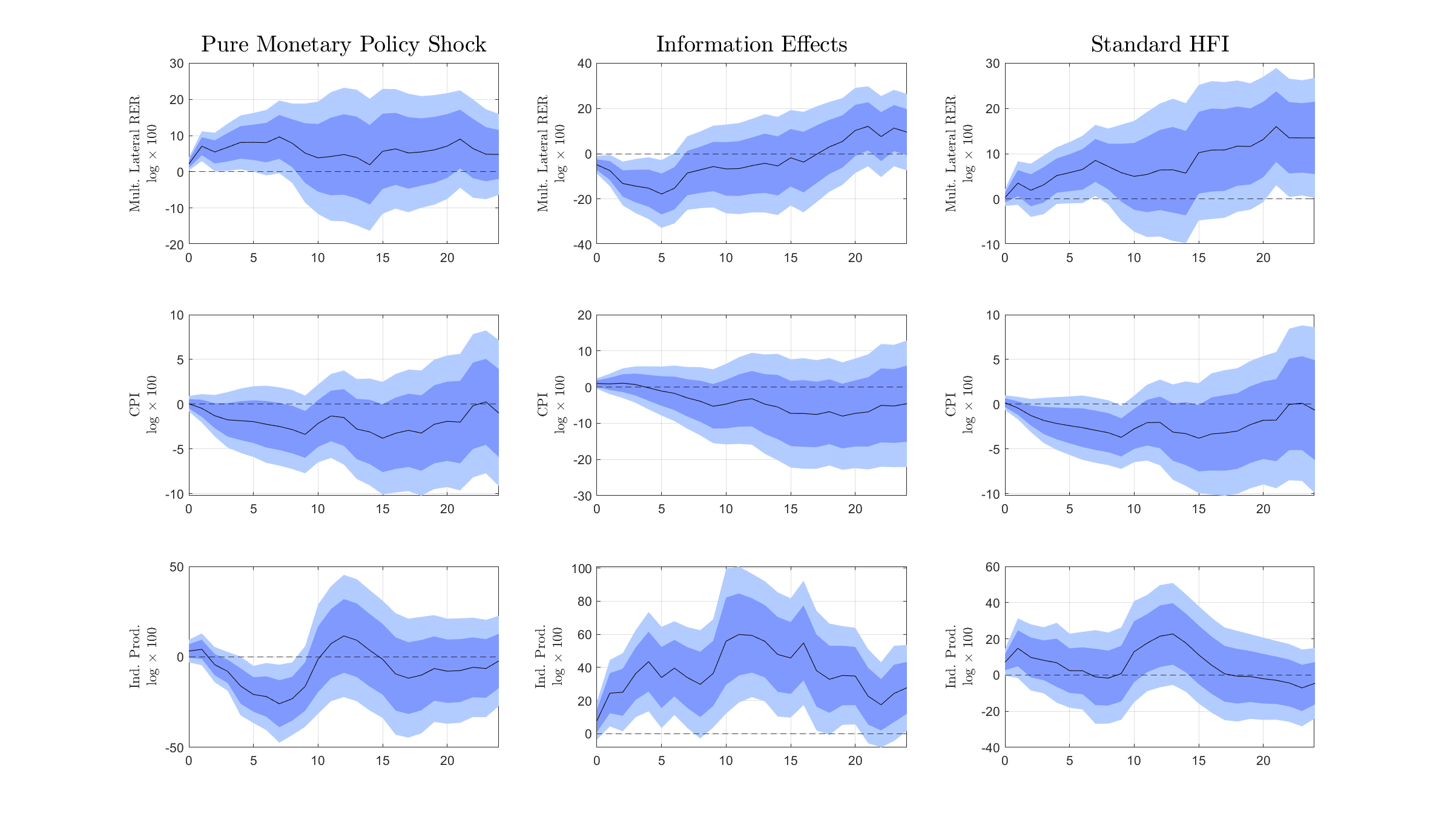}
    \caption{Impulse Response Functions - Alternative Variables \\ Multi. REER Sample - Removing LT Rates \& Equity Index - Starting Jan 2008}
    \label{fig:Removing_LT_Equity_REER_08}
    \floatfoot{\textbf{Note:} The figure is comprised of 9 sub-figures ordered in three columns and three rows. The left column relates to the estimates of $\beta^{MP}$ in Equation \ref{eq:LP_pooled}, the middle column relates to the estimate of $\beta^{FIE}$ in Equation \ref{eq:LP_pooled}, while the right column relates to estimating Equation \ref{eq:LP_pooled}, replacing the MP and FIE components with the un-orthogonalized monetary policy surprise. The rows represent the impact on (i) the multilateral trade weighted real exchange rate index (in logs times 100); (ii) the consumer price index (in logs times 100); (iii) the industrial production index (in logs times 100). The solid black line represents the point estimate, the dark blue area represents the 68\% confidence interval, and the light blue area represents the 90\% confidence interval. In the text, when referring to Panel $(i,j)$, $i$ refers to the row and $j$ to the column of the figure. Each variable, in its own transformation, is demeaned at the country level.}
\end{figure}

\newpage
\begin{figure}
    \centering
    \includegraphics[scale=0.4]{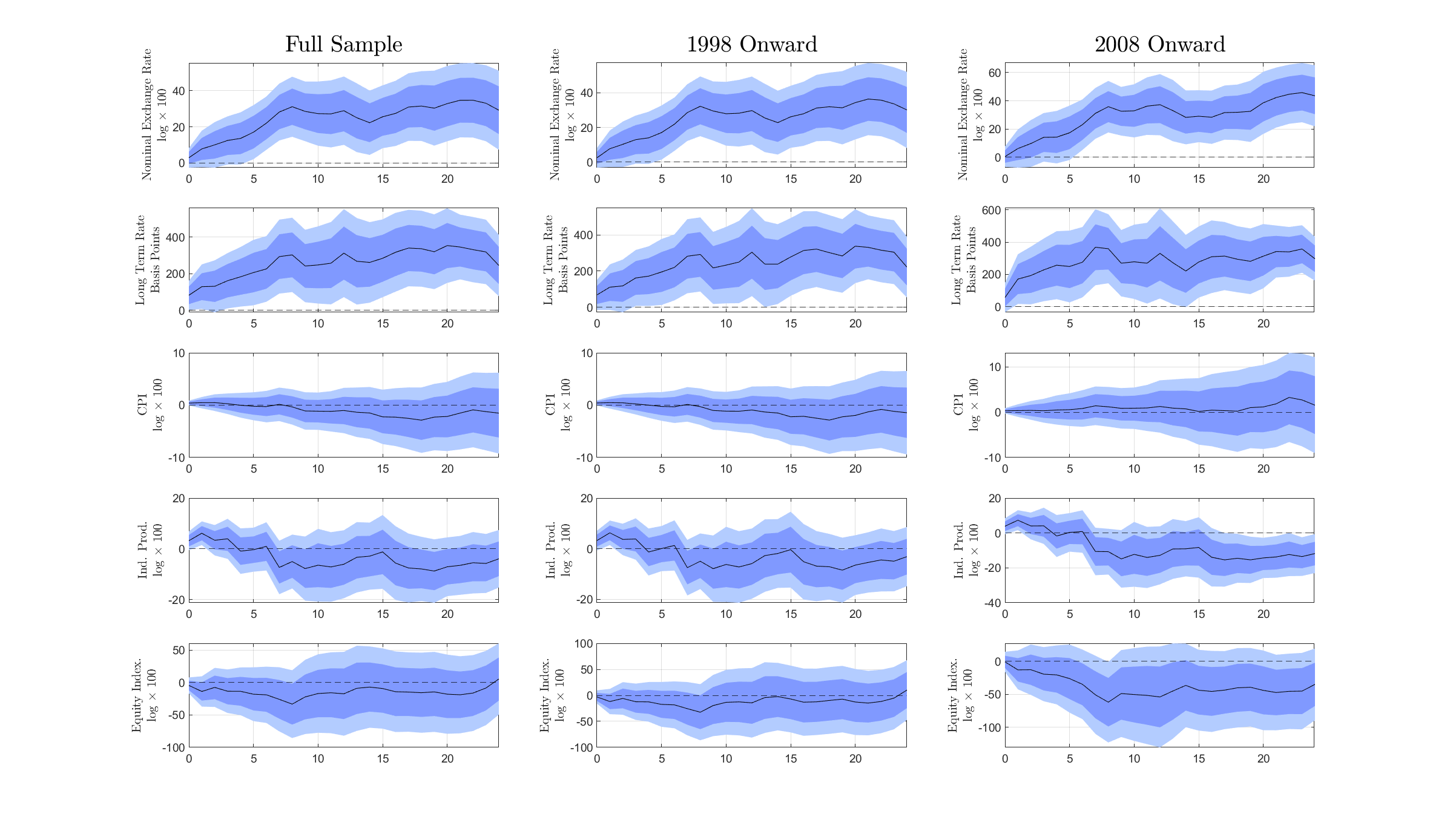}
    \caption{Impulse Response Functions \\ \cite{bu2021unified}}
    \label{fig:BenchmarkNER_BU}
    \floatfoot{\textbf{Note:} The figure is comprised of 15 sub-figures ordered in three columns and five rows. The left column relates to the estimates of $\beta^{MP}$ in Equation \ref{eq:LP_pooled} which replaces the benchmark $MP$ and $FIE$ components with the monetary policy shock introduced by \cite{bu2021unified} for the full sample of January 1994 to September 2019, the middle column for the sample of January 1998 to September 2019, while the right column the sample January 2008 to September 2019. The rows represent the impact on (i) the nominal exchange rate with the US dollar (in logs times 100); (ii) long term interest rates in basis points; (iii) the consumer price index (in logs times 100); (iv) the industrial production index (in logs times 100); (v) the equity index (in logs times 100). The solid black line represents the point estimate, the dark blue area represents the 68\% confidence interval, and the light blue area represents the 90\% confidence interval. In the text, when referring to Panel $(i,j)$, $i$ refers to the row and $j$ to the column of the figure. Each variable, in its own transformation, is demeaned at the country level.}
\end{figure}

\newpage
\begin{figure}
    \centering
    \includegraphics[scale=0.4]{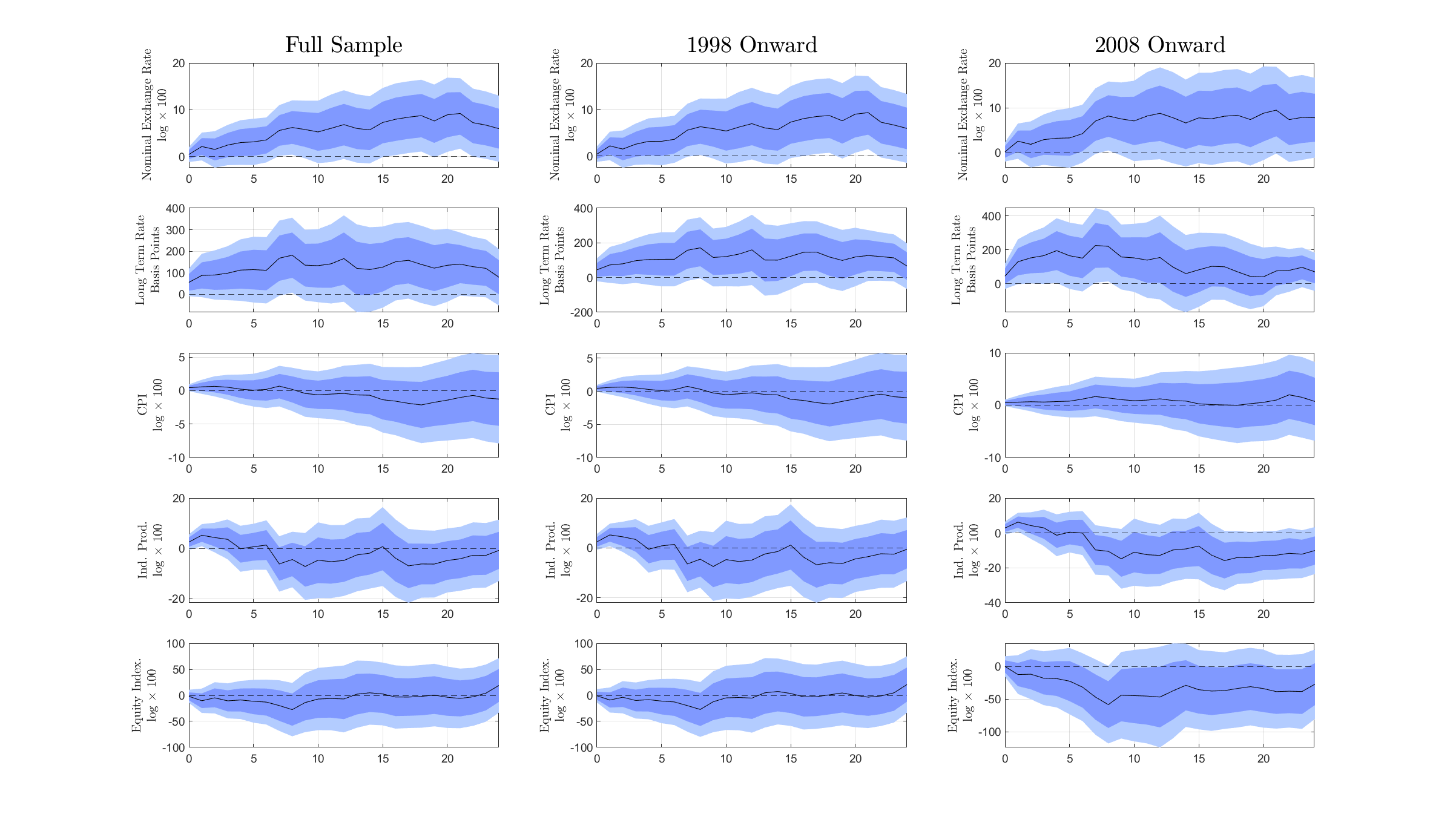}
    \caption{Impulse Response Functions \\ Multi. REER Sample \cite{bu2021unified}}
    \label{fig:BenchmarkREER_BU}
    \floatfoot{\textbf{Note:} The figure is comprised of 15 sub-figures ordered in three columns and five rows. The left column relates to the estimates of $\beta^{MP}$ in Equation \ref{eq:LP_pooled} which replaces the benchmark $MP$ and $FIE$ components with the monetary policy shock introduced by \cite{bu2021unified} for the full sample of January 1994 to September 2019, the middle column for the sample of January 1998 to September 2019, while the right column the sample January 2008 to September 2019. The rows represent the impact on (i) the trade weighted multilateral real exchange rate (in logs times 100); (ii) long term interest rates in basis points; (iii) the consumer price index (in logs times 100); (iv) the industrial production index (in logs times 100); (v) the equity index (in logs times 100). The solid black line represents the point estimate, the dark blue area represents the 68\% confidence interval, and the light blue area represents the 90\% confidence interval. In the text, when referring to Panel $(i,j)$, $i$ refers to the row and $j$ to the column of the figure. Each variable, in its own transformation, is demeaned at the country level.}
\end{figure}

\newpage
\begin{figure}
    \centering
    \includegraphics[scale=0.4]{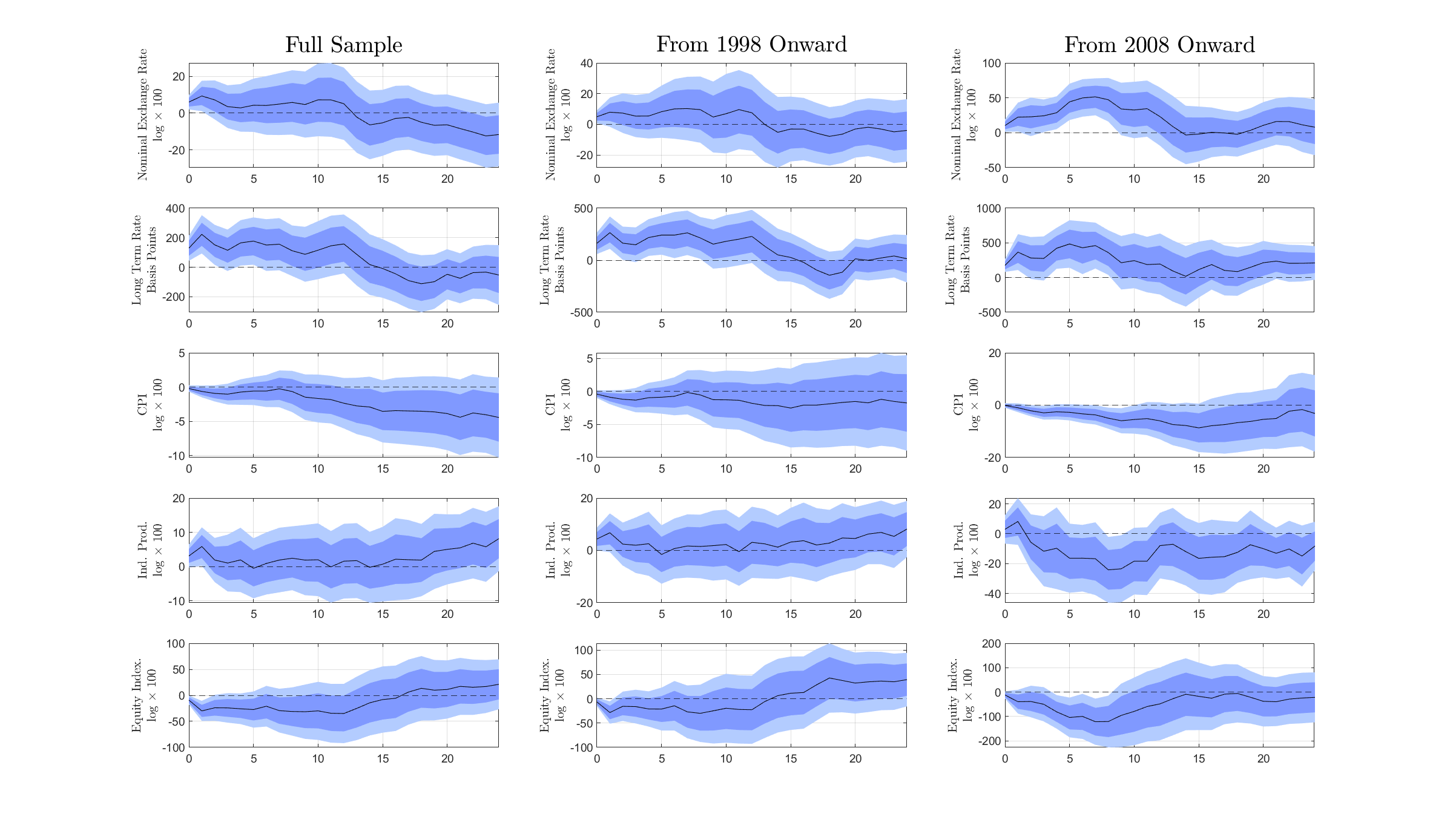}
    \caption{Impulse Response Functions - Pure Monetary Policy Shock \\ \cite{jarocinski2020deconstructing}}
    \label{fig:JK_Comparison_MP}
    \floatfoot{\textbf{Note:} The figure is comprised of 15 sub-figures ordered in three columns and five rows. To estimate the international spillovers I replace the $MP$ and $FIE$ components in the benchmark regression in Equation \ref{eq:LP_pooled} with the pure monetary policy shock and the information disclosure shock constructed by \cite{jarocinski2020deconstructing}. This figure presents the results for the impact of the pure monetary policy shock. The left column presents the results for the sample January 1990 to December 2016, the middle column for the sample January 1998 to December 2016, and the right column for the sample January 2008 to December 2016. The rows represent the impact on (i) the nominal exchange rate with the US dollar (in logs times 100); (ii) long term interest rates in basis points; (iii) the consumer price index (in logs times 100); (iv) the industrial production index (in logs times 100); (v) the equity index (in logs times 100). The solid black line represents the point estimate, the dark blue area represents the 68\% confidence interval, and the light blue area represents the 90\% confidence interval. In the text, when referring to Panel $(i,j)$, $i$ refers to the row and $j$ to the column of the figure. Each variable, in its own transformation, is demeaned at the country level.}
\end{figure}
\newpage
\begin{figure}
    \centering
    \includegraphics[scale=0.4]{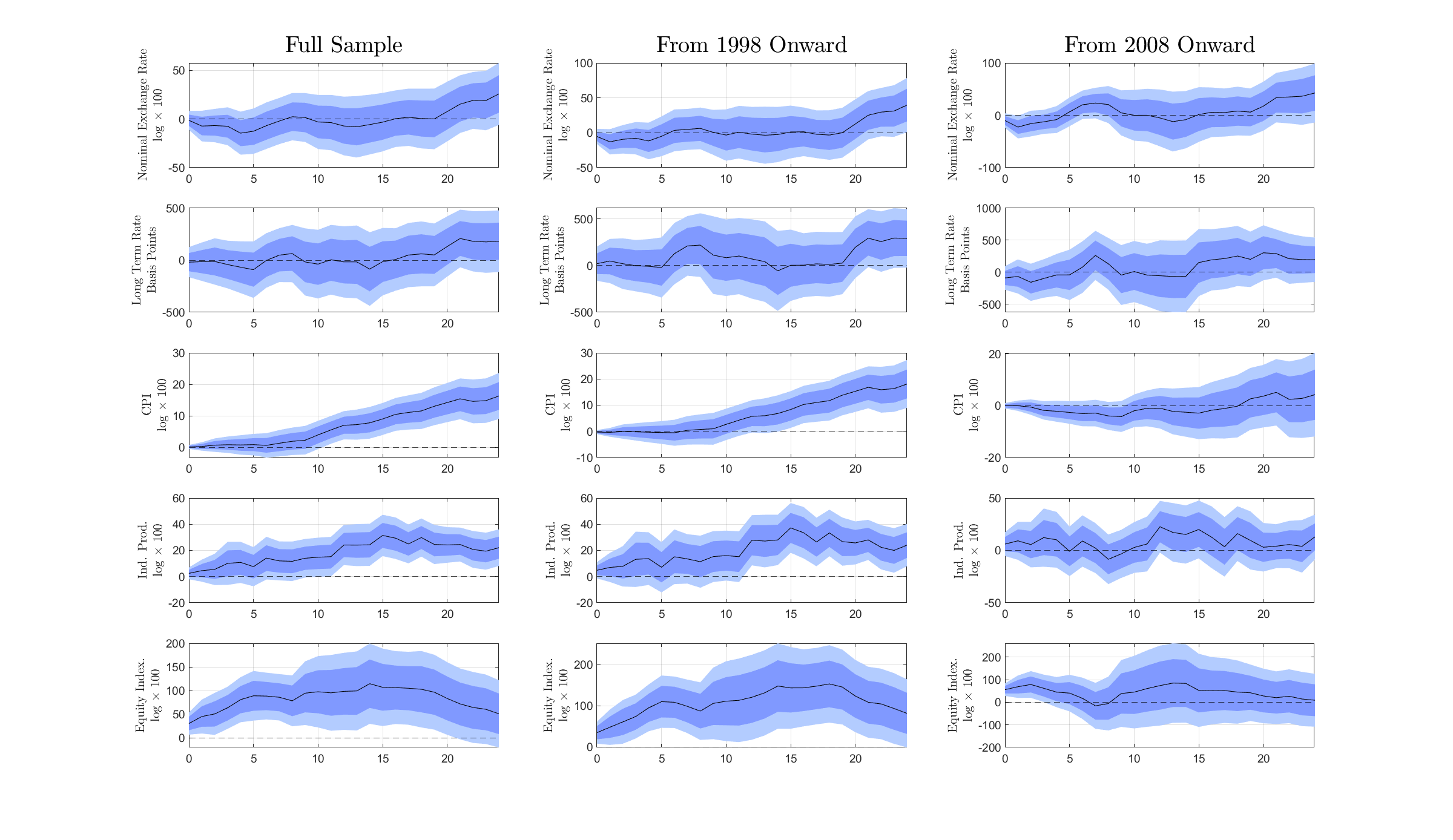}
    \caption{Impulse Response Functions - Information Disclosure Shock \\ \cite{jarocinski2020deconstructing}}
    \label{fig:JK_Comparison_CBI}
    \floatfoot{\textbf{Note:} The figure is comprised of 15 sub-figures ordered in three columns and five rows. To estimate the international spillovers I replace the $MP$ and $FIE$ components in the benchmark regression in Equation \ref{eq:LP_pooled} with the pure monetary policy shock and the information disclosure shock constructed by \cite{jarocinski2020deconstructing}. This figure presents the results for the impact of the information disclosure shock. The left column presents the results for the sample January 1990 to December 2016, the middle column for the sample January 1998 to December 2016, and the right column for the sample January 2008 to December 2016. The rows represent the impact on (i) the nominal exchange rate with the US dollar (in logs times 100); (ii) long term interest rates in basis points; (iii) the consumer price index (in logs times 100); (iv) the industrial production index (in logs times 100); (v) the equity index (in logs times 100). The solid black line represents the point estimate, the dark blue area represents the 68\% confidence interval, and the light blue area represents the 90\% confidence interval. In the text, when referring to Panel $(i,j)$, $i$ refers to the row and $j$ to the column of the figure. Each variable, in its own transformation, is demeaned at the country level.}
\end{figure}
\newpage
\begin{figure}
    \centering
    \includegraphics[scale=0.4]{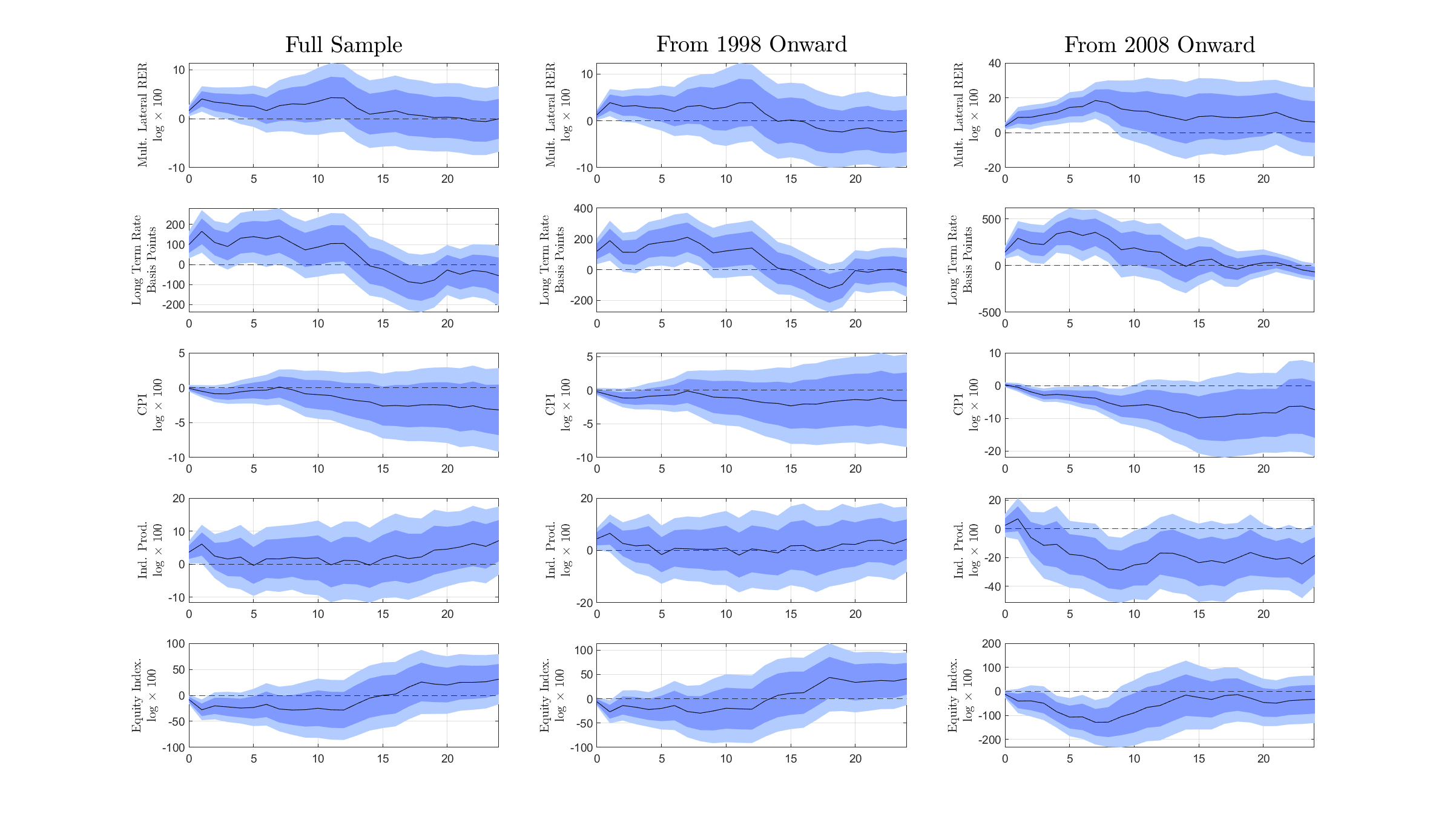}
    \caption{Impulse Response Functions - Pure Monetary Policy Shock \\ Multi. REER Sample \cite{jarocinski2020deconstructing}}
    \label{fig:JK_Comparison_MP_REER}
    \floatfoot{\textbf{Note:} The figure is comprised of 15 sub-figures ordered in three columns and five rows. To estimate the international spillovers I replace the $MP$ and $FIE$ components in the benchmark regression in Equation \ref{eq:LP_pooled} with the pure monetary policy shock and the information disclosure shock constructed by \cite{jarocinski2020deconstructing}.  This figure presents the results for the impact of the pure monetary policy shock.   
    The left column presents the results for the sample January 1990 to December 2016, the middle column for the sample January 1998 to December 2016, and the right column for the sample January 2008 to December 2016. The rows represent the impact on (i) the trade weighted multilateral real exchange rate (in logs times 100); (ii) long term interest rates in basis points; (iii) the consumer price index (in logs times 100); (iv) the industrial production index (in logs times 100); (v) the equity index (in logs times 100). The solid black line represents the point estimate, the dark blue area represents the 68\% confidence interval, and the light blue area represents the 90\% confidence interval. In the text, when referring to Panel $(i,j)$, $i$ refers to the row and $j$ to the column of the figure. Each variable, in its own transformation, is demeaned at the country level.}
\end{figure}
\newpage
\begin{figure}
    \centering
    \includegraphics[scale=0.4]{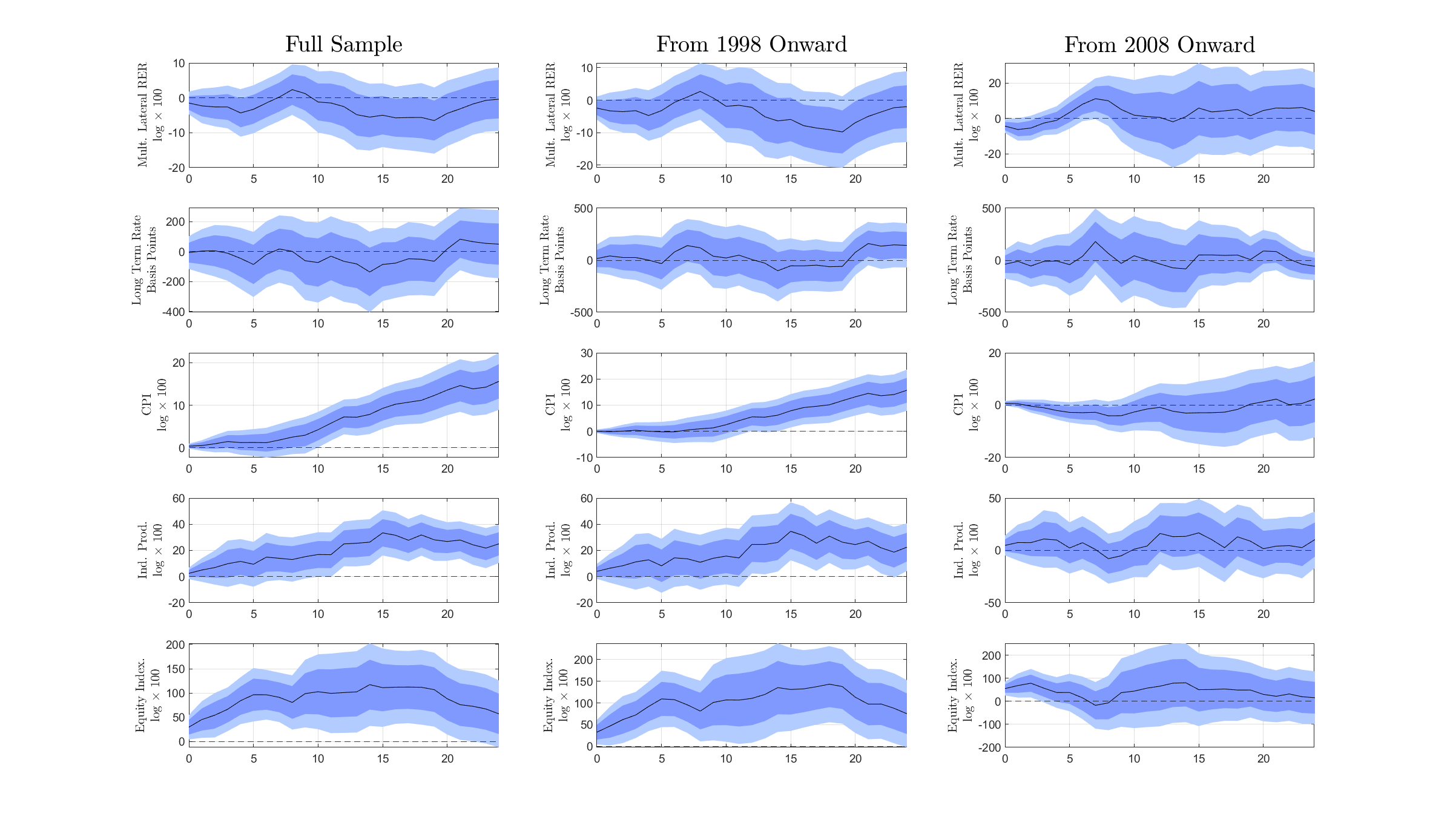}
    \caption{Impulse Response Functions - Information Disclosure Shock \\ Multi. REER Sample \cite{jarocinski2020deconstructing}}
    \label{fig:JK_Comparison_CBI_REER}
    \floatfoot{\textbf{Note:} The figure is comprised of 15 sub-figures ordered in three columns and five rows. To estimate the international spillovers I replace the $MP$ and $FIE$ components in the benchmark regression in Equation \ref{eq:LP_pooled} with the pure monetary policy shock and the information disclosure shock constructed by \cite{jarocinski2020deconstructing}. This figure presents the results for the impact of the information disclosure shock. The left column presents the results for the sample January 1990 to December 2016, the middle column for the sample January 1998 to December 2016, and the right column for the sample January 2008 to December 2016. The rows represent the impact on (i) the trade weighted multilateral real exchange rate (in logs times 100); (ii) long term interest rates in basis points; (iii) the consumer price index (in logs times 100); (iv) the industrial production index (in logs times 100); (v) the equity index (in logs times 100). The solid black line represents the point estimate, the dark blue area represents the 68\% confidence interval, and the light blue area represents the 90\% confidence interval. In the text, when referring to Panel $(i,j)$, $i$ refers to the row and $j$ to the column of the figure. Each variable, in its own transformation, is demeaned at the country level.}
\end{figure}



\newpage
\begin{figure}
    \centering
    \includegraphics[scale=0.4]{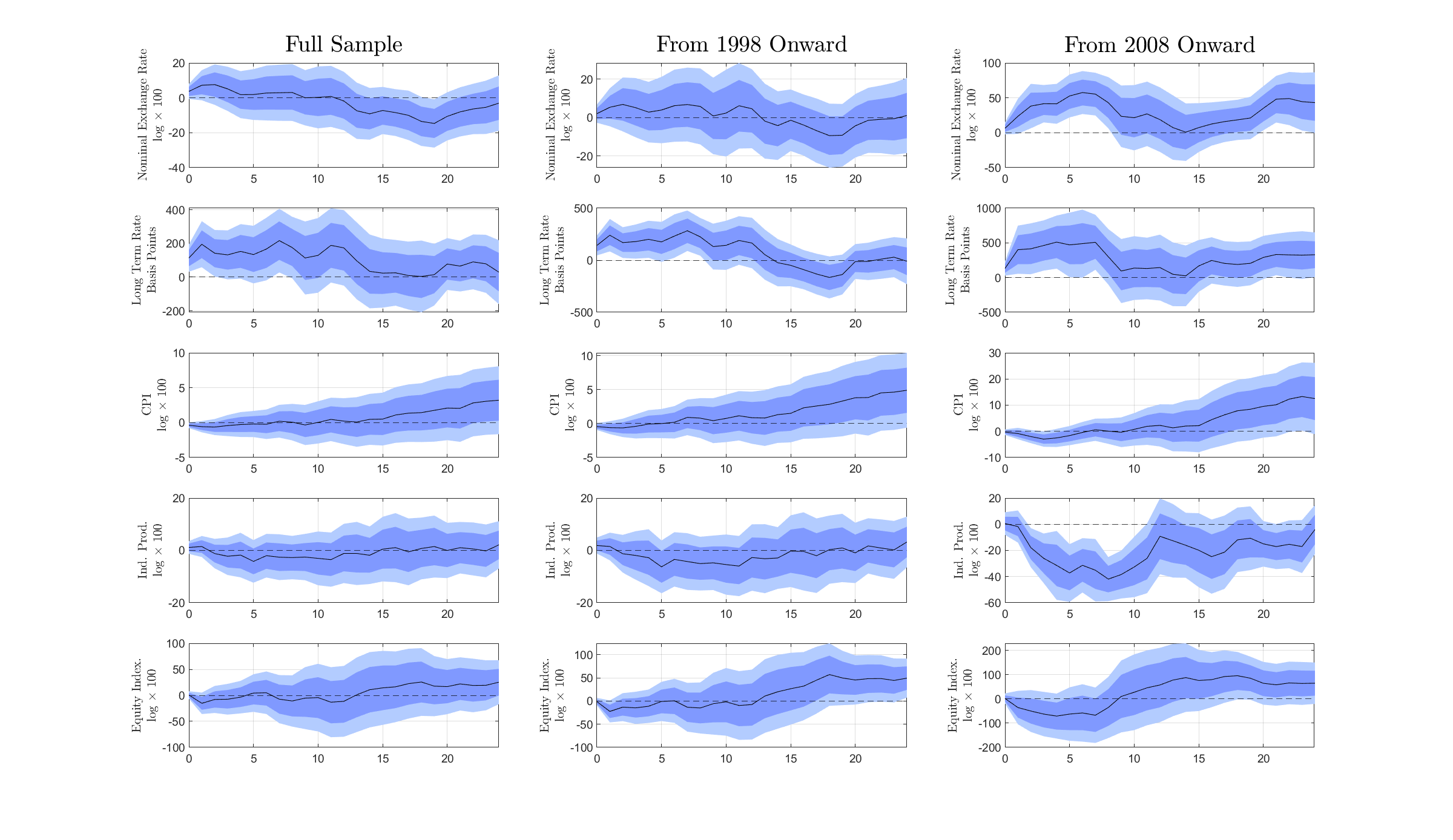}
    \caption{Impulse Response Functions - Pure Monetary Policy Shock \\ \cite{miranda2021transmission}}
    \label{fig:MAR_Comparison_MP}
    \floatfoot{\textbf{Note:} The figure is comprised of 15 sub-figures ordered in three columns and five rows. To estimate the international spillovers I replace the $MP$ and $FIE$ components in the benchmark regression in Equation \ref{eq:LP_pooled} with the pure monetary policy shock and the information component or ``signalling effect'' constructed by \cite{miranda2021transmission}. This figure presents the results for the impact of the pure monetary policy shock. The left column presents the results for the sample January 1990 to December 2016, the middle column for the sample January 1998 to December 2016, and the right column for the sample January 2008 to December 2016. The rows represent the impact on (i) the nominal exchange rate with the US dollar (in logs times 100); (ii) long term interest rates in basis points; (iii) the consumer price index (in logs times 100); (iv) the industrial production index (in logs times 100); (v) the equity index (in logs times 100). The solid black line represents the point estimate, the dark blue area represents the 68\% confidence interval, and the light blue area represents the 90\% confidence interval. In the text, when referring to Panel $(i,j)$, $i$ refers to the row and $j$ to the column of the figure. Each variable, in its own transformation, is demeaned at the country level.}
\end{figure}
\newpage
\begin{figure}
    \centering
    \includegraphics[scale=0.4]{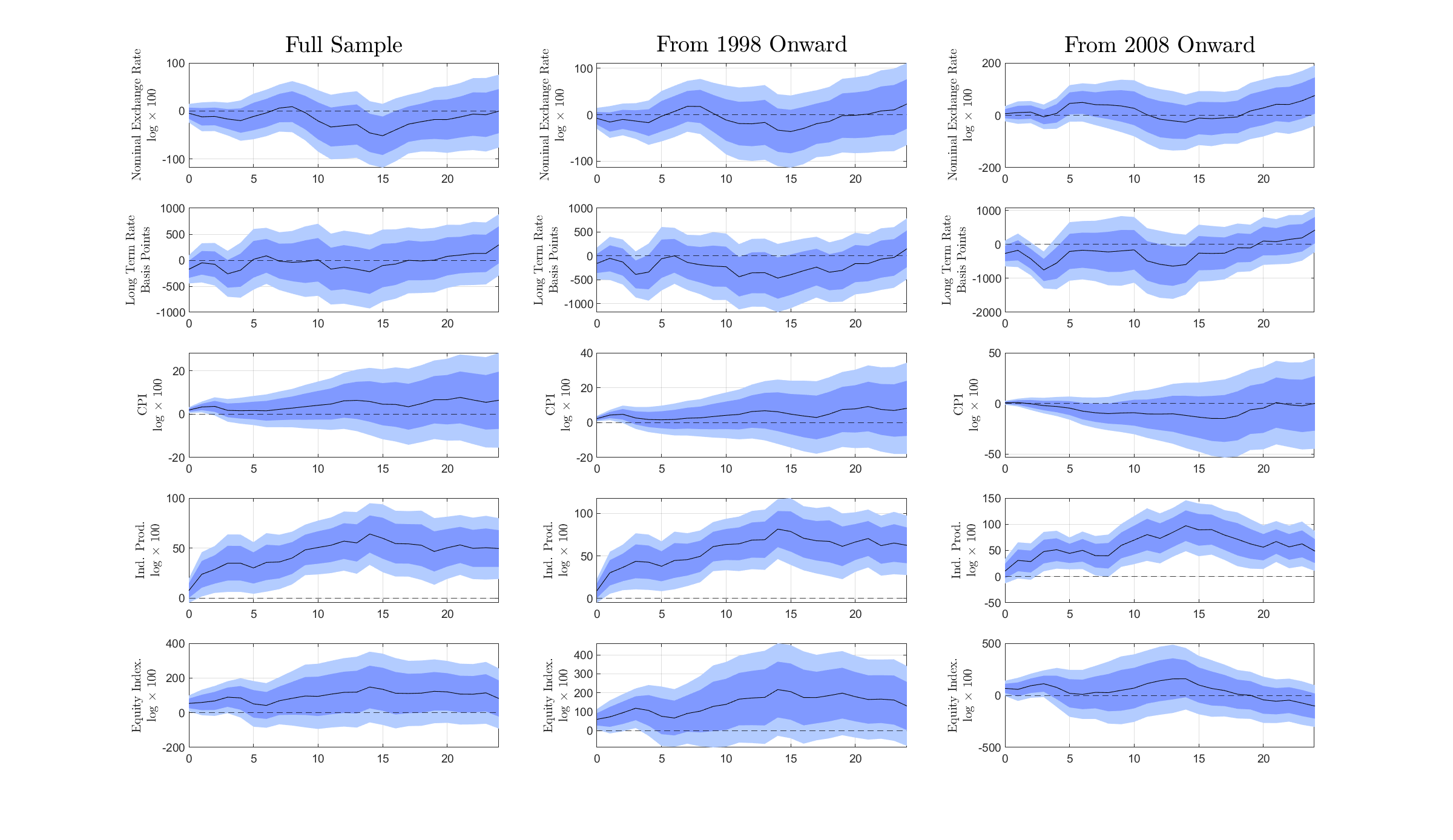}
    \caption{Impulse Response Functions - Information Component \\ \cite{miranda2021transmission}}
    \label{fig:MAR_Comparison_CBI}
    \floatfoot{\textbf{Note:} The figure is comprised of 15 sub-figures ordered in three columns and five rows. I replace the $MP$ and $FIE$ components in the benchmark regression in Equation \ref{eq:LP_pooled} with the pure monetary policy shock and the information component or ``signalling effect'' constructed by \cite{miranda2021transmission}. This figure presents the results for the impact of the information component or ``signalling effect'' shock. The left column presents the results for the sample January 1990 to December 2016, the middle column for the sample January 1998 to December 2016, and the right column for the sample January 2008 to December 2016. The rows represent the impact on (i) the nominal exchange rate with the US dollar (in logs times 100); (ii) long term interest rates in basis points; (iii) the consumer price index (in logs times 100); (iv) the industrial production index (in logs times 100); (v) the equity index (in logs times 100). The solid black line represents the point estimate, the dark blue area represents the 68\% confidence interval, and the light blue area represents the 90\% confidence interval. In the text, when referring to Panel $(i,j)$, $i$ refers to the row and $j$ to the column of the figure. Each variable, in its own transformation, is demeaned at the country level.}
\end{figure}
\newpage
\begin{figure}
    \centering
    \includegraphics[scale=0.4]{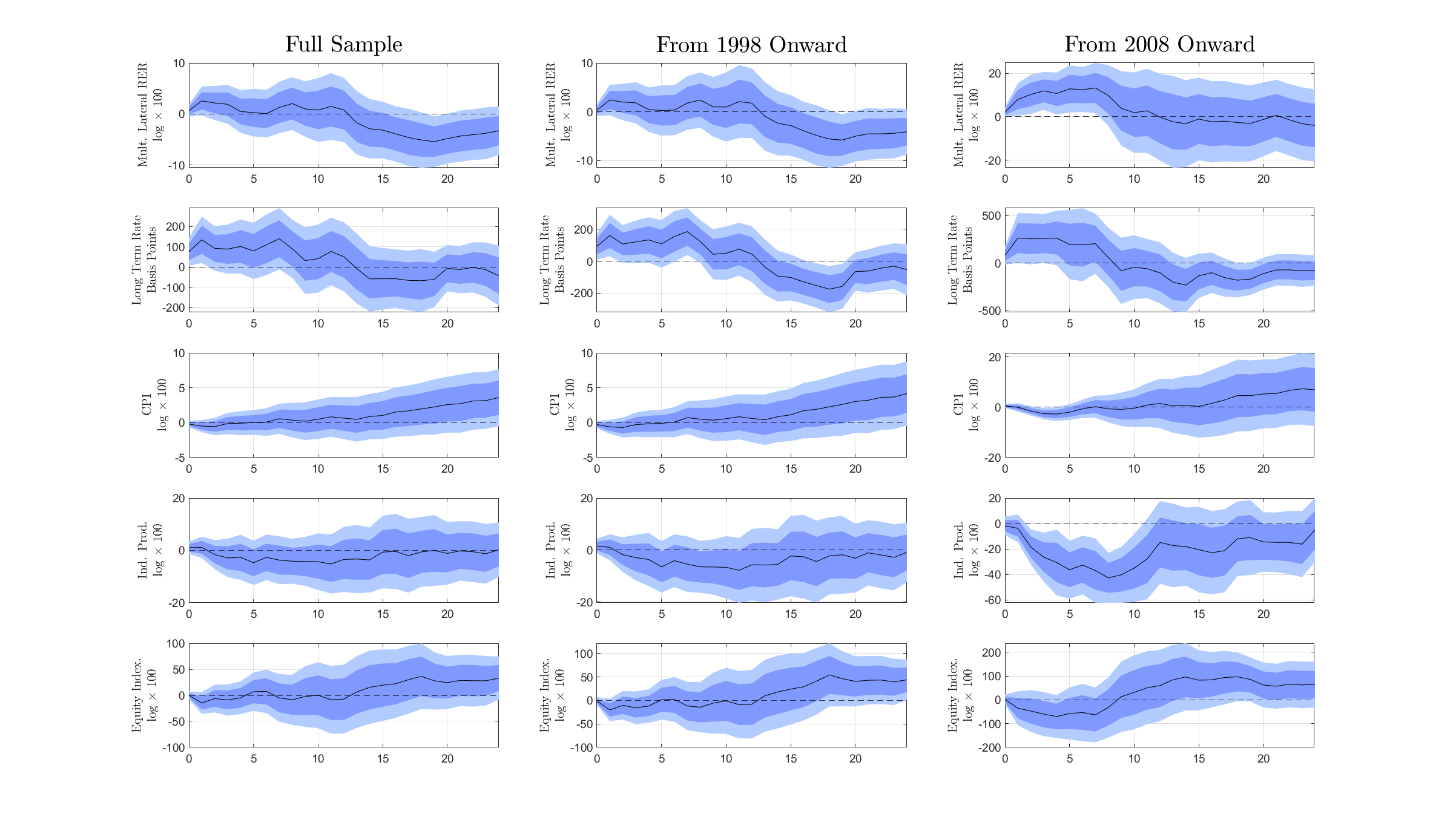}
    \caption{Impulse Response Functions - Pure Monetary Policy Shock \\ Multi. REER Sample \cite{miranda2021transmission}}
    \label{fig:MAR_Comparison_MP_REER}
    \floatfoot{\textbf{Note:} The figure is comprised of 15 sub-figures ordered in three columns and five rows. I replace the $MP$ and $FIE$ components in the benchmark regression in Equation \ref{eq:LP_pooled} with the pure monetary policy shock and the information component or ``signalling effect'' constructed by \cite{miranda2021transmission}. This figure presents the results for the impact of the pure monetary policy shock. The left column presents the results for the sample January 1990 to December 2016, the middle column for the sample January 1998 to December 2016, and the right column for the sample January 2008 to December 2016. The rows represent the impact on (i) the trade weighted multilateral real exchange rate (in logs times 100); (ii) long term interest rates in basis points; (iii) the consumer price index (in logs times 100); (iv) the industrial production index (in logs times 100); (v) the equity index (in logs times 100). The solid black line represents the point estimate, the dark blue area represents the 68\% confidence interval, and the light blue area represents the 90\% confidence interval. In the text, when referring to Panel $(i,j)$, $i$ refers to the row and $j$ to the column of the figure. Each variable, in its own transformation, is demeaned at the country level.}
\end{figure}

\newpage
\begin{figure}
    \centering
    \includegraphics[scale=0.4]{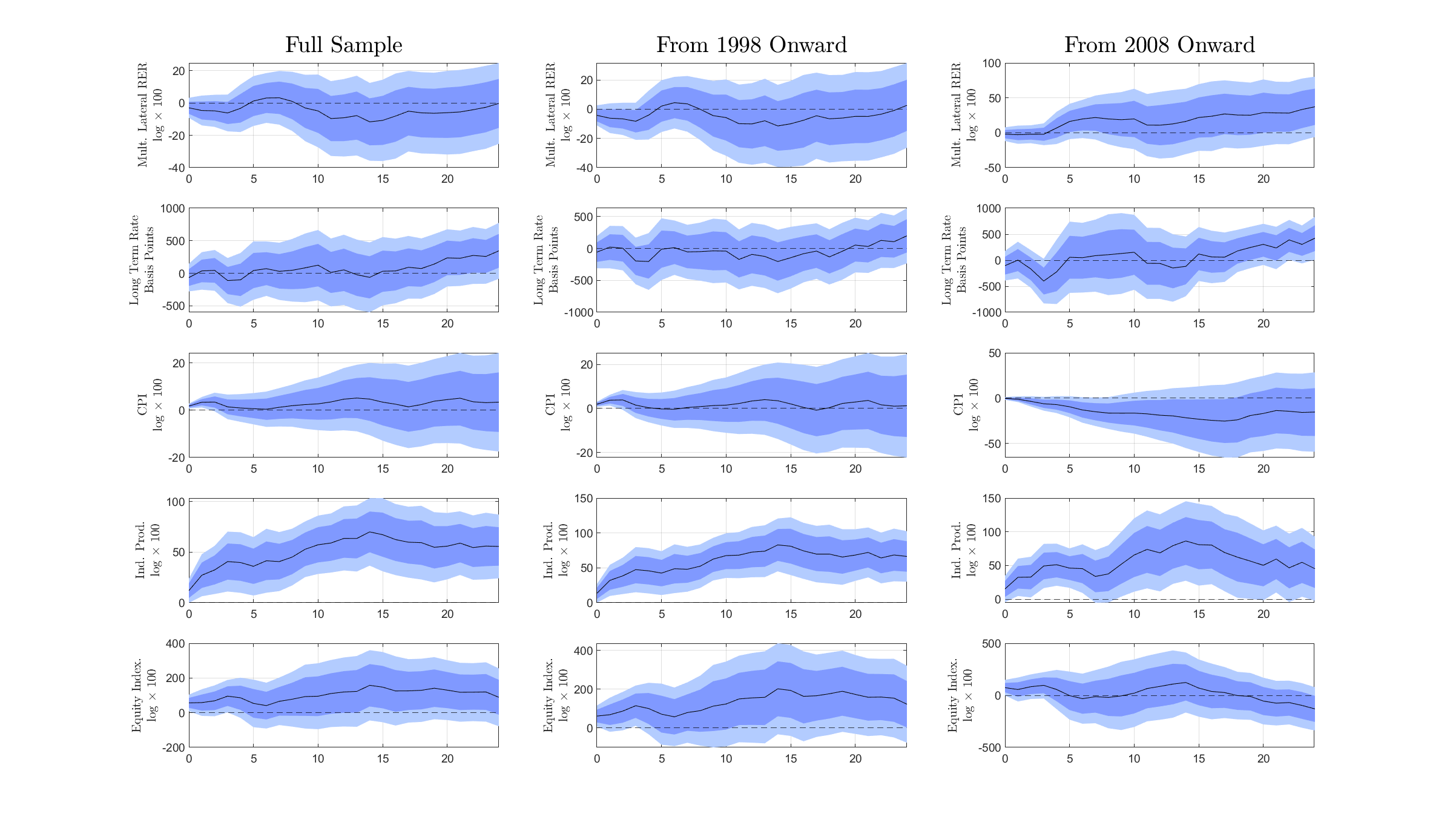}
    \caption{Impulse Response Functions - Information Component \\ Multi. REER Sample - \cite{miranda2021transmission}}
    \label{fig:MAR_Comparison_CBI_REER}
    \floatfoot{\textbf{Note:} The figure is comprised of 15 sub-figures ordered in three columns and five rows. I replace the $MP$ and $FIE$ components in the benchmark regression in Equation \ref{eq:LP_pooled} with the pure monetary policy shock and the information component or ``signalling effect'' constructed by \cite{miranda2021transmission}. This figure presents the results for the impact of the information component or ``signalling effect'' shock. The left column presents the results for the sample January 1990 to December 2016, the middle column for the sample January 1998 to December 2016, and the right column for the sample January 2008 to December 2016. The rows represent the impact on (i) the trade weighted multilateral real exchange rate (in logs times 100); (ii) long term interest rates in basis points; (iii) the consumer price index (in logs times 100); (iv) the industrial production index (in logs times 100); (v) the equity index (in logs times 100). The solid black line represents the point estimate, the dark blue area represents the 68\% confidence interval, and the light blue area represents the 90\% confidence interval. In the text, when referring to Panel $(i,j)$, $i$ refers to the row and $j$ to the column of the figure. Each variable, in its own transformation, is demeaned at the country level.}
\end{figure}


\newpage
\begin{figure}
    \centering
    \includegraphics[scale=0.4]{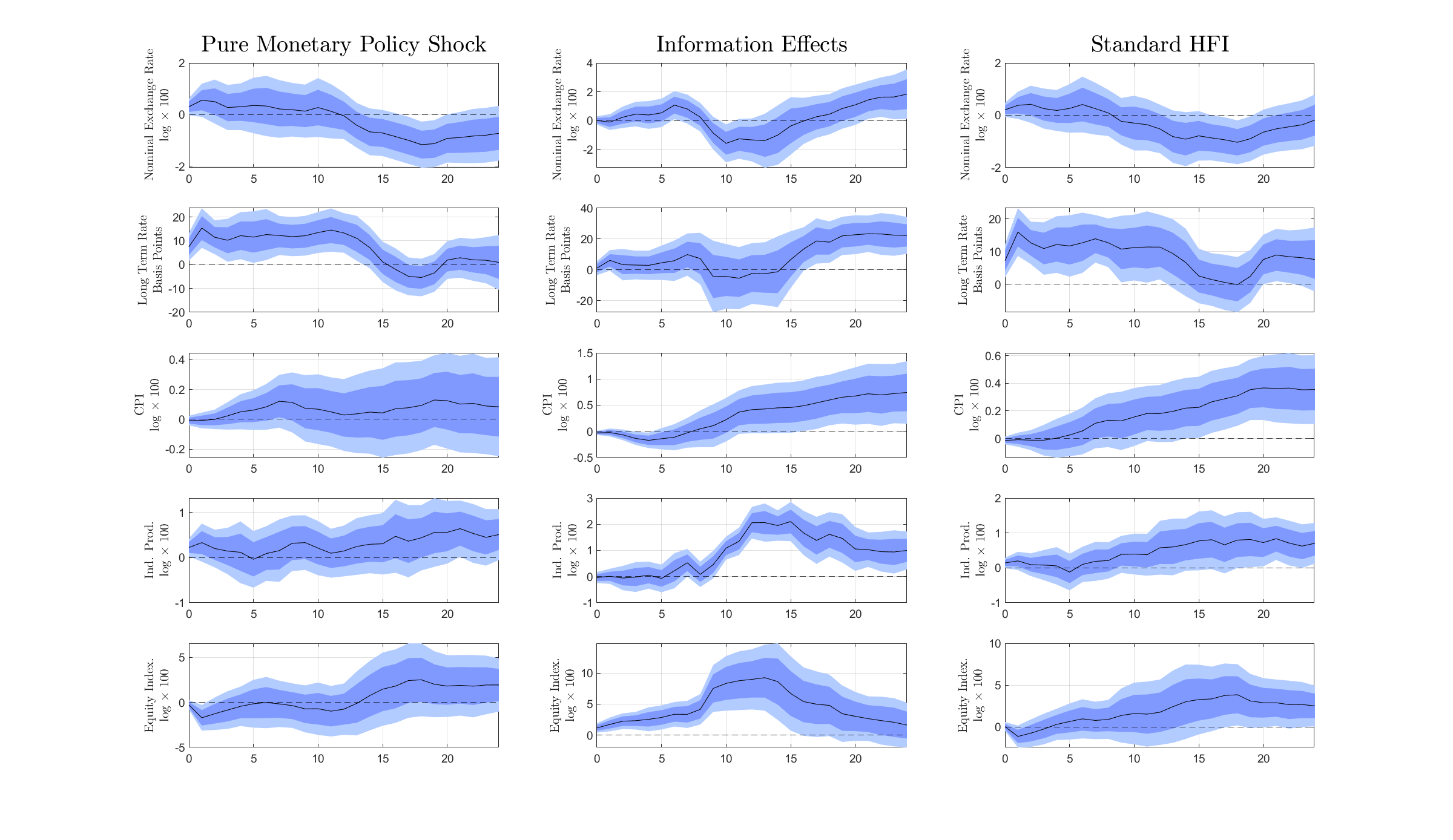}
    \caption{Impulse Response Functions \\ Target Shock - \cite{miranda2022tale} }
    \label{fig:Target_BenchmarkNER}
    \floatfoot{\textbf{Note:} The figure is comprised of 15 sub-figures ordered in three columns and five rows. The left column relates to the estimates of $\beta^{MP}$ in Equation \ref{eq:LP_pooled}, the middle column relates to the estimate of $\beta^{FIE}$ in Equation \ref{eq:LP_pooled}, while the right column relates to estimating Equation \ref{eq:LP_pooled}, replacing the MP and FIE components with the un-orthogonalized monetary policy surprise. Each component is constructed for the target shock following the identification strategy in \cite{miranda2022tale}. The rows represent the impact on (i) the nominal exchange rate with the US dollar (in logs times 100); (ii) long term interest rates in basis points; (iii) the consumer price index (in logs times 100); (iv) the industrial production index (in logs times 100); (v) the equity index (in logs times 100). The solid black line represents the point estimate, the dark blue area represents the 68\% confidence interval, and the light blue area represents the 90\% confidence interval. In the text, when referring to Panel $(i,j)$, $i$ refers to the row and $j$ to the column of the figure. Each variable, in its own transformation, is demeaned at the country level. }
\end{figure}

\newpage
\begin{figure}
    \centering
    \includegraphics[scale=0.4]{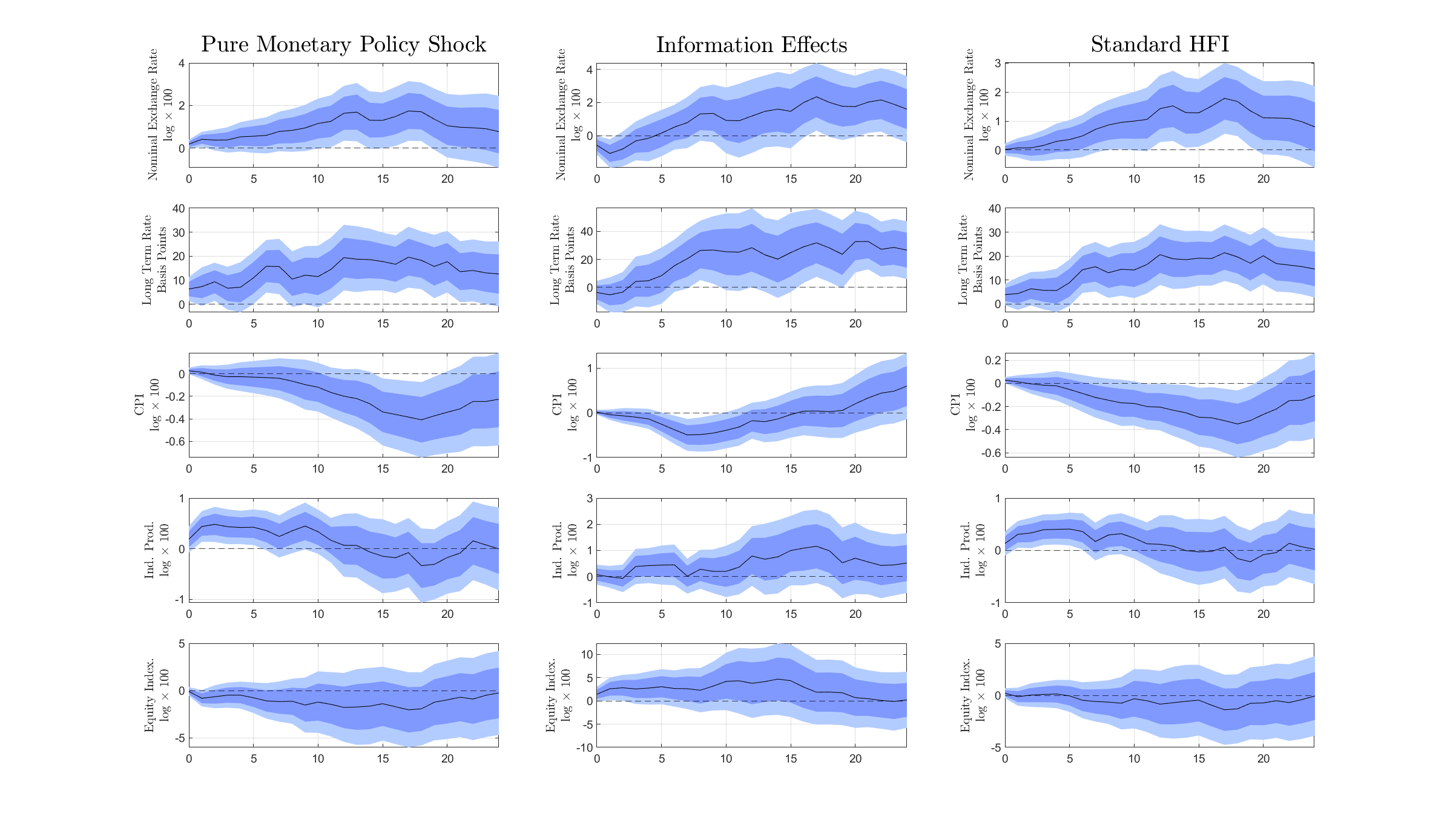}
    \caption{Impulse Response Functions \\ Path Shock - \cite{miranda2022tale} }
    \label{fig:Path_BenchmarkNER}
    \floatfoot{\textbf{Note:} The figure is comprised of 15 sub-figures ordered in three columns and five rows. The left column relates to the estimates of $\beta^{MP}$ in Equation \ref{eq:LP_pooled}, the middle column relates to the estimate of $\beta^{FIE}$ in Equation \ref{eq:LP_pooled}, while the right column relates to estimating Equation \ref{eq:LP_pooled}, replacing the MP and FIE components with the un-orthogonalized monetary policy surprise. Each component is constructed for the path shock following the identification strategy in \cite{miranda2022tale}. The rows represent the impact on (i) the nominal exchange rate with the US dollar (in logs times 100); (ii) long term interest rates in basis points; (iii) the consumer price index (in logs times 100); (iv) the industrial production index (in logs times 100); (v) the equity index (in logs times 100). The solid black line represents the point estimate, the dark blue area represents the 68\% confidence interval, and the light blue area represents the 90\% confidence interval. In the text, when referring to Panel $(i,j)$, $i$ refers to the row and $j$ to the column of the figure. Each variable, in its own transformation, is demeaned at the country level. }
\end{figure}

\newpage
\begin{figure}
    \centering
    \includegraphics[scale=0.4]{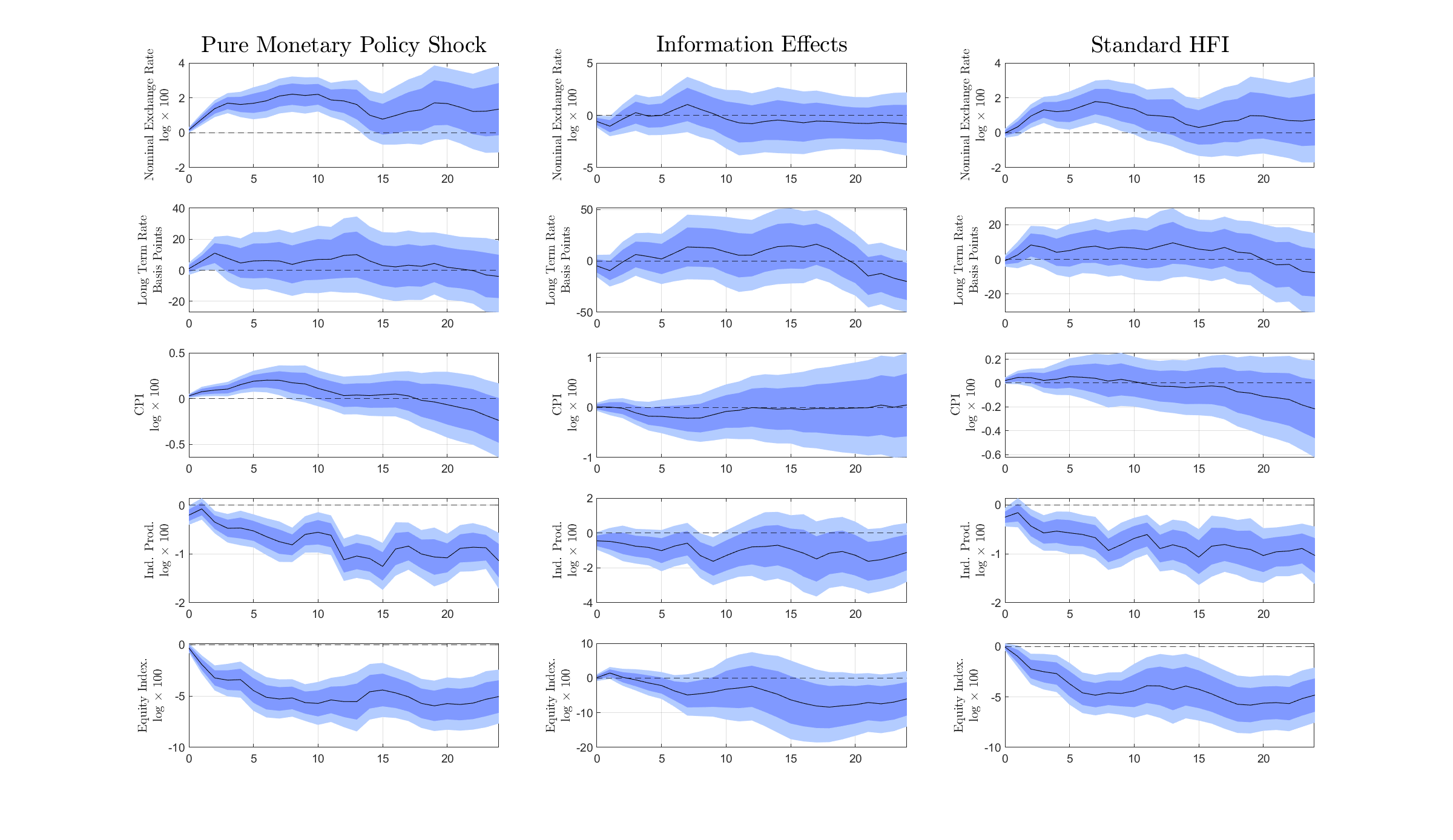}
    \caption{Impulse Response Functions \\ QE Shock - \cite{miranda2022tale} }
    \label{fig:QE_BenchmarkNER}
    \floatfoot{\textbf{Note:} The figure is comprised of 15 sub-figures ordered in three columns and five rows. The left column relates to the estimates of $\beta^{MP}$ in Equation \ref{eq:LP_pooled}, the middle column relates to the estimate of $\beta^{FIE}$ in Equation \ref{eq:LP_pooled}, while the right column relates to estimating Equation \ref{eq:LP_pooled}, replacing the MP and FIE components with the un-orthogonalized monetary policy surprise. Each component is constructed for the QE shock following the identification strategy in \cite{miranda2022tale}. The rows represent the impact on (i) the nominal exchange rate with the US dollar (in logs times 100); (ii) long term interest rates in basis points; (iii) the consumer price index (in logs times 100); (iv) the industrial production index (in logs times 100); (v) the equity index (in logs times 100). The solid black line represents the point estimate, the dark blue area represents the 68\% confidence interval, and the light blue area represents the 90\% confidence interval. In the text, when referring to Panel $(i,j)$, $i$ refers to the row and $j$ to the column of the figure. Each variable, in its own transformation, is demeaned at the country level. }
\end{figure}

\newpage
\begin{figure}
    \centering
    \includegraphics[scale=0.4]{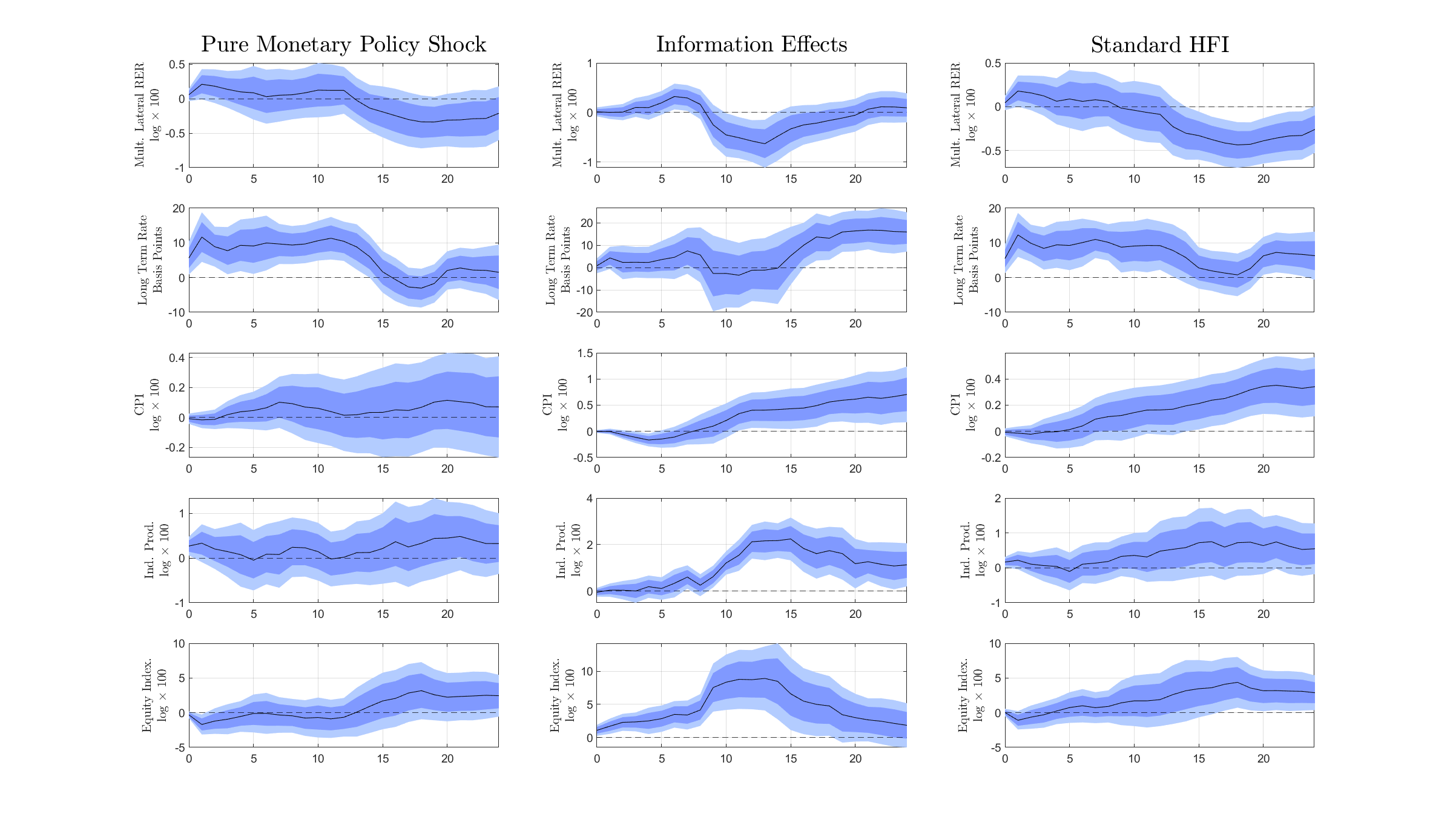}
    \caption{Impulse Response Functions \\ Target Shock - Multi. REER Sample - \cite{miranda2022tale} }
    \label{fig:Target_BenchmarkREER}
    \floatfoot{\textbf{Note:} The figure is comprised of 15 sub-figures ordered in three columns and five rows. The left column relates to the estimates of $\beta^{MP}$ in Equation \ref{eq:LP_pooled}, the middle column relates to the estimate of $\beta^{FIE}$ in Equation \ref{eq:LP_pooled}, while the right column relates to estimating Equation \ref{eq:LP_pooled}, replacing the MP and FIE components with the un-orthogonalized monetary policy surprise. Each component is constructed for the target shock following the identification strategy in \cite{miranda2022tale}. The rows represent the impact on (i) the trade weighted multilateral real exchange rate (in logs times 100); (ii) long term interest rates in basis points; (iii) the consumer price index (in logs times 100); (iv) the industrial production index (in logs times 100); (v) the equity index (in logs times 100). The solid black line represents the point estimate, the dark blue area represents the 68\% confidence interval, and the light blue area represents the 90\% confidence interval. In the text, when referring to Panel $(i,j)$, $i$ refers to the row and $j$ to the column of the figure. Each variable, in its own transformation, is demeaned at the country level. }
\end{figure}

\newpage
\begin{figure}
    \centering
    \includegraphics[scale=0.4]{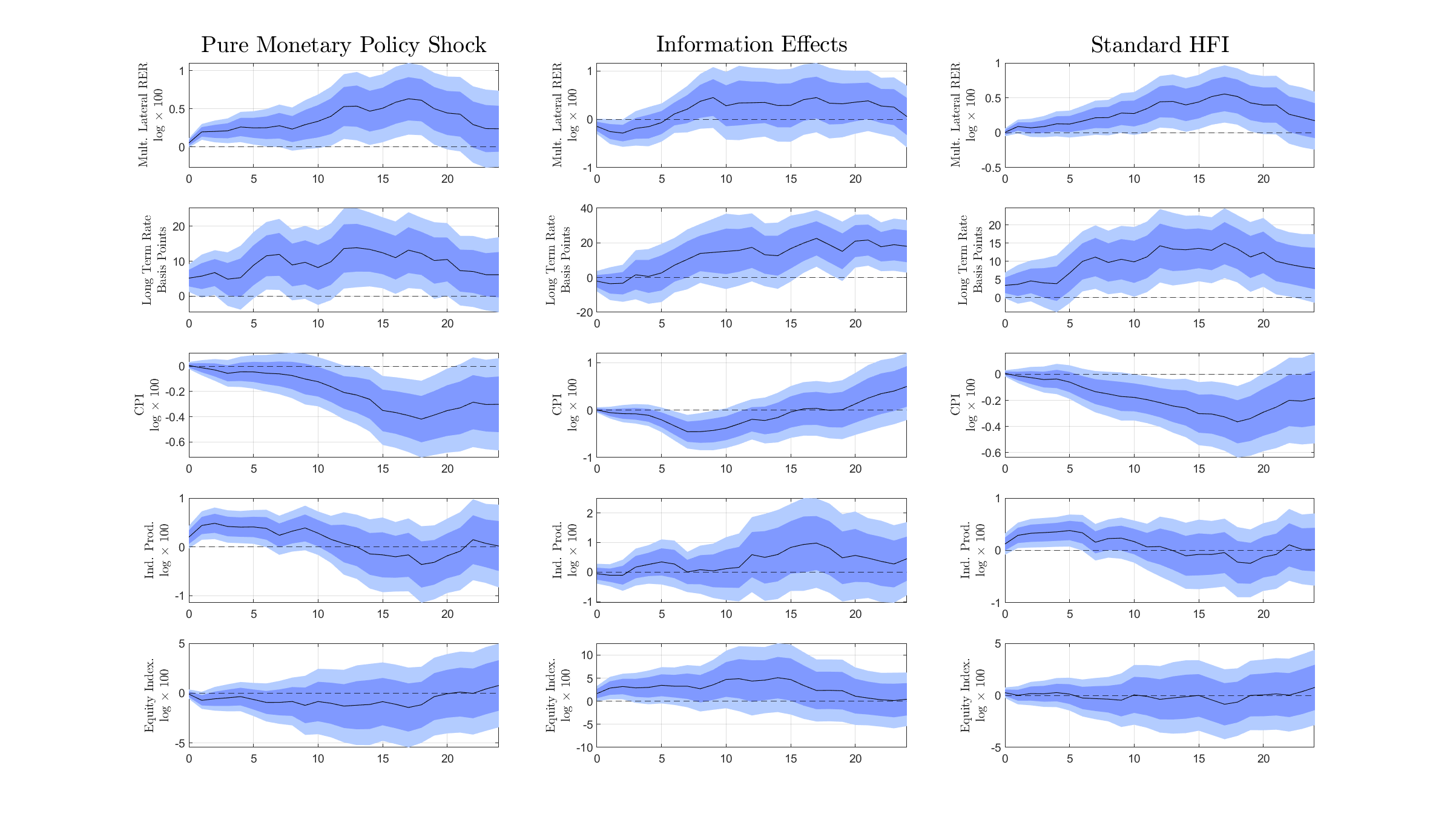}
    \caption{Impulse Response Functions \\ Path Shock - Multi. REER Sample - \cite{miranda2022tale} }
    \label{fig:Path_BenchmarkREER}
    \floatfoot{\textbf{Note:} The figure is comprised of 15 sub-figures ordered in three columns and five rows. The left column relates to the estimates of $\beta^{MP}$ in Equation \ref{eq:LP_pooled}, the middle column relates to the estimate of $\beta^{FIE}$ in Equation \ref{eq:LP_pooled}, while the right column relates to estimating Equation \ref{eq:LP_pooled}, replacing the MP and FIE components with the un-orthogonalized monetary policy surprise. Each component is constructed for the path shock following the identification strategy in \cite{miranda2022tale}. The rows represent the impact on (i) the trade weighted multilateral real exchange rate (in logs times 100); (ii) long term interest rates in basis points; (iii) the consumer price index (in logs times 100); (iv) the industrial production index (in logs times 100); (v) the equity index (in logs times 100). The solid black line represents the point estimate, the dark blue area represents the 68\% confidence interval, and the light blue area represents the 90\% confidence interval. In the text, when referring to Panel $(i,j)$, $i$ refers to the row and $j$ to the column of the figure. Each variable, in its own transformation, is demeaned at the country level. }
\end{figure}

\newpage
\begin{figure}
    \centering
    \includegraphics[scale=0.4]{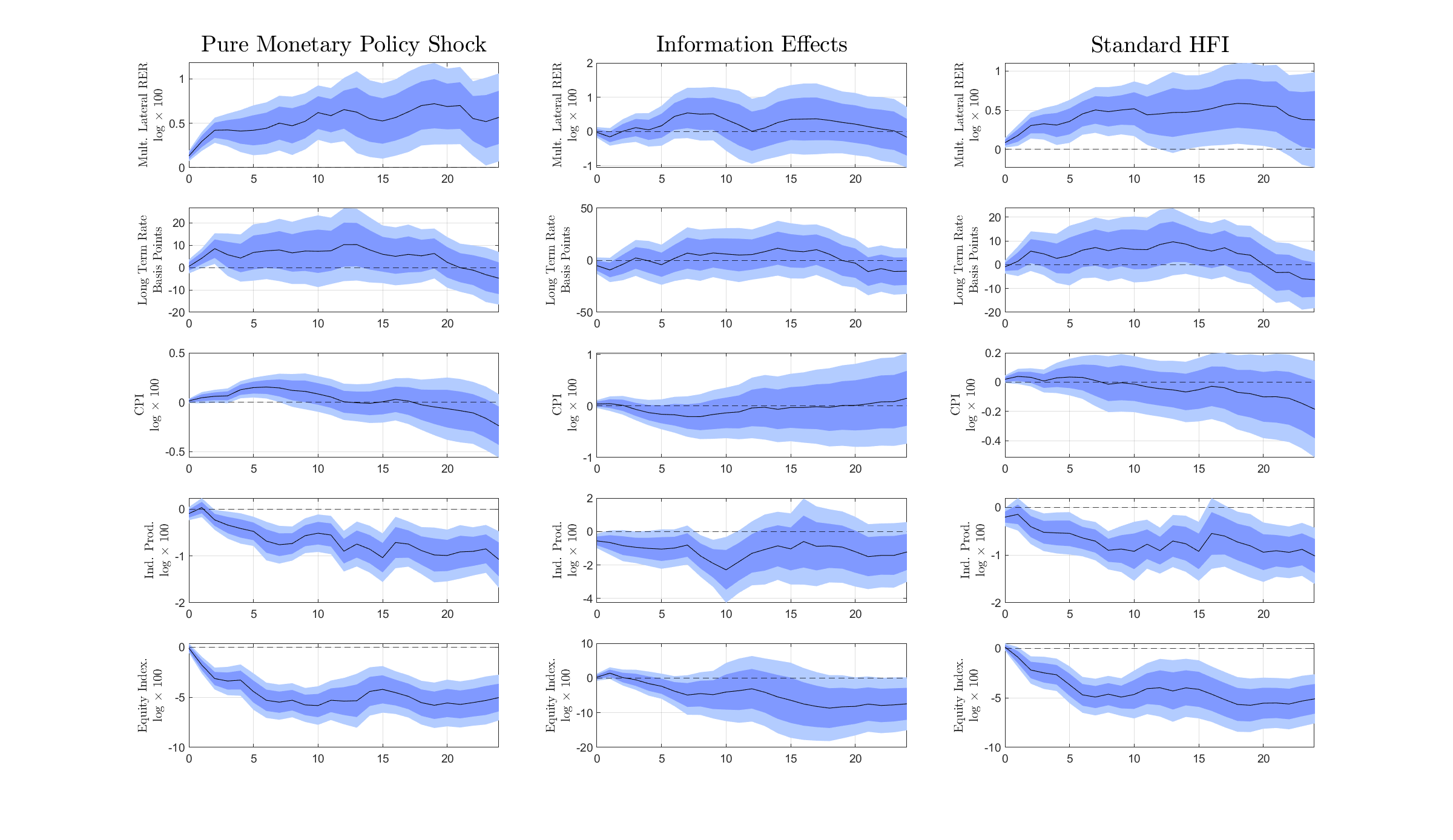}
    \caption{Impulse Response Functions \\ QE Shock - Multi. REER Sample - \cite{miranda2022tale} }
    \label{fig:QE_BenchmarkREER}
    \floatfoot{\textbf{Note:} The figure is comprised of 15 sub-figures ordered in three columns and five rows. The left column relates to the estimates of $\beta^{MP}$ in Equation \ref{eq:LP_pooled}, the middle column relates to the estimate of $\beta^{FIE}$ in Equation \ref{eq:LP_pooled}, while the right column relates to estimating Equation \ref{eq:LP_pooled}, replacing the MP and FIE components with the un-orthogonalized monetary policy surprise. Each component is constructed for the QE shock following the identification strategy in \cite{miranda2022tale}. The rows represent the impact on (i) the trade weighted multilateral real exchange rate (in logs times 100); (ii) long term interest rates in basis points; (iii) the consumer price index (in logs times 100); (iv) the industrial production index (in logs times 100); (v) the equity index (in logs times 100). The solid black line represents the point estimate, the dark blue area represents the 68\% confidence interval, and the light blue area represents the 90\% confidence interval. In the text, when referring to Panel $(i,j)$, $i$ refers to the row and $j$ to the column of the figure. Each variable, in its own transformation, is demeaned at the country level. }
\end{figure}

\newpage
\begin{figure}
    \centering
    \includegraphics[scale=0.4]{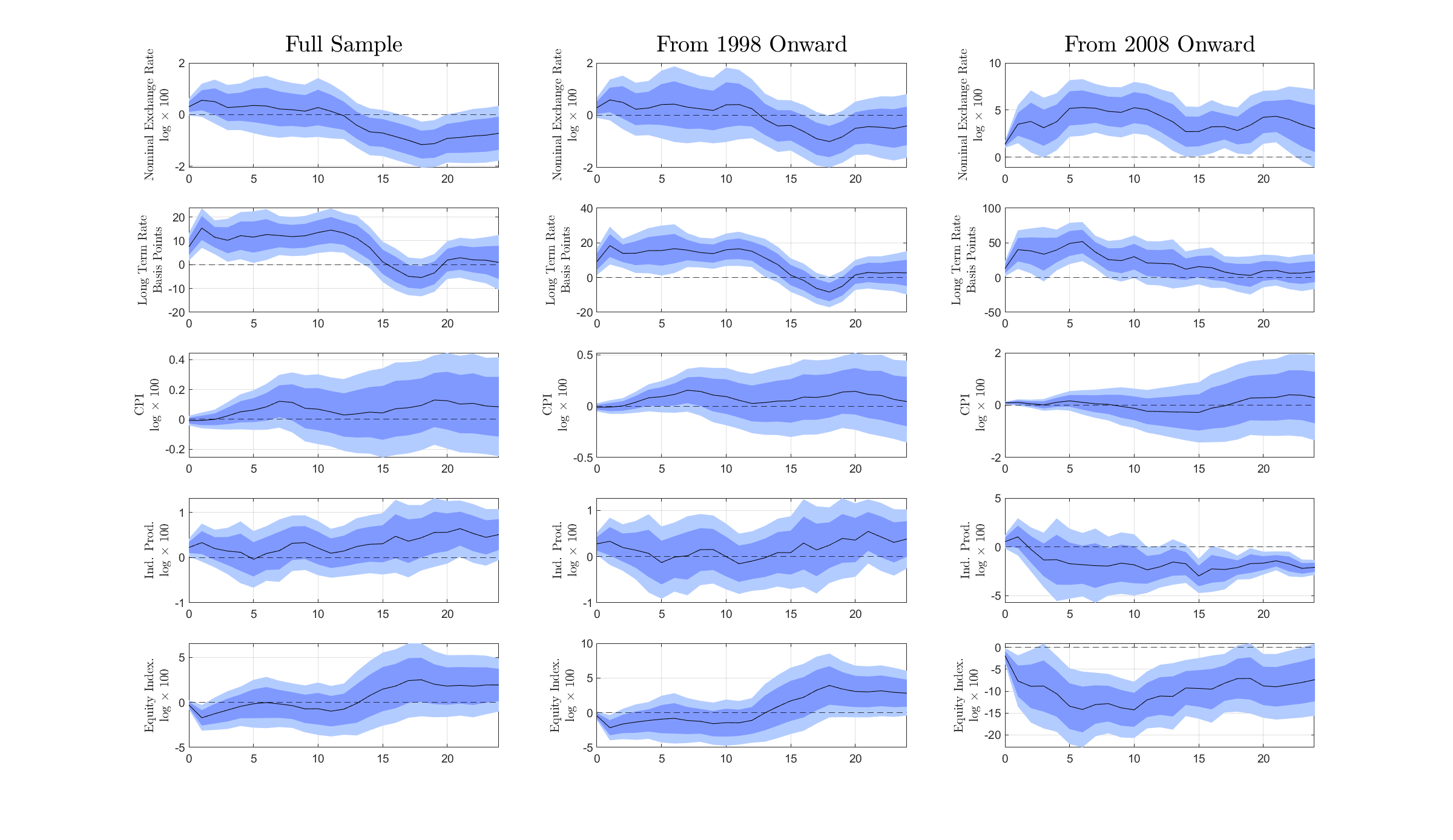}
    \caption{Impulse Response Functions - Pure Monetary Policy Shock \\ Target Shock - \cite{miranda2022tale}}
    \label{fig:Target_Comparison_MP}
    \floatfoot{\textbf{Note:} The figure is comprised of 15 sub-figures ordered in three columns and five rows. To estimate the international spillovers I replace the $MP$ and $FIE$ components in the benchmark regression in Equation \ref{eq:LP_pooled} with the pure monetary policy shock and the information component constructed by \cite{miranda2022tale}. This figure presents the results for the impact of the pure monetary policy shock. The left column presents the results for the sample July 1991 to June 2019, the middle column for the sample January 1998 to June 2019, and the right column for the sample January 2008 to June 2019. The rows represent the impact on (i) the nominal exchange rate with the US dollar (in logs times 100); (ii) long term interest rates in basis points; (iii) the consumer price index (in logs times 100); (iv) the industrial production index (in logs times 100); (v) the equity index (in logs times 100). The solid black line represents the point estimate, the dark blue area represents the 68\% confidence interval, and the light blue area represents the 90\% confidence interval. In the text, when referring to Panel $(i,j)$, $i$ refers to the row and $j$ to the column of the figure. Each variable, in its own transformation, is demeaned at the country level.}
\end{figure}

\newpage
\begin{figure}
    \centering
    \includegraphics[scale=0.4]{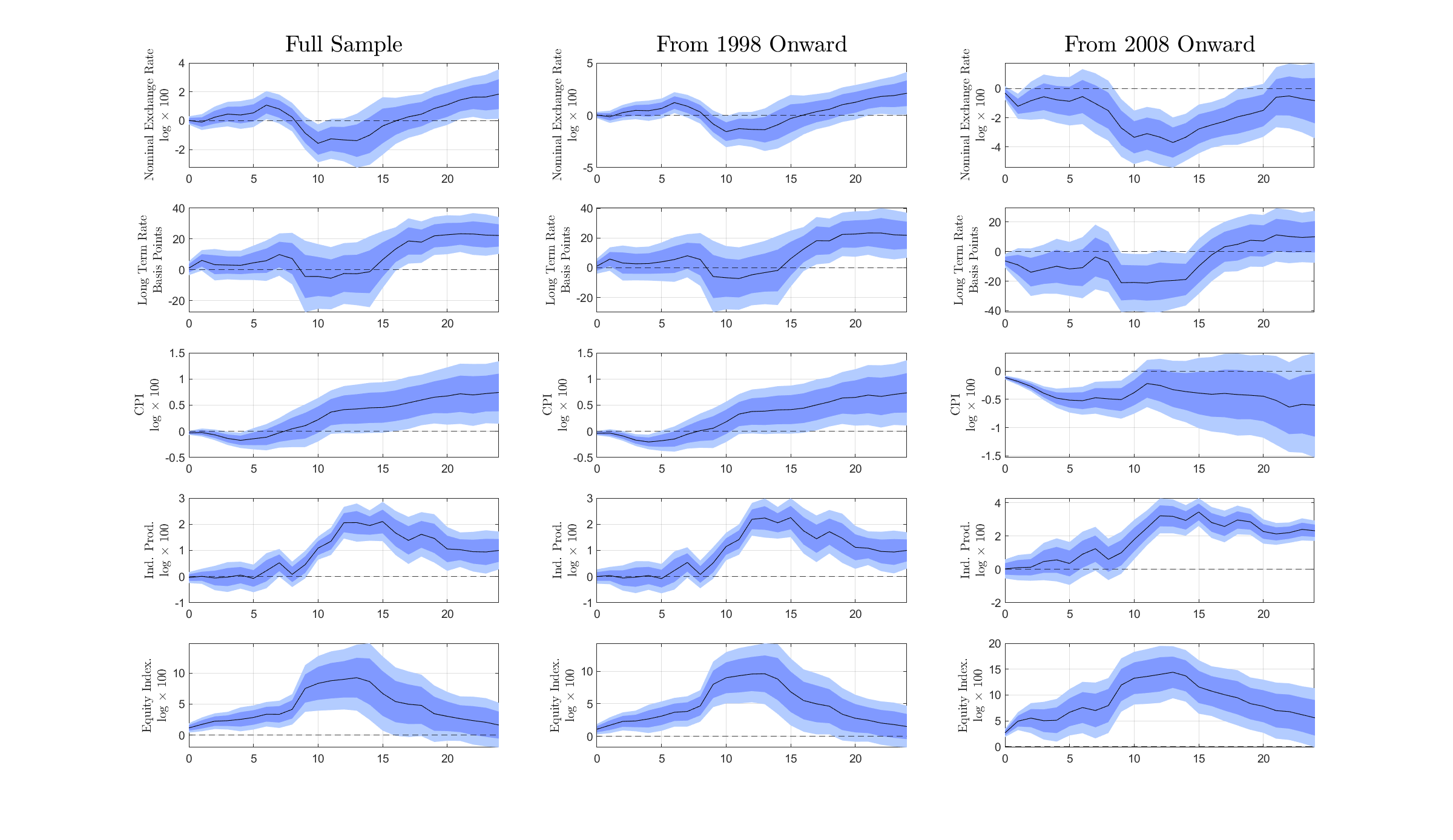}
    \caption{Impulse Response Functions - Information Component \\  Target Shock - \cite{miranda2022tale}}
    \label{fig:Target_Comparison_CBI}
    \floatfoot{\textbf{Note:} The figure is comprised of 15 sub-figures ordered in three columns and five rows. To estimate the international spillovers I replace the $MP$ and $FIE$ components in the benchmark regression in Equation \ref{eq:LP_pooled} with the pure monetary policy shock and the information component constructed by \cite{miranda2022tale}. This figure presents the results for the information component. The left column presents the results for the sample July 1991 to June 2019, the middle column for the sample January 1998 to June 2019, and the right column for the sample January 2008 to June 2019. The rows represent the impact on (i) the nominal exchange rate with the US dollar (in logs times 100); (ii) long term interest rates in basis points; (iii) the consumer price index (in logs times 100); (iv) the industrial production index (in logs times 100); (v) the equity index (in logs times 100). The solid black line represents the point estimate, the dark blue area represents the 68\% confidence interval, and the light blue area represents the 90\% confidence interval. In the text, when referring to Panel $(i,j)$, $i$ refers to the row and $j$ to the column of the figure. Each variable, in its own transformation, is demeaned at the country level.}
\end{figure}

\newpage
\begin{figure}
    \centering
    \includegraphics[scale=0.4]{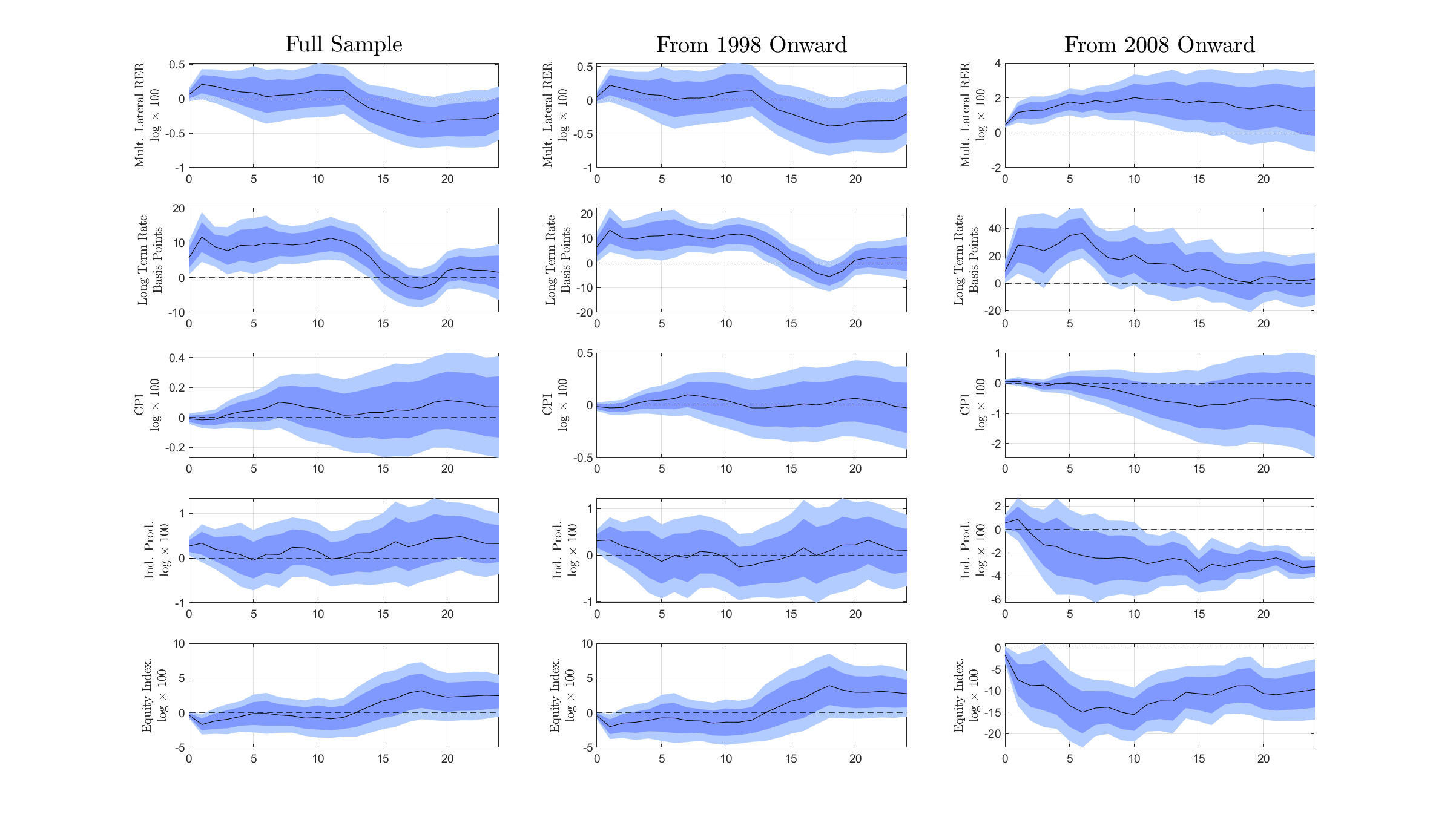}
    \caption{Impulse Response Functions - Pure Monetary Policy Shock \\ Multi. REER Sample - Target Shock - \cite{miranda2022tale}}
    \label{fig:Target_Comparison_MP_REER}
    \floatfoot{\textbf{Note:} The figure is comprised of 15 sub-figures ordered in three columns and five rows. To estimate the international spillovers I replace the $MP$ and $FIE$ components in the benchmark regression in Equation \ref{eq:LP_pooled} with the pure monetary policy shock and the information component constructed by \cite{miranda2022tale}. This figure presents the results for the impact of the pure monetary policy shock. The left column presents the results for the sample July 1991 to June 2019, the middle column for the sample January 1998 to June 2019, and the right column for the sample January 2008 to June 2019. The rows represent the impact on (i) the trade weighted multilateral real exchange rate (in logs times 100); (ii) long term interest rates in basis points; (iii) the consumer price index (in logs times 100); (iv) the industrial production index (in logs times 100); (v) the equity index (in logs times 100). The solid black line represents the point estimate, the dark blue area represents the 68\% confidence interval, and the light blue area represents the 90\% confidence interval. In the text, when referring to Panel $(i,j)$, $i$ refers to the row and $j$ to the column of the figure. Each variable, in its own transformation, is demeaned at the country level.}
\end{figure}

\newpage
\begin{figure}
    \centering
    \includegraphics[scale=0.4]{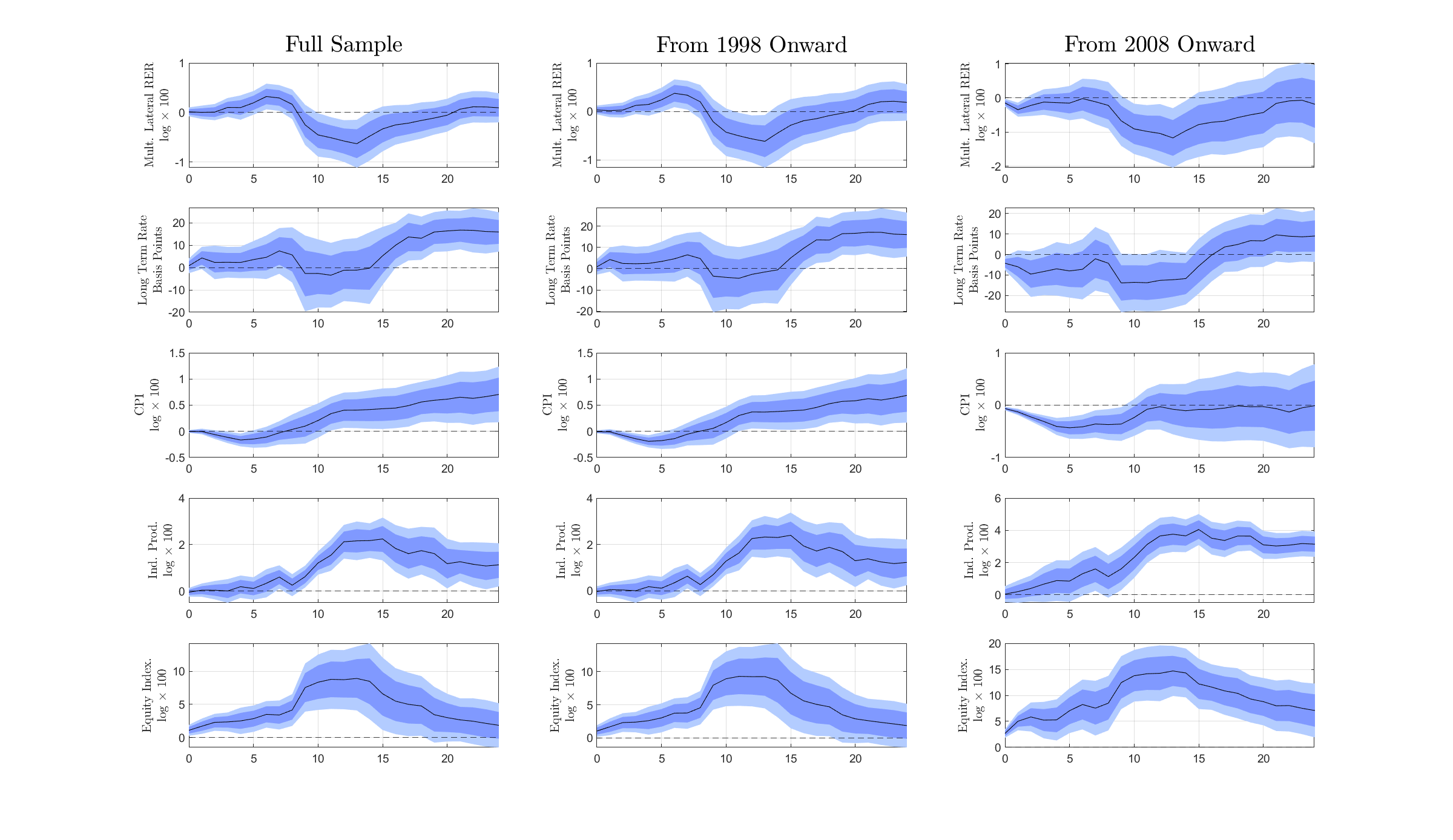}
    \caption{Impulse Response Functions - Information Component \\ Multi. REER Sample - Target Shock -  \cite{miranda2022tale}}
    \label{fig:Target_Comparison_CBI_REER}
    \floatfoot{\textbf{Note:} The figure is comprised of 15 sub-figures ordered in three columns and five rows. To estimate the international spillovers I replace the $MP$ and $FIE$ components in the benchmark regression in Equation \ref{eq:LP_pooled} with the pure monetary policy shock and the information component constructed by \cite{miranda2022tale}. This figure presents the results for the impact of the information component. The left column presents the results for the sample July 1991 to June 2019, the middle column for the sample January 1998 to June 2019, and the right column for the sample January 2008 to June 2019. The rows represent the impact on (i) the trade weighted multilateral real exchange rate (in logs times 100); (ii) long term interest rates in basis points; (iii) the consumer price index (in logs times 100); (iv) the industrial production index (in logs times 100); (v) the equity index (in logs times 100). The solid black line represents the point estimate, the dark blue area represents the 68\% confidence interval, and the light blue area represents the 90\% confidence interval. In the text, when referring to Panel $(i,j)$, $i$ refers to the row and $j$ to the column of the figure. Each variable, in its own transformation, is demeaned at the country level.}
\end{figure}

\newpage
\begin{figure}
    \centering
    \includegraphics[scale=0.4]{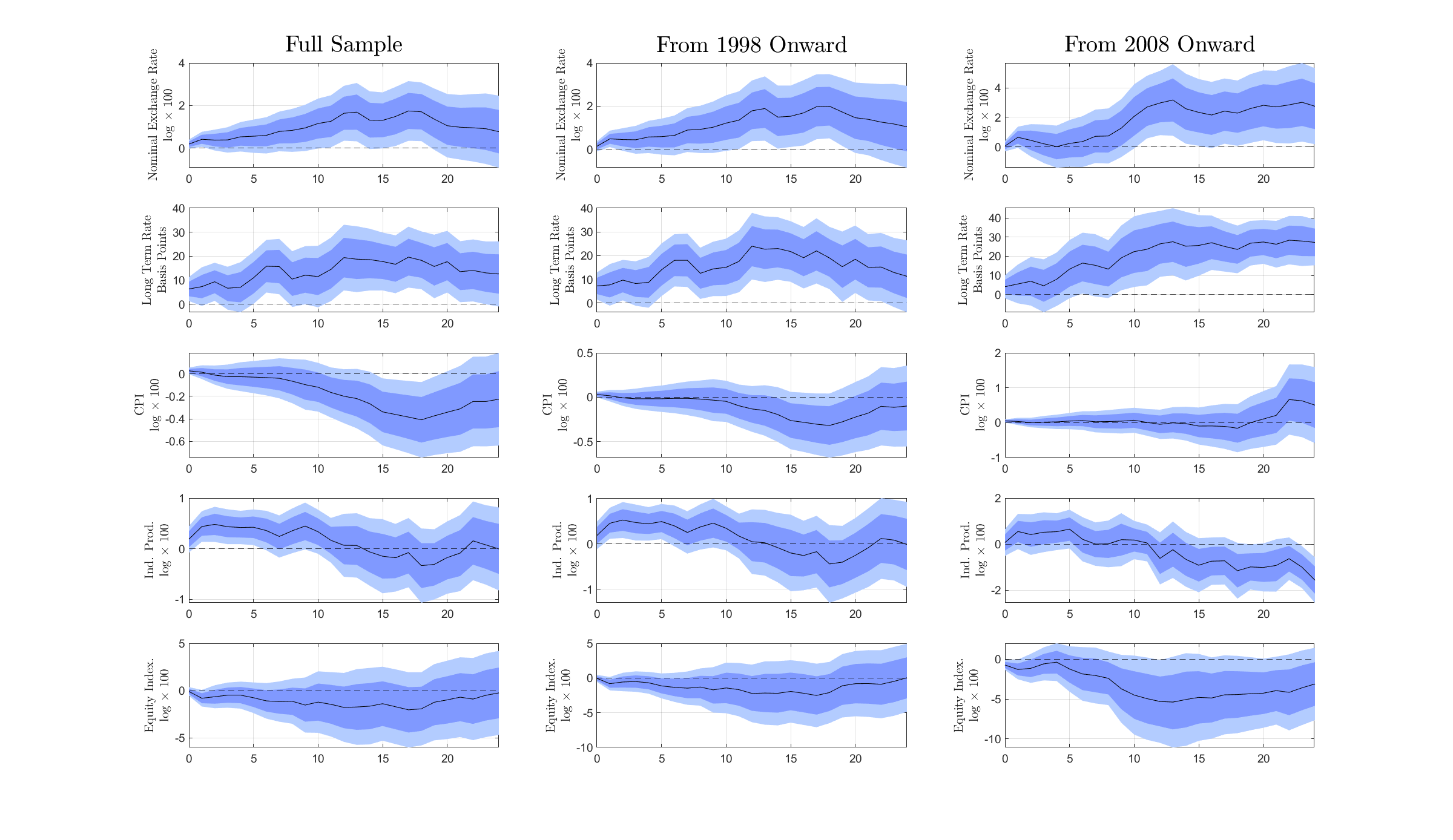}
    \caption{Impulse Response Functions - Pure Monetary Policy Shock \\ Path Shock - \cite{miranda2022tale}}
    \label{fig:Path_Comparison_MP}
    \floatfoot{\textbf{Note:} The figure is comprised of 15 sub-figures ordered in three columns and five rows. To estimate the international spillovers I replace the $MP$ and $FIE$ components in the benchmark regression in Equation \ref{eq:LP_pooled} with the pure monetary policy shock and the information component constructed by \cite{miranda2022tale}. This figure presents the results for the impact of the pure monetary policy shock. The left column presents the results for the sample July 1991 to June 2019, the middle column for the sample January 1998 to June 2019, and the right column for the sample January 2008 to June 2019. The rows represent the impact on (i) the nominal exchange rate with the US dollar (in logs times 100); (ii) long term interest rates in basis points; (iii) the consumer price index (in logs times 100); (iv) the industrial production index (in logs times 100); (v) the equity index (in logs times 100). The solid black line represents the point estimate, the dark blue area represents the 68\% confidence interval, and the light blue area represents the 90\% confidence interval. In the text, when referring to Panel $(i,j)$, $i$ refers to the row and $j$ to the column of the figure. Each variable, in its own transformation, is demeaned at the country level.}
\end{figure}

\newpage
\begin{figure}
    \centering
    \includegraphics[scale=0.4]{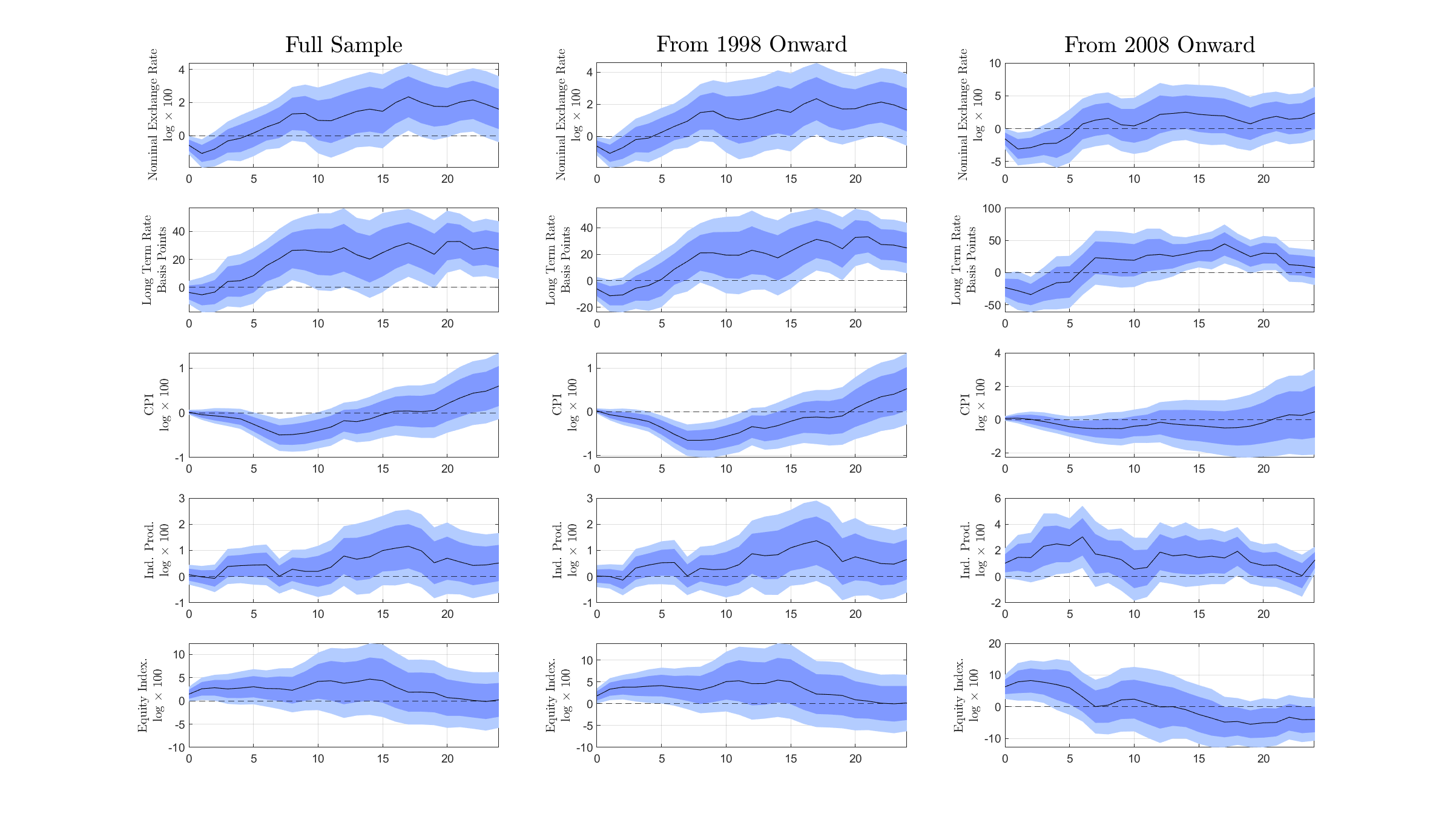}
    \caption{Impulse Response Functions - Information Component \\  Path Shock - \cite{miranda2022tale}}
    \label{fig:Path_Comparison_CBI}
    \floatfoot{\textbf{Note:} The figure is comprised of 15 sub-figures ordered in three columns and five rows. To estimate the international spillovers I replace the $MP$ and $FIE$ components in the benchmark regression in Equation \ref{eq:LP_pooled} with the pure monetary policy shock and the information component constructed by \cite{miranda2022tale}. This figure presents the results for the information component. The left column presents the results for the sample July 1991 to June 2019, the middle column for the sample January 1998 to June 2019, and the right column for the sample January 2008 to June 2019. The rows represent the impact on (i) the nominal exchange rate with the US dollar (in logs times 100); (ii) long term interest rates in basis points; (iii) the consumer price index (in logs times 100); (iv) the industrial production index (in logs times 100); (v) the equity index (in logs times 100). The solid black line represents the point estimate, the dark blue area represents the 68\% confidence interval, and the light blue area represents the 90\% confidence interval. In the text, when referring to Panel $(i,j)$, $i$ refers to the row and $j$ to the column of the figure. Each variable, in its own transformation, is demeaned at the country level.}
\end{figure}

\newpage
\begin{figure}
    \centering
    \includegraphics[scale=0.4]{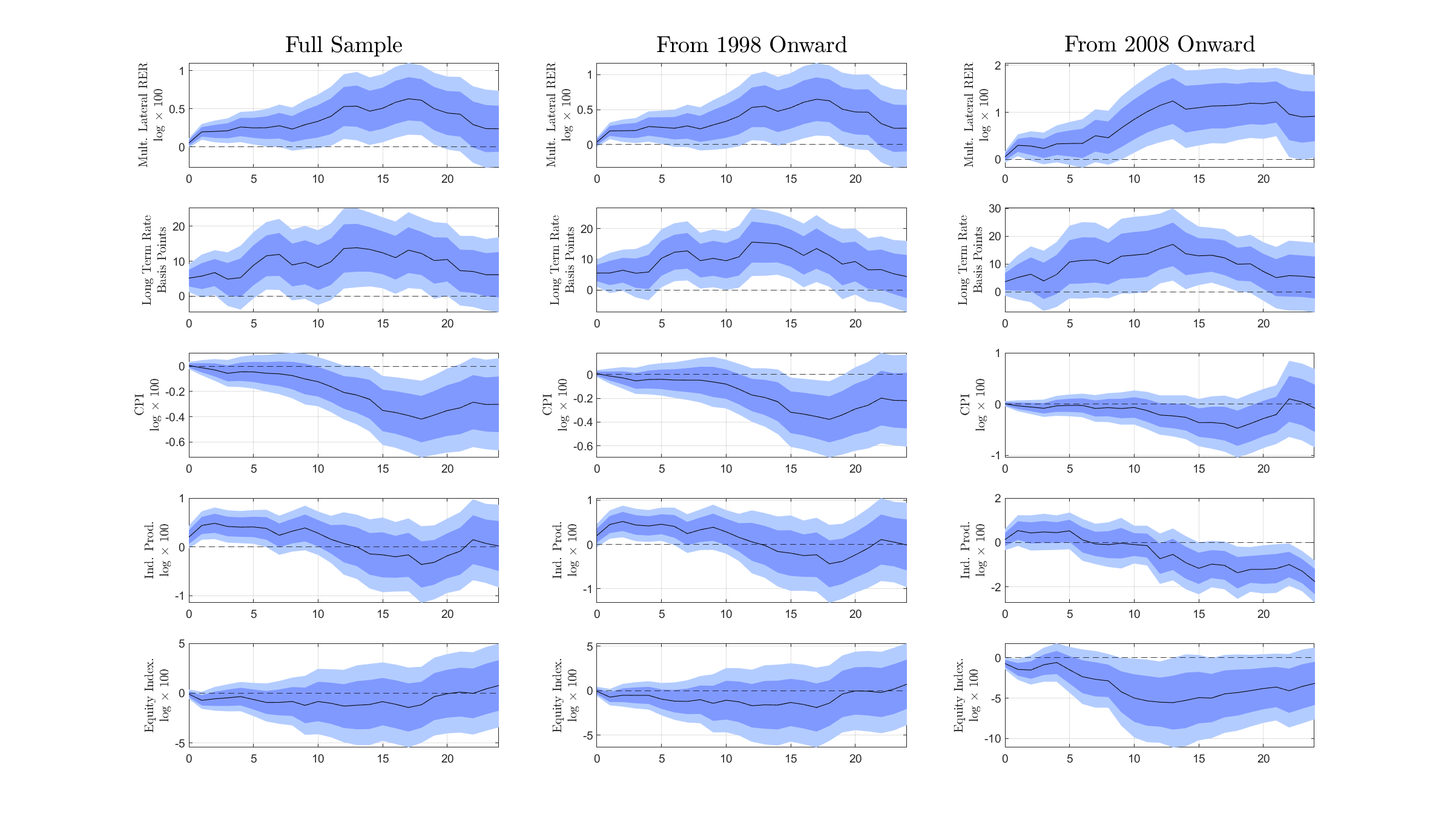}
    \caption{Impulse Response Functions - Pure Monetary Policy Shock \\ Multi. REER Sample - Path Shock - \cite{miranda2022tale}}
    \label{fig:Path_Comparison_MP_REER}
    \floatfoot{\textbf{Note:} The figure is comprised of 15 sub-figures ordered in three columns and five rows. To estimate the international spillovers I replace the $MP$ and $FIE$ components in the benchmark regression in Equation \ref{eq:LP_pooled} with the pure monetary policy shock and the information component constructed by \cite{miranda2022tale}. This figure presents the results for the impact of the pure monetary policy shock. The left column presents the results for the sample July 1991 to June 2019, the middle column for the sample January 1998 to June 2019, and the right column for the sample January 2008 to June 2019. The rows represent the impact on (i) the trade weighted multilateral real exchange rate (in logs times 100); (ii) long term interest rates in basis points; (iii) the consumer price index (in logs times 100); (iv) the industrial production index (in logs times 100); (v) the equity index (in logs times 100). The solid black line represents the point estimate, the dark blue area represents the 68\% confidence interval, and the light blue area represents the 90\% confidence interval. In the text, when referring to Panel $(i,j)$, $i$ refers to the row and $j$ to the column of the figure. Each variable, in its own transformation, is demeaned at the country level.}
\end{figure}

\newpage
\begin{figure}
    \centering
    \includegraphics[scale=0.4]{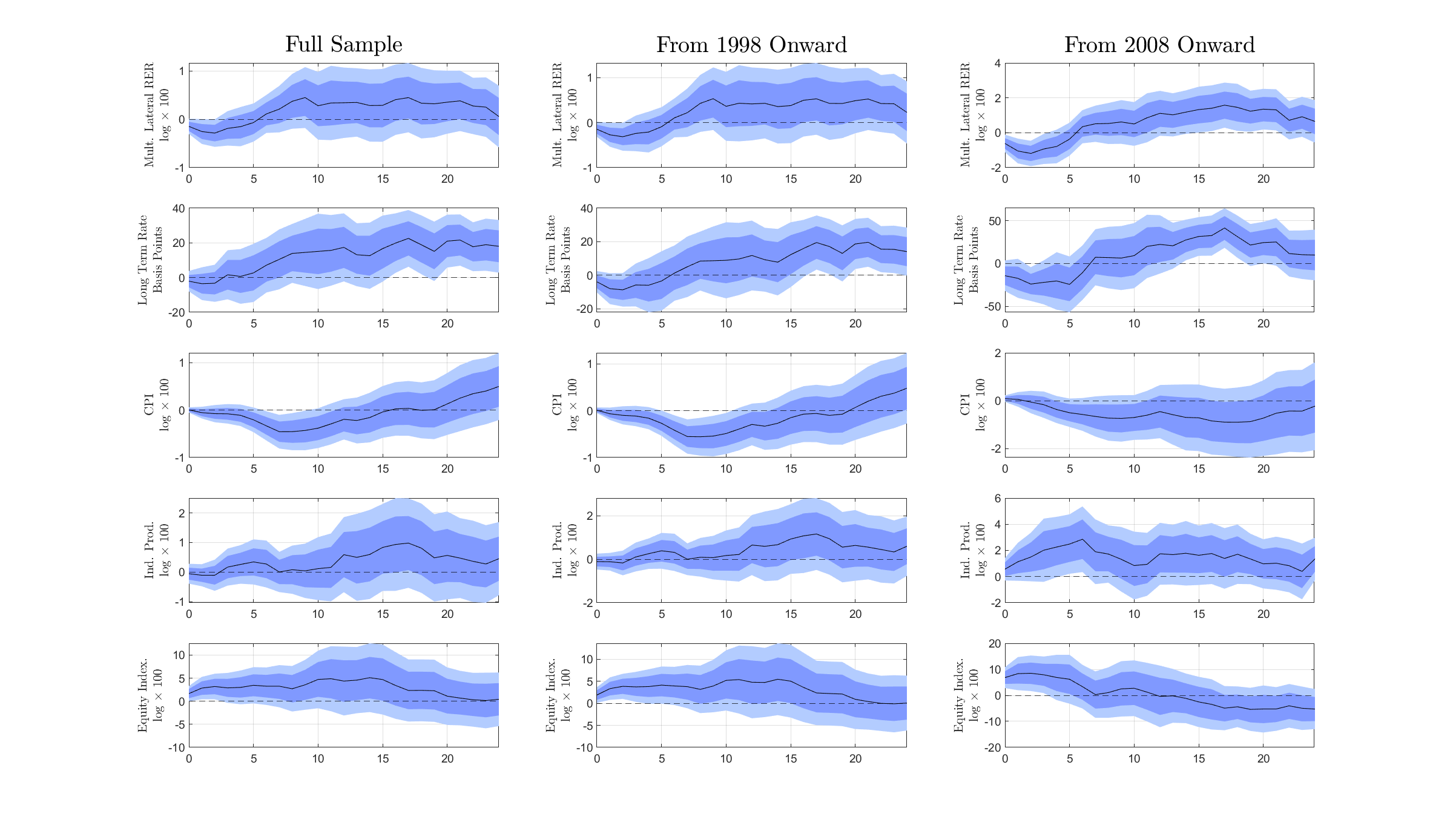}
    \caption{Impulse Response Functions - Information Component \\ Multi. REER Sample - Path Shock -  \cite{miranda2022tale}}
    \label{fig:Path_Comparison_CBI_REER}
    \floatfoot{\textbf{Note:} The figure is comprised of 15 sub-figures ordered in three columns and five rows. To estimate the international spillovers I replace the $MP$ and $FIE$ components in the benchmark regression in Equation \ref{eq:LP_pooled} with the pure monetary policy shock and the information component constructed by \cite{miranda2022tale}. This figure presents the results for the impact of the information component. The left column presents the results for the sample July 1991 to June 2019, the middle column for the sample January 1998 to June 2019, and the right column for the sample January 2008 to June 2019. The rows represent the impact on (i) the trade weighted multilateral real exchange rate (in logs times 100); (ii) long term interest rates in basis points; (iii) the consumer price index (in logs times 100); (iv) the industrial production index (in logs times 100); (v) the equity index (in logs times 100). The solid black line represents the point estimate, the dark blue area represents the 68\% confidence interval, and the light blue area represents the 90\% confidence interval. In the text, when referring to Panel $(i,j)$, $i$ refers to the row and $j$ to the column of the figure. Each variable, in its own transformation, is demeaned at the country level.}
\end{figure}

\newpage
\begin{figure}
    \centering
    \includegraphics[scale=0.4]{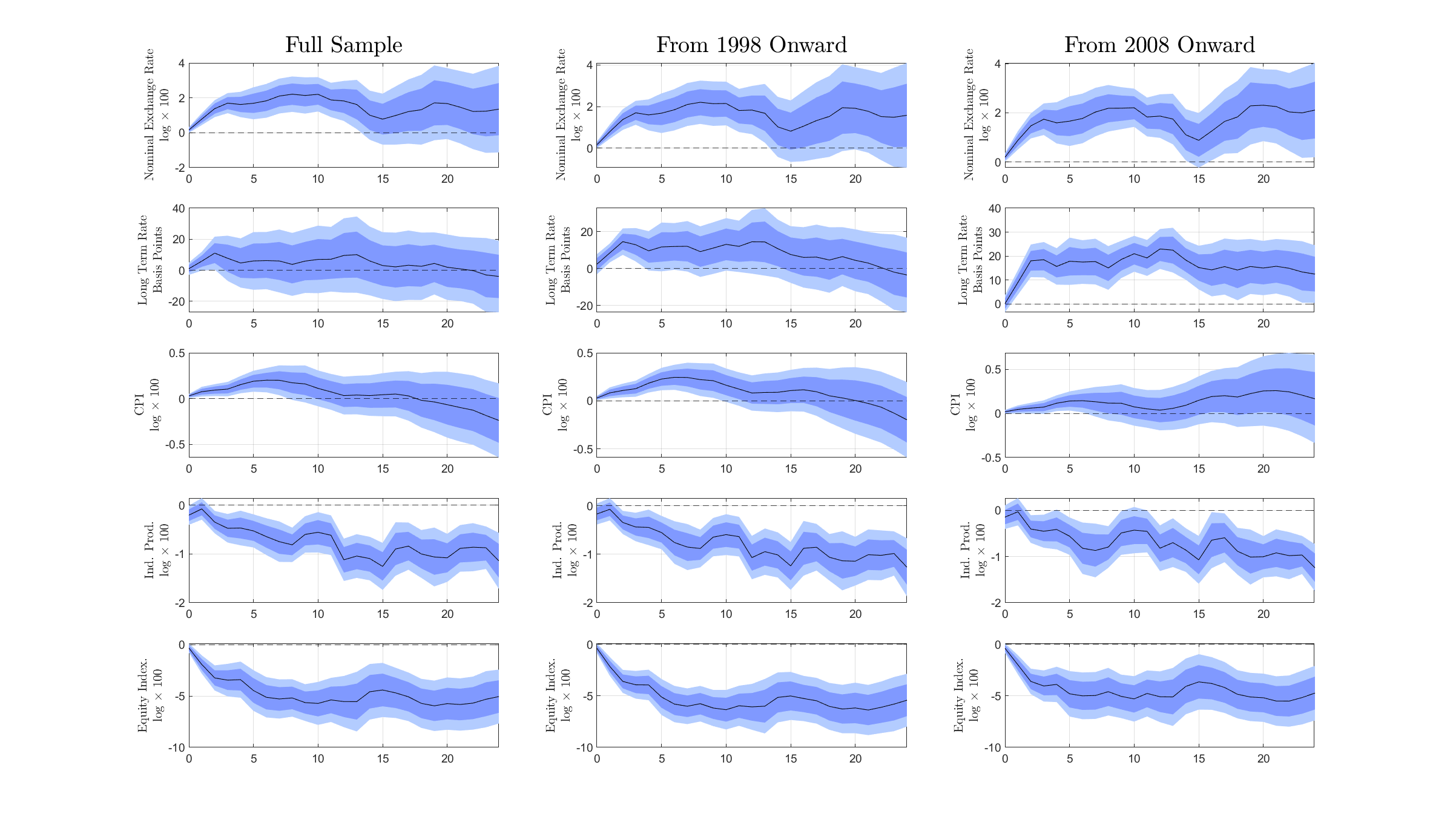}
    \caption{Impulse Response Functions - Pure Monetary Policy Shock \\ QE Shock - \cite{miranda2022tale}}
    \label{fig:QE_Comparison_MP}
    \floatfoot{\textbf{Note:} The figure is comprised of 15 sub-figures ordered in three columns and five rows. To estimate the international spillovers I replace the $MP$ and $FIE$ components in the benchmark regression in Equation \ref{eq:LP_pooled} with the pure monetary policy shock and the information component constructed by \cite{miranda2022tale}. This figure presents the results for the impact of the pure monetary policy shock. The left column presents the results for the sample July 1991 to June 2019, the middle column for the sample January 1998 to June 2019, and the right column for the sample January 2008 to June 2019. The rows represent the impact on (i) the nominal exchange rate with the US dollar (in logs times 100); (ii) long term interest rates in basis points; (iii) the consumer price index (in logs times 100); (iv) the industrial production index (in logs times 100); (v) the equity index (in logs times 100). The solid black line represents the point estimate, the dark blue area represents the 68\% confidence interval, and the light blue area represents the 90\% confidence interval. In the text, when referring to Panel $(i,j)$, $i$ refers to the row and $j$ to the column of the figure. Each variable, in its own transformation, is demeaned at the country level.}
\end{figure}

\newpage
\begin{figure}
    \centering
    \includegraphics[scale=0.4]{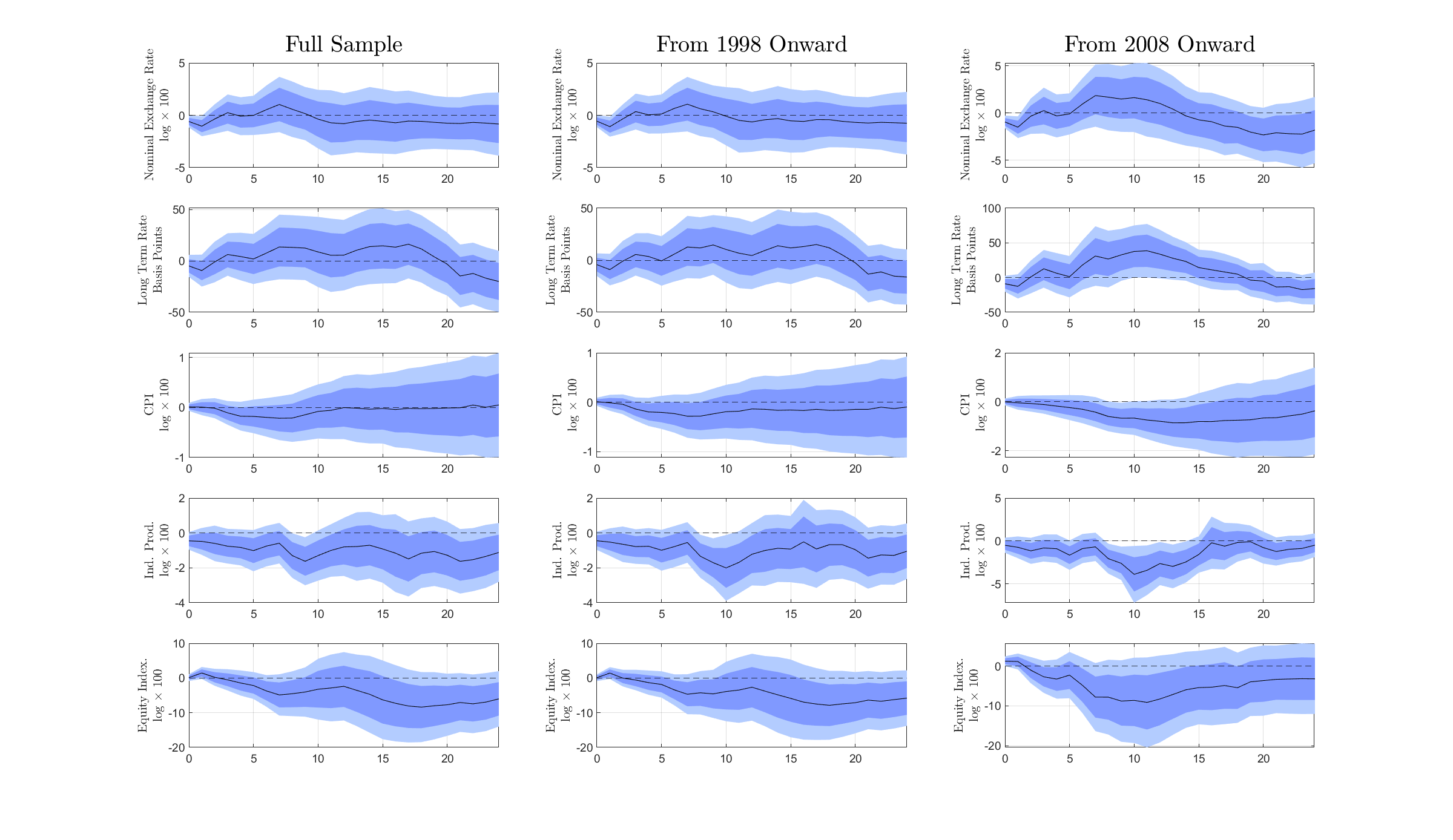}
    \caption{Impulse Response Functions - Information Component \\  QE Shock - \cite{miranda2022tale}}
    \label{fig:QE_Comparison_CBI}
    \floatfoot{\textbf{Note:} The figure is comprised of 15 sub-figures ordered in three columns and five rows. To estimate the international spillovers I replace the $MP$ and $FIE$ components in the benchmark regression in Equation \ref{eq:LP_pooled} with the pure monetary policy shock and the information component constructed by \cite{miranda2022tale}. This figure presents the results for the information component. The left column presents the results for the sample July 1991 to June 2019, the middle column for the sample January 1998 to June 2019, and the right column for the sample January 2008 to June 2019. The rows represent the impact on (i) the nominal exchange rate with the US dollar (in logs times 100); (ii) long term interest rates in basis points; (iii) the consumer price index (in logs times 100); (iv) the industrial production index (in logs times 100); (v) the equity index (in logs times 100). The solid black line represents the point estimate, the dark blue area represents the 68\% confidence interval, and the light blue area represents the 90\% confidence interval. In the text, when referring to Panel $(i,j)$, $i$ refers to the row and $j$ to the column of the figure. Each variable, in its own transformation, is demeaned at the country level.}
\end{figure}

\newpage
\begin{figure}
    \centering
    \includegraphics[scale=0.4]{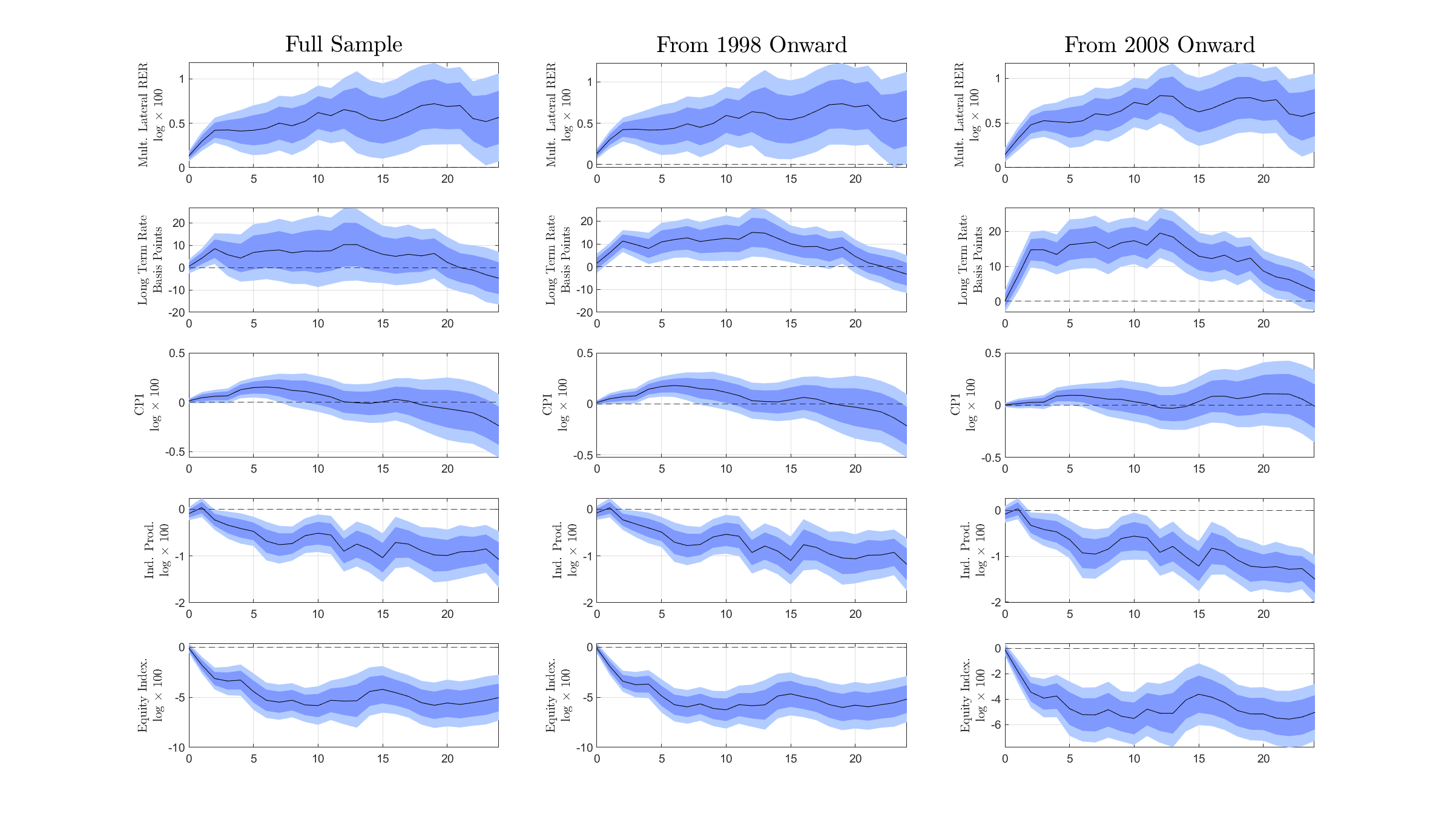}
    \caption{Impulse Response Functions - Pure Monetary Policy Shock \\ Multi. REER Sample - QE Shock - \cite{miranda2022tale}}
    \label{fig:QE_Comparison_MP_REER}
    \floatfoot{\textbf{Note:} The figure is comprised of 15 sub-figures ordered in three columns and five rows. To estimate the international spillovers I replace the $MP$ and $FIE$ components in the benchmark regression in Equation \ref{eq:LP_pooled} with the pure monetary policy shock and the information component constructed by \cite{miranda2022tale}. This figure presents the results for the impact of the pure monetary policy shock. The left column presents the results for the sample July 1991 to June 2019, the middle column for the sample January 1998 to June 2019, and the right column for the sample January 2008 to June 2019. The rows represent the impact on (i) the trade weighted multilateral real exchange rate (in logs times 100); (ii) long term interest rates in basis points; (iii) the consumer price index (in logs times 100); (iv) the industrial production index (in logs times 100); (v) the equity index (in logs times 100). The solid black line represents the point estimate, the dark blue area represents the 68\% confidence interval, and the light blue area represents the 90\% confidence interval. In the text, when referring to Panel $(i,j)$, $i$ refers to the row and $j$ to the column of the figure. Each variable, in its own transformation, is demeaned at the country level.}
\end{figure}

\newpage
\begin{figure}
    \centering
    \includegraphics[scale=0.4]{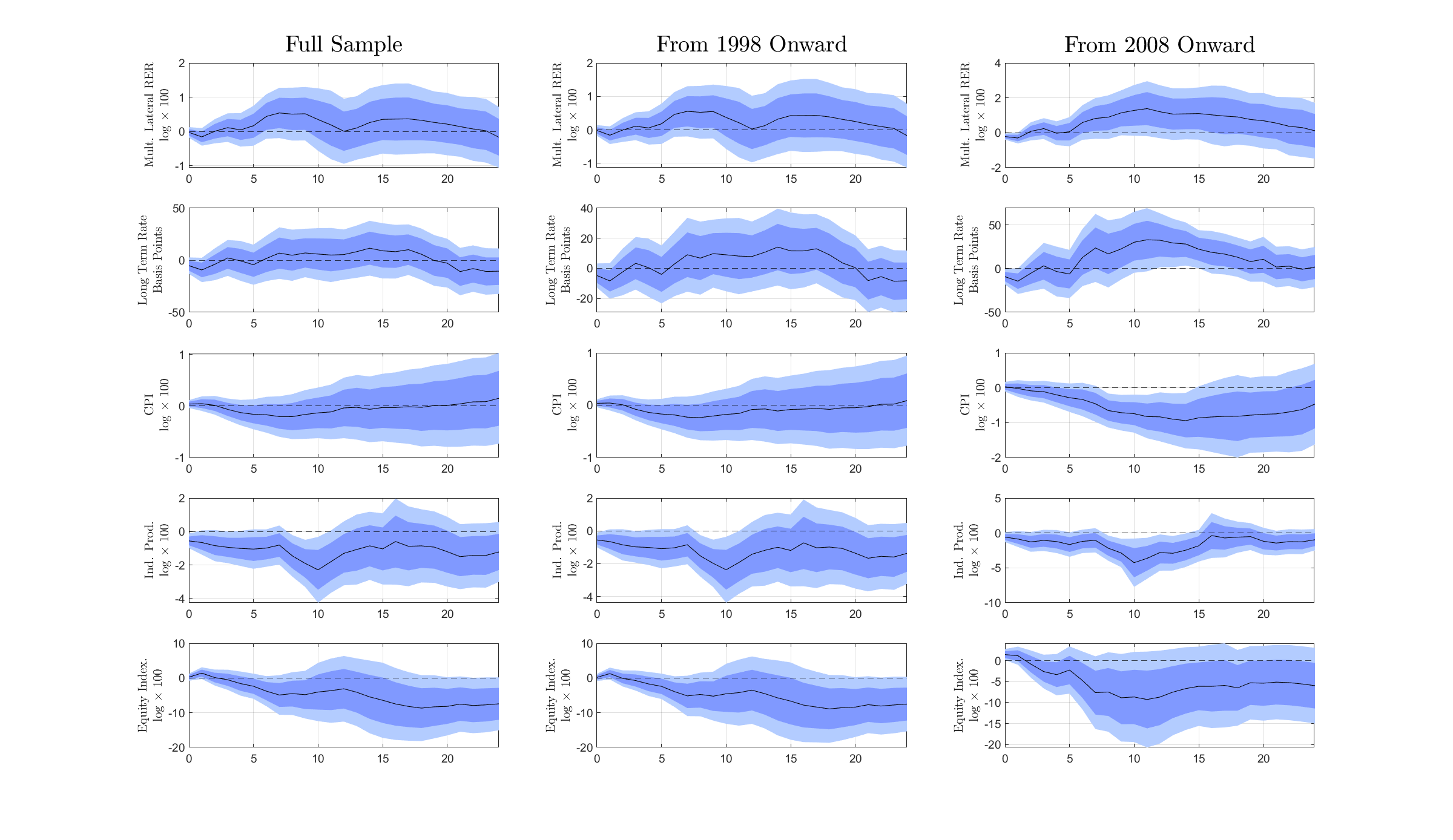}
    \caption{Impulse Response Functions - Information Component \\ Multi. REER Sample - QE Shock -  \cite{miranda2022tale}}
    \label{fig:QE_Comparison_CBI_REER}
    \floatfoot{\textbf{Note:} The figure is comprised of 15 sub-figures ordered in three columns and five rows. To estimate the international spillovers I replace the $MP$ and $FIE$ components in the benchmark regression in Equation \ref{eq:LP_pooled} with the pure monetary policy shock and the information component constructed by \cite{miranda2022tale}. This figure presents the results for the impact of the information component. The left column presents the results for the sample July 1991 to June 2019, the middle column for the sample January 1998 to June 2019, and the right column for the sample January 2008 to June 2019. The rows represent the impact on (i) the trade weighted multilateral real exchange rate (in logs times 100); (ii) long term interest rates in basis points; (iii) the consumer price index (in logs times 100); (iv) the industrial production index (in logs times 100); (v) the equity index (in logs times 100). The solid black line represents the point estimate, the dark blue area represents the 68\% confidence interval, and the light blue area represents the 90\% confidence interval. In the text, when referring to Panel $(i,j)$, $i$ refers to the row and $j$ to the column of the figure. Each variable, in its own transformation, is demeaned at the country level.}
\end{figure}

\newpage
\begin{figure}
    \centering
    \includegraphics[scale=0.4]{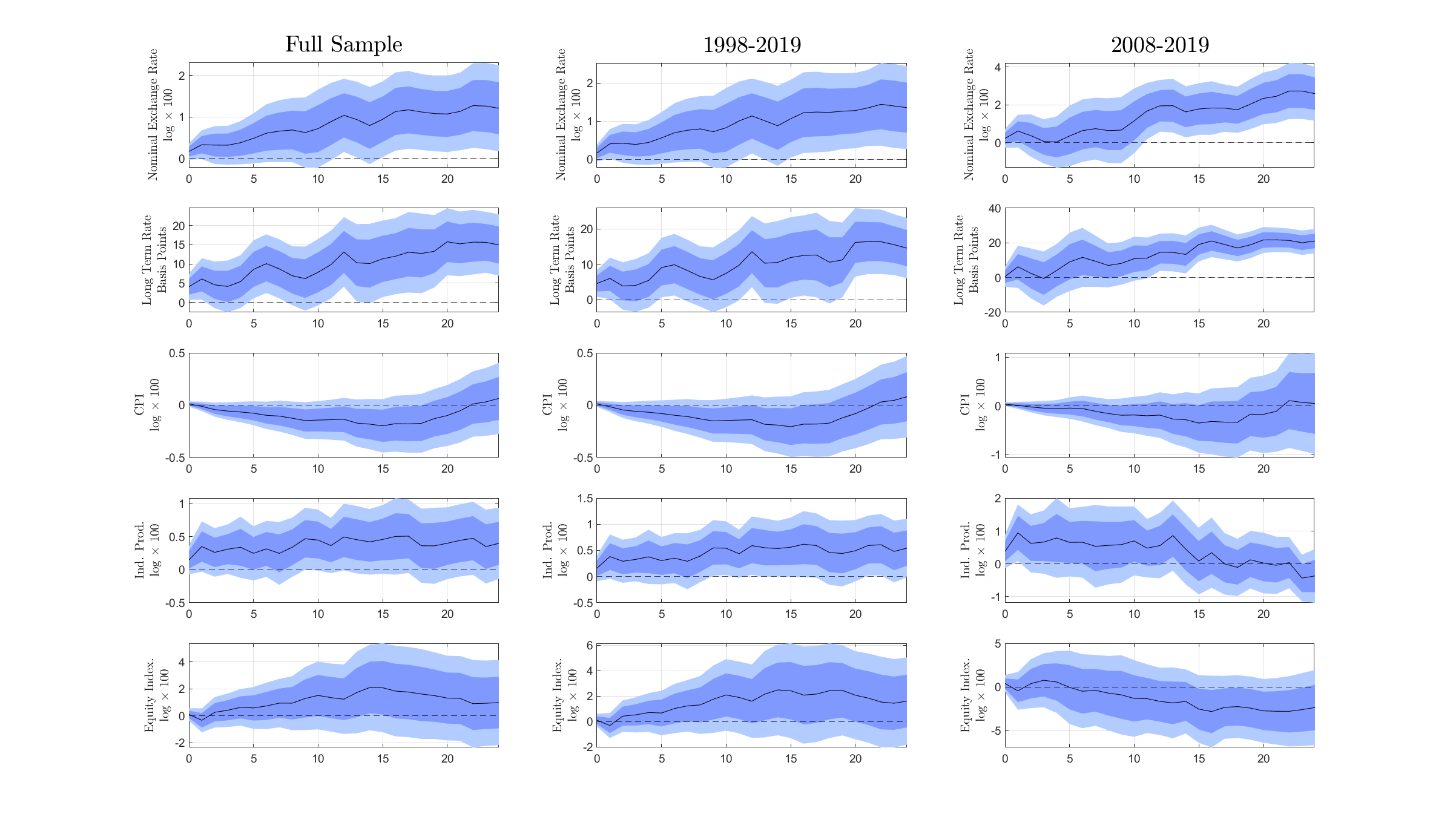}
    \caption{Impulse Response Functions - \cite{nakamura2018identification}}
    \label{fig:BenchmarkNER_NS}
    \floatfoot{\textbf{Note:} The figure is comprised of 15 sub-figures ordered in three columns and five rows. To estimate the international spillovers I replace the $MP$ and $FIE$ components in the benchmark regression in Equation \ref{eq:LP_pooled} with the US monetary policy shock constructed by \cite{nakamura2018identification}. The left column presents the results for the sample July 1991 to June 2019, the middle column for the sample January 1998 to June 2019, and the right column for the sample January 2008 to June 2019. The rows represent the impact on (i) the nominal exchange rate with the US dollar (in logs times 100); (ii) long term interest rates in basis points; (iii) the consumer price index (in logs times 100); (iv) the industrial production index (in logs times 100); (v) the equity index (in logs times 100). The solid black line represents the point estimate, the dark blue area represents the 68\% confidence interval, and the light blue area represents the 90\% confidence interval. In the text, when referring to Panel $(i,j)$, $i$ refers to the row and $j$ to the column of the figure. Each variable, in its own transformation, is demeaned at the country level.}
\end{figure}

\newpage
\begin{figure}
    \centering
    \includegraphics[scale=0.4]{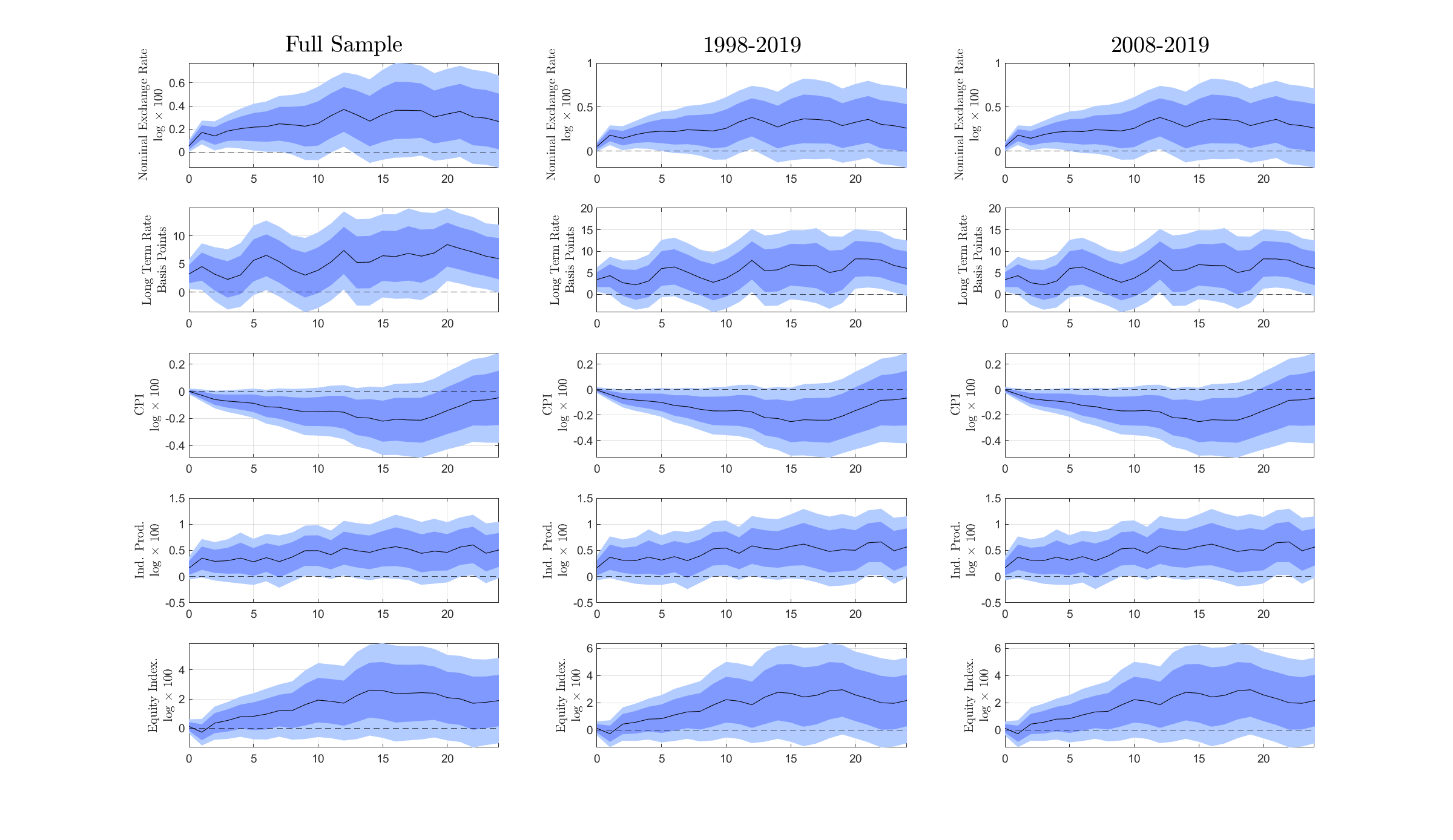}
    \caption{Impulse Response Functions - \cite{nakamura2018identification} \\ Multi. REER Sample}
    \label{fig:BenchmarkREER_NS}
    \floatfoot{\textbf{Note:} The figure is comprised of 15 sub-figures ordered in three columns and five rows. To estimate the international spillovers I replace the $MP$ and $FIE$ components in the benchmark regression in Equation \ref{eq:LP_pooled} with the US monetary policy shock constructed by \cite{nakamura2018identification}. The left column presents the results for the sample July 1991 to June 2019, the middle column for the sample January 1998 to June 2019, and the right column for the sample January 2008 to June 2019. The rows represent the impact on (i) the trade weighted real exchange rate (in logs times 100); (ii) long term interest rates in basis points; (iii) the consumer price index (in logs times 100); (iv) the industrial production index (in logs times 100); (v) the equity index (in logs times 100). The solid black line represents the point estimate, the dark blue area represents the 68\% confidence interval, and the light blue area represents the 90\% confidence interval. In the text, when referring to Panel $(i,j)$, $i$ refers to the row and $j$ to the column of the figure. Each variable, in its own transformation, is demeaned at the country level.}
\end{figure}


\begin{landscape}
\begin{figure}
    \centering
    \includegraphics[scale=0.4]{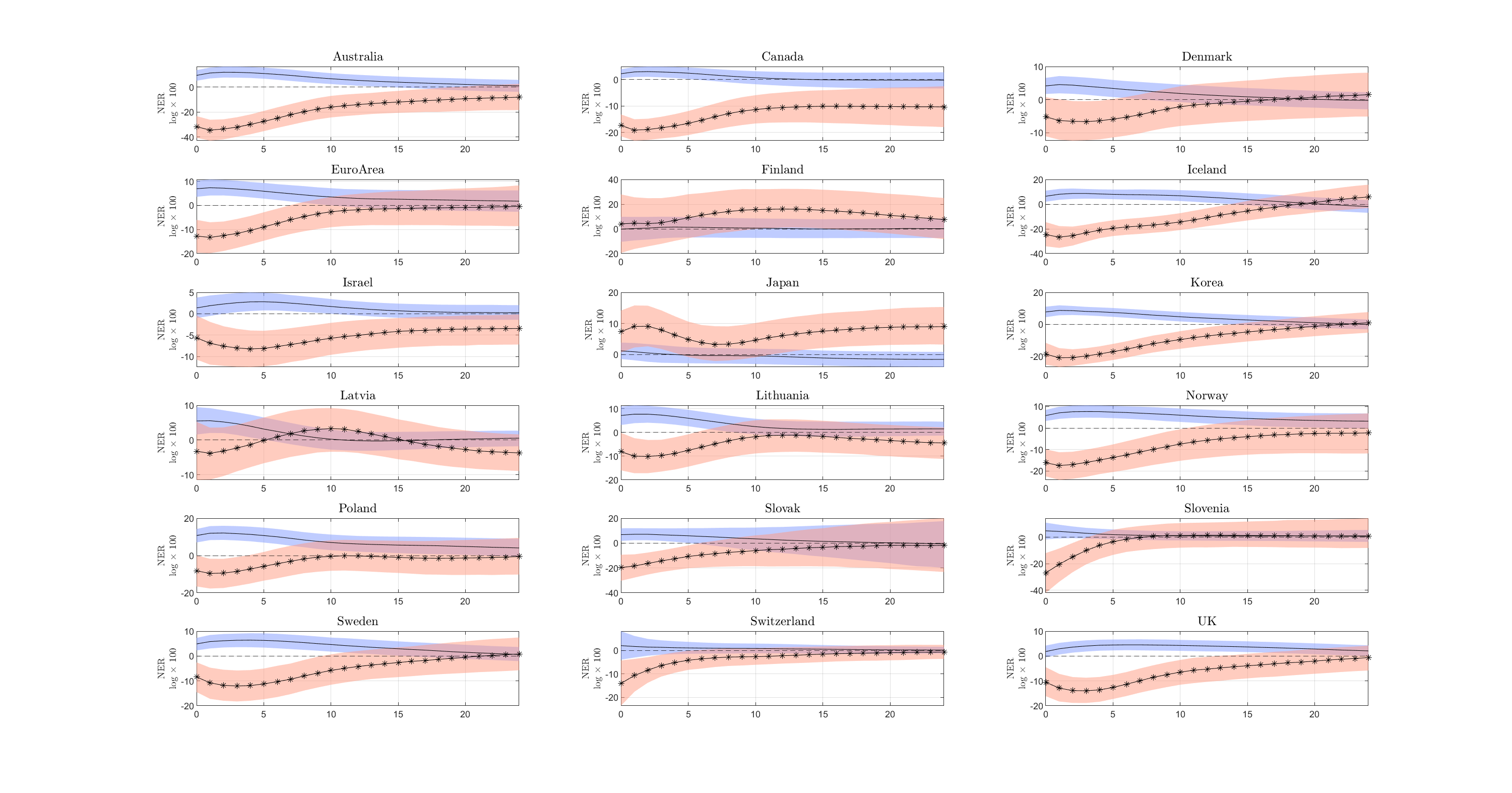}
    \caption{Impulse Response Functions - country-by-country \\ Advanced Economies - NER}
    \label{fig:CbC_NER_AE_NER}
    \floatfoot{\textbf{Note:} The figure is comprised of 18 sub figures ordered in three columns and six rows. The figure represents the response of the nominal exchange rate with respect to the US dollar (log times 100) for the 18 Advanced Economies. The full specification is as specified in Section \ref{subsec:additional_results}. The model is estimated for each country for its longest possible sample. See Appendix \ref{sec:appendix_data_details}. Note that I replace the countries that enter the Euro Area early with an Euro Area unit. The solid black line represents the median impulse response function of the MP component. The blue area represents the 68\% confidence interval for the MP component. The asterisk-line represents the median impulse response function of the FIE component. The red area represents the 68\% confidence interval for the MP component. In the text, when referring to Panel $(i,j)$, $i$ refers to the row and $j$ to the column of the figure.}
\end{figure}
\end{landscape}

\begin{landscape}
\begin{figure}
    \centering
    \includegraphics[scale=0.4]{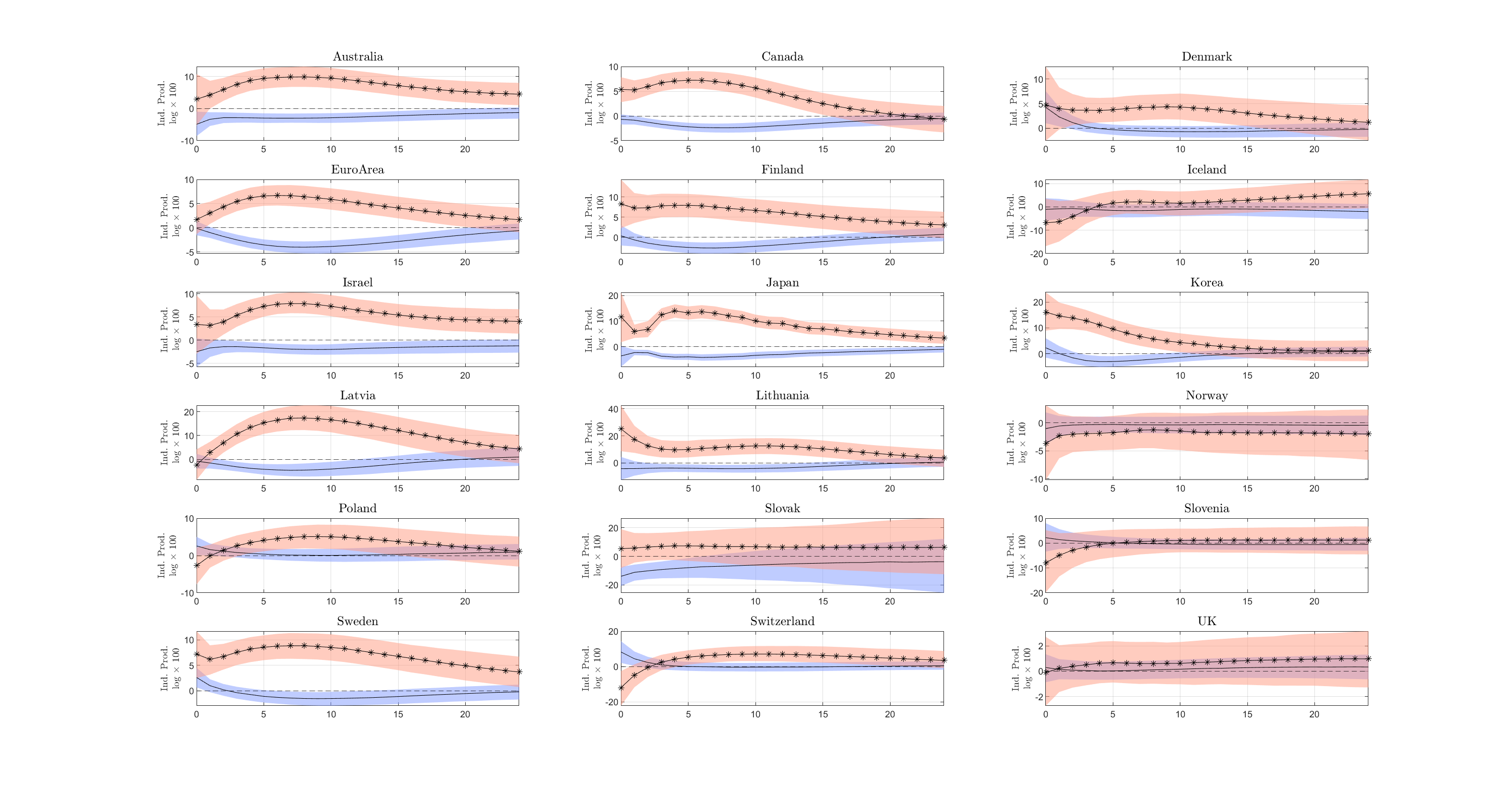}
    \caption{Impulse Response Functions - country-by-country \\ Advanced Economies - Industrial Production}
    \label{fig:CbC_NER_AE_IND}
    \floatfoot{\textbf{Note:} The figure is comprised of 18 sub figures ordered in three columns and six rows. The figure represents the response of the industrial production index (log times 100) for the 18 Advanced Economies. The full specification is as specified in Section \ref{subsec:additional_results}. The model is estimated for each country for its longest possible sample. See Appendix \ref{sec:appendix_data_details}. Note that I replace the countries that enter the Euro Area early with an Euro Area unit. The solid black line represents the median impulse response function of the MP component. The blue area represents the 68\% confidence interval for the MP component. The asterisk-line represents the median impulse response function of the FIE component. The red area represents the 68\% confidence interval for the MP component. In the text, when referring to Panel $(i,j)$, $i$ refers to the row and $j$ to the column of the figure.}
\end{figure}
\end{landscape}

\begin{landscape}
\begin{figure}
    \centering
    \includegraphics[scale=0.4]{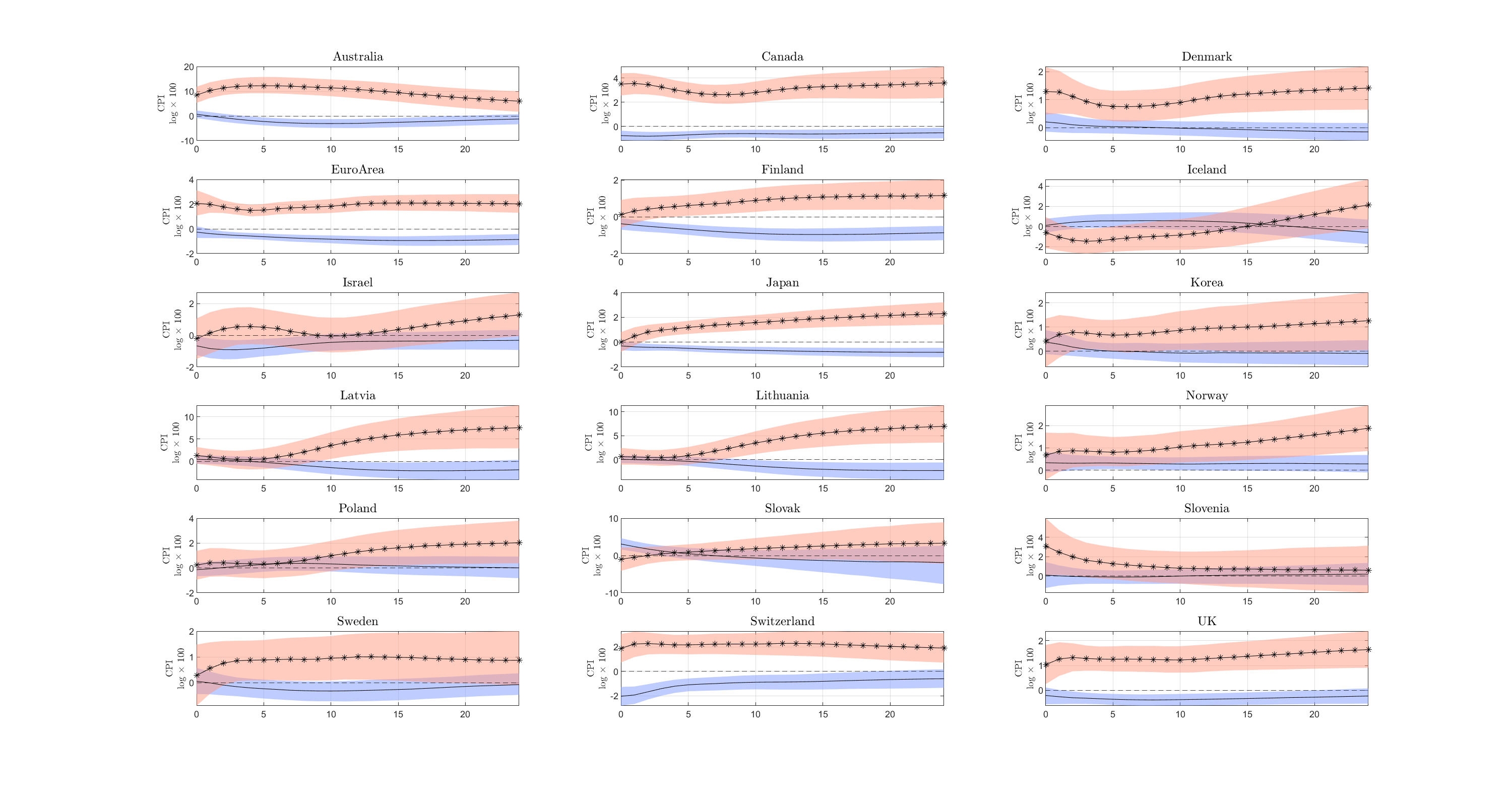}
    \caption{Impulse Response Functions - country-by-country \\ Advanced Economies - CPI}
    \label{fig:CbC_NER_AE_CPI}
    \floatfoot{\textbf{Note:} The figure is comprised of 18 sub figures ordered in three columns and six rows. The figure represents the response of the consumer price index (log times 100) for the 18 Advanced Economies. The full specification is as specified in Section \ref{subsec:additional_results}. The model is estimated for each country for its longest possible sample. See Appendix \ref{sec:appendix_data_details}. Note that I replace the countries that enter the Euro Area early with an Euro Area unit. The solid black line represents the median impulse response function of the MP component. The blue area represents the 68\% confidence interval for the MP component. The asterisk-line represents the median impulse response function of the FIE component. The red area represents the 68\% confidence interval for the MP component. In the text, when referring to Panel $(i,j)$, $i$ refers to the row and $j$ to the column of the figure.}
\end{figure}
\end{landscape}

\begin{landscape}
\begin{figure}
    \centering
    \includegraphics[scale=0.4]{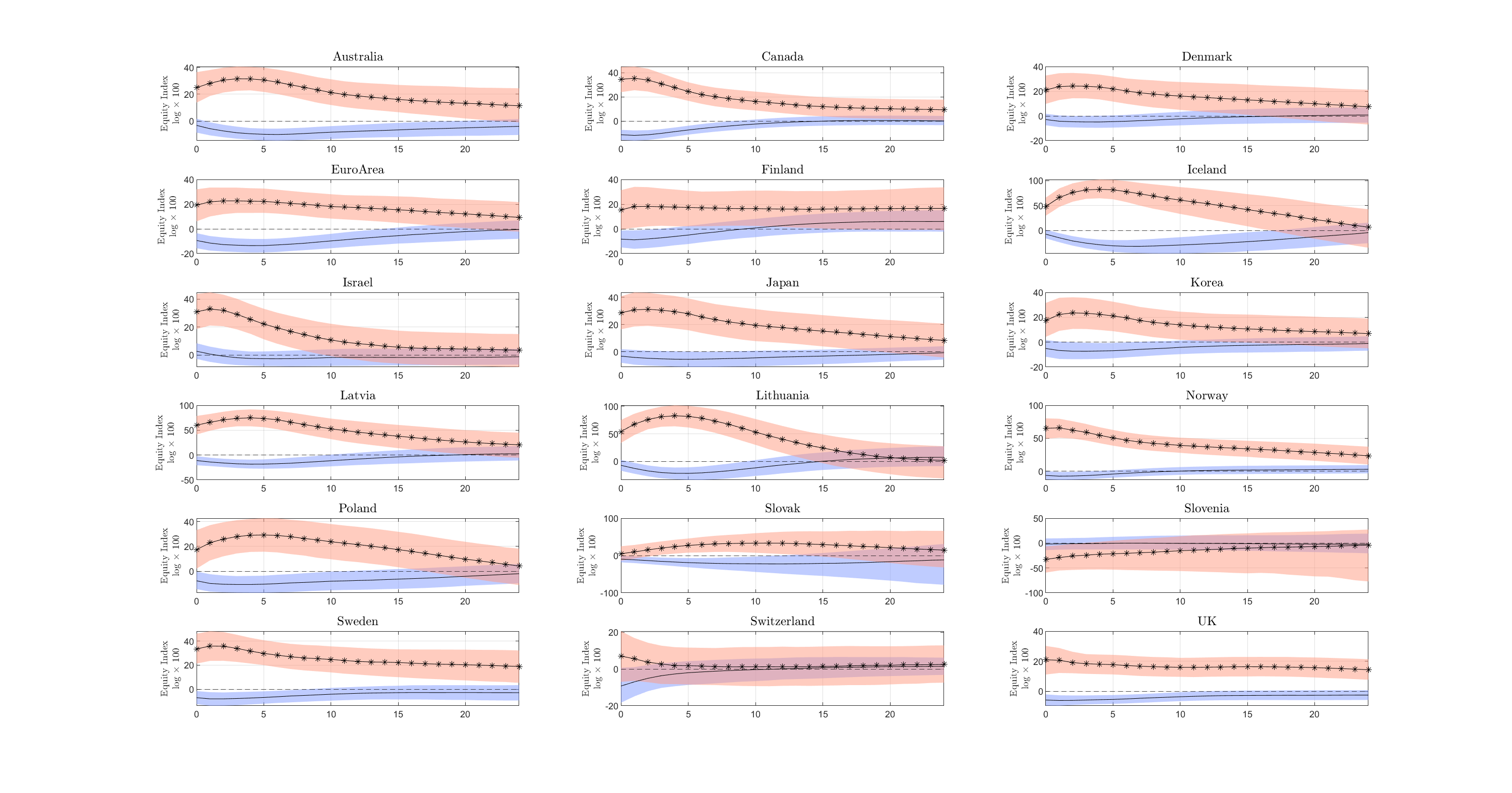}
    \caption{Impulse Response Functions - country-by-country \\ Advanced Economies - Equity Index}
    \label{fig:CbC_NER_AE_EQUITY}
    \floatfoot{\textbf{Note:} The figure is comprised of 18 sub figures ordered in three columns and six rows. The figure represents the response of the equity index (log times 100) for the 18 Advanced Economies. The full specification is as specified in Section \ref{subsec:additional_results}. The model is estimated for each country for its longest possible sample. See Appendix \ref{sec:appendix_data_details}. Note that I replace the countries that enter the Euro Area early with an Euro Area unit. The solid black line represents the median impulse response function of the MP component. The blue area represents the 68\% confidence interval for the MP component. The asterisk-line represents the median impulse response function of the FIE component. The red area represents the 68\% confidence interval for the MP component. In the text, when referring to Panel $(i,j)$, $i$ refers to the row and $j$ to the column of the figure.}
\end{figure}
\end{landscape}

\begin{landscape}
\begin{figure}
    \centering
    \includegraphics[scale=0.4]{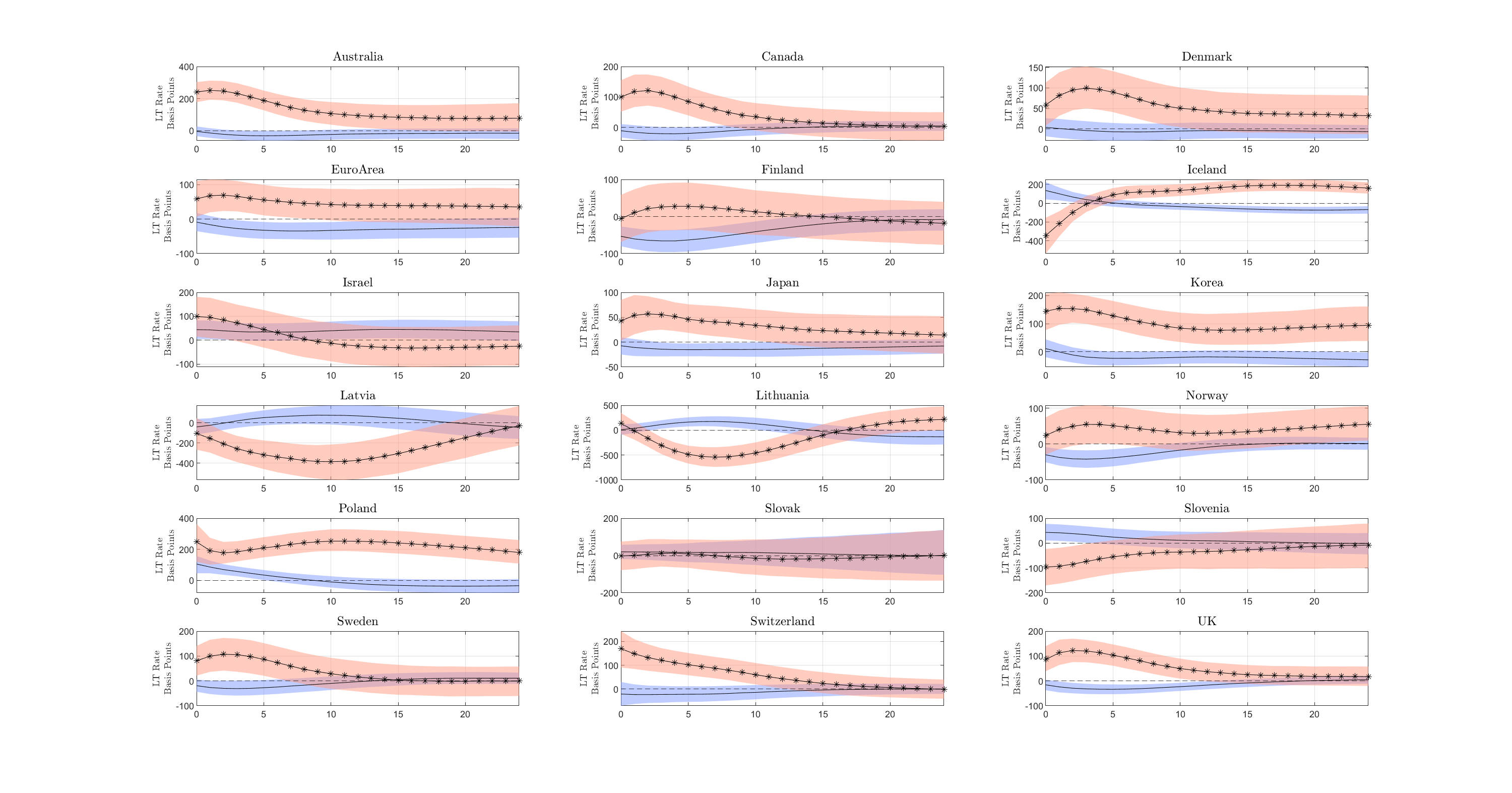}
    \caption{Impulse Response Functions - country-by-country \\ Advanced Economies - Long Term Rate}
    \label{CbC_NER_AE_LT}
    \floatfoot{\textbf{Note:} The figure is comprised of 18 sub figures ordered in three columns and six rows. The figure represents the response of the long term rate (in basis points) for the 18 Advanced Economies. The full specification is as specified in Section \ref{subsec:additional_results}. The model is estimated for each country for its longest possible sample. See Appendix \ref{sec:appendix_data_details}. Note that I replace the countries that enter the Euro Area early with an Euro Area unit. The solid black line represents the median impulse response function of the MP component. The blue area represents the 68\% confidence interval for the MP component. The asterisk-line represents the median impulse response function of the FIE component. The red area represents the 68\% confidence interval for the MP component. In the text, when referring to Panel $(i,j)$, $i$ refers to the row and $j$ to the column of the figure.}
\end{figure}
\end{landscape}

\begin{landscape}
\begin{figure}
    \centering
    \includegraphics[scale=0.4]{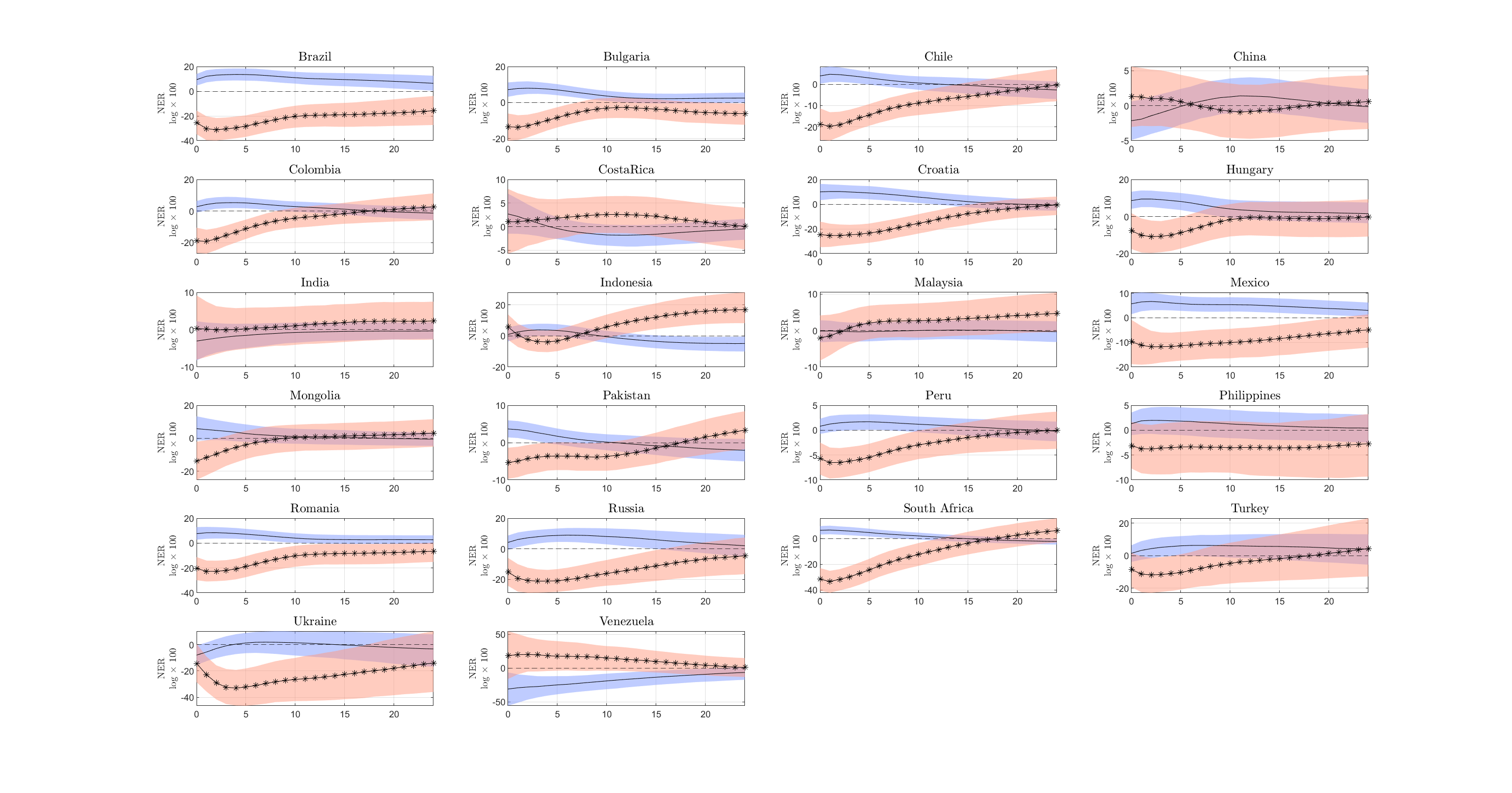}
    \caption{Impulse Response Functions - country-by-country \\ Emerging Market Economies - NER}
    \label{fig:CbC_NER_EM_NER}
    \floatfoot{\textbf{Note:} The figure is comprised of 22 sub figures ordered in four columns and six rows. The figure represents the response of the nominal exchange rate with respect to the US dollar (log times 100) for the 22 Emerging Market Economies. The full specification is as specified in Section \ref{subsec:additional_results}. The model is estimated for each country for its longest possible sample. See Appendix \ref{sec:appendix_data_details}. The solid black line represents the median impulse response function of the MP component. The blue area represents the 68\% confidence interval for the MP component. The asterisk-line represents the median impulse response function of the FIE component. The red area represents the 68\% confidence interval for the MP component. In the text, when referring to Panel $(i,j)$, $i$ refers to the row and $j$ to the column of the figure.}
\end{figure}
\end{landscape}

\begin{landscape}
\begin{figure}
    \centering
    \includegraphics[scale=0.4]{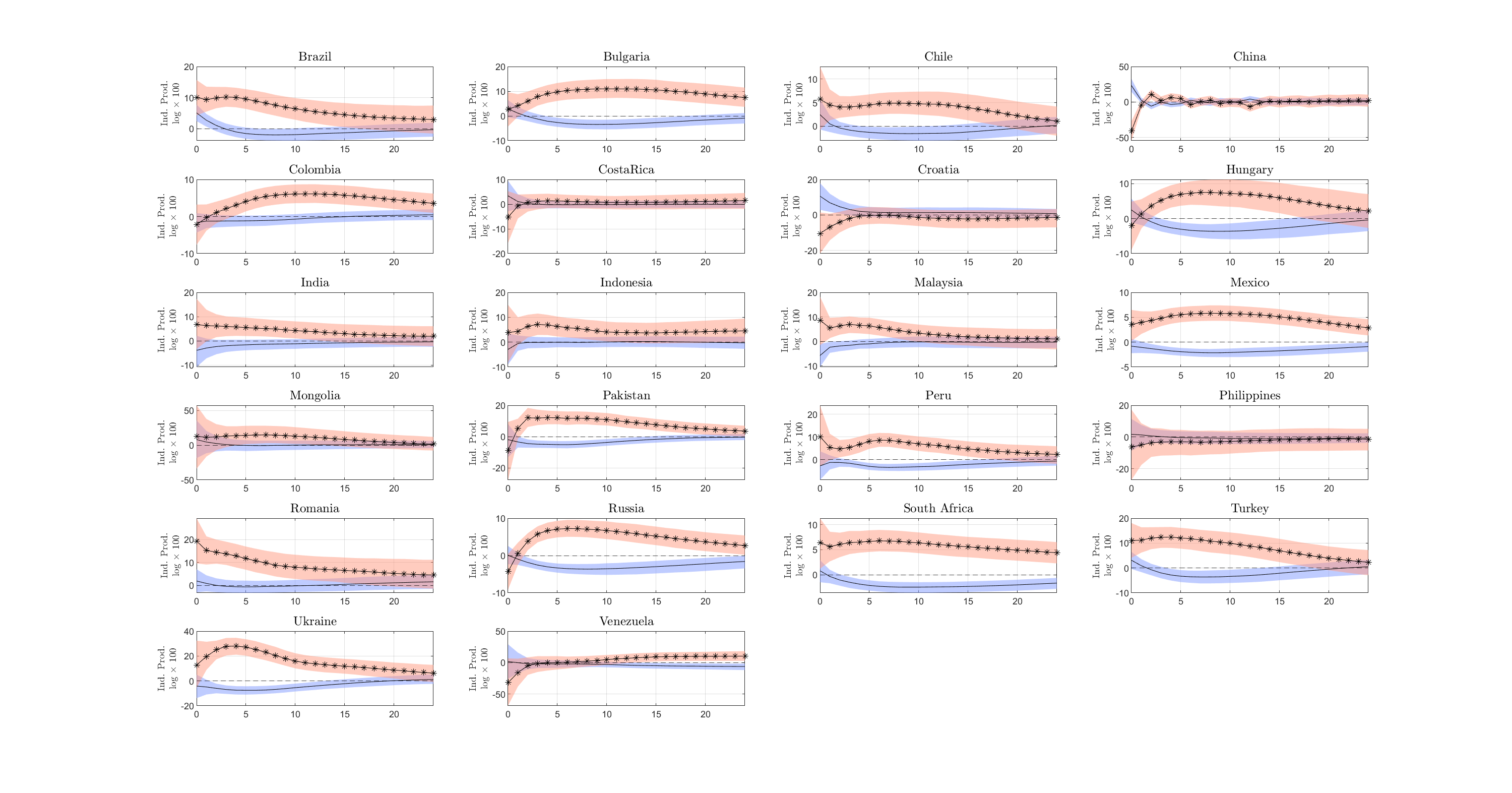}
    \caption{Impulse Response Functions - country-by-country \\ Emerging Market Economies - Ind. Prod.}
    \label{fig:CbC_NER_EM_IND}
    \floatfoot{\textbf{Note:} The figure is comprised of 22 sub figures ordered in four columns and six rows. The figure represents the response of the industrial production index (log times 100) for the 22 Emerging Market Economies. The full specification is as specified in Section \ref{subsec:additional_results}. The model is estimated for each country for its longest possible sample. See Appendix \ref{sec:appendix_data_details}. The solid black line represents the median impulse response function of the MP component. The blue area represents the 68\% confidence interval for the MP component. The asterisk-line represents the median impulse response function of the FIE component. The red area represents the 68\% confidence interval for the MP component. In the text, when referring to Panel $(i,j)$, $i$ refers to the row and $j$ to the column of the figure.}
\end{figure}
\end{landscape}

\begin{landscape}
\begin{figure}
    \centering
    \includegraphics[scale=0.4]{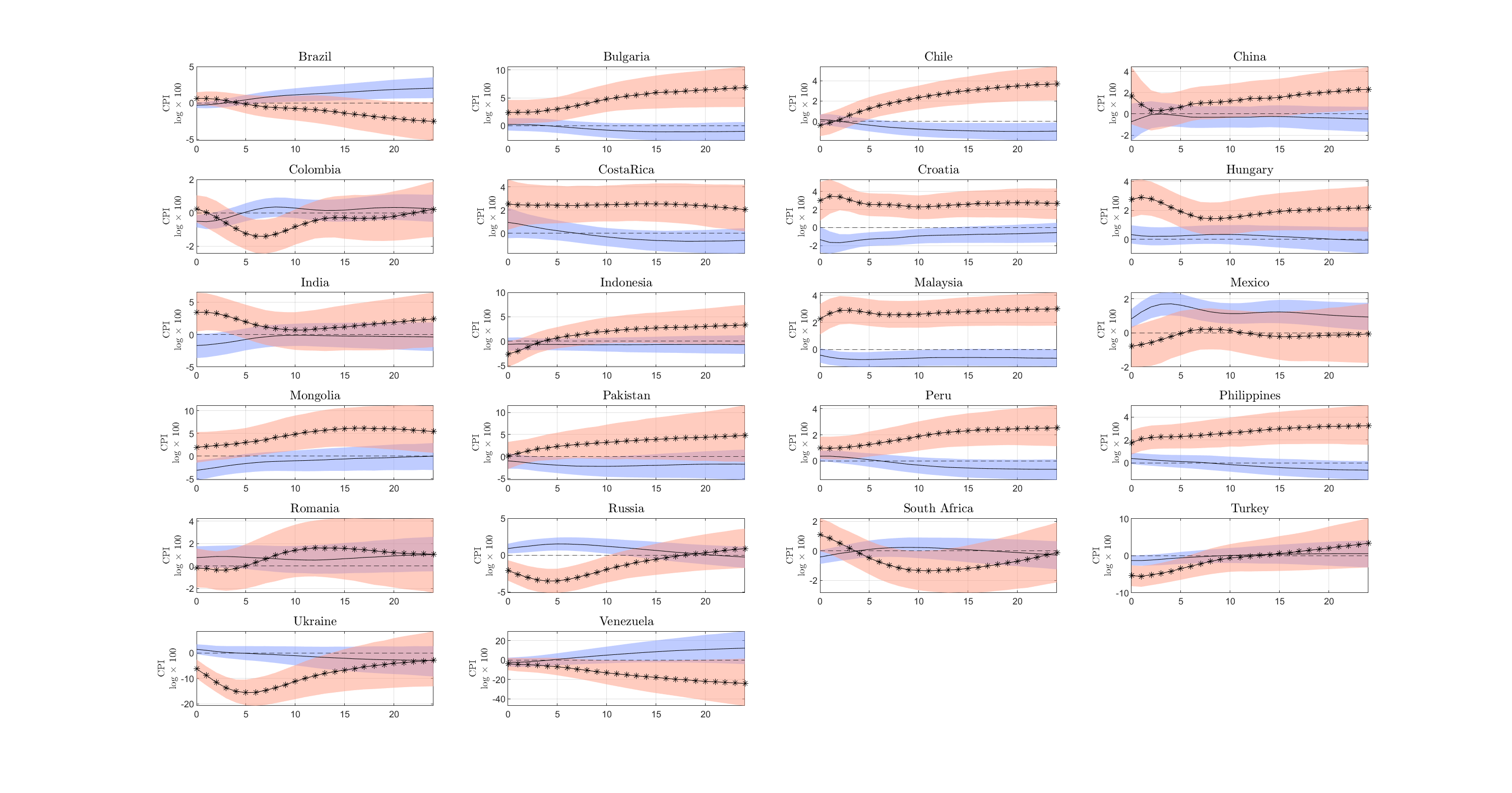}
    \caption{Impulse Response Functions - country-by-country \\ Emerging Market Economies - CPI}
    \label{fig:CbC_NER_EM_CPI}
    \floatfoot{\textbf{Note:} The figure is comprised of 22 sub figures ordered in four columns and six rows. The figure represents the response of the consumer price index (log times 100) for the 22 Emerging Market Economies. The full specification is as specified in Section \ref{subsec:additional_results}. The model is estimated for each country for its longest possible sample. See Appendix \ref{sec:appendix_data_details}. The solid black line represents the median impulse response function of the MP component. The blue area represents the 68\% confidence interval for the MP component. The asterisk-line represents the median impulse response function of the FIE component. The red area represents the 68\% confidence interval for the MP component. In the text, when referring to Panel $(i,j)$, $i$ refers to the row and $j$ to the column of the figure.}
\end{figure}
\end{landscape}

\begin{landscape}
\begin{figure}
    \centering
    \includegraphics[scale=0.4]{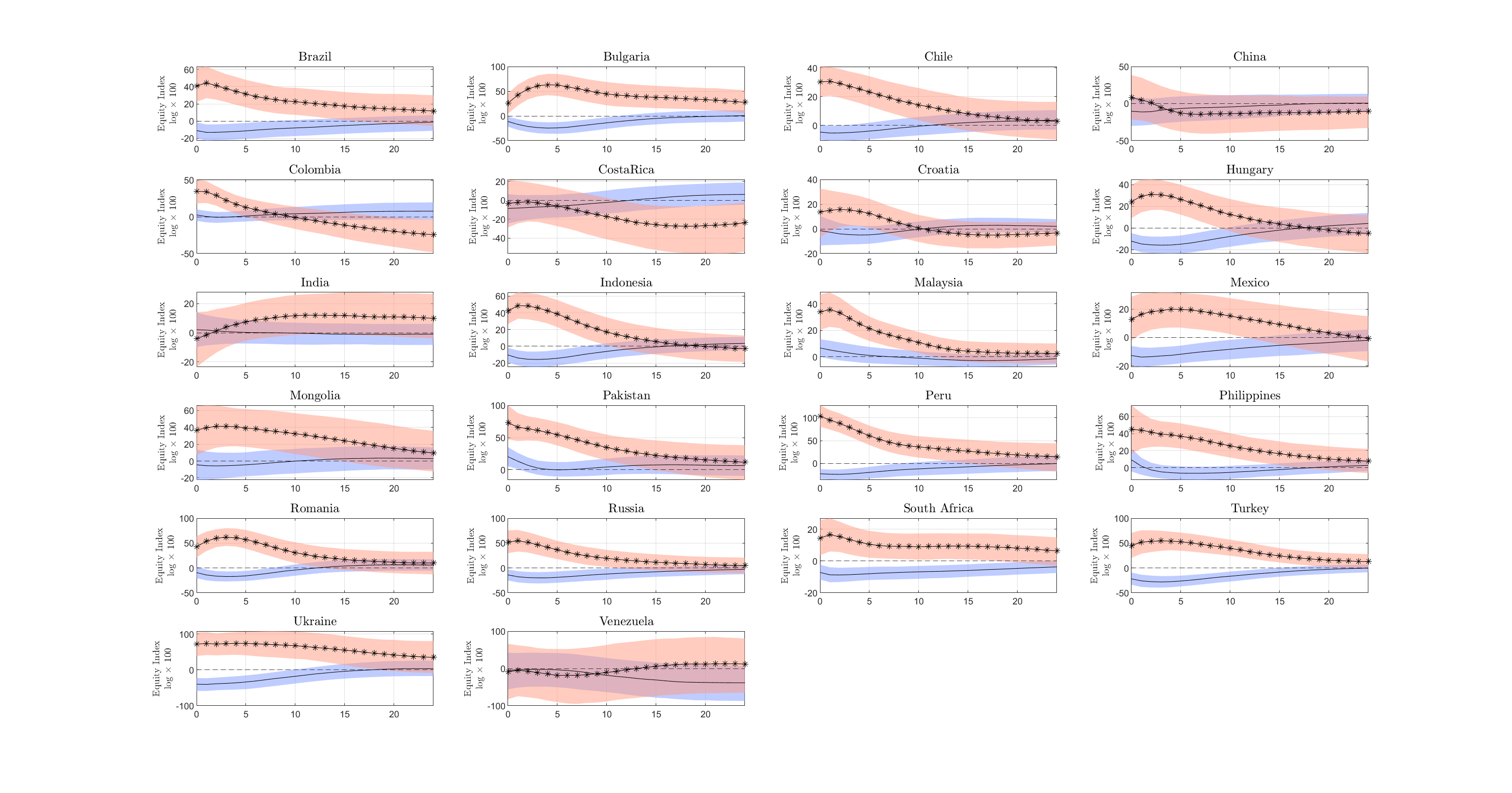}
    \caption{Impulse Response Functions - country-by-country \\ Emerging Market Economies - Equity Index}
    \label{fig:CbC_NER_EM_Equity}
    \floatfoot{\textbf{Note:} The figure is comprised of 22 sub figures ordered in four columns and six rows. The figure represents the response of the equity index (log times 100) for the 22 Emerging Market Economies. The full specification is as specified in Section \ref{subsec:additional_results}. The model is estimated for each country for its longest possible sample. See Appendix \ref{sec:appendix_data_details}. The solid black line represents the median impulse response function of the MP component. The blue area represents the 68\% confidence interval for the MP component. The asterisk-line represents the median impulse response function of the FIE component. The red area represents the 68\% confidence interval for the MP component. In the text, when referring to Panel $(i,j)$, $i$ refers to the row and $j$ to the column of the figure.}
\end{figure}
\end{landscape}

\begin{landscape}
\begin{figure}
    \centering
    \includegraphics[scale=0.4]{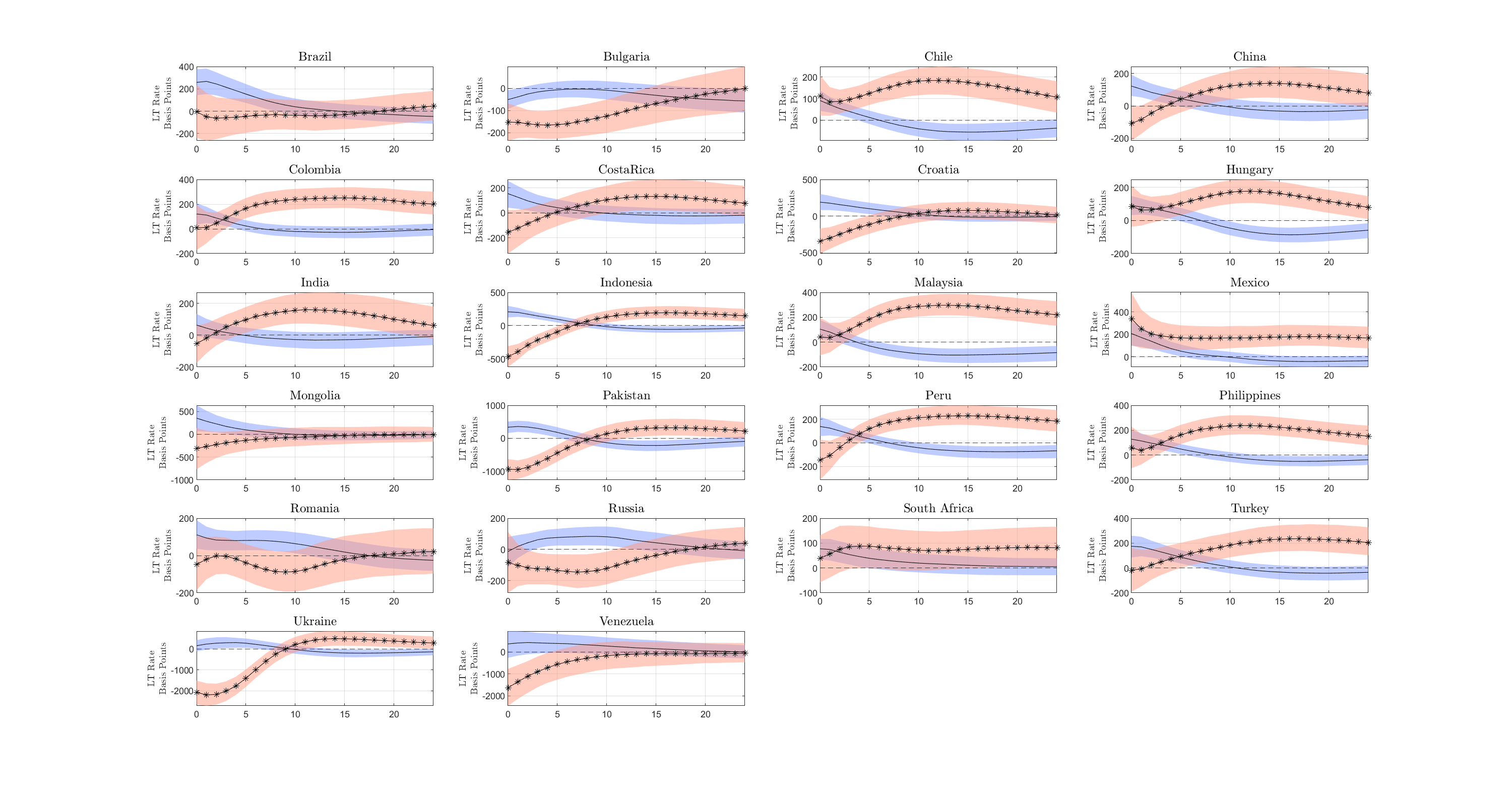}
    \caption{Impulse Response Functions - country-by-country \\ Emerging Market Economies - LT Rate}
    \label{fig:CbC_NER_EM_LT}
    \floatfoot{\textbf{Note:} The figure is comprised of 22 sub figures ordered in four columns and six rows. The figure represents the response of the long term rates (basis points) for the 22 Emerging Market Economies. The full specification is as specified in Section \ref{subsec:additional_results}. The model is estimated for each country for its longest possible sample. See Appendix \ref{sec:appendix_data_details}. The solid black line represents the median impulse response function of the MP component. The blue area represents the 68\% confidence interval for the MP component. The asterisk-line represents the median impulse response function of the FIE component. The red area represents the 68\% confidence interval for the MP component. In the text, when referring to Panel $(i,j)$, $i$ refers to the row and $j$ to the column of the figure.}
\end{figure}
\end{landscape}

\begin{figure}
    \centering
    \includegraphics[scale=0.4]{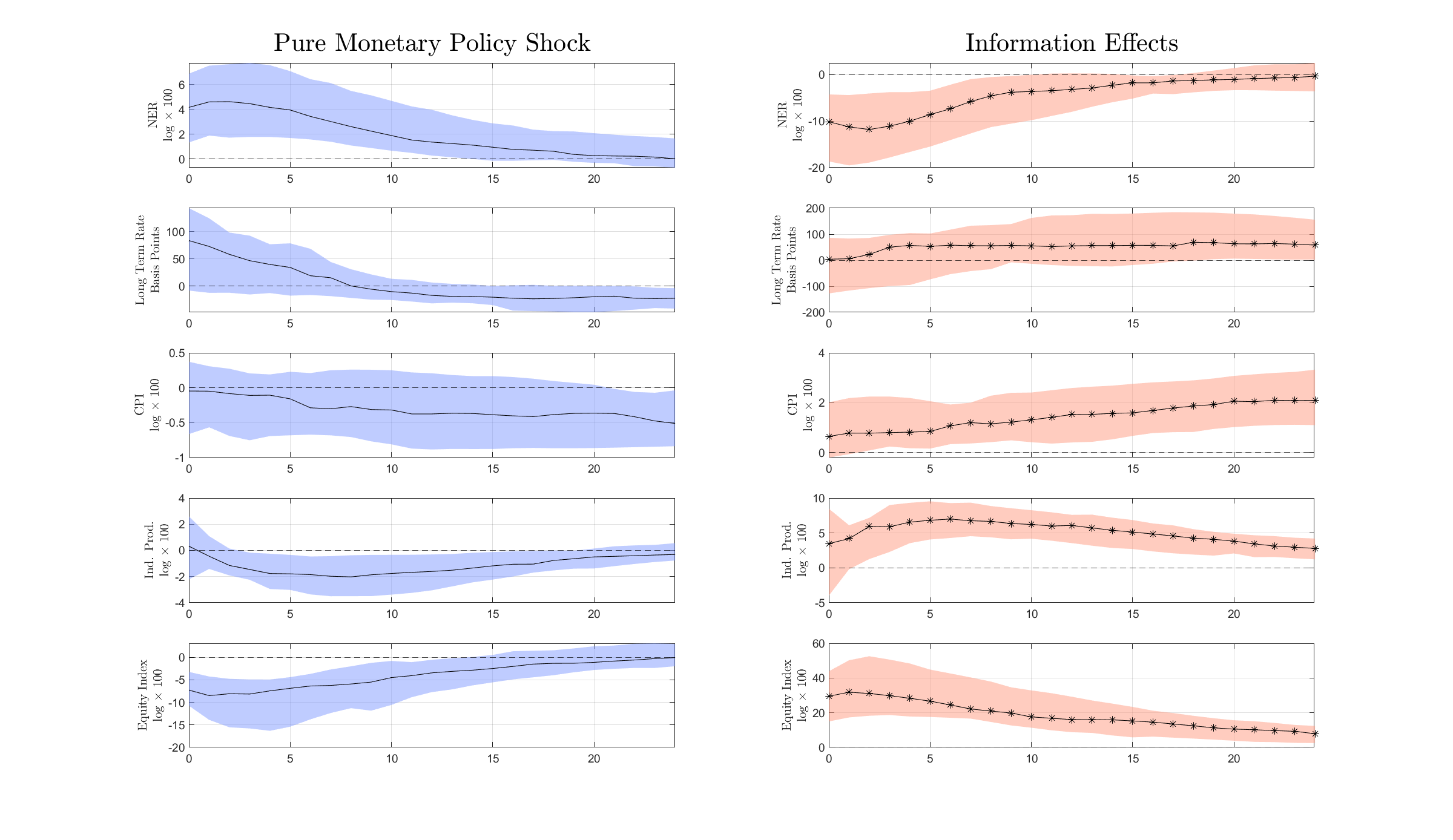}
    \caption{Impulse Response Functions - Median Responses across Countries}
    \label{fig:NER_MedianResponse}
    \floatfoot{\textbf{Note:} The figure is comprised of 10 sub-figures ordered in two columns and five rows. The left column presents the impulse response functions to the MP component while the right column present the impulse response functions to the FIE component. The rows represent the impact on (i) the nominal exchange rate with the US dollar (in logs times 100); (ii) long term interest rates in basis points; (iii) the consumer price index (in logs times 100); (iv) the industrial production index (in logs times 100); (v) the equity index (in logs times 100). I keep the median impulse response function for every country in the sample. The solid black line represents the median response from the set of impulse response functions for each country to a MP component. The light blue area represents the interquartile range of impulse response functions to a MP component. The asterisk black line represents the median response from the set of impulse response functions for each country to a FIE component. The light red area represents the interquartile range of impulse response functions to a FIE component. In the text, when referring to Panel $(i,j)$, $i$ refers to the row and $j$ to the column of the figure.}
\end{figure}

\begin{landscape}
\begin{figure}
    \centering
    \includegraphics[scale=0.4]{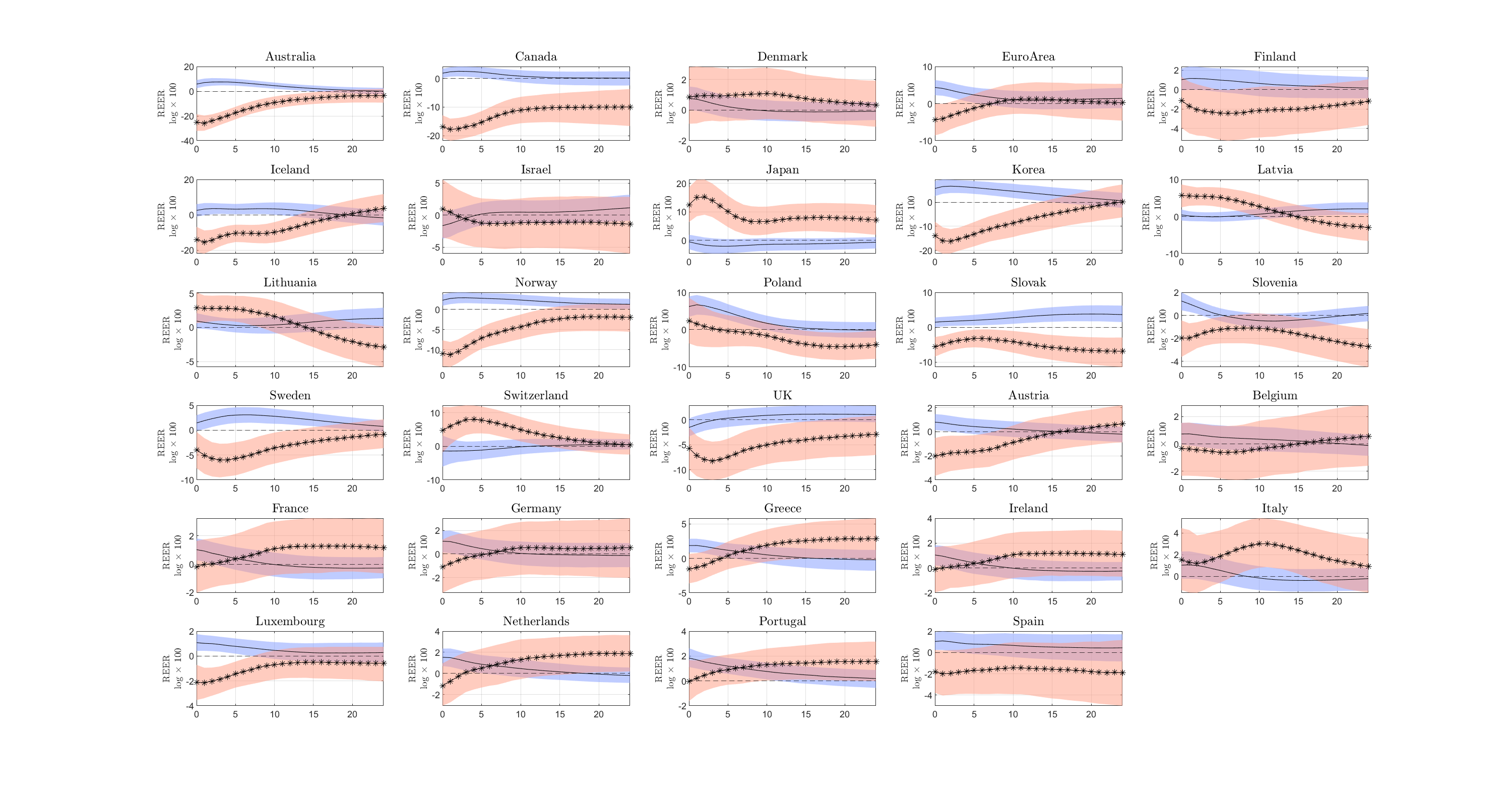}
    \caption{Impulse Response Functions - country-by-country \\ Multi. REER - AE - REER}
    \label{fig:CbC_REER_AE_REER}
    \floatfoot{\textbf{Note:} The figure is comprised of 29 sub figures ordered in five columns and six rows. To estimate the figures I replace the nominal exchange rate with the US with the trade weighted multilateral real exchange rate. The figure represents the response of the trade weighted real exchange rate (log times 100) for the 28 Advanced Economies plus the Euro Area. The full specification is as specified in Section \ref{subsec:additional_results}. The model is estimated for each country for its longest possible sample. See Appendix \ref{sec:appendix_data_details}. The solid black line represents the median impulse response function of the MP component. The blue area represents the 68\% confidence interval for the MP component. The asterisk-line represents the median impulse response function of the FIE component. The red area represents the 68\% confidence interval for the MP component. In the text, when referring to Panel $(i,j)$, $i$ refers to the row and $j$ to the column of the figure.}
\end{figure}
\end{landscape}

\begin{landscape}
\begin{figure}
    \centering
    \includegraphics[scale=0.4]{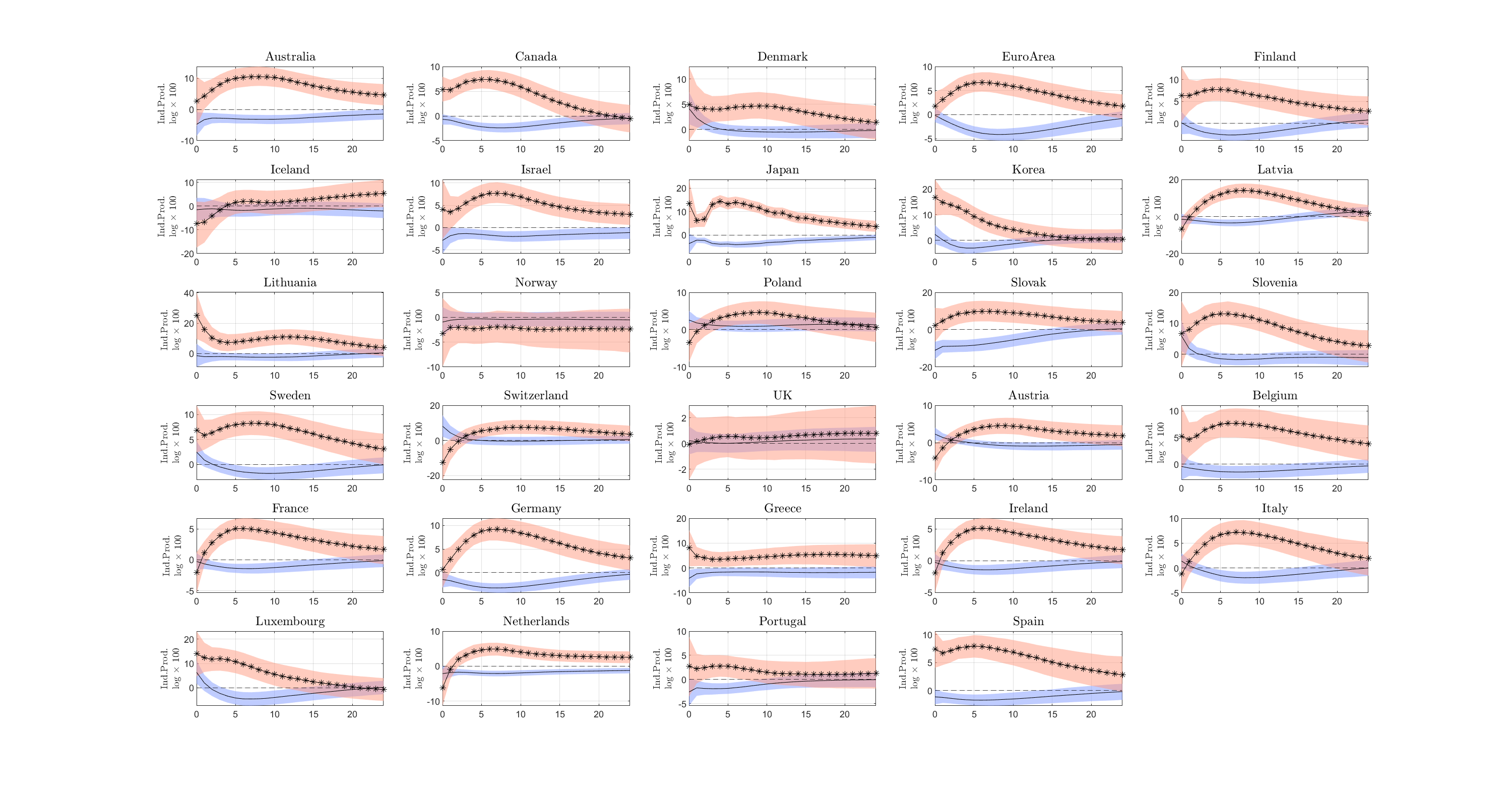}
    \caption{Impulse Response Functions - country-by-country \\ Multi. REER - AE - Ind. Prod.}
    \label{fig:CbC_REER_AE_Ind}
    \floatfoot{\textbf{Note:} The figure is comprised of 29 sub figures ordered in five columns and six rows. To estimate the figures I replace the nominal exchange rate with the US with the trade weighted multilateral real exchange rate. The figure represents the response of the industrial production index (log times 100) for the 28 Advanced Economies plus the Euro Area. The full specification is as specified in Section \ref{subsec:additional_results}. The model is estimated for each country for its longest possible sample. See Appendix \ref{sec:appendix_data_details}. The solid black line represents the median impulse response function of the MP component. The blue area represents the 68\% confidence interval for the MP component. The asterisk-line represents the median impulse response function of the FIE component. The red area represents the 68\% confidence interval for the MP component. In the text, when referring to Panel $(i,j)$, $i$ refers to the row and $j$ to the column of the figure.}
\end{figure}
\end{landscape}

\begin{landscape}
\begin{figure}
    \centering
    \includegraphics[scale=0.4]{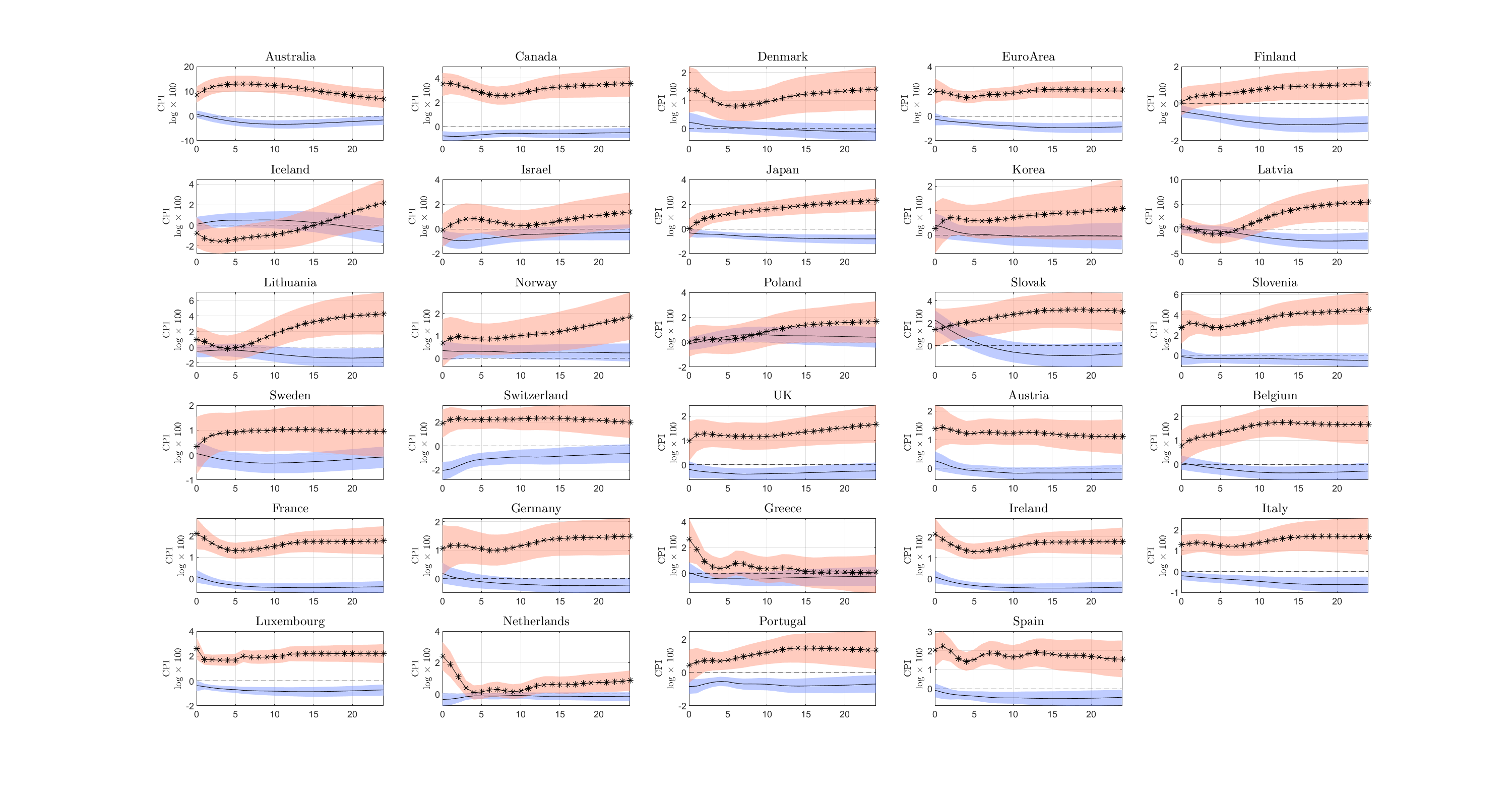}
    \caption{Impulse Response Functions - country-by-country \\ Multi. REER - AE - CPI}
    \label{fig:CbC_REER_AE_CPI}
    \floatfoot{\textbf{Note:} The figure is comprised of 29 sub figures ordered in five columns and six rows. To estimate the figures I replace the nominal exchange rate with the US with the trade weighted multilateral real exchange rate. The figure represents the response of the consumer price index (log times 100) for the 28 Advanced Economies plus the Euro Area. The full specification is as specified in Section \ref{subsec:additional_results}. The model is estimated for each country for its longest possible sample. See Appendix \ref{sec:appendix_data_details}. The solid black line represents the median impulse response function of the MP component. The blue area represents the 68\% confidence interval for the MP component. The asterisk-line represents the median impulse response function of the FIE component. The red area represents the 68\% confidence interval for the MP component. In the text, when referring to Panel $(i,j)$, $i$ refers to the row and $j$ to the column of the figure.}
\end{figure}
\end{landscape}

\begin{landscape}
\begin{figure}
    \centering
    \includegraphics[scale=0.4]{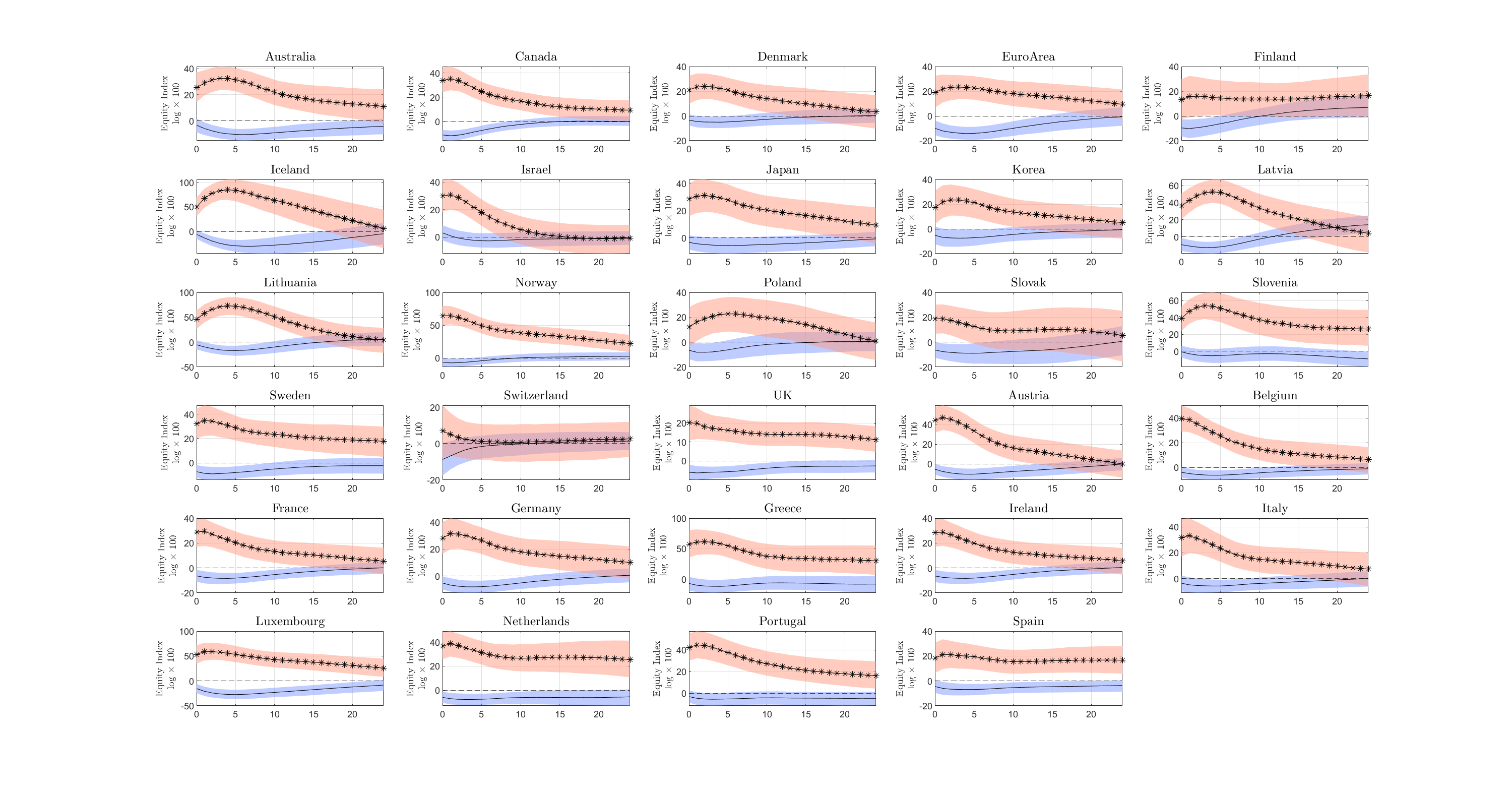}
    \caption{Impulse Response Functions - country-by-country \\ Multi. REER - AE - Equity Index}
    \label{fig:CbC_REER_AE_Equity}
    \floatfoot{\textbf{Note:} The figure is comprised of 29 sub figures ordered in five columns and six rows. To estimate the figures I replace the nominal exchange rate with the US with the trade weighted multilateral real exchange rate. The figure represents the response of the equity index (log times 100) for the 28 Advanced Economies plus the Euro Area. The full specification is as specified in Section \ref{subsec:additional_results}. The model is estimated for each country for its longest possible sample. See Appendix \ref{sec:appendix_data_details}. The solid black line represents the median impulse response function of the MP component. The blue area represents the 68\% confidence interval for the MP component. The asterisk-line represents the median impulse response function of the FIE component. The red area represents the 68\% confidence interval for the MP component. In the text, when referring to Panel $(i,j)$, $i$ refers to the row and $j$ to the column of the figure.}
\end{figure}
\end{landscape}

\begin{landscape}
\begin{figure}
    \centering
    \includegraphics[scale=0.4]{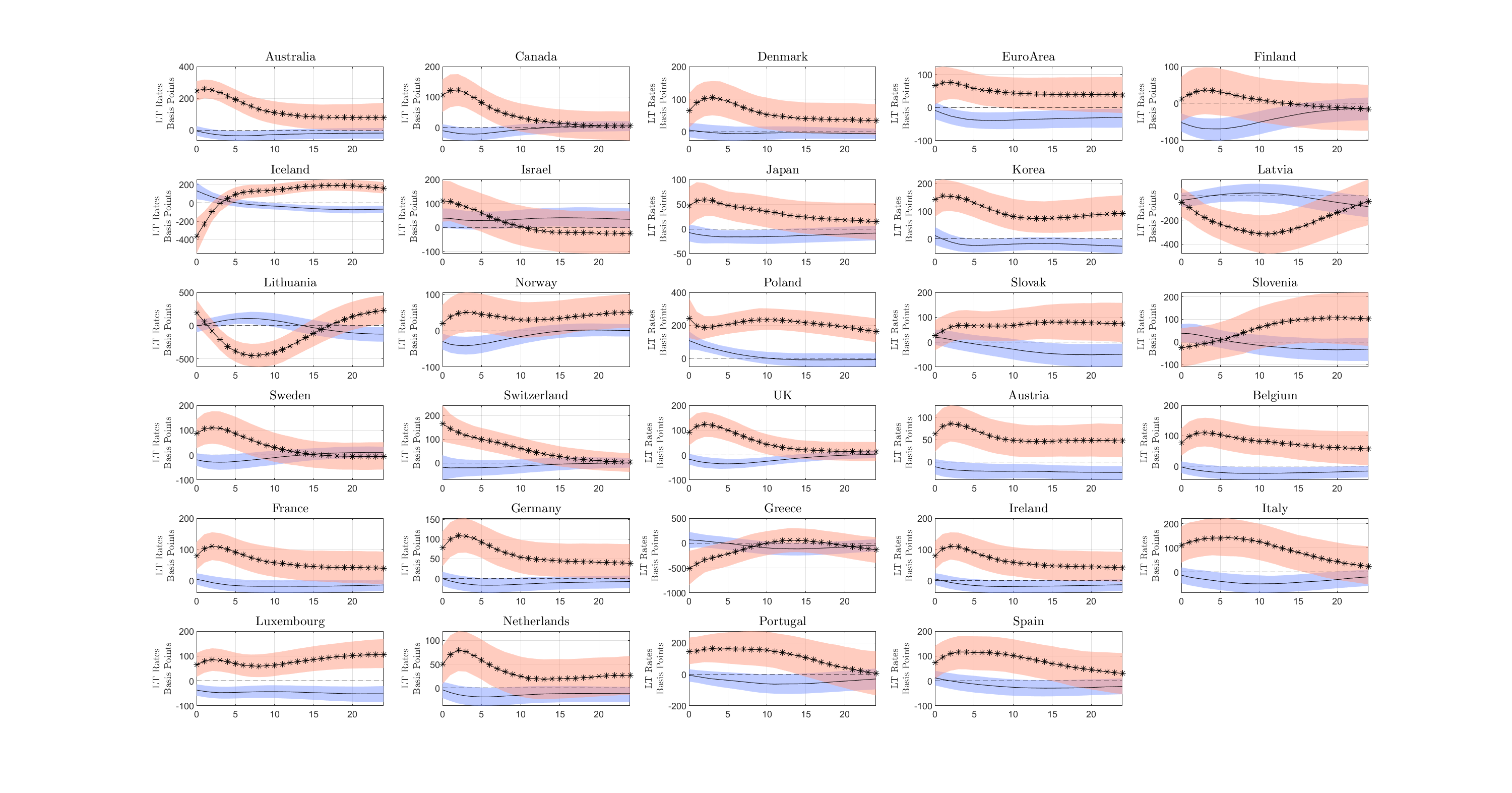}
    \caption{Impulse Response Functions - country-by-country \\ Multi. REER - AE - LT Rate}
    \label{fig:CbC_REER_AE_LT}
    \floatfoot{\textbf{Note:} The figure is comprised of 29 sub figures ordered in five columns and six rows. To estimate the figures I replace the nominal exchange rate with the US with the trade weighted multilateral real exchange rate. The figure represents the response of the long term rate (basis points) for the 28 Advanced Economies plus the Euro Area. The full specification is as specified in Section \ref{subsec:additional_results}. The model is estimated for each country for its longest possible sample. See Appendix \ref{sec:appendix_data_details}. The solid black line represents the median impulse response function of the MP component. The blue area represents the 68\% confidence interval for the MP component. The asterisk-line represents the median impulse response function of the FIE component. The red area represents the 68\% confidence interval for the MP component. In the text, when referring to Panel $(i,j)$, $i$ refers to the row and $j$ to the column of the figure.}
\end{figure}
\end{landscape}

\begin{landscape}
\begin{figure}
    \centering
    \includegraphics[scale=0.4]{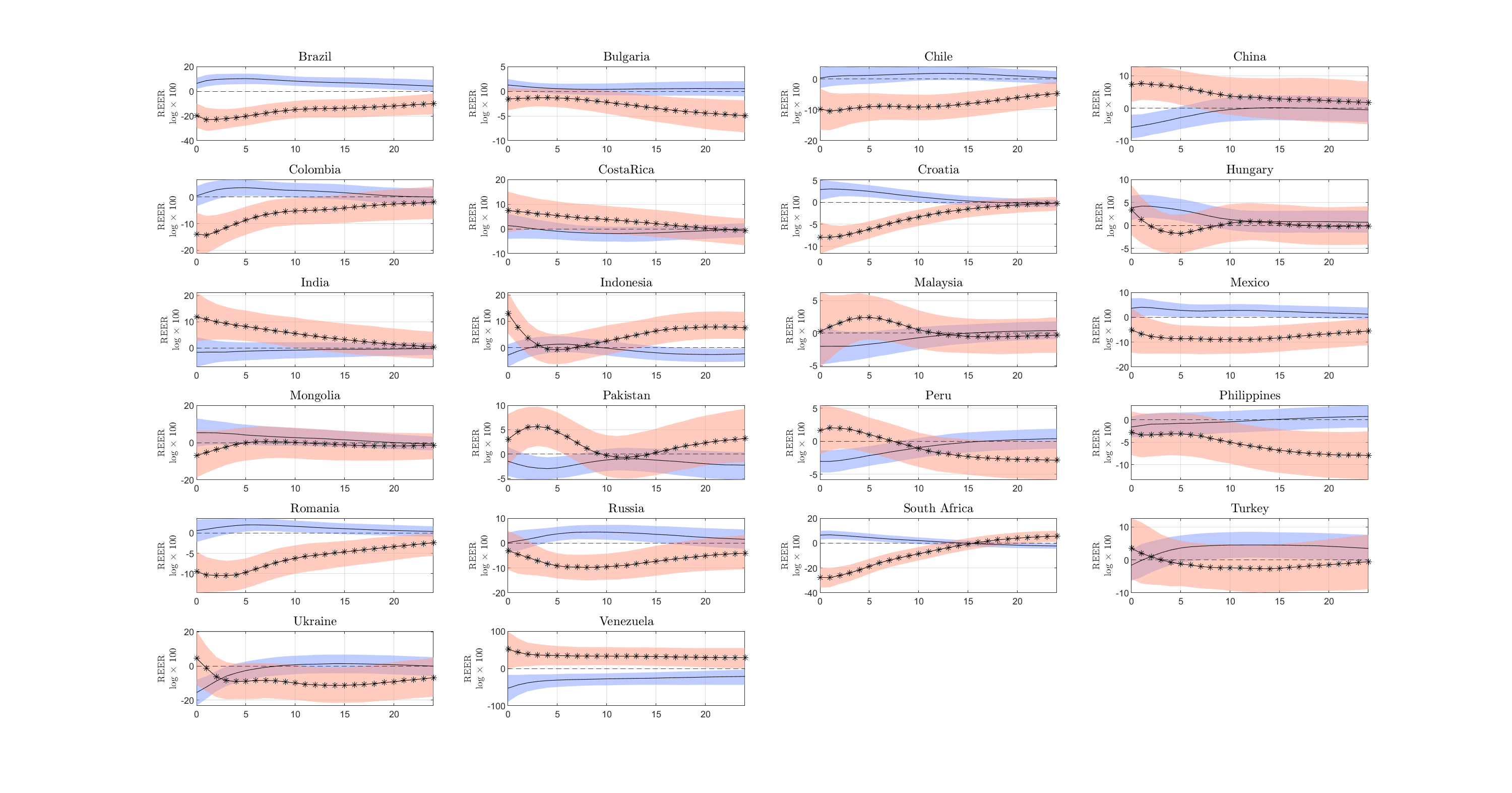}
    \caption{Impulse Response Functions - country-by-country \\ Multi. REER- EME - REER}
    \label{fig:CbC_REER_EM_REER}
    \floatfoot{\textbf{Note:} The figure is comprised of 22 sub figures ordered in four columns and six rows. To estimate the figures I replace the nominal exchange rate with the US with the trade weighted multilateral real exchange rate. The figure represents the response of the nominal exchange rate with respect to the US dollar (log times 100) for the 22 Emerging Market Economies. The full specification is as specified in Section \ref{subsec:additional_results}. The model is estimated for each country for its longest possible sample. See Appendix \ref{sec:appendix_data_details}. The solid black line represents the median impulse response function of the MP component. The blue area represents the 68\% confidence interval for the MP component. The asterisk-line represents the median impulse response function of the FIE component. The red area represents the 68\% confidence interval for the MP component. In the text, when referring to Panel $(i,j)$, $i$ refers to the row and $j$ to the column of the figure.}
\end{figure}
\end{landscape}

\begin{landscape}
\begin{figure}
    \centering
    \includegraphics[scale=0.4]{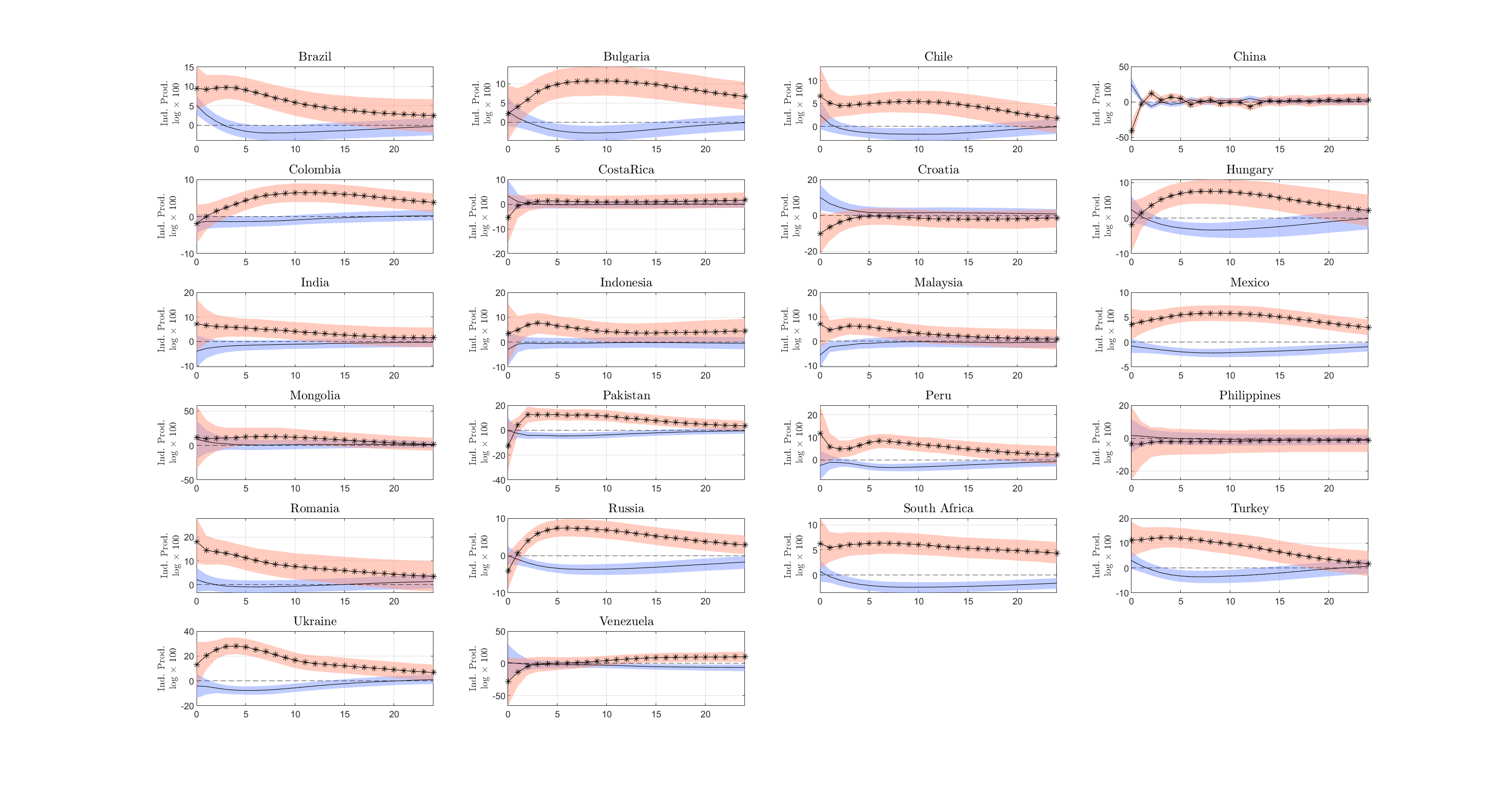}
    \caption{Impulse Response Functions - country-by-country \\ Multi. REER- EME - Ind. Prod.}
    \label{fig:CbC_REER_EM_IND}
    \floatfoot{\textbf{Note:} The figure is comprised of 22 sub figures ordered in four columns and six rows. To estimate the figures I replace the nominal exchange rate with the US with the trade weighted multilateral real exchange rate. The figure represents the response of the industrial production index (log times 100) for the 22 Emerging Market Economies. The full specification is as specified in Section \ref{subsec:additional_results}. The model is estimated for each country for its longest possible sample. See Appendix \ref{sec:appendix_data_details}. The solid black line represents the median impulse response function of the MP component. The blue area represents the 68\% confidence interval for the MP component. The asterisk-line represents the median impulse response function of the FIE component. The red area represents the 68\% confidence interval for the MP component. In the text, when referring to Panel $(i,j)$, $i$ refers to the row and $j$ to the column of the figure.}
\end{figure}
\end{landscape}

\begin{landscape}
\begin{figure}
    \centering
    \includegraphics[scale=0.4]{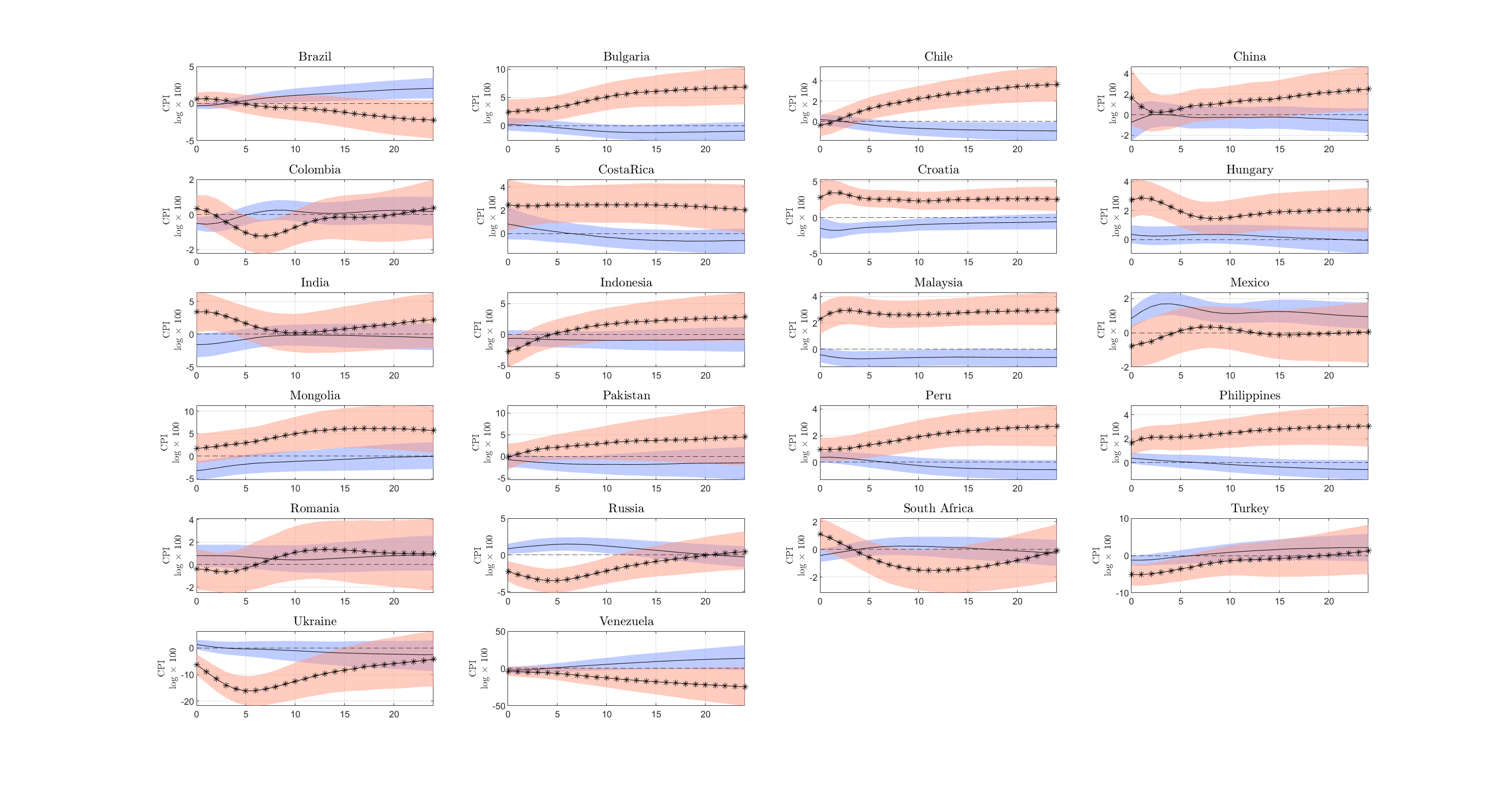}
    \caption{Impulse Response Functions - country-by-country \\ Multi. REER- EME - CPI}
    \label{fig:CbC_REER_EM_CPI}
    \floatfoot{\textbf{Note:} The figure is comprised of 22 sub figures ordered in four columns and six rows. To estimate the figures I replace the nominal exchange rate with the US with the trade weighted multilateral real exchange rate. The figure represents the response of the consumer price index (log times 100) for the 22 Emerging Market Economies. The full specification is as specified in Section \ref{subsec:additional_results}. The model is estimated for each country for its longest possible sample. See Appendix \ref{sec:appendix_data_details}. The solid black line represents the median impulse response function of the MP component. The blue area represents the 68\% confidence interval for the MP component. The asterisk-line represents the median impulse response function of the FIE component. The red area represents the 68\% confidence interval for the MP component. In the text, when referring to Panel $(i,j)$, $i$ refers to the row and $j$ to the column of the figure.}
\end{figure}
\end{landscape}

\begin{landscape}
\begin{figure}
    \centering
    \includegraphics[scale=0.4]{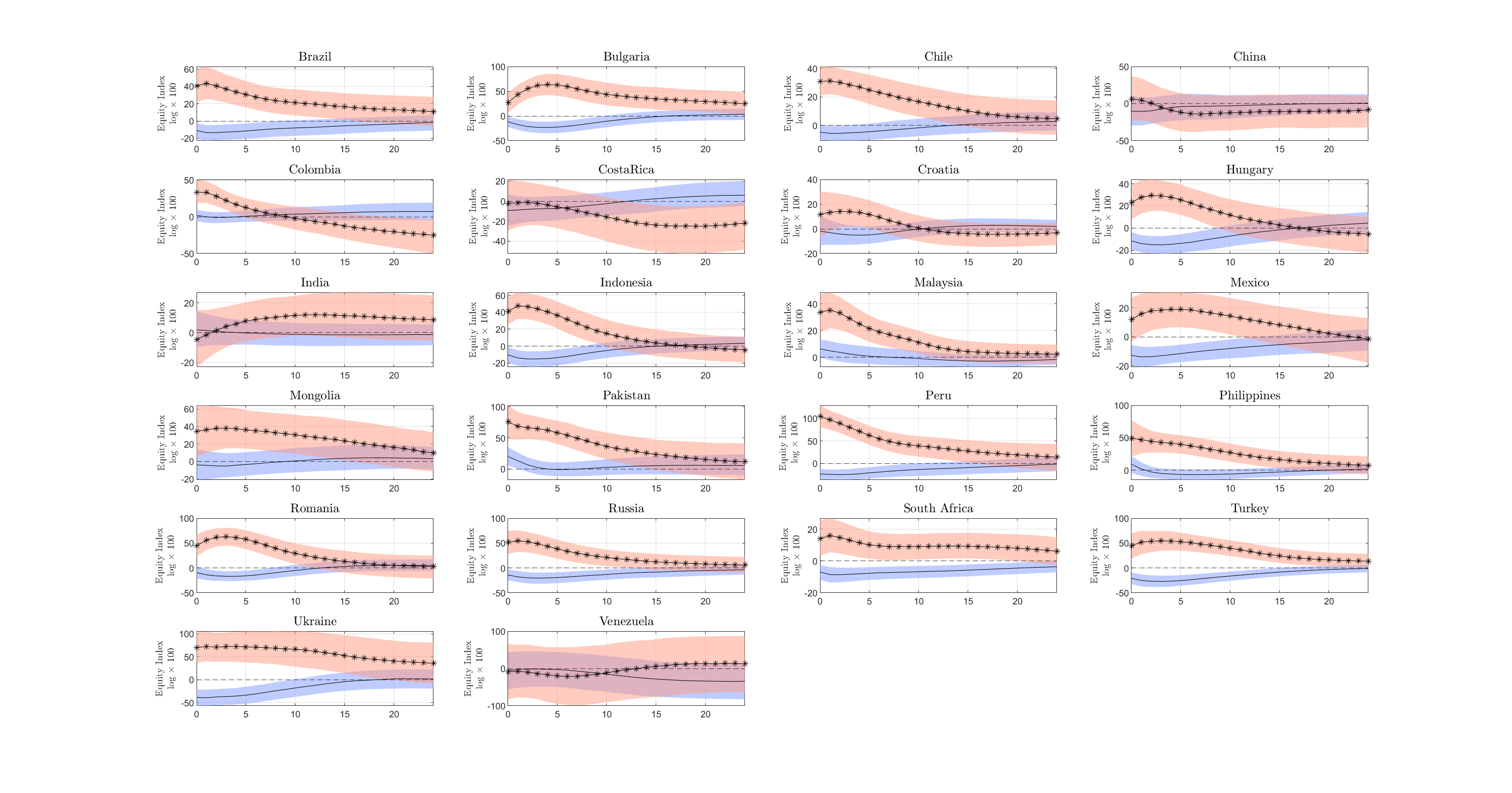}
    \caption{Impulse Response Functions - country-by-country \\ Multi. REER- EME - Equity Index}
    \label{fig:CbC_REER_EM_Equity}
    \floatfoot{\textbf{Note:} The figure is comprised of 22 sub figures ordered in four columns and six rows. To estimate the figures I replace the nominal exchange rate with the US with the trade weighted multilateral real exchange rate. The figure represents the response of the equity index (log times 100) for the 22 Emerging Market Economies. The full specification is as specified in Section \ref{subsec:additional_results}. The model is estimated for each country for its longest possible sample. See Appendix \ref{sec:appendix_data_details}. The solid black line represents the median impulse response function of the MP component. The blue area represents the 68\% confidence interval for the MP component. The asterisk-line represents the median impulse response function of the FIE component. The red area represents the 68\% confidence interval for the MP component. In the text, when referring to Panel $(i,j)$, $i$ refers to the row and $j$ to the column of the figure.}
\end{figure}
\end{landscape}

\begin{landscape}
\begin{figure}
    \centering
    \includegraphics[scale=0.4]{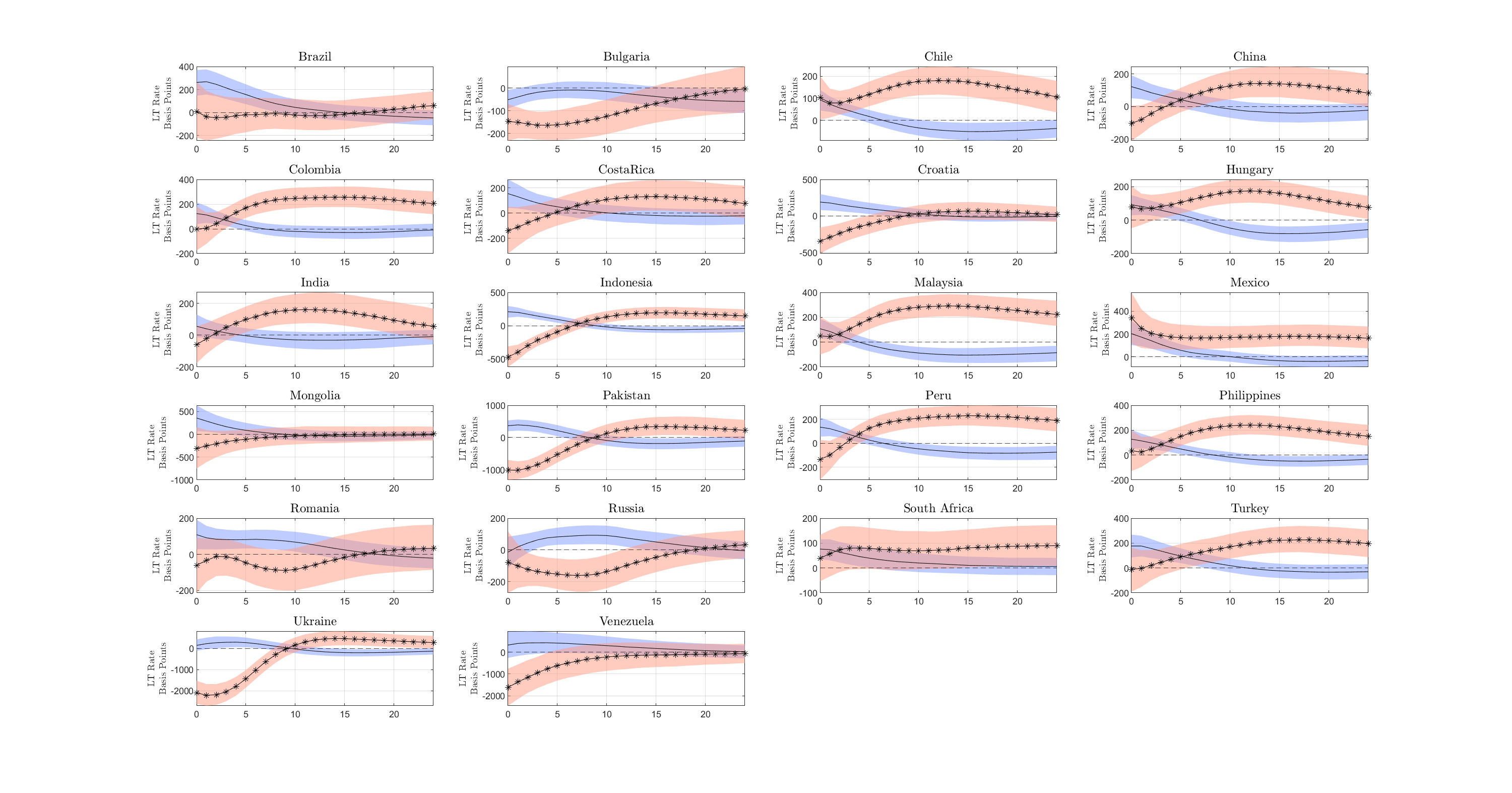}
    \caption{Impulse Response Functions - country-by-country \\ Multi. REER- EME - LT Rate}
    \label{fig:CbC_REER_EM_LT}
    \floatfoot{\textbf{Note:} The figure is comprised of 22 sub figures ordered in four columns and six rows. To estimate the figures I replace the nominal exchange rate with the US with the trade weighted multilateral real exchange rate. The figure represents the response of the long term rate  (basis points) for the 22 Emerging Market Economies. The full specification is as specified in Section \ref{subsec:additional_results}. The model is estimated for each country for its longest possible sample. See Appendix \ref{sec:appendix_data_details}. The solid black line represents the median impulse response function of the MP component. The blue area represents the 68\% confidence interval for the MP component. The asterisk-line represents the median impulse response function of the FIE component. The red area represents the 68\% confidence interval for the MP component. In the text, when referring to Panel $(i,j)$, $i$ refers to the row and $j$ to the column of the figure.}
\end{figure}
\end{landscape}

\begin{figure}
    \centering
    \includegraphics[scale=0.4]{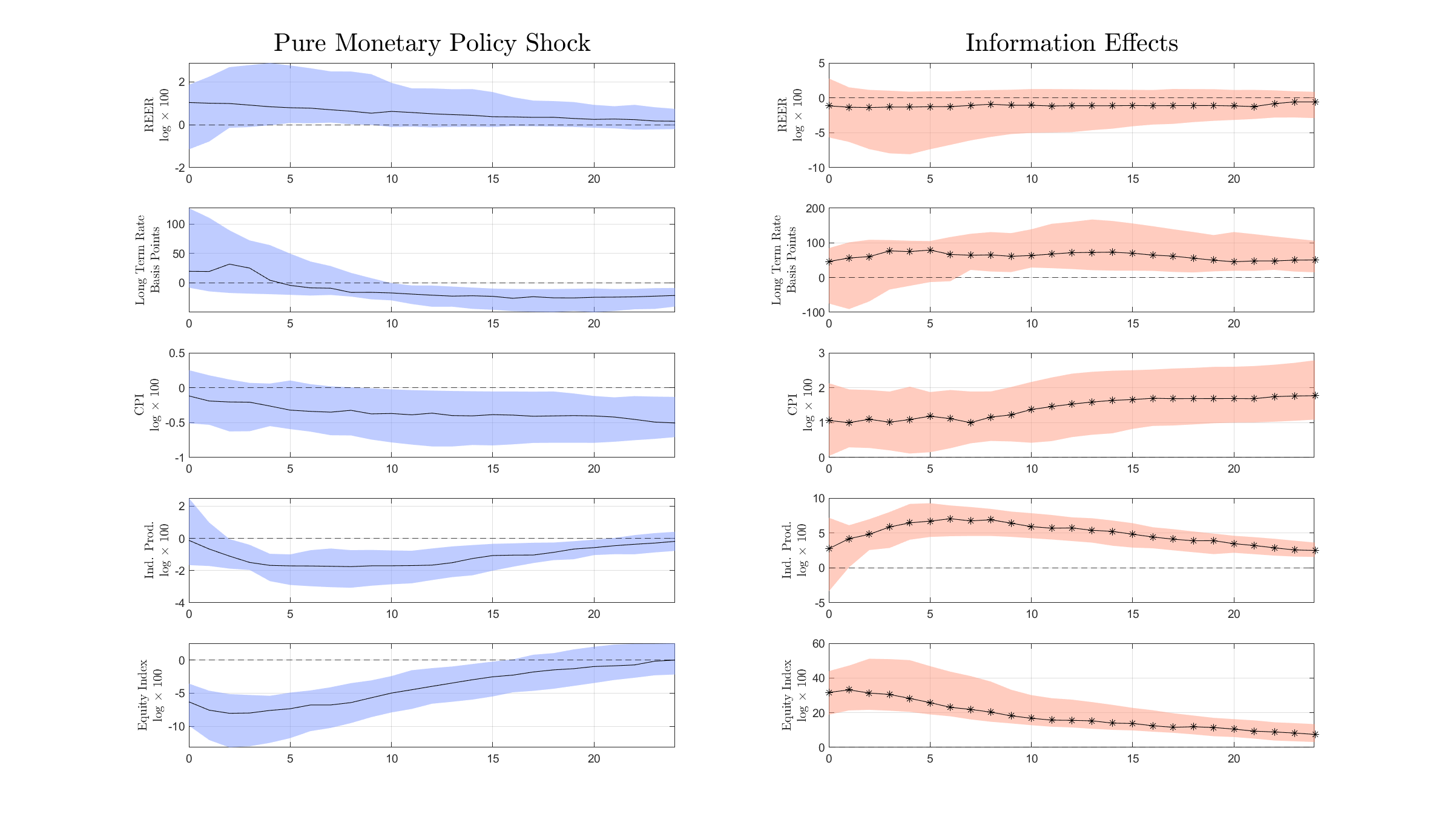}
    \caption{Impulse Response Functions - Median Responses across Countries \\ Multi. REER Sample}
    \label{fig:REER_MedianResponse}
    \floatfoot{\textbf{Note:} The figure is comprised of 10 sub-figures ordered in two columns and five rows. The left column presents the impulse response functions to the MP component while the right column present the impulse response functions to the FIE component. Impulse response functions are estimated for the sample of countries that replace the nominal exchange rate with the trade weighted multilateral real exchange rate. The rows represent the impact on (i) the trade weighted multilateral real exchange rate (in logs times 100); (ii) long term interest rates in basis points; (iii) the consumer price index (in logs times 100); (iv) the industrial production index (in logs times 100); (v) the equity index (in logs times 100). I keep the median impulse response function for every country in the sample. The solid black line represents the median response from the set of impulse response functions for each country to a MP component. The light blue area represents the interquartile range of impulse response functions to a MP component. The asterisk black line represents the median response from the set of impulse response functions for each country to a FIE component. The light red area represents the interquartile range of impulse response functions to a FIE component. In the text, when referring to Panel $(i,j)$, $i$ refers to the row and $j$ to the column of the figure.}
\end{figure}

\newpage
\begin{figure}[ht]
    \centering
    \includegraphics[scale=0.4]{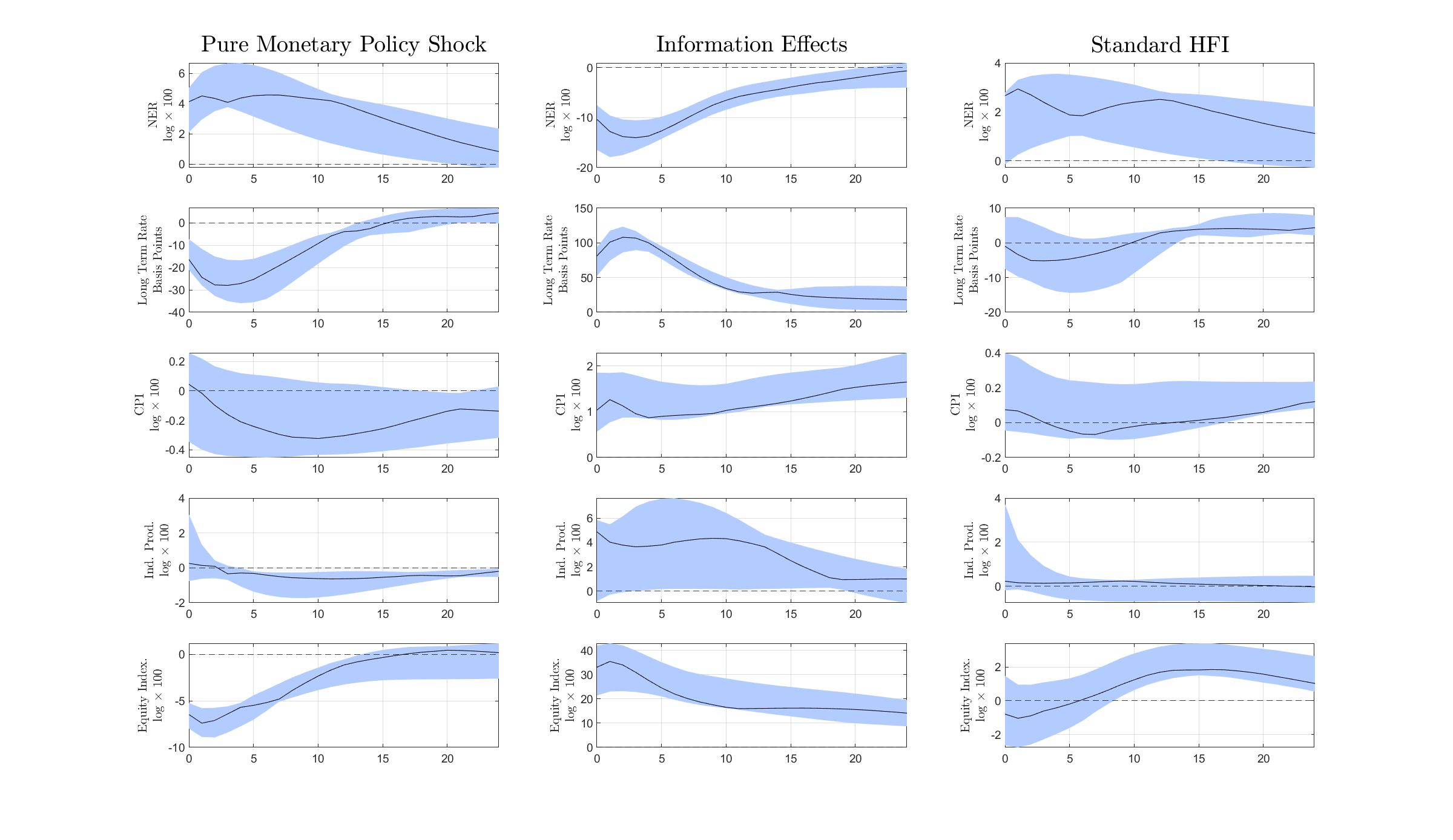}
    \caption{Impulse Response Functions - country-by-country \\ Full Sample}
    \label{fig:MeanMedian_88_NER_Modify}
    \floatfoot{\textbf{Note:} The figure is comprised of 15 sub-figures ordered in three columns and five rows. Each sub-figure is comprised by pooling the median impulse response function estimated for each country in the time sample January 1988 to December 2019. See Appendix \ref{sec:appendix_data_details} for details on this sample. The left panel presents the impulse response functions to the MP component, the middle column to the FIE component, and the right column following the Standard HFI approach. The rows represent the impact on (i) the nominal exchange rate with the US dollar (in logs times 100); (ii) long term interest rates in basis points; (iii) the consumer price index (in logs times 100); (iv) the industrial production index (in logs times 100); (v) the equity index (in logs times 100). The solid black line represents the median response from the set of impulse response functions for each country. The light blue area represents the interquartile range of impulse response functions. In the text, when referring to Panel $(i,j)$, $i$ refers to the row and $j$ to the column of the figure.  }
\end{figure}

\newpage
\begin{figure}[ht]
    \centering
    \includegraphics[scale=0.4]{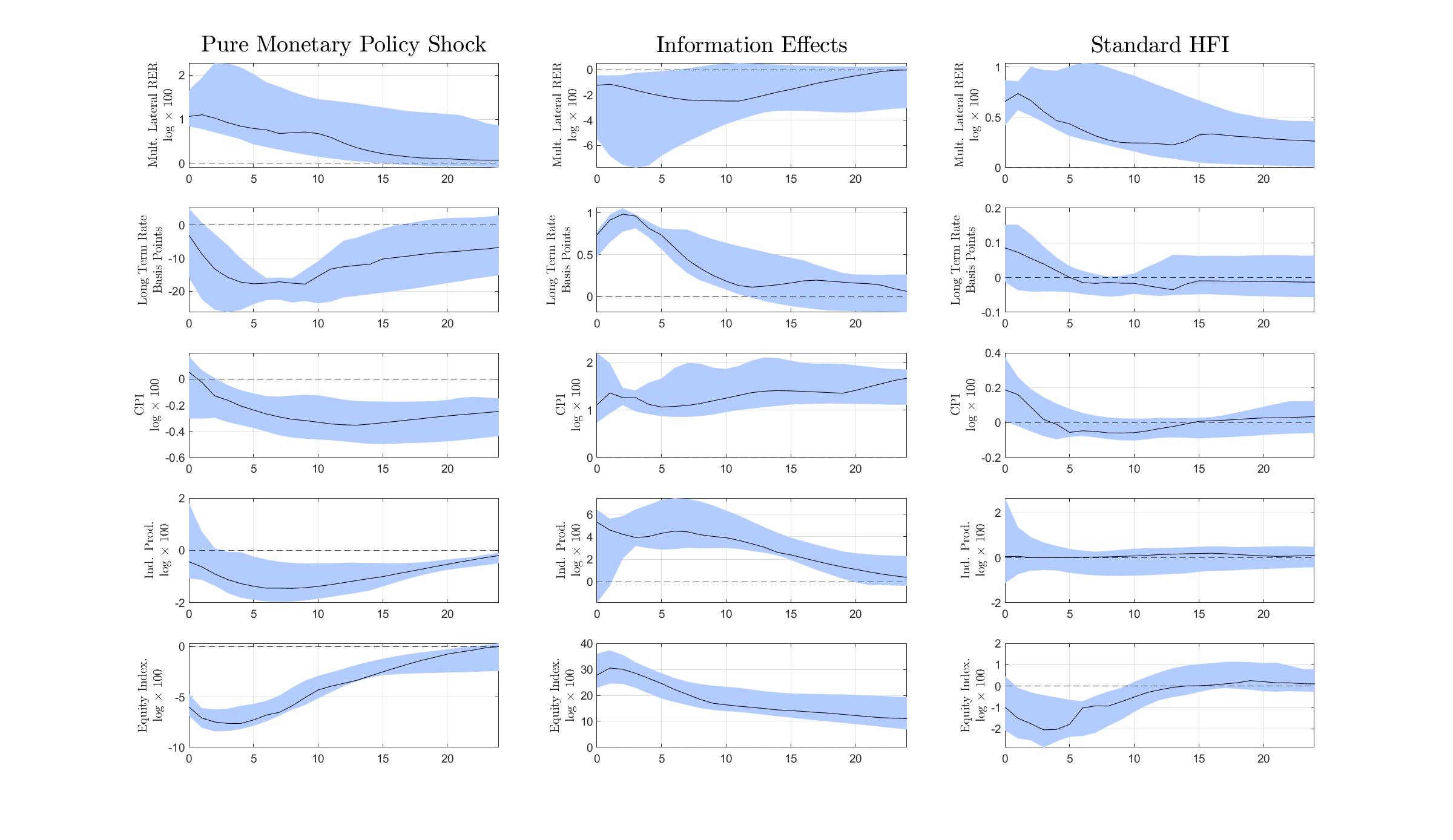}
    \caption{Impulse Response Functions - country-by-country \\ Multi. REER Sample - Full Sample}
    \label{fig:MeanMedian_88_REER_Modify}
    \floatfoot{\textbf{Note:} The figure is comprised of 15 sub-figures ordered in three columns and five rows. Each sub-figure is comprised by pooling the median impulse response function estimated for each country in the time sample January 1988 to December 2019 for the sample that replaces the exchange rate with respect to the US dollar with the trade weighted multilateral real exchange rate. See Appendix \ref{sec:appendix_data_details} for details on this sample. The left panel presents the impulse response functions to the MP component, the middle column to the FIE component, and the right column following the Standard HFI approach. The rows represent the impact on (i) the trade weighted multilateral real exchange rate (in logs times 100); (ii) long term interest rates in basis points; (iii) the consumer price index (in logs times 100); (iv) the industrial production index (in logs times 100); (v) the equity index (in logs times 100). The solid black line represents the median response from the set of impulse response functions for each country. The light blue area represents the interquartile range of impulse response functions. In the text, when referring to Panel $(i,j)$, $i$ refers to the row and $j$ to the column of the figure.  }
\end{figure}

\newpage
\begin{figure}[ht]
    \centering
    \includegraphics[scale=0.4]{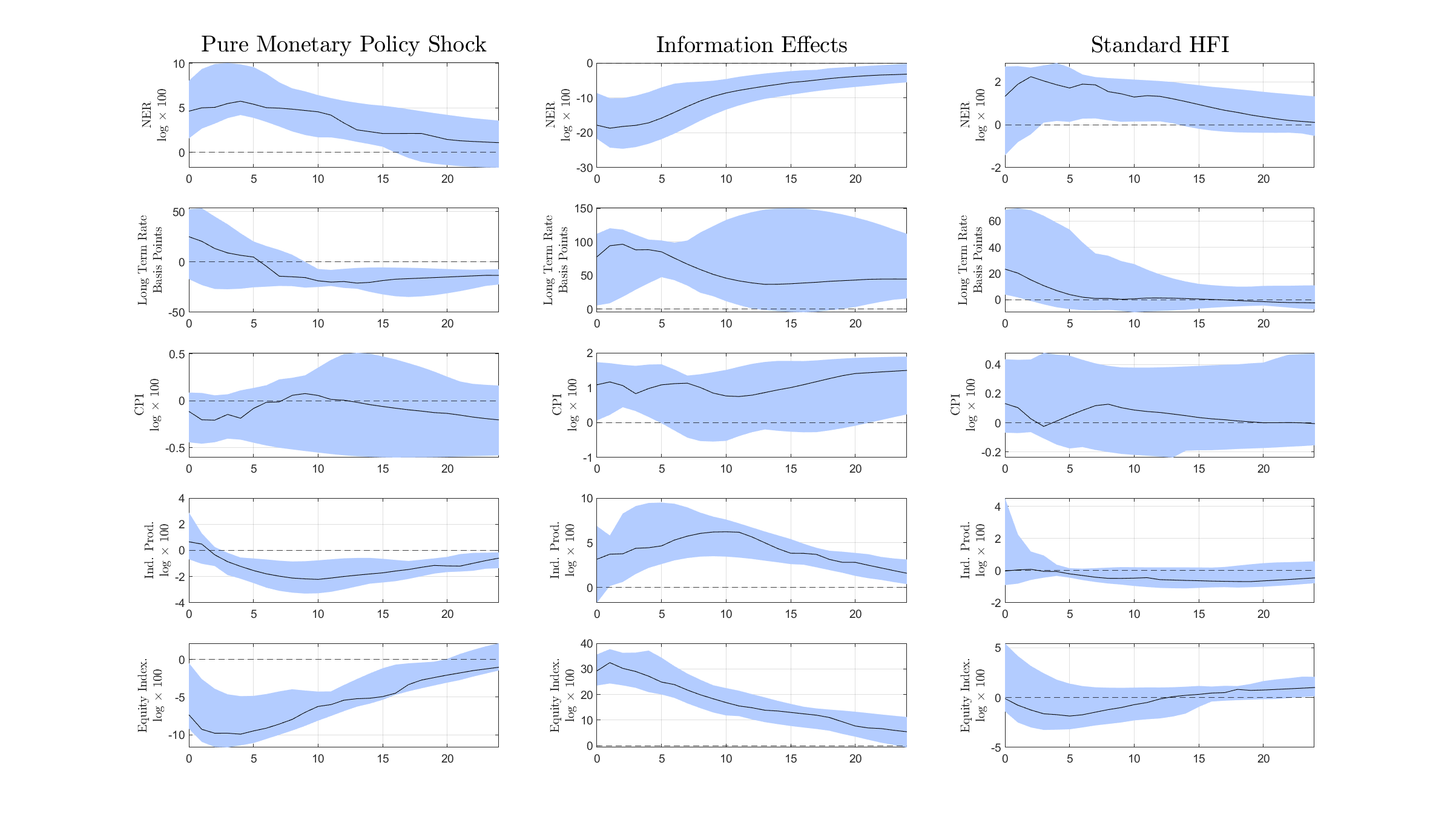}
    \caption{Impulse Response Functions - country-by-country \\ 1998 to 2019}
    \label{fig:MeanMedian_98_NER_Modify}
    \floatfoot{\textbf{Note:} The figure is comprised of 15 sub-figures ordered in three columns and five rows. Each sub-figure is comprised by pooling the median impulse response function estimated for each country in the time sample January 1998 to December 2019. See Appendix \ref{sec:appendix_data_details} for details on this sample. The left panel presents the impulse response functions to the MP component, the middle column to the FIE component, and the right column following the Standard HFI approach. The rows represent the impact on (i) the nominal exchange rate with the US dollar (in logs times 100); (ii) long term interest rates in basis points; (iii) the consumer price index (in logs times 100); (iv) the industrial production index (in logs times 100); (v) the equity index (in logs times 100). The solid black line represents the median response from the set of impulse response functions for each country. The light blue area represents the interquartile range of impulse response functions. In the text, when referring to Panel $(i,j)$, $i$ refers to the row and $j$ to the column of the figure.  }
\end{figure}

\newpage
\begin{figure}[ht]
    \centering
    \includegraphics[scale=0.4]{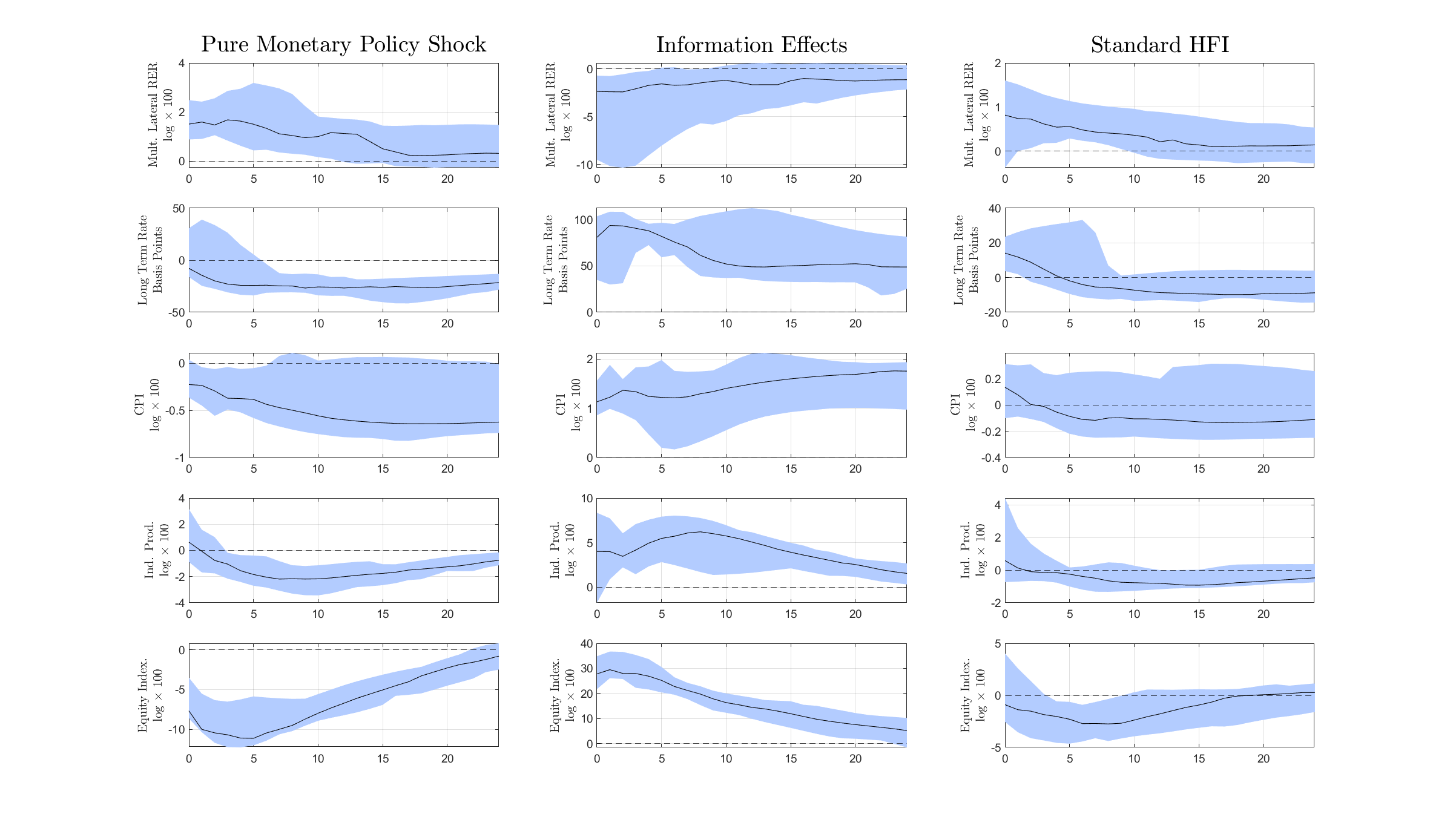}
    \caption{Impulse Response Functions - country-by-country \\ Multi. REER Sample - 1998 to 2019}
    \label{fig:MeanMedian_98_REER_Modify}
    \floatfoot{\textbf{Note:} The figure is comprised of 15 sub-figures ordered in three columns and five rows. Each sub-figure is comprised by pooling the median impulse response function estimated for each country in the time sample January 1998 to December 2019 for the sample that replaces the exchange rate with respect to the US dollar with the trade weighted multilateral real exchange rate. See Appendix \ref{sec:appendix_data_details} for details on this sample. The left panel presents the impulse response functions to the MP component, the middle column to the FIE component, and the right column following the Standard HFI approach. The rows represent the impact on (i) the trade weighted multilateral real exchange rate (in logs times 100); (ii) long term interest rates in basis points; (iii) the consumer price index (in logs times 100); (iv) the industrial production index (in logs times 100); (v) the equity index (in logs times 100). The solid black line represents the median response from the set of impulse response functions for each country. The light blue area represents the interquartile range of impulse response functions. In the text, when referring to Panel $(i,j)$, $i$ refers to the row and $j$ to the column of the figure.  }
\end{figure}

\newpage
\begin{figure}[ht]
    \centering
    \includegraphics[scale=0.4]{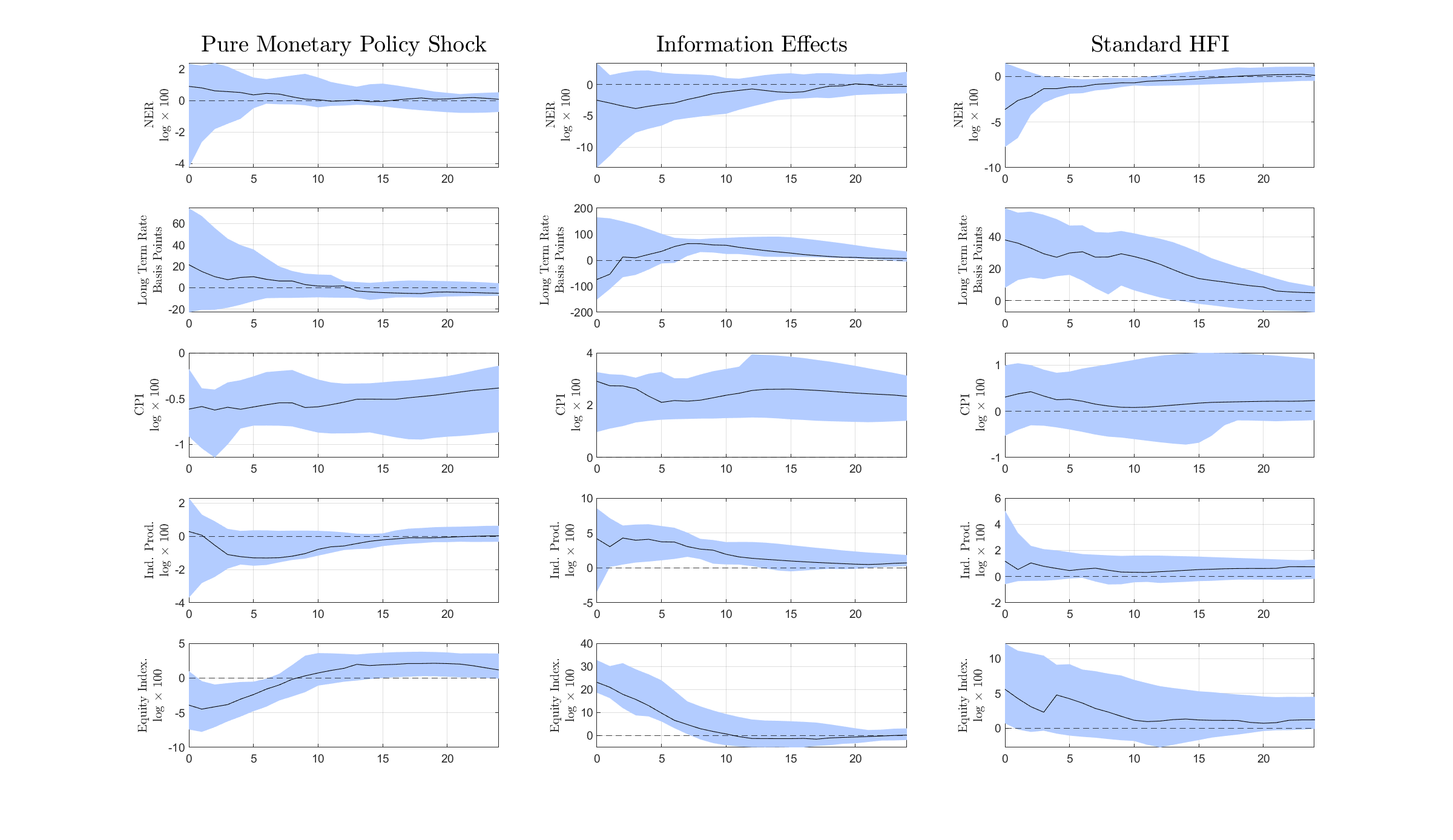}
    \caption{Impulse Response Functions - country-by-country \\ 2008 to 2019}
    \label{fig:MeanMedian_08_NER_Modify}
    \floatfoot{\textbf{Note:} The figure is comprised of 15 sub-figures ordered in three columns and five rows. Each sub-figure is comprised by pooling the median impulse response function estimated for each country in the time sample January 2008 to December 2019. See Appendix \ref{sec:appendix_data_details} for details on this sample. The left panel presents the impulse response functions to the MP component, the middle column to the FIE component, and the right column following the Standard HFI approach. The rows represent the impact on (i) the nominal exchange rate with the US dollar (in logs times 100); (ii) long term interest rates in basis points; (iii) the consumer price index (in logs times 100); (iv) the industrial production index (in logs times 100); (v) the equity index (in logs times 100). The solid black line represents the median response from the set of impulse response functions for each country. The light blue area represents the interquartile range of impulse response functions. In the text, when referring to Panel $(i,j)$, $i$ refers to the row and $j$ to the column of the figure.  }
\end{figure}

\newpage
\begin{figure}[ht]
    \centering
    \includegraphics[scale=0.4]{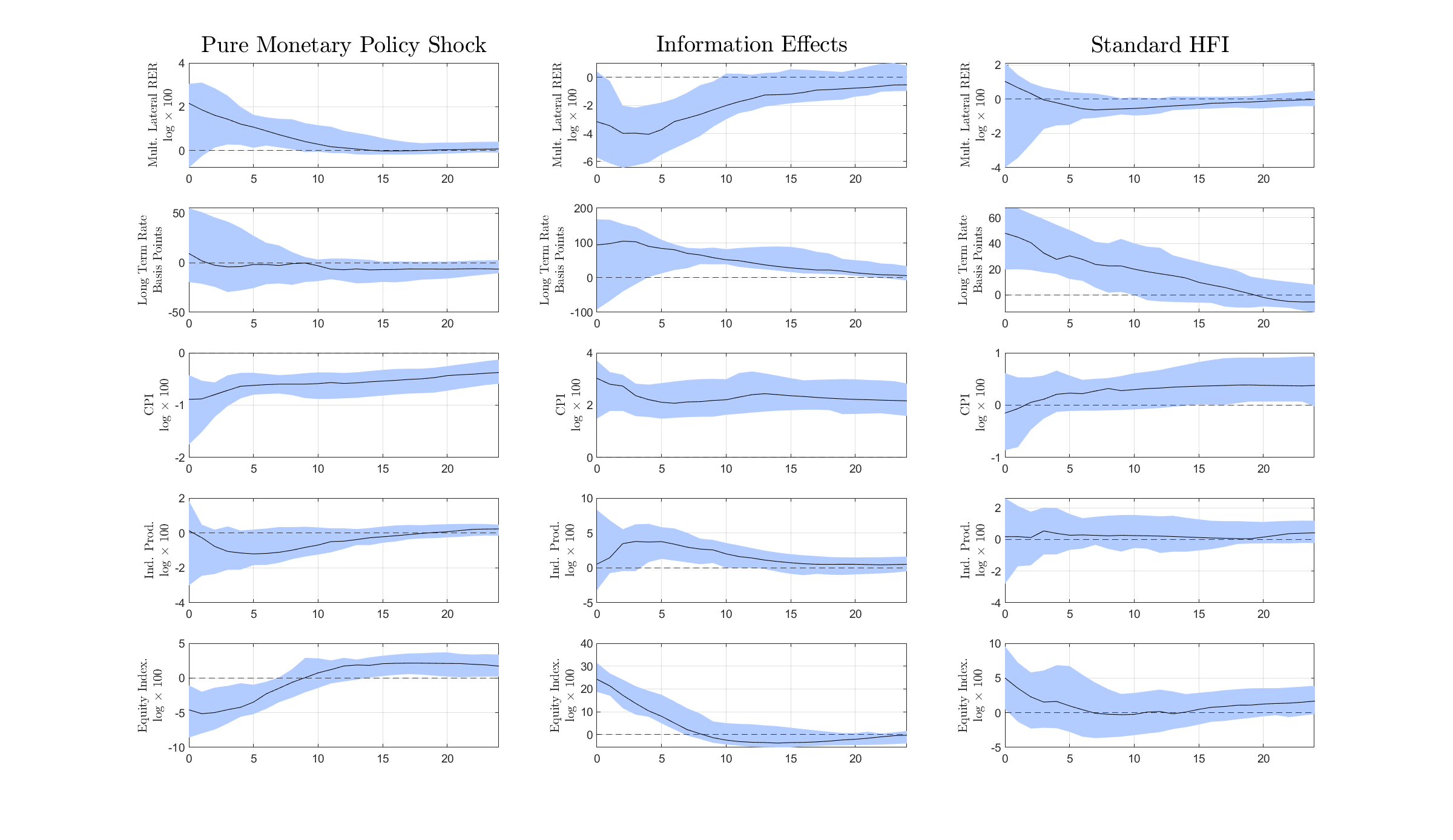}
    \caption{Impulse Response Functions - country-by-country \\ Multi. REER Sample - 2008 to 2019}
    \label{fig:MeanMedian_08_REER_Modify}
    \floatfoot{\textbf{Note:} The figure is comprised of 15 sub-figures ordered in three columns and five rows. Each sub-figure is comprised by pooling the median impulse response function estimated for each country in the time sample January 2008 to December 2019 for the sample that replaces the exchange rate with respect to the US dollar with the trade weighted multilateral real exchange rate. See Appendix \ref{sec:appendix_data_details} for details on this sample. The left panel presents the impulse response functions to the MP component, the middle column to the FIE component, and the right column following the Standard HFI approach. The rows represent the impact on (i) the trade weighted multilateral real exchange rate (in logs times 100); (ii) long term interest rates in basis points; (iii) the consumer price index (in logs times 100); (iv) the industrial production index (in logs times 100); (v) the equity index (in logs times 100). The solid black line represents the median response from the set of impulse response functions for each country. The light blue area represents the interquartile range of impulse response functions. In the text, when referring to Panel $(i,j)$, $i$ refers to the row and $j$ to the column of the figure.  }
\end{figure}


\subsection{Robustness Checks Figures}
\newpage
\begin{figure}
    \centering
    \includegraphics[scale=0.4]{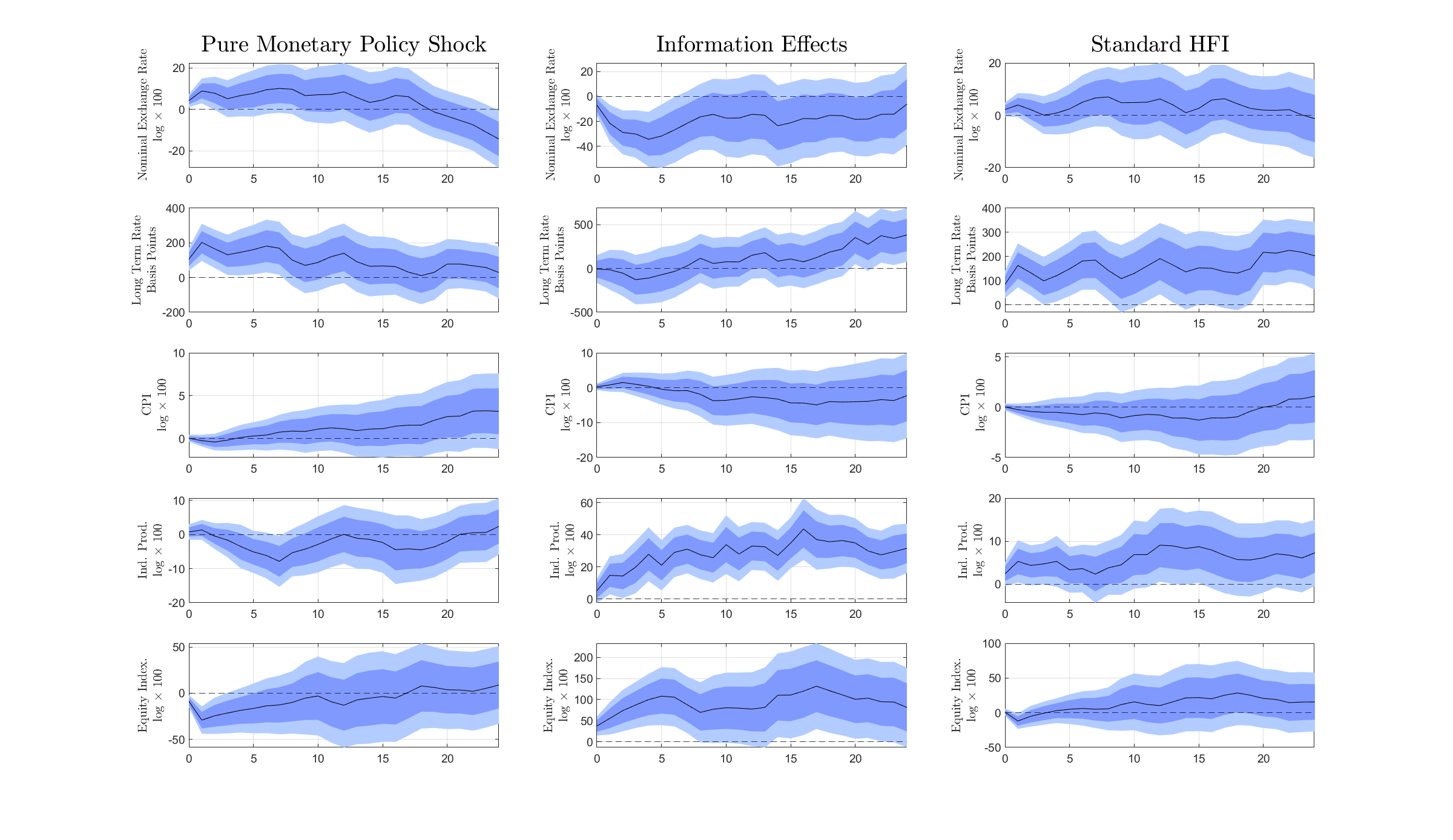}
    \caption{Impulse Response Functions \\ Alternative Regression Specification with Time Trend}
    \label{fig:Trend_NER}
    \floatfoot{\textbf{Note:} The figure is comprised of 15 sub-figures ordered in three columns and five rows. The left column relates to the estimates of $\beta^{MP}$ in Equation \ref{eq:LP_Trend}, the middle column relates to the estimate of $\beta^{FIE}$ in Equation \ref{eq:LP_Trend}, while the right column relates to estimating Equation \ref{eq:LP_Trend}, replacing the MP and FIE components with the un-orthogonalized monetary policy surprise. The rows represent the impact on (i) the nominal exchange rate with the US dollar (in logs times 100); (ii) long term interest rates in basis points; (iii) the consumer price index (in logs times 100); (iv) the industrial production index (in logs times 100); (v) the equity index (in logs times 100). The solid black line represents the point estimate, the dark blue area represents the 68\% confidence interval, and the light blue area represents the 90\% confidence interval. In the text, when referring to Panel $(i,j)$, $i$ refers to the row and $j$ to the column of the figure. Each variable, in its own transformation, is demeaned at the country level. }
\end{figure}

\newpage
\begin{figure}
    \centering
    \includegraphics[scale=0.4]{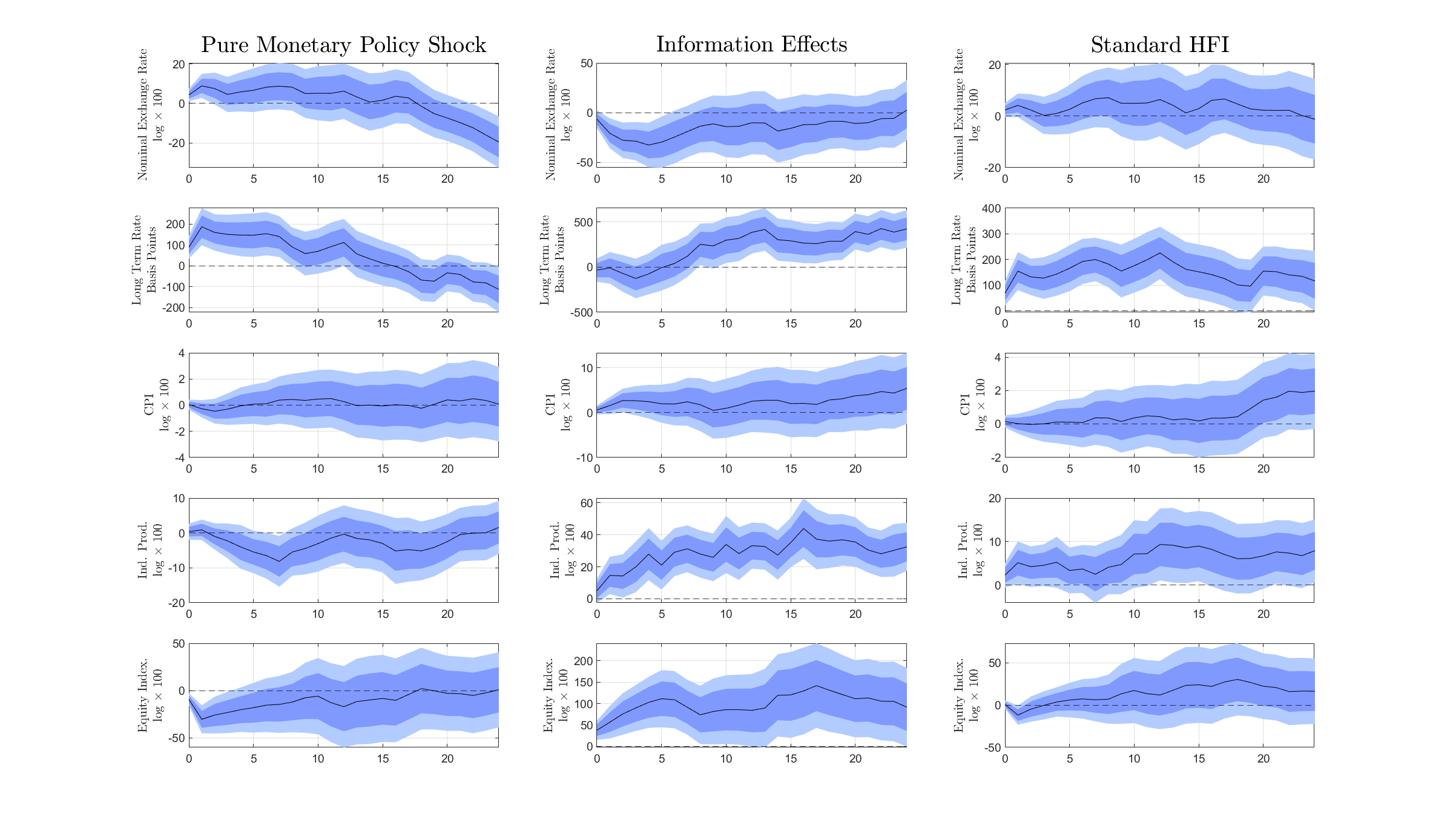}
    \caption{Impulse Response Functions \\ Alternative Regression Specification with Time Trend + F.E.}
    \label{fig:TrendFE_NER}
    \floatfoot{\textbf{Note:} The figure is comprised of 15 sub-figures ordered in three columns and five rows. The left column relates to the estimates of $\beta^{MP}$ in Equation \ref{eq:LP_Trend_FE}, the middle column relates to the estimate of $\beta^{FIE}$ in Equation \ref{eq:LP_Trend_FE}, while the right column relates to estimating Equation \ref{eq:LP_Trend_FE}, replacing the MP and FIE components with the un-orthogonalized monetary policy surprise. The rows represent the impact on (i) the nominal exchange rate with the US dollar (in logs times 100); (ii) long term interest rates in basis points; (iii) the consumer price index (in logs times 100); (iv) the industrial production index (in logs times 100); (v) the equity index (in logs times 100). The solid black line represents the point estimate, the dark blue area represents the 68\% confidence interval, and the light blue area represents the 90\% confidence interval. In the text, when referring to Panel $(i,j)$, $i$ refers to the row and $j$ to the column of the figure. Each variable, in its own transformation, is demeaned at the country level. }
\end{figure}

\newpage
\begin{figure}
    \centering
    \includegraphics[scale=0.4]{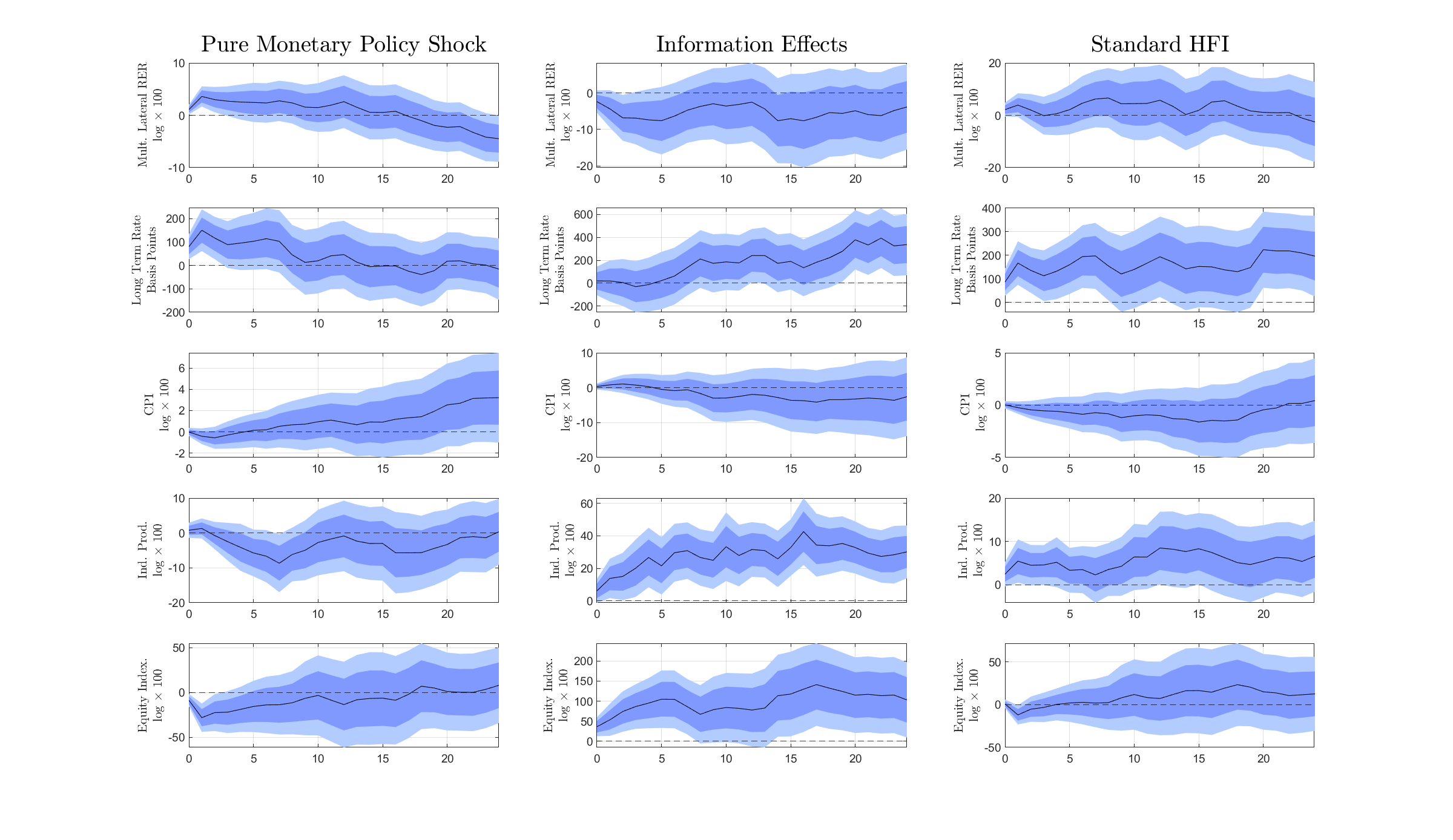}
    \caption{Impulse Response Functions \\ Multi. REER Sample - Alternative Regression Specification with Time Trend}
    \label{fig:Trend_REER}
    \floatfoot{\textbf{Note:} The figure is comprised of 15 sub-figures ordered in three columns and five rows. The left column relates to the estimates of $\beta^{MP}$ in Equation \ref{eq:LP_Trend}, the middle column relates to the estimate of $\beta^{FIE}$ in Equation \ref{eq:LP_Trend}, while the right column relates to estimating Equation \ref{eq:LP_Trend}, replacing the MP and FIE components with the un-orthogonalized monetary policy surprise. The rows represent the impact on (i) the trade weighted multilateral real exchange rate (in logs times 100); (ii) long term interest rates in basis points; (iii) the consumer price index (in logs times 100); (iv) the industrial production index (in logs times 100); (v) the equity index (in logs times 100). The solid black line represents the point estimate, the dark blue area represents the 68\% confidence interval, and the light blue area represents the 90\% confidence interval. In the text, when referring to Panel $(i,j)$, $i$ refers to the row and $j$ to the column of the figure. Each variable, in its own transformation, is demeaned at the country level. }
\end{figure}

\newpage
\begin{figure}
    \centering
    \includegraphics[scale=0.4]{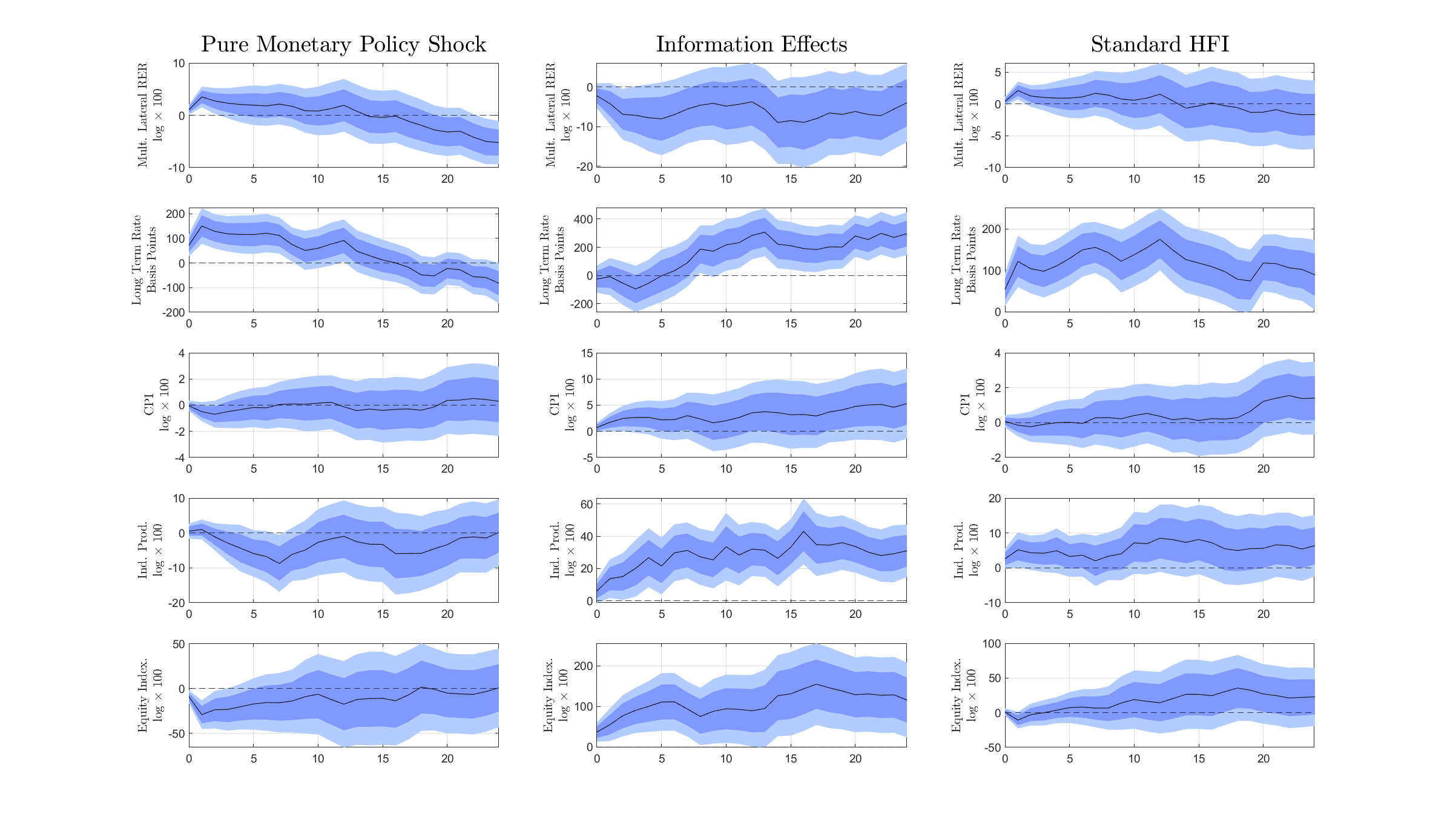}
    \caption{Impulse Response Functions \\ Multi. REER Sample - Alternative Regression Specification with Time Trend + F.E.}
    \label{fig:TrendFE_REER}
    \floatfoot{\textbf{Note:} The figure is comprised of 15 sub-figures ordered in three columns and five rows. The left column relates to the estimates of $\beta^{MP}$ in Equation \ref{eq:LP_Trend_FE}, the middle column relates to the estimate of $\beta^{FIE}$ in Equation \ref{eq:LP_Trend_FE}, while the right column relates to estimating Equation \ref{eq:LP_Trend_FE}, replacing the MP and FIE components with the un-orthogonalized monetary policy surprise. The rows represent the impact on (i) the trade weighted multilateral real exchange rate (in logs times 100); (ii) long term interest rates in basis points; (iii) the consumer price index (in logs times 100); (iv) the industrial production index (in logs times 100); (v) the equity index (in logs times 100). The solid black line represents the point estimate, the dark blue area represents the 68\% confidence interval, and the light blue area represents the 90\% confidence interval. In the text, when referring to Panel $(i,j)$, $i$ refers to the row and $j$ to the column of the figure. Each variable, in its own transformation, is demeaned at the country level. }
\end{figure}

\newpage
\newpage
\begin{figure}[ht]
    \centering
    \includegraphics[scale=0.4]{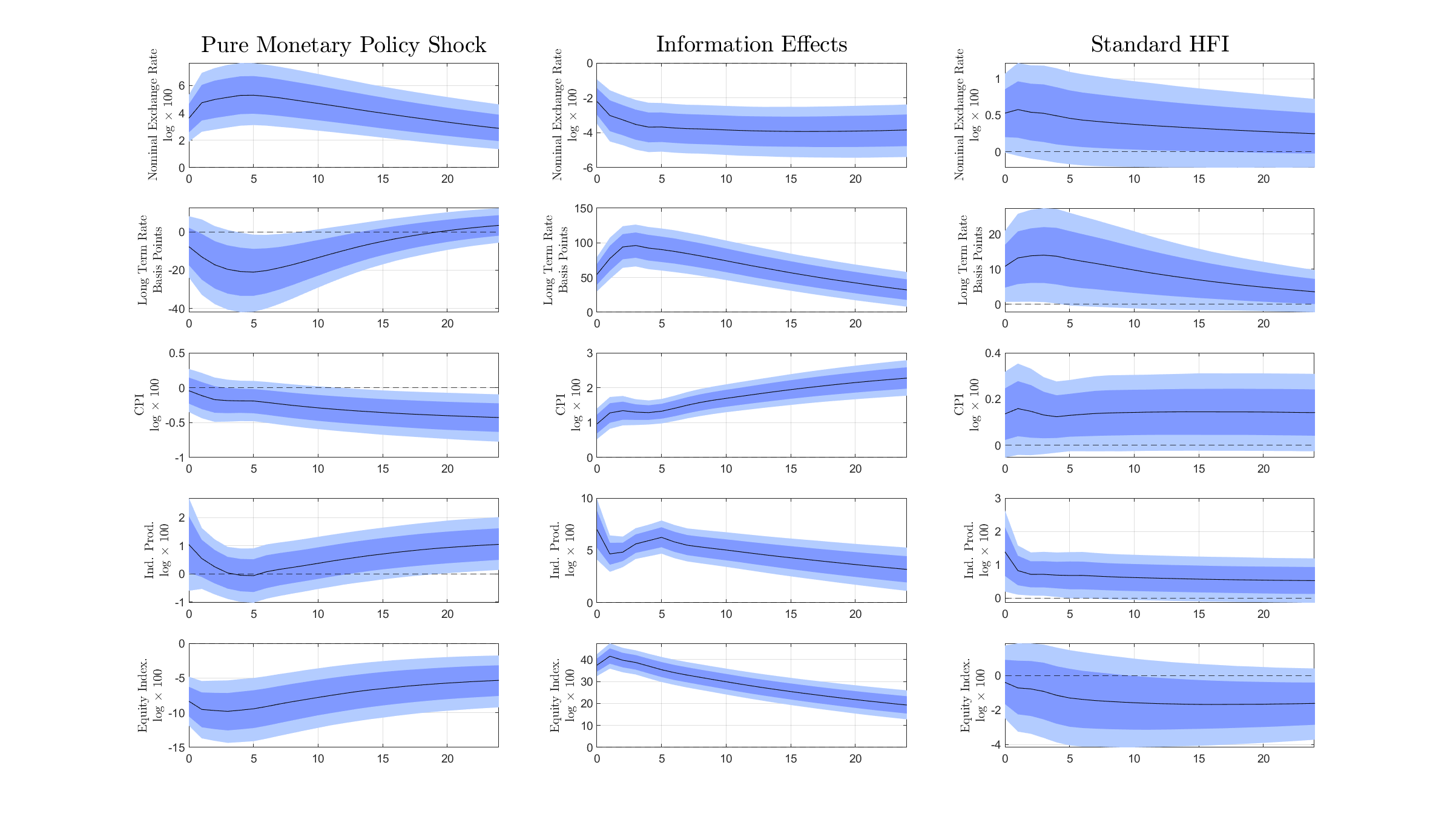}
    \caption{IRFs - Pooled Panel SVAR \\ January 1988 to December 2019}
    \label{fig:Pool_88_NER}
    \floatfoot{\textbf{Note:} The figure is comprised of 15 sub-figures ordered in three columns and five rows. The impulse response functions are estimated following a panel SVAR model, as described in Appendix \ref{sec:appendix_model_details} and for the sample of countries with a valid nominal exchange rate for the period January 1988 to December 2019. The left panel presents the impulse response functions to the MP component, the middle column to the FIE component, and the right column following the Standard HFI approach. The rows represent the impact on (i) the nominal exchange rate with the US dollar (in logs times 100); (ii) long term interest rates in basis points; (iii) the consumer price index (in logs times 100); (iv) the industrial production index (in logs times 100); (v) the equity index (in logs times 100). The solid black line represents the median response from the set of impulse response functions for each country. The light blue area represents the interquartile range of impulse response functions. In the text, when referring to Panel $(i,j)$, $i$ refers to the row and $j$ to the column of the figure. Variables are demeaned at the country level. }
\end{figure}

\newpage
\begin{figure}[ht]
    \centering
    \includegraphics[scale=0.4]{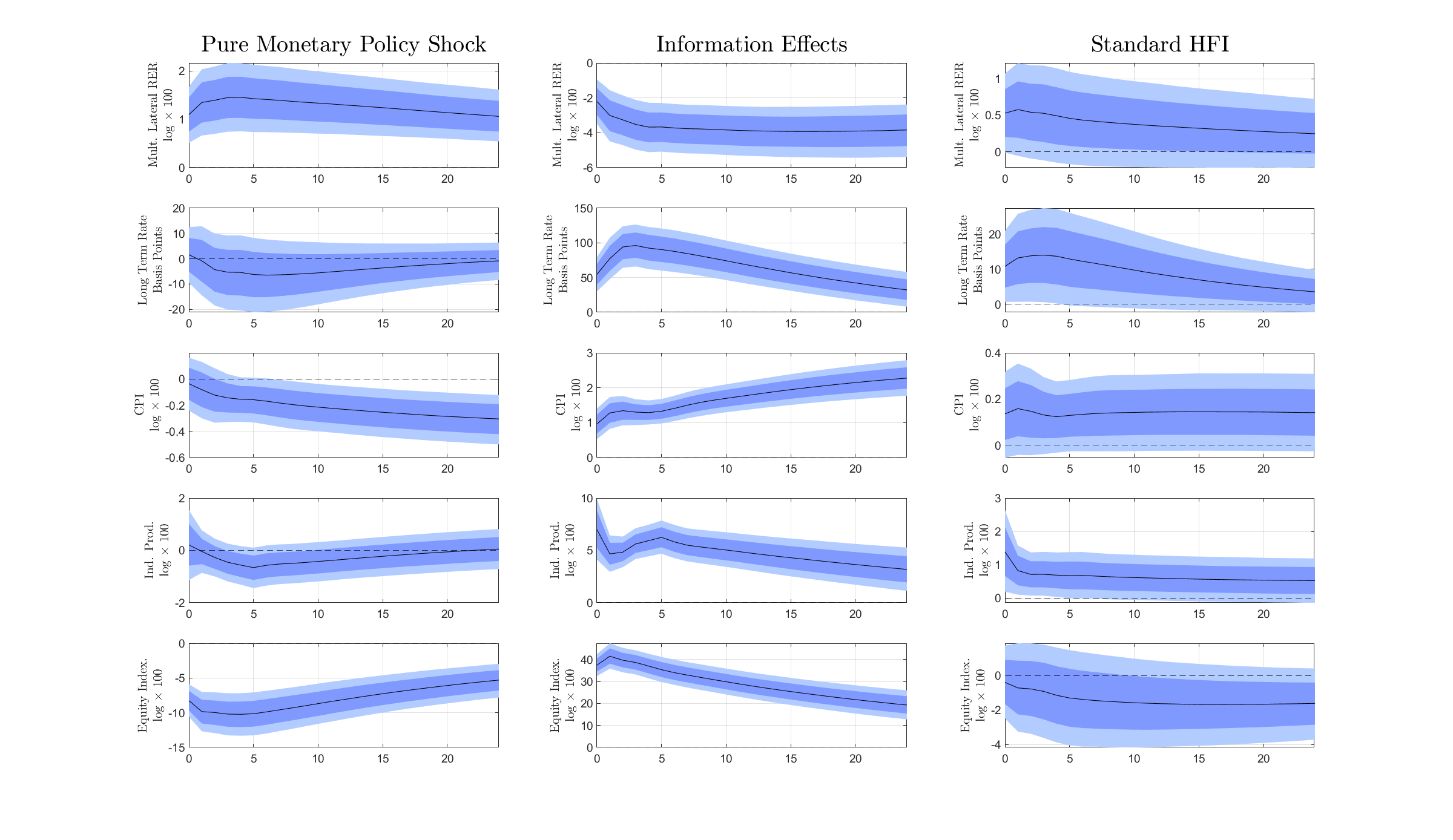}
    \caption{IRFs - Pooled Panel SVAR \\ Multi. REER Sample -  Jan 1988 to Dec 2019}
    \label{fig:Pool_88_REER}
    \floatfoot{\textbf{Note:} The figure is comprised of 15 sub-figures ordered in three columns and five rows. The impulse response functions are estimated following a panel SVAR model, as described in Appendix \ref{sec:appendix_model_details} and for the sample of countries with a valid multilateral trade weighted real exchange rate for the January 1988 to December 2019. The left panel presents the impulse response functions to the MP component, the middle column to the FIE component, and the right column following the Standard HFI approach. The rows represent the impact on (i) the multilateral trade weighted real exchange rate (in logs times 100); (ii) long term interest rates in basis points; (iii) the consumer price index (in logs times 100); (iv) the industrial production index (in logs times 100); (v) the equity index (in logs times 100). The solid black line represents the median response from the set of impulse response functions for each country. The light blue area represents the interquartile range of impulse response functions. In the text, when referring to Panel $(i,j)$, $i$ refers to the row and $j$ to the column of the figure. Variables are demeaned at the country level. }
\end{figure}

\newpage
\newpage
\begin{figure}[ht]
    \centering
    \includegraphics[scale=0.4]{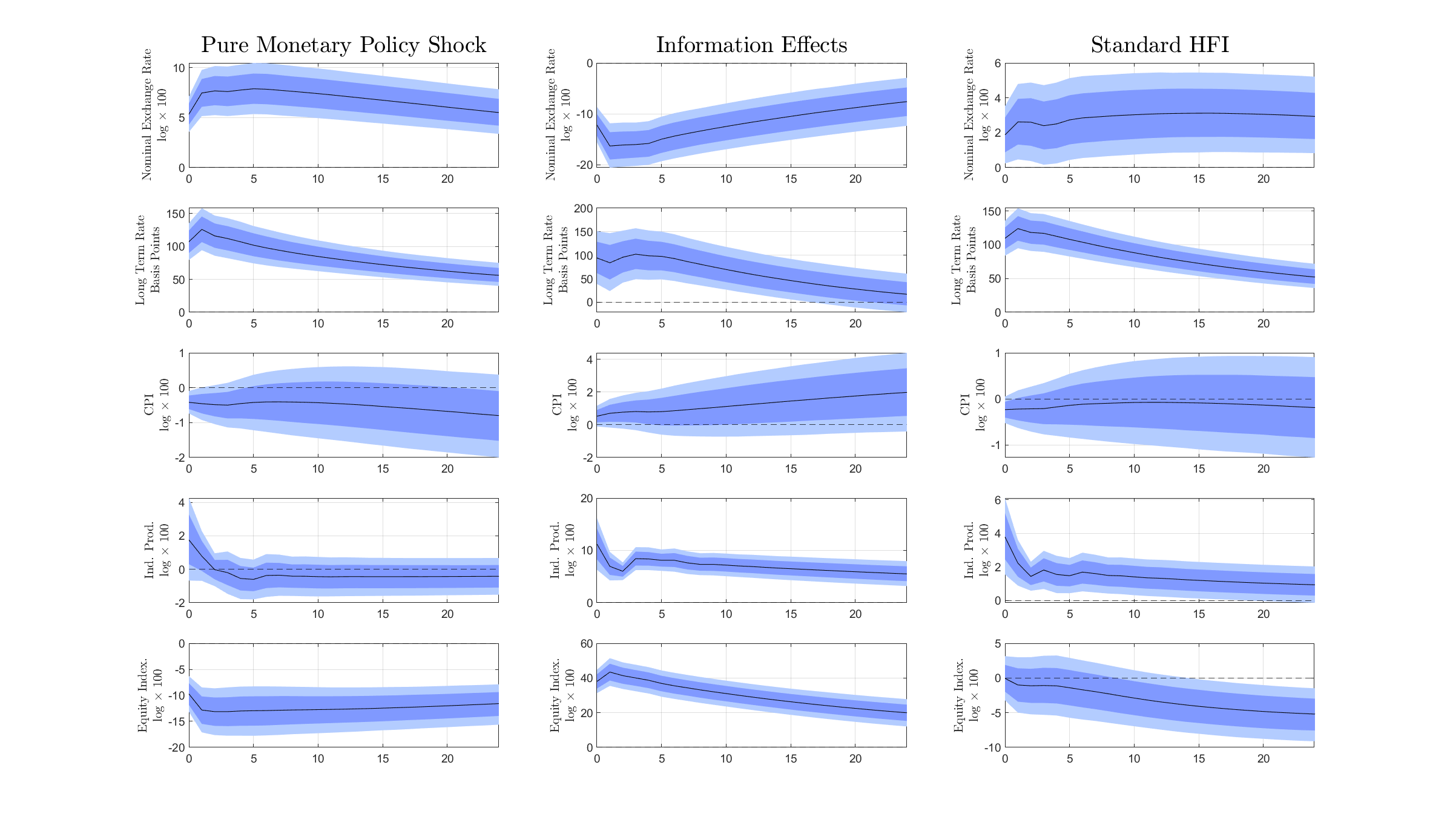}
    \caption{IRFs - Pooled Panel SVAR \\ January 1998 to December 2019}
    \label{fig:Pool_98_NER}
    \floatfoot{\textbf{Note:} The figure is comprised of 15 sub-figures ordered in three columns and five rows. The impulse response functions are estimated following a panel SVAR model, as described in Appendix \ref{sec:appendix_model_details} and for the sample of countries with a valid nominal exchange rate for the period January 1998 to December 2019. The left panel presents the impulse response functions to the MP component, the middle column to the FIE component, and the right column following the Standard HFI approach. The rows represent the impact on (i) the nominal exchange rate with the US dollar (in logs times 100); (ii) long term interest rates in basis points; (iii) the consumer price index (in logs times 100); (iv) the industrial production index (in logs times 100); (v) the equity index (in logs times 100). The solid black line represents the median response from the set of impulse response functions for each country. The light blue area represents the interquartile range of impulse response functions. In the text, when referring to Panel $(i,j)$, $i$ refers to the row and $j$ to the column of the figure. Variables are demeaned at the country level. }
\end{figure}

\newpage
\begin{figure}[ht]
    \centering
    \includegraphics[scale=0.4]{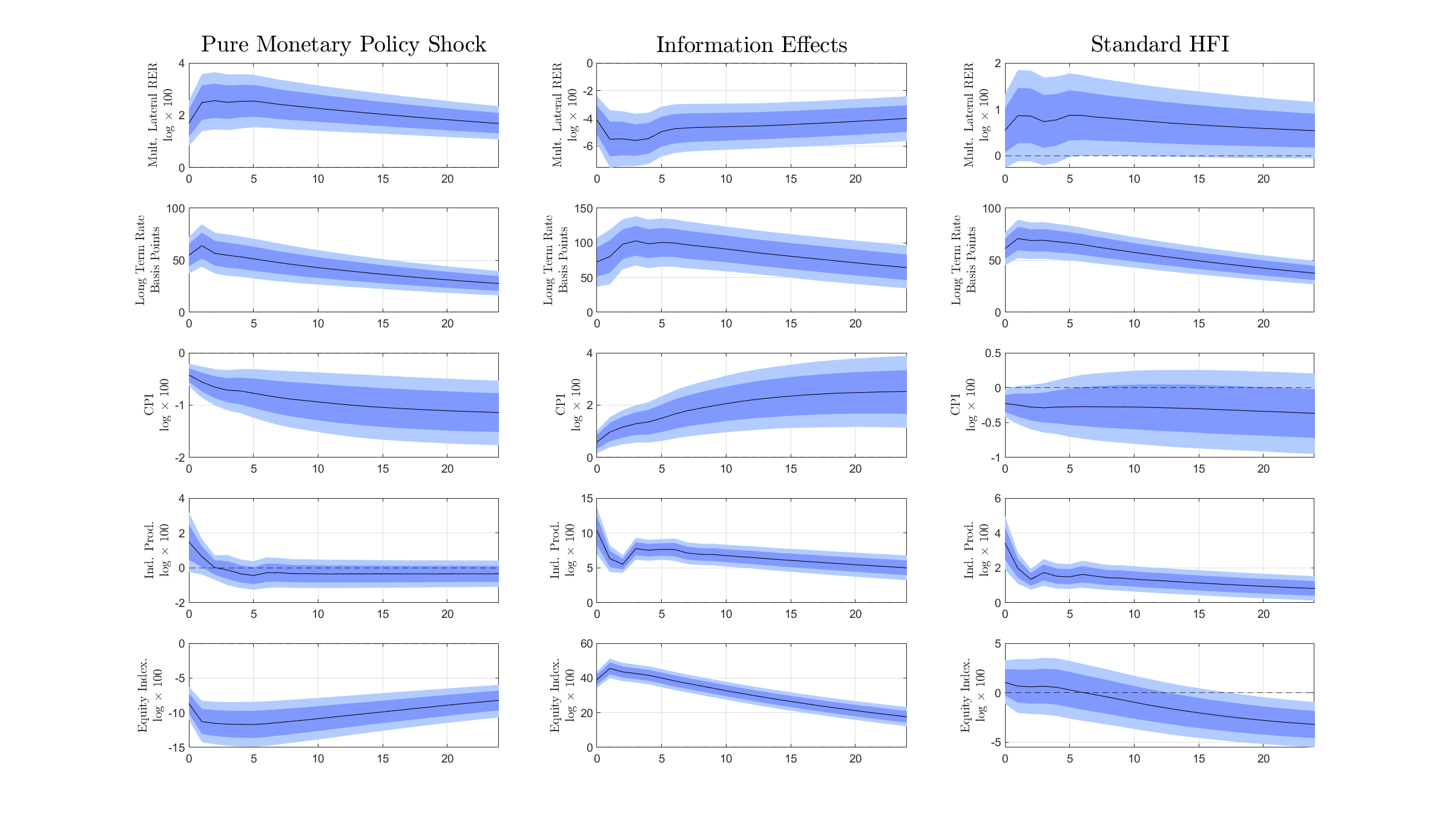}
    \caption{IRFs - Pooled Panel SVAR \\ Multi. REER Sample -  Jan 1998 to Dec 2019}
    \label{fig:Pool_98_REER}
    \floatfoot{\textbf{Note:} The figure is comprised of 15 sub-figures ordered in three columns and five rows. The impulse response functions are estimated following a panel SVAR model, as described in Appendix \ref{sec:appendix_model_details} and for the sample of countries with a valid multilateral trade weighted real exchange rate for the January 1998 to December 2019. The left panel presents the impulse response functions to the MP component, the middle column to the FIE component, and the right column following the Standard HFI approach. The rows represent the impact on (i) the multilateral trade weighted real exchange rate (in logs times 100); (ii) long term interest rates in basis points; (iii) the consumer price index (in logs times 100); (iv) the industrial production index (in logs times 100); (v) the equity index (in logs times 100). The solid black line represents the median response from the set of impulse response functions for each country. The light blue area represents the interquartile range of impulse response functions. In the text, when referring to Panel $(i,j)$, $i$ refers to the row and $j$ to the column of the figure. Variables are demeaned at the country level. }
\end{figure}

\newpage
\newpage
\begin{figure}[ht]
    \centering
    \includegraphics[scale=0.4]{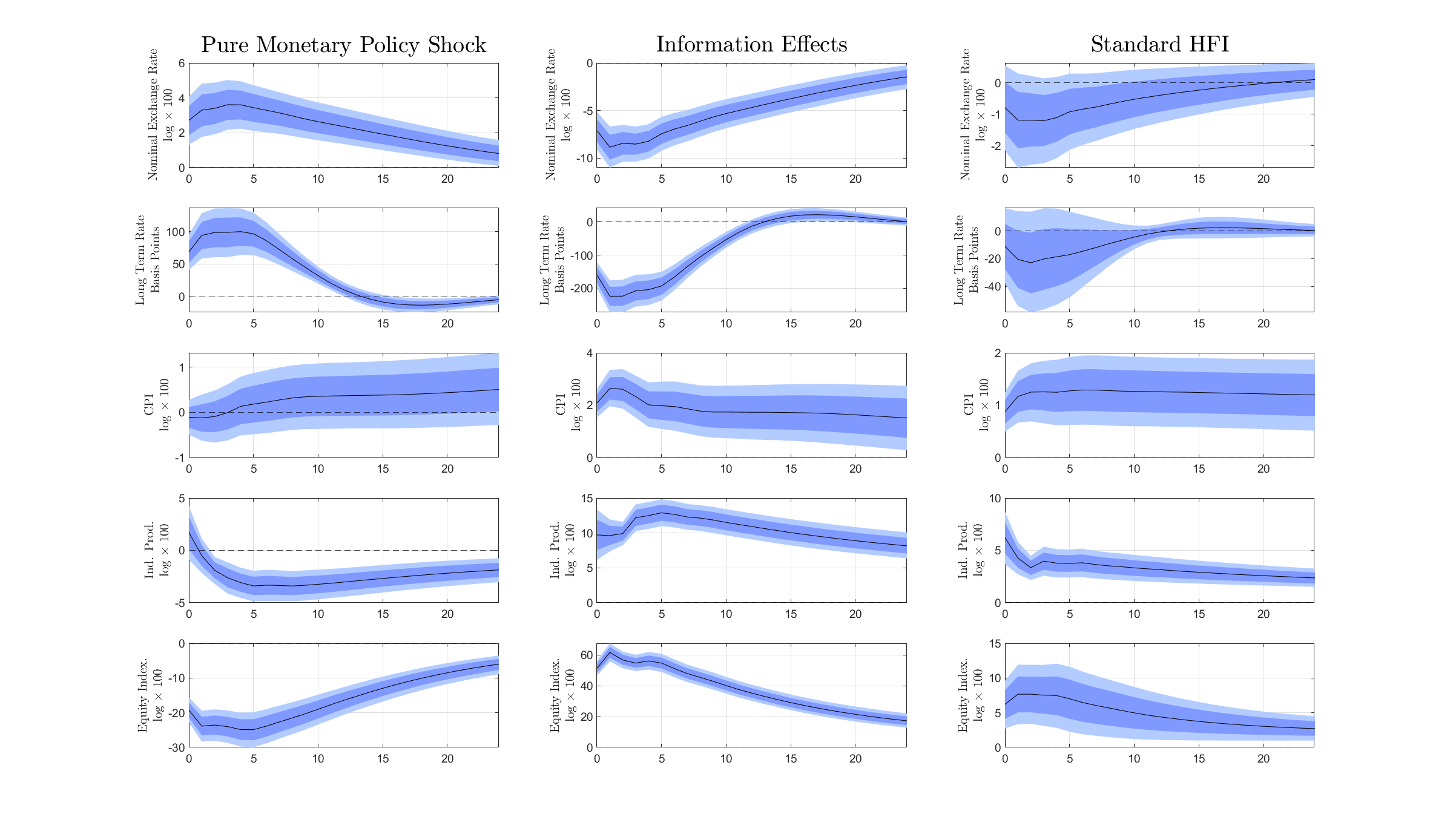}
    \caption{IRFs - Pooled Panel SVAR \\ January 2008 to December 2019}
    \label{fig:Pool_08_NER}
    \floatfoot{\textbf{Note:} The figure is comprised of 15 sub-figures ordered in three columns and five rows. The impulse response functions are estimated following a panel SVAR model, as described in Appendix \ref{sec:appendix_model_details} and for the sample of countries with a valid nominal exchange rate for the period January 2008 to December 2019. The left panel presents the impulse response functions to the MP component, the middle column to the FIE component, and the right column following the Standard HFI approach. The rows represent the impact on (i) the nominal exchange rate with the US dollar (in logs times 100); (ii) long term interest rates in basis points; (iii) the consumer price index (in logs times 100); (iv) the industrial production index (in logs times 100); (v) the equity index (in logs times 100). The solid black line represents the median response from the set of impulse response functions for each country. The light blue area represents the interquartile range of impulse response functions. In the text, when referring to Panel $(i,j)$, $i$ refers to the row and $j$ to the column of the figure. Variables are demeaned at the country level. }
\end{figure}

\newpage
\begin{figure}[ht]
    \centering
    \includegraphics[scale=0.4]{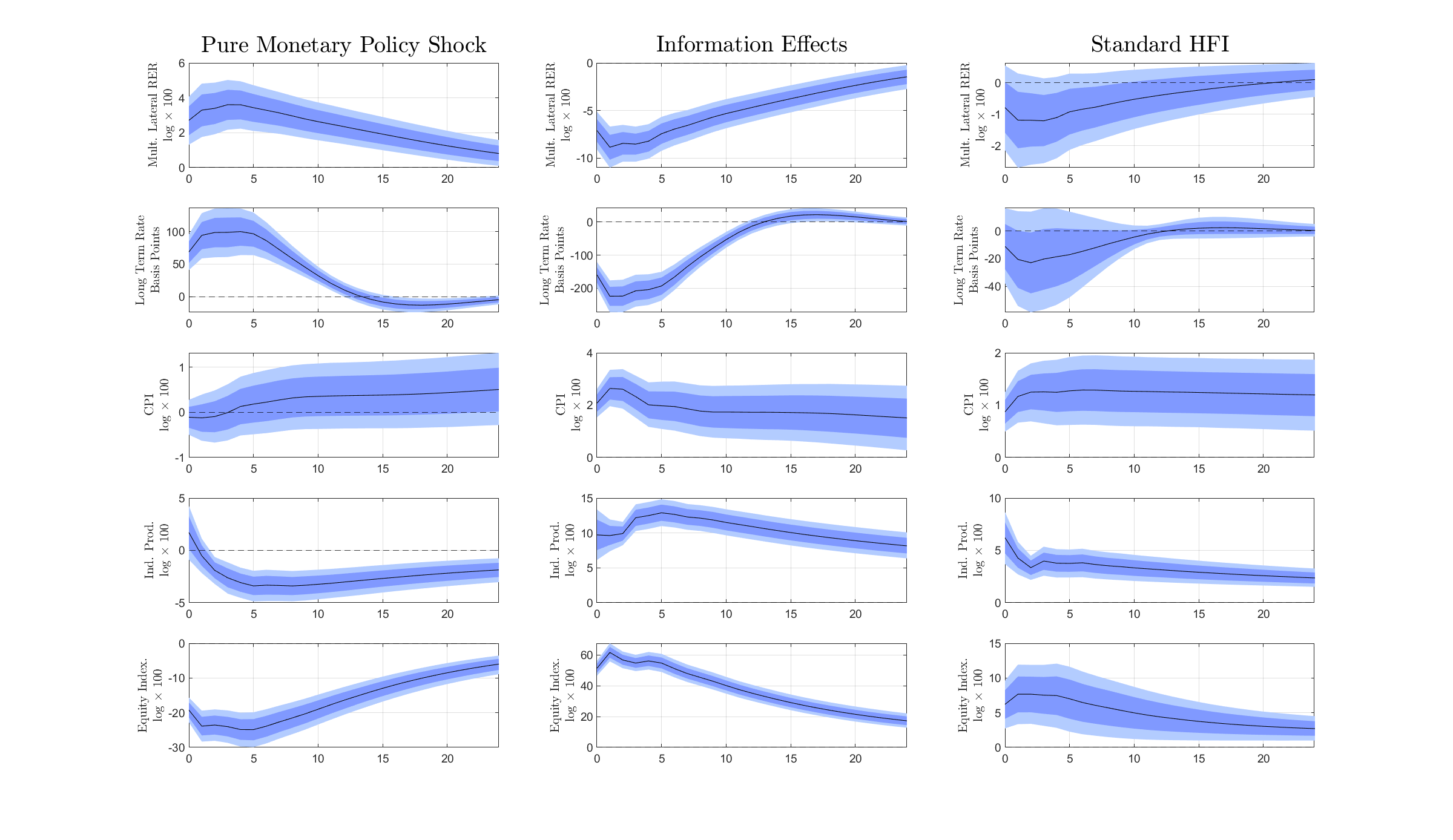}
    \caption{IRFs - Pooled Panel SVAR \\ Multi. REER Sample -  Jan 2008 to Dec 2019}
    \label{fig:Pool_08_REER}
    \floatfoot{\textbf{Note:} The figure is comprised of 15 sub-figures ordered in three columns and five rows. The impulse response functions are estimated following a panel SVAR model, as described in Appendix \ref{sec:appendix_model_details} and for the sample of countries with a valid multilateral trade weighted real exchange rate for the January 2008 to December 2019. The left panel presents the impulse response functions to the MP component, the middle column to the FIE component, and the right column following the Standard HFI approach. The rows represent the impact on (i) the multilateral trade weighted real exchange rate (in logs times 100); (ii) long term interest rates in basis points; (iii) the consumer price index (in logs times 100); (iv) the industrial production index (in logs times 100); (v) the equity index (in logs times 100). The solid black line represents the median response from the set of impulse response functions for each country. The light blue area represents the interquartile range of impulse response functions. In the text, when referring to Panel $(i,j)$, $i$ refers to the row and $j$ to the column of the figure. Variables are demeaned at the country level. }
\end{figure}


\newpage
\newpage
\begin{figure}[ht]
    \centering
    \includegraphics[scale=0.4]{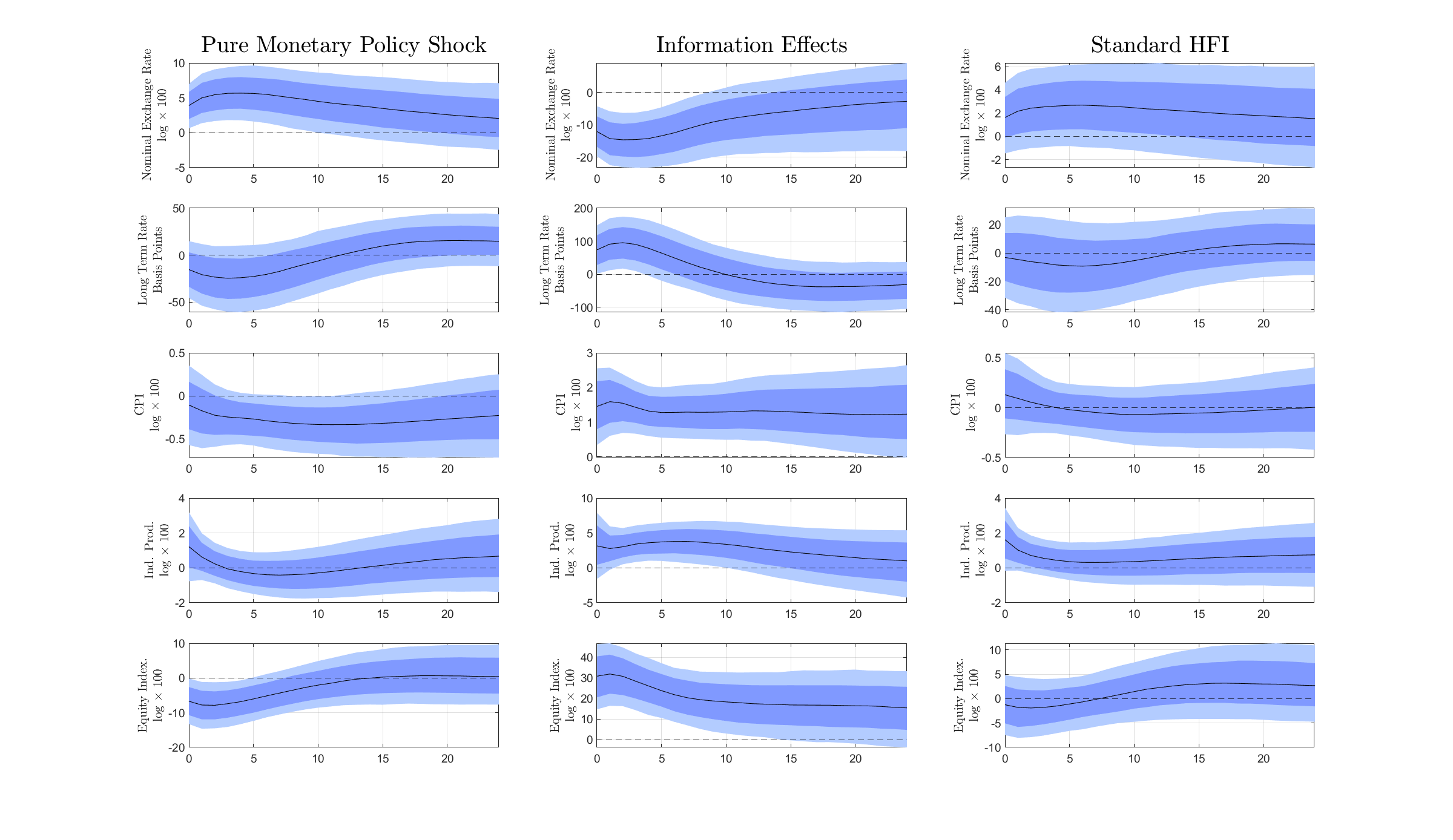}
    \caption{IRFs - Mean Group Panel SVAR \\ January 1988 to December 2019}
    \label{fig:Average_1988_NER}
    \floatfoot{\textbf{Note:} The figure is comprised of 15 sub-figures ordered in three columns and five rows. The impulse response functions are estimated following a panel SVAR model, as described in Appendix \ref{sec:appendix_model_details} and for the sample of countries with a valid nominal exchange rate for the period January 1988 to December 2019. The left panel presents the impulse response functions to the MP component, the middle column to the FIE component, and the right column following the Standard HFI approach. The rows represent the impact on (i) the nominal exchange rate with the US dollar (in logs times 100); (ii) long term interest rates in basis points; (iii) the consumer price index (in logs times 100); (iv) the industrial production index (in logs times 100); (v) the equity index (in logs times 100). The solid black line represents the median response from the set of impulse response functions for each country. The light blue area represents the interquartile range of impulse response functions. In the text, when referring to Panel $(i,j)$, $i$ refers to the row and $j$ to the column of the figure. Variables are demeaned at the country level and then averaged across countries. }
\end{figure}

\newpage
\begin{figure}[ht]
    \centering
    \includegraphics[scale=0.4]{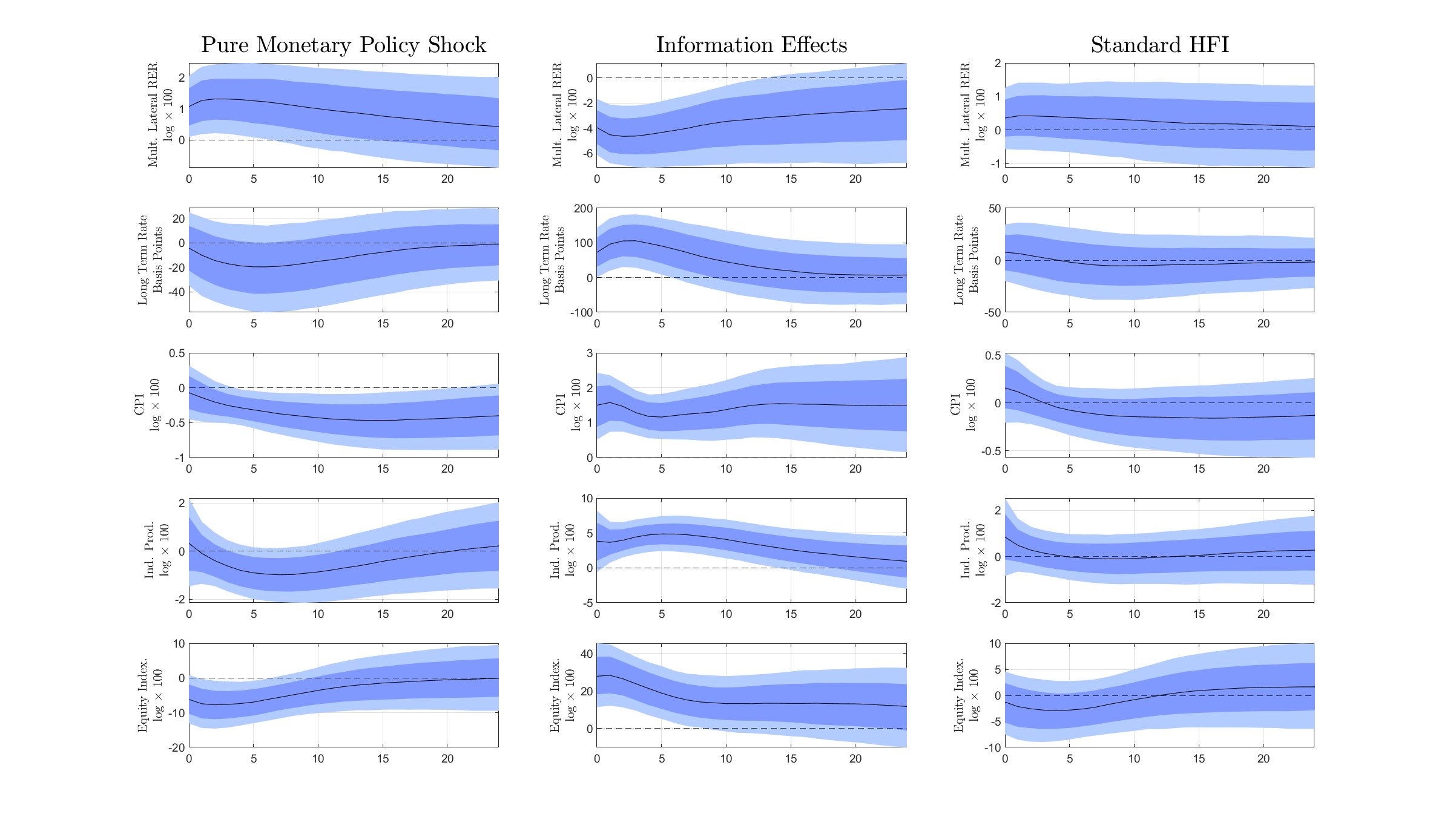}
    \caption{IRFs - Mean Group Panel SVAR \\ Multi. REER Sample -  Jan 1988 to Dec 2019}
    \label{fig:Average_1988_REER}
    \floatfoot{\textbf{Note:} The figure is comprised of 15 sub-figures ordered in three columns and five rows. The impulse response functions are estimated following a panel SVAR model, as described in Appendix \ref{sec:appendix_model_details} and for the sample of countries with a valid multilateral trade weighted real exchange rate for the January 1988 to December 2019. The left panel presents the impulse response functions to the MP component, the middle column to the FIE component, and the right column following the Standard HFI approach. The rows represent the impact on (i) the multilateral trade weighted real exchange rate (in logs times 100); (ii) long term interest rates in basis points; (iii) the consumer price index (in logs times 100); (iv) the industrial production index (in logs times 100); (v) the equity index (in logs times 100). The solid black line represents the median response from the set of impulse response functions for each country. The light blue area represents the interquartile range of impulse response functions. In the text, when referring to Panel $(i,j)$, $i$ refers to the row and $j$ to the column of the figure. Variables are demeaned at the country level and then averaged across countries.}
\end{figure}

\newpage
\newpage
\begin{figure}[ht]
    \centering
    \includegraphics[scale=0.4]{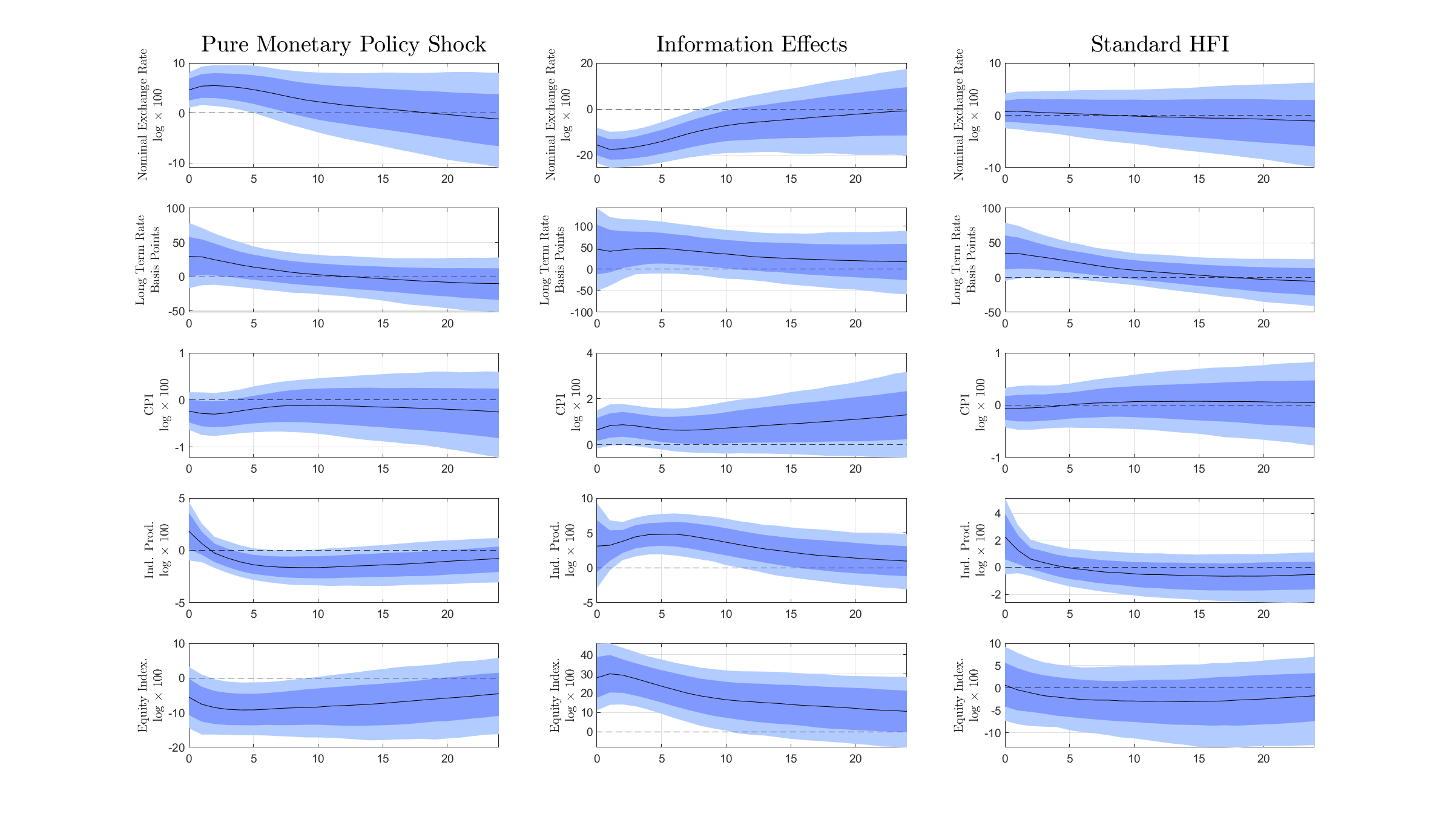}
    \caption{IRFs - Mean Group Panel SVAR \\ January 1998 to December 2019}
    \label{fig:Average_1998_NER}
    \floatfoot{\textbf{Note:} The figure is comprised of 15 sub-figures ordered in three columns and five rows. The impulse response functions are estimated following a panel SVAR model, as described in Appendix \ref{sec:appendix_model_details} and for the sample of countries with a valid nominal exchange rate for the period January 1998 to December 2019. The left panel presents the impulse response functions to the MP component, the middle column to the FIE component, and the right column following the Standard HFI approach. The rows represent the impact on (i) the nominal exchange rate with the US dollar (in logs times 100); (ii) long term interest rates in basis points; (iii) the consumer price index (in logs times 100); (iv) the industrial production index (in logs times 100); (v) the equity index (in logs times 100). The solid black line represents the median response from the set of impulse response functions for each country. The light blue area represents the interquartile range of impulse response functions. In the text, when referring to Panel $(i,j)$, $i$ refers to the row and $j$ to the column of the figure. Variables are demeaned at the country level and then averaged across countries.}
\end{figure}

\newpage
\begin{figure}[ht]
    \centering
    \includegraphics[scale=0.4]{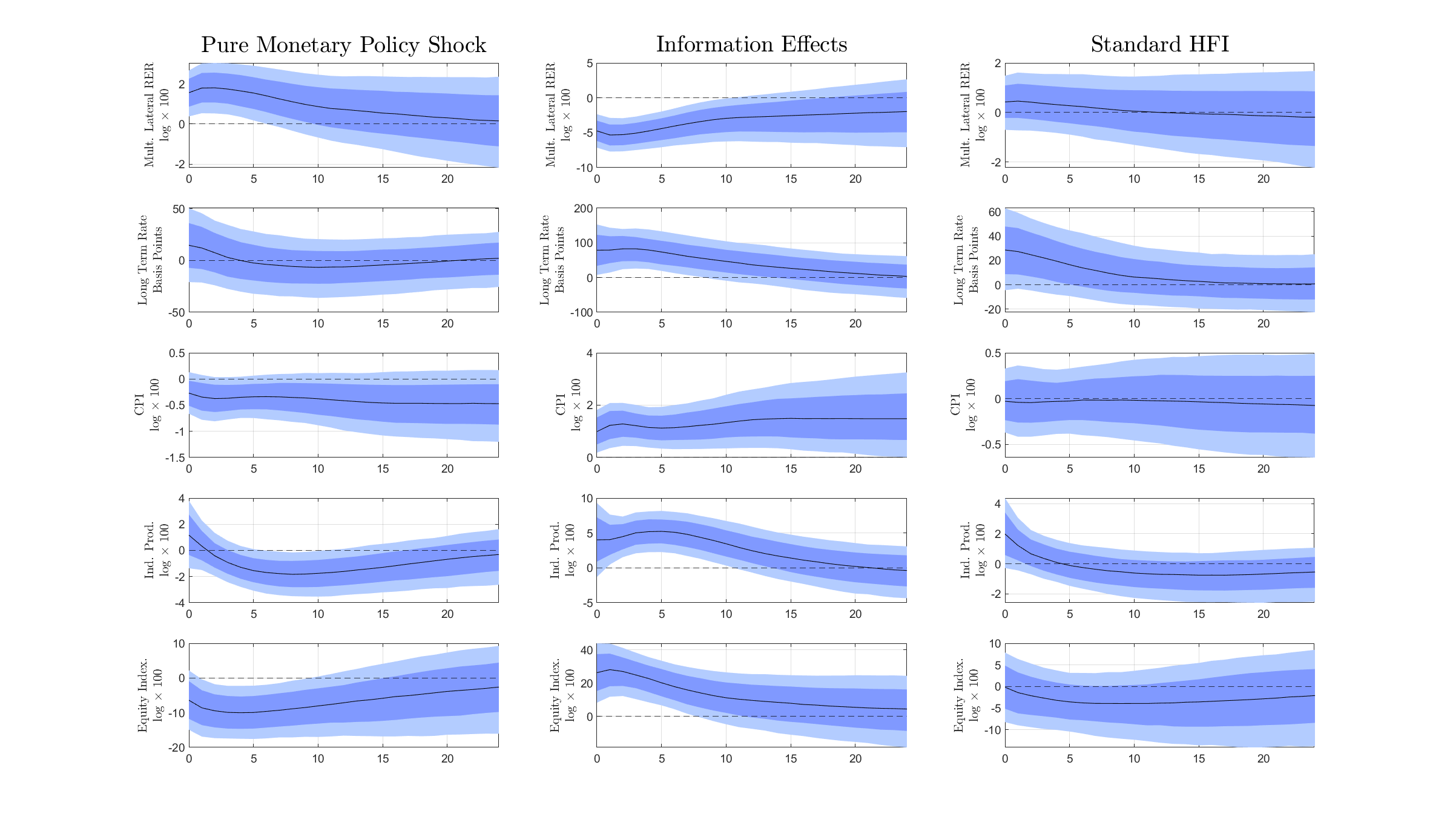}
    \caption{IRFs - Mean Group Panel SVAR \\ Multi. REER Sample -  Jan 1998 to Dec 2019}
    \label{fig:Average_1998_REER}
    \floatfoot{\textbf{Note:} The figure is comprised of 15 sub-figures ordered in three columns and five rows. The impulse response functions are estimated following a panel SVAR model, as described in Appendix \ref{sec:appendix_model_details} and for the sample of countries with a valid multilateral trade weighted real exchange rate for the January 1998 to December 2019. The left panel presents the impulse response functions to the MP component, the middle column to the FIE component, and the right column following the Standard HFI approach. The rows represent the impact on (i) the multilateral trade weighted real exchange rate (in logs times 100); (ii) long term interest rates in basis points; (iii) the consumer price index (in logs times 100); (iv) the industrial production index (in logs times 100); (v) the equity index (in logs times 100). The solid black line represents the median response from the set of impulse response functions for each country. The light blue area represents the interquartile range of impulse response functions. In the text, when referring to Panel $(i,j)$, $i$ refers to the row and $j$ to the column of the figure. Variables are demeaned at the country level and then averaged across countries.}
\end{figure}

\newpage
\newpage
\begin{figure}[ht]
    \centering
    \includegraphics[scale=0.4]{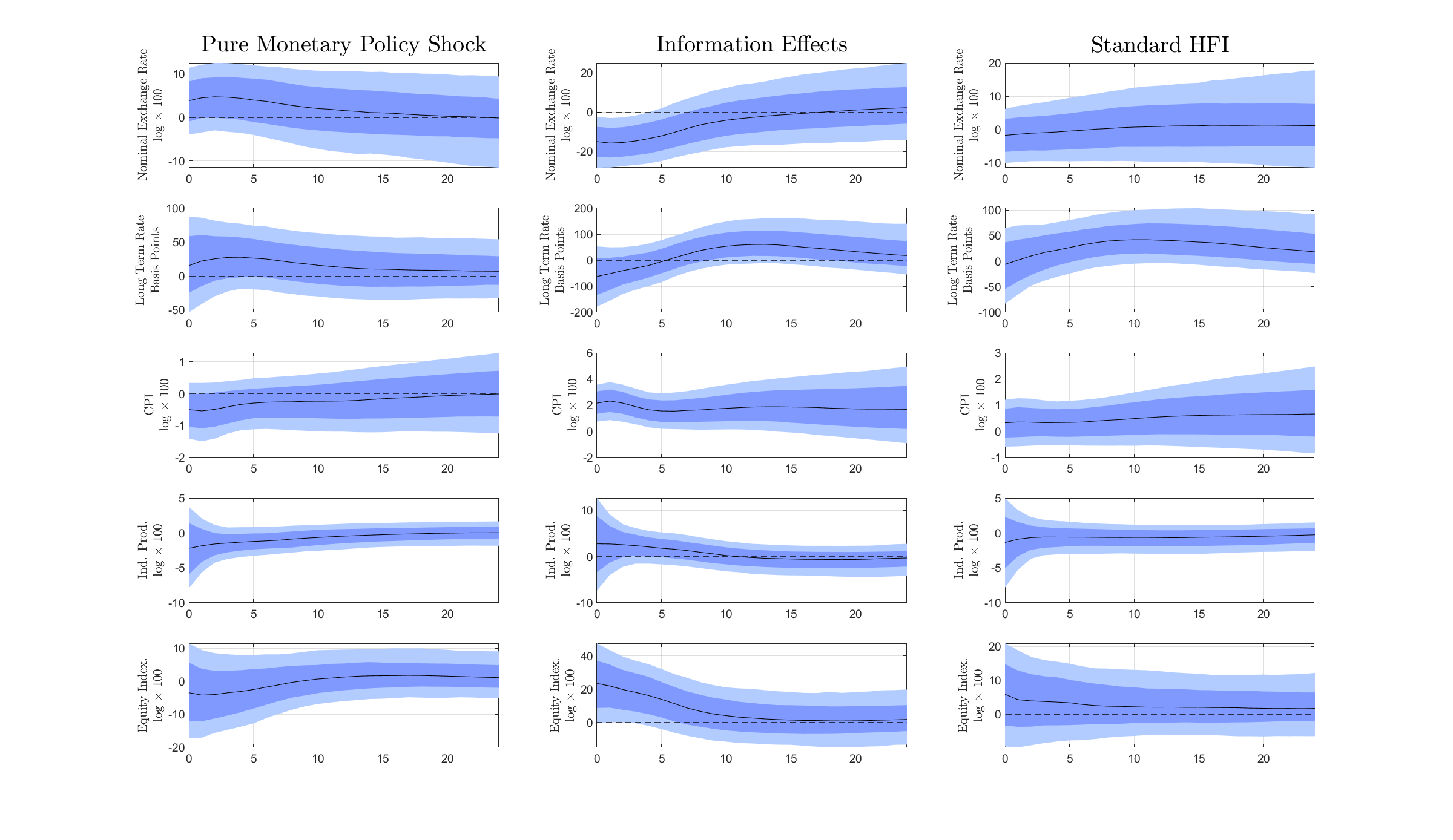}
    \caption{IRFs - Mean Group Panel SVAR \\ January 2008 to December 2019}
    \label{fig:Average_2008_NER}
    \floatfoot{\textbf{Note:} The figure is comprised of 15 sub-figures ordered in three columns and five rows. The impulse response functions are estimated following a panel SVAR model, as described in Appendix \ref{sec:appendix_model_details} and for the sample of countries with a valid nominal exchange rate for the period January 2008 to December 2019. The left panel presents the impulse response functions to the MP component, the middle column to the FIE component, and the right column following the Standard HFI approach. The rows represent the impact on (i) the nominal exchange rate with the US dollar (in logs times 100); (ii) long term interest rates in basis points; (iii) the consumer price index (in logs times 100); (iv) the industrial production index (in logs times 100); (v) the equity index (in logs times 100). The solid black line represents the median response from the set of impulse response functions for each country. The light blue area represents the interquartile range of impulse response functions. In the text, when referring to Panel $(i,j)$, $i$ refers to the row and $j$ to the column of the figure. Variables are demeaned at the country level and then averaged across countries.}
\end{figure}

\newpage
\begin{figure}[ht]
    \centering
    \includegraphics[scale=0.4]{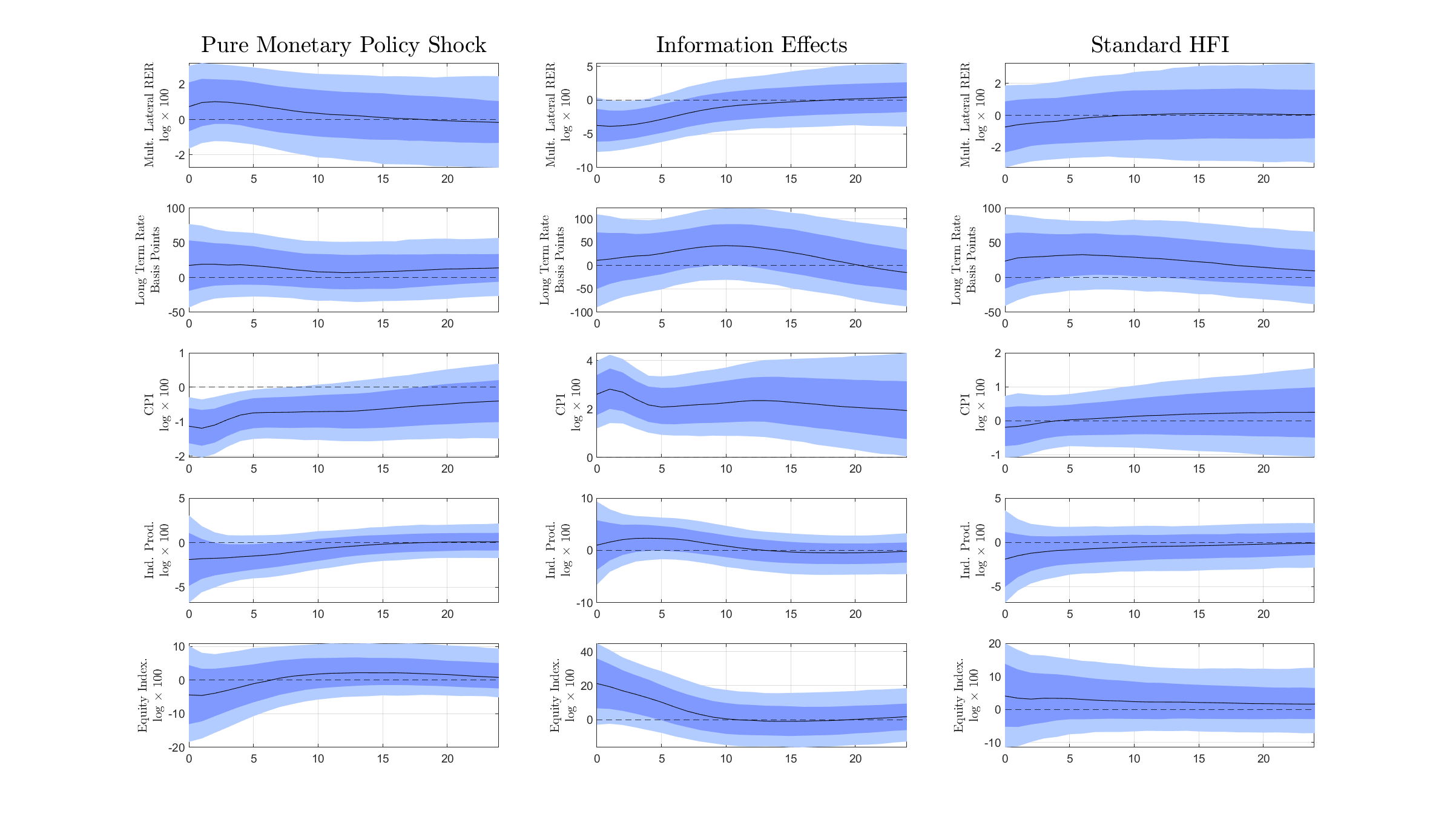}
    \caption{IRFs - Mean Group Panel SVAR \\ Multi. REER Sample -  Jan 2008 to Dec 2019}
    \label{fig:Average_2008_REER}
    \floatfoot{\textbf{Note:} The figure is comprised of 15 sub-figures ordered in three columns and five rows. The impulse response functions are estimated following a panel SVAR model, as described in Appendix \ref{sec:appendix_model_details} and for the sample of countries with a valid multilateral trade weighted real exchange rate for the January 2008 to December 2019. The left panel presents the impulse response functions to the MP component, the middle column to the FIE component, and the right column following the Standard HFI approach. The rows represent the impact on (i) the multilateral trade weighted real exchange rate (in logs times 100); (ii) long term interest rates in basis points; (iii) the consumer price index (in logs times 100); (iv) the industrial production index (in logs times 100); (v) the equity index (in logs times 100). The solid black line represents the median response from the set of impulse response functions for each country. The light blue area represents the interquartile range of impulse response functions. In the text, when referring to Panel $(i,j)$, $i$ refers to the row and $j$ to the column of the figure. Variables are demeaned at the country level and then averaged across countries.}
\end{figure}


\newpage
\begin{figure}
    \centering
    \includegraphics[scale=0.4]{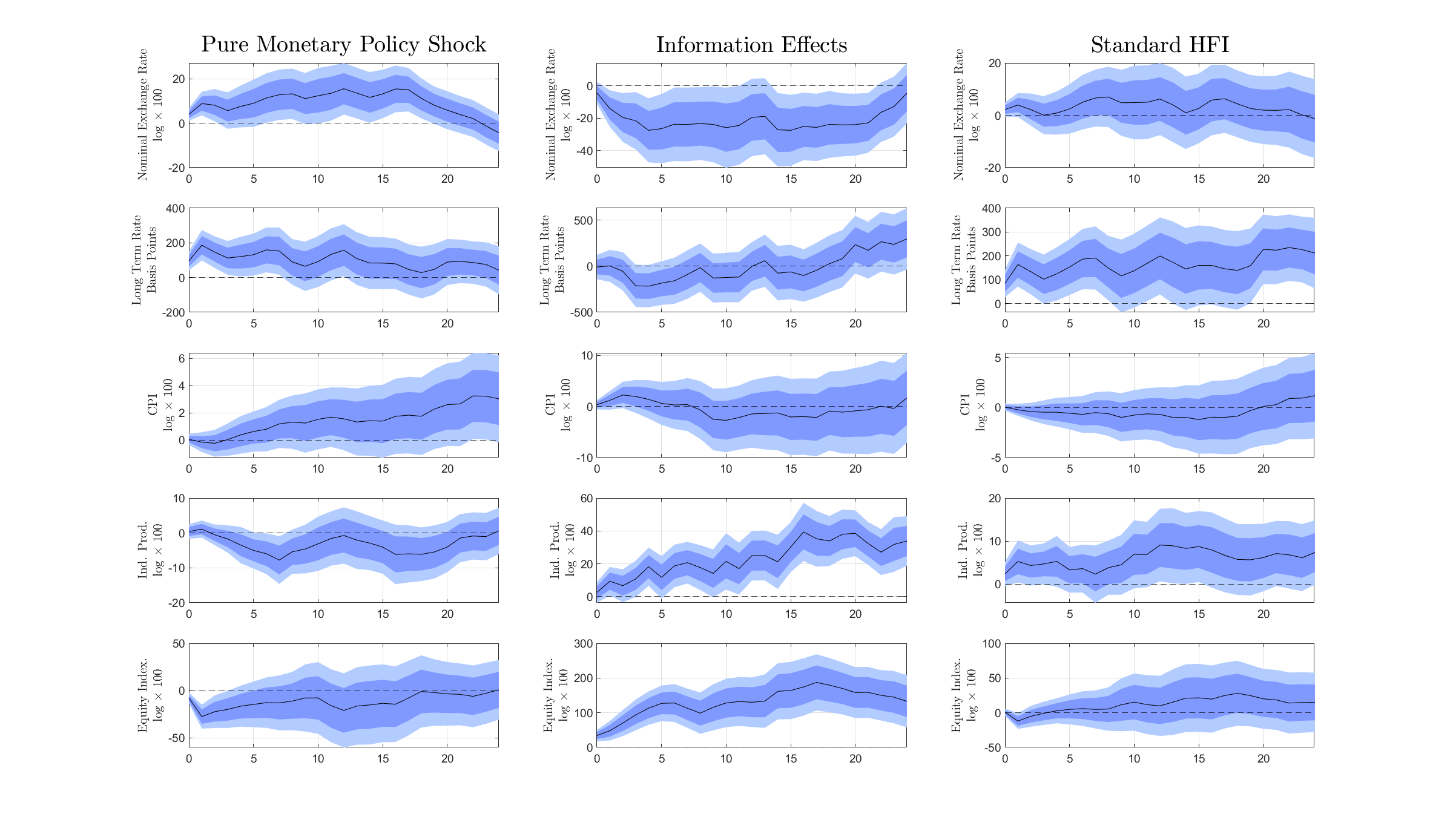}
    \caption{Impulse Response Functions \\ Introducing US Control Variables}
    \label{fig:US_BenchmarkNER}
    \floatfoot{\textbf{Note:} The figure is comprised of 15 sub-figures ordered in three columns and five rows. The left column relates to the estimates of $\beta^{MP}$ in Equation \ref{eq:LP_pooled_USA}, the middle column relates to the estimate of $\beta^{FIE}$ in Equation \ref{eq:LP_pooled_USA}, while the right column relates to estimating Equation \ref{eq:LP_pooled_USA}, replacing the MP and FIE components with the un-orthogonalized monetary policy surprise. The rows represent the impact on (i) the nominal exchange rate with the US dollar (in logs times 100); (ii) long term interest rates in basis points; (iii) the consumer price index (in logs times 100); (iv) the industrial production index (in logs times 100); (v) the equity index (in logs times 100). The solid black line represents the point estimate, the dark blue area represents the 68\% confidence interval, and the light blue area represents the 90\% confidence interval. In the text, when referring to Panel $(i,j)$, $i$ refers to the row and $j$ to the column of the figure. Each variable, in its own transformation, is demeaned at the country level. }
\end{figure}

\newpage
\begin{figure}
    \centering
    \includegraphics[scale=0.4]{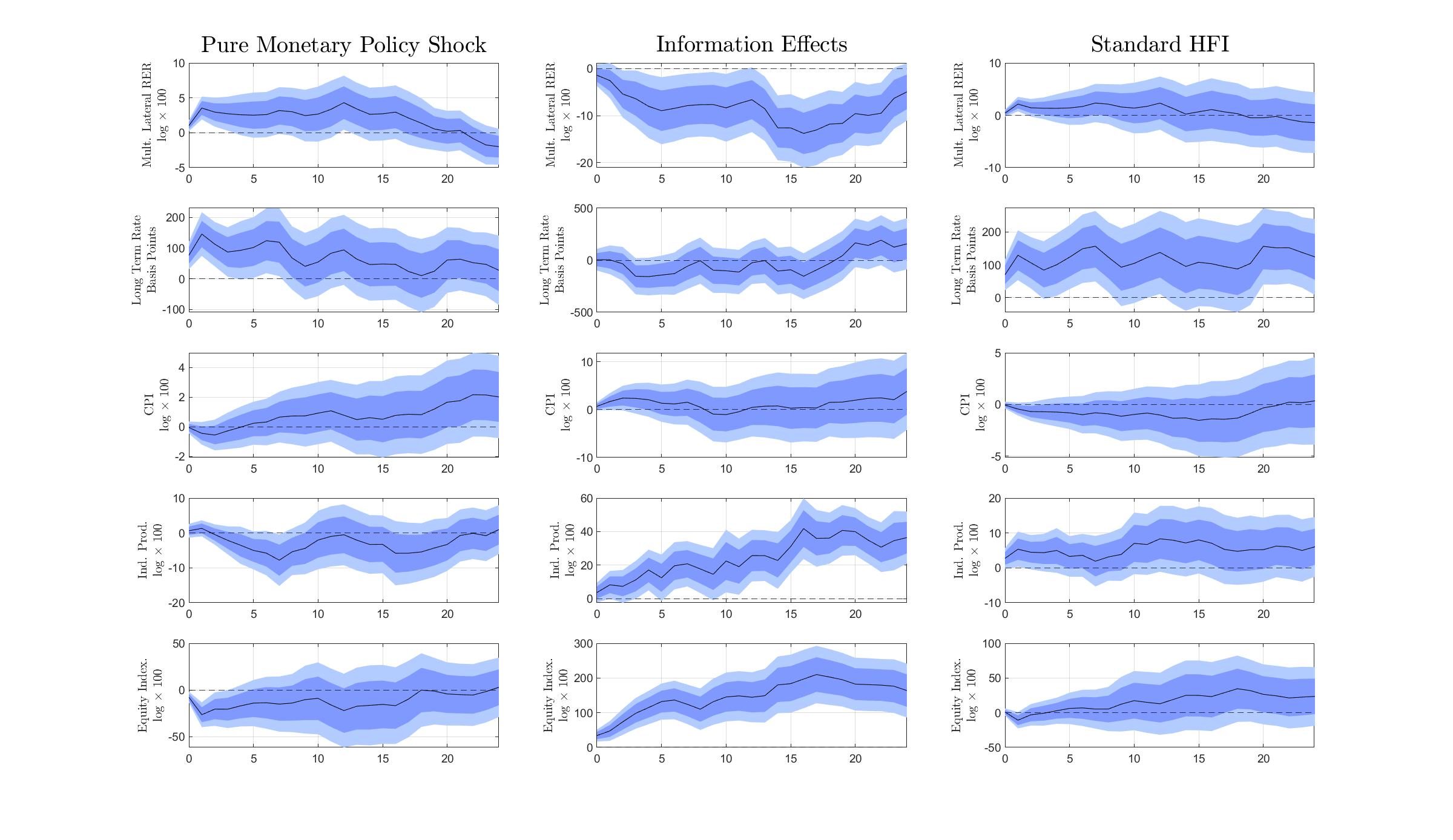}
    \caption{Impulse Response Functions \\ US Variables - Multi. REER Specification}
    \label{fig:US_BenchmarkREER}
    \floatfoot{\textbf{Note:} The figure is comprised of 15 sub-figures ordered in three columns and five rows. The left column relates to the estimates of $\beta^{MP}$ in Equation \ref{eq:LP_pooled_USA}, the middle column relates to the estimate of $\beta^{FIE}$ in Equation \ref{eq:LP_pooled_USA}, while the right column relates to estimating Equation \ref{eq:LP_pooled_USA}, replacing the MP and FIE components with the un-orthogonalized monetary policy surprise. The rows represent the impact on (i) the multilateral trade weighted real exchange rate index (in logs times 100); (ii) long term interest rates in basis points; (iii) the consumer price index (in logs times 100); (iv) the industrial production index (in logs times 100); (v) the equity index (in logs times 100). The solid black line represents the point estimate, the dark blue area represents the 68\% confidence interval, and the light blue area represents the 90\% confidence interval. In the text, when referring to Panel $(i,j)$, $i$ refers to the row and $j$ to the column of the figure. Each variable, in its own transformation, is demeaned at the country level. }
\end{figure}

\newpage
\begin{figure}
    \centering
    \includegraphics[scale=0.4]{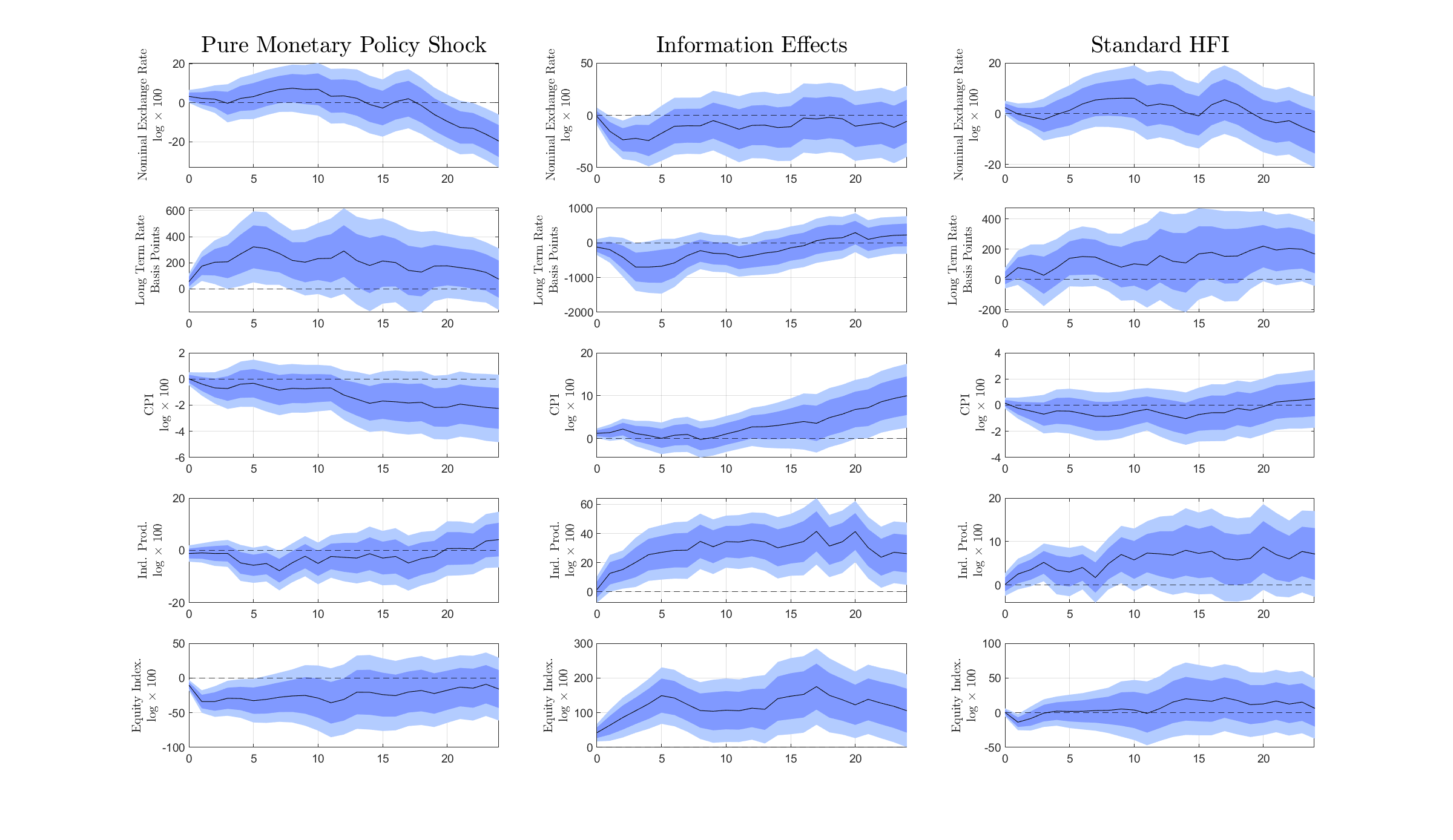}
    \caption{Impulse Response Functions \\ De Facto Peg \cite{ilzetzki2017country}}
    \label{fig:BenchmarkNER_ERA1}
    \floatfoot{\textbf{Note:} The figure is comprised of 15 sub-figures ordered in three columns and five rows. The left column relates to the estimates of $\beta^{MP}$ in Equation \ref{eq:LP_pooled}, the middle column relates to the estimate of $\beta^{FIE}$ in Equation \ref{eq:LP_pooled}, while the right column relates to estimating Equation \ref{eq:LP_pooled}, replacing the MP and FIE components with the un-orthogonalized monetary policy surprise. This figure presents the results for countries with classification of 1 by \cite{ilzetzki2017country} in month $t$.  The rows represent the impact on (i) the nominal exchange rate with the US dollar (in logs times 100); (ii) long term interest rates in basis points; (iii) the consumer price index (in logs times 100); (iv) the industrial production index (in logs times 100); (v) the equity index (in logs times 100). The solid black line represents the point estimate, the dark blue area represents the 68\% confidence interval, and the light blue area represents the 90\% confidence interval. In the text, when referring to Panel $(i,j)$, $i$ refers to the row and $j$ to the column of the figure. Each variable, in its own transformation, is demeaned at the country level. }
\end{figure}

\newpage
\begin{figure}
    \centering
    \includegraphics[scale=0.4]{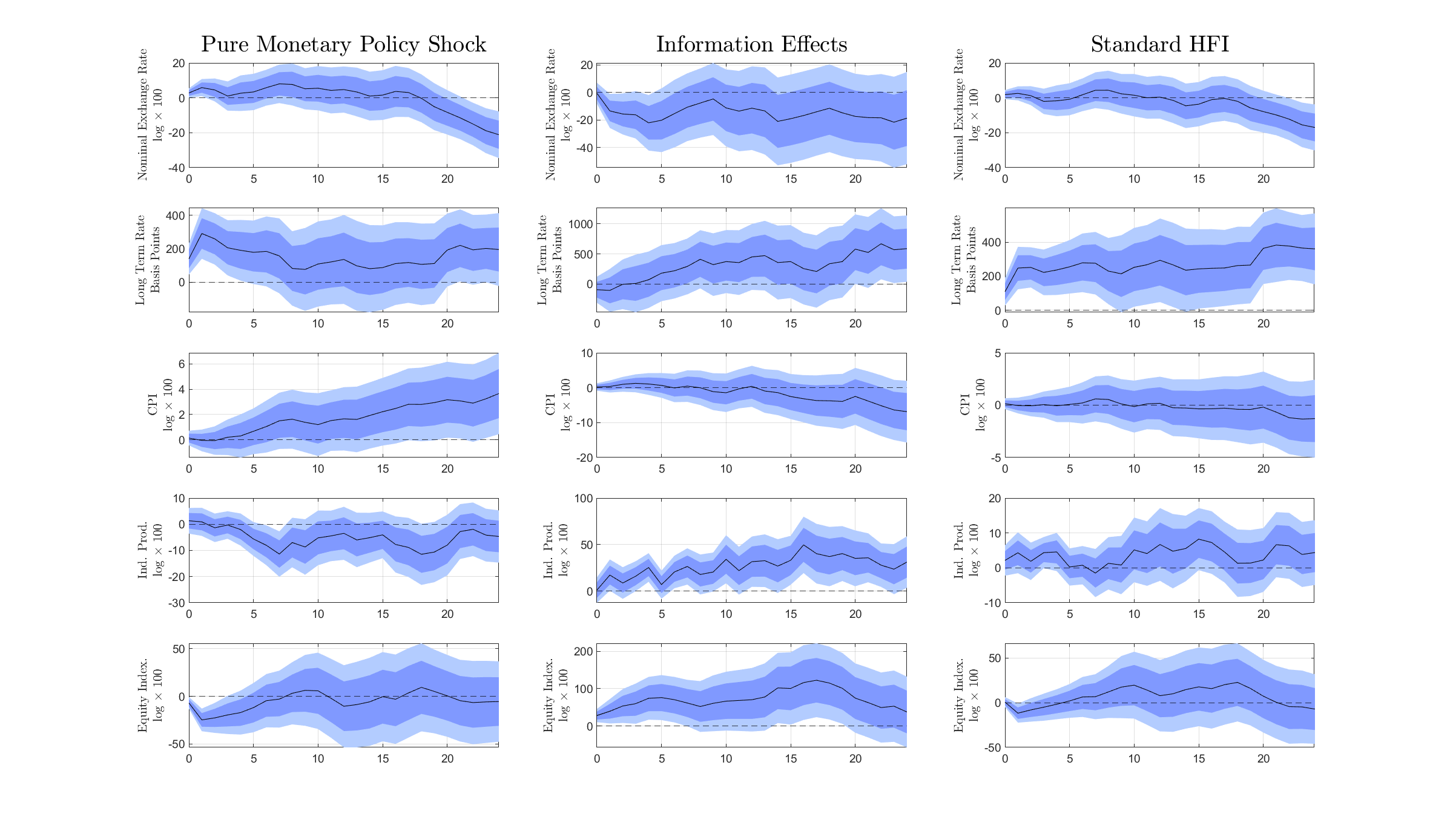}
    \caption{Impulse Response Functions \\ Pre-Announced Crawling Peg \cite{ilzetzki2017country}}
    \label{fig:BenchmarkNER_ERA2}
    \floatfoot{\textbf{Note:} The figure is comprised of 15 sub-figures ordered in three columns and five rows. The left column relates to the estimates of $\beta^{MP}$ in Equation \ref{eq:LP_pooled}, the middle column relates to the estimate of $\beta^{FIE}$ in Equation \ref{eq:LP_pooled}, while the right column relates to estimating Equation \ref{eq:LP_pooled}, replacing the MP and FIE components with the un-orthogonalized monetary policy surprise. This figure presents the results for countries with classification of 2 by \cite{ilzetzki2017country} in month $t$.  The rows represent the impact on (i) the nominal exchange rate with the US dollar (in logs times 100); (ii) long term interest rates in basis points; (iii) the consumer price index (in logs times 100); (iv) the industrial production index (in logs times 100); (v) the equity index (in logs times 100). The solid black line represents the point estimate, the dark blue area represents the 68\% confidence interval, and the light blue area represents the 90\% confidence interval. In the text, when referring to Panel $(i,j)$, $i$ refers to the row and $j$ to the column of the figure. Each variable, in its own transformation, is demeaned at the country level. }
\end{figure}

\newpage
\begin{figure}
    \centering
    \includegraphics[scale=0.4]{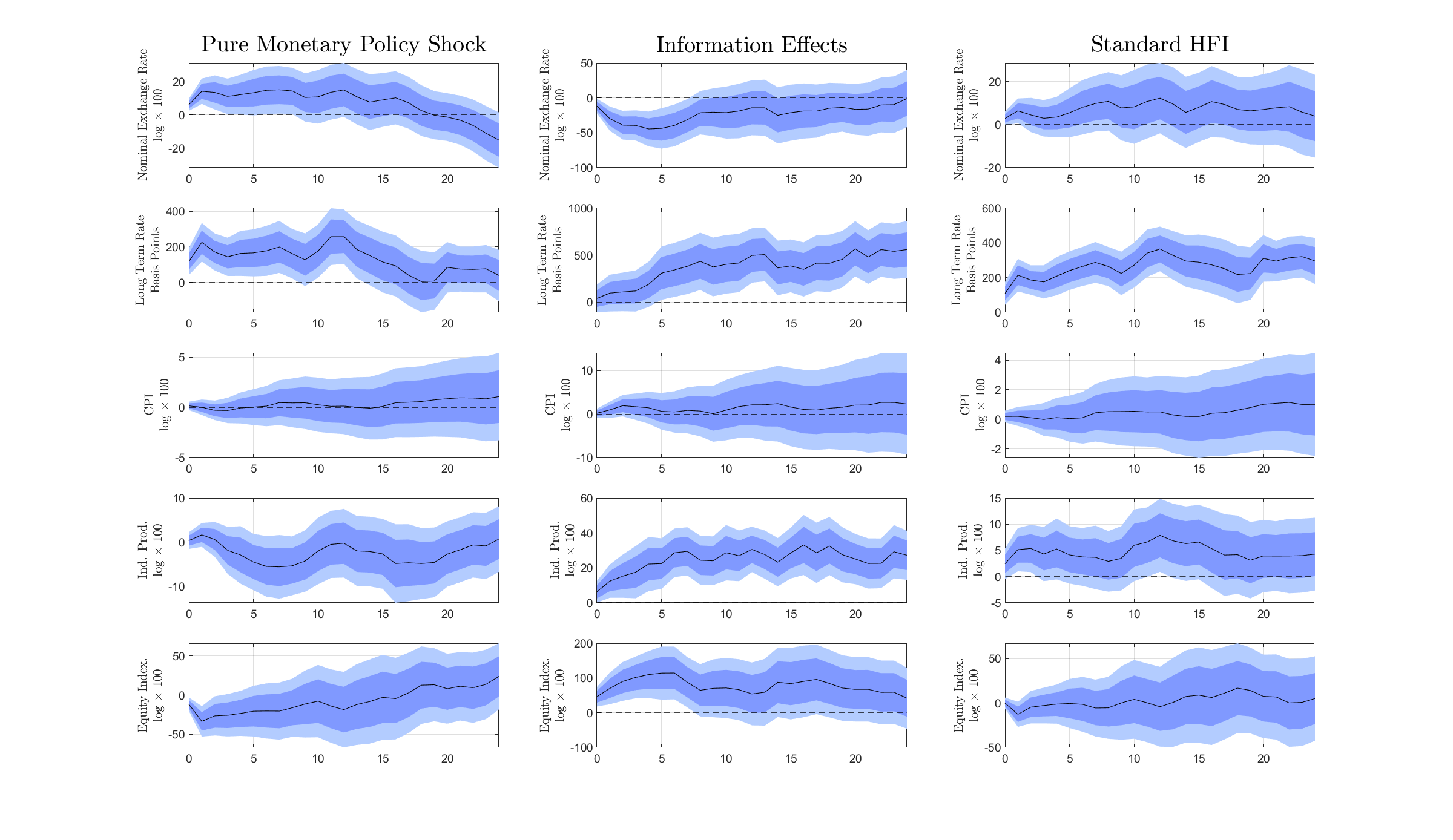}
    \caption{Impulse Response Functions \\ Managed Floating \cite{ilzetzki2017country}}
    \label{fig:BenchmarkNER_ERA3}
    \floatfoot{\textbf{Note:} The figure is comprised of 15 sub-figures ordered in three columns and five rows. The left column relates to the estimates of $\beta^{MP}$ in Equation \ref{eq:LP_pooled}, the middle column relates to the estimate of $\beta^{FIE}$ in Equation \ref{eq:LP_pooled}, while the right column relates to estimating Equation \ref{eq:LP_pooled}, replacing the MP and FIE components with the un-orthogonalized monetary policy surprise. This figure presents the results for countries with classification of 3 by \cite{ilzetzki2017country} in month $t$.  The rows represent the impact on (i) the nominal exchange rate with the US dollar (in logs times 100); (ii) long term interest rates in basis points; (iii) the consumer price index (in logs times 100); (iv) the industrial production index (in logs times 100); (v) the equity index (in logs times 100). The solid black line represents the point estimate, the dark blue area represents the 68\% confidence interval, and the light blue area represents the 90\% confidence interval. In the text, when referring to Panel $(i,j)$, $i$ refers to the row and $j$ to the column of the figure. Each variable, in its own transformation, is demeaned at the country level. }
\end{figure}

\newpage
\begin{figure}
    \centering
    \includegraphics[scale=0.4]{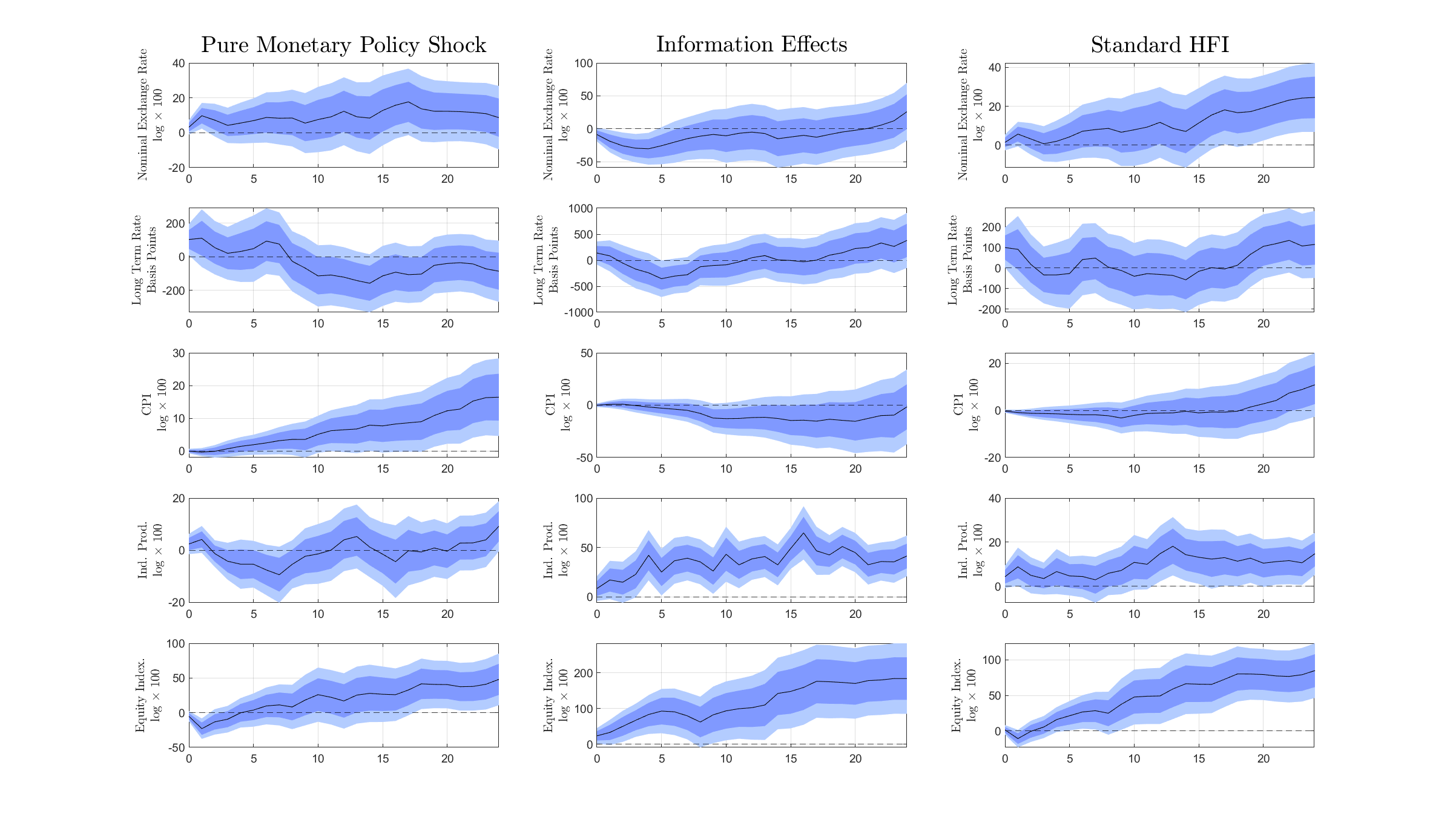}
    \caption{Impulse Response Functions \\ Freely Floating to Dual Market \cite{ilzetzki2017country}}
    \label{fig:BenchmarkNER_ERA4}
    \floatfoot{\textbf{Note:} The figure is comprised of 15 sub-figures ordered in three columns and five rows. The left column relates to the estimates of $\beta^{MP}$ in Equation \ref{eq:LP_pooled}, the middle column relates to the estimate of $\beta^{FIE}$ in Equation \ref{eq:LP_pooled}, while the right column relates to estimating Equation \ref{eq:LP_pooled}, replacing the MP and FIE components with the un-orthogonalized monetary policy surprise. This figure presents the results for countries with classification of 4 or higher by \cite{ilzetzki2017country} in month $t$.  The rows represent the impact on (i) the nominal exchange rate with the US dollar (in logs times 100); (ii) long term interest rates in basis points; (iii) the consumer price index (in logs times 100); (iv) the industrial production index (in logs times 100); (v) the equity index (in logs times 100). The solid black line represents the point estimate, the dark blue area represents the 68\% confidence interval, and the light blue area represents the 90\% confidence interval. In the text, when referring to Panel $(i,j)$, $i$ refers to the row and $j$ to the column of the figure. Each variable, in its own transformation, is demeaned at the country level. }
\end{figure}

\newpage
\begin{figure}
    \centering
    \includegraphics[scale=0.4]{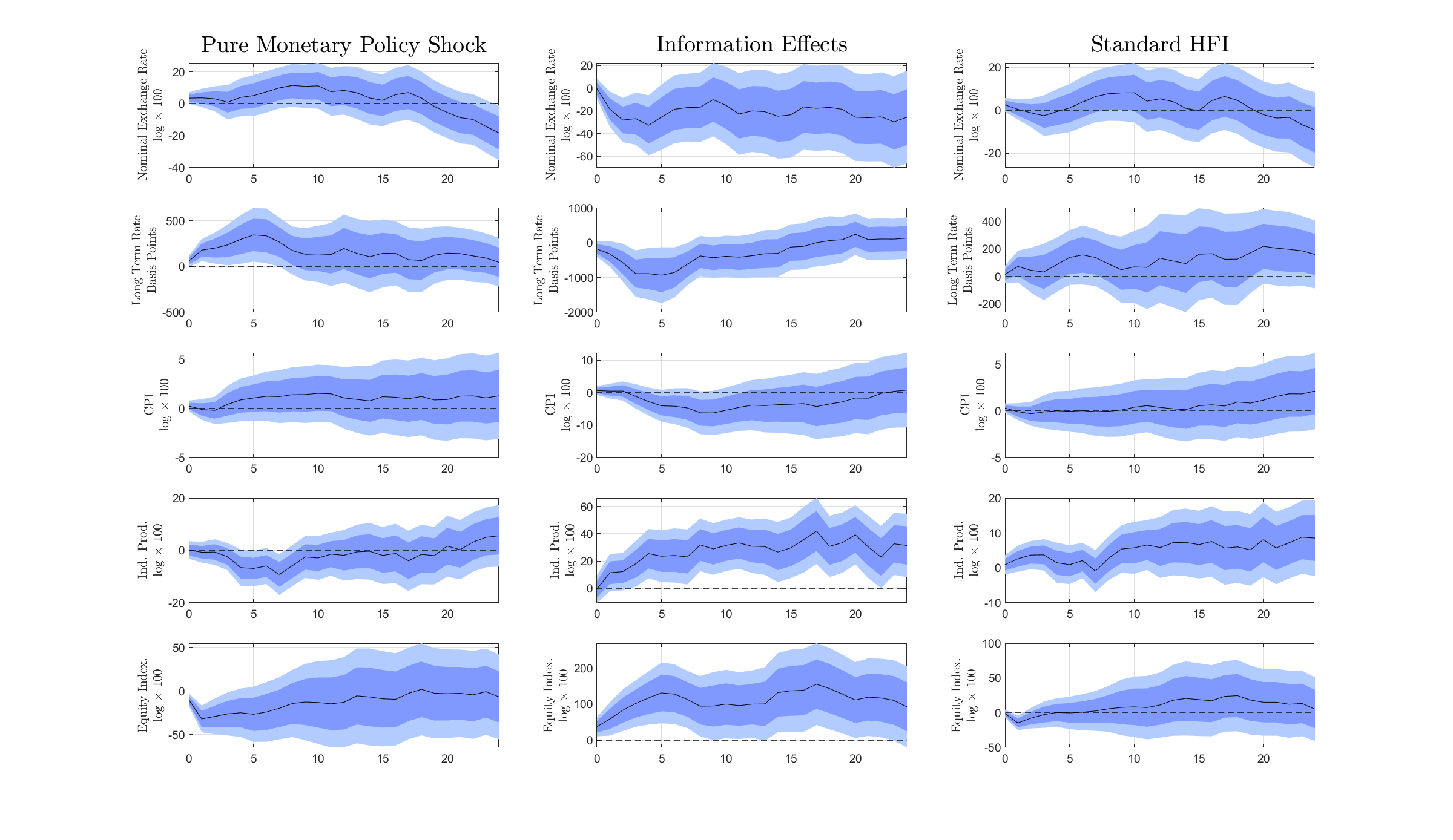}
    \caption{Impulse Response Functions \\ Median De Facto Peg \cite{ilzetzki2017country}}
    \label{fig:BenchmarkNER_Median_ERA1}
    \floatfoot{\textbf{Note:} The figure is comprised of 15 sub-figures ordered in three columns and five rows. The left column relates to the estimates of $\beta^{MP}$ in Equation \ref{eq:LP_pooled}, the middle column relates to the estimate of $\beta^{FIE}$ in Equation \ref{eq:LP_pooled}, while the right column relates to estimating Equation \ref{eq:LP_pooled}, replacing the MP and FIE components with the un-orthogonalized monetary policy surprise. This figure presents the results for countries with a median classification of 1 by \cite{ilzetzki2017country} in month $t$.  The rows represent the impact on (i) the nominal exchange rate with the US dollar (in logs times 100); (ii) long term interest rates in basis points; (iii) the consumer price index (in logs times 100); (iv) the industrial production index (in logs times 100); (v) the equity index (in logs times 100). The solid black line represents the point estimate, the dark blue area represents the 68\% confidence interval, and the light blue area represents the 90\% confidence interval. In the text, when referring to Panel $(i,j)$, $i$ refers to the row and $j$ to the column of the figure. Each variable, in its own transformation, is demeaned at the country level. }
\end{figure}

\newpage
\begin{figure}
    \centering
    \includegraphics[scale=0.4]{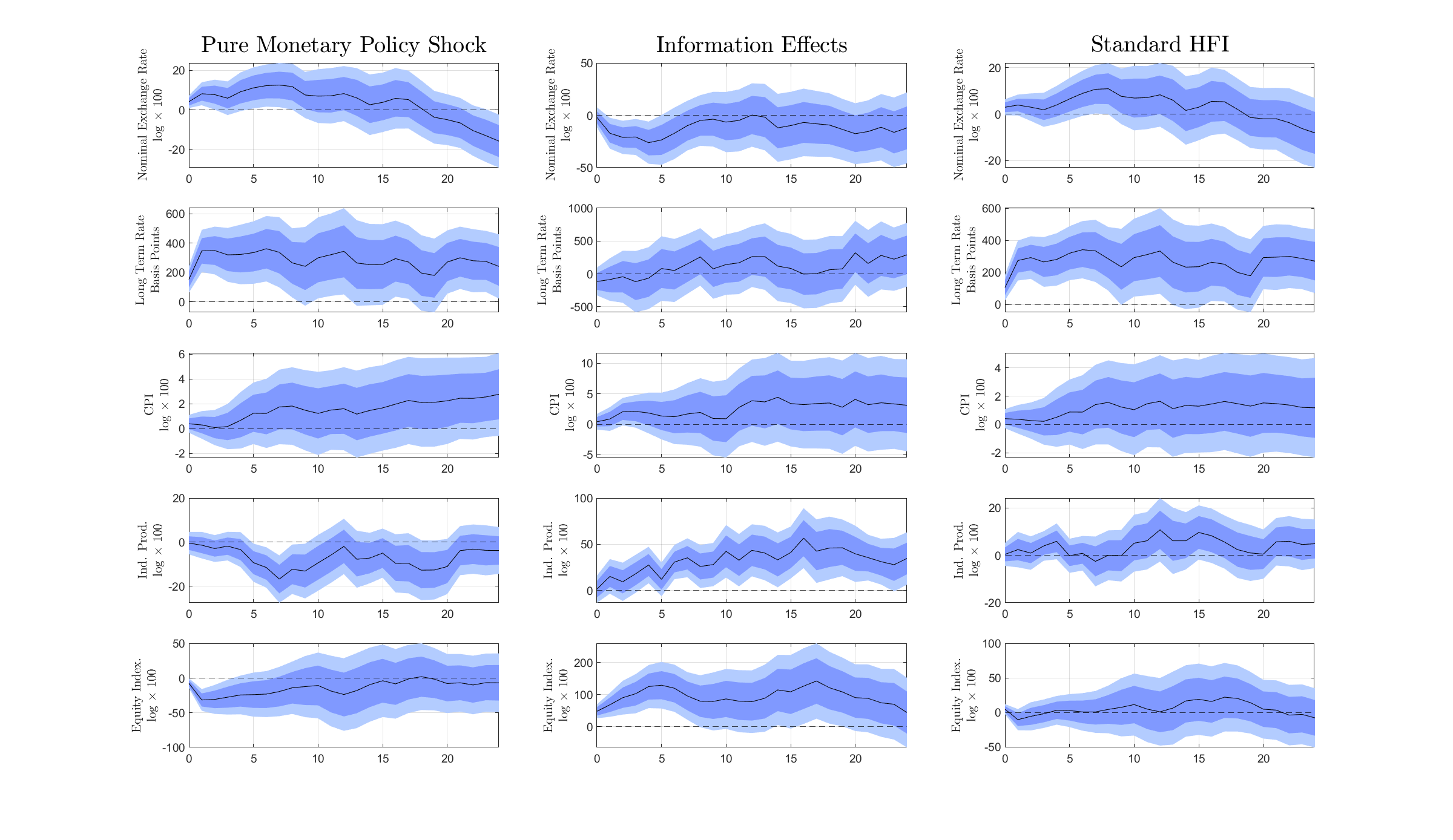}
    \caption{Impulse Response Functions \\ Median Pre-Announced Crawling Peg \cite{ilzetzki2017country}}
    \label{fig:BenchmarkNER_Median_ERA2}
    \floatfoot{\textbf{Note:} The figure is comprised of 15 sub-figures ordered in three columns and five rows. The left column relates to the estimates of $\beta^{MP}$ in Equation \ref{eq:LP_pooled}, the middle column relates to the estimate of $\beta^{FIE}$ in Equation \ref{eq:LP_pooled}, while the right column relates to estimating Equation \ref{eq:LP_pooled}, replacing the MP and FIE components with the un-orthogonalized monetary policy surprise. This figure presents the results for countries with a median classification of 2 by \cite{ilzetzki2017country} in month $t$.  The rows represent the impact on (i) the nominal exchange rate with the US dollar (in logs times 100); (ii) long term interest rates in basis points; (iii) the consumer price index (in logs times 100); (iv) the industrial production index (in logs times 100); (v) the equity index (in logs times 100). The solid black line represents the point estimate, the dark blue area represents the 68\% confidence interval, and the light blue area represents the 90\% confidence interval. In the text, when referring to Panel $(i,j)$, $i$ refers to the row and $j$ to the column of the figure. Each variable, in its own transformation, is demeaned at the country level. }
\end{figure}

\newpage
\begin{figure}
    \centering
    \includegraphics[scale=0.4]{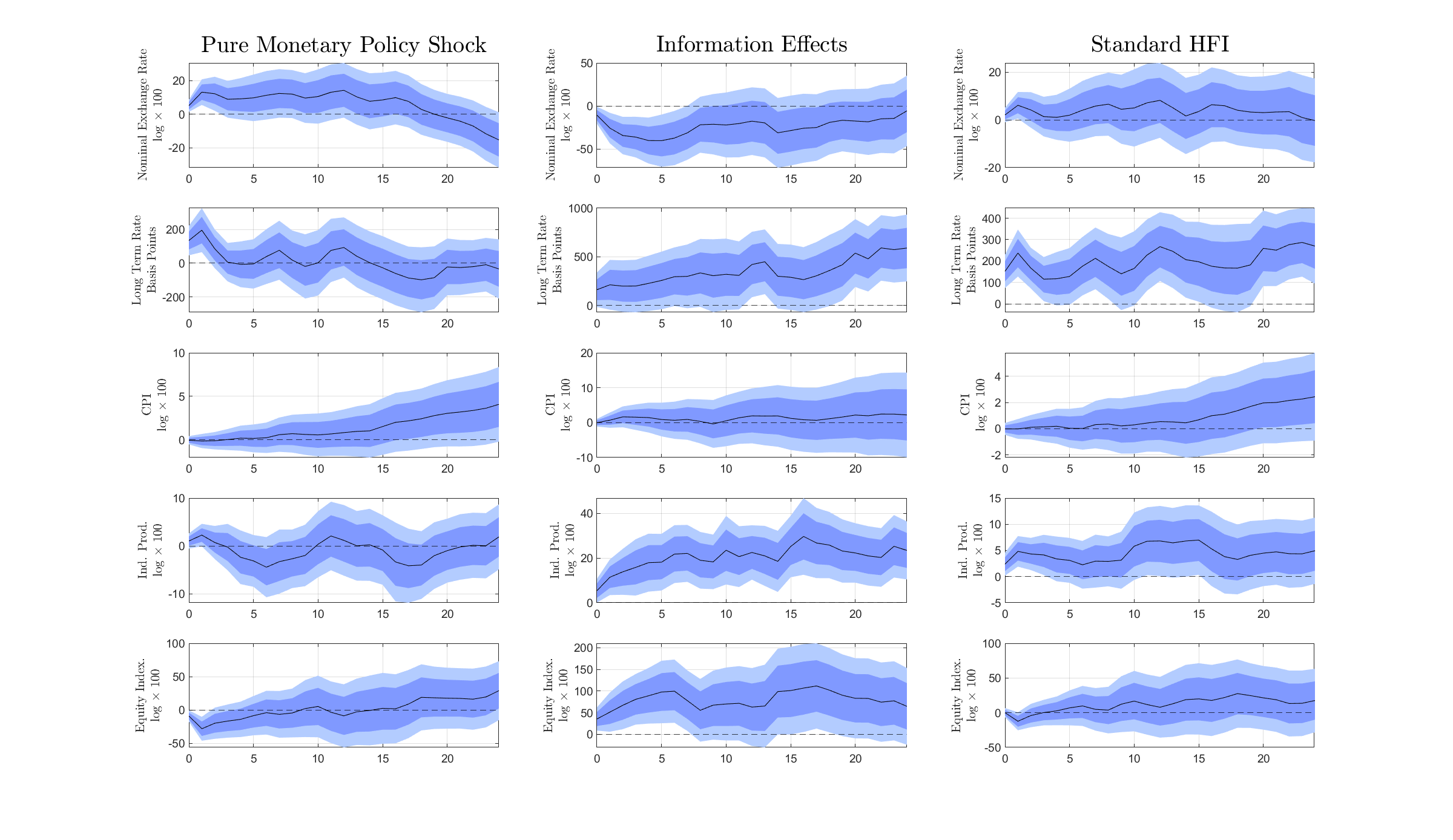}
    \caption{Impulse Response Functions \\ Median Managed Floating \cite{ilzetzki2017country}}
    \label{fig:BenchmarkNER_Median_ERA3}
    \floatfoot{\textbf{Note:} The figure is comprised of 15 sub-figures ordered in three columns and five rows. The left column relates to the estimates of $\beta^{MP}$ in Equation \ref{eq:LP_pooled}, the middle column relates to the estimate of $\beta^{FIE}$ in Equation \ref{eq:LP_pooled}, while the right column relates to estimating Equation \ref{eq:LP_pooled}, replacing the MP and FIE components with the un-orthogonalized monetary policy surprise. This figure presents the results for countries with a median classification of 3 by \cite{ilzetzki2017country} in month $t$.  The rows represent the impact on (i) the nominal exchange rate with the US dollar (in logs times 100); (ii) long term interest rates in basis points; (iii) the consumer price index (in logs times 100); (iv) the industrial production index (in logs times 100); (v) the equity index (in logs times 100). The solid black line represents the point estimate, the dark blue area represents the 68\% confidence interval, and the light blue area represents the 90\% confidence interval. In the text, when referring to Panel $(i,j)$, $i$ refers to the row and $j$ to the column of the figure. Each variable, in its own transformation, is demeaned at the country level. }
\end{figure}

\newpage
\begin{figure}
    \centering
    \includegraphics[scale=0.4]{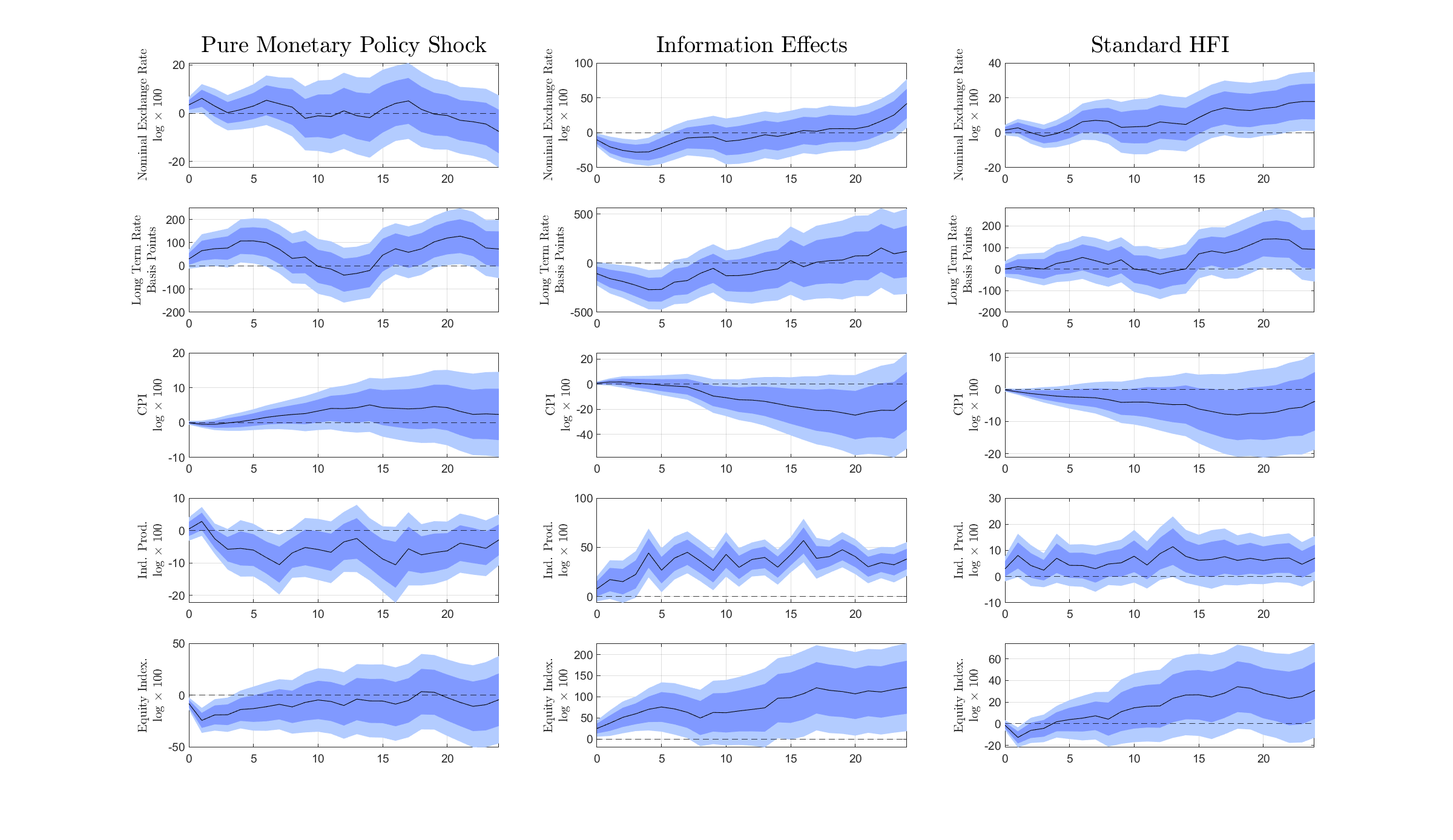}
    \caption{Impulse Response Functions \\ Median Freely Floating to Dual Market \cite{ilzetzki2017country}}
    \label{fig:BenchmarkNER_Median_ERA4}
    \floatfoot{\textbf{Note:} The figure is comprised of 15 sub-figures ordered in three columns and five rows. The left column relates to the estimates of $\beta^{MP}$ in Equation \ref{eq:LP_pooled}, the middle column relates to the estimate of $\beta^{FIE}$ in Equation \ref{eq:LP_pooled}, while the right column relates to estimating Equation \ref{eq:LP_pooled}, replacing the MP and FIE components with the un-orthogonalized monetary policy surprise. This figure presents the results for countries with a median classification of 4 or higher by \cite{ilzetzki2017country} in month $t$.  The rows represent the impact on (i) the nominal exchange rate with the US dollar (in logs times 100); (ii) long term interest rates in basis points; (iii) the consumer price index (in logs times 100); (iv) the industrial production index (in logs times 100); (v) the equity index (in logs times 100). The solid black line represents the point estimate, the dark blue area represents the 68\% confidence interval, and the light blue area represents the 90\% confidence interval. In the text, when referring to Panel $(i,j)$, $i$ refers to the row and $j$ to the column of the figure. Each variable, in its own transformation, is demeaned at the country level. }
\end{figure}

\newpage
\begin{figure}
    \centering
    \includegraphics[scale=0.4]{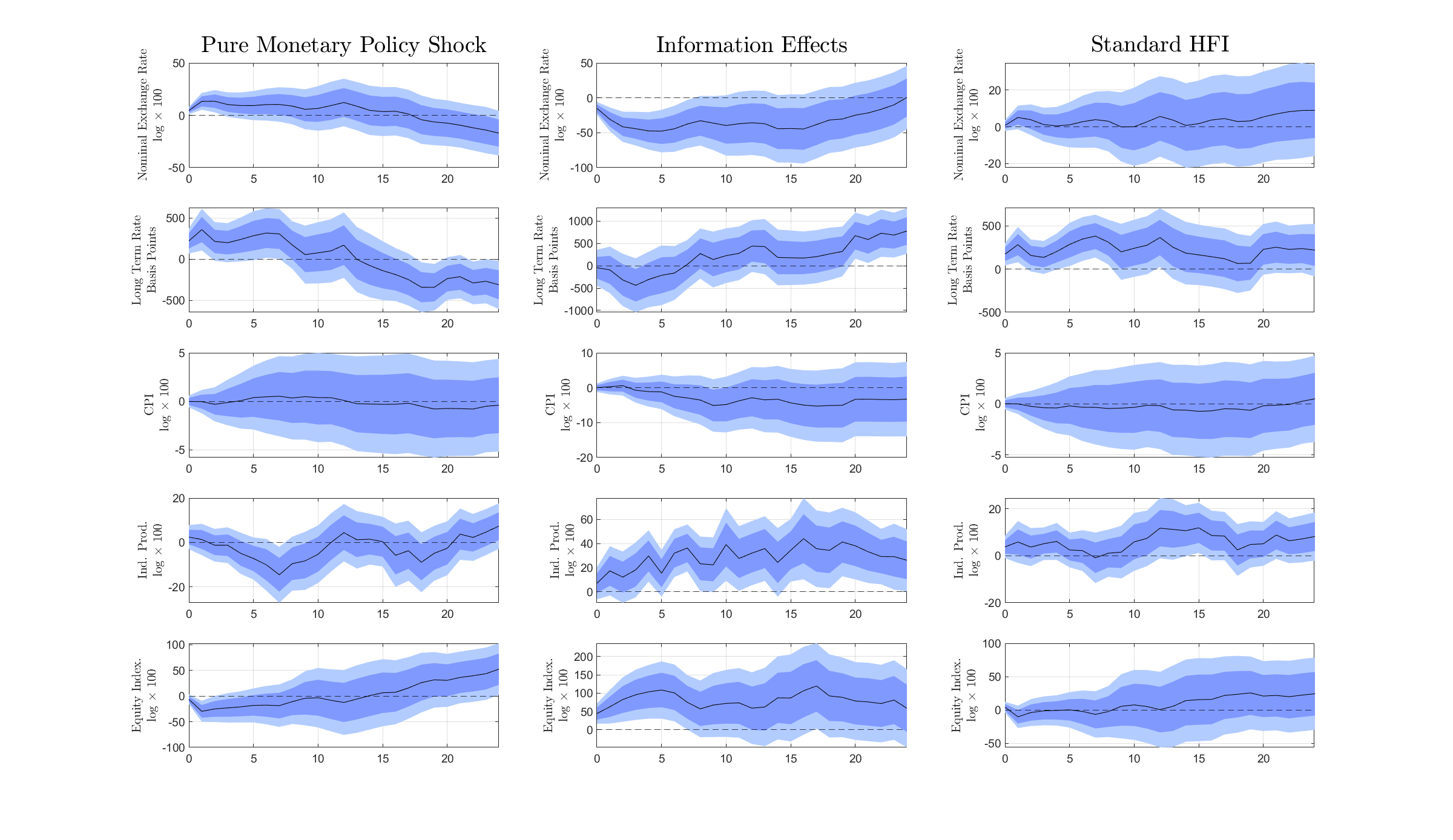}
    \caption{Impulse Response Functions \\ Group 1 Current Account Openness  \cite{chinn2008new}}
    \label{fig:BenchmarkNER_KA1}
    \floatfoot{\textbf{Note:} The figure is comprised of 15 sub-figures ordered in three columns and five rows. The left column relates to the estimates of $\beta^{MP}$ in Equation \ref{eq:LP_pooled}, the middle column relates to the estimate of $\beta^{FIE}$ in Equation \ref{eq:LP_pooled}, while the right column relates to estimating Equation \ref{eq:LP_pooled}, replacing the MP and FIE components with the un-orthogonalized monetary policy surprise. This figure presents the results for the group of countries in the lowest quartile of mean current account openness in our sample, based on \cite{chinn2008new}. The rows represent the impact on (i) the nominal exchange rate with the US dollar (in logs times 100); (ii) long term interest rates in basis points; (iii) the consumer price index (in logs times 100); (iv) the industrial production index (in logs times 100); (v) the equity index (in logs times 100). The solid black line represents the point estimate, the dark blue area represents the 68\% confidence interval, and the light blue area represents the 90\% confidence interval. In the text, when referring to Panel $(i,j)$, $i$ refers to the row and $j$ to the column of the figure. Each variable, in its own transformation, is demeaned at the country level. }
\end{figure}

\newpage
\begin{figure}
    \centering
    \includegraphics[scale=0.4]{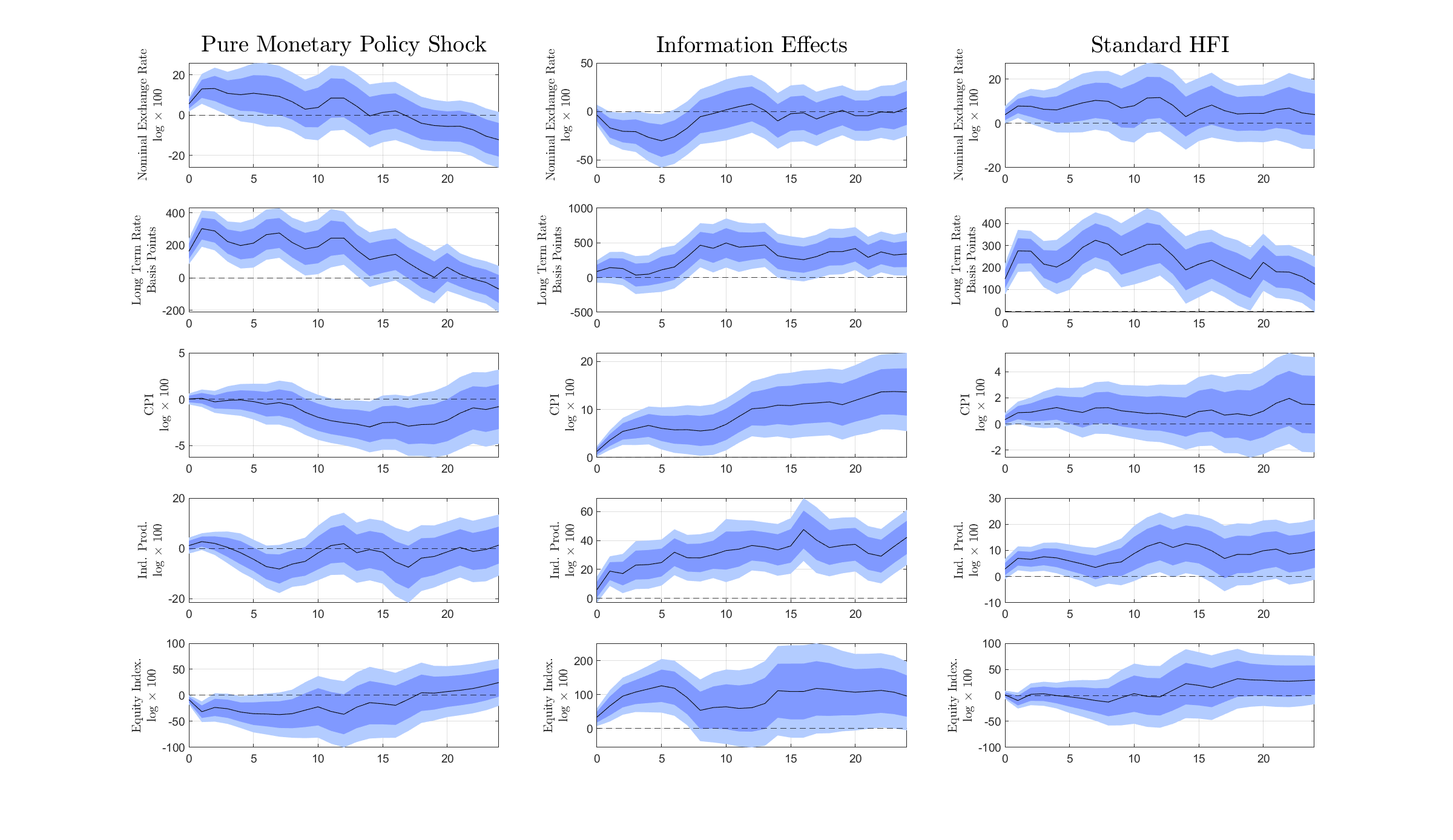}
    \caption{Impulse Response Functions \\ Group 2 Current Account Openness  \cite{chinn2008new}}
    \label{fig:BenchmarkNER_KA2}
    \floatfoot{\textbf{Note:} The figure is comprised of 15 sub-figures ordered in three columns and five rows. The left column relates to the estimates of $\beta^{MP}$ in Equation \ref{eq:LP_pooled}, the middle column relates to the estimate of $\beta^{FIE}$ in Equation \ref{eq:LP_pooled}, while the right column relates to estimating Equation \ref{eq:LP_pooled}, replacing the MP and FIE components with the un-orthogonalized monetary policy surprise. This figure presents the results for the group of countries in the second lowest quartile of mean current account openness in our sample, based on \cite{chinn2008new}. The rows represent the impact on (i) the nominal exchange rate with the US dollar (in logs times 100); (ii) long term interest rates in basis points; (iii) the consumer price index (in logs times 100); (iv) the industrial production index (in logs times 100); (v) the equity index (in logs times 100). The solid black line represents the point estimate, the dark blue area represents the 68\% confidence interval, and the light blue area represents the 90\% confidence interval. In the text, when referring to Panel $(i,j)$, $i$ refers to the row and $j$ to the column of the figure. Each variable, in its own transformation, is demeaned at the country level. }
\end{figure}

\newpage
\begin{figure}
    \centering
    \includegraphics[scale=0.4]{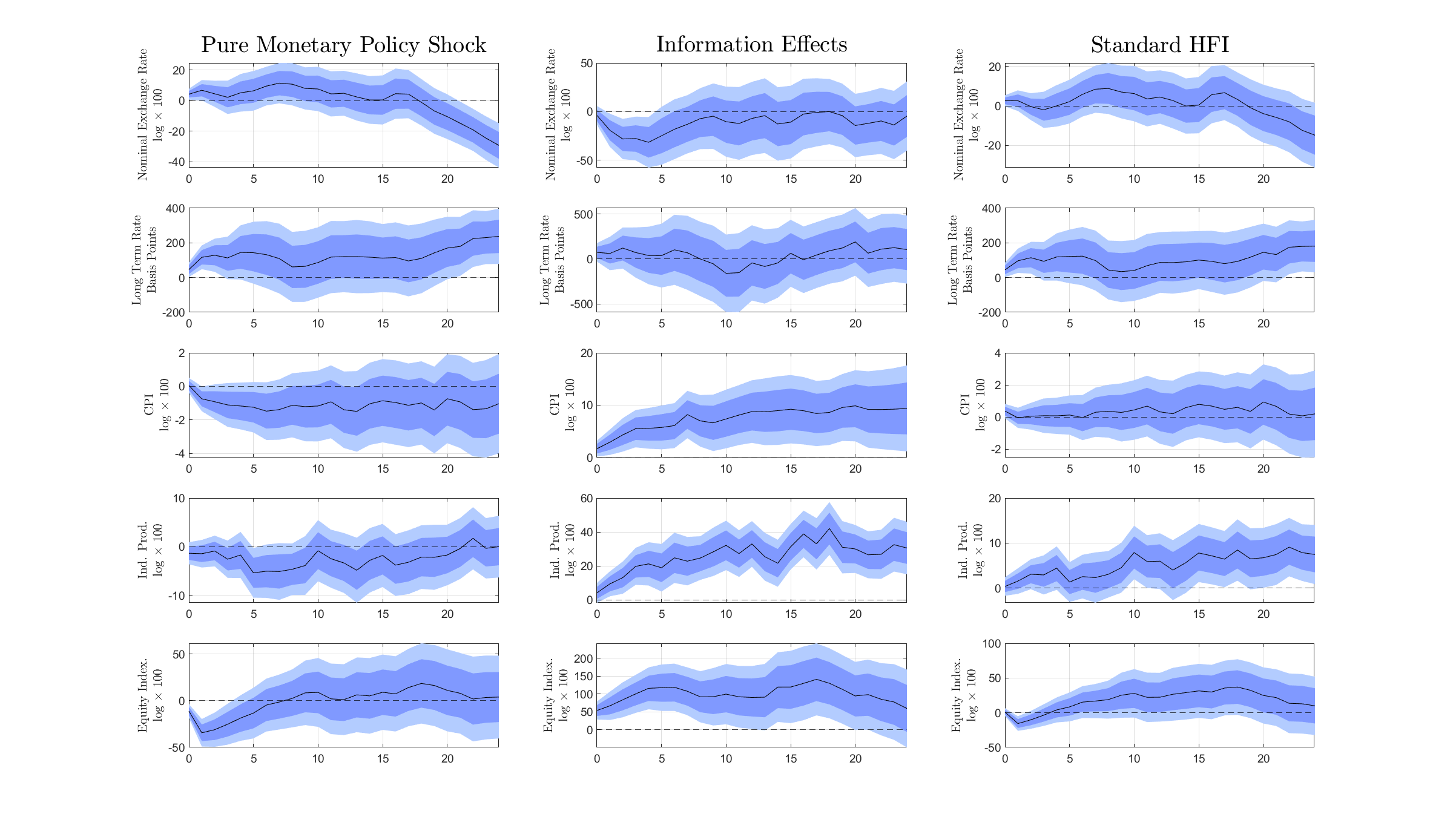}
    \caption{Impulse Response Functions \\ Group 3 Current Account Openness \cite{chinn2008new}}
    \label{fig:BenchmarkNER_KA3}
    \floatfoot{\textbf{Note:} The figure is comprised of 15 sub-figures ordered in three columns and five rows. The left column relates to the estimates of $\beta^{MP}$ in Equation \ref{eq:LP_pooled}, the middle column relates to the estimate of $\beta^{FIE}$ in Equation \ref{eq:LP_pooled}, while the right column relates to estimating Equation \ref{eq:LP_pooled}, replacing the MP and FIE components with the un-orthogonalized monetary policy surprise. This figure presents the results for the group of countries in the second highest quartile of mean current account openness in our sample, based on \cite{chinn2008new}. The rows represent the impact on (i) the nominal exchange rate with the US dollar (in logs times 100); (ii) long term interest rates in basis points; (iii) the consumer price index (in logs times 100); (iv) the industrial production index (in logs times 100); (v) the equity index (in logs times 100). The solid black line represents the point estimate, the dark blue area represents the 68\% confidence interval, and the light blue area represents the 90\% confidence interval. In the text, when referring to Panel $(i,j)$, $i$ refers to the row and $j$ to the column of the figure. Each variable, in its own transformation, is demeaned at the country level. }
\end{figure}

\newpage
\begin{figure}
    \centering
    \includegraphics[scale=0.4]{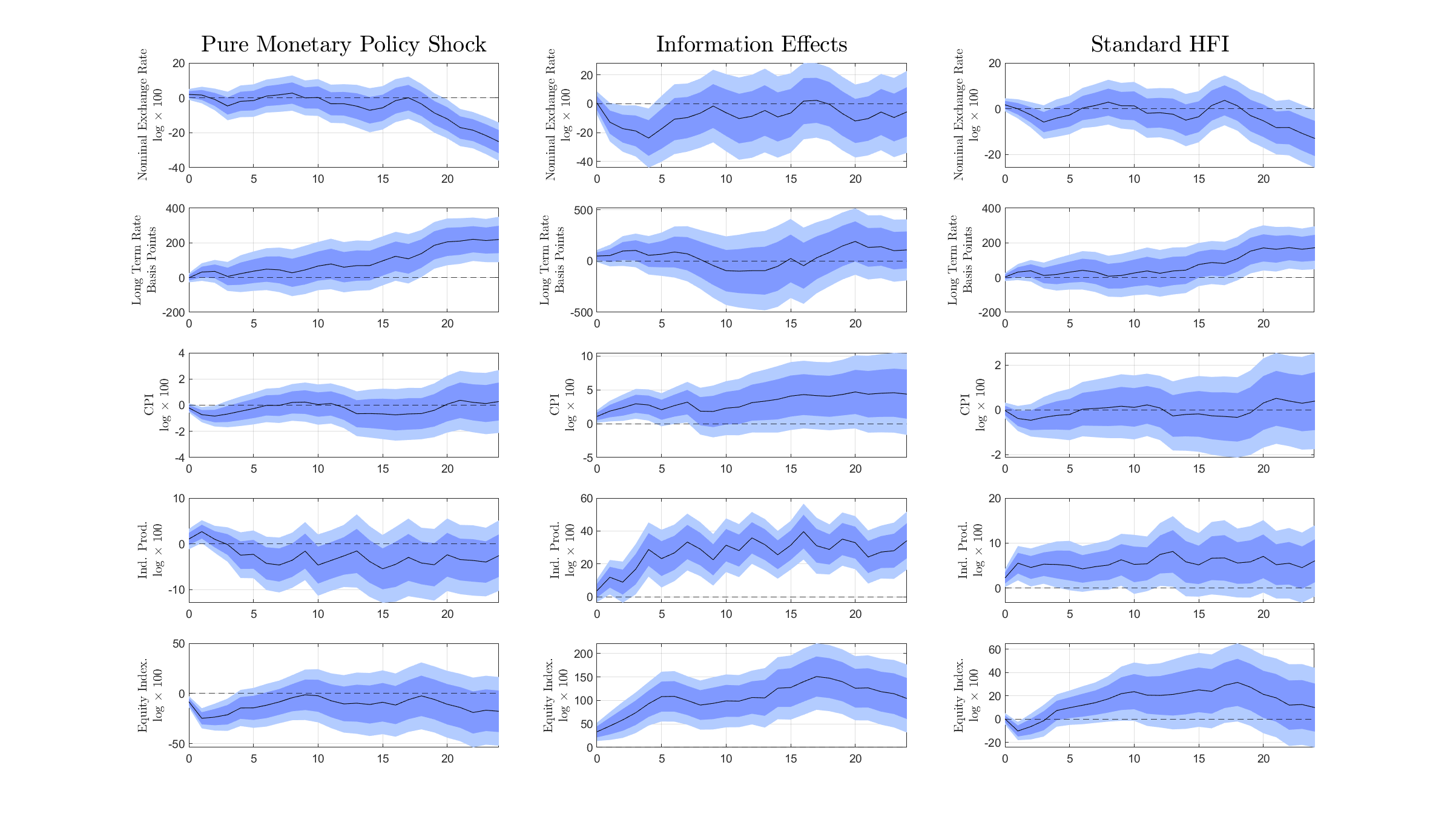}
    \caption{Impulse Response Functions \\ Group 4 Current Account Openness \cite{chinn2008new}}
    \label{fig:BenchmarkNER_KA4}
    \floatfoot{\textbf{Note:} The figure is comprised of 15 sub-figures ordered in three columns and five rows. The left column relates to the estimates of $\beta^{MP}$ in Equation \ref{eq:LP_pooled}, the middle column relates to the estimate of $\beta^{FIE}$ in Equation \ref{eq:LP_pooled}, while the right column relates to estimating Equation \ref{eq:LP_pooled}, replacing the MP and FIE components with the un-orthogonalized monetary policy surprise. This figure presents the results for the group of countries in the highest quartile of mean current account openness in our sample, based on \cite{chinn2008new}. The rows represent the impact on (i) the nominal exchange rate with the US dollar (in logs times 100); (ii) long term interest rates in basis points; (iii) the consumer price index (in logs times 100); (iv) the industrial production index (in logs times 100); (v) the equity index (in logs times 100). The solid black line represents the point estimate, the dark blue area represents the 68\% confidence interval, and the light blue area represents the 90\% confidence interval. In the text, when referring to Panel $(i,j)$, $i$ refers to the row and $j$ to the column of the figure. Each variable, in its own transformation, is demeaned at the country level. }
\end{figure}

\newpage
\begin{figure}
    \centering
    \includegraphics[scale=0.4]{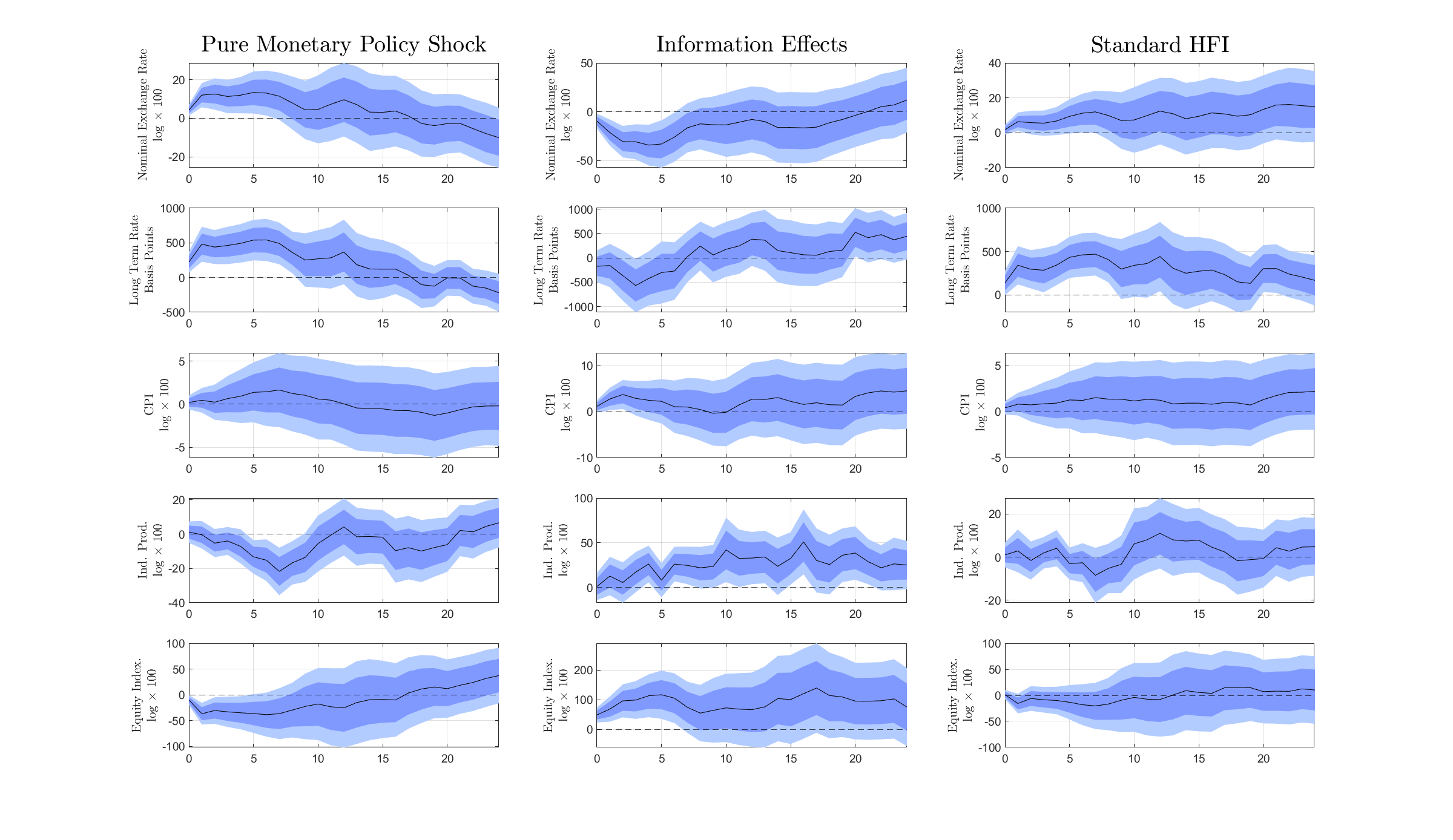}
    \caption{Impulse Response Functions \\ Lowest Level of GDP per Capita}
    \label{fig:BenchmarkNER_GDPpc_1}
    \floatfoot{\textbf{Note:} The figure is comprised of 15 sub-figures ordered in three columns and five rows. The left column relates to the estimates of $\beta^{MP}$ in Equation \ref{eq:LP_pooled}, the middle column relates to the estimate of $\beta^{FIE}$ in Equation \ref{eq:LP_pooled}, while the right column relates to estimating Equation \ref{eq:LP_pooled}, replacing the MP and FIE components with the un-orthogonalized monetary policy surprise. This figure presents the results for the group of countries in the lowest quartile of GDP per Capita in the sample for the year 2004. The rows represent the impact on (i) the nominal exchange rate with the US dollar (in logs times 100); (ii) long term interest rates in basis points; (iii) the consumer price index (in logs times 100); (iv) the industrial production index (in logs times 100); (v) the equity index (in logs times 100). The solid black line represents the point estimate, the dark blue area represents the 68\% confidence interval, and the light blue area represents the 90\% confidence interval. In the text, when referring to Panel $(i,j)$, $i$ refers to the row and $j$ to the column of the figure. Each variable, in its own transformation, is demeaned at the country level. }
\end{figure}

\newpage
\begin{figure}
    \centering
    \includegraphics[scale=0.4]{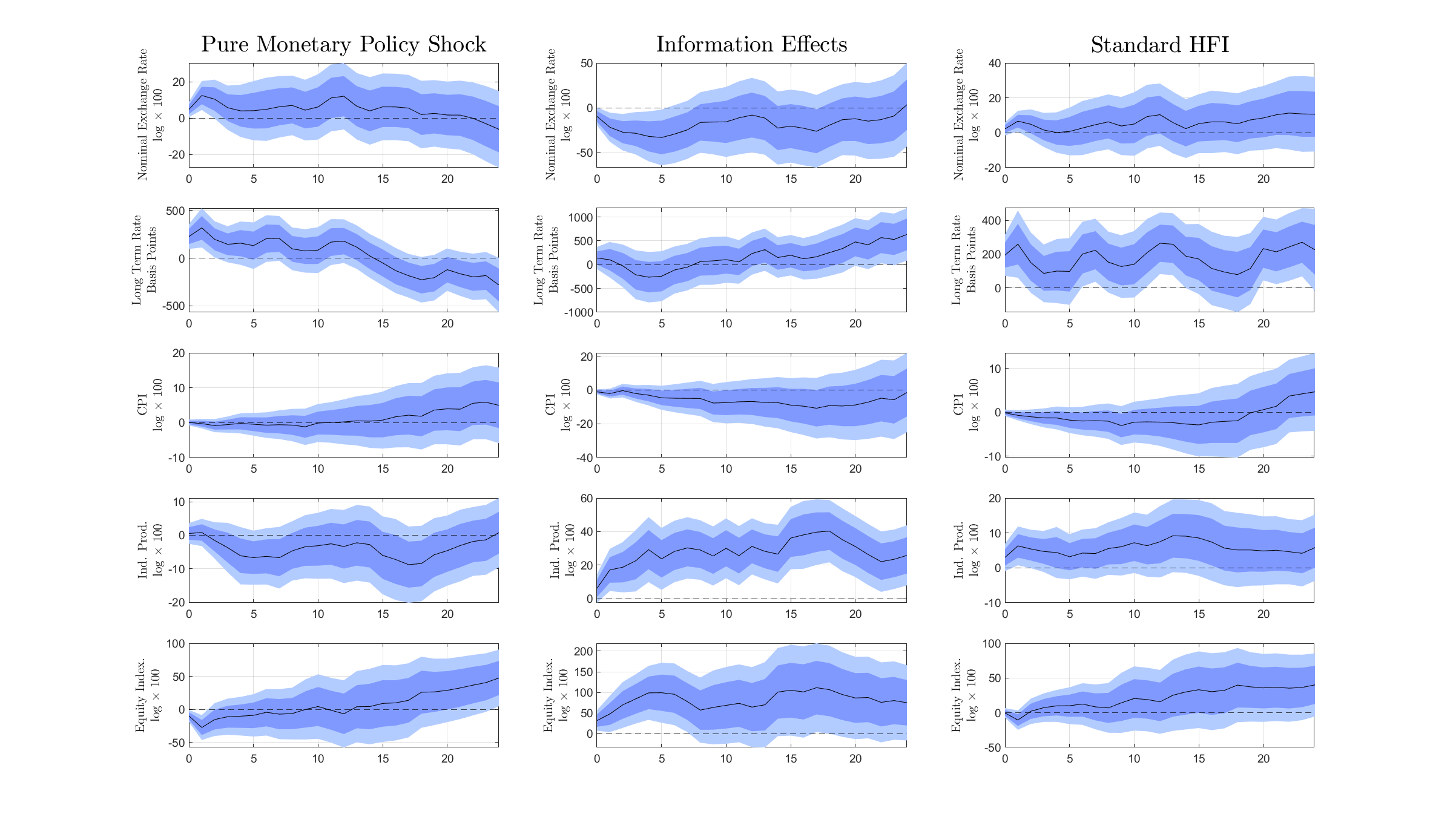}
    \caption{Impulse Response Functions \\ Second Lowest Level of GDP per Capita}
    \label{fig:BenchmarkNER_GDPpc_2}
    \floatfoot{\textbf{Note:} The figure is comprised of 15 sub-figures ordered in three columns and five rows. The left column relates to the estimates of $\beta^{MP}$ in Equation \ref{eq:LP_pooled}, the middle column relates to the estimate of $\beta^{FIE}$ in Equation \ref{eq:LP_pooled}, while the right column relates to estimating Equation \ref{eq:LP_pooled}, replacing the MP and FIE components with the un-orthogonalized monetary policy surprise. This figure presents the results for the group of countries in the second lowest quartile of GDP per Capita in the sample for the year 2004. The rows represent the impact on (i) the nominal exchange rate with the US dollar (in logs times 100); (ii) long term interest rates in basis points; (iii) the consumer price index (in logs times 100); (iv) the industrial production index (in logs times 100); (v) the equity index (in logs times 100). The solid black line represents the point estimate, the dark blue area represents the 68\% confidence interval, and the light blue area represents the 90\% confidence interval. In the text, when referring to Panel $(i,j)$, $i$ refers to the row and $j$ to the column of the figure. Each variable, in its own transformation, is demeaned at the country level. }
\end{figure}

\newpage
\begin{figure}
    \centering
    \includegraphics[scale=0.4]{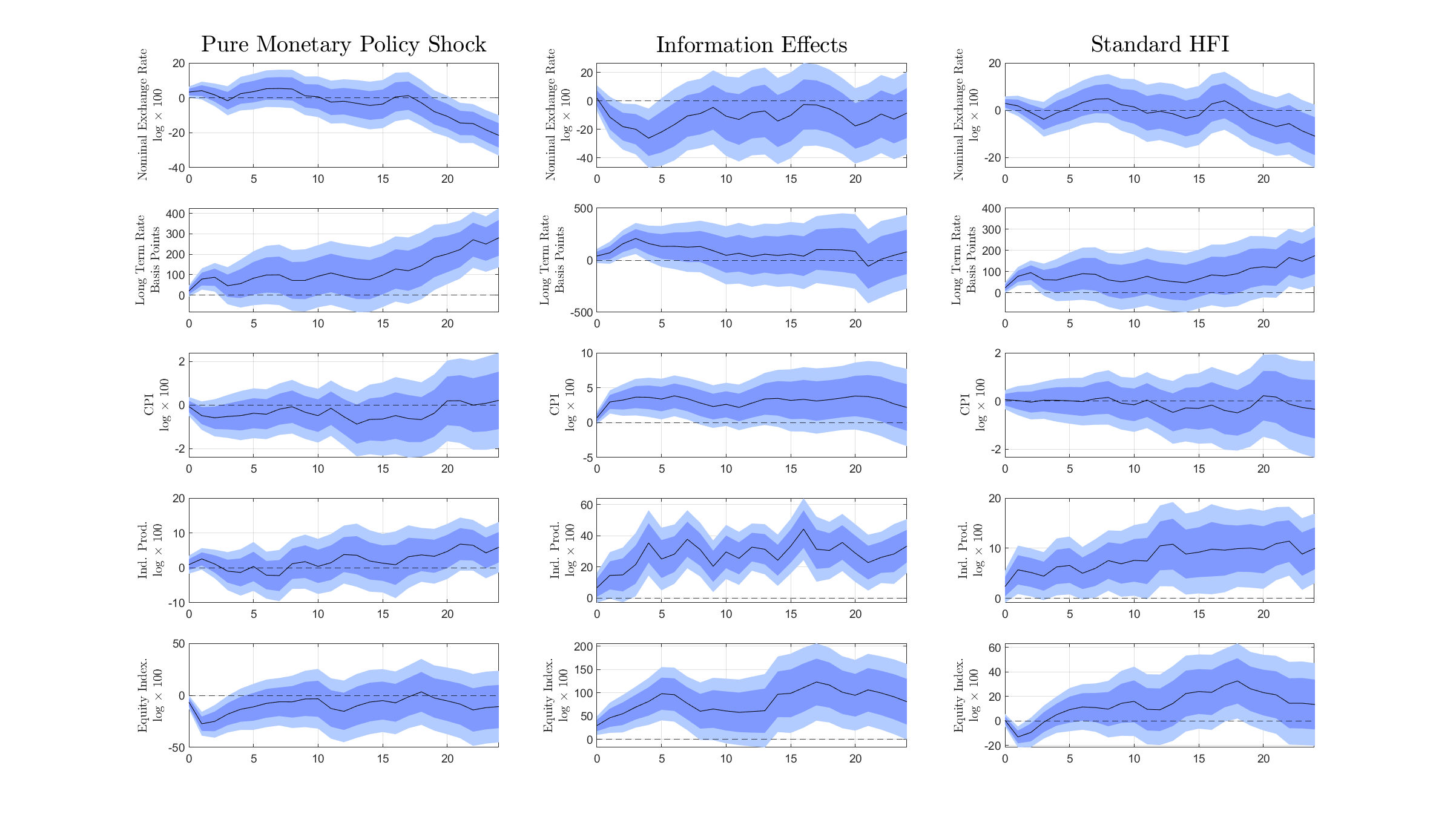}
    \caption{Impulse Response Functions \\ Second Highest Level of GDP per Capita}
    \label{fig:BenchmarkNER_GDPpc_3}
    \floatfoot{\textbf{Note:} The figure is comprised of 15 sub-figures ordered in three columns and five rows. The left column relates to the estimates of $\beta^{MP}$ in Equation \ref{eq:LP_pooled}, the middle column relates to the estimate of $\beta^{FIE}$ in Equation \ref{eq:LP_pooled}, while the right column relates to estimating Equation \ref{eq:LP_pooled}, replacing the MP and FIE components with the un-orthogonalized monetary policy surprise. This figure presents the results for the group of countries in the second highest quartile of GDP per Capita in the sample for the year 2004. The rows represent the impact on (i) the nominal exchange rate with the US dollar (in logs times 100); (ii) long term interest rates in basis points; (iii) the consumer price index (in logs times 100); (iv) the industrial production index (in logs times 100); (v) the equity index (in logs times 100). The solid black line represents the point estimate, the dark blue area represents the 68\% confidence interval, and the light blue area represents the 90\% confidence interval. In the text, when referring to Panel $(i,j)$, $i$ refers to the row and $j$ to the column of the figure. Each variable, in its own transformation, is demeaned at the country level. }
\end{figure}

\newpage
\begin{figure}
    \centering
    \includegraphics[scale=0.4]{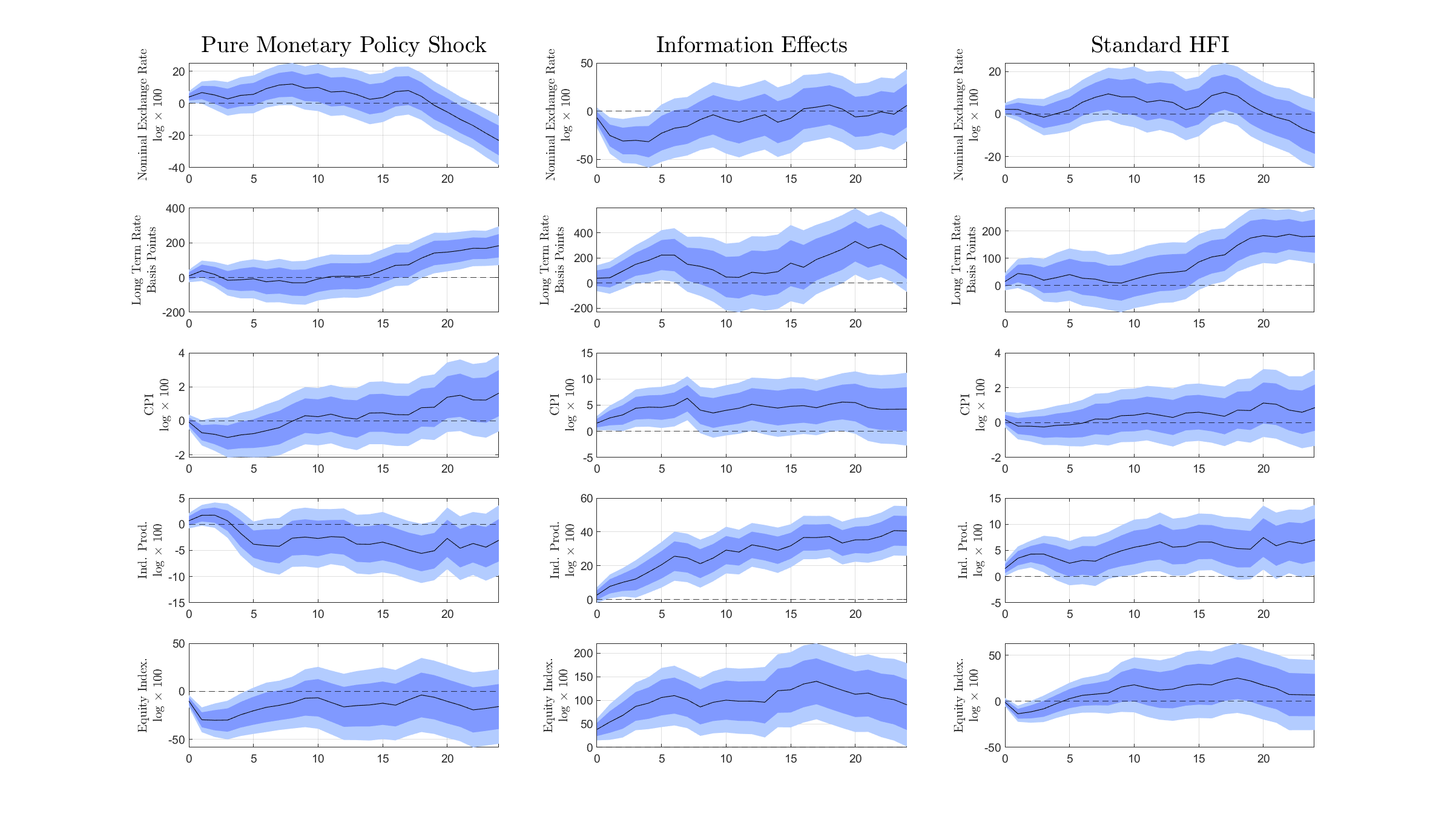}
    \caption{Impulse Response Functions \\ Second Highest Level of GDP per Capita}
    \label{fig:BenchmarkNER_GDPpc_4}
    \floatfoot{\textbf{Note:} The figure is comprised of 15 sub-figures ordered in three columns and five rows. The left column relates to the estimates of $\beta^{MP}$ in Equation \ref{eq:LP_pooled}, the middle column relates to the estimate of $\beta^{FIE}$ in Equation \ref{eq:LP_pooled}, while the right column relates to estimating Equation \ref{eq:LP_pooled}, replacing the MP and FIE components with the un-orthogonalized monetary policy surprise. This figure presents the results for the group of countries in the highest quartile of GDP per Capita in the sample for the year 2004. The rows represent the impact on (i) the nominal exchange rate with the US dollar (in logs times 100); (ii) long term interest rates in basis points; (iii) the consumer price index (in logs times 100); (iv) the industrial production index (in logs times 100); (v) the equity index (in logs times 100). The solid black line represents the point estimate, the dark blue area represents the 68\% confidence interval, and the light blue area represents the 90\% confidence interval. In the text, when referring to Panel $(i,j)$, $i$ refers to the row and $j$ to the column of the figure. Each variable, in its own transformation, is demeaned at the country level. }
\end{figure}

\newpage
\begin{figure}
    \centering
    \includegraphics[scale=0.4]{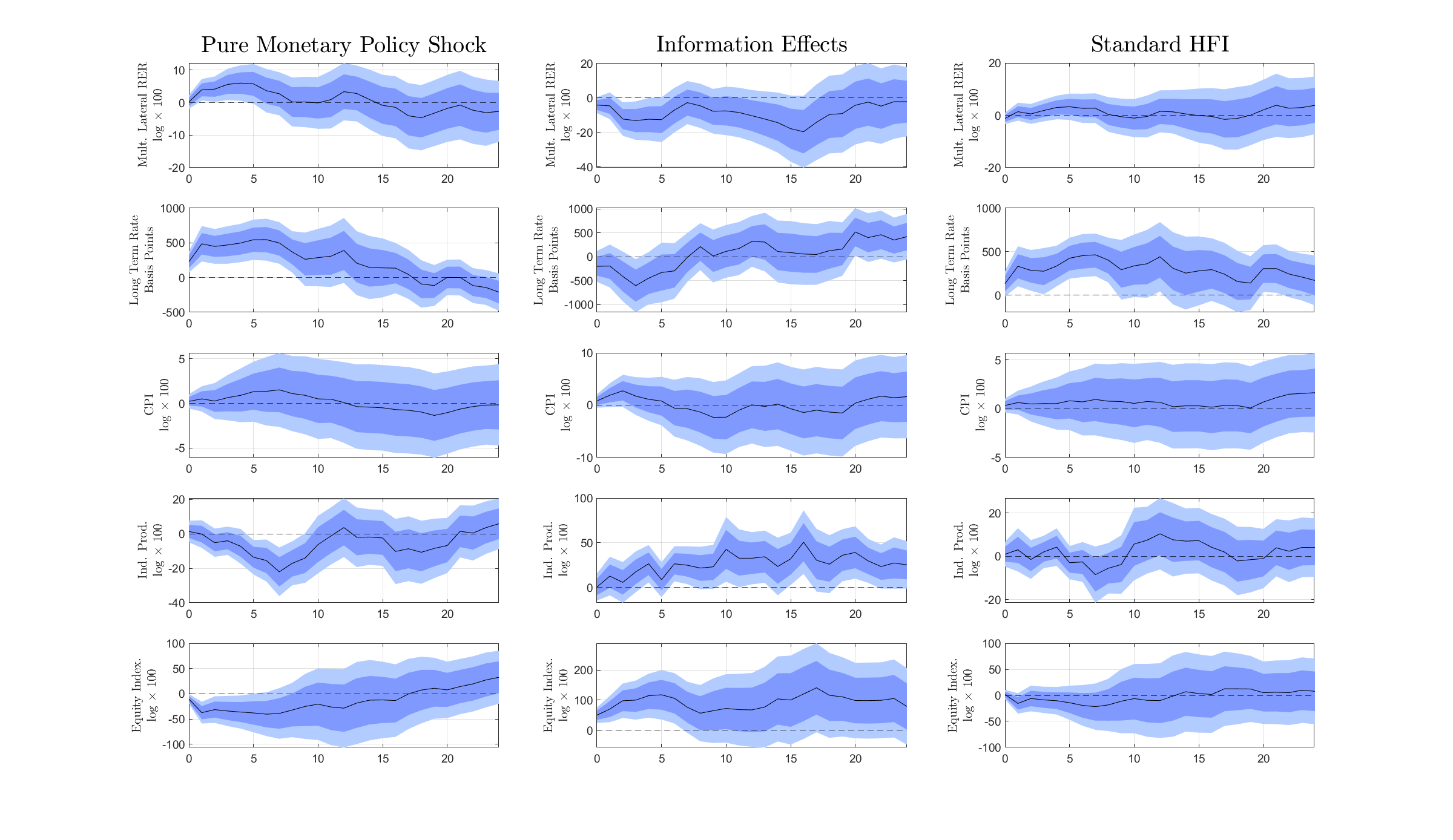}
    \caption{Impulse Response Functions \\ Multi. REER Sample - Lowest Level of GDP per Capita}
    \label{fig:BenchmarkREER_GDPpc_1}
    \floatfoot{\textbf{Note:} The figure is comprised of 15 sub-figures ordered in three columns and five rows. The left column relates to the estimates of $\beta^{MP}$ in Equation \ref{eq:LP_pooled}, the middle column relates to the estimate of $\beta^{FIE}$ in Equation \ref{eq:LP_pooled}, while the right column relates to estimating Equation \ref{eq:LP_pooled}, replacing the MP and FIE components with the un-orthogonalized monetary policy surprise. This figure presents the results for the group of countries in the lowest quartile of GDP per Capita in the sample for the year 2004, replacing the nominal exchange rate with the US dollar with the multilateral trade weighted real exchange rate. The rows represent the impact on (i) the nominal exchange rate with the US dollar (in logs times 100); (ii) long term interest rates in basis points; (iii) the consumer price index (in logs times 100); (iv) the industrial production index (in logs times 100); (v) the equity index (in logs times 100). The solid black line represents the point estimate, the dark blue area represents the 68\% confidence interval, and the light blue area represents the 90\% confidence interval. In the text, when referring to Panel $(i,j)$, $i$ refers to the row and $j$ to the column of the figure. Each variable, in its own transformation, is demeaned at the country level. }
\end{figure}

\newpage
\begin{figure}
    \centering
    \includegraphics[scale=0.4]{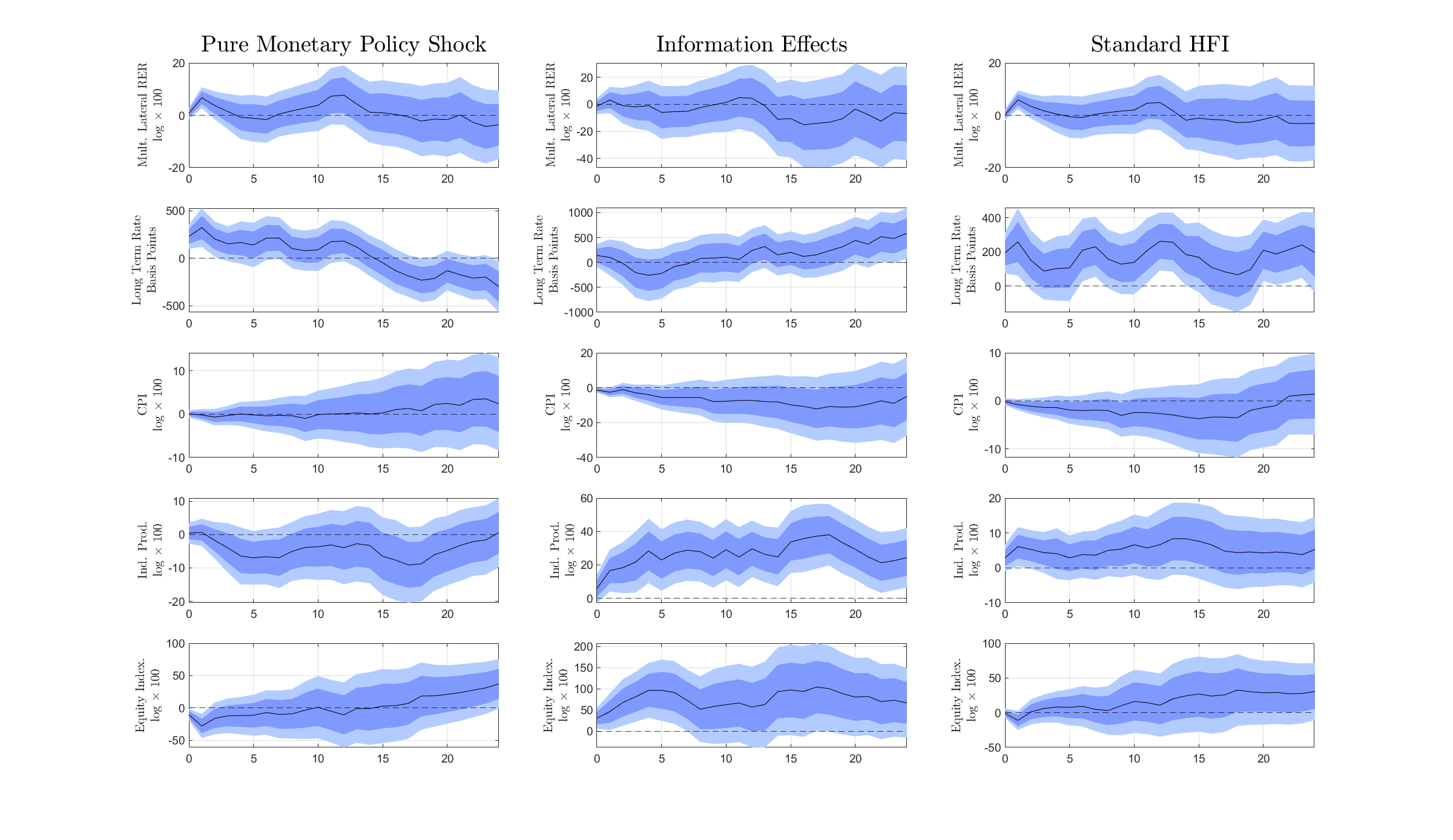}
    \caption{Impulse Response Functions \\ Multi. REER Sample - Second Lowest Level of GDP per Capita}
    \label{fig:BenchmarkREER_GDPpc_2}
    \floatfoot{\textbf{Note:} The figure is comprised of 15 sub-figures ordered in three columns and five rows. The left column relates to the estimates of $\beta^{MP}$ in Equation \ref{eq:LP_pooled}, the middle column relates to the estimate of $\beta^{FIE}$ in Equation \ref{eq:LP_pooled}, while the right column relates to estimating Equation \ref{eq:LP_pooled}, replacing the MP and FIE components with the un-orthogonalized monetary policy surprise. This figure presents the results for the group of countries in the second lowest quartile of GDP per Capita in the sample for the year 2004, replacing the nominal exchange rate with the US dollar with the multilateral trade weighted real exchange rate. The rows represent the impact on (i) the nominal exchange rate with the US dollar (in logs times 100); (ii) long term interest rates in basis points; (iii) the consumer price index (in logs times 100); (iv) the industrial production index (in logs times 100); (v) the equity index (in logs times 100). The solid black line represents the point estimate, the dark blue area represents the 68\% confidence interval, and the light blue area represents the 90\% confidence interval. In the text, when referring to Panel $(i,j)$, $i$ refers to the row and $j$ to the column of the figure. Each variable, in its own transformation, is demeaned at the country level. }
\end{figure}

\newpage
\begin{figure}
    \centering
    \includegraphics[scale=0.4]{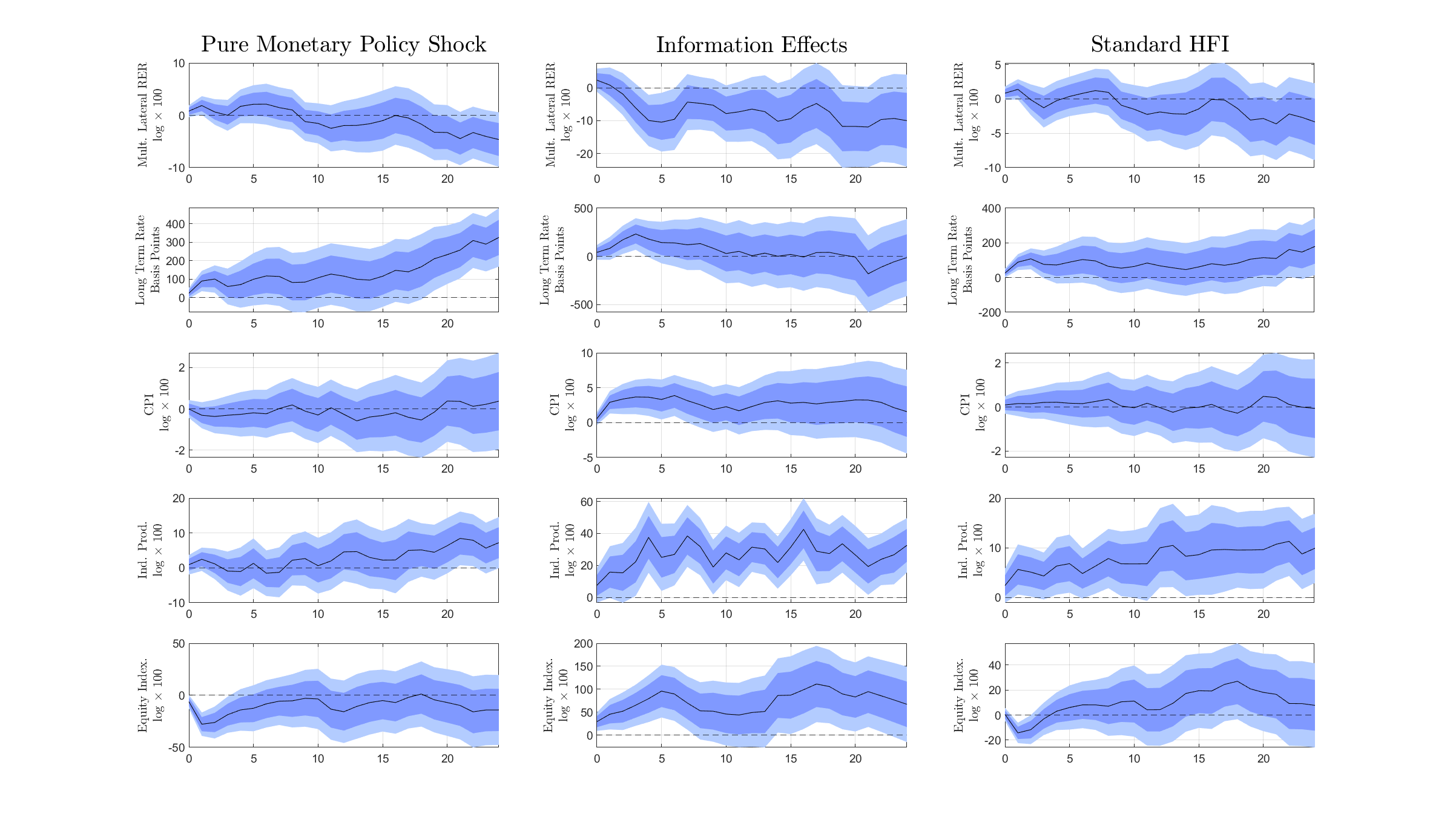}
    \caption{Impulse Response Functions \\ Multi. REER Sample - Second Highest Level of GDP per Capita}
    \label{fig:BenchmarkREER_GDPpc_3}
    \floatfoot{\textbf{Note:} The figure is comprised of 15 sub-figures ordered in three columns and five rows. The left column relates to the estimates of $\beta^{MP}$ in Equation \ref{eq:LP_pooled}, the middle column relates to the estimate of $\beta^{FIE}$ in Equation \ref{eq:LP_pooled}, while the right column relates to estimating Equation \ref{eq:LP_pooled}, replacing the MP and FIE components with the un-orthogonalized monetary policy surprise. This figure presents the results for the group of countries in the second highest quartile of GDP per Capita in the sample for the year 2004, replacing the nominal exchange rate with the US dollar with the multilateral trade weighted real exchange rate. The rows represent the impact on (i) the nominal exchange rate with the US dollar (in logs times 100); (ii) long term interest rates in basis points; (iii) the consumer price index (in logs times 100); (iv) the industrial production index (in logs times 100); (v) the equity index (in logs times 100). The solid black line represents the point estimate, the dark blue area represents the 68\% confidence interval, and the light blue area represents the 90\% confidence interval. In the text, when referring to Panel $(i,j)$, $i$ refers to the row and $j$ to the column of the figure. Each variable, in its own transformation, is demeaned at the country level. }
\end{figure}

\newpage
\begin{figure}
    \centering
    \includegraphics[scale=0.4]{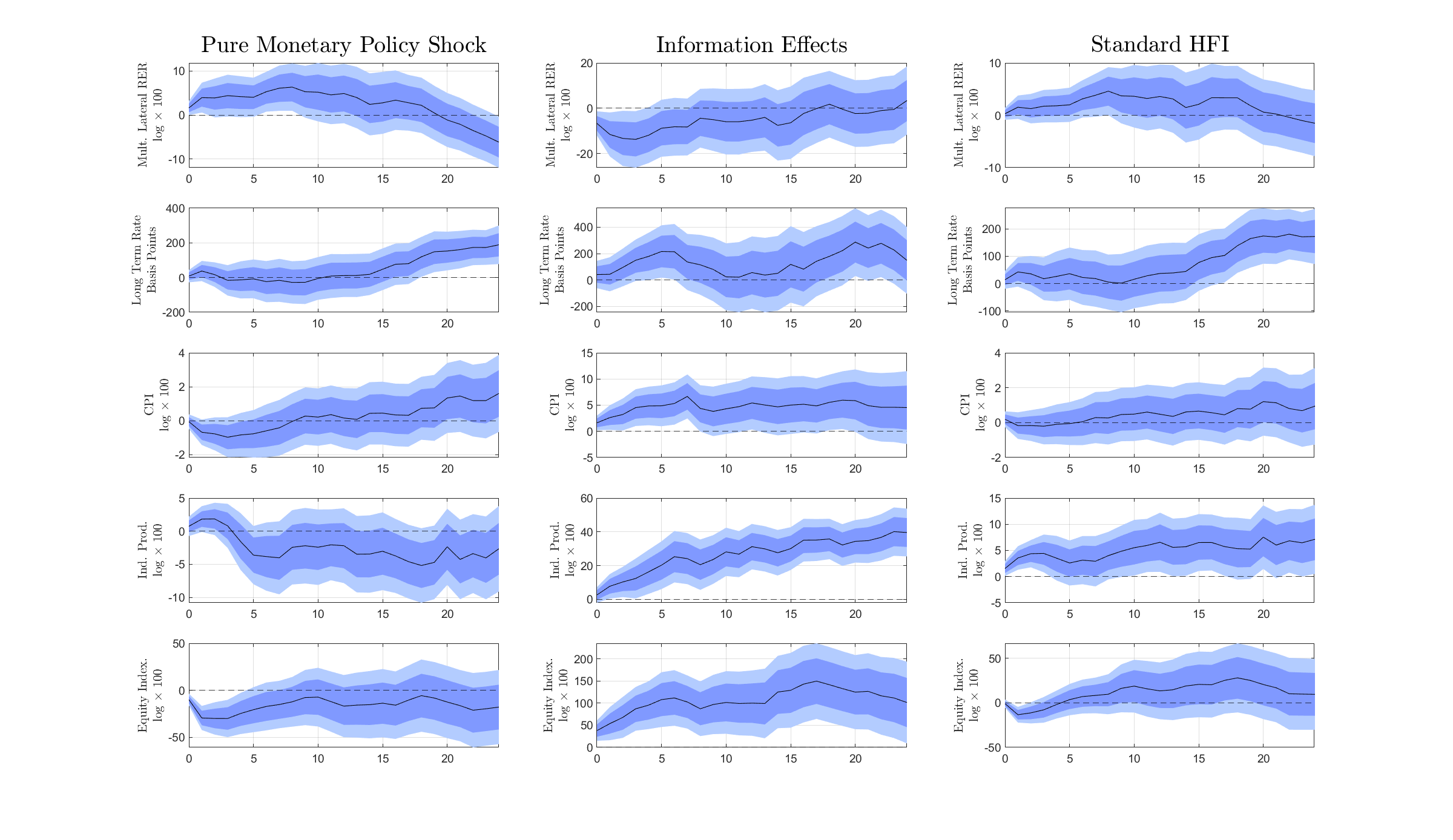}
    \caption{Impulse Response Functions \\ Multi. REER Sample -  Highest Level of GDP per Capita}
    \label{fig:BenchmarkREER_GDPpc_4}
    \floatfoot{\textbf{Note:} The figure is comprised of 15 sub-figures ordered in three columns and five rows. The left column relates to the estimates of $\beta^{MP}$ in Equation \ref{eq:LP_pooled}, the middle column relates to the estimate of $\beta^{FIE}$ in Equation \ref{eq:LP_pooled}, while the right column relates to estimating Equation \ref{eq:LP_pooled}, replacing the MP and FIE components with the un-orthogonalized monetary policy surprise. This figure presents the results for the group of countries in the highest quartile of GDP per Capita in the sample for the year 2004, replacing the nominal exchange rate with the US dollar with the multilateral trade weighted real exchange rate. The rows represent the impact on (i) the nominal exchange rate with the US dollar (in logs times 100); (ii) long term interest rates in basis points; (iii) the consumer price index (in logs times 100); (iv) the industrial production index (in logs times 100); (v) the equity index (in logs times 100). The solid black line represents the point estimate, the dark blue area represents the 68\% confidence interval, and the light blue area represents the 90\% confidence interval. In the text, when referring to Panel $(i,j)$, $i$ refers to the row and $j$ to the column of the figure. Each variable, in its own transformation, is demeaned at the country level. }
\end{figure}

\newpage
\begin{figure}
    \centering
    \includegraphics[scale=0.4]{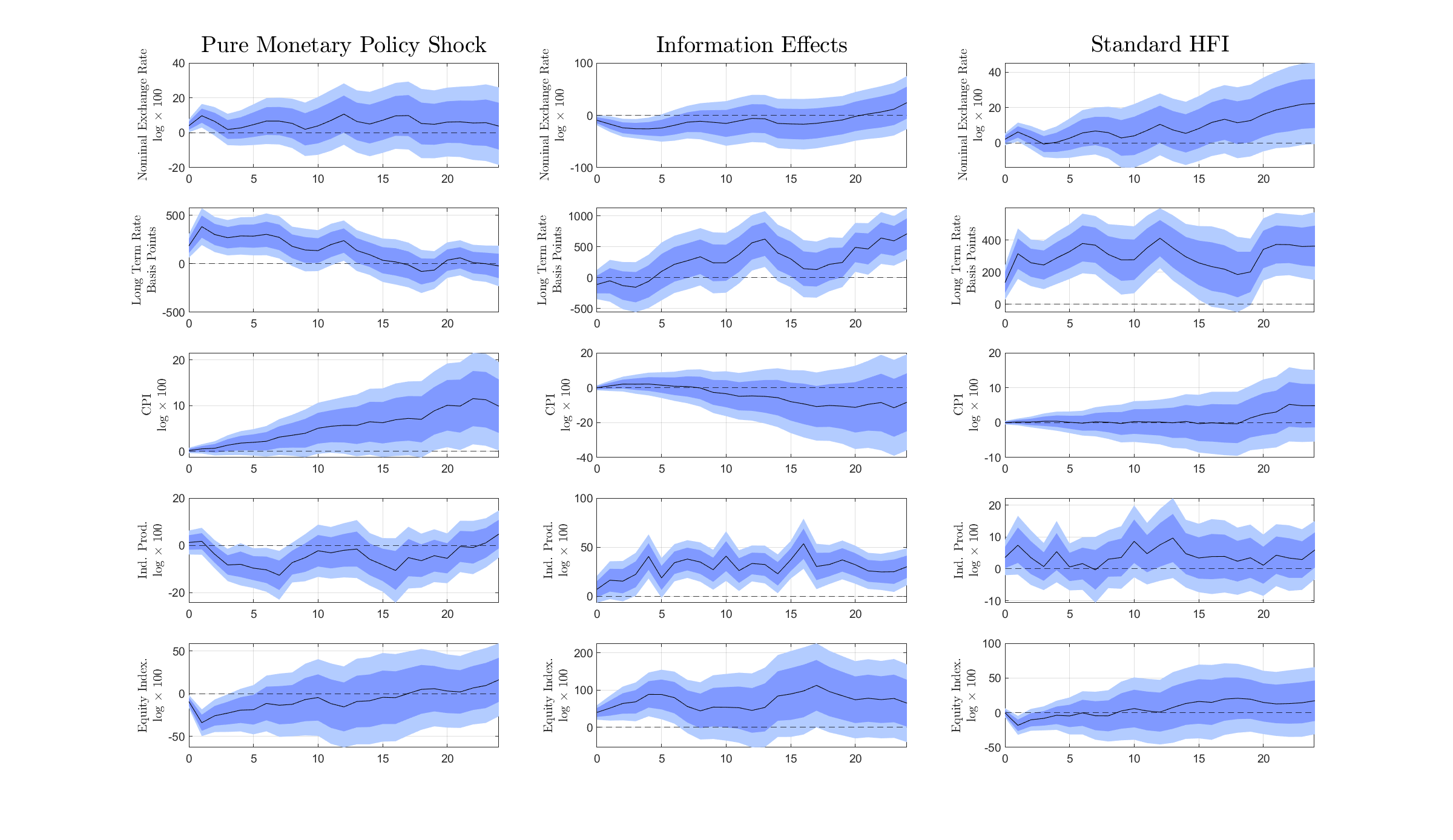}
    \caption{Impulse Response Functions \\ Lowest Level of Total Trade to GDP}
    \label{fig:BenchmarkNER_TT_1}
    \floatfoot{\textbf{Note:} The figure is comprised of 15 sub-figures ordered in three columns and five rows. The left column relates to the estimates of $\beta^{MP}$ in Equation \ref{eq:LP_pooled}, the middle column relates to the estimate of $\beta^{FIE}$ in Equation \ref{eq:LP_pooled}, while the right column relates to estimating Equation \ref{eq:LP_pooled}, replacing the MP and FIE components with the un-orthogonalized monetary policy surprise. This figure presents the results for the group of countries in the lowest quartile of total trade to GDP. The rows represent the impact on (i) the nominal exchange rate with the US dollar (in logs times 100); (ii) long term interest rates in basis points; (iii) the consumer price index (in logs times 100); (iv) the industrial production index (in logs times 100); (v) the equity index (in logs times 100). The solid black line represents the point estimate, the dark blue area represents the 68\% confidence interval, and the light blue area represents the 90\% confidence interval. In the text, when referring to Panel $(i,j)$, $i$ refers to the row and $j$ to the column of the figure. Each variable, in its own transformation, is demeaned at the country level. }
\end{figure}

\newpage
\begin{figure}
    \centering
    \includegraphics[scale=0.4]{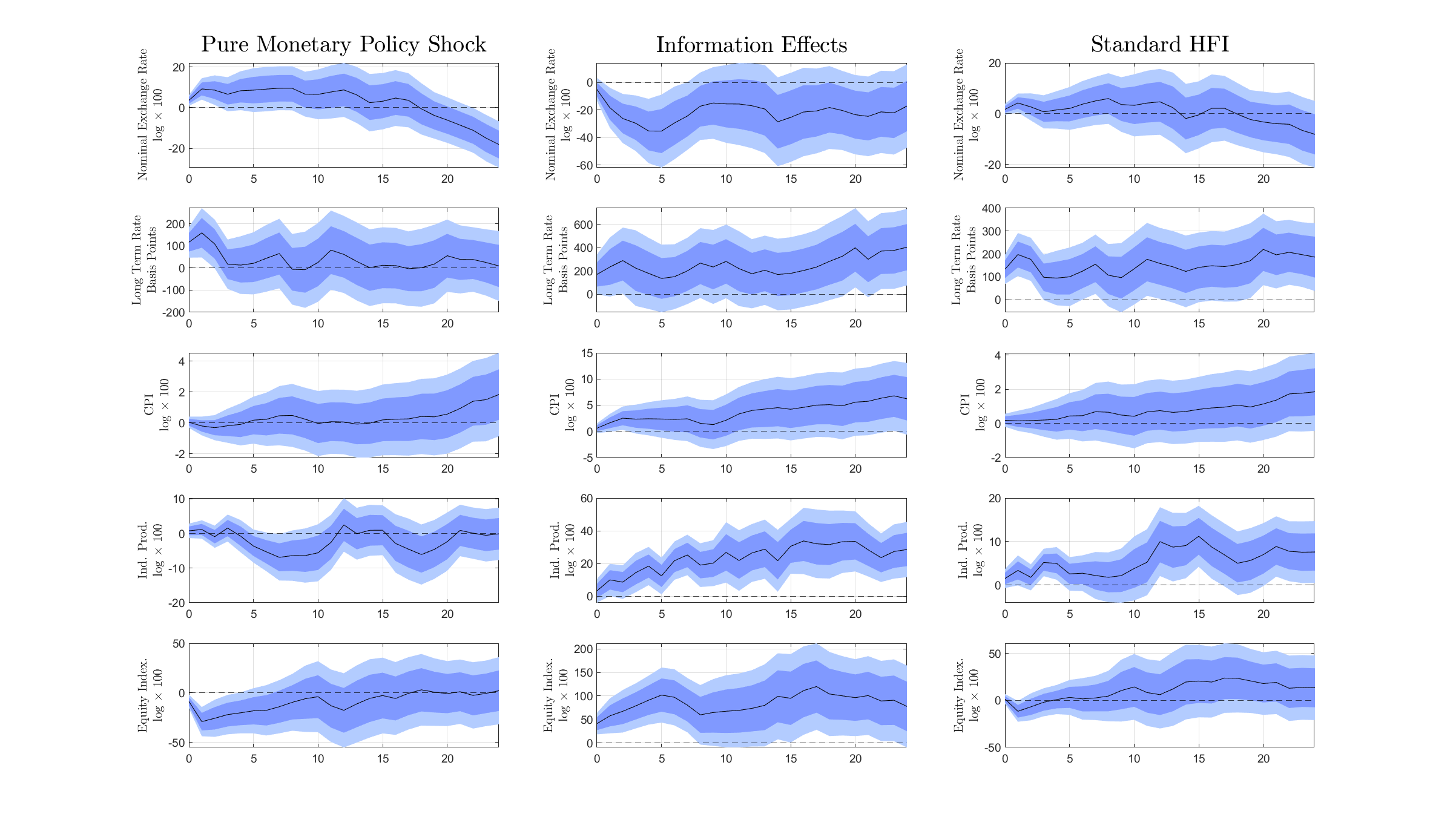}
    \caption{Impulse Response Functions \\ Second Lowest Level of Total Trade to GDP}
    \label{fig:BenchmarkNER_TT_2}
    \floatfoot{\textbf{Note:} The figure is comprised of 15 sub-figures ordered in three columns and five rows. The left column relates to the estimates of $\beta^{MP}$ in Equation \ref{eq:LP_pooled}, the middle column relates to the estimate of $\beta^{FIE}$ in Equation \ref{eq:LP_pooled}, while the right column relates to estimating Equation \ref{eq:LP_pooled}, replacing the MP and FIE components with the un-orthogonalized monetary policy surprise. This figure presents the results for the group of countries in the second lowest quartile of total trade to GDP. The rows represent the impact on (i) the nominal exchange rate with the US dollar (in logs times 100); (ii) long term interest rates in basis points; (iii) the consumer price index (in logs times 100); (iv) the industrial production index (in logs times 100); (v) the equity index (in logs times 100). The solid black line represents the point estimate, the dark blue area represents the 68\% confidence interval, and the light blue area represents the 90\% confidence interval. In the text, when referring to Panel $(i,j)$, $i$ refers to the row and $j$ to the column of the figure. Each variable, in its own transformation, is demeaned at the country level. }
\end{figure}

\newpage
\begin{figure}
    \centering
    \includegraphics[scale=0.4]{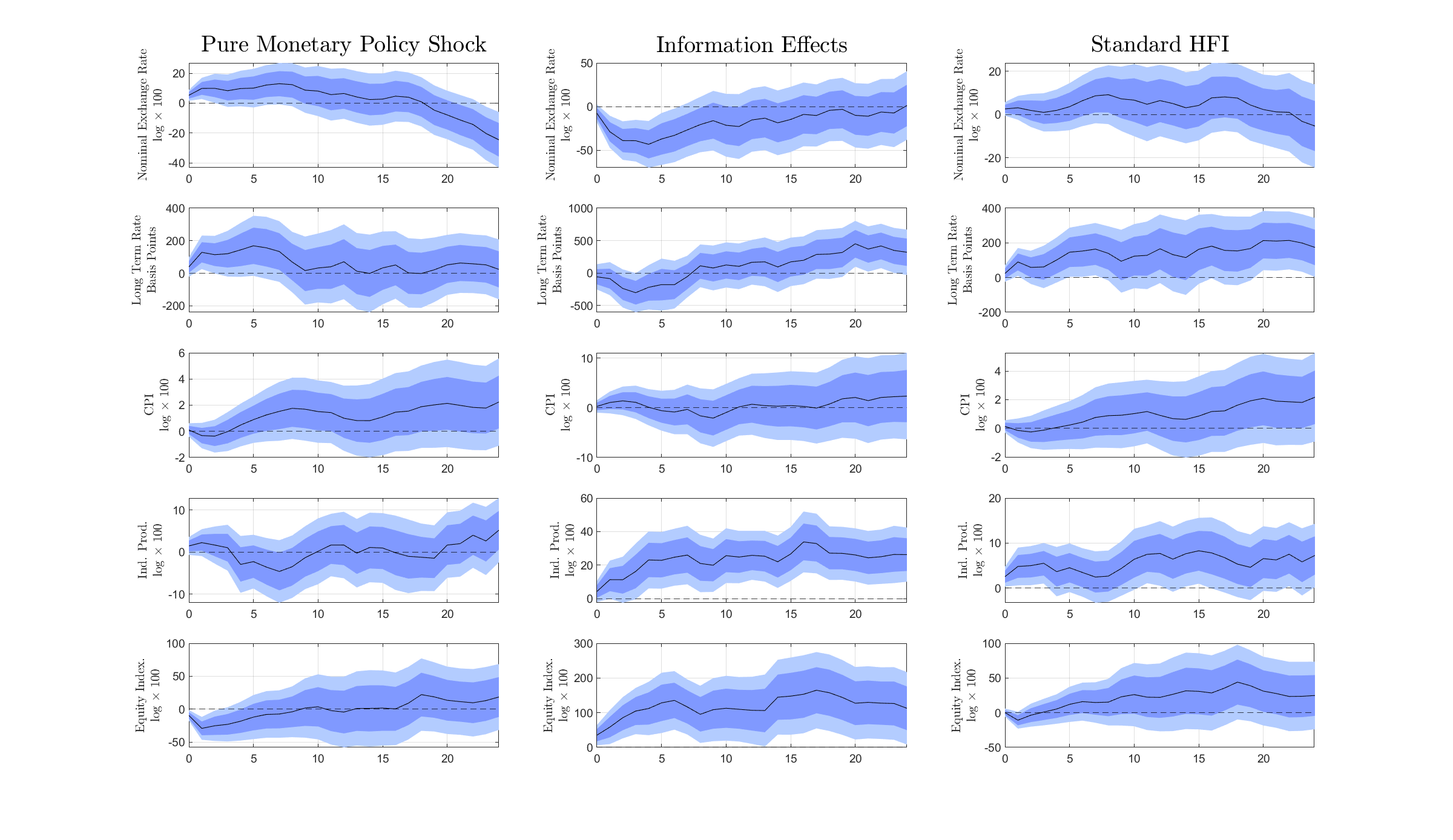}
    \caption{Impulse Response Functions \\ Second Highest Level of Total Trade to GDP}
    \label{fig:BenchmarkNER_TT_3}
    \floatfoot{\textbf{Note:} The figure is comprised of 15 sub-figures ordered in three columns and five rows. The left column relates to the estimates of $\beta^{MP}$ in Equation \ref{eq:LP_pooled}, the middle column relates to the estimate of $\beta^{FIE}$ in Equation \ref{eq:LP_pooled}, while the right column relates to estimating Equation \ref{eq:LP_pooled}, replacing the MP and FIE components with the un-orthogonalized monetary policy surprise. This figure presents the results for the group of countries in the second highest quartile of total trade to GDP. The rows represent the impact on (i) the nominal exchange rate with the US dollar (in logs times 100); (ii) long term interest rates in basis points; (iii) the consumer price index (in logs times 100); (iv) the industrial production index (in logs times 100); (v) the equity index (in logs times 100). The solid black line represents the point estimate, the dark blue area represents the 68\% confidence interval, and the light blue area represents the 90\% confidence interval. In the text, when referring to Panel $(i,j)$, $i$ refers to the row and $j$ to the column of the figure. Each variable, in its own transformation, is demeaned at the country level. }
\end{figure}

\newpage
\begin{figure}
    \centering
    \includegraphics[scale=0.4]{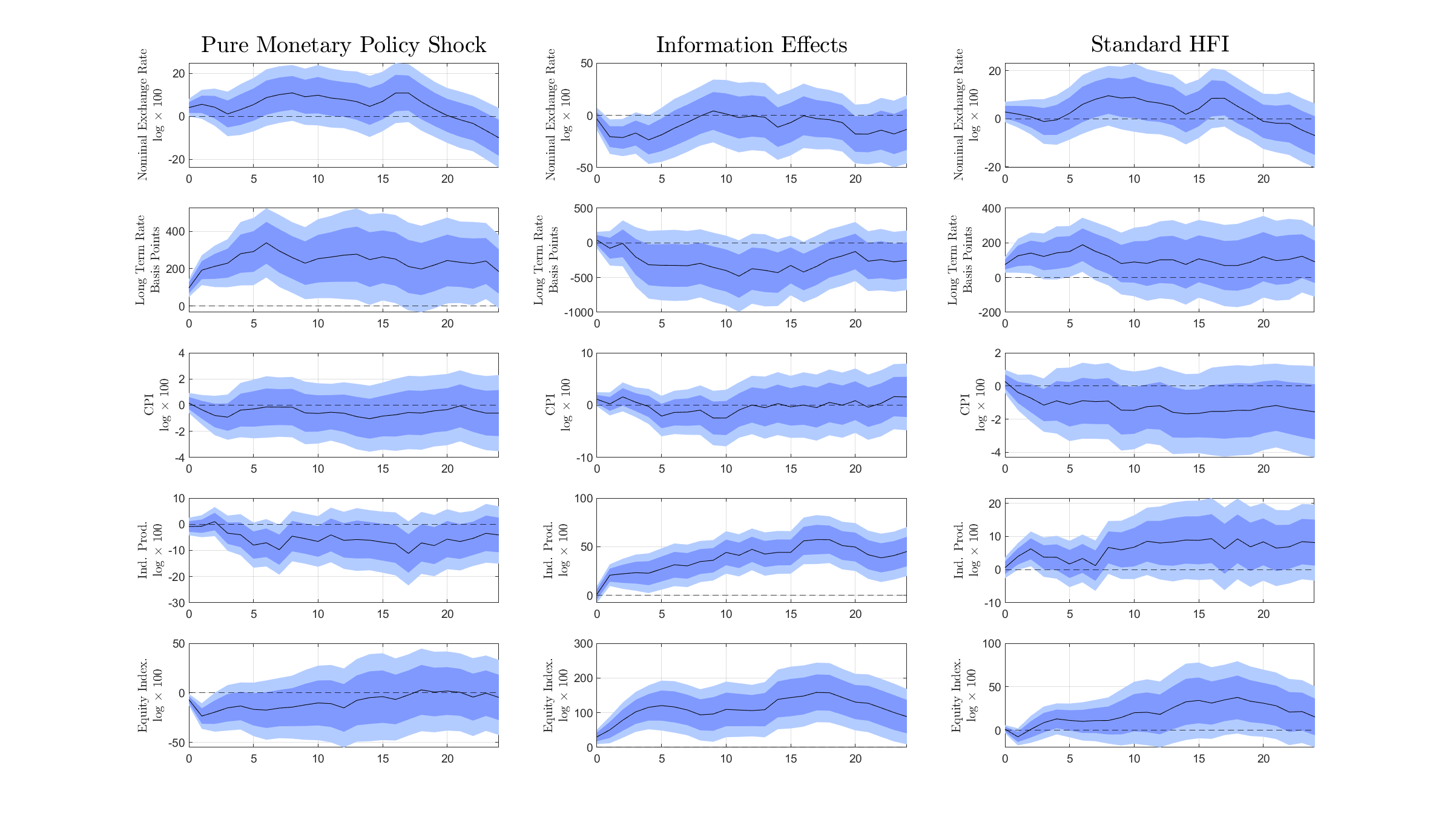}
    \caption{Impulse Response Functions \\ Highest Level of Total Trade to GDP}
    \label{fig:BenchmarkNER_TT_4}
    \floatfoot{\textbf{Note:} The figure is comprised of 15 sub-figures ordered in three columns and five rows. The left column relates to the estimates of $\beta^{MP}$ in Equation \ref{eq:LP_pooled}, the middle column relates to the estimate of $\beta^{FIE}$ in Equation \ref{eq:LP_pooled}, while the right column relates to estimating Equation \ref{eq:LP_pooled}, replacing the MP and FIE components with the un-orthogonalized monetary policy surprise. This figure presents the results for the group of countries in the second highest quartile of total trade to GDP. The rows represent the impact on (i) the nominal exchange rate with the US dollar (in logs times 100); (ii) long term interest rates in basis points; (iii) the consumer price index (in logs times 100); (iv) the industrial production index (in logs times 100); (v) the equity index (in logs times 100). The solid black line represents the point estimate, the dark blue area represents the 68\% confidence interval, and the light blue area represents the 90\% confidence interval. In the text, when referring to Panel $(i,j)$, $i$ refers to the row and $j$ to the column of the figure. Each variable, in its own transformation, is demeaned at the country level. }
\end{figure}

\newpage
\begin{figure}
    \centering
    \includegraphics[scale=0.4]{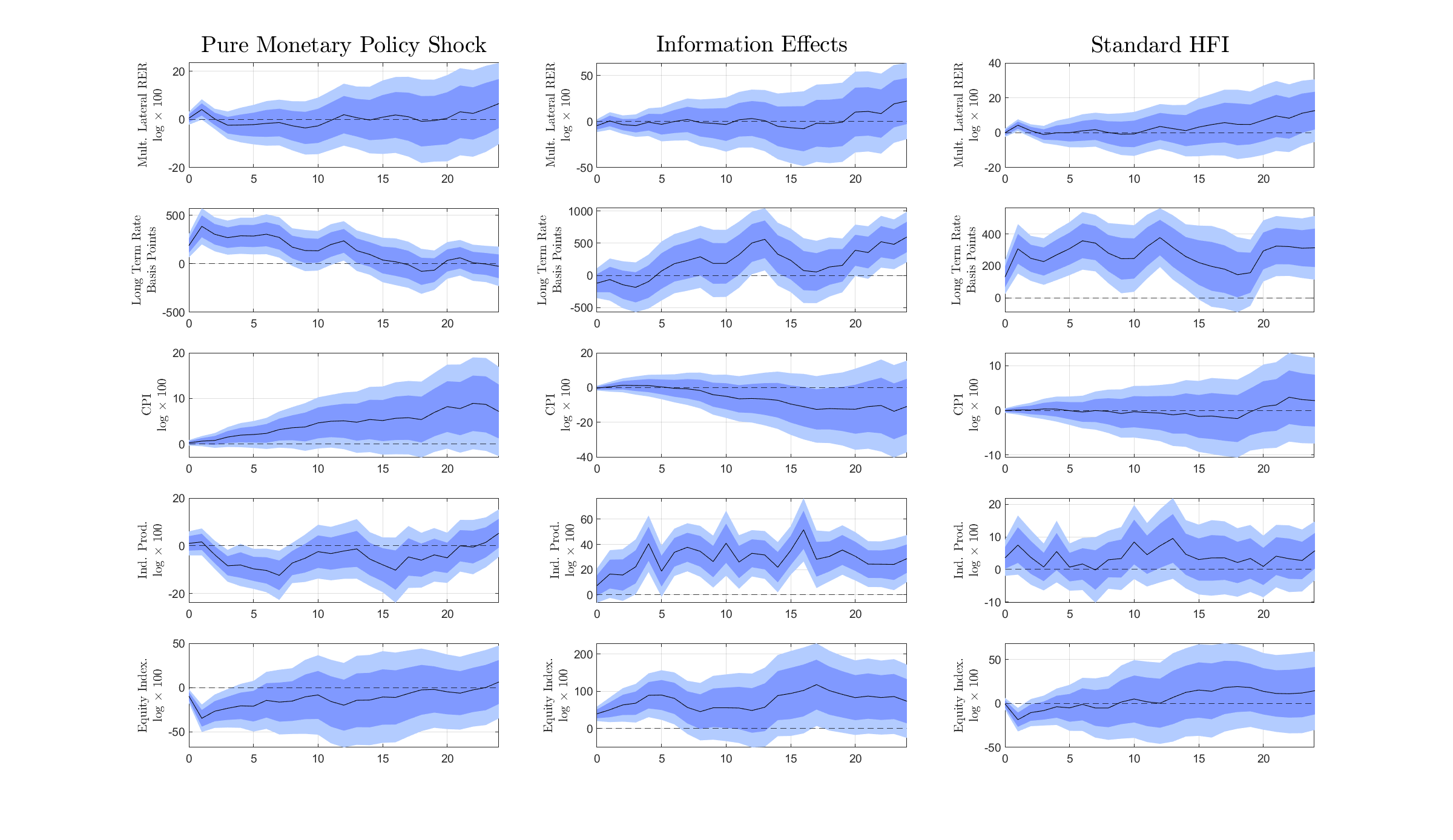}
    \caption{Impulse Response Functions \\ Multi. REER Sample -  Lowest Level of Total Trade to GDP}
    \label{fig:BenchmarkREER_TT_1}
    \floatfoot{\textbf{Note:} The figure is comprised of 15 sub-figures ordered in three columns and five rows. The left column relates to the estimates of $\beta^{MP}$ in Equation \ref{eq:LP_pooled}, the middle column relates to the estimate of $\beta^{FIE}$ in Equation \ref{eq:LP_pooled}, while the right column relates to estimating Equation \ref{eq:LP_pooled}, replacing the MP and FIE components with the un-orthogonalized monetary policy surprise. This figure presents the results for the group of countries in the lowest quartile of total trade to GDP, replacing the nominal exchange rate with the US dollar with the multilateral trade weighted real exchange rate. The rows represent the impact on (i) the nominal exchange rate with the US dollar (in logs times 100); (ii) long term interest rates in basis points; (iii) the consumer price index (in logs times 100); (iv) the industrial production index (in logs times 100); (v) the equity index (in logs times 100). The solid black line represents the point estimate, the dark blue area represents the 68\% confidence interval, and the light blue area represents the 90\% confidence interval. In the text, when referring to Panel $(i,j)$, $i$ refers to the row and $j$ to the column of the figure. Each variable, in its own transformation, is demeaned at the country level. }
\end{figure}

\newpage
\begin{figure}
    \centering
    \includegraphics[scale=0.4]{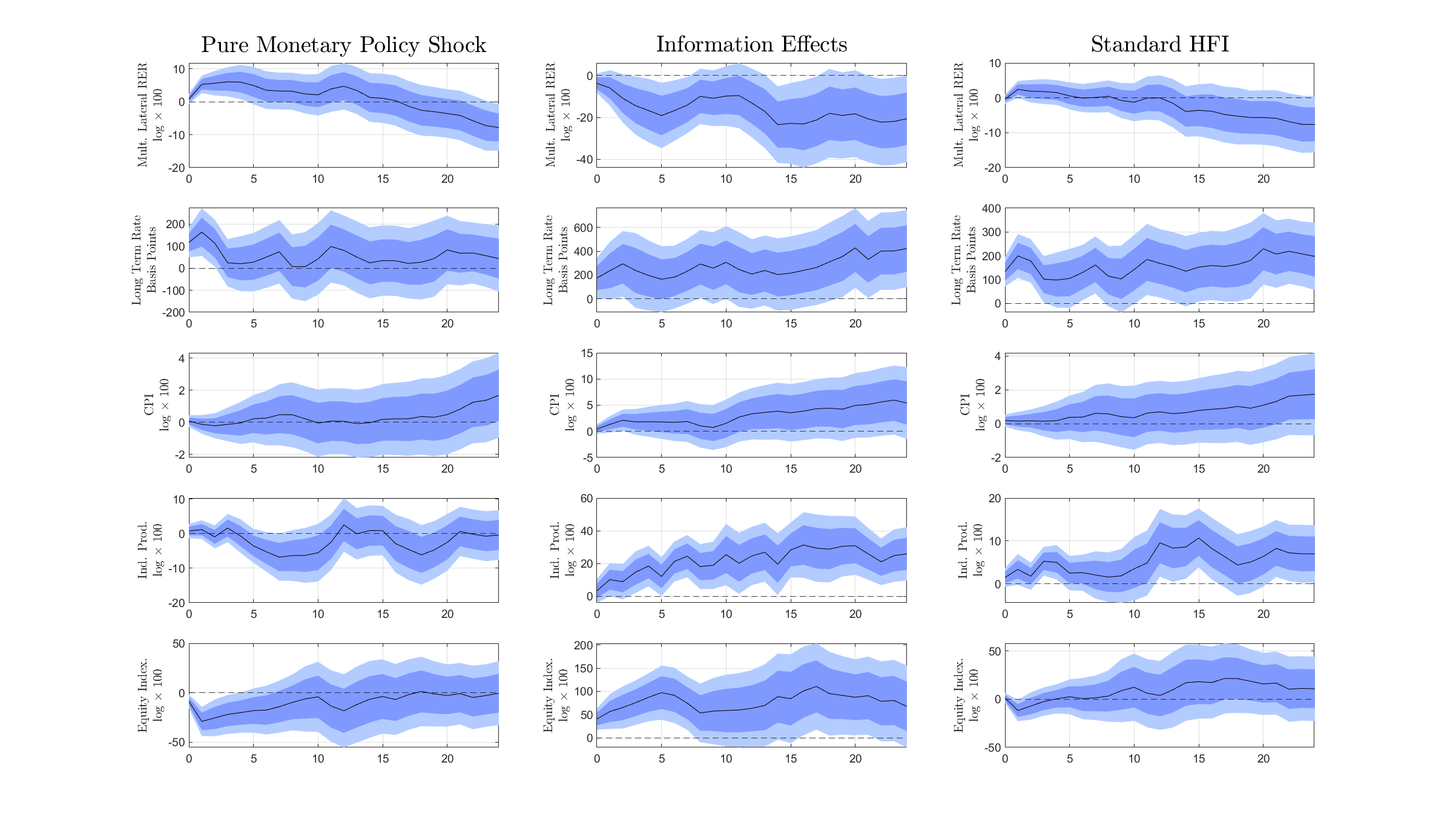}
    \caption{Impulse Response Functions \\ Multi. REER Sample - Second Lowest Level of Total Trade to GDP}
    \label{fig:BenchmarkREER_TT_2}
    \floatfoot{\textbf{Note:} The figure is comprised of 15 sub-figures ordered in three columns and five rows. The left column relates to the estimates of $\beta^{MP}$ in Equation \ref{eq:LP_pooled}, the middle column relates to the estimate of $\beta^{FIE}$ in Equation \ref{eq:LP_pooled}, while the right column relates to estimating Equation \ref{eq:LP_pooled}, replacing the MP and FIE components with the un-orthogonalized monetary policy surprise. This figure presents the results for the group of countries in the second lowest quartile of total trade to GDP, replacing the nominal exchange rate with the US dollar with the multilateral trade weighted real exchange rate. The rows represent the impact on (i) the nominal exchange rate with the US dollar (in logs times 100); (ii) long term interest rates in basis points; (iii) the consumer price index (in logs times 100); (iv) the industrial production index (in logs times 100); (v) the equity index (in logs times 100). The solid black line represents the point estimate, the dark blue area represents the 68\% confidence interval, and the light blue area represents the 90\% confidence interval. In the text, when referring to Panel $(i,j)$, $i$ refers to the row and $j$ to the column of the figure. Each variable, in its own transformation, is demeaned at the country level. }
\end{figure}

\newpage
\begin{figure}
    \centering
    \includegraphics[scale=0.4]{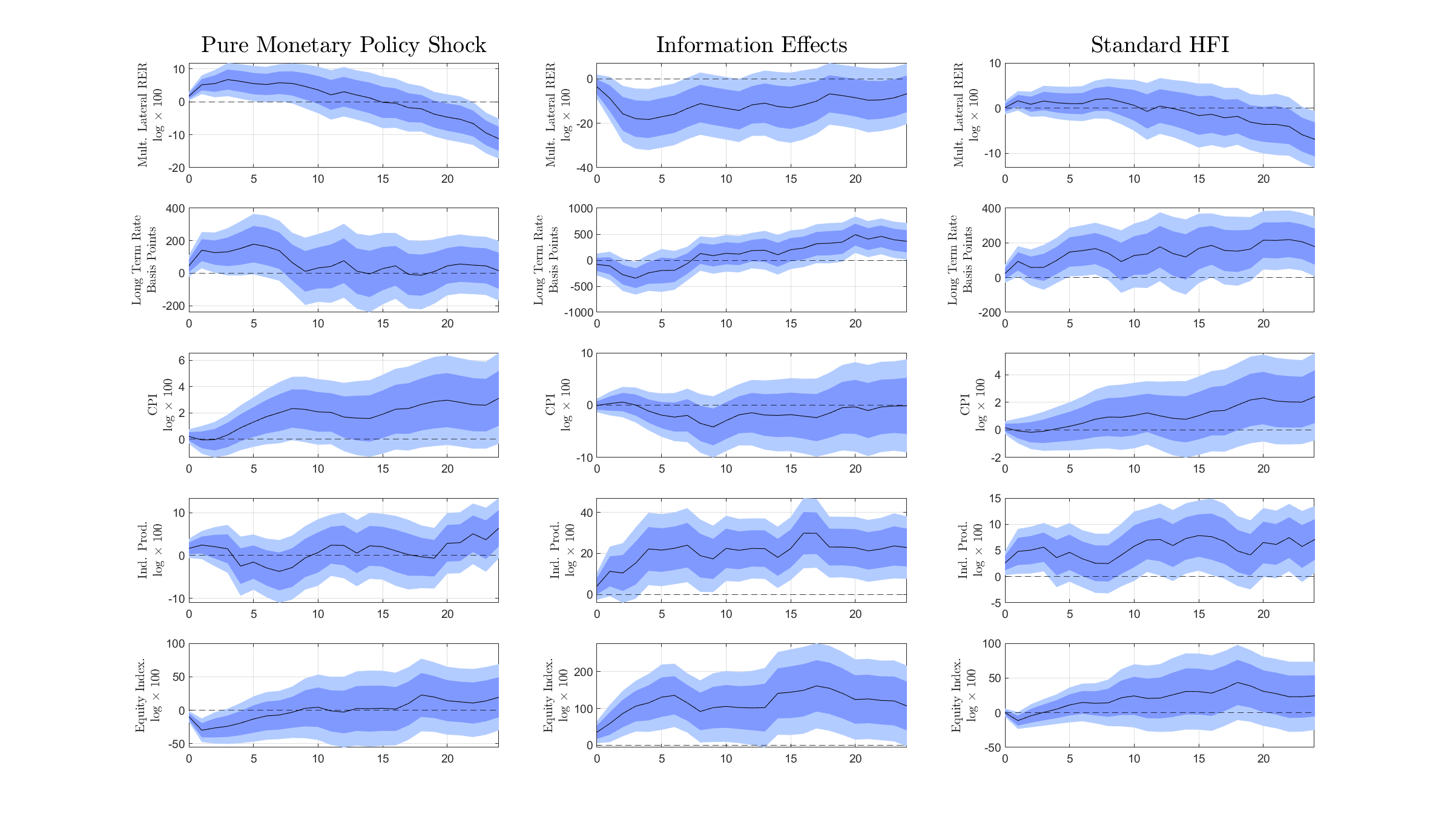}
    \caption{Impulse Response Functions \\ Multi. REER Sample - Second Highest Level of Total Trade to GDP}
    \label{fig:BenchmarkREER_TT_3}
    \floatfoot{\textbf{Note:} The figure is comprised of 15 sub-figures ordered in three columns and five rows. The left column relates to the estimates of $\beta^{MP}$ in Equation \ref{eq:LP_pooled}, the middle column relates to the estimate of $\beta^{FIE}$ in Equation \ref{eq:LP_pooled}, while the right column relates to estimating Equation \ref{eq:LP_pooled}, replacing the MP and FIE components with the un-orthogonalized monetary policy surprise. This figure presents the results for the group of countries in the second highest quartile of total trade to GDP, replacing the nominal exchange rate with the US dollar with the multilateral trade weighted real exchange rate. The rows represent the impact on (i) the nominal exchange rate with the US dollar (in logs times 100); (ii) long term interest rates in basis points; (iii) the consumer price index (in logs times 100); (iv) the industrial production index (in logs times 100); (v) the equity index (in logs times 100). The solid black line represents the point estimate, the dark blue area represents the 68\% confidence interval, and the light blue area represents the 90\% confidence interval. In the text, when referring to Panel $(i,j)$, $i$ refers to the row and $j$ to the column of the figure. Each variable, in its own transformation, is demeaned at the country level. }
\end{figure}

\newpage
\begin{figure}
    \centering
    \includegraphics[scale=0.4]{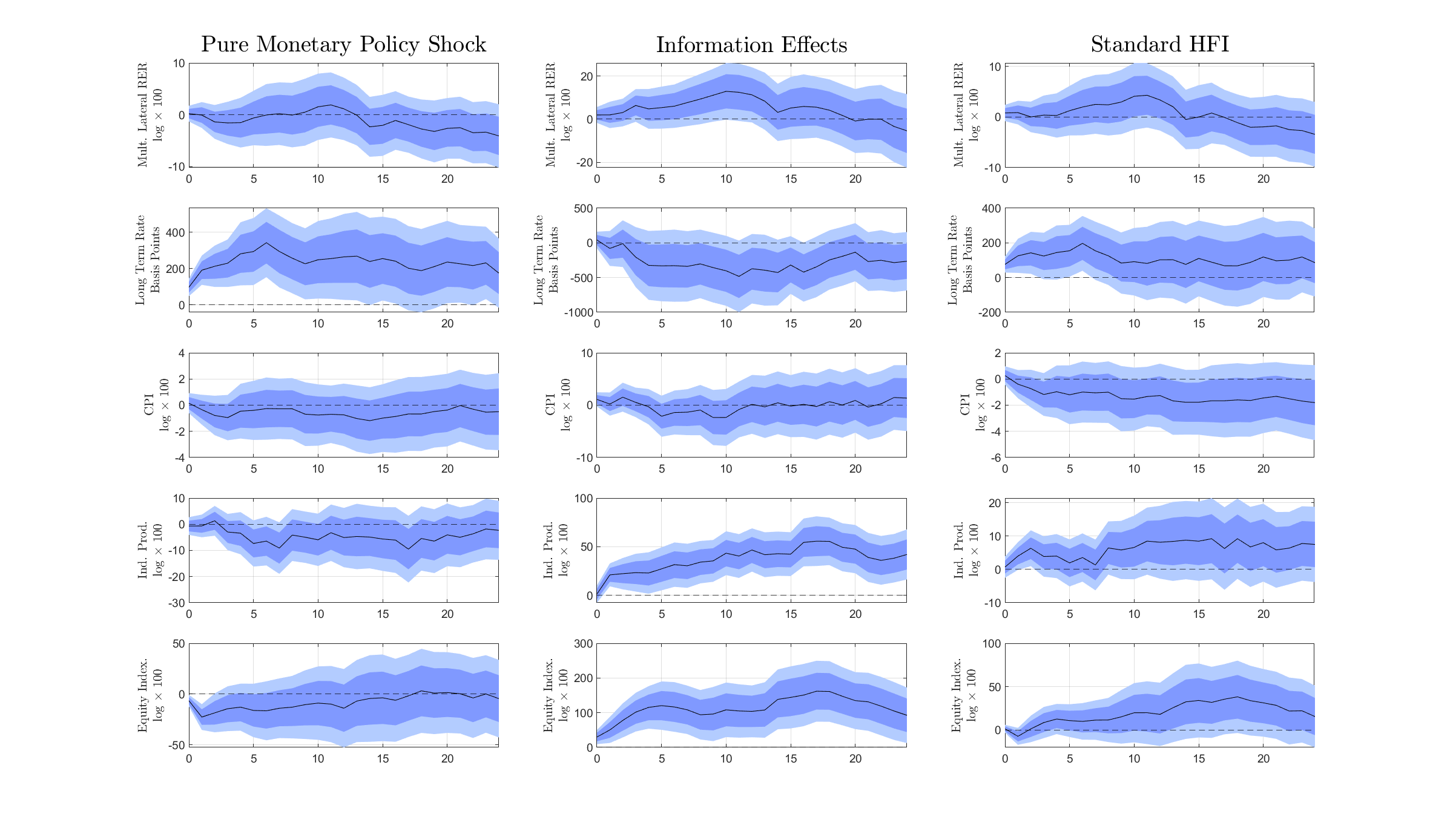}
    \caption{Impulse Response Functions \\ Multi. REER Sample - Highest Level of Total Trade to GDP}
    \label{fig:BenchmarkREER_TT_4}
    \floatfoot{\textbf{Note:} The figure is comprised of 15 sub-figures ordered in three columns and five rows. The left column relates to the estimates of $\beta^{MP}$ in Equation \ref{eq:LP_pooled}, the middle column relates to the estimate of $\beta^{FIE}$ in Equation \ref{eq:LP_pooled}, while the right column relates to estimating Equation \ref{eq:LP_pooled}, replacing the MP and FIE components with the un-orthogonalized monetary policy surprise. This figure presents the results for the group of countries in the highest quartile of total trade to GDP, replacing the nominal exchange rate with the US dollar with the multilateral trade weighted real exchange rate. The rows represent the impact on (i) the nominal exchange rate with the US dollar (in logs times 100); (ii) long term interest rates in basis points; (iii) the consumer price index (in logs times 100); (iv) the industrial production index (in logs times 100); (v) the equity index (in logs times 100). The solid black line represents the point estimate, the dark blue area represents the 68\% confidence interval, and the light blue area represents the 90\% confidence interval. In the text, when referring to Panel $(i,j)$, $i$ refers to the row and $j$ to the column of the figure. Each variable, in its own transformation, is demeaned at the country level. }
\end{figure}

\newpage
\begin{figure}[ht]
    \centering
    \includegraphics[scale=0.4]{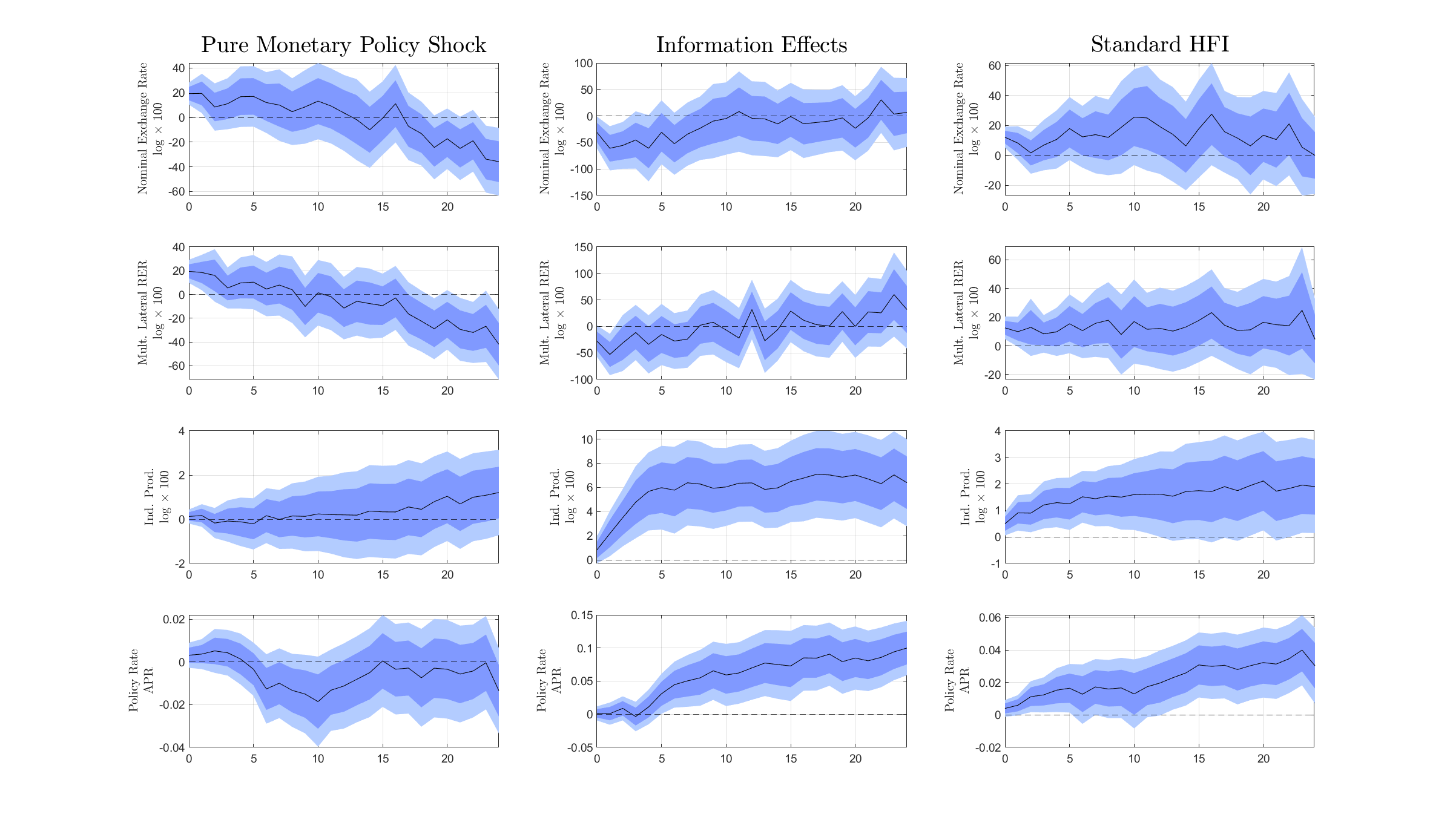}
    \caption{Impulse Response Functions \\ Dataset \cite{ilzetzki2021puzzling} - Eq. \ref{eq:LP_pooled} }
    \label{fig:Results_1}
    \floatfoot{\textbf{Note:} The figure is comprised of 12 sub-figures ordered in three columns and four rows. The left column relates to the estimates of $\beta^{MP}$ in Equation \ref{eq:LP_pooled}, the middle column relates to the estimate of $\beta^{FIE}$ in Equation \ref{eq:LP_pooled}, while the right column relates to estimating Equation \ref{eq:LP_pooled}, replacing the MP and FIE components with the un-orthogonalized monetary policy surprise. The rows represent the impact on (i) the nominal exchange rate with the US dollar; (ii) the real exchange rate with the US; (iii) the industrial production index; (iv) monetary policy rate. I estimate the impulse response functions for the nominal exchange rate, the industrial production and the monetary policy rate using these three variables as controls. For the impulse response function for the real exchange rate, I replace the nominal exchange rate with the real exchange, given the strong correlation between the nominal and real exchange rate with the US dollar. The solid black line represents the point estimate, the dark blue area represents the 68\% confidence interval, and the light blue area represents the 90\% confidence interval. In the text, when referring to Panel $(i,j)$, $i$ refers to the row and $j$ to the column of the figure. Each variable is as defined in \cite{ilzetzki2021puzzling}. }
\end{figure}

\newpage
\begin{figure}[ht]
    \centering
    \includegraphics[scale=0.4]{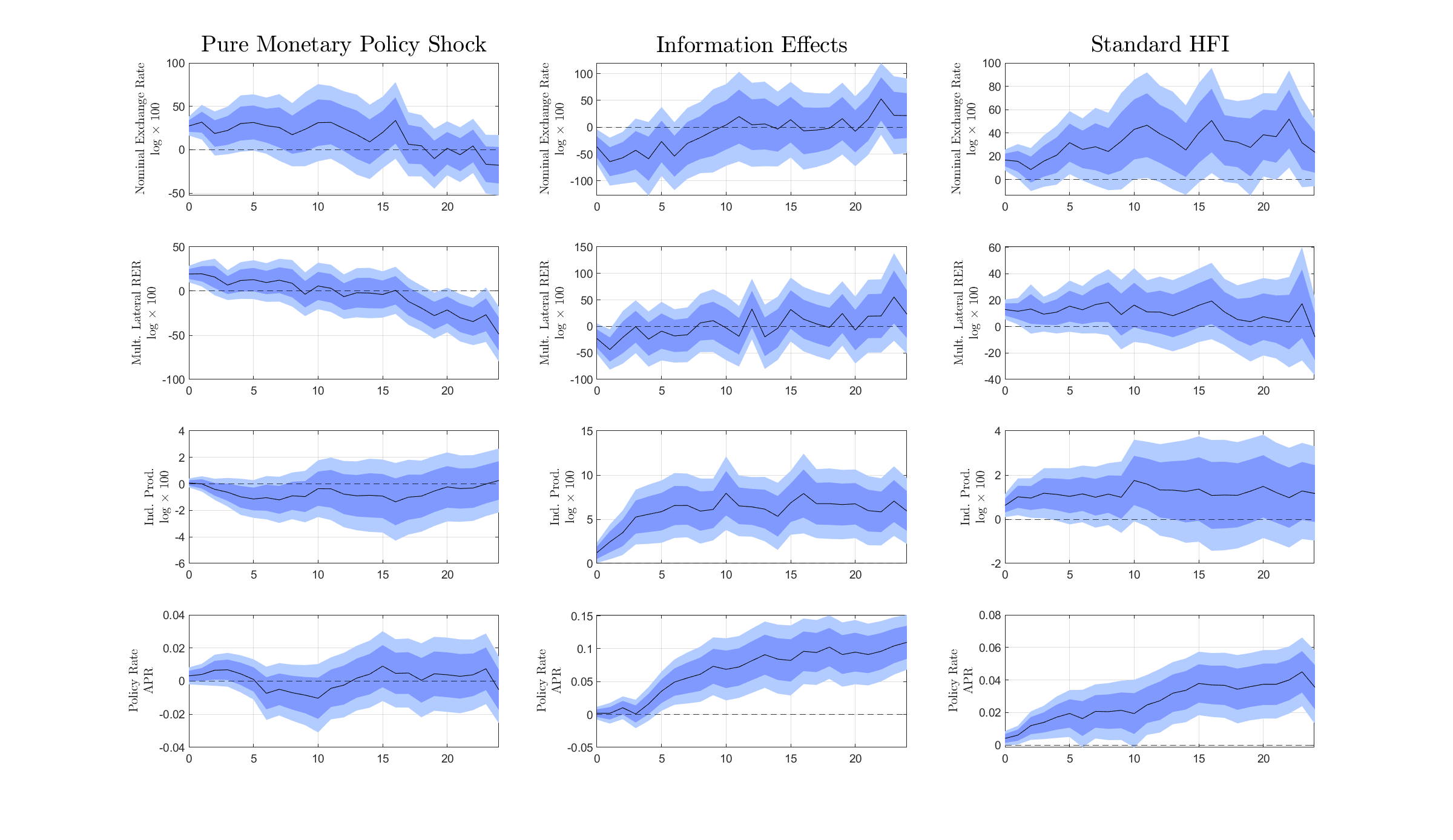}
    \caption{Impulse Response Functions \\ Dataset \cite{ilzetzki2021puzzling} 1998 - Onward - Eq. \ref{eq:LP_pooled}}
    \label{fig:Results_1_98}
    \floatfoot{\textbf{Note:} The figure is comprised of 12 sub-figures ordered in three columns and four rows. The left column relates to the estimates of $\beta^{MP}$ in Equation \ref{eq:LP_pooled}, the middle column relates to the estimate of $\beta^{FIE}$ in Equation \ref{eq:LP_pooled}, while the right column relates to estimating Equation \ref{eq:LP_pooled}, replacing the MP and FIE components with the un-orthogonalized monetary policy surprise. The figure presents results for the sample starting in January 1998 onward. The rows represent the impact on (i) the nominal exchange rate with the US dollar; (ii) the real exchange rate with the US; (iii) the industrial production index; (iv) monetary policy rate. I estimate the impulse response functions for the nominal exchange rate, the industrial production and the monetary policy rate using these three variables as controls. For the impulse response function for the real exchange rate, I replace the nominal exchange rate with the real exchange, given the strong correlation between the nominal and real exchange rate with the US dollar.  The solid black line represents the point estimate, the dark blue area represents the 68\% confidence interval, and the light blue area represents the 90\% confidence interval. In the text, when referring to Panel $(i,j)$, $i$ refers to the row and $j$ to the column of the figure. Each variable is as defined in \cite{ilzetzki2021puzzling}. }
\end{figure}

\newpage
\begin{figure}[ht]
    \centering
    \includegraphics[scale=0.4]{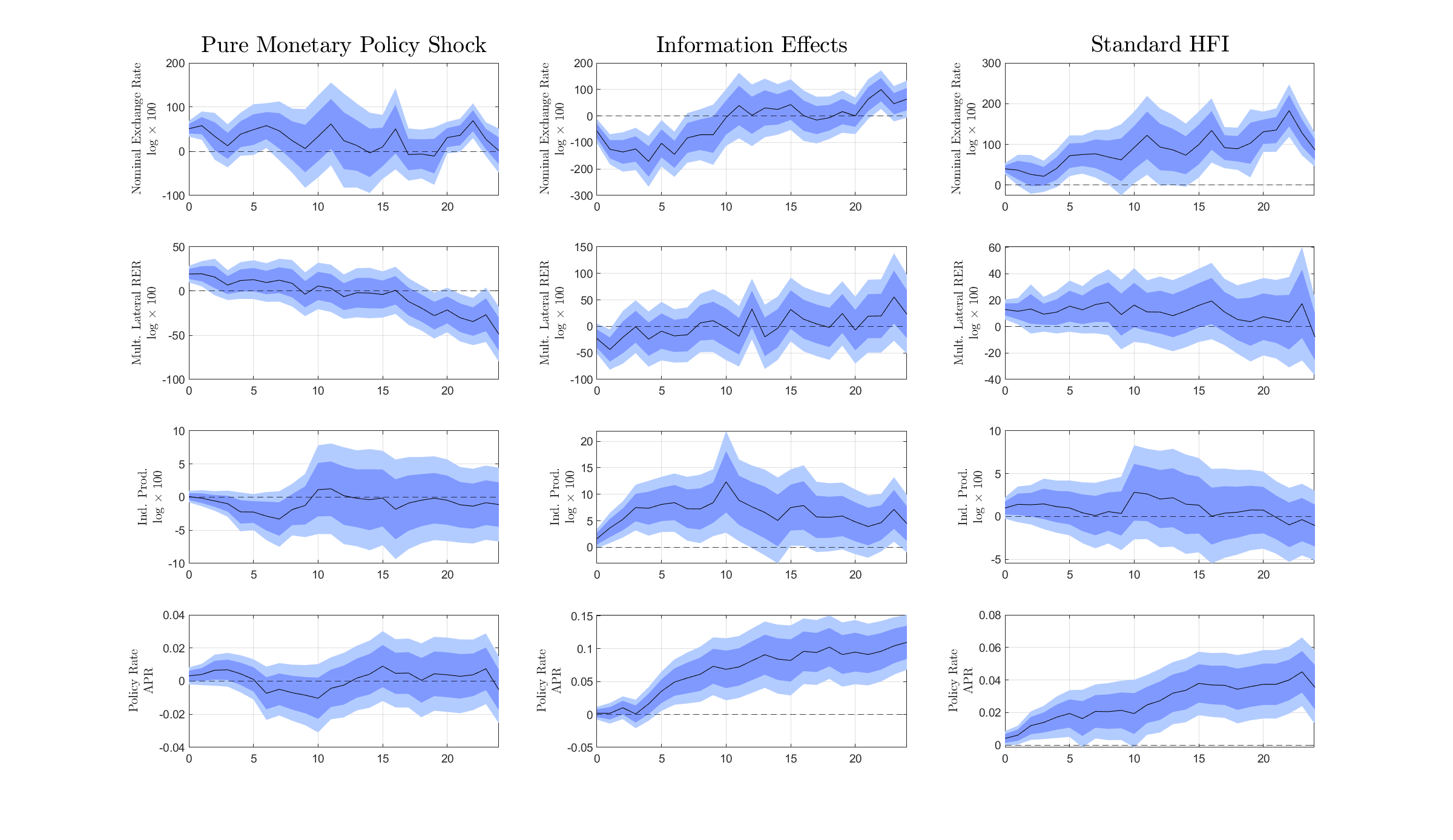}
    \caption{Impulse Response Functions \\ Dataset \cite{ilzetzki2021puzzling} 2008 - Onward - Eq. \ref{eq:LP_pooled}}
    \label{fig:Results_1_08}
    \floatfoot{\textbf{Note:} The figure is comprised of 12 sub-figures ordered in three columns and four rows. The left column relates to the estimates of $\beta^{MP}$ in Equation \ref{eq:LP_pooled}, the middle column relates to the estimate of $\beta^{FIE}$ in Equation \ref{eq:LP_pooled}, while the right column relates to estimating Equation \ref{eq:LP_pooled}, replacing the MP and FIE components with the un-orthogonalized monetary policy surprise. The figure presents results for the sample starting in January 2008 onward. The rows represent the impact on (i) the nominal exchange rate with the US dollar; (ii) the real exchange rate with the US; (iii) the industrial production index; (iv) monetary policy rate. I estimate the impulse response functions for the nominal exchange rate, the industrial production and the monetary policy rate using these three variables as controls. For the impulse response function for the real exchange rate, I replace the nominal exchange rate with the real exchange, given the strong correlation between the nominal and real exchange rate with the US dollar. The solid black line represents the point estimate, the dark blue area represents the 68\% confidence interval, and the light blue area represents the 90\% confidence interval. In the text, when referring to Panel $(i,j)$, $i$ refers to the row and $j$ to the column of the figure. Each variable is as defined in \cite{ilzetzki2021puzzling}. }
\end{figure}

\newpage
\begin{figure}[ht]
    \centering
    \includegraphics[scale=0.4]{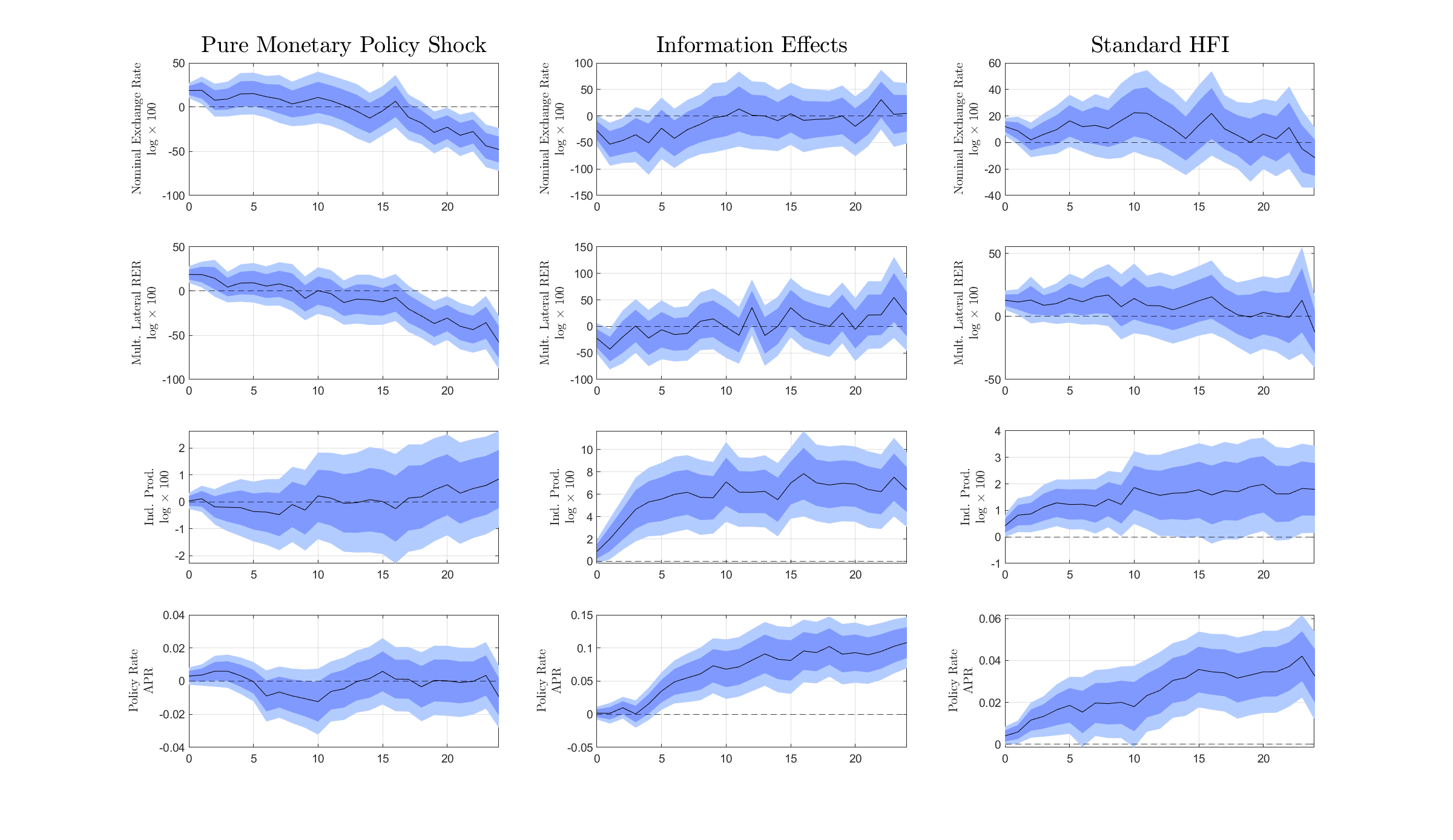}
    \caption{Impulse Response Functions \\ Dataset \cite{ilzetzki2021puzzling} - Eq. \ref{eq:LP_Trend} }
    \label{fig:Results_Trend}
    \floatfoot{\textbf{Note:} The figure is comprised of 12 sub-figures ordered in three columns and four rows. The left column relates to the estimates of $\beta^{MP}$ in Equation \ref{eq:LP_Trend}, the middle column relates to the estimate of $\beta^{FIE}$ in Equation \ref{eq:LP_Trend}, while the right column relates to estimating Equation \ref{eq:LP_Trend}, replacing the MP and FIE components with the un-orthogonalized monetary policy surprise. The rows represent the impact on (i) the nominal exchange rate with the US dollar; (ii) the real exchange rate with the US; (iii) the industrial production index; (iv) monetary policy rate. I estimate the impulse response functions for the nominal exchange rate, the industrial production and the monetary policy rate using these three variables as controls. For the impulse response function for the real exchange rate, I replace the nominal exchange rate with the real exchange, given the strong correlation between the nominal and real exchange rate with the US dollar. The solid black line represents the point estimate, the dark blue area represents the 68\% confidence interval, and the light blue area represents the 90\% confidence interval. In the text, when referring to Panel $(i,j)$, $i$ refers to the row and $j$ to the column of the figure. Each variable is as defined in \cite{ilzetzki2021puzzling}. }
\end{figure}

\newpage
\begin{figure}[ht]
    \centering
    \includegraphics[scale=0.4]{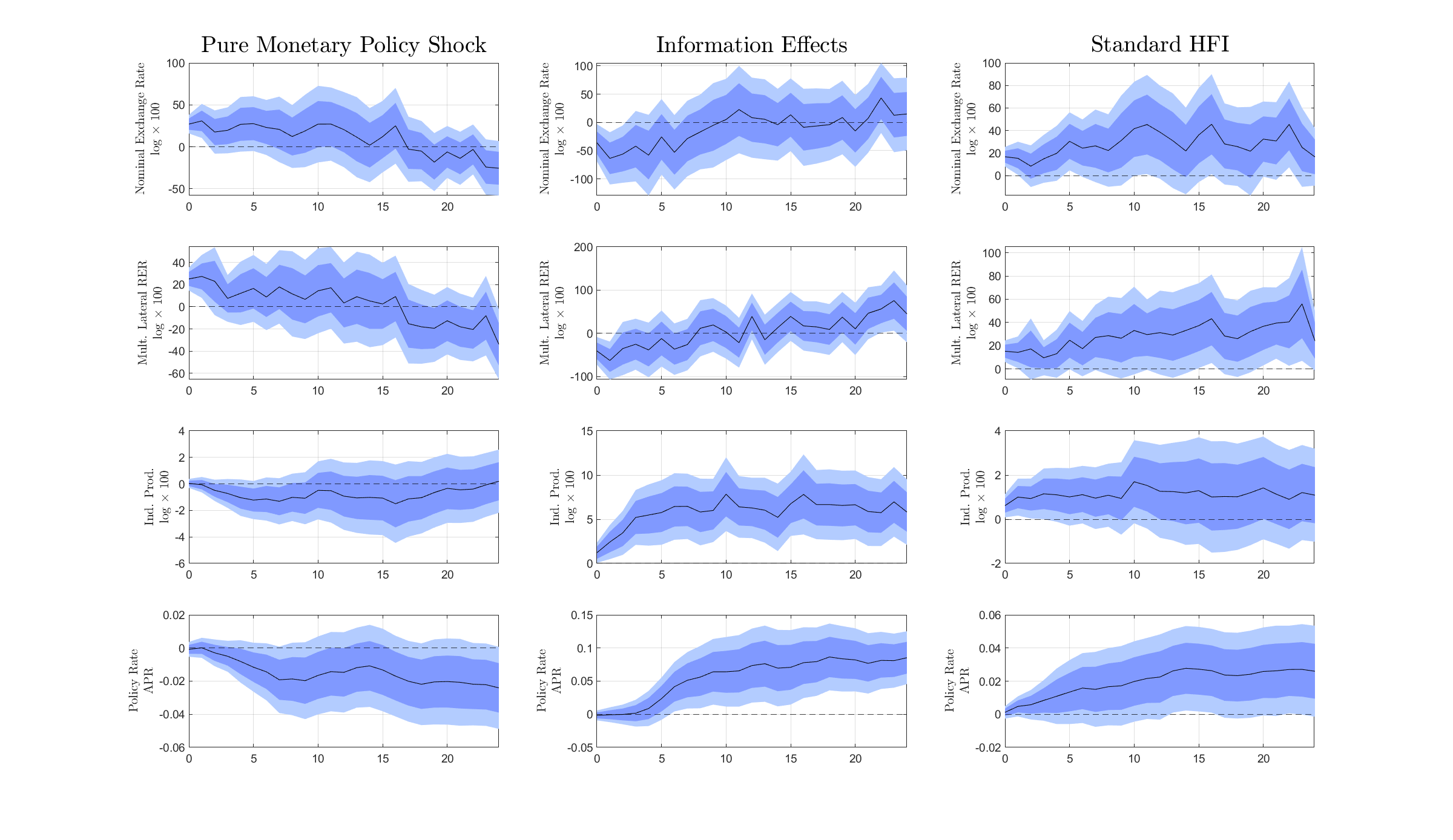}
    \caption{Impulse Response Functions \\ Dataset \cite{ilzetzki2021puzzling} 1998 - Onward - Eq. \ref{eq:LP_Trend}}
    \label{fig:Results_Trend_98}
    \floatfoot{\textbf{Note:} The figure is comprised of 12 sub-figures ordered in three columns and four rows. The left column relates to the estimates of $\beta^{MP}$ in Equation \ref{eq:LP_Trend}, the middle column relates to the estimate of $\beta^{FIE}$ in Equation \ref{eq:LP_Trend}, while the right column relates to estimating Equation \ref{eq:LP_Trend}, replacing the MP and FIE components with the un-orthogonalized monetary policy surprise. The figure presents results for the sample starting in January 1998 onward. The rows represent the impact on (i) the nominal exchange rate with the US dollar; (ii) the real exchange rate with the US; (iii) the industrial production index; (iv) monetary policy rate. I estimate the impulse response functions for the nominal exchange rate, the industrial production and the monetary policy rate using these three variables as controls. For the impulse response function for the real exchange rate, I replace the nominal exchange rate with the real exchange, given the strong correlation between the nominal and real exchange rate with the US dollar.  The solid black line represents the point estimate, the dark blue area represents the 68\% confidence interval, and the light blue area represents the 90\% confidence interval. In the text, when referring to Panel $(i,j)$, $i$ refers to the row and $j$ to the column of the figure. Each variable is as defined in \cite{ilzetzki2021puzzling}. }
\end{figure}

\newpage
\begin{figure}[ht]
    \centering
    \includegraphics[scale=0.4]{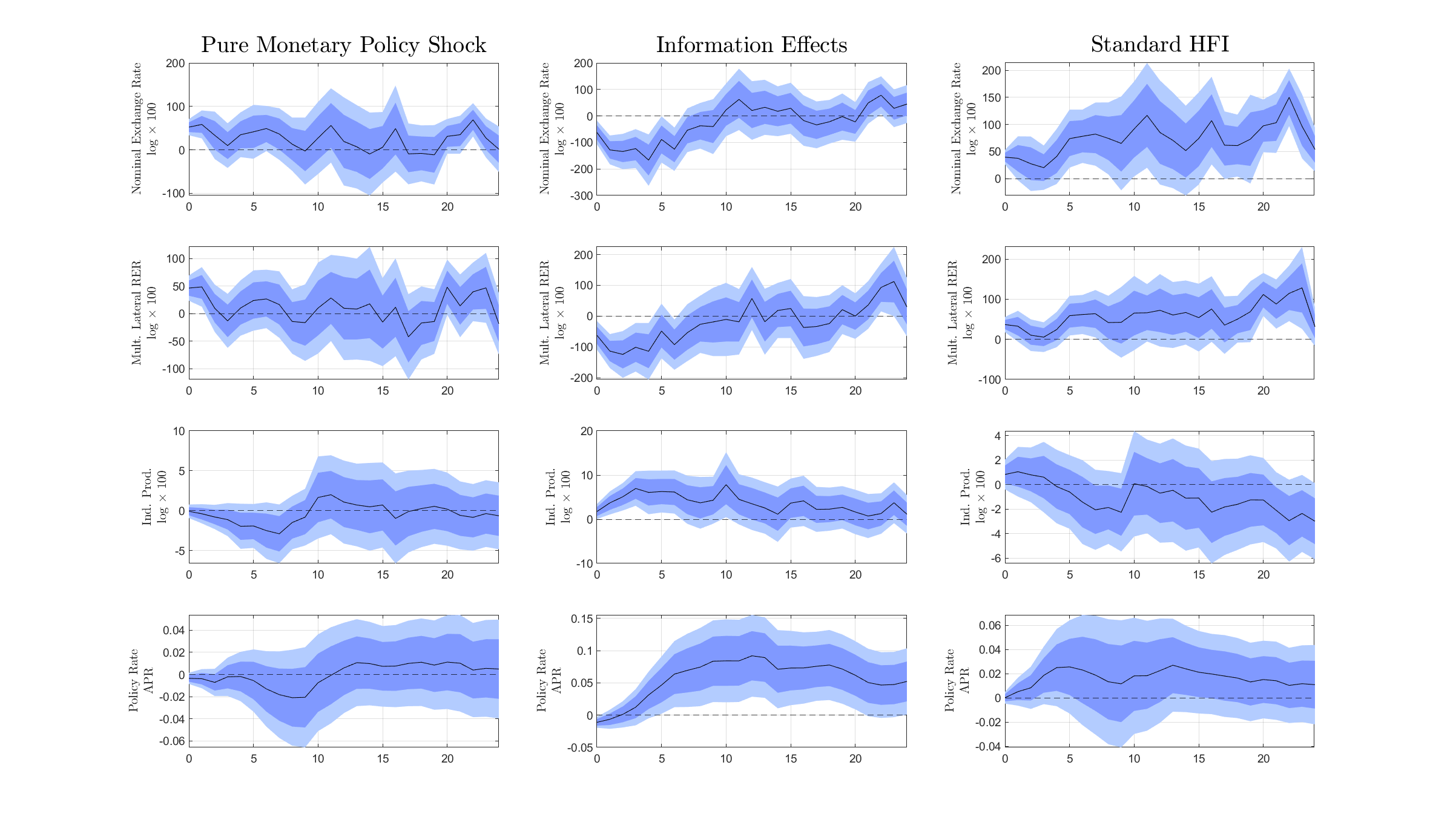}
    \caption{Impulse Response Functions \\ Dataset \cite{ilzetzki2021puzzling} 2008 - Onward - Eq. \ref{eq:LP_Trend}}
    \label{fig:Results_Trend_08}
    \floatfoot{\textbf{Note:} The figure is comprised of 12 sub-figures ordered in three columns and four rows. The left column relates to the estimates of $\beta^{MP}$ in Equation \ref{eq:LP_Trend}, the middle column relates to the estimate of $\beta^{FIE}$ in Equation \ref{eq:LP_Trend}, while the right column relates to estimating Equation \ref{eq:LP_Trend}, replacing the MP and FIE components with the un-orthogonalized monetary policy surprise. The figure presents results for the sample starting in January 2008 onward. The rows represent the impact on (i) the nominal exchange rate with the US dollar; (ii) the real exchange rate with the US; (iii) the industrial production index; (iv) monetary policy rate. I estimate the impulse response functions for the nominal exchange rate, the industrial production and the monetary policy rate using these three variables as controls. For the impulse response function for the real exchange rate, I replace the nominal exchange rate with the real exchange, given the strong correlation between the nominal and real exchange rate with the US dollar.  The solid black line represents the point estimate, the dark blue area represents the 68\% confidence interval, and the light blue area represents the 90\% confidence interval. In the text, when referring to Panel $(i,j)$, $i$ refers to the row and $j$ to the column of the figure. Each variable is as defined in \cite{ilzetzki2021puzzling}. }
\end{figure}

\newpage
\begin{figure}[ht]
    \centering
    \includegraphics[scale=0.4]{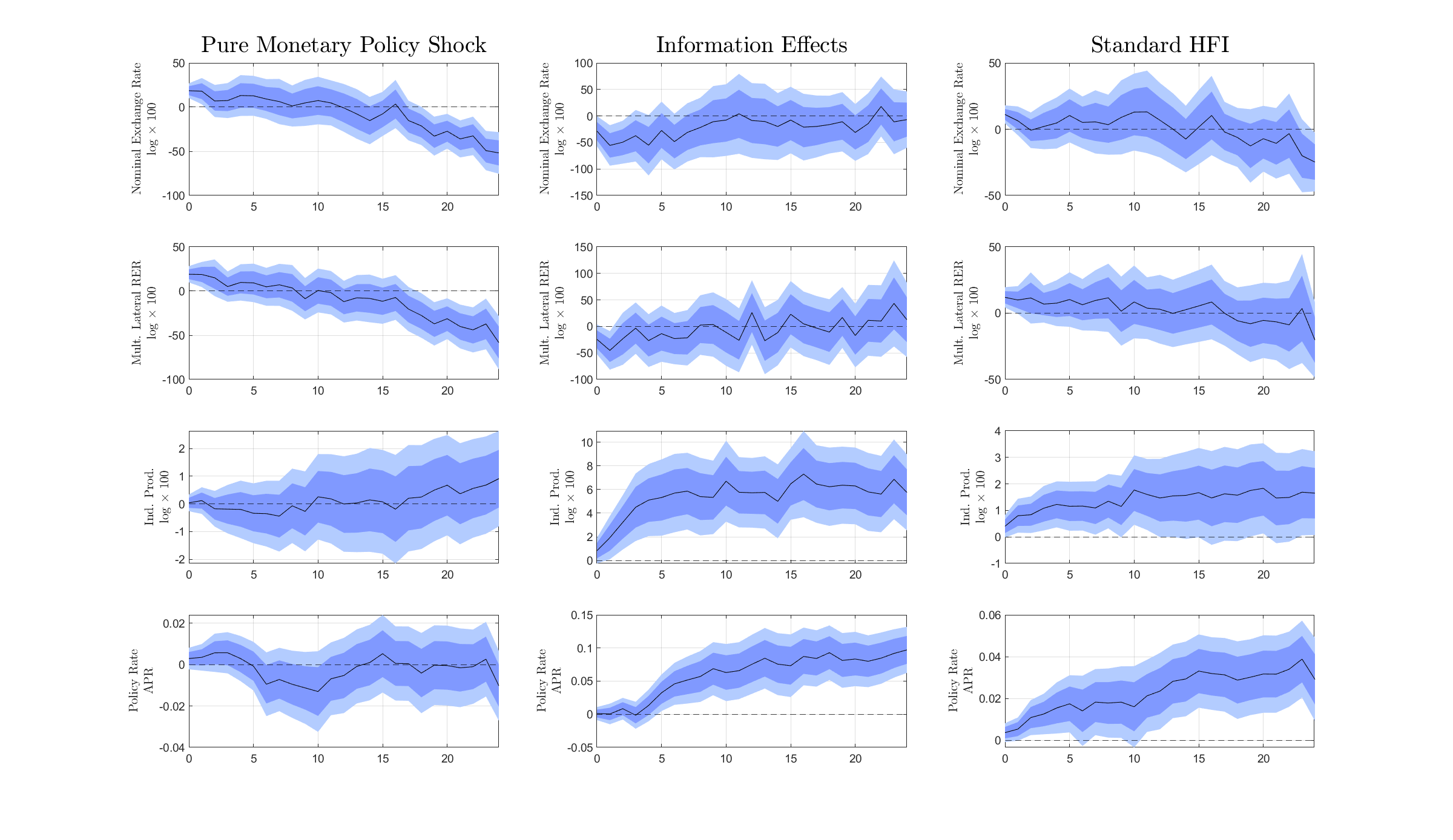}
    \caption{Impulse Response Functions \\ Dataset \cite{ilzetzki2021puzzling} - Eq. \ref{eq:LP_Trend_FE} }
    \label{fig:Results_Trend_FE}
    \floatfoot{\textbf{Note:} The figure is comprised of 12 sub-figures ordered in three columns and four rows. The left column relates to the estimates of $\beta^{MP}$ in Equation \ref{eq:LP_Trend_FE}, the middle column relates to the estimate of $\beta^{FIE}$ in Equation \ref{eq:LP_Trend_FE}, while the right column relates to estimating Equation \ref{eq:LP_Trend_FE}, replacing the MP and FIE components with the un-orthogonalized monetary policy surprise. The rows represent the impact on (i) the nominal exchange rate with the US dollar; (ii) the real exchange rate with the US; (iii) the industrial production index; (iv) monetary policy rate. I estimate the impulse response functions for the nominal exchange rate, the industrial production and the monetary policy rate using these three variables as controls. For the impulse response function for the real exchange rate, I replace the nominal exchange rate with the real exchange, given the strong correlation between the nominal and real exchange rate with the US dollar. The solid black line represents the point estimate, the dark blue area represents the 68\% confidence interval, and the light blue area represents the 90\% confidence interval. In the text, when referring to Panel $(i,j)$, $i$ refers to the row and $j$ to the column of the figure. Each variable is as defined in \cite{ilzetzki2021puzzling}. }
\end{figure}

\newpage
\begin{figure}[ht]
    \centering
    \includegraphics[scale=0.4]{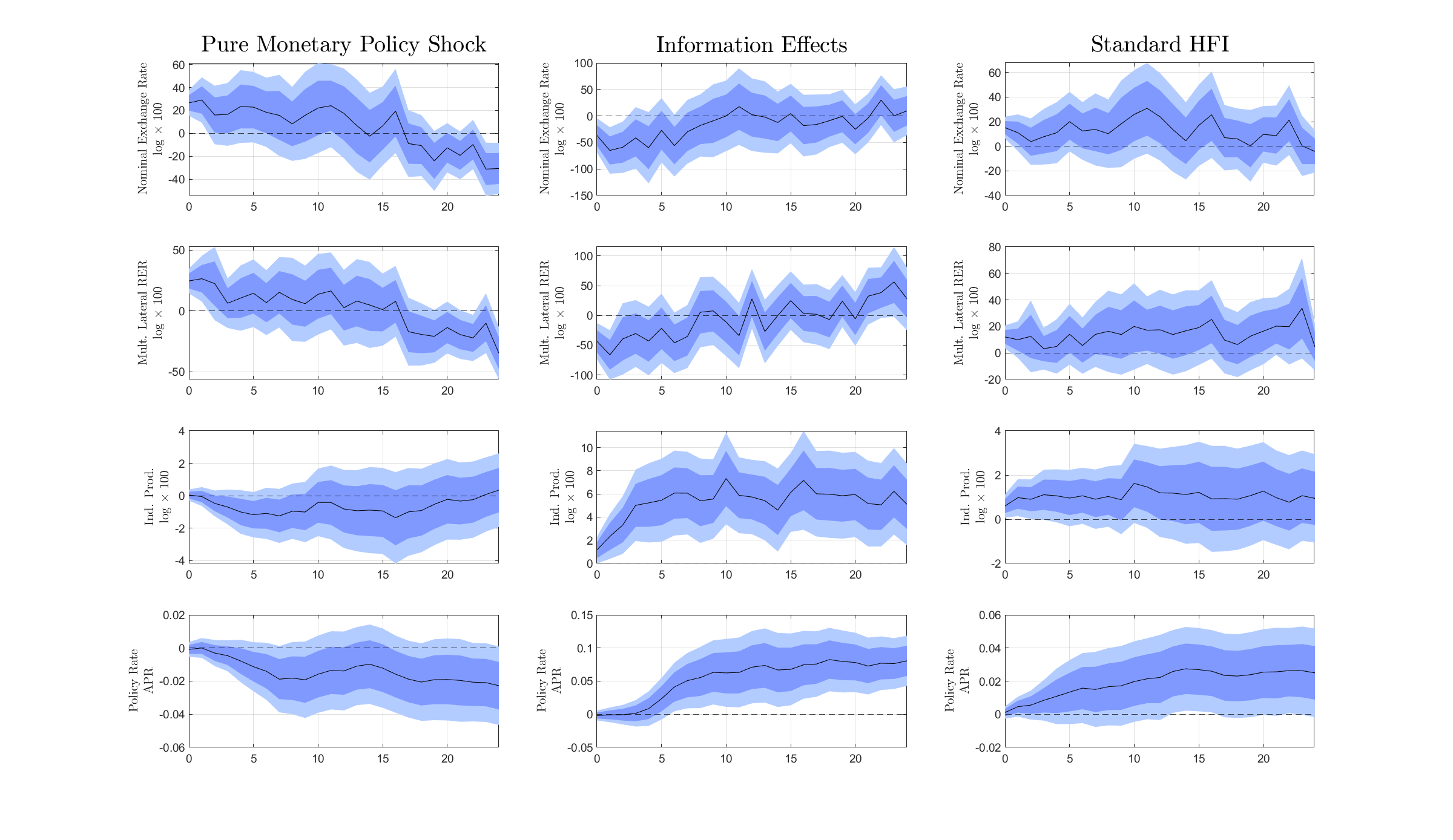}
    \caption{Impulse Response Functions \\ Dataset \cite{ilzetzki2021puzzling} 1998 - Onward - Eq. \ref{eq:LP_Trend_FE}}
    \label{fig:Results_Trend_FE_98}
    \floatfoot{\textbf{Note:} The figure is comprised of 12 sub-figures ordered in three columns and four rows. The left column relates to the estimates of $\beta^{MP}$ in Equation \ref{eq:LP_Trend_FE}, the middle column relates to the estimate of $\beta^{FIE}$ in Equation \ref{eq:LP_Trend_FE}, while the right column relates to estimating Equation \ref{eq:LP_Trend_FE}, replacing the MP and FIE components with the un-orthogonalized monetary policy surprise. The figure presents results for the sample starting in January 1998 onward. The rows represent the impact on (i) the nominal exchange rate with the US dollar; (ii) the real exchange rate with the US; (iii) the industrial production index; (iv) monetary policy rate. I estimate the impulse response functions for the nominal exchange rate, the industrial production and the monetary policy rate using these three variables as controls. For the impulse response function for the real exchange rate, I replace the nominal exchange rate with the real exchange, given the strong correlation between the nominal and real exchange rate with the US dollar. The solid black line represents the point estimate, the dark blue area represents the 68\% confidence interval, and the light blue area represents the 90\% confidence interval. In the text, when referring to Panel $(i,j)$, $i$ refers to the row and $j$ to the column of the figure. Each variable is as defined in \cite{ilzetzki2021puzzling}. }
\end{figure}

\newpage
\begin{figure}[ht]
    \centering
    \includegraphics[scale=0.4]{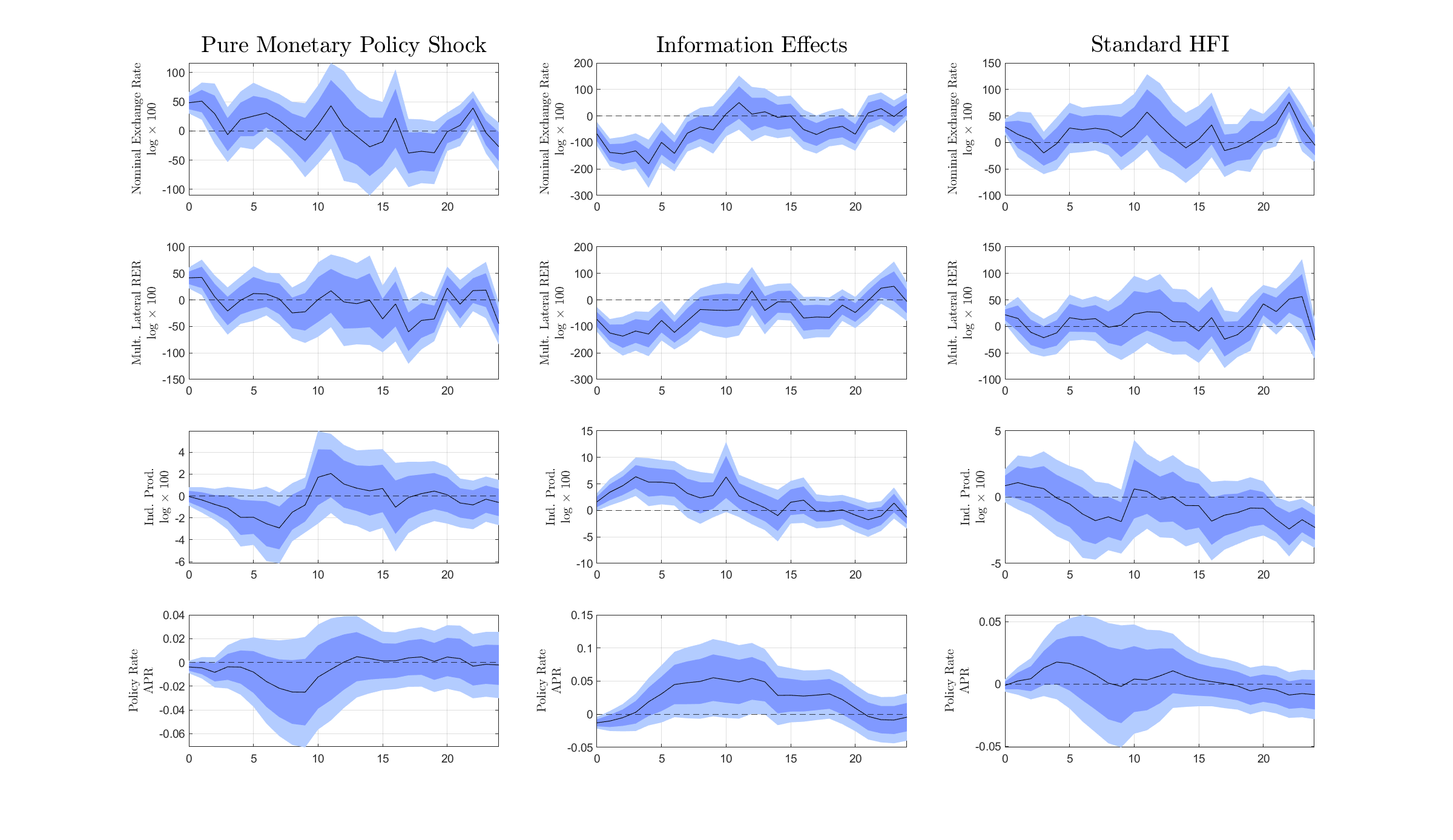}
    \caption{Impulse Response Functions \\ Dataset \cite{ilzetzki2021puzzling} 2008 - Onward - Eq. \ref{eq:LP_Trend_FE}}
    \label{fig:Results_Trend_FE_08}
    \floatfoot{\textbf{Note:} The figure is comprised of 12 sub-figures ordered in three columns and four rows. The left column relates to the estimates of $\beta^{MP}$ in Equation \ref{eq:LP_Trend_FE}, the middle column relates to the estimate of $\beta^{FIE}$ in Equation \ref{eq:LP_Trend_FE}, while the right column relates to estimating Equation \ref{eq:LP_Trend_FE}, replacing the MP and FIE components with the un-orthogonalized monetary policy surprise. The figure presents results for the sample starting in January 2008 onward. The rows represent the impact on (i) the nominal exchange rate with the US dollar; (ii) the real exchange rate with the US; (iii) the industrial production index; (iv) monetary policy rate. I estimate the impulse response functions for the nominal exchange rate, the industrial production and the monetary policy rate using these three variables as controls. For the impulse response function for the real exchange rate, I replace the nominal exchange rate with the real exchange, given the strong correlation between the nominal and real exchange rate with the US dollar. The solid black line represents the point estimate, the dark blue area represents the 68\% confidence interval, and the light blue area represents the 90\% confidence interval. In the text, when referring to Panel $(i,j)$, $i$ refers to the row and $j$ to the column of the figure. Each variable is as defined in \cite{ilzetzki2021puzzling}. }
\end{figure}

\newpage
\begin{figure}
    \centering
    \includegraphics[scale=0.4]{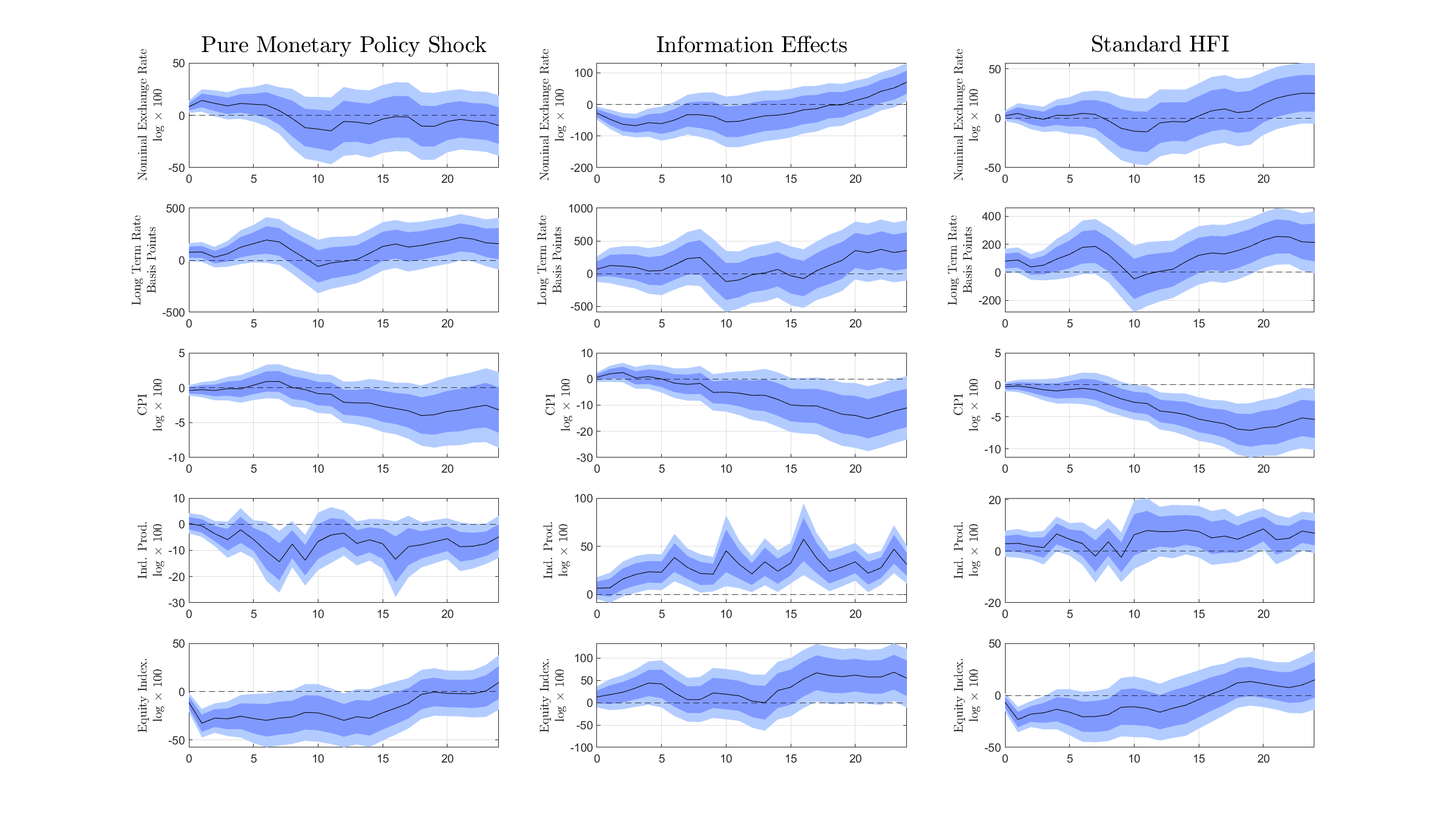}
    \caption{Impulse Response Functions \\ Results for Africa}
    \label{fig:BenchmarkNER_AF}
    \floatfoot{\textbf{Note:} The figure is comprised of 15 sub-figures ordered in three columns and five rows. The left column relates to the estimates of $\beta^{MP}$ in Equation \ref{eq:LP_pooled}, the middle column relates to the estimate of $\beta^{FIE}$ in Equation \ref{eq:LP_pooled}, while the right column relates to estimating Equation \ref{eq:LP_pooled}, replacing the MP and FIE components with the un-orthogonalized monetary policy surprise. The rows represent the impact on (i) the nominal exchange rate with the US dollar (in logs times 100); (ii) long term interest rates in basis points; (iii) the consumer price index (in logs times 100); (iv) the industrial production index (in logs times 100); (v) the equity index (in logs times 100). The solid black line represents the point estimate, the dark blue area represents the 68\% confidence interval, and the light blue area represents the 90\% confidence interval. In the text, when referring to Panel $(i,j)$, $i$ refers to the row and $j$ to the column of the figure. Each variable, in its own transformation, is demeaned at the country level. Results for countries in Africa alone.}
\end{figure}

\newpage
\begin{figure}
    \centering
    \includegraphics[scale=0.4]{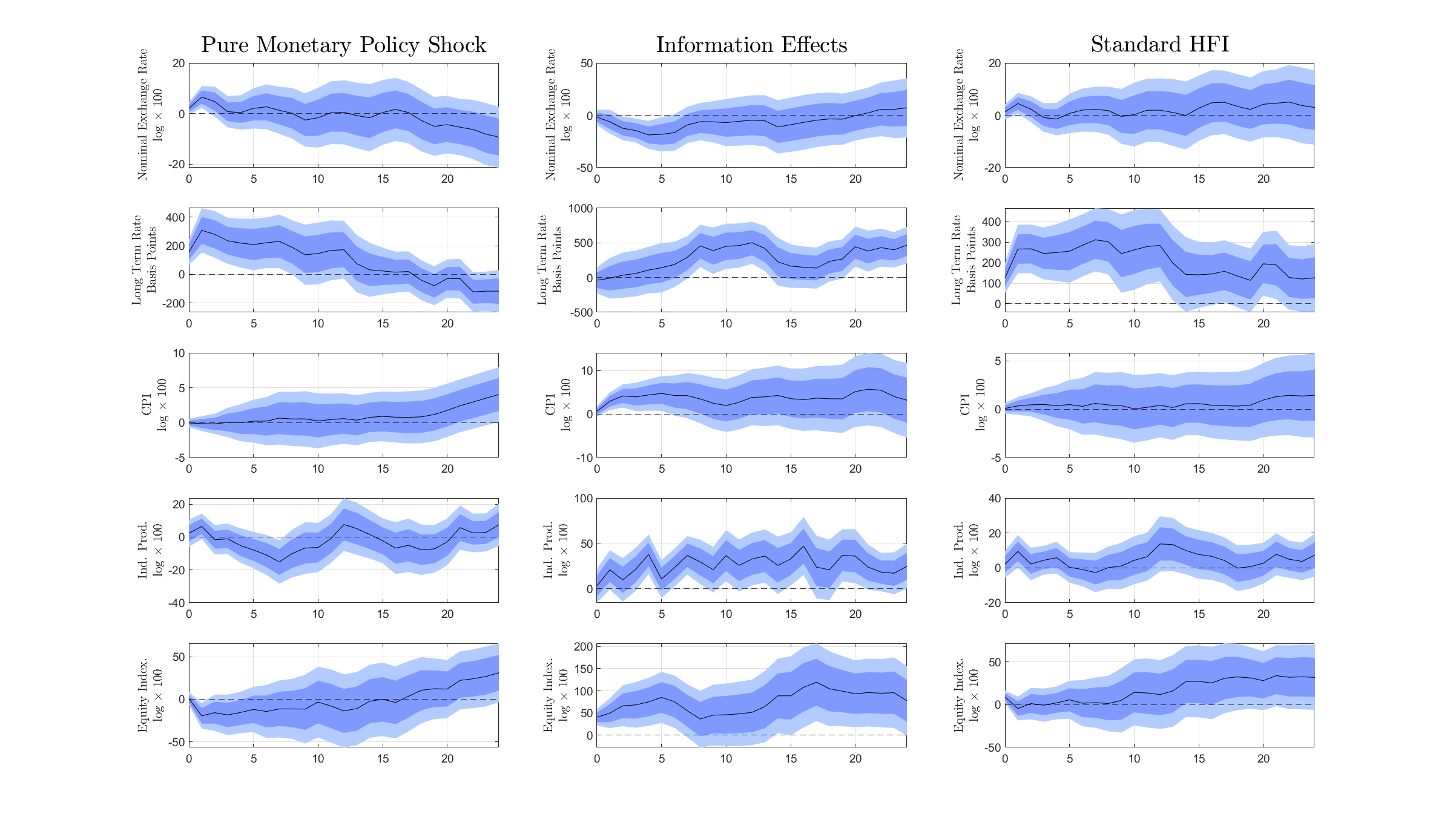}
    \caption{Impulse Response Functions \\ Results for Asia}
    \label{fig:BenchmarkNER_AS}
    \floatfoot{\textbf{Note:} The figure is comprised of 15 sub-figures ordered in three columns and five rows. The left column relates to the estimates of $\beta^{MP}$ in Equation \ref{eq:LP_pooled}, the middle column relates to the estimate of $\beta^{FIE}$ in Equation \ref{eq:LP_pooled}, while the right column relates to estimating Equation \ref{eq:LP_pooled}, replacing the MP and FIE components with the un-orthogonalized monetary policy surprise. The rows represent the impact on (i) the nominal exchange rate with the US dollar (in logs times 100); (ii) long term interest rates in basis points; (iii) the consumer price index (in logs times 100); (iv) the industrial production index (in logs times 100); (v) the equity index (in logs times 100). The solid black line represents the point estimate, the dark blue area represents the 68\% confidence interval, and the light blue area represents the 90\% confidence interval. In the text, when referring to Panel $(i,j)$, $i$ refers to the row and $j$ to the column of the figure. Each variable, in its own transformation, is demeaned at the country level. Results for countries in Asia alone.}
\end{figure}

\newpage
\begin{figure}
    \centering
    \includegraphics[scale=0.4]{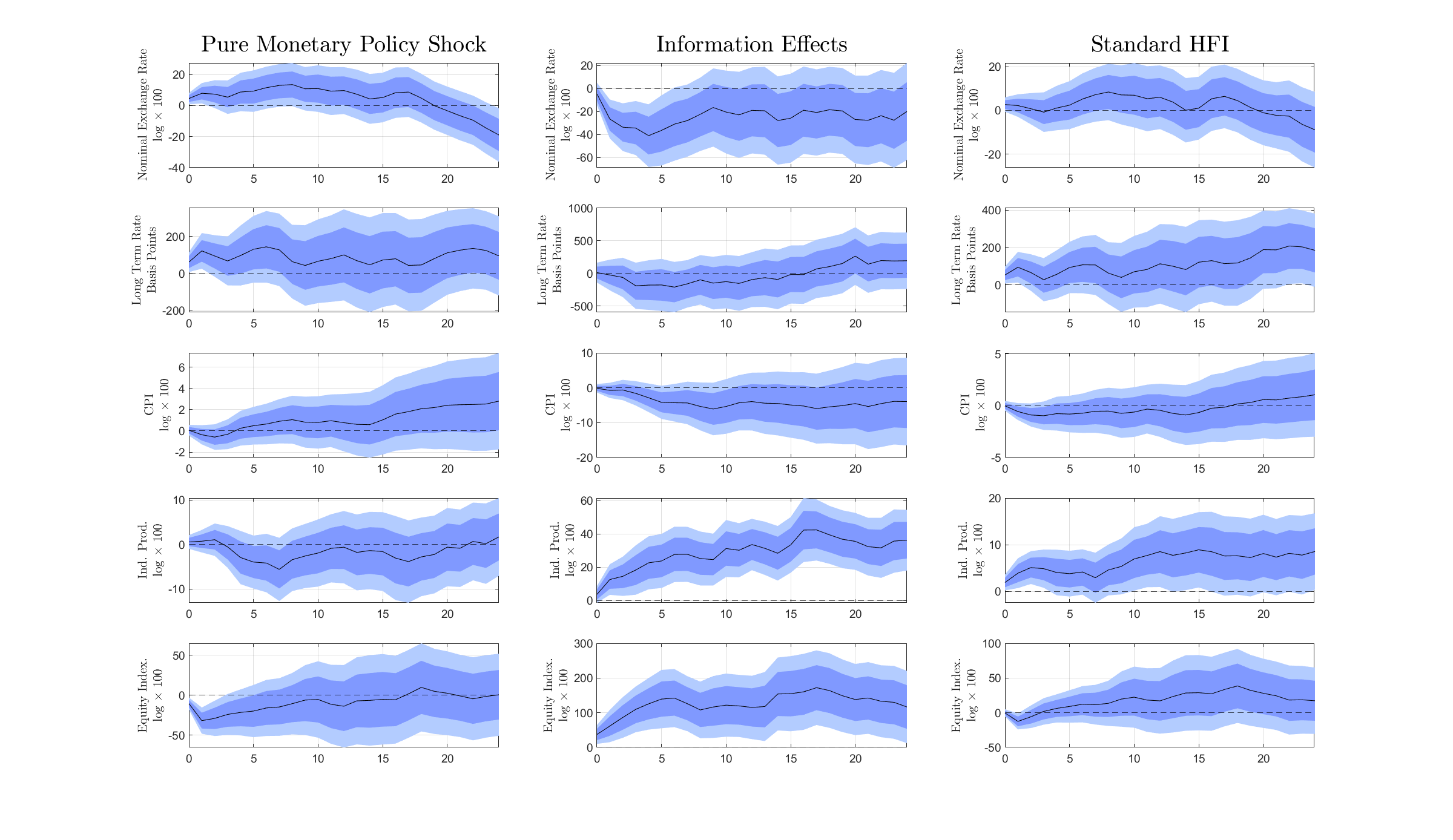}
    \caption{Impulse Response Functions \\ Results for Europe}
    \label{fig:BenchmarkNER_EU}
    \floatfoot{\textbf{Note:} The figure is comprised of 15 sub-figures ordered in three columns and five rows. The left column relates to the estimates of $\beta^{MP}$ in Equation \ref{eq:LP_pooled}, the middle column relates to the estimate of $\beta^{FIE}$ in Equation \ref{eq:LP_pooled}, while the right column relates to estimating Equation \ref{eq:LP_pooled}, replacing the MP and FIE components with the un-orthogonalized monetary policy surprise. The rows represent the impact on (i) the nominal exchange rate with the US dollar (in logs times 100); (ii) long term interest rates in basis points; (iii) the consumer price index (in logs times 100); (iv) the industrial production index (in logs times 100); (v) the equity index (in logs times 100). The solid black line represents the point estimate, the dark blue area represents the 68\% confidence interval, and the light blue area represents the 90\% confidence interval. In the text, when referring to Panel $(i,j)$, $i$ refers to the row and $j$ to the column of the figure. Each variable, in its own transformation, is demeaned at the country level. Results for countries in Europe alone.}
\end{figure}

\newpage
\begin{figure}
    \centering
    \includegraphics[scale=0.4]{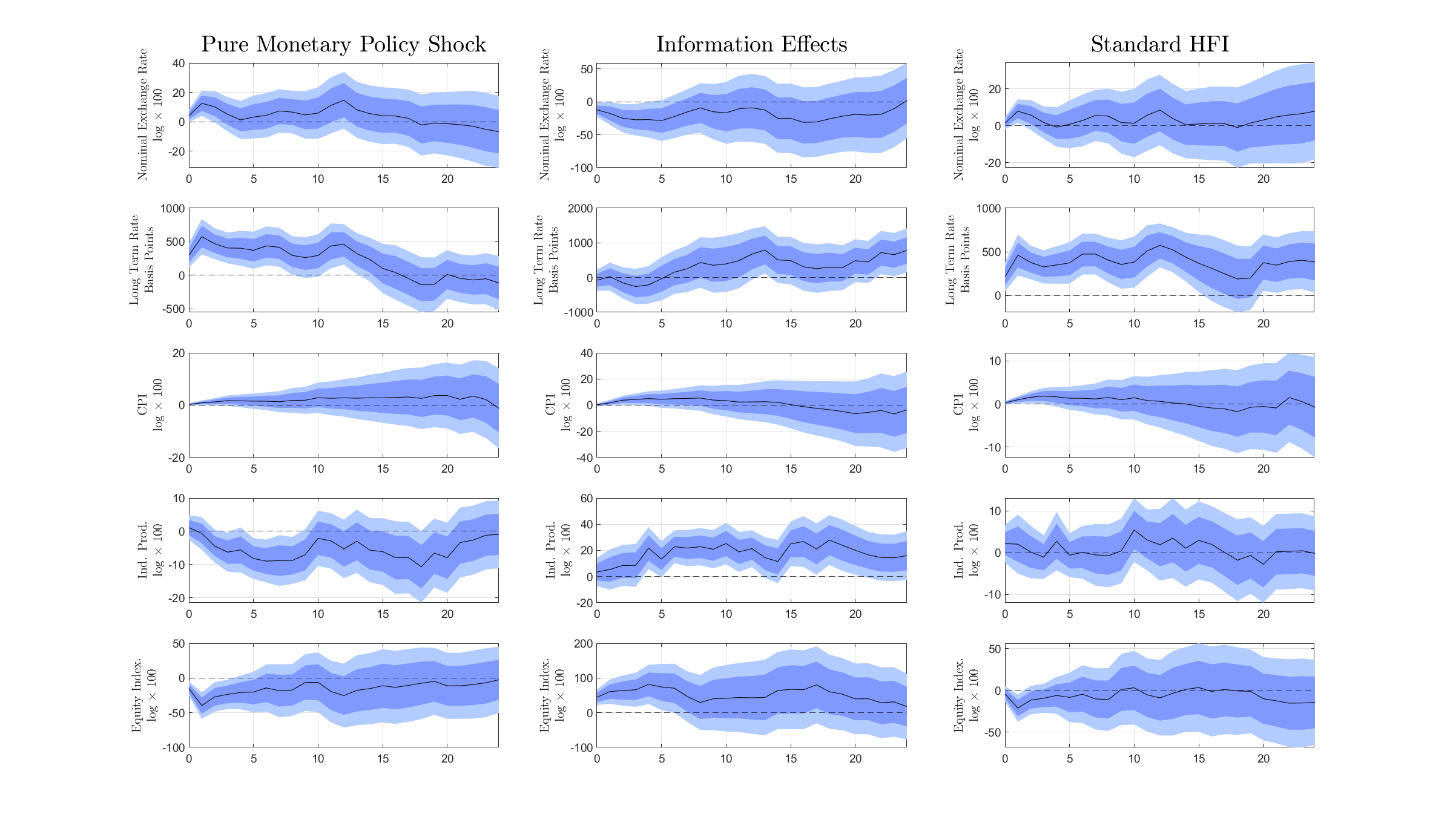}
    \caption{Impulse Response Functions \\ Results for Latin America}
    \label{fig:BenchmarkNER_LA}
    \floatfoot{\textbf{Note:} The figure is comprised of 15 sub-figures ordered in three columns and five rows. The left column relates to the estimates of $\beta^{MP}$ in Equation \ref{eq:LP_pooled}, the middle column relates to the estimate of $\beta^{FIE}$ in Equation \ref{eq:LP_pooled}, while the right column relates to estimating Equation \ref{eq:LP_pooled}, replacing the MP and FIE components with the un-orthogonalized monetary policy surprise. The rows represent the impact on (i) the nominal exchange rate with the US dollar (in logs times 100); (ii) long term interest rates in basis points; (iii) the consumer price index (in logs times 100); (iv) the industrial production index (in logs times 100); (v) the equity index (in logs times 100). The solid black line represents the point estimate, the dark blue area represents the 68\% confidence interval, and the light blue area represents the 90\% confidence interval. In the text, when referring to Panel $(i,j)$, $i$ refers to the row and $j$ to the column of the figure. Each variable, in its own transformation, is demeaned at the country level. Results for countries in Latin America alone.}
\end{figure}

\newpage
\begin{figure}
    \centering
    \includegraphics[scale=0.4]{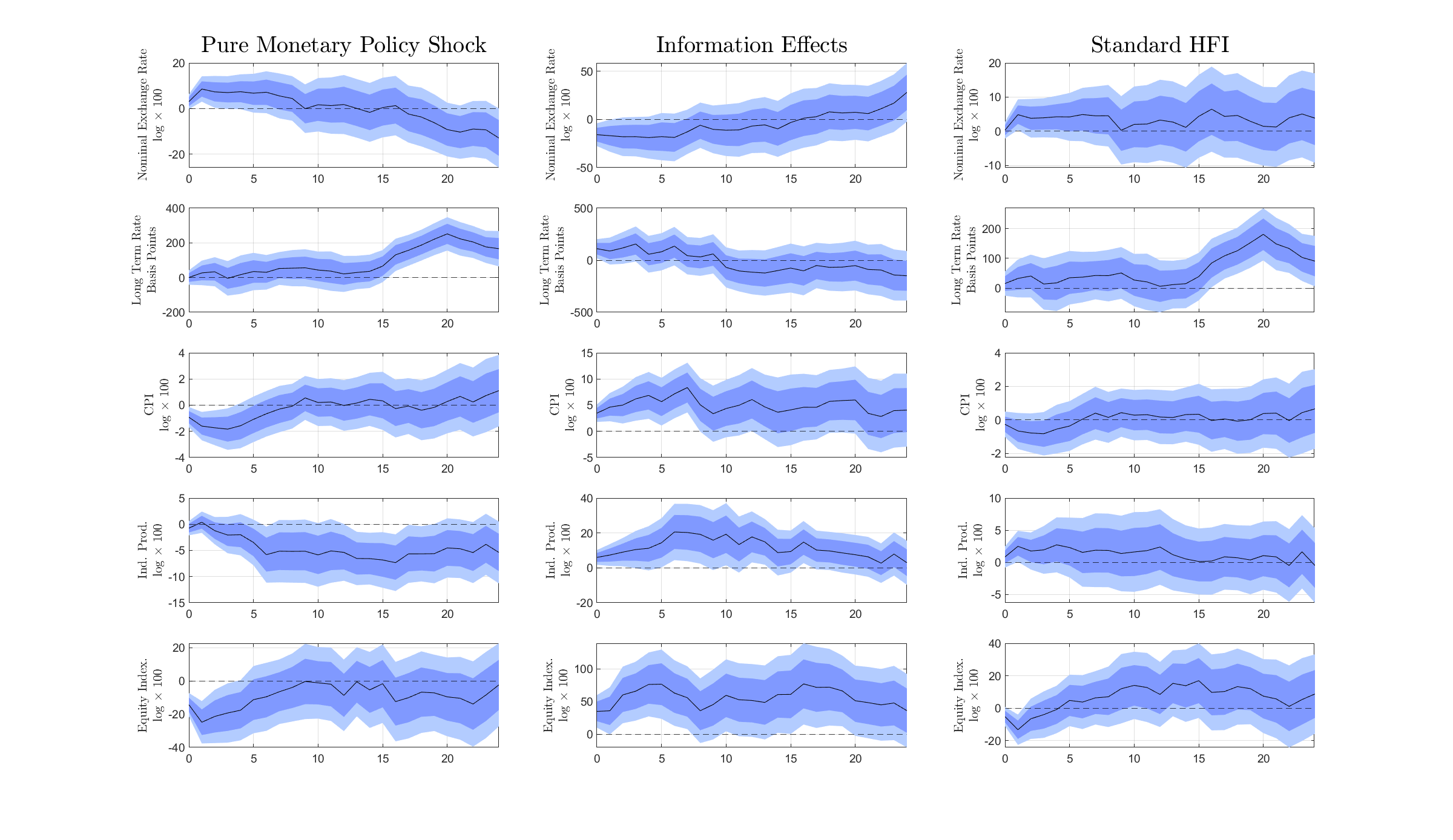}
    \caption{Impulse Response Functions \\ Results for North America}
    \label{fig:BenchmarkNER_NA}
    \floatfoot{\textbf{Note:} The figure is comprised of 15 sub-figures ordered in three columns and five rows. The left column relates to the estimates of $\beta^{MP}$ in Equation \ref{eq:LP_pooled}, the middle column relates to the estimate of $\beta^{FIE}$ in Equation \ref{eq:LP_pooled}, while the right column relates to estimating Equation \ref{eq:LP_pooled}, replacing the MP and FIE components with the un-orthogonalized monetary policy surprise. The rows represent the impact on (i) the nominal exchange rate with the US dollar (in logs times 100); (ii) long term interest rates in basis points; (iii) the consumer price index (in logs times 100); (iv) the industrial production index (in logs times 100); (v) the equity index (in logs times 100). The solid black line represents the point estimate, the dark blue area represents the 68\% confidence interval, and the light blue area represents the 90\% confidence interval. In the text, when referring to Panel $(i,j)$, $i$ refers to the row and $j$ to the column of the figure. Each variable, in its own transformation, is demeaned at the country level. Results for countries in North America alone.}
\end{figure}

\newpage
\begin{figure}
    \centering
    \includegraphics[scale=0.4]{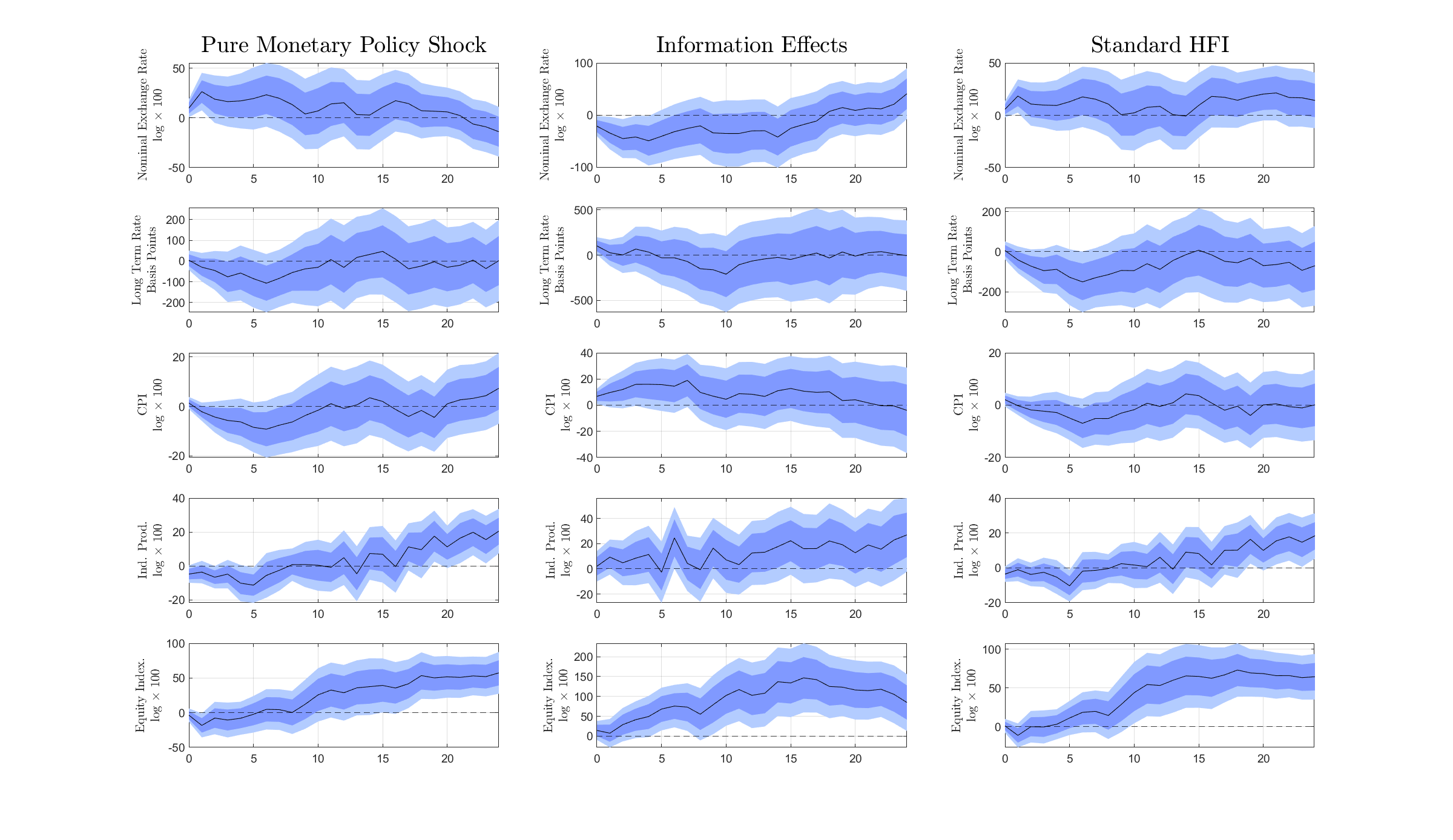}
    \caption{Impulse Response Functions \\ Results for Oceania }
    \label{fig:BenchmarkNER_OC}
    \floatfoot{\textbf{Note:} The figure is comprised of 15 sub-figures ordered in three columns and five rows. The left column relates to the estimates of $\beta^{MP}$ in Equation \ref{eq:LP_pooled}, the middle column relates to the estimate of $\beta^{FIE}$ in Equation \ref{eq:LP_pooled}, while the right column relates to estimating Equation \ref{eq:LP_pooled}, replacing the MP and FIE components with the un-orthogonalized monetary policy surprise. The rows represent the impact on (i) the nominal exchange rate with the US dollar (in logs times 100); (ii) long term interest rates in basis points; (iii) the consumer price index (in logs times 100); (iv) the industrial production index (in logs times 100); (v) the equity index (in logs times 100). The solid black line represents the point estimate, the dark blue area represents the 68\% confidence interval, and the light blue area represents the 90\% confidence interval. In the text, when referring to Panel $(i,j)$, $i$ refers to the row and $j$ to the column of the figure. Each variable, in its own transformation, is demeaned at the country level. Results for countries in Oceania alone.}
\end{figure}

\newpage
\begin{figure}
    \centering
    \includegraphics[scale=0.4]{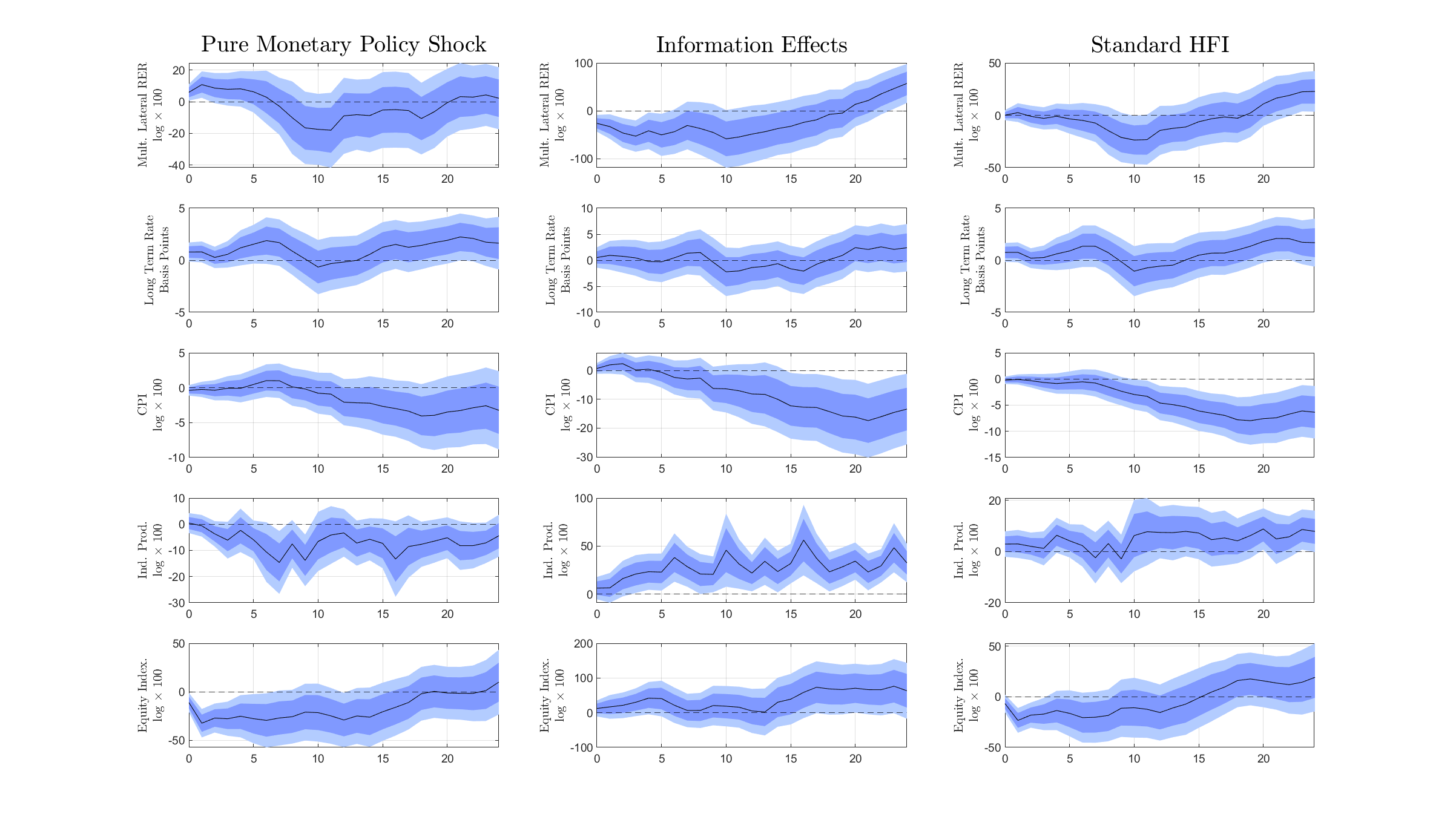}
    \caption{Impulse Response Functions \\ Multi. REER Sample -Results for Africa}
    \label{fig:BenchmarkREER_AF}
    \floatfoot{\textbf{Note:} The figure is comprised of 15 sub-figures ordered in three columns and five rows. The left column relates to the estimates of $\beta^{MP}$ in Equation \ref{eq:LP_pooled}, the middle column relates to the estimate of $\beta^{FIE}$ in Equation \ref{eq:LP_pooled}, while the right column relates to estimating Equation \ref{eq:LP_pooled}, replacing the MP and FIE components with the un-orthogonalized monetary policy surprise. The rows represent the impact on (i) the multilateral trade weighted real exchange rate (in logs times 100); (ii) long term interest rates in basis points; (iii) the consumer price index (in logs times 100); (iv) the industrial production index (in logs times 100); (v) the equity index (in logs times 100). The solid black line represents the point estimate, the dark blue area represents the 68\% confidence interval, and the light blue area represents the 90\% confidence interval. In the text, when referring to Panel $(i,j)$, $i$ refers to the row and $j$ to the column of the figure. Each variable, in its own transformation, is demeaned at the country level. Results for countries in Africa alone.}
\end{figure}

\newpage
\begin{figure}
    \centering
    \includegraphics[scale=0.4]{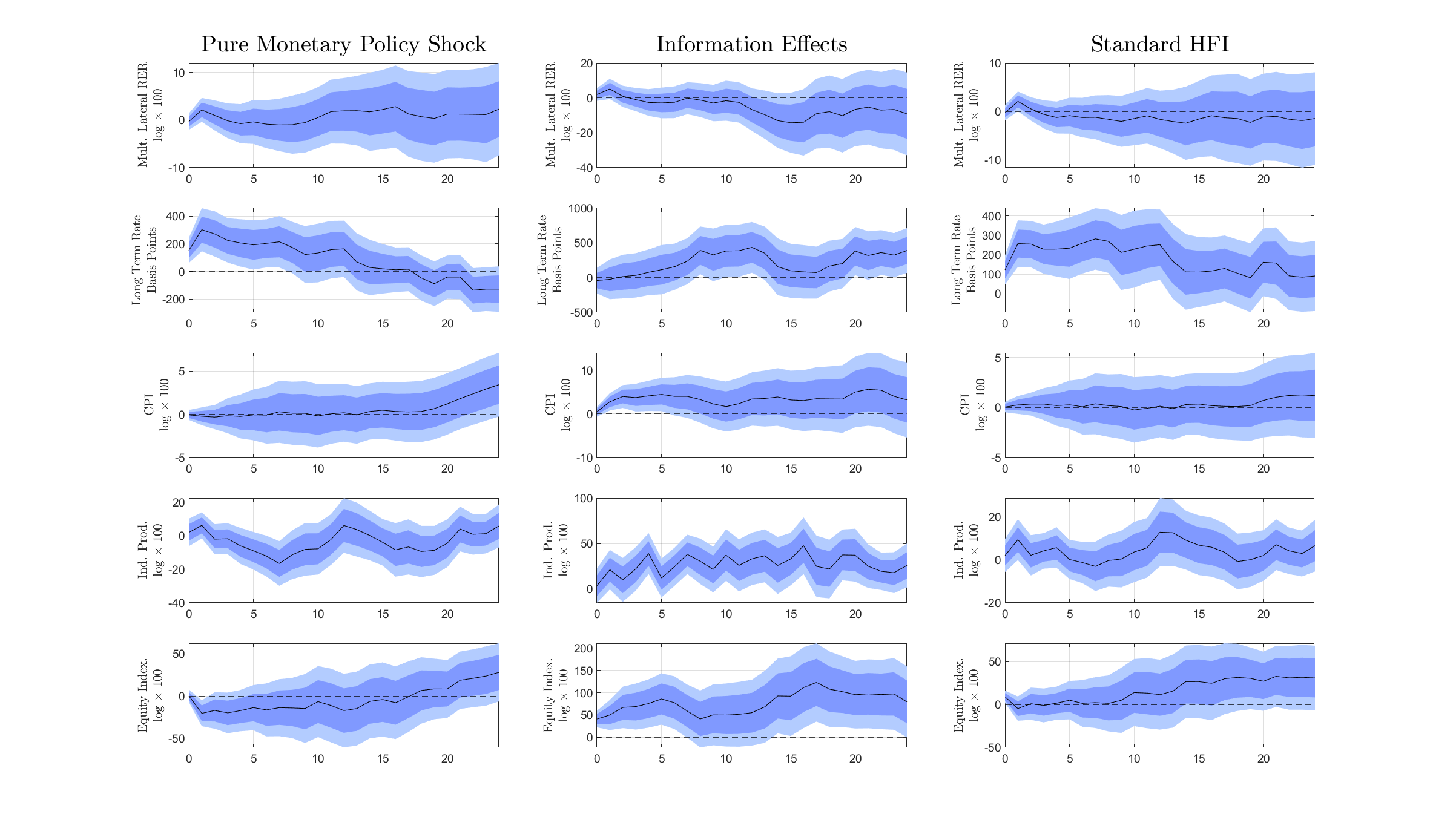}
    \caption{Impulse Response Functions \\ Multi. REER Sample -Results for Asia}
    \label{fig:BenchmarkREER_AS}
    \floatfoot{\textbf{Note:} The figure is comprised of 15 sub-figures ordered in three columns and five rows. The left column relates to the estimates of $\beta^{MP}$ in Equation \ref{eq:LP_pooled}, the middle column relates to the estimate of $\beta^{FIE}$ in Equation \ref{eq:LP_pooled}, while the right column relates to estimating Equation \ref{eq:LP_pooled}, replacing the MP and FIE components with the un-orthogonalized monetary policy surprise. The rows represent the impact on (i) the multilateral trade weighted real exchange rate (in logs times 100); (ii) long term interest rates in basis points; (iii) the consumer price index (in logs times 100); (iv) the industrial production index (in logs times 100); (v) the equity index (in logs times 100). The solid black line represents the point estimate, the dark blue area represents the 68\% confidence interval, and the light blue area represents the 90\% confidence interval. In the text, when referring to Panel $(i,j)$, $i$ refers to the row and $j$ to the column of the figure. Each variable, in its own transformation, is demeaned at the country level. Results for countries in Asia alone.}
\end{figure}

\newpage
\begin{figure}
    \centering
    \includegraphics[scale=0.4]{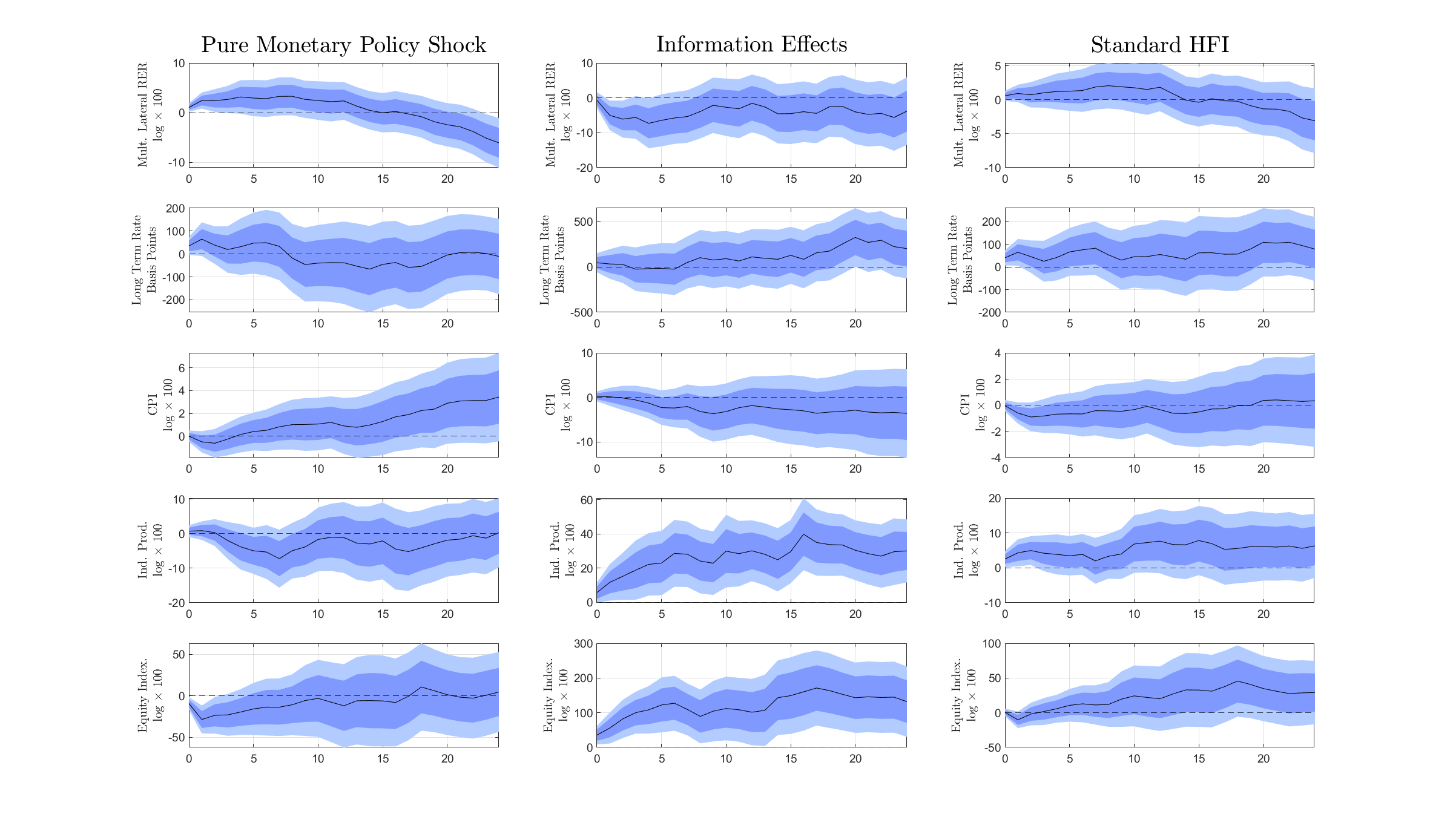}
    \caption{Impulse Response Functions \\ Multi. REER Sample -Results for Europe}
    \label{fig:BenchmarkREER_EU}
    \floatfoot{\textbf{Note:} The figure is comprised of 15 sub-figures ordered in three columns and five rows. The left column relates to the estimates of $\beta^{MP}$ in Equation \ref{eq:LP_pooled}, the middle column relates to the estimate of $\beta^{FIE}$ in Equation \ref{eq:LP_pooled}, while the right column relates to estimating Equation \ref{eq:LP_pooled}, replacing the MP and FIE components with the un-orthogonalized monetary policy surprise. The rows represent the impact on (i) the multilateral trade weighted real exchange rate (in logs times 100); (ii) long term interest rates in basis points; (iii) the consumer price index (in logs times 100); (iv) the industrial production index (in logs times 100); (v) the equity index (in logs times 100). The solid black line represents the point estimate, the dark blue area represents the 68\% confidence interval, and the light blue area represents the 90\% confidence interval. In the text, when referring to Panel $(i,j)$, $i$ refers to the row and $j$ to the column of the figure. Each variable, in its own transformation, is demeaned at the country level. Results for countries in Europe alone.}
\end{figure}

\newpage
\begin{figure}
    \centering
    \includegraphics[scale=0.4]{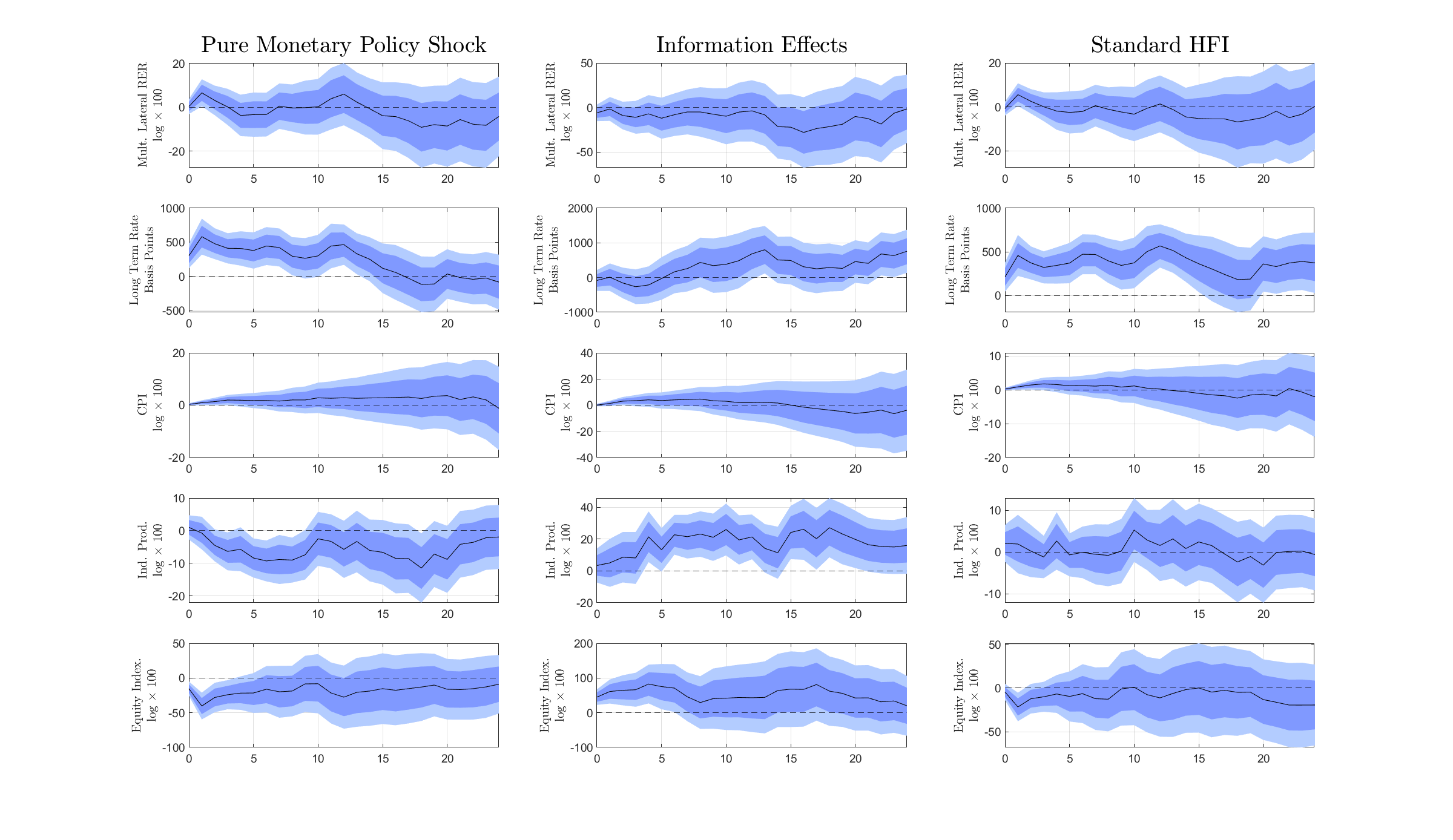}
    \caption{Impulse Response Functions \\ Multi. REER Sample -Results for Latin America}
    \label{fig:BenchmarkREER_LA}
    \floatfoot{\textbf{Note:} The figure is comprised of 15 sub-figures ordered in three columns and five rows. The left column relates to the estimates of $\beta^{MP}$ in Equation \ref{eq:LP_pooled}, the middle column relates to the estimate of $\beta^{FIE}$ in Equation \ref{eq:LP_pooled}, while the right column relates to estimating Equation \ref{eq:LP_pooled}, replacing the MP and FIE components with the un-orthogonalized monetary policy surprise. The rows represent the impact on (i) the multilateral trade weighted real exchange rate (in logs times 100); (ii) long term interest rates in basis points; (iii) the consumer price index (in logs times 100); (iv) the industrial production index (in logs times 100); (v) the equity index (in logs times 100). The solid black line represents the point estimate, the dark blue area represents the 68\% confidence interval, and the light blue area represents the 90\% confidence interval. In the text, when referring to Panel $(i,j)$, $i$ refers to the row and $j$ to the column of the figure. Each variable, in its own transformation, is demeaned at the country level. Results for countries in Latin America alone.}
\end{figure}

\newpage
\begin{figure}
    \centering
    \includegraphics[scale=0.4]{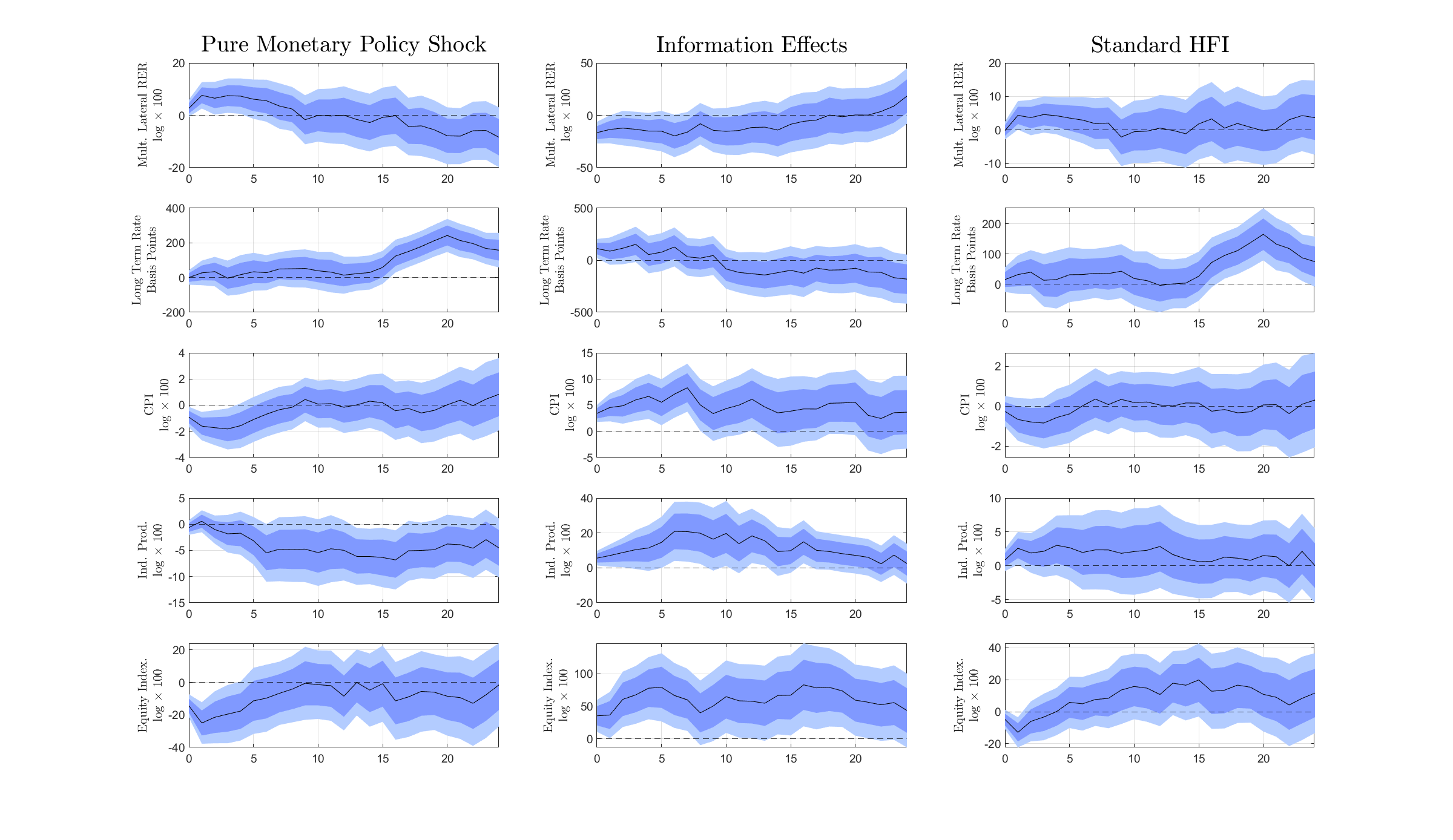}
    \caption{Impulse Response Functions \\ Multi. REER Sample -Results for North America}
    \label{fig:BenchmarkREER_NA}
    \floatfoot{\textbf{Note:} The figure is comprised of 15 sub-figures ordered in three columns and five rows. The left column relates to the estimates of $\beta^{MP}$ in Equation \ref{eq:LP_pooled}, the middle column relates to the estimate of $\beta^{FIE}$ in Equation \ref{eq:LP_pooled}, while the right column relates to estimating Equation \ref{eq:LP_pooled}, replacing the MP and FIE components with the un-orthogonalized monetary policy surprise. The rows represent the impact on (i) the multilateral trade weighted real exchange rate (in logs times 100); (ii) long term interest rates in basis points; (iii) the consumer price index (in logs times 100); (iv) the industrial production index (in logs times 100); (v) the equity index (in logs times 100). The solid black line represents the point estimate, the dark blue area represents the 68\% confidence interval, and the light blue area represents the 90\% confidence interval. In the text, when referring to Panel $(i,j)$, $i$ refers to the row and $j$ to the column of the figure. Each variable, in its own transformation, is demeaned at the country level. Results for countries in North America alone.}
\end{figure}

\newpage
\begin{figure}
    \centering
    \includegraphics[scale=0.4]{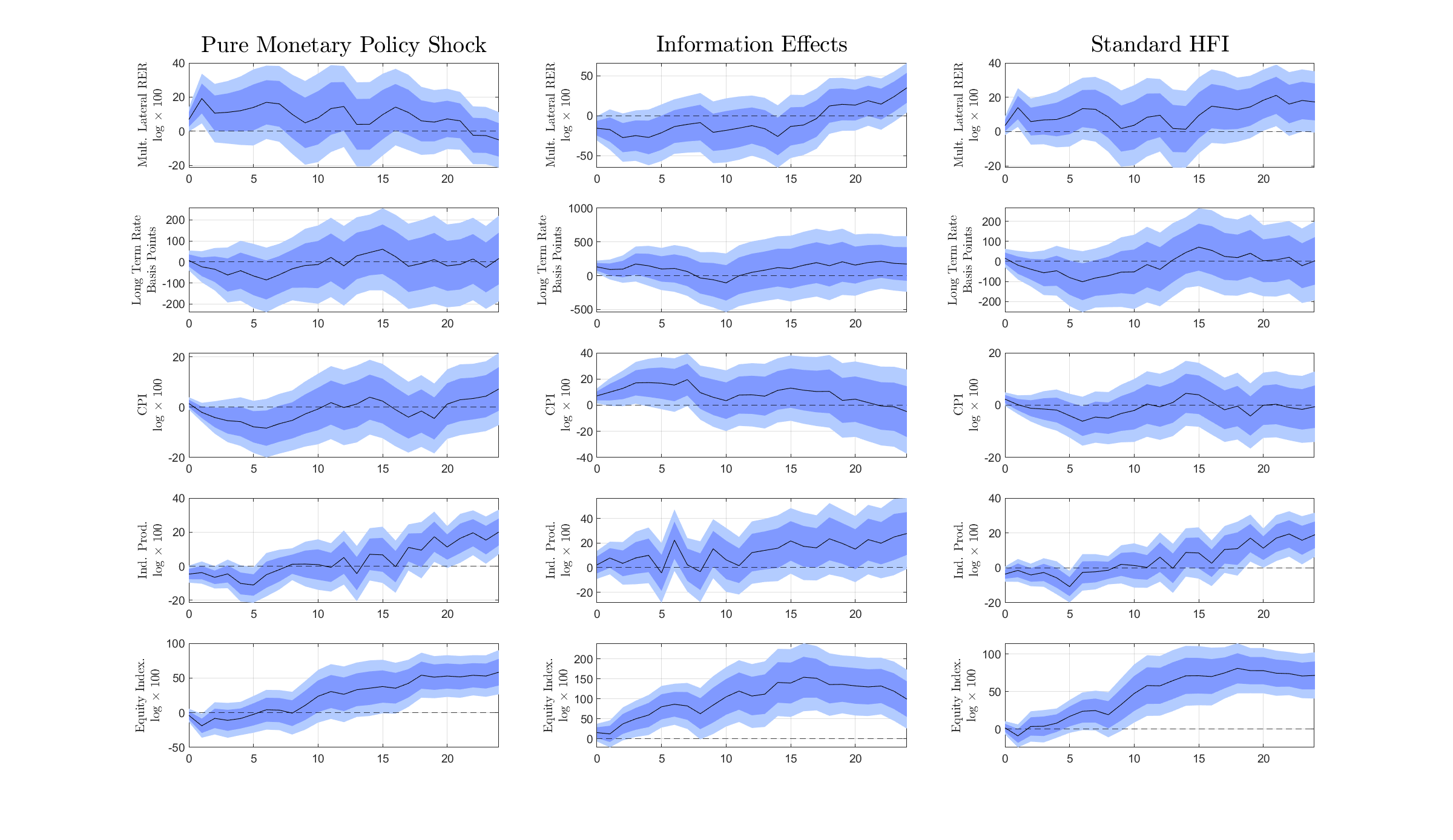}
    \caption{Impulse Response Functions \\ Multi. REER Sample -Results for Oceania}
    \label{fig:BenchmarkREER_OC}
    \floatfoot{\textbf{Note:} The figure is comprised of 15 sub-figures ordered in three columns and five rows. The left column relates to the estimates of $\beta^{MP}$ in Equation \ref{eq:LP_pooled}, the middle column relates to the estimate of $\beta^{FIE}$ in Equation \ref{eq:LP_pooled}, while the right column relates to estimating Equation \ref{eq:LP_pooled}, replacing the MP and FIE components with the un-orthogonalized monetary policy surprise. The rows represent the impact on (i) the multilateral trade weighted real exchange rate (in logs times 100); (ii) long term interest rates in basis points; (iii) the consumer price index (in logs times 100); (iv) the industrial production index (in logs times 100); (v) the equity index (in logs times 100). The solid black line represents the point estimate, the dark blue area represents the 68\% confidence interval, and the light blue area represents the 90\% confidence interval. In the text, when referring to Panel $(i,j)$, $i$ refers to the row and $j$ to the column of the figure. Each variable, in its own transformation, is demeaned at the country level. Results for countries in Oceania alone.}
\end{figure}

\newpage
\begin{figure}
    \centering
    \includegraphics[scale=0.4]{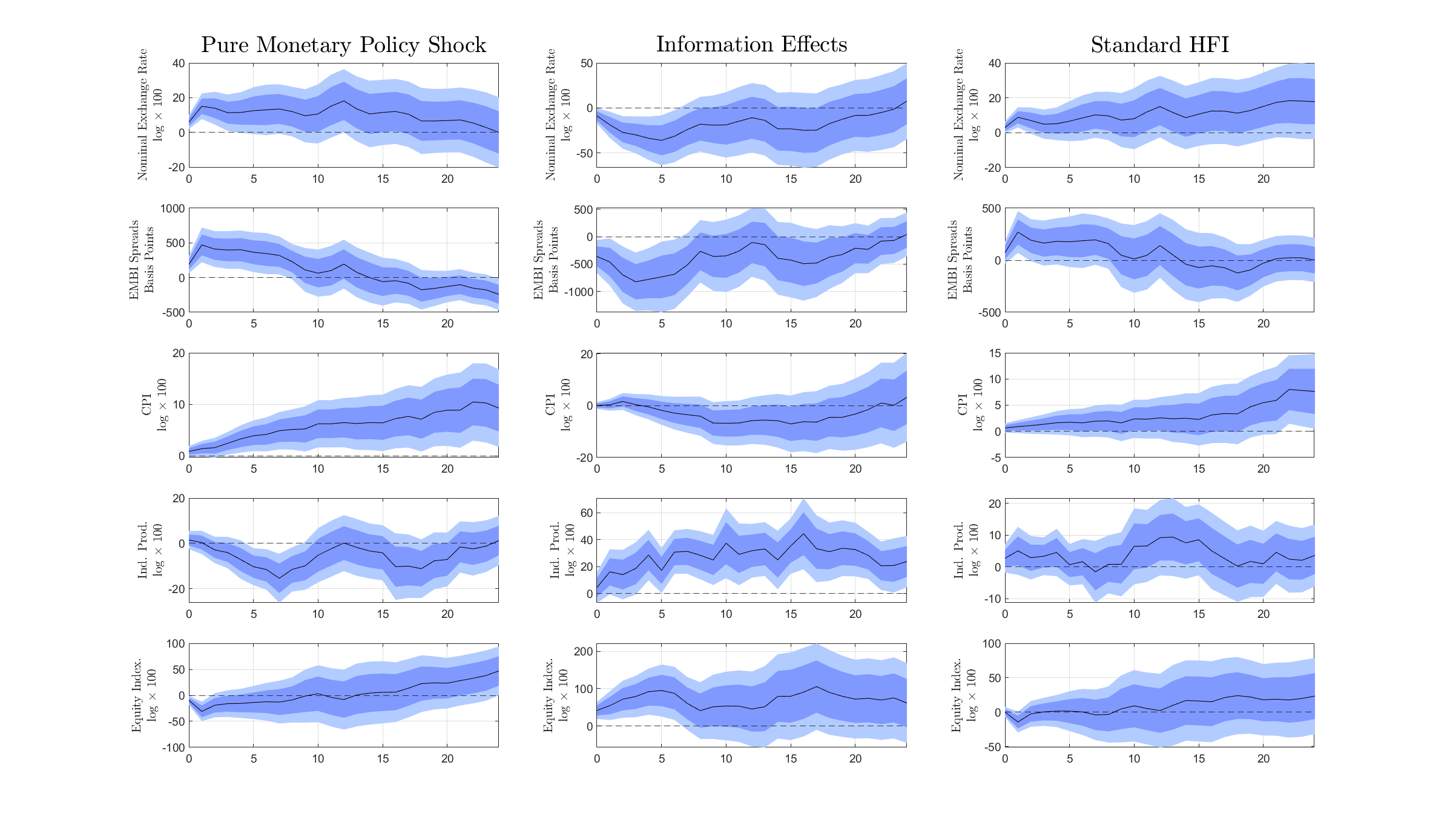}
    \caption{Impulse Response Functions \\ EMBI Specification}
    \label{fig:EMBI_BenchmarkNER}
    \floatfoot{\textbf{Note:} The figure is comprised of 15 sub-figures ordered in three columns and five rows. The left column relates to the estimates of $\beta^{MP}$ in Equation \ref{eq:LP_pooled}, the middle column relates to the estimate of $\beta^{FIE}$ in Equation \ref{eq:LP_pooled}, while the right column relates to estimating Equation \ref{eq:LP_pooled}, replacing the MP and FIE components with the un-orthogonalized monetary policy surprise. The rows represent the impact on (i) the nominal exchange rate with the US dollar (in logs times 100); (ii) EMBI spreads in basis points; (iii) the consumer price index (in logs times 100); (iv) the industrial production index (in logs times 100); (v) the equity index (in logs times 100). The solid black line represents the point estimate, the dark blue area represents the 68\% confidence interval, and the light blue area represents the 90\% confidence interval. In the text, when referring to Panel $(i,j)$, $i$ refers to the row and $j$ to the column of the figure. Each variable, in its own transformation, is demeaned at the country level. The list of countries in this sample are: Brazil, Bulgaria, Chile,  China, P.R.: Mainland, Colombia Costa Rica, Croatia, Hungary, India, Indonesia,  Latvia, Lithuania, Malaysia, Mexico, Mongolia, Pakistan, Peru, Philippines, Poland, Romania, Russian Federation, South Africa, Türkiye, Ukraine, Venezuela.}
\end{figure}

\newpage
\begin{figure}
    \centering
    \includegraphics[scale=0.4]{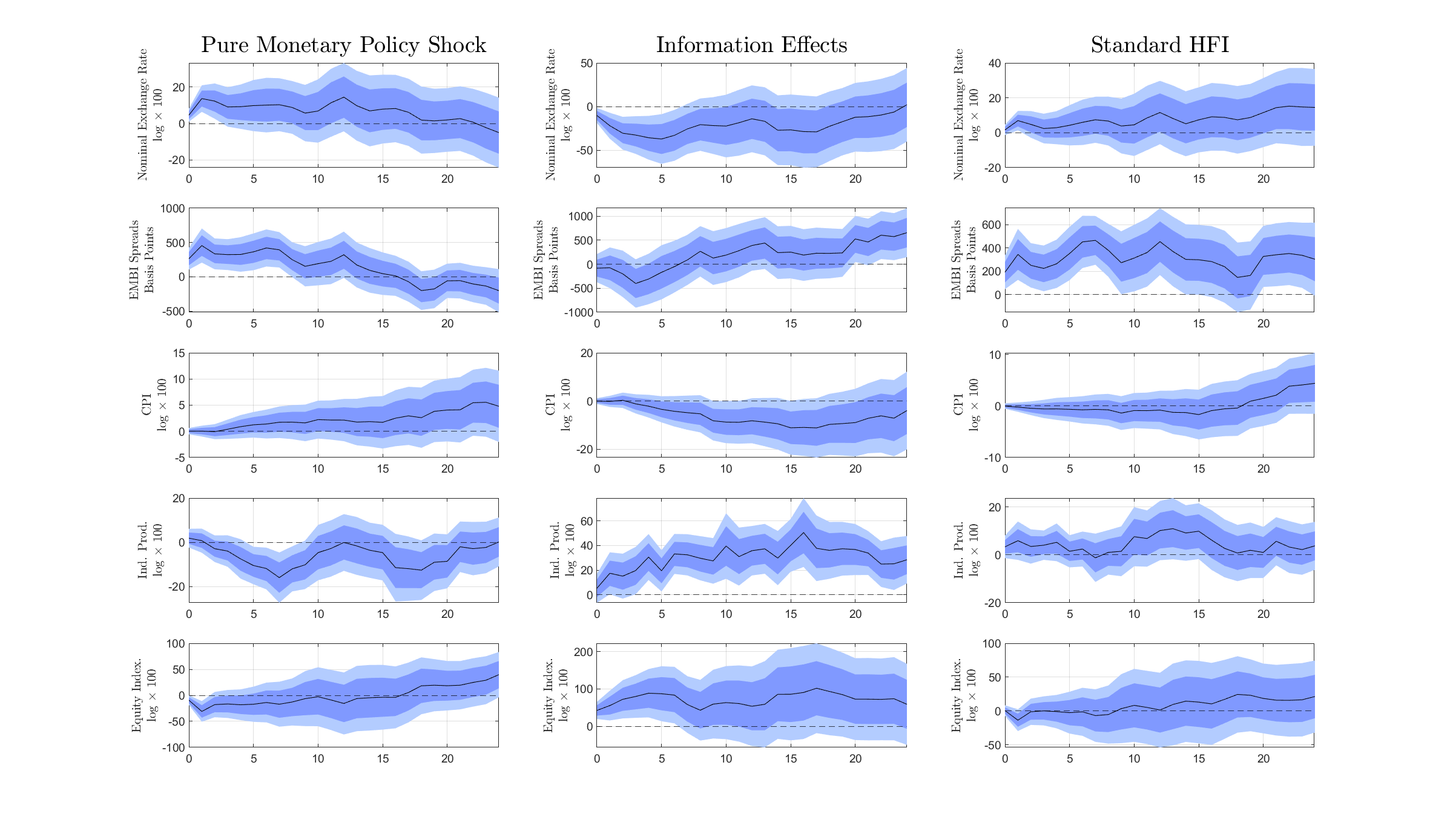}
    \caption{Impulse Response Functions \\ EMBI Specification - 1998 Onward}
    \label{fig:EMBI_Regressions_98NER}
    \floatfoot{\textbf{Note:} The figure is comprised of 15 sub-figures ordered in three columns and five rows. The left column relates to the estimates of $\beta^{MP}$ in Equation \ref{eq:LP_pooled}, the middle column relates to the estimate of $\beta^{FIE}$ in Equation \ref{eq:LP_pooled}, while the right column relates to estimating Equation \ref{eq:LP_pooled}, replacing the MP and FIE components with the un-orthogonalized monetary policy surprise. The rows represent the impact on (i) the nominal exchange rate with the US dollar (in logs times 100); (ii) EMBI spreads in basis points; (iii) the consumer price index (in logs times 100); (iv) the industrial production index (in logs times 100); (v) the equity index (in logs times 100). The solid black line represents the point estimate, the dark blue area represents the 68\% confidence interval, and the light blue area represents the 90\% confidence interval. In the text, when referring to Panel $(i,j)$, $i$ refers to the row and $j$ to the column of the figure. Each variable, in its own transformation, is demeaned at the country level. The list of countries in this sample are: Brazil, Bulgaria, Chile,  China, P.R.: Mainland, Colombia Costa Rica, Croatia, Hungary, India, Indonesia,  Latvia, Lithuania, Malaysia, Mexico, Mongolia, Pakistan, Peru, Philippines, Poland, Romania, Russian Federation, South Africa, Türkiye, Ukraine, Venezuela. Sample starts in January 1998 and ends in December 2019.}
\end{figure}

\newpage
\begin{figure}
    \centering
    \includegraphics[scale=0.4]{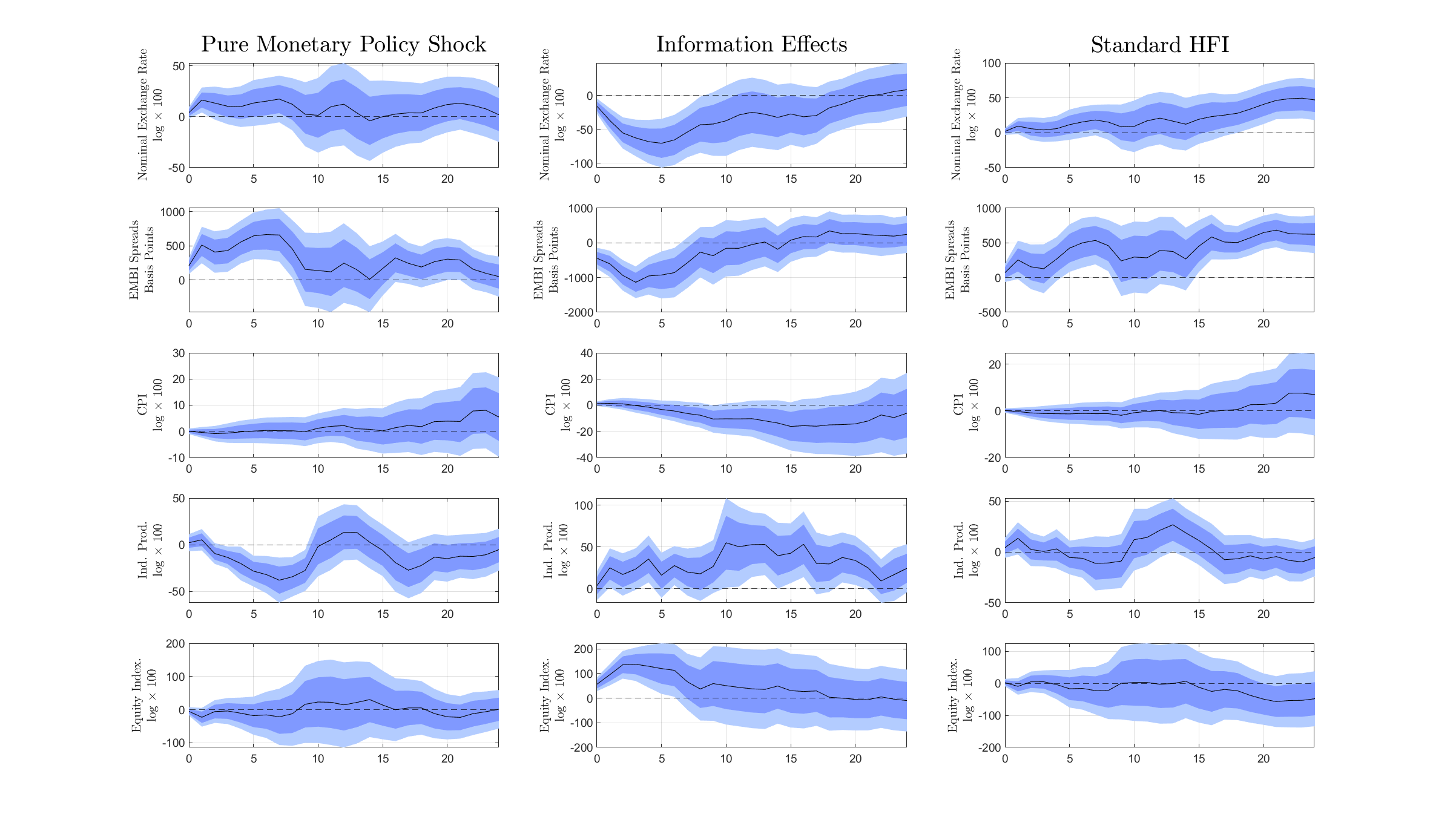}
    \caption{Impulse Response Functions \\ EMBI Specification - 2008 Onward}
    \label{fig:EMBI_Regressions_08NER}
    \floatfoot{\textbf{Note:} The figure is comprised of 15 sub-figures ordered in three columns and five rows. The left column relates to the estimates of $\beta^{MP}$ in Equation \ref{eq:LP_pooled}, the middle column relates to the estimate of $\beta^{FIE}$ in Equation \ref{eq:LP_pooled}, while the right column relates to estimating Equation \ref{eq:LP_pooled}, replacing the MP and FIE components with the un-orthogonalized monetary policy surprise. The rows represent the impact on (i) the nominal exchange rate with the US dollar (in logs times 100); (ii) EMBI spreads in basis points; (iii) the consumer price index (in logs times 100); (iv) the industrial production index (in logs times 100); (v) the equity index (in logs times 100). The solid black line represents the point estimate, the dark blue area represents the 68\% confidence interval, and the light blue area represents the 90\% confidence interval. In the text, when referring to Panel $(i,j)$, $i$ refers to the row and $j$ to the column of the figure. Each variable, in its own transformation, is demeaned at the country level. The list of countries in this sample are: Brazil, Bulgaria, Chile,  China, P.R.: Mainland, Colombia Costa Rica, Croatia, Hungary, India, Indonesia,  Latvia, Lithuania, Malaysia, Mexico, Mongolia, Pakistan, Peru, Philippines, Poland, Romania, Russian Federation, South Africa, Türkiye, Ukraine, Venezuela. Sample starts in January 2008 and ends in December 2019.}
\end{figure}

\newpage
\begin{figure}
    \centering
    \includegraphics[scale=0.4]{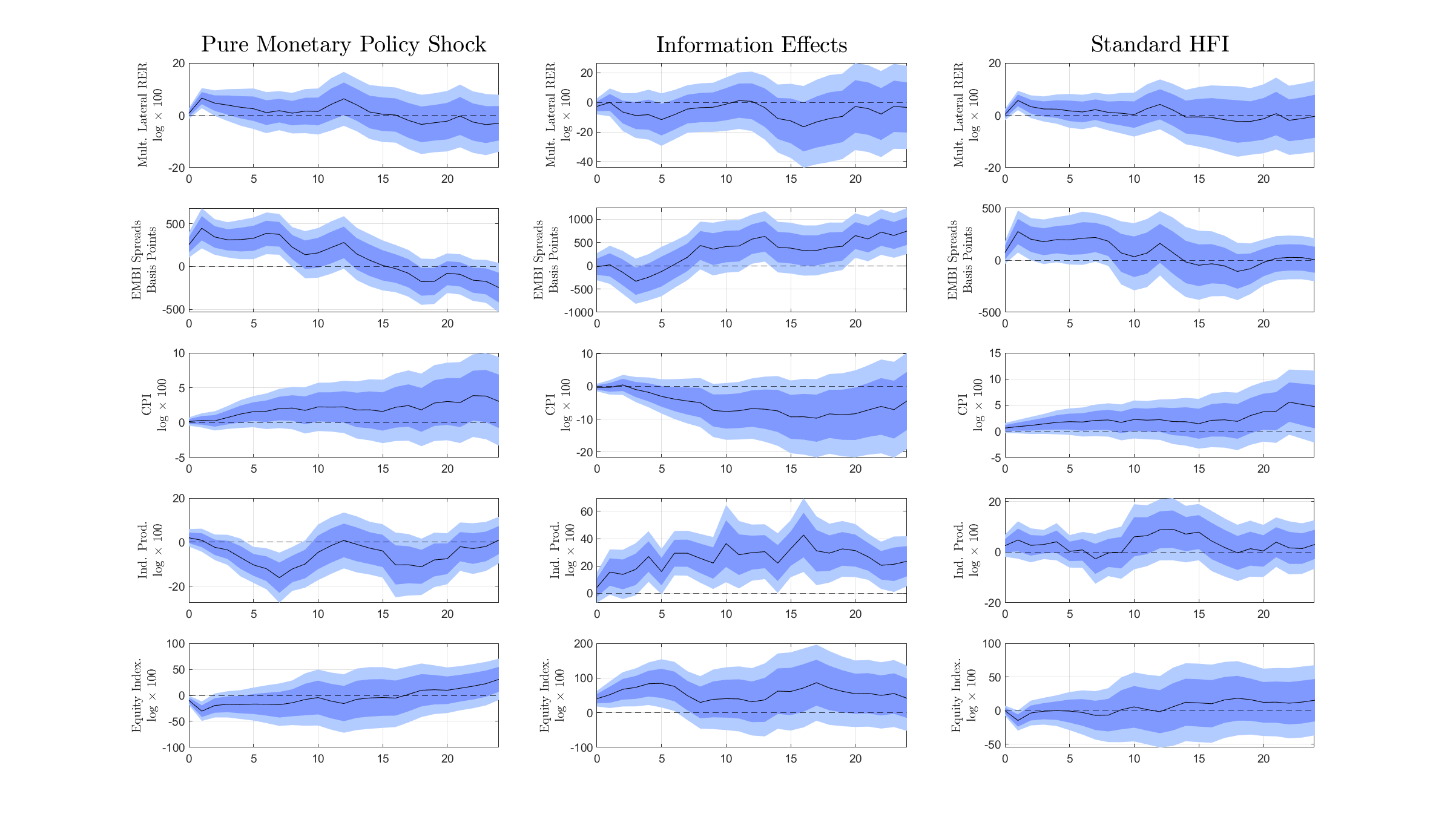}
    \caption{Impulse Response Functions \\ EMBI Multi. REER Specification}
    \label{fig:EMBI_BenchmarkREER}
    \floatfoot{\textbf{Note:} The figure is comprised of 15 sub-figures ordered in three columns and five rows. The left column relates to the estimates of $\beta^{MP}$ in Equation \ref{eq:LP_pooled}, the middle column relates to the estimate of $\beta^{FIE}$ in Equation \ref{eq:LP_pooled}, while the right column relates to estimating Equation \ref{eq:LP_pooled}, replacing the MP and FIE components with the un-orthogonalized monetary policy surprise. The rows represent the impact on (i) the trade weighted multilateral real exchange rate (in logs times 100); (ii) EMBI spreads in basis points; (iii) the consumer price index (in logs times 100); (iv) the industrial production index (in logs times 100); (v) the equity index (in logs times 100). The solid black line represents the point estimate, the dark blue area represents the 68\% confidence interval, and the light blue area represents the 90\% confidence interval. In the text, when referring to Panel $(i,j)$, $i$ refers to the row and $j$ to the column of the figure. Each variable, in its own transformation, is demeaned at the country level. The list of countries in this sample are: Brazil, Bulgaria, Chile, China, P.R.: Mainland, Colombia, Costa Rica, Croatia, Hungary, India, Indonesia, Latvia, Lithuania, Malaysia, Mexico, Mongolia, Pakistan, Peru, Philippines, Poland, Portugal, Romania, Russian Federation, Slovak Rep., South Africa, Türkiye, Rep of, Ukraine, Venezuela. }
\end{figure}

\newpage
\begin{figure}
    \centering
    \includegraphics[scale=0.4]{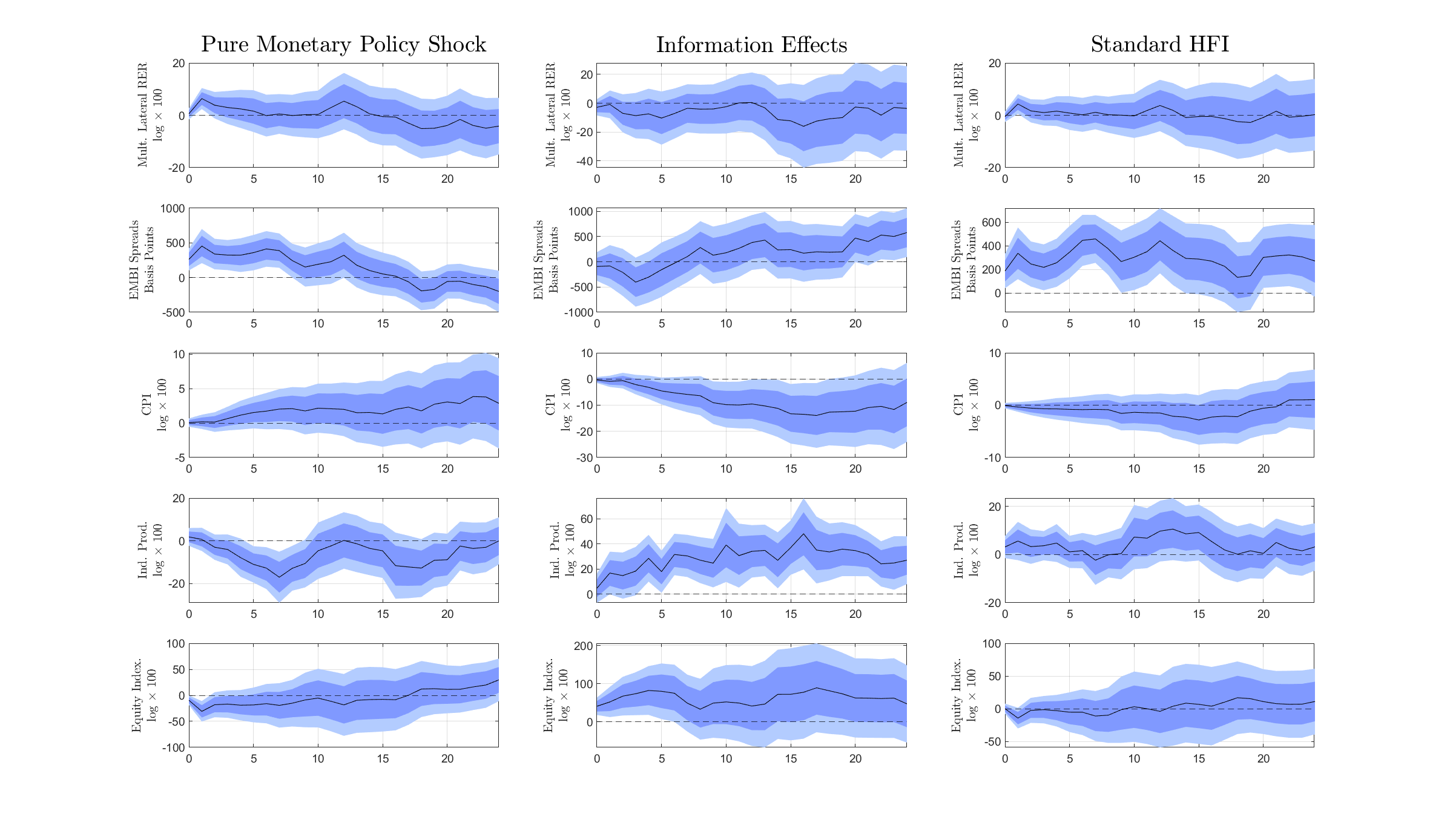}
    \caption{Impulse Response Functions \\ EMBI Multi. REER Specification - 1998 Onward}
    \label{fig:EMBI_Regressions_98REER}
    \floatfoot{\textbf{Note:} The figure is comprised of 15 sub-figures ordered in three columns and five rows. The left column relates to the estimates of $\beta^{MP}$ in Equation \ref{eq:LP_pooled}, the middle column relates to the estimate of $\beta^{FIE}$ in Equation \ref{eq:LP_pooled}, while the right column relates to estimating Equation \ref{eq:LP_pooled}, replacing the MP and FIE components with the un-orthogonalized monetary policy surprise. The rows represent the impact on (i) the trade weighted multilateral real exchange rate (in logs times 100); (ii) EMBI spreads in basis points; (iii) the consumer price index (in logs times 100); (iv) the industrial production index (in logs times 100); (v) the equity index (in logs times 100). The solid black line represents the point estimate, the dark blue area represents the 68\% confidence interval, and the light blue area represents the 90\% confidence interval. In the text, when referring to Panel $(i,j)$, $i$ refers to the row and $j$ to the column of the figure. Each variable, in its own transformation, is demeaned at the country level. The list of countries in this sample are: Brazil, Bulgaria, Chile, China, P.R.: Mainland, Colombia, Costa Rica, Croatia, Hungary, India, Indonesia, Latvia, Lithuania, Malaysia, Mexico, Mongolia, Pakistan, Peru, Philippines, Poland, Portugal, Romania, Russian Federation, Slovak Rep., South Africa, Türkiye, Rep of, Ukraine, Venezuela. Sample starts in January 1998 and ends in December 2019.}
\end{figure}

\newpage
\begin{figure}
    \centering
    \includegraphics[scale=0.4]{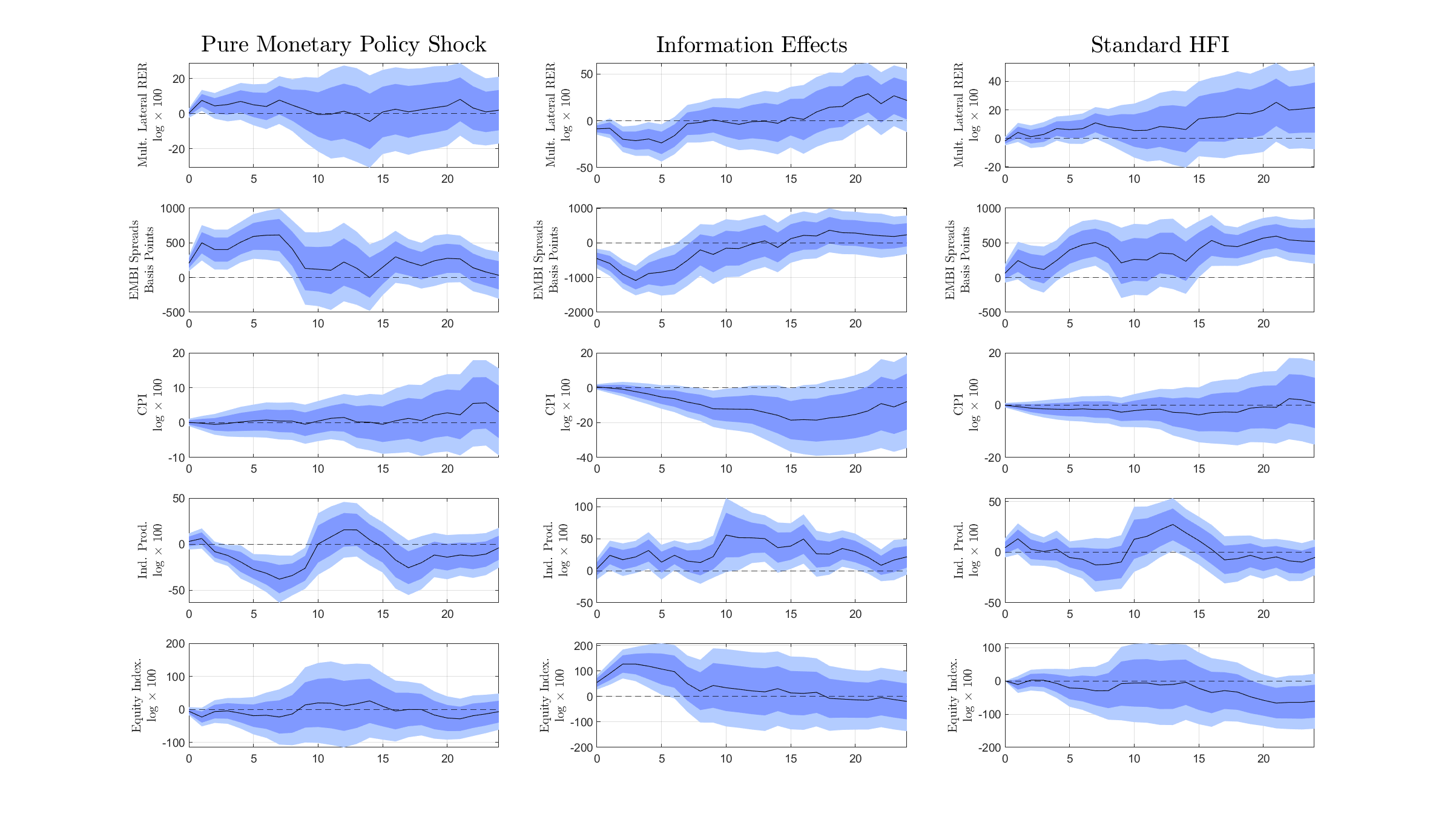}
    \caption{Impulse Response Functions \\ EMBI Multi. REER Specification - 2008 Onward}
    \label{fig:EMBI_Regressions_08REER}
    \floatfoot{\textbf{Note:} The figure is comprised of 15 sub-figures ordered in three columns and five rows. The left column relates to the estimates of $\beta^{MP}$ in Equation \ref{eq:LP_pooled}, the middle column relates to the estimate of $\beta^{FIE}$ in Equation \ref{eq:LP_pooled}, while the right column relates to estimating Equation \ref{eq:LP_pooled}, replacing the MP and FIE components with the un-orthogonalized monetary policy surprise. The rows represent the impact on (i) the trade weighted multilateral real exchange rate (in logs times 100); (ii) EMBI spreads in basis points; (iii) the consumer price index (in logs times 100); (iv) the industrial production index (in logs times 100); (v) the equity index (in logs times 100). The solid black line represents the point estimate, the dark blue area represents the 68\% confidence interval, and the light blue area represents the 90\% confidence interval. In the text, when referring to Panel $(i,j)$, $i$ refers to the row and $j$ to the column of the figure. Each variable, in its own transformation, is demeaned at the country level. The list of countries in this sample are: Brazil, Bulgaria, Chile, China, P.R.: Mainland, Colombia, Costa Rica, Croatia, Hungary, India, Indonesia, Latvia, Lithuania, Malaysia, Mexico, Mongolia, Pakistan, Peru, Philippines, Poland, Portugal, Romania, Russian Federation, Slovak Rep., South Africa, Türkiye, Rep of, Ukraine, Venezuela. Sample starts in January 2008 and ends in December 2019.}
\end{figure}

\newpage
\begin{figure}[ht]
    \centering
    \includegraphics[scale=0.4]{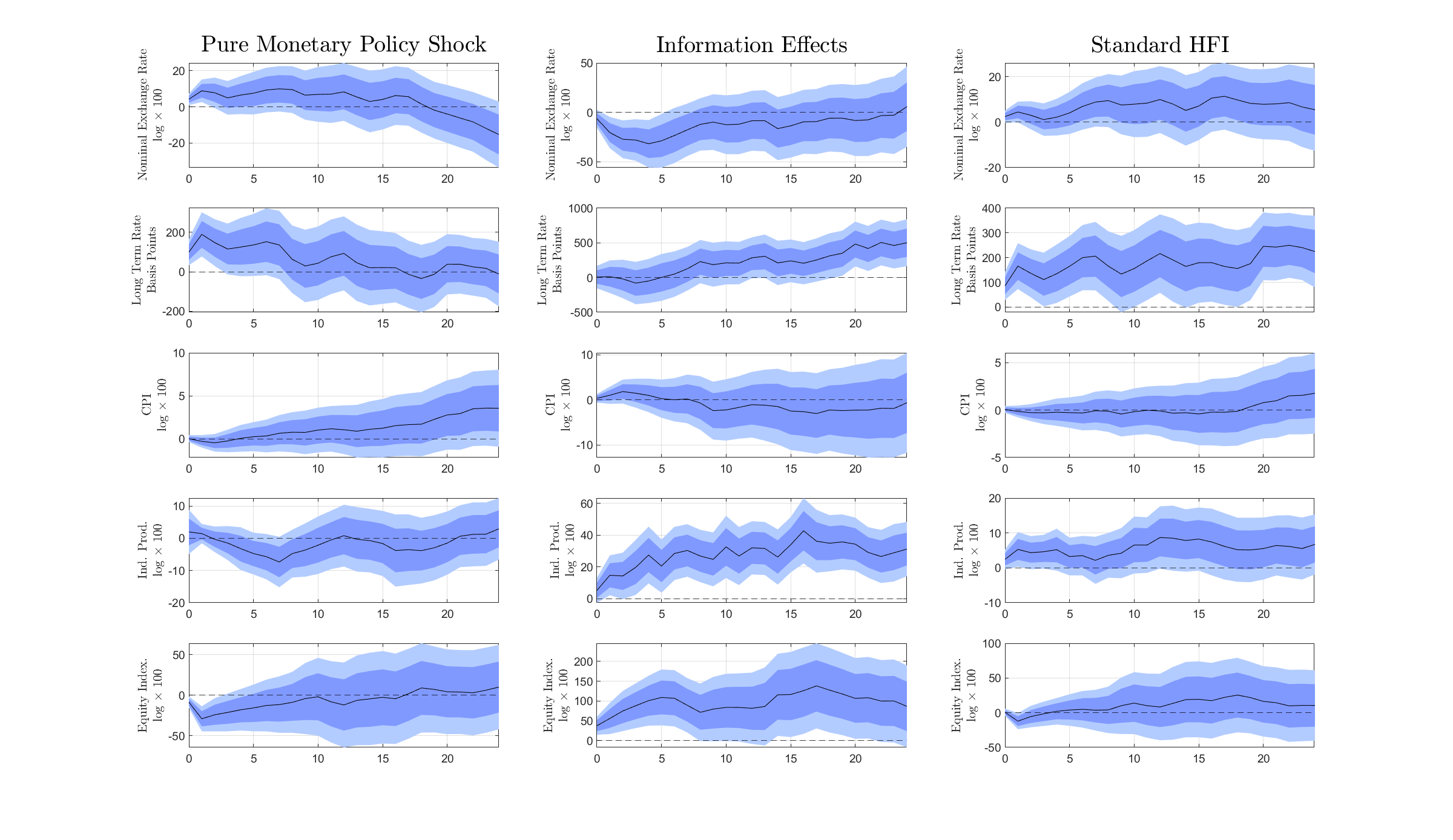}
    \caption{Impulse Response Functions \\ Variables not Demeaned}
    \label{fig:Levels_BenchmarkNER}
    \floatfoot{\textbf{Note:} The figure is comprised of 15 sub-figures ordered in three columns and five rows. The left column relates to the estimates of $\beta^{MP}$ in Equation \ref{eq:LP_pooled}, the middle column relates to the estimate of $\beta^{FIE}$ in Equation \ref{eq:LP_pooled}, while the right column relates to estimating Equation \ref{eq:LP_pooled}, replacing the MP and FIE components with the un-orthogonalized monetary policy surprise. The rows represent the impact on (i) the nominal exchange rate with the US dollar (in logs times 100); (ii) long term interest rates in basis points; (iii) the consumer price index (in logs times 100); (iv) the industrial production index (in logs times 100); (v) the equity index (in logs times 100). The solid black line represents the point estimate, the dark blue area represents the 68\% confidence interval, and the light blue area represents the 90\% confidence interval. In the text, when referring to Panel $(i,j)$, $i$ refers to the row and $j$ to the column of the figure. Variables are not demeaned.}
\end{figure}

\newpage
\begin{figure}[ht]
    \centering
    \includegraphics[scale=0.4]{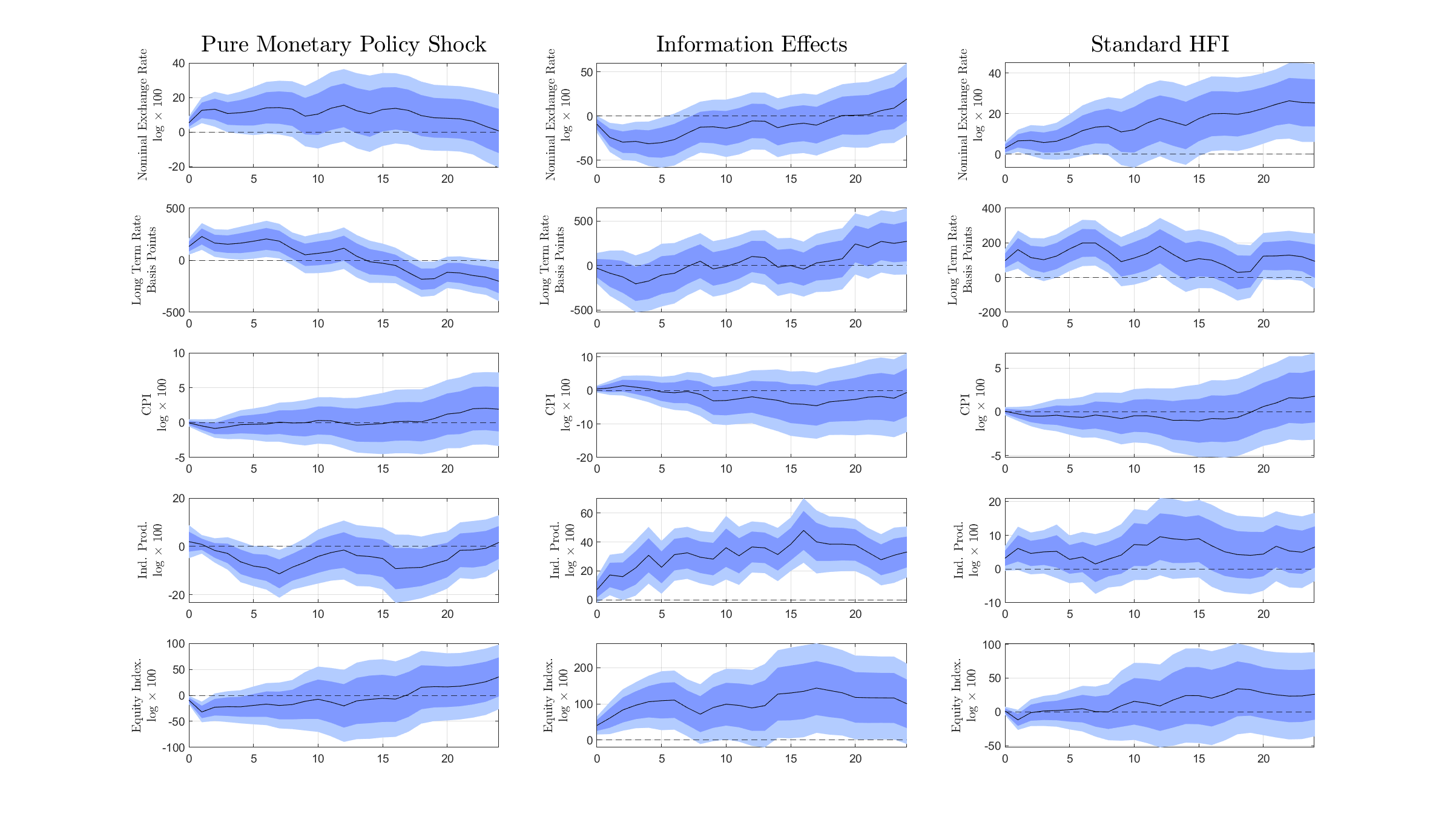}
    \caption{Impulse Response Functions \\ Variables not Demeaned - January 1998 onward}
    \label{fig:Levels_BenchmarkNER_98}
    \floatfoot{\textbf{Note:} The figure is comprised of 15 sub-figures ordered in three columns and five rows. The left column relates to the estimates of $\beta^{MP}$ in Equation \ref{eq:LP_pooled}, the middle column relates to the estimate of $\beta^{FIE}$ in Equation \ref{eq:LP_pooled}, while the right column relates to estimating Equation \ref{eq:LP_pooled}, replacing the MP and FIE components with the un-orthogonalized monetary policy surprise. The rows represent the impact on (i) the nominal exchange rate with the US dollar (in logs times 100); (ii) long term interest rates in basis points; (iii) the consumer price index (in logs times 100); (iv) the industrial production index (in logs times 100); (v) the equity index (in logs times 100). The solid black line represents the point estimate, the dark blue area represents the 68\% confidence interval, and the light blue area represents the 90\% confidence interval. In the text, when referring to Panel $(i,j)$, $i$ refers to the row and $j$ to the column of the figure. Variables are not demeaned. Sample of January 1998 to December 2019.}
\end{figure}

\newpage
\begin{figure}[ht]
    \centering
    \includegraphics[scale=0.4]{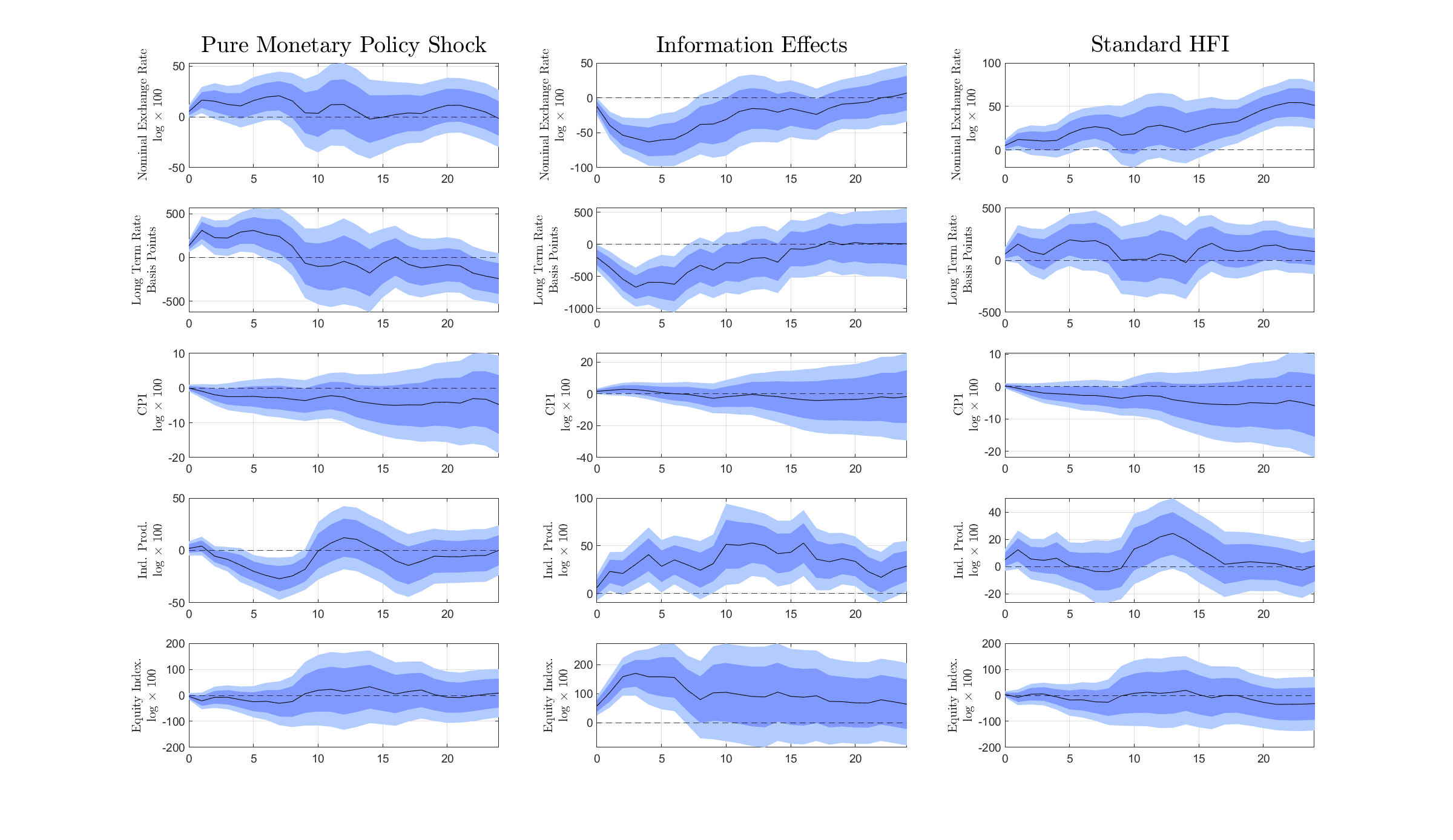}
    \caption{Impulse Response Functions \\ Variables not Demeaned - January 2008 onward}
    \label{fig:Levels_BenchmarkNER_08}
    \floatfoot{\textbf{Note:} The figure is comprised of 15 sub-figures ordered in three columns and five rows. The left column relates to the estimates of $\beta^{MP}$ in Equation \ref{eq:LP_pooled}, the middle column relates to the estimate of $\beta^{FIE}$ in Equation \ref{eq:LP_pooled}, while the right column relates to estimating Equation \ref{eq:LP_pooled}, replacing the MP and FIE components with the un-orthogonalized monetary policy surprise. The rows represent the impact on (i) the nominal exchange rate with the US dollar (in logs times 100); (ii) long term interest rates in basis points; (iii) the consumer price index (in logs times 100); (iv) the industrial production index (in logs times 100); (v) the equity index (in logs times 100). The solid black line represents the point estimate, the dark blue area represents the 68\% confidence interval, and the light blue area represents the 90\% confidence interval. In the text, when referring to Panel $(i,j)$, $i$ refers to the row and $j$ to the column of the figure. Variables are not demeaned. Sample of January 2008 to December 2019.}
\end{figure}

\newpage
\begin{figure}[ht]
    \centering
    \includegraphics[scale=0.4]{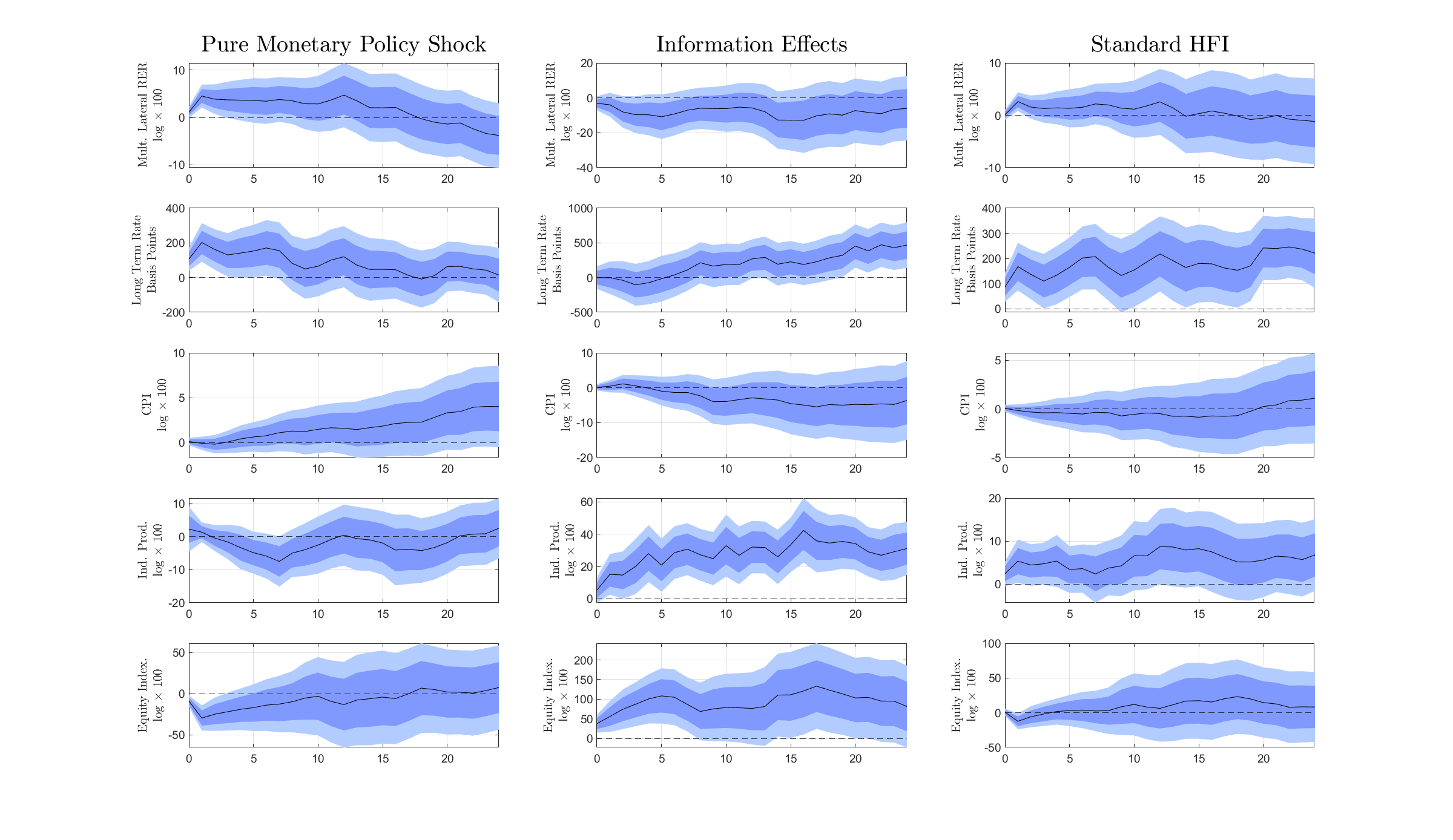}
    \caption{Impulse Response Functions \\ Multi. REER Sample - Variables not Demeaned}
    \label{fig:Levels_BenchmarkREER}
    \floatfoot{\textbf{Note:} The figure is comprised of 15 sub-figures ordered in three columns and five rows. The left column relates to the estimates of $\beta^{MP}$ in Equation \ref{eq:LP_pooled}, the middle column relates to the estimate of $\beta^{FIE}$ in Equation \ref{eq:LP_pooled}, while the right column relates to estimating Equation \ref{eq:LP_pooled}, replacing the MP and FIE components with the un-orthogonalized monetary policy surprise. The rows represent the impact on (i) the multilateral trade weighted real exchange rate (in logs times 100); (ii) long term interest rates in basis points; (iii) the consumer price index (in logs times 100); (iv) the industrial production index (in logs times 100); (v) the equity index (in logs times 100). The solid black line represents the point estimate, the dark blue area represents the 68\% confidence interval, and the light blue area represents the 90\% confidence interval. In the text, when referring to Panel $(i,j)$, $i$ refers to the row and $j$ to the column of the figure. Variables are not demeaned.}
\end{figure}

\newpage
\begin{figure}[ht]
    \centering
    \includegraphics[scale=0.4]{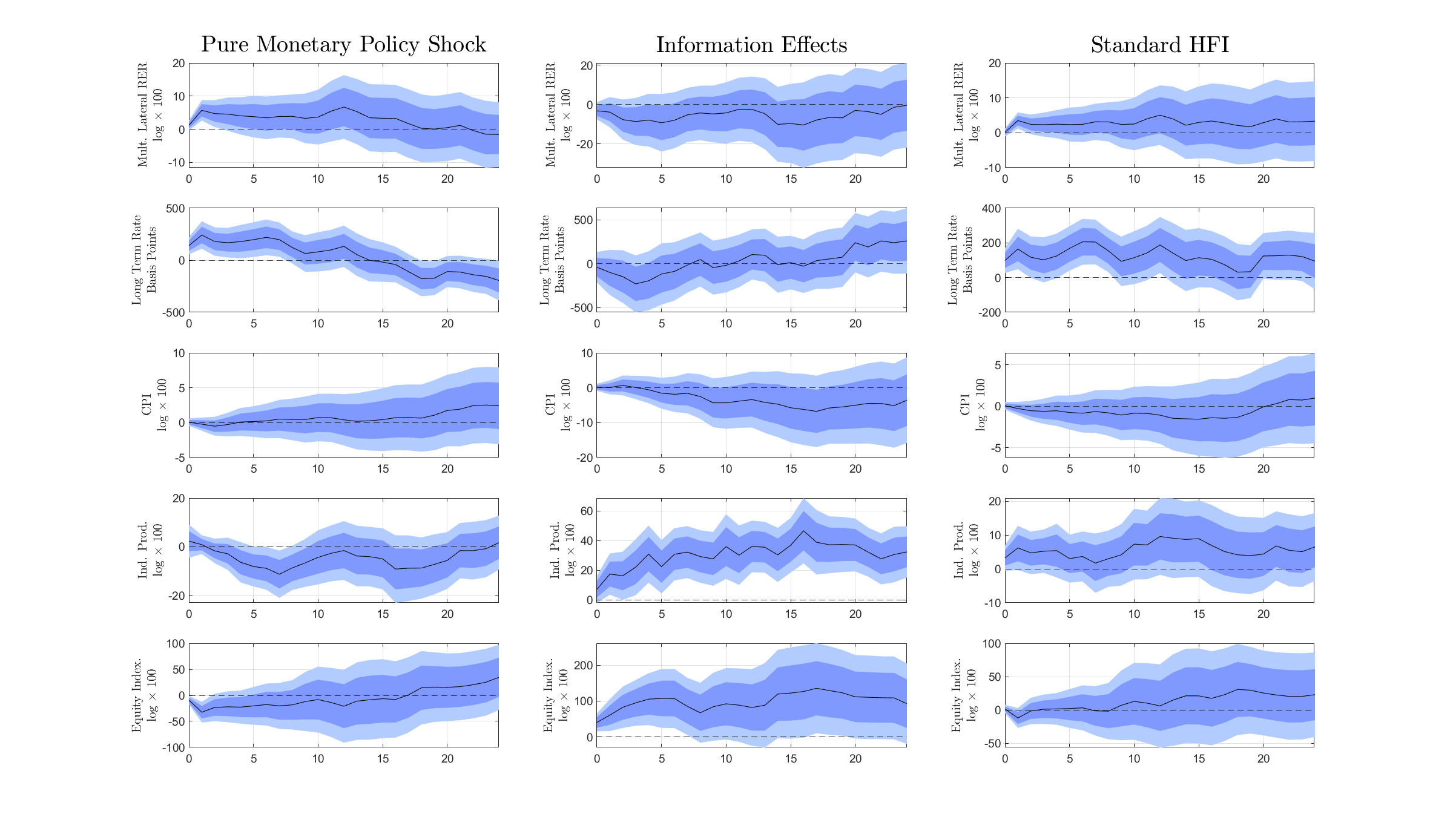}
    \caption{Impulse Response Functions \\ Multi. REER Sample - Variables not Demeaned - 1998 onward}
    \label{fig:Levels_BenchmarkREER_98}
    \floatfoot{\textbf{Note:} The figure is comprised of 15 sub-figures ordered in three columns and five rows. The left column relates to the estimates of $\beta^{MP}$ in Equation \ref{eq:LP_pooled}, the middle column relates to the estimate of $\beta^{FIE}$ in Equation \ref{eq:LP_pooled}, while the right column relates to estimating Equation \ref{eq:LP_pooled}, replacing the MP and FIE components with the un-orthogonalized monetary policy surprise. The rows represent the impact on (i) the multilateral trade weighted real exchange rate (in logs times 100); (ii) long term interest rates in basis points; (iii) the consumer price index (in logs times 100); (iv) the industrial production index (in logs times 100); (v) the equity index (in logs times 100). The solid black line represents the point estimate, the dark blue area represents the 68\% confidence interval, and the light blue area represents the 90\% confidence interval. In the text, when referring to Panel $(i,j)$, $i$ refers to the row and $j$ to the column of the figure. Variables are not demeaned. Sample of January 1998 to December 2019}
\end{figure}

\newpage
\begin{figure}[ht]
    \centering
    \includegraphics[scale=0.4]{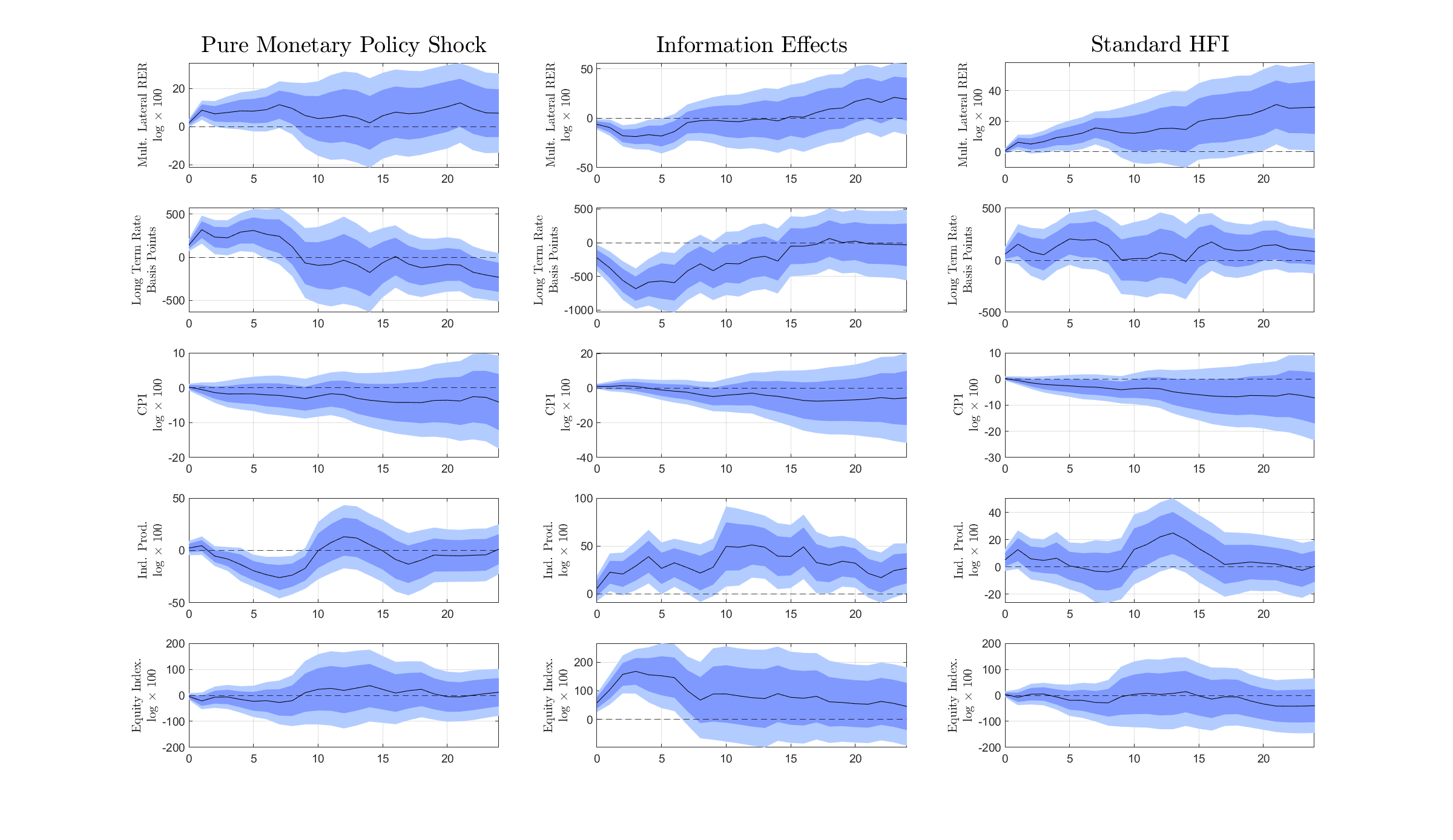}
    \caption{Impulse Response Functions \\ Multi. REER Sample - Variables not Demeaned - 2008 onward}
    \label{fig:Levels_BenchmarkREER_08}
    \floatfoot{\textbf{Note:} The figure is comprised of 15 sub-figures ordered in three columns and five rows. The left column relates to the estimates of $\beta^{MP}$ in Equation \ref{eq:LP_pooled}, the middle column relates to the estimate of $\beta^{FIE}$ in Equation \ref{eq:LP_pooled}, while the right column relates to estimating Equation \ref{eq:LP_pooled}, replacing the MP and FIE components with the un-orthogonalized monetary policy surprise. The rows represent the impact on (i) the multilateral trade weighted real exchange rate (in logs times 100); (ii) long term interest rates in basis points; (iii) the consumer price index (in logs times 100); (iv) the industrial production index (in logs times 100); (v) the equity index (in logs times 100). The solid black line represents the point estimate, the dark blue area represents the 68\% confidence interval, and the light blue area represents the 90\% confidence interval. In the text, when referring to Panel $(i,j)$, $i$ refers to the row and $j$ to the column of the figure. Variables are not demeaned. Sample of January 2008 to December 2019}
\end{figure}

\newpage
\begin{figure}[ht]
    \centering
    \includegraphics[scale=0.4]{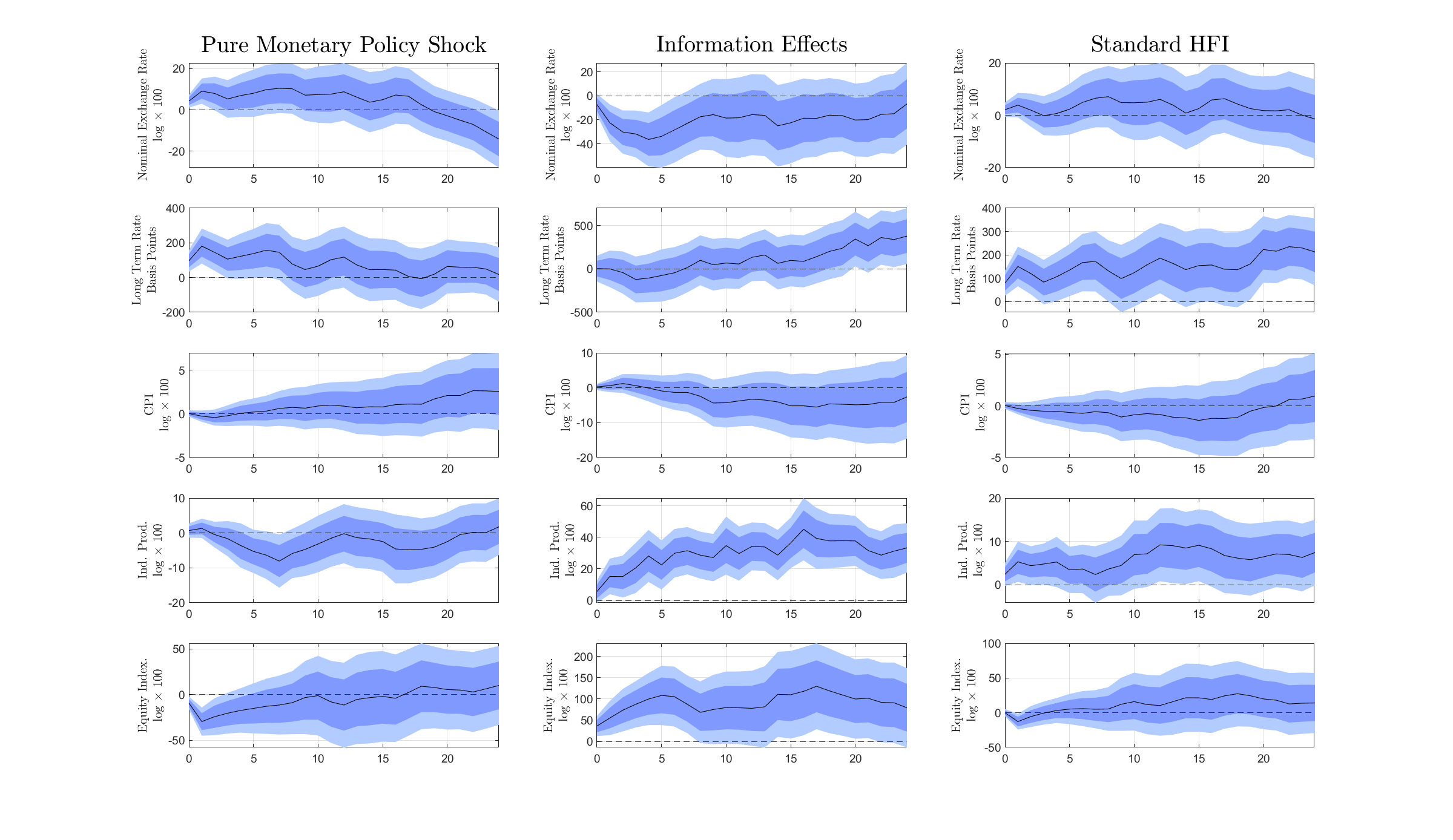}
    \caption{Impulse Response Functions \\ Sample without countries that exit \& re-enter sample}
    \label{fig:Exiters_BenchmarkNER}
    \floatfoot{\textbf{Note:} The figure is comprised of 15 sub-figures ordered in three columns and five rows. The left column relates to the estimates of $\beta^{MP}$ in Equation \ref{eq:LP_pooled}, the middle column relates to the estimate of $\beta^{FIE}$ in Equation \ref{eq:LP_pooled}, while the right column relates to estimating Equation \ref{eq:LP_pooled}, replacing the MP and FIE components with the un-orthogonalized monetary policy surprise. The rows represent the impact on (i) the nominal exchange rate with the US dollar (in logs times 100); (ii) long term interest rates in basis points; (iii) the consumer price index (in logs times 100); (iv) the industrial production index (in logs times 100); (v) the equity index (in logs times 100). The solid black line represents the point estimate, the dark blue area represents the 68\% confidence interval, and the light blue area represents the 90\% confidence interval. In the text, when referring to Panel $(i,j)$, $i$ refers to the row and $j$ to the column of the figure. Each variable, in its own transformation, is demeaned at the country level. }
\end{figure}

\newpage
\begin{figure}[ht]
    \centering
    \includegraphics[scale=0.4]{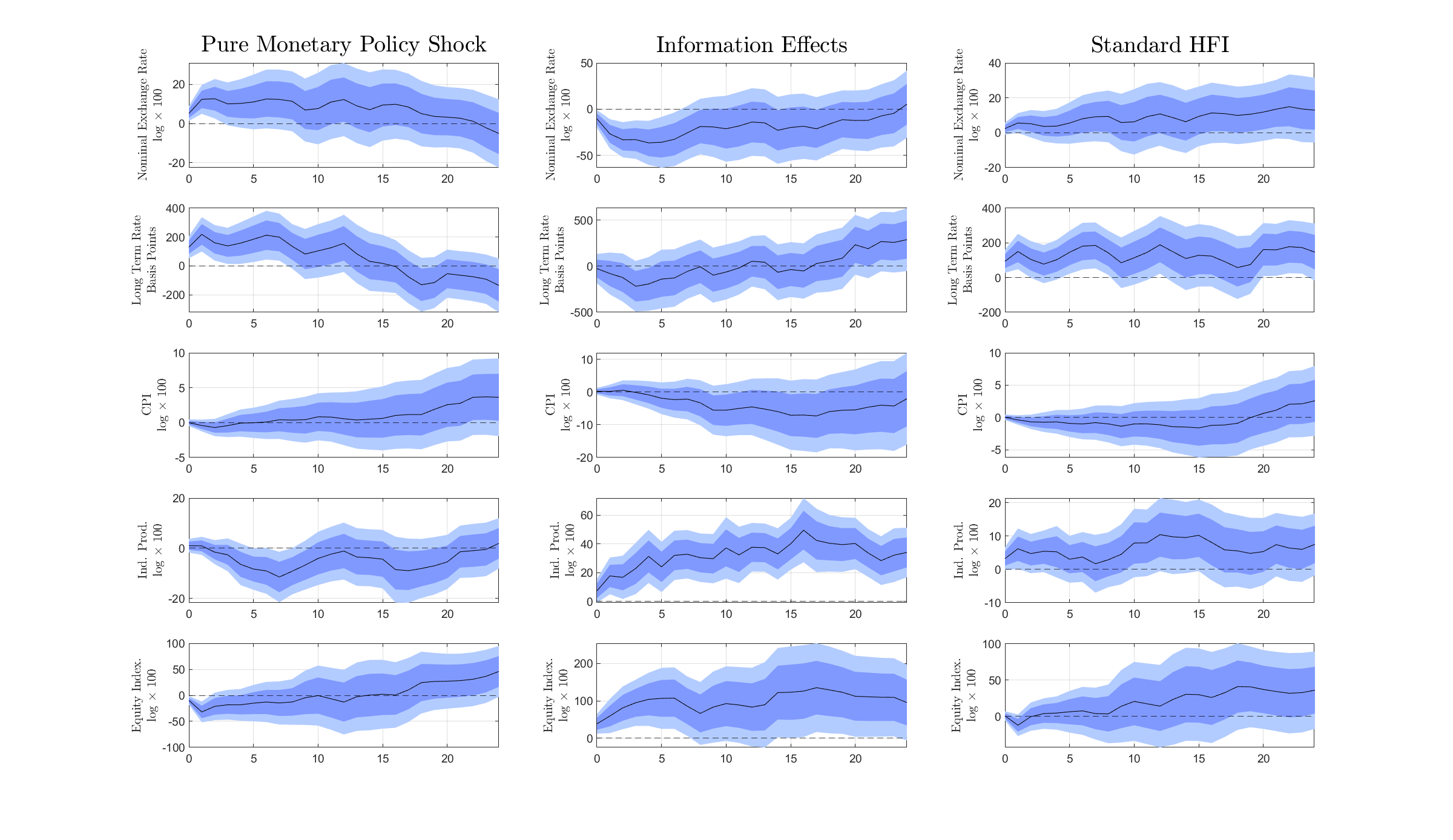}
    \caption{Impulse Response Functions \\ Sample without countries that exit \& re-enter sample}
    \label{fig:Exiters_BenchmarkNER_98}
    \floatfoot{\textbf{Note:} The figure is comprised of 15 sub-figures ordered in three columns and five rows. The left column relates to the estimates of $\beta^{MP}$ in Equation \ref{eq:LP_pooled}, the middle column relates to the estimate of $\beta^{FIE}$ in Equation \ref{eq:LP_pooled}, while the right column relates to estimating Equation \ref{eq:LP_pooled}, replacing the MP and FIE components with the un-orthogonalized monetary policy surprise. The rows represent the impact on (i) the nominal exchange rate with the US dollar (in logs times 100); (ii) long term interest rates in basis points; (iii) the consumer price index (in logs times 100); (iv) the industrial production index (in logs times 100); (v) the equity index (in logs times 100). The solid black line represents the point estimate, the dark blue area represents the 68\% confidence interval, and the light blue area represents the 90\% confidence interval. In the text, when referring to Panel $(i,j)$, $i$ refers to the row and $j$ to the column of the figure. Each variable, in its own transformation, is demeaned at the country level. Sample from January 1998 to December 2019.}
\end{figure}

\newpage
\begin{figure}[ht]
    \centering
    \includegraphics[scale=0.4]{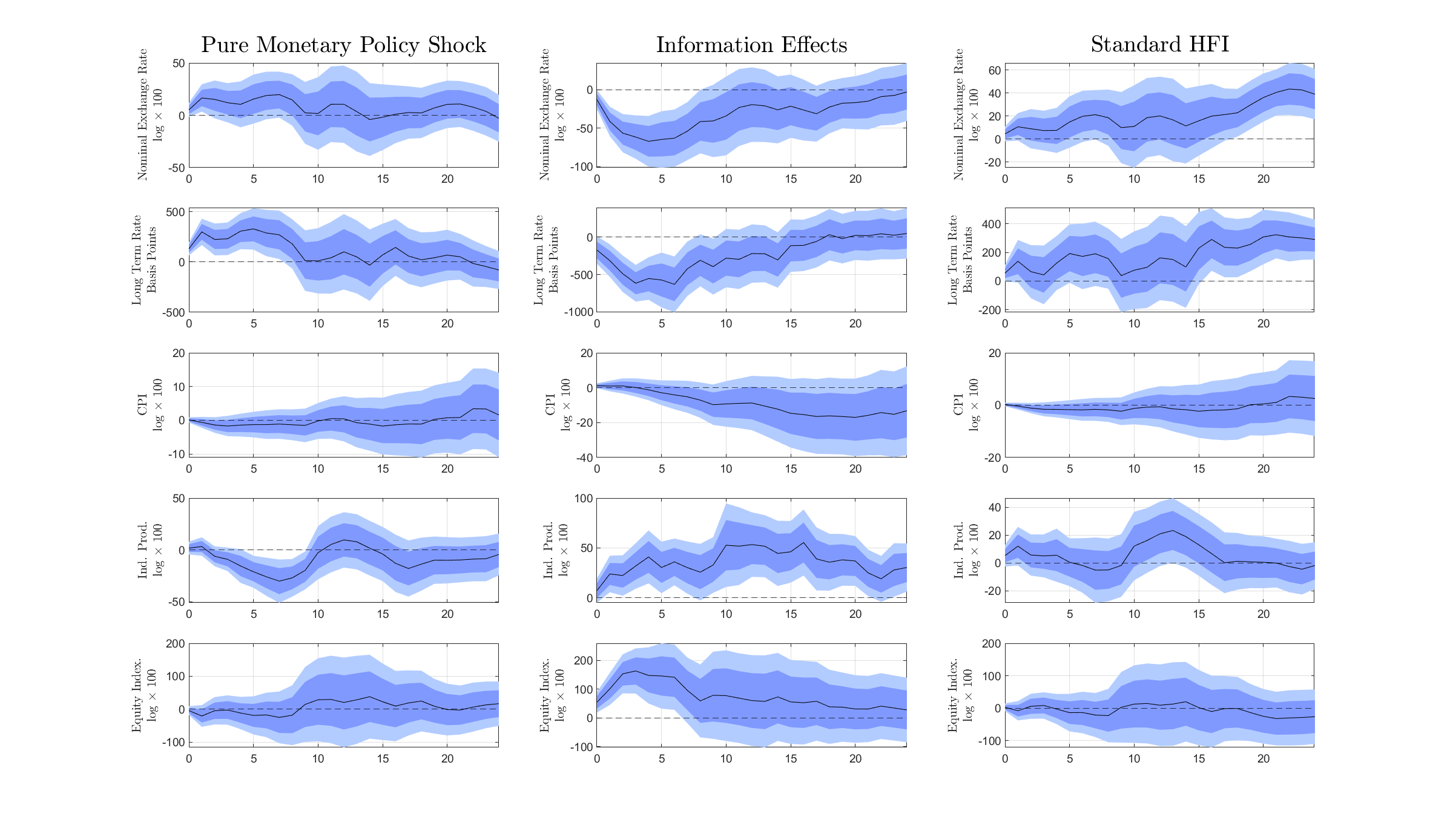}
    \caption{Impulse Response Functions \\ Sample without countries that exit \& re-enter sample}
    \label{fig:Exiters_BenchmarkNER_08}
    \floatfoot{\textbf{Note:} The figure is comprised of 15 sub-figures ordered in three columns and five rows. The left column relates to the estimates of $\beta^{MP}$ in Equation \ref{eq:LP_pooled}, the middle column relates to the estimate of $\beta^{FIE}$ in Equation \ref{eq:LP_pooled}, while the right column relates to estimating Equation \ref{eq:LP_pooled}, replacing the MP and FIE components with the un-orthogonalized monetary policy surprise. The rows represent the impact on (i) the nominal exchange rate with the US dollar (in logs times 100); (ii) long term interest rates in basis points; (iii) the consumer price index (in logs times 100); (iv) the industrial production index (in logs times 100); (v) the equity index (in logs times 100). The solid black line represents the point estimate, the dark blue area represents the 68\% confidence interval, and the light blue area represents the 90\% confidence interval. In the text, when referring to Panel $(i,j)$, $i$ refers to the row and $j$ to the column of the figure. Each variable, in its own transformation, is demeaned at the country level. Sample from January 2008 to December 2019.}
\end{figure}

\newpage
\begin{figure}[ht]
    \centering
    \includegraphics[scale=0.4]{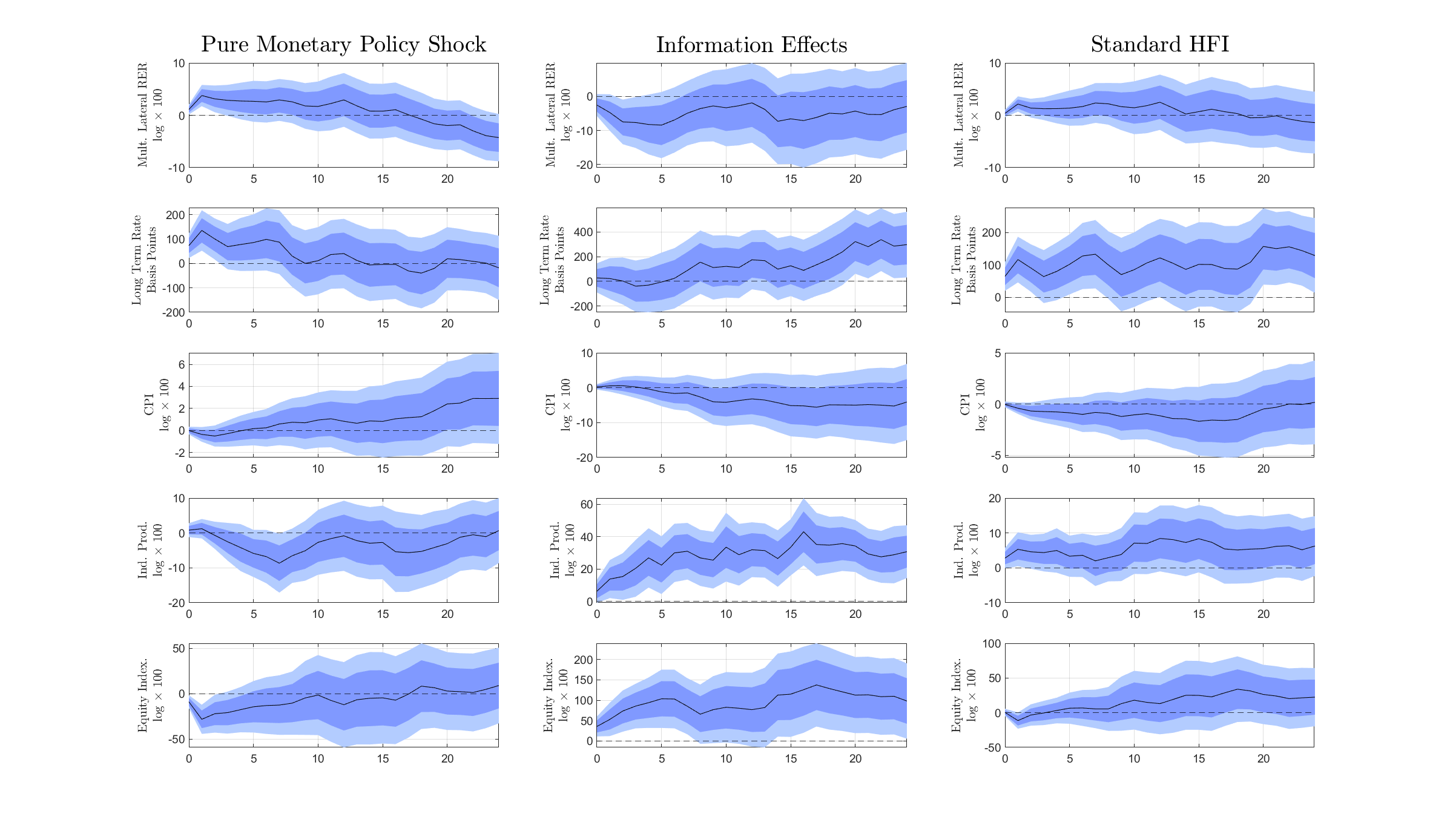}
    \caption{Impulse Response Functions \\  Multi. REER Sample. Without countries that exit \& re-enter sample}
    \label{fig:Exiters_BenchmarkREER}
    \floatfoot{\textbf{Note:} The figure is comprised of 15 sub-figures ordered in three columns and five rows. The left column relates to the estimates of $\beta^{MP}$ in Equation \ref{eq:LP_pooled}, the middle column relates to the estimate of $\beta^{FIE}$ in Equation \ref{eq:LP_pooled}, while the right column relates to estimating Equation \ref{eq:LP_pooled}, replacing the MP and FIE components with the un-orthogonalized monetary policy surprise. The rows represent the impact on (i) the trade weighted multilateral real exchange rate (in logs times 100); (ii) long term interest rates in basis points; (iii) the consumer price index (in logs times 100); (iv) the industrial production index (in logs times 100); (v) the equity index (in logs times 100). The solid black line represents the point estimate, the dark blue area represents the 68\% confidence interval, and the light blue area represents the 90\% confidence interval. In the text, when referring to Panel $(i,j)$, $i$ refers to the row and $j$ to the column of the figure. Each variable, in its own transformation, is demeaned at the country level. }
\end{figure}

\newpage
\begin{figure}[ht]
    \centering
    \includegraphics[scale=0.4]{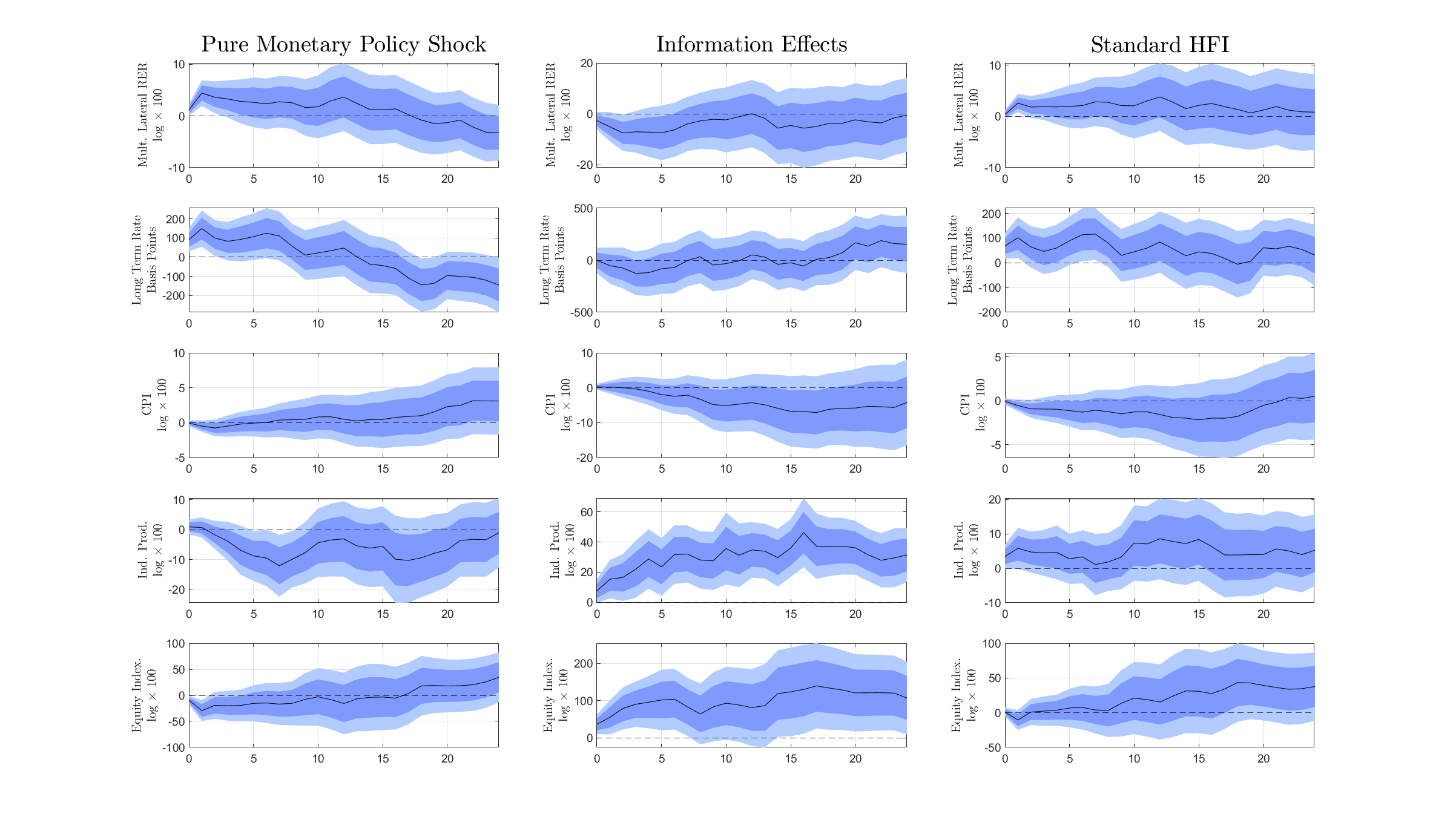}
    \caption{Impulse Response Functions \\ Multi. REER Sample. Without countries that exit \& re-enter sample - 1998 Onward}
    \label{fig:Exiters_BenchmarkREER_98}
    \floatfoot{\textbf{Note:} The figure is comprised of 15 sub-figures ordered in three columns and five rows. The left column relates to the estimates of $\beta^{MP}$ in Equation \ref{eq:LP_pooled}, the middle column relates to the estimate of $\beta^{FIE}$ in Equation \ref{eq:LP_pooled}, while the right column relates to estimating Equation \ref{eq:LP_pooled}, replacing the MP and FIE components with the un-orthogonalized monetary policy surprise. The rows represent the impact on (i) the trade weighted multilateral real exchange rate (in logs times 100); (ii) long term interest rates in basis points; (iii) the consumer price index (in logs times 100); (iv) the industrial production index (in logs times 100); (v) the equity index (in logs times 100). The solid black line represents the point estimate, the dark blue area represents the 68\% confidence interval, and the light blue area represents the 90\% confidence interval. In the text, when referring to Panel $(i,j)$, $i$ refers to the row and $j$ to the column of the figure. Each variable, in its own transformation, is demeaned at the country level. Sample from January 1998 to December 2019.}
\end{figure}

\newpage
\begin{figure}[ht]
    \centering
    \includegraphics[scale=0.4]{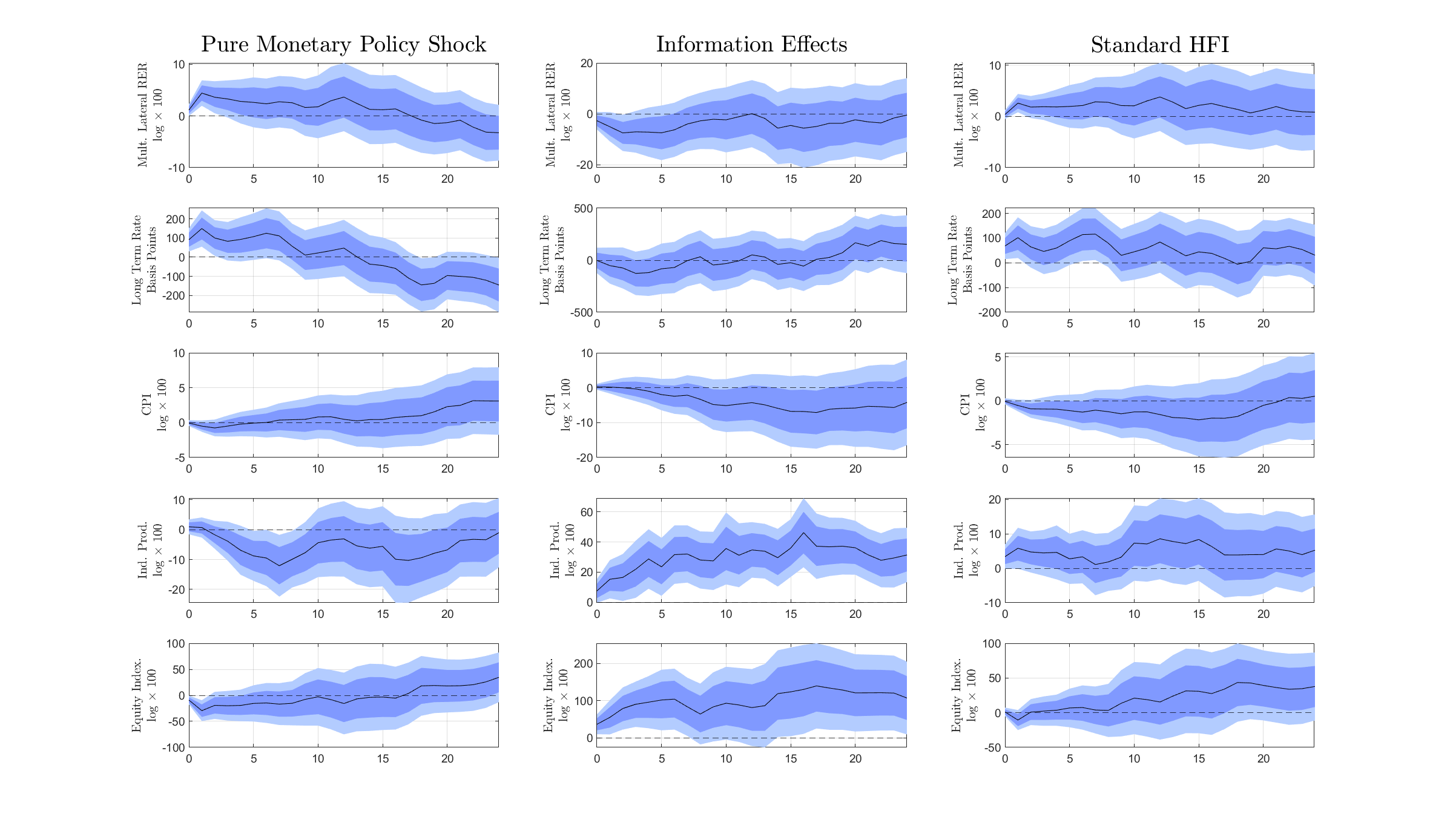}
    \caption{Impulse Response Functions \\ Multi. REER Sample. Without countries that exit \& re-enter sample  - 2008 Onward}
    \label{fig:Exiters_BenchmarkREER_08}
    \floatfoot{\textbf{Note:} The figure is comprised of 15 sub-figures ordered in three columns and five rows. The left column relates to the estimates of $\beta^{MP}$ in Equation \ref{eq:LP_pooled}, the middle column relates to the estimate of $\beta^{FIE}$ in Equation \ref{eq:LP_pooled}, while the right column relates to estimating Equation \ref{eq:LP_pooled}, replacing the MP and FIE components with the un-orthogonalized monetary policy surprise. The rows represent the impact on (i) the trade weighted multilateral real exchange rate (in logs times 100); (ii) long term interest rates in basis points; (iii) the consumer price index (in logs times 100); (iv) the industrial production index (in logs times 100); (v) the equity index (in logs times 100). The solid black line represents the point estimate, the dark blue area represents the 68\% confidence interval, and the light blue area represents the 90\% confidence interval. In the text, when referring to Panel $(i,j)$, $i$ refers to the row and $j$ to the column of the figure. Each variable, in its own transformation, is demeaned at the country level. Sample from January 2008 to December 2019.}
\end{figure}

\newpage
\begin{figure}[ht]
    \centering
    \includegraphics[scale=0.4]{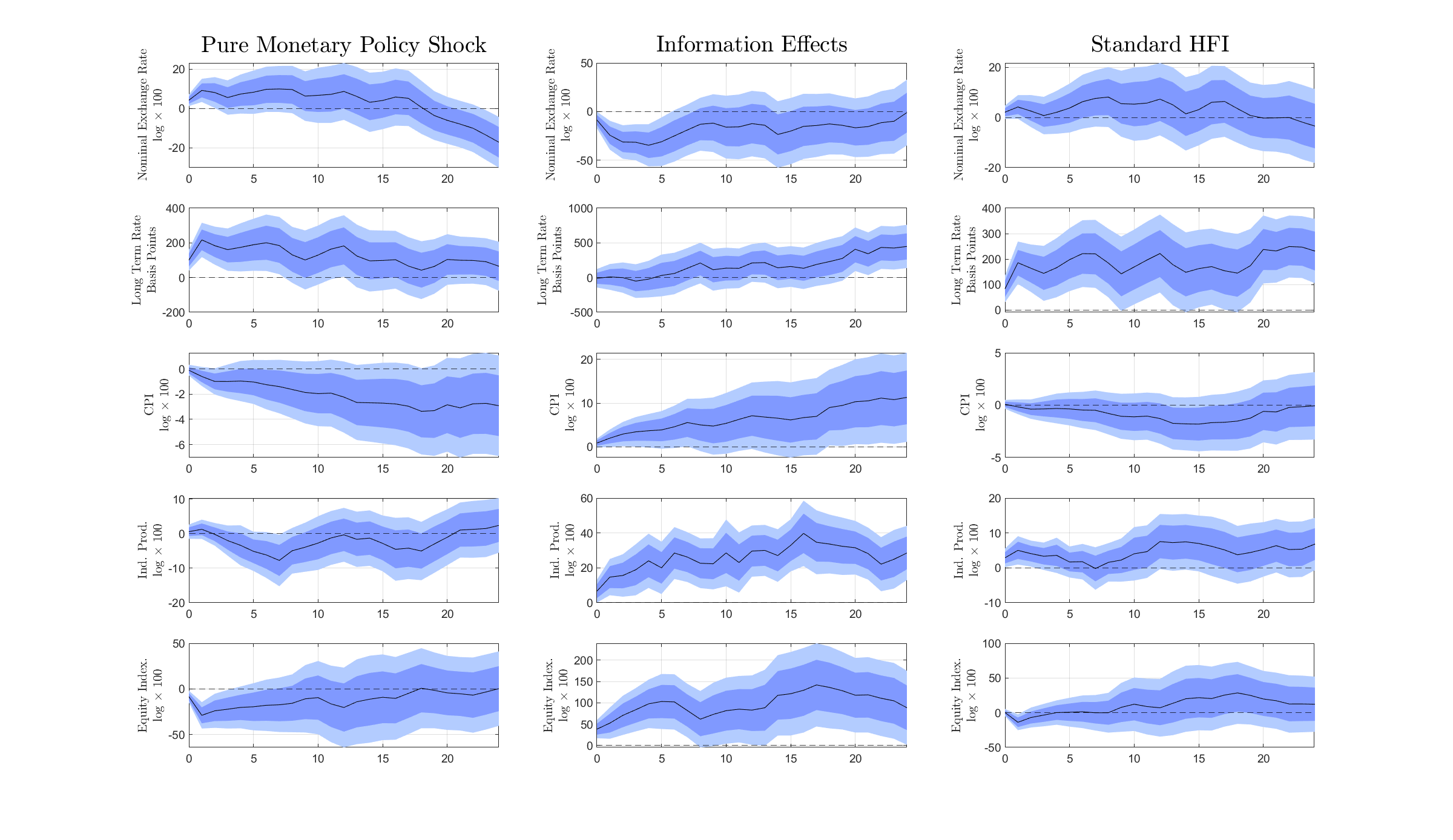}
    \caption{Impulse Response Functions \\ $J_y = J_x = 12$  }
    \label{fig:Regressions_NER_deplag12}
    \floatfoot{\textbf{Note:} The figure is comprised of 15 sub-figures ordered in three columns and five rows. The left column relates to the estimates of $\beta^{MP}$ in Equation \ref{eq:LP_pooled}, the middle column relates to the estimate of $\beta^{FIE}$ in Equation \ref{eq:LP_pooled}, while the right column relates to estimating Equation \ref{eq:LP_pooled}, replacing the MP and FIE components with the un-orthogonalized monetary policy surprise. The rows represent the impact on (i) the nominal exchange rate with the US dollar (in logs times 100); (ii) long term interest rates in basis points; (iii) the consumer price index (in logs times 100); (iv) the industrial production index (in logs times 100); (v) the equity index (in logs times 100). The solid black line represents the point estimate, the dark blue area represents the 68\% confidence interval, and the light blue area represents the 90\% confidence interval. In the text, when referring to Panel $(i,j)$, $i$ refers to the row and $j$ to the column of the figure. Each variable, in its own transformation, is demeaned at the country level. }
\end{figure}

\newpage
\begin{figure}[ht]
    \centering
    \includegraphics[scale=0.4]{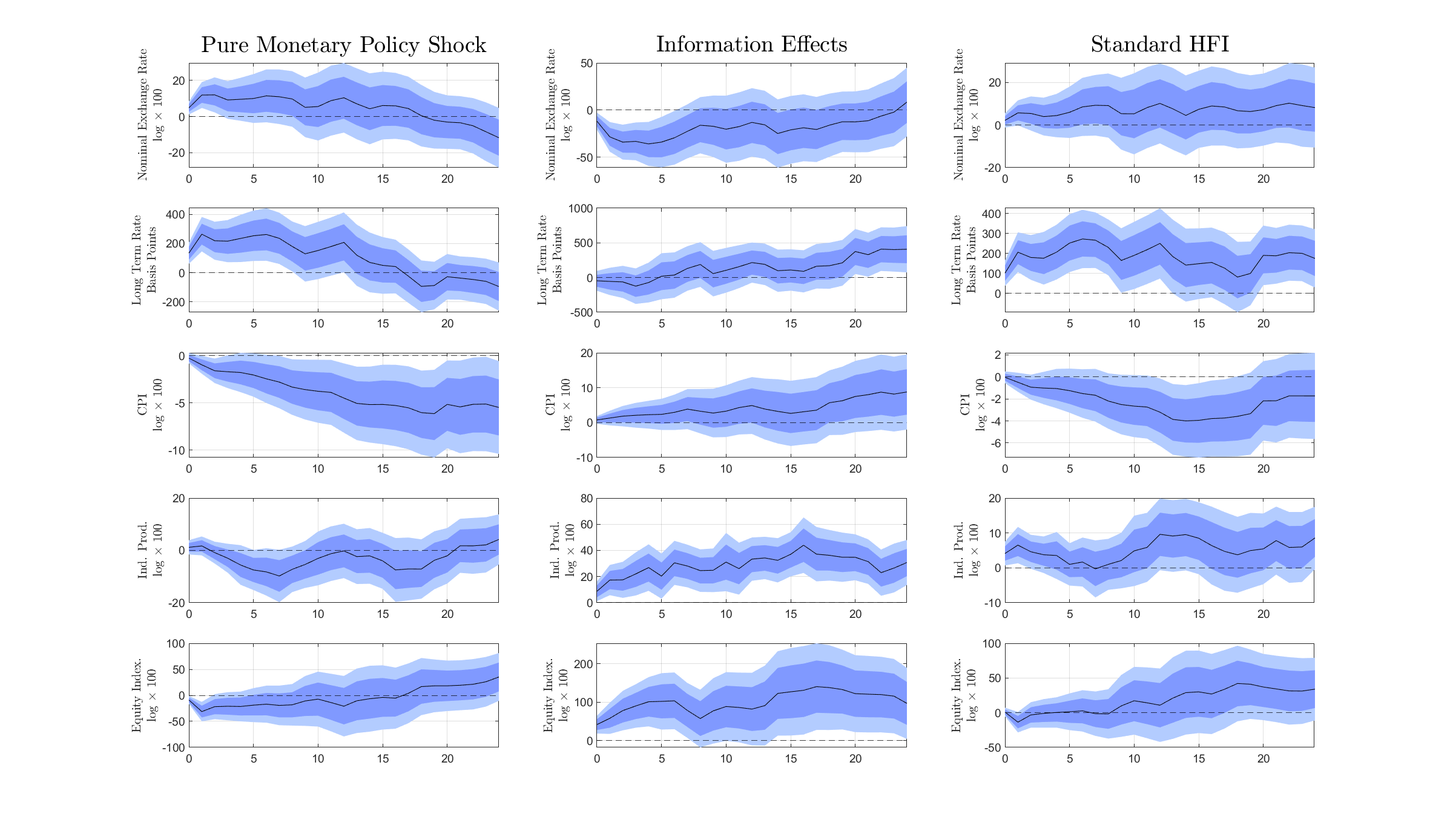}
    \caption{Impulse Response Functions \\ $J_y = J_x = 12$  }
    \label{fig:Regressions_98NER_deplag12}
    \floatfoot{\textbf{Note:} The figure is comprised of 15 sub-figures ordered in three columns and five rows. The left column relates to the estimates of $\beta^{MP}$ in Equation \ref{eq:LP_pooled}, the middle column relates to the estimate of $\beta^{FIE}$ in Equation \ref{eq:LP_pooled}, while the right column relates to estimating Equation \ref{eq:LP_pooled}, replacing the MP and FIE components with the un-orthogonalized monetary policy surprise. The rows represent the impact on (i) the nominal exchange rate with the US dollar (in logs times 100); (ii) long term interest rates in basis points; (iii) the consumer price index (in logs times 100); (iv) the industrial production index (in logs times 100); (v) the equity index (in logs times 100). The solid black line represents the point estimate, the dark blue area represents the 68\% confidence interval, and the light blue area represents the 90\% confidence interval. In the text, when referring to Panel $(i,j)$, $i$ refers to the row and $j$ to the column of the figure. Each variable, in its own transformation, is demeaned at the country level. Sample from January 1998 to December 2019.}
\end{figure}

\newpage
\begin{figure}[ht]
    \centering
    \includegraphics[scale=0.4]{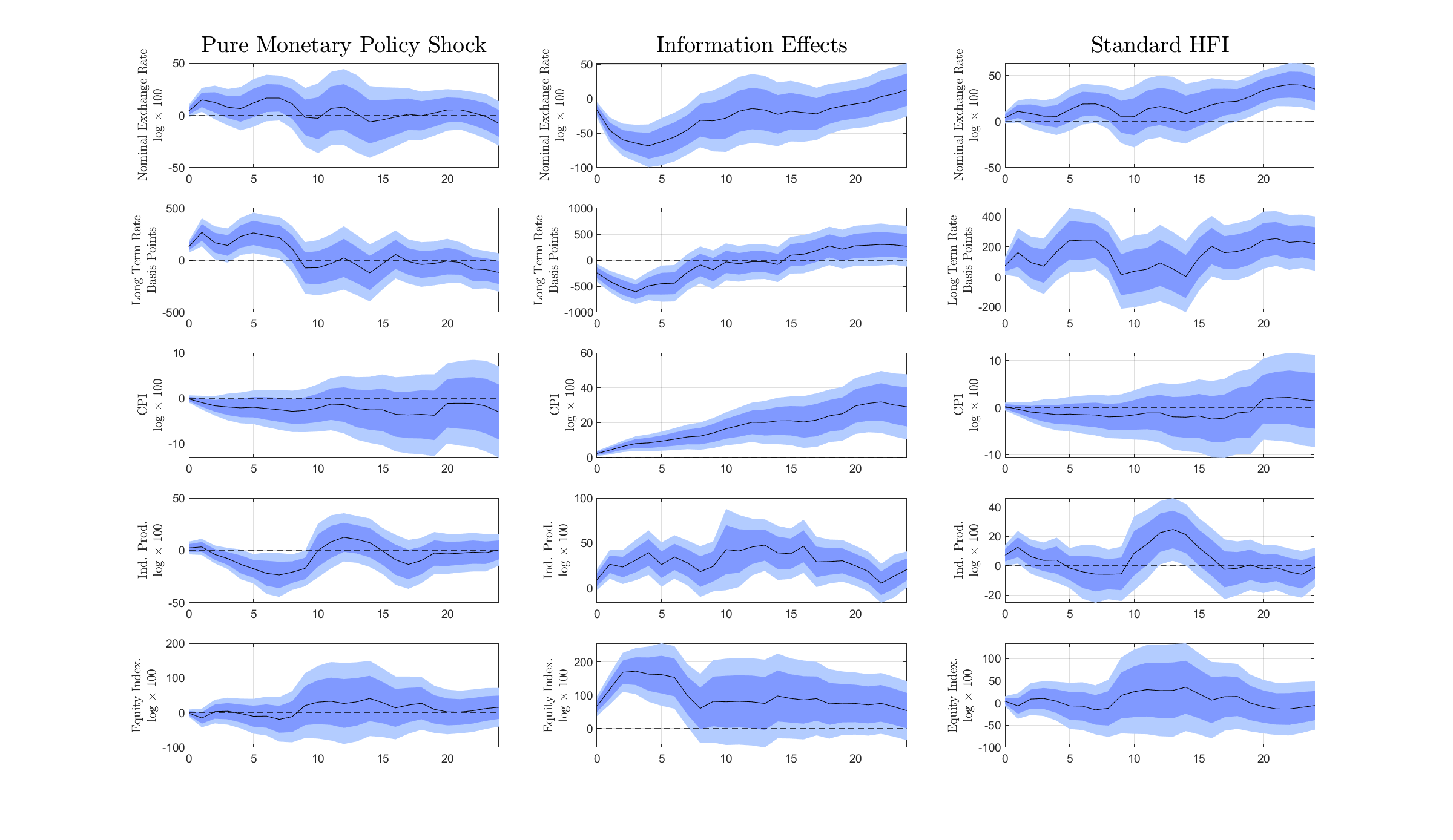}
    \caption{Impulse Response Functions \\ $J_y = J_x = 12$  }
    \label{fig:Regressions_08NER_deplag12}
    \floatfoot{\textbf{Note:} The figure is comprised of 15 sub-figures ordered in three columns and five rows. The left column relates to the estimates of $\beta^{MP}$ in Equation \ref{eq:LP_pooled}, the middle column relates to the estimate of $\beta^{FIE}$ in Equation \ref{eq:LP_pooled}, while the right column relates to estimating Equation \ref{eq:LP_pooled}, replacing the MP and FIE components with the un-orthogonalized monetary policy surprise. The rows represent the impact on (i) the nominal exchange rate with the US dollar (in logs times 100); (ii) long term interest rates in basis points; (iii) the consumer price index (in logs times 100); (iv) the industrial production index (in logs times 100); (v) the equity index (in logs times 100). The solid black line represents the point estimate, the dark blue area represents the 68\% confidence interval, and the light blue area represents the 90\% confidence interval. In the text, when referring to Panel $(i,j)$, $i$ refers to the row and $j$ to the column of the figure. Each variable, in its own transformation, is demeaned at the country level. Sample from January 2008 to December 2019.}
\end{figure}

\newpage
\begin{figure}[ht]
    \centering
    \includegraphics[scale=0.4]{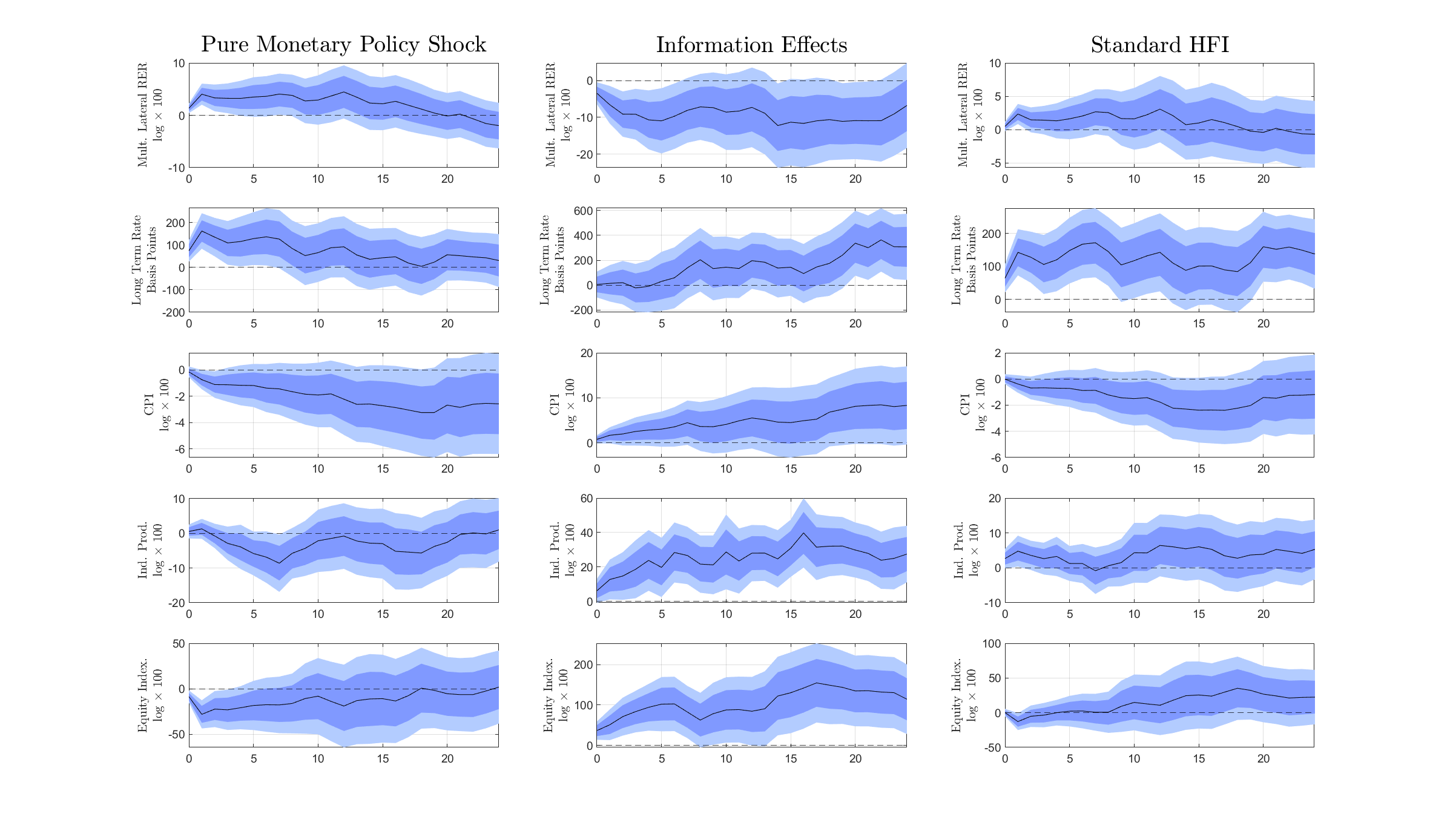}
    \caption{Impulse Response Functions \\  Multi. REER Sample - $J_y = J_x = 12$  }
    \label{fig:Regressions_REER_deplag12}
    \floatfoot{\textbf{Note:} The figure is comprised of 15 sub-figures ordered in three columns and five rows. The left column relates to the estimates of $\beta^{MP}$ in Equation \ref{eq:LP_pooled}, the middle column relates to the estimate of $\beta^{FIE}$ in Equation \ref{eq:LP_pooled}, while the right column relates to estimating Equation \ref{eq:LP_pooled}, replacing the MP and FIE components with the un-orthogonalized monetary policy surprise. The rows represent the impact on (i) the trade weighted multilateral real exchange rate (in logs times 100); (ii) long term interest rates in basis points; (iii) the consumer price index (in logs times 100); (iv) the industrial production index (in logs times 100); (v) the equity index (in logs times 100). The solid black line represents the point estimate, the dark blue area represents the 68\% confidence interval, and the light blue area represents the 90\% confidence interval. In the text, when referring to Panel $(i,j)$, $i$ refers to the row and $j$ to the column of the figure. Each variable, in its own transformation, is demeaned at the country level. }
\end{figure}

\newpage
\begin{figure}[ht]
    \centering
    \includegraphics[scale=0.4]{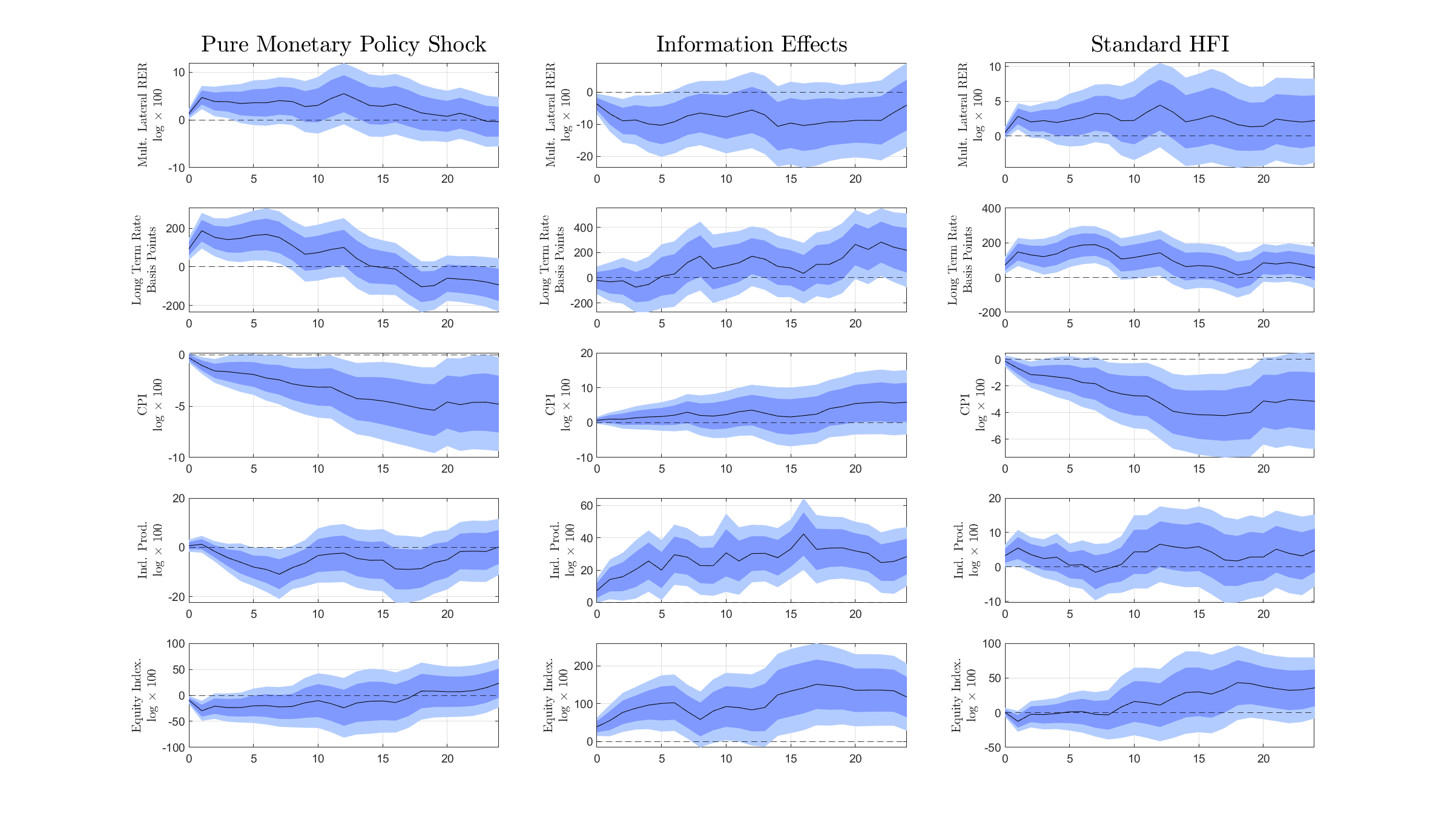}
    \caption{Impulse Response Functions \\ Multi. REER Sample - $J_y = J_x = 12$   - 1998 Onward}
    \label{fig:Regressions_98REER_deplag12}
    \floatfoot{\textbf{Note:} The figure is comprised of 15 sub-figures ordered in three columns and five rows. The left column relates to the estimates of $\beta^{MP}$ in Equation \ref{eq:LP_pooled}, the middle column relates to the estimate of $\beta^{FIE}$ in Equation \ref{eq:LP_pooled}, while the right column relates to estimating Equation \ref{eq:LP_pooled}, replacing the MP and FIE components with the un-orthogonalized monetary policy surprise. The rows represent the impact on (i) the trade weighted multilateral real exchange rate (in logs times 100); (ii) long term interest rates in basis points; (iii) the consumer price index (in logs times 100); (iv) the industrial production index (in logs times 100); (v) the equity index (in logs times 100). The solid black line represents the point estimate, the dark blue area represents the 68\% confidence interval, and the light blue area represents the 90\% confidence interval. In the text, when referring to Panel $(i,j)$, $i$ refers to the row and $j$ to the column of the figure. Each variable, in its own transformation, is demeaned at the country level. Sample from January 1998 to December 2019.}
\end{figure}

\newpage
\begin{figure}[ht]
    \centering
    \includegraphics[scale=0.4]{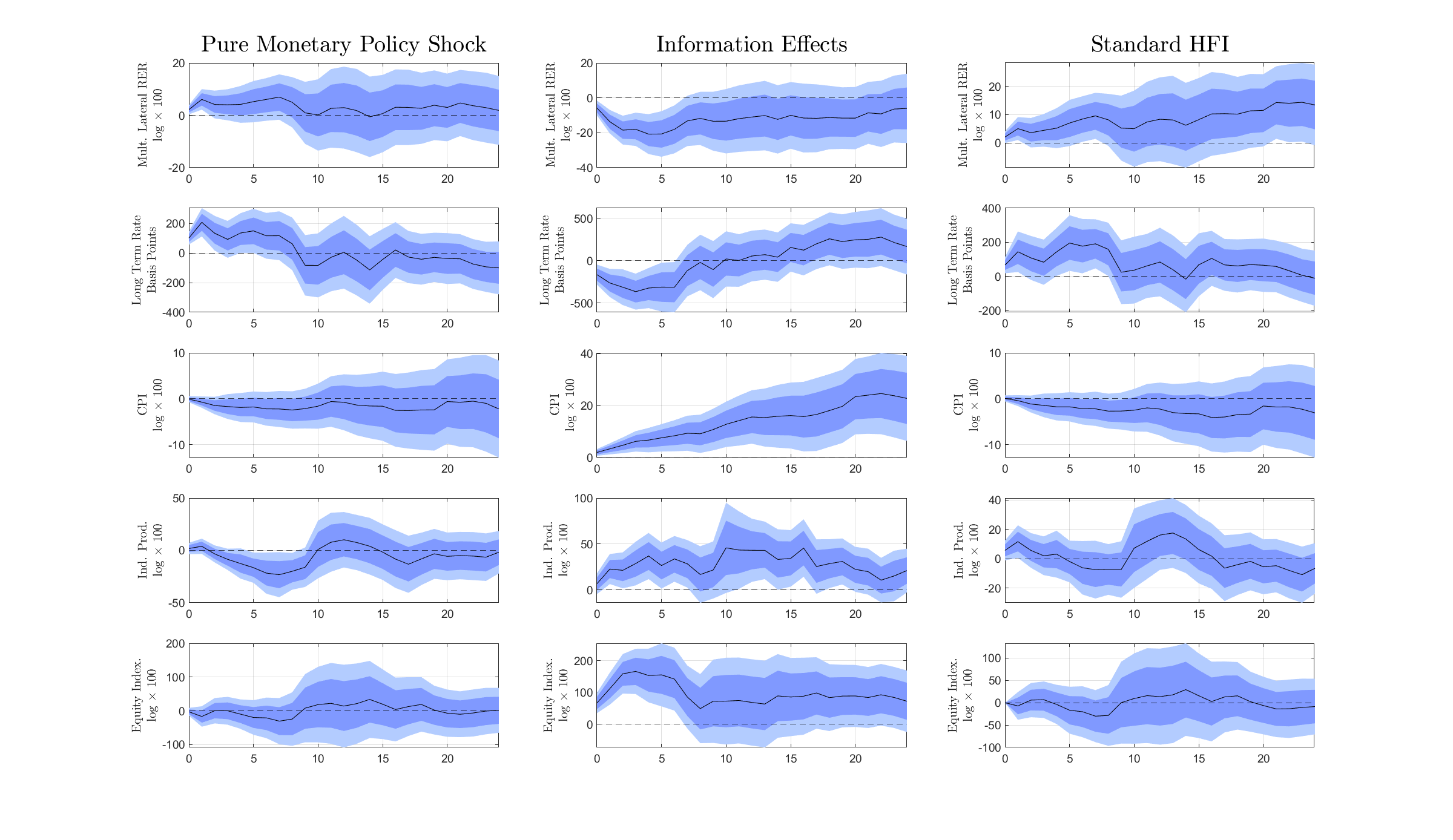}
    \caption{Impulse Response Functions \\ Multi. REER Sample - $J_y = J_x = 12$    - 2008 Onward}
    \label{fig:Regressions_08REER_deplag12}
    \floatfoot{\textbf{Note:} The figure is comprised of 15 sub-figures ordered in three columns and five rows. The left column relates to the estimates of $\beta^{MP}$ in Equation \ref{eq:LP_pooled}, the middle column relates to the estimate of $\beta^{FIE}$ in Equation \ref{eq:LP_pooled}, while the right column relates to estimating Equation \ref{eq:LP_pooled}, replacing the MP and FIE components with the un-orthogonalized monetary policy surprise. The rows represent the impact on (i) the trade weighted multilateral real exchange rate (in logs times 100); (ii) long term interest rates in basis points; (iii) the consumer price index (in logs times 100); (iv) the industrial production index (in logs times 100); (v) the equity index (in logs times 100). The solid black line represents the point estimate, the dark blue area represents the 68\% confidence interval, and the light blue area represents the 90\% confidence interval. In the text, when referring to Panel $(i,j)$, $i$ refers to the row and $j$ to the column of the figure. Each variable, in its own transformation, is demeaned at the country level. Sample from January 2008 to December 2019.}
\end{figure}


\newpage
\begin{figure}[ht]
    \centering
    \includegraphics[scale=0.4]{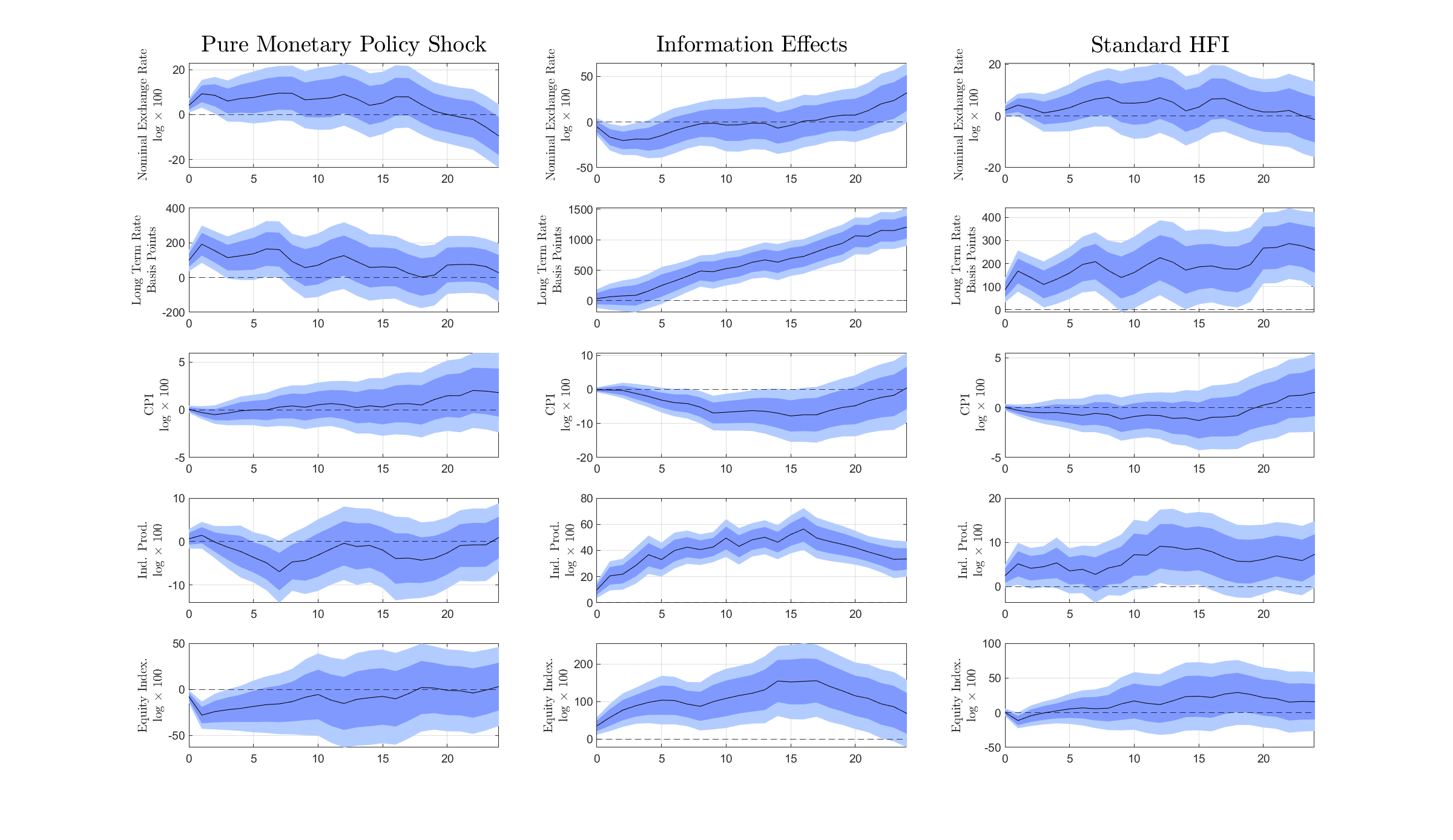}
    \caption{Impulse Response Functions \\ $J_i= 0$  }
    \label{fig:Regressions_NER_shock0lag}
    \floatfoot{\textbf{Note:} The figure is comprised of 15 sub-figures ordered in three columns and five rows. The left column relates to the estimates of $\beta^{MP}$ in Equation \ref{eq:LP_pooled}, the middle column relates to the estimate of $\beta^{FIE}$ in Equation \ref{eq:LP_pooled}, while the right column relates to estimating Equation \ref{eq:LP_pooled}, replacing the MP and FIE components with the un-orthogonalized monetary policy surprise. The rows represent the impact on (i) the nominal exchange rate with the US dollar (in logs times 100); (ii) long term interest rates in basis points; (iii) the consumer price index (in logs times 100); (iv) the industrial production index (in logs times 100); (v) the equity index (in logs times 100). The solid black line represents the point estimate, the dark blue area represents the 68\% confidence interval, and the light blue area represents the 90\% confidence interval. In the text, when referring to Panel $(i,j)$, $i$ refers to the row and $j$ to the column of the figure. Each variable, in its own transformation, is demeaned at the country level. }
\end{figure}

\newpage
\begin{figure}[ht]
    \centering
    \includegraphics[scale=0.4]{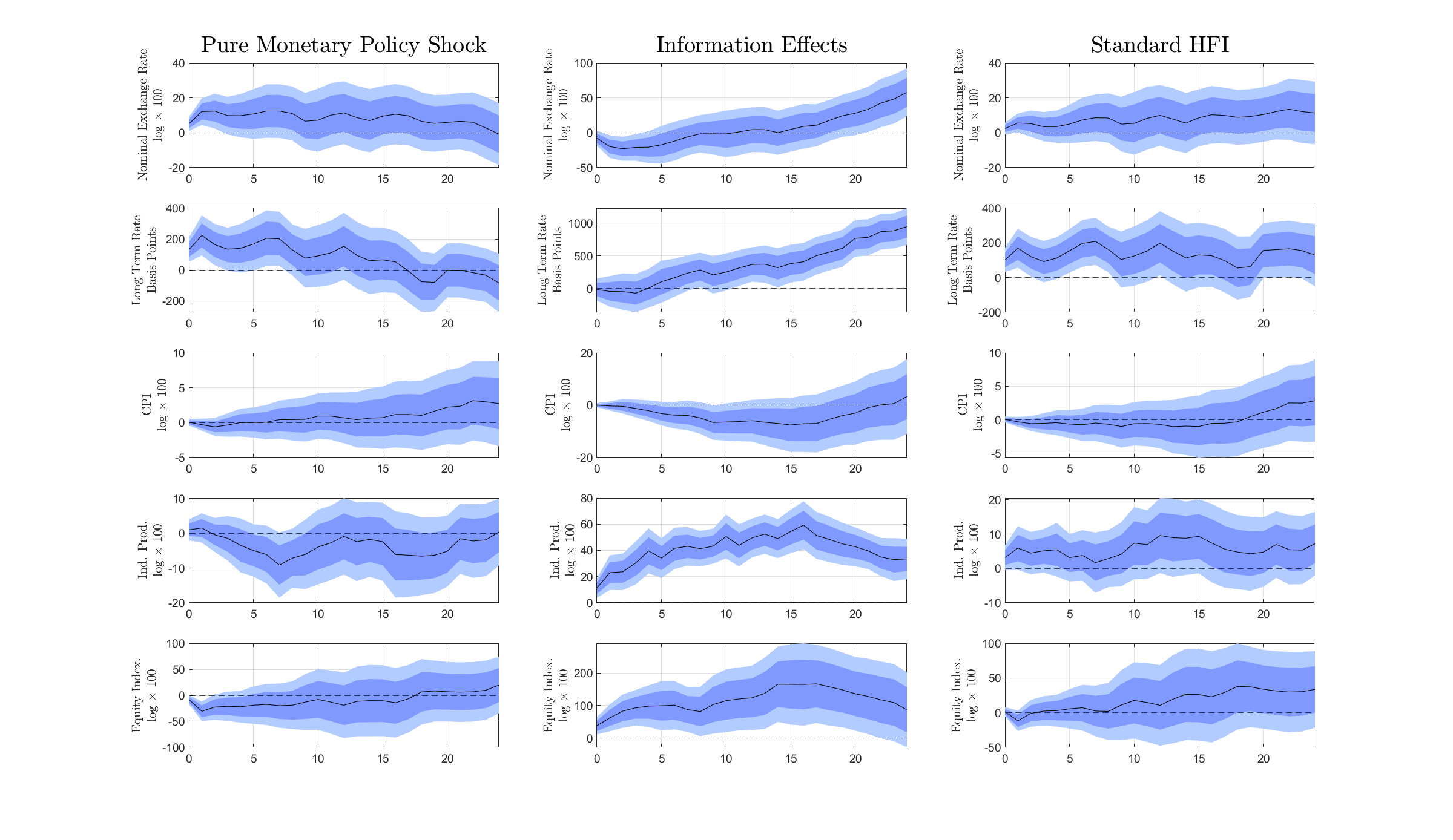}
    \caption{Impulse Response Functions \\ $J_i= 0$  }
    \label{fig:Regressions_98NER_shock0lag}
    \floatfoot{\textbf{Note:} The figure is comprised of 15 sub-figures ordered in three columns and five rows. The left column relates to the estimates of $\beta^{MP}$ in Equation \ref{eq:LP_pooled}, the middle column relates to the estimate of $\beta^{FIE}$ in Equation \ref{eq:LP_pooled}, while the right column relates to estimating Equation \ref{eq:LP_pooled}, replacing the MP and FIE components with the un-orthogonalized monetary policy surprise. The rows represent the impact on (i) the nominal exchange rate with the US dollar (in logs times 100); (ii) long term interest rates in basis points; (iii) the consumer price index (in logs times 100); (iv) the industrial production index (in logs times 100); (v) the equity index (in logs times 100). The solid black line represents the point estimate, the dark blue area represents the 68\% confidence interval, and the light blue area represents the 90\% confidence interval. In the text, when referring to Panel $(i,j)$, $i$ refers to the row and $j$ to the column of the figure. Each variable, in its own transformation, is demeaned at the country level. Sample from January 1998 to December 2019.}
\end{figure}

\newpage
\begin{figure}[ht]
    \centering
    \includegraphics[scale=0.4]{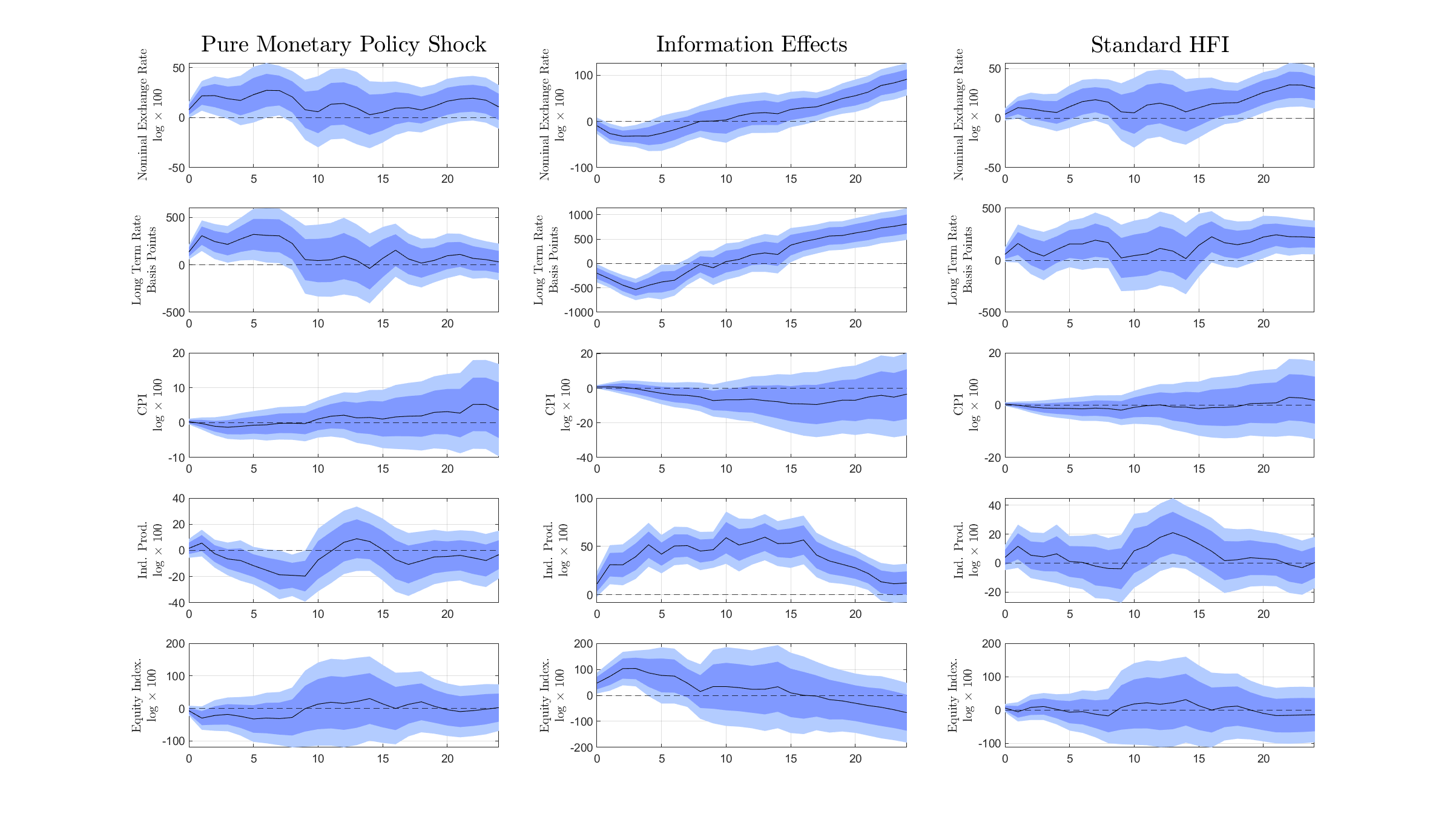}
    \caption{Impulse Response Functions \\ $J_i= 0$  }
    \label{fig:Regressions_08NER_shock0lag}
    \floatfoot{\textbf{Note:} The figure is comprised of 15 sub-figures ordered in three columns and five rows. The left column relates to the estimates of $\beta^{MP}$ in Equation \ref{eq:LP_pooled}, the middle column relates to the estimate of $\beta^{FIE}$ in Equation \ref{eq:LP_pooled}, while the right column relates to estimating Equation \ref{eq:LP_pooled}, replacing the MP and FIE components with the un-orthogonalized monetary policy surprise. The rows represent the impact on (i) the nominal exchange rate with the US dollar (in logs times 100); (ii) long term interest rates in basis points; (iii) the consumer price index (in logs times 100); (iv) the industrial production index (in logs times 100); (v) the equity index (in logs times 100). The solid black line represents the point estimate, the dark blue area represents the 68\% confidence interval, and the light blue area represents the 90\% confidence interval. In the text, when referring to Panel $(i,j)$, $i$ refers to the row and $j$ to the column of the figure. Each variable, in its own transformation, is demeaned at the country level. Sample from January 2008 to December 2019.}
\end{figure}

\newpage
\begin{figure}[ht]
    \centering
    \includegraphics[scale=0.4]{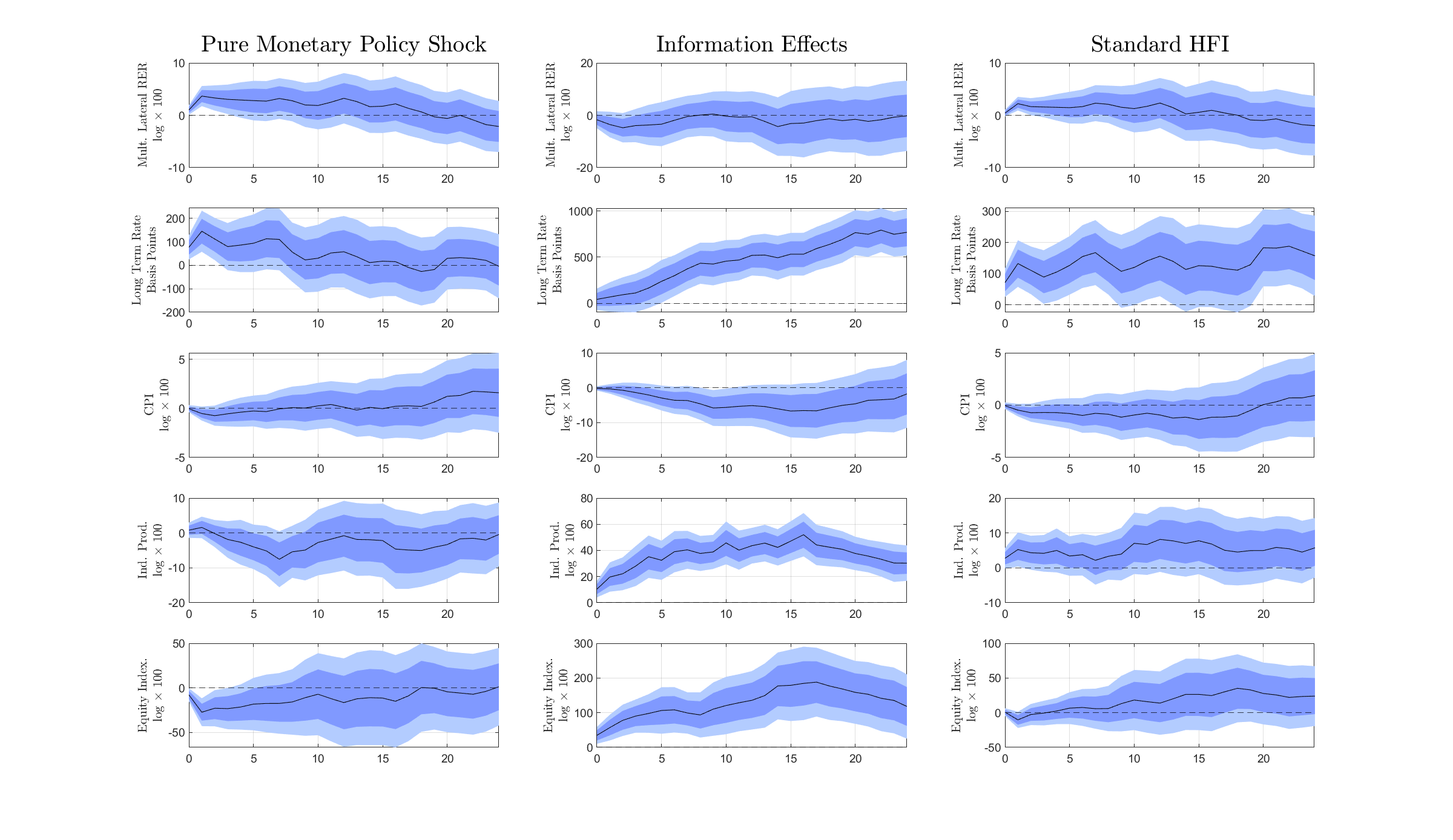}
    \caption{Impulse Response Functions \\  Multi. REER Sample - $J_i= 0$  }
    \label{fig:Regressions_REER_shock0lag}
    \floatfoot{\textbf{Note:} The figure is comprised of 15 sub-figures ordered in three columns and five rows. The left column relates to the estimates of $\beta^{MP}$ in Equation \ref{eq:LP_pooled}, the middle column relates to the estimate of $\beta^{FIE}$ in Equation \ref{eq:LP_pooled}, while the right column relates to estimating Equation \ref{eq:LP_pooled}, replacing the MP and FIE components with the un-orthogonalized monetary policy surprise. The rows represent the impact on (i) the trade weighted multilateral real exchange rate (in logs times 100); (ii) long term interest rates in basis points; (iii) the consumer price index (in logs times 100); (iv) the industrial production index (in logs times 100); (v) the equity index (in logs times 100). The solid black line represents the point estimate, the dark blue area represents the 68\% confidence interval, and the light blue area represents the 90\% confidence interval. In the text, when referring to Panel $(i,j)$, $i$ refers to the row and $j$ to the column of the figure. Each variable, in its own transformation, is demeaned at the country level. }
\end{figure}

\newpage
\begin{figure}[ht]
    \centering
    \includegraphics[scale=0.4]{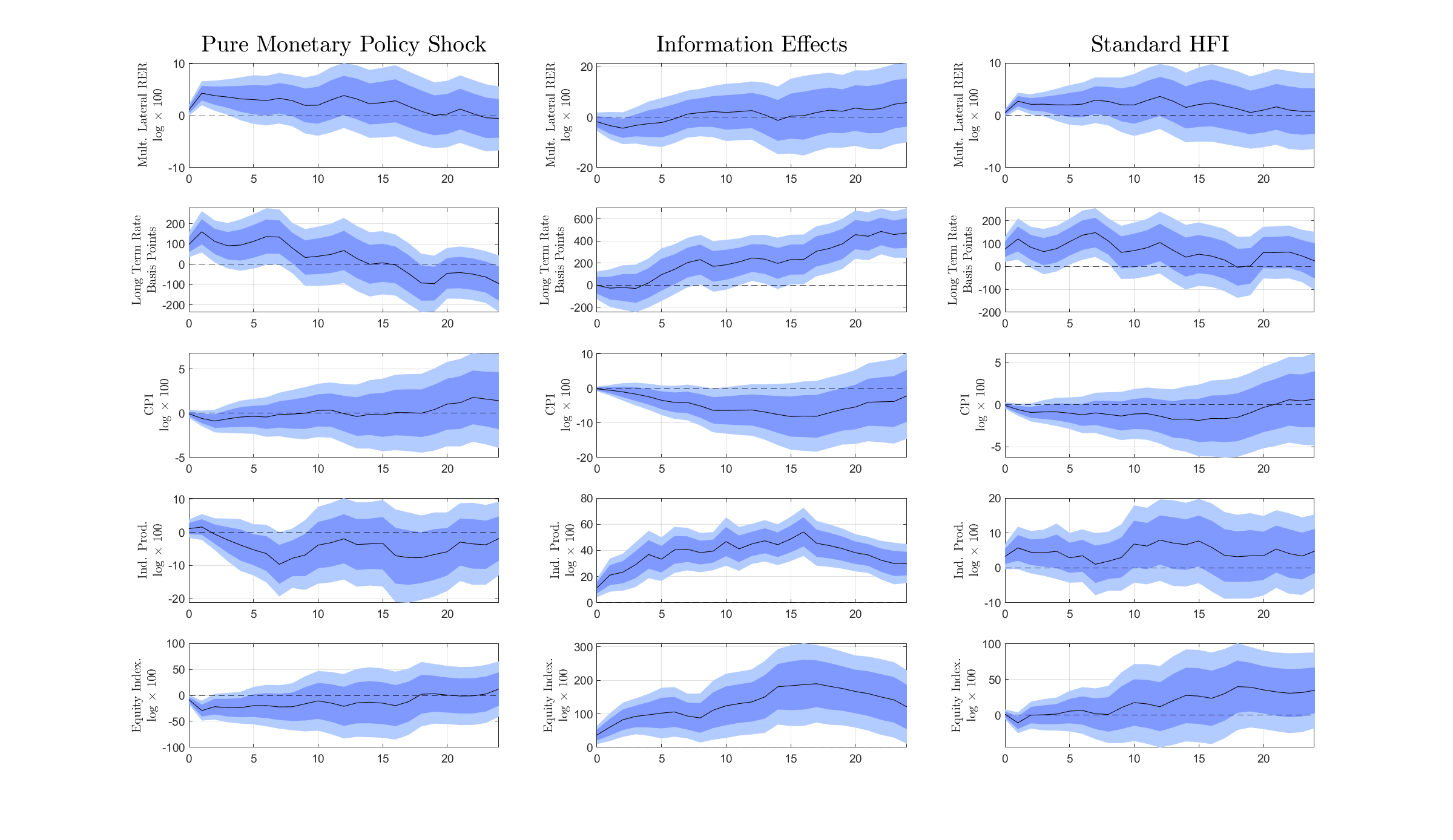}
    \caption{Impulse Response Functions \\ Multi. REER Sample - $J_i= 0$   - 1998 Onward}
    \label{fig:Regressions_98REER_shock0lag}
    \floatfoot{\textbf{Note:} The figure is comprised of 15 sub-figures ordered in three columns and five rows. The left column relates to the estimates of $\beta^{MP}$ in Equation \ref{eq:LP_pooled}, the middle column relates to the estimate of $\beta^{FIE}$ in Equation \ref{eq:LP_pooled}, while the right column relates to estimating Equation \ref{eq:LP_pooled}, replacing the MP and FIE components with the un-orthogonalized monetary policy surprise. The rows represent the impact on (i) the trade weighted multilateral real exchange rate (in logs times 100); (ii) long term interest rates in basis points; (iii) the consumer price index (in logs times 100); (iv) the industrial production index (in logs times 100); (v) the equity index (in logs times 100). The solid black line represents the point estimate, the dark blue area represents the 68\% confidence interval, and the light blue area represents the 90\% confidence interval. In the text, when referring to Panel $(i,j)$, $i$ refers to the row and $j$ to the column of the figure. Each variable, in its own transformation, is demeaned at the country level. Sample from January 1998 to December 2019.}
\end{figure}

\newpage
\newpage
\begin{figure}[ht]
    \centering
    \includegraphics[scale=0.4]{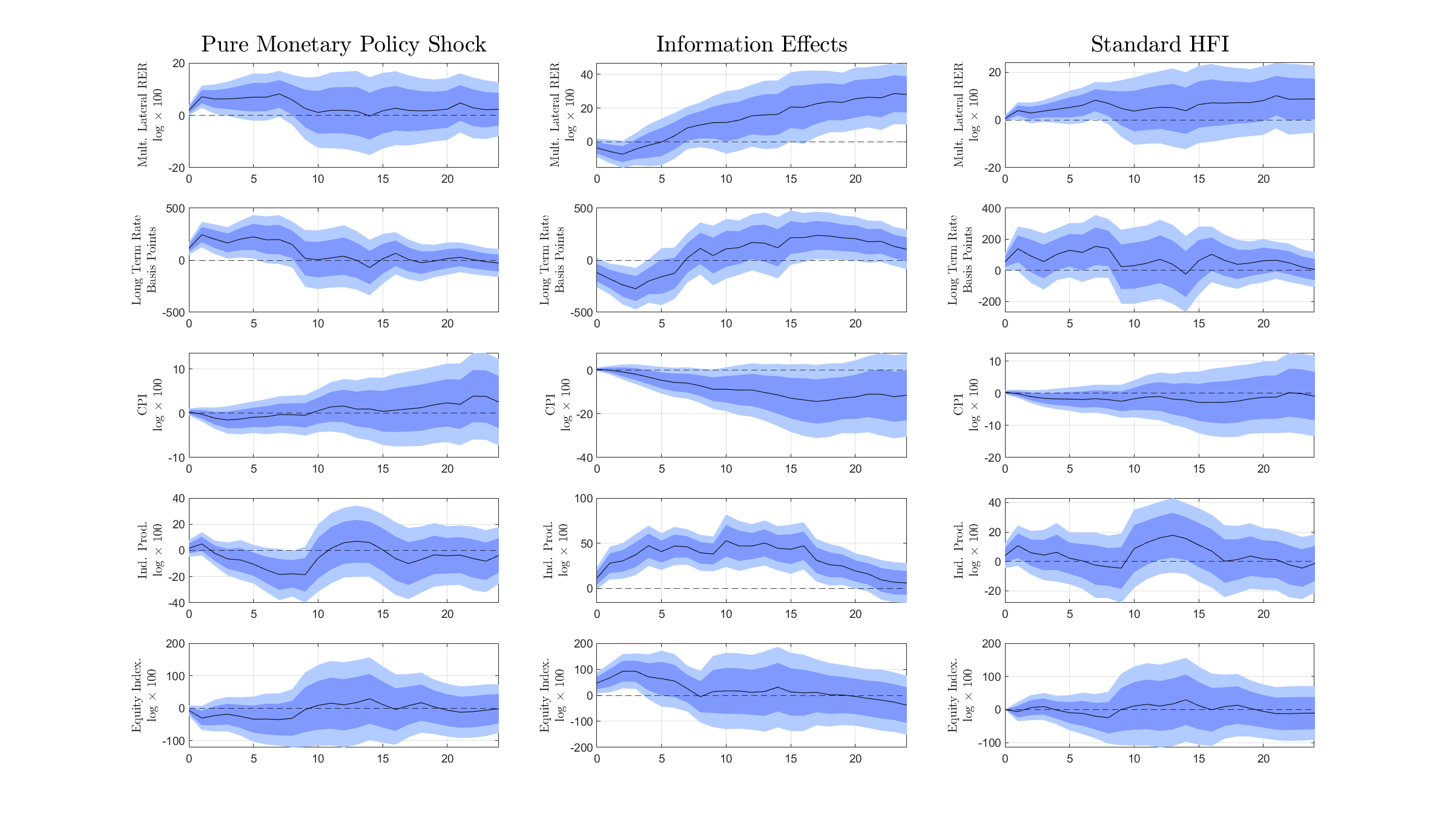}
    \caption{Impulse Response Functions \\ Multi. REER Sample - $J_i= 0$    - 2008 Onward}
    \label{fig:Regressions_08REER_shock0lag}
    \floatfoot{\textbf{Note:} The figure is comprised of 15 sub-figures ordered in three columns and five rows. The left column relates to the estimates of $\beta^{MP}$ in Equation \ref{eq:LP_pooled}, the middle column relates to the estimate of $\beta^{FIE}$ in Equation \ref{eq:LP_pooled}, while the right column relates to estimating Equation \ref{eq:LP_pooled}, replacing the MP and FIE components with the un-orthogonalized monetary policy surprise. The rows represent the impact on (i) the trade weighted multilateral real exchange rate (in logs times 100); (ii) long term interest rates in basis points; (iii) the consumer price index (in logs times 100); (iv) the industrial production index (in logs times 100); (v) the equity index (in logs times 100). The solid black line represents the point estimate, the dark blue area represents the 68\% confidence interval, and the light blue area represents the 90\% confidence interval. In the text, when referring to Panel $(i,j)$, $i$ refers to the row and $j$ to the column of the figure. Each variable, in its own transformation, is demeaned at the country level. Sample from January 2008 to December 2019.}
\end{figure}

\end{document}